\documentclass[twocolumn]{aastex61}

\pdfoutput=1

\usepackage{enumerate}
\usepackage{natbib,ifthen}
\usepackage{fancyhdr}
\usepackage{amssymb,amsmath}
\usepackage[T1]{fontenc}

\shorttitle{YSOs in Canis Major}
\shortauthors{Sewi{\l}o et al.}

\newenvironment{packed_enum}{
\begin{enumerate}
  \setlength{\itemsep}{6.5pt}
  \setlength{\parskip}{0pt}
  \setlength{\parsep}{0pt}
}{\end{enumerate}}

\begin{document}


\title{Identifying Young Stellar Objects in the outer Galaxy: \MakeLowercase{{\it l}} = {\rm 224$^{\circ}$ Region in Canis Major}}


\correspondingauthor{Marta Sewi{\l}o}
\email{marta.m.sewilo@nasa.gov}

\author{Marta Sewi{\l}o}
\affiliation{Department of Astronomy, University of Maryland, College Park, MD 20742, USA}
\affiliation{CRESST II and Exoplanets and Stellar Astrophysics Laboratory, NASA Goddard Space Flight Center, Greenbelt, MD 20771, USA}

\author{Barbara A. Whitney}
\affiliation{Space Science Institute,  4750 Walnut St. Suite 205, Boulder, CO 80301, USA}
\affiliation{Department of Astronomy, University of Wisconsin -- Madison, 475 North Charter St., Madison, WI 53706, USA}

\author{Bosco H. K. Yung}
\affiliation{N. Copernicus Astronomical Centre, PAS, Rabia\'nska 8, 87-100 Toru\'n, Poland}
\affiliation{Astronomical Observatory of the Jagiellonian University, ul. Orla 171, 30-244 Krak{\'o}w, Poland}

\author{Thomas P. Robitaille}
\affiliation{Freelance Consultant, Headingley Enterprise and Arts Centre, Bennett Road Headingley, Leeds LS6 3HN, United Kingdom}

\author{Davide Elia}
\affiliation{Istituto di Astrofisica e Planetologia Spaziali, INAF-IAPS, Via Fosso del Cavaliere 100, I-00133 Roma, Italy}

\author{Remy Indebetouw}
\affiliation{Department of Astronomy, University of Virginia, PO Box 3818, Charlottesville, VA 22903, USA}
\affiliation{National Radio Astronomy Observatory, 520 Edgemont Rd, Charlottesville, VA 22903, USA}

\author{Eugenio Schisano}
\affiliation{Istituto di Astrofisica e Planetologia Spaziali, INAF-IAPS, Via Fosso del Cavaliere 100, 00133 Roma, Italy}
\affiliation{Istituto di Radio Astronomia, INAF-IRA, Via Pietro Gobetti 101, 40126, Bologna, Italy}

\author{Ryszard Szczerba}
\affiliation{N. Copernicus Astronomical Centre, PAS, Rabia\'nska 8, 87-100 Toru\'n, Poland}

\author{Agata Karska}
\affiliation{Centre for Astronomy, Faculty of Physics, Astronomy and Informatics,
 Nicolaus Copernicus University, Grudzi\k{a}dzka 5, 87-100 Toru\'n, Poland}
 
\author{Jennifer Wiseman}
\affiliation{NASA Goddard Space Flight Center, 8800 Greenbelt Rd, Greenbelt, MD 20771, USA}

\author{Brian Babler}
\affiliation{Department of Astronomy, University of Wisconsin -- Madison, 475 North Charter St., Madison, WI 53706, USA}

\author{Martha Boyer}
\affiliation{Space Telescope Science Institute,  3700 San Martin Dr., Baltimore, MD 21218, USA}

\author{William J. Fischer} 
\affiliation{Space Telescope Science Institute,  3700 San Martin Dr., Baltimore, MD 21218, USA}

\author{Marilyn Meade}
\affiliation{Department of Astronomy, University of Wisconsin -- Madison, 475 North Charter St., Madison, WI 53706, USA}

\author{Luca Olmi}
\affiliation{INAF, Osservatorio Astrofisico di Arcetri, Largo E. Fermi 5, I-50125 Firenze, Italy}

\author{Deborah Padgett} 
\affiliation{NASA Jet Propulsion Laboratory, California Institute of Technology, 4800 Oak Grove Dr., Pasadena, CA 91109, USA}

\author{Natasza Si\'odmiak}
\affiliation{N. Copernicus Astronomical Centre, PAS, Rabia\'nska 8, 87-100 Toru\'n, Poland}


\begin{abstract}
We study a very young star-forming region in the outer Galaxy that is the most concentrated source of outflows in the {\it Spitzer Space Telescope} GLIMPSE360 survey. This  region, dubbed CMa--$l224$, is located in the Canis Major OB1 association. CMa--$l224$ is relatively faint in the mid-infrared, but it shines brightly at the far-infrared wavelengths as revealed by the {\it Herschel Space Observatory} data from the Hi-GAL survey. Using the 3.6 and 4.5 $\mu$m data from the {\it Spitzer}/GLIMPSE360 survey, combined with the {\it JHK$_s$} 2MASS and the 70--500 $\mu$m {\it Herschel}/Hi-GAL data, we develop a young stellar object (YSO) selection criteria based on color-color cuts and fitting of the YSO candidates' spectral energy distributions with YSO 2D radiative transfer models.  We identify 293 YSO candidates and estimate physical parameters for 210 sources well-fit with YSO models.  We select an additional 47 sources with GLIMPSE360-only photometry as `possible YSO candidates'. The vast majority of these sources are associated with high H$_2$ column density regions and are good targets for follow-up studies. The distribution of YSO candidates at different evolutionary stages with respect to {\it Herschel} filaments supports the idea that stars are formed in the filaments and become more dispersed with time.  Both the supernova-induced and spontaneous star formation scenarios are plausible in the environmental context of CMa--$l224$.  However, our results indicate that a spontaneous gravitational collapse of filaments is a more likely scenario. The methods developed for CMa--$l224$ can be used for larger regions in the Galactic plane where the same set of photometry is available.
\end{abstract}



\section{Introduction}
\label{s:intro}

Since the advent of the {\it Spitzer Space Telescope} ({\it Spitzer}; \citealt{werner2004}), young stellar objects (YSOs) in the Galaxy have been studied extensively (e.g., \citealt{gutermuth2009}; \citealt{megeath2012}; \citealt{povich2011}; \citealt{dunham2015}). Most of these studies, however, have concentrated on individual, well-known star formation regions in the inner Galaxy, leaving the larger environments outside the central regions and in the outer Galaxy unstudied. However, star formation in the outer Galactic disk has increasingly been gaining attention in the last few years (e.g., \citealt{elia2013}; \citealt{schisano2014}; \citealt{fischer2016}; \citealt{olmi2016}). 

The Warm {\it Spitzer Space Telescope} Exploration Science Program ``GLIMPSE360: Completing the Spitzer Galactic Plane Survey'' (3.6 and 4.5 $\mu$m; \citealt{gldoc}) is a large-scale survey covering the outer Galactic disk, enabling star formation studies with high sensitivity and angular resolution in these relatively uncharted territories, in  environments different than in the inner Galaxy (e.g., reduced metallicity at larger Galactocentric distances, lower cosmic ray flux: \citealt{balser2011}; \citealt{bloemen1984}). The GLIMPSE360 survey can detect low-mass star formation throughout the Perseus spiral arm (at a distance of $\sim$3.6 kpc toward {\it l} = 225$^{\circ}$) and higher mass star formation beyond. 

One of the most interesting regions uncovered by the GLIMPSE360 survey is located at {\it l} $\sim$ 224$^{\circ}$. It contains a large concentration of sources associated with extended 4.5 $\mu$m emission that resemble the Extended Green Objects (EGOs) identified by \citet{cyganowski2008} in the inner Galaxy based on the images from the {\it Spitzer} ``Galactic Legacy Infrared Mid-Plane Survey Extraordinaire'' (GLIMPSE; \citealt{benjamin2003}). The 4.5 $\mu$m emission is likely dominated by H$_2$ emission from outflow shocks, a scenario that is supported by high-angular resolution SiO detections (\citealt{cyganowski2011}). Maser surveys show that EGOs possess dense gas and outflow activity characteristic for the earliest, deeply-embedded YSOs (e.g., \citealt{cyganowski2013}; \citealt{he2012}). Thus, the region at  {\it l} $\sim$ 224$^{\circ}$ is the site of a very young population of protostars, actively feeding back on their surroundings.

\begin{figure*}
\includegraphics[width=\textwidth]{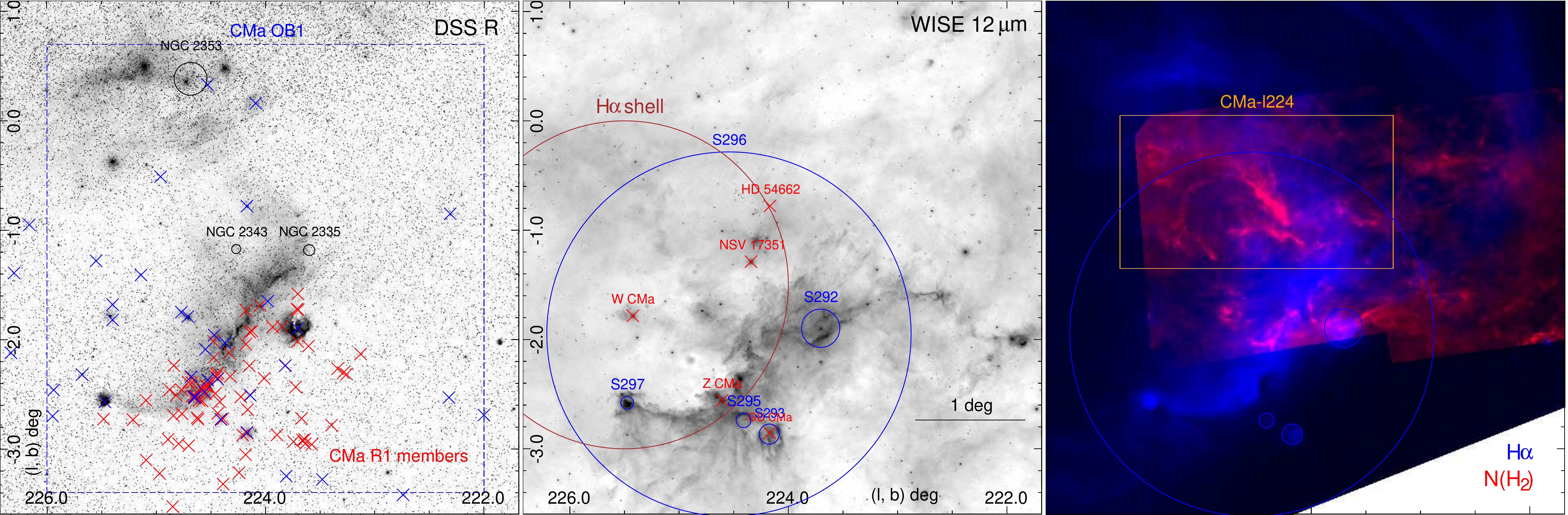}
\caption{The Digital Sky Survey (DSS) R ({\it left}), WISE 12 $\mu$m ({\it middle}), and the combined SHASSA H$\alpha$ (``The Southern H$\alpha$ Sky Survey Atlas'', \citealt{gaustad2001}) and H$_{2}$ column density (based on the {\it Herschel} data, \citealt{elia2013}; {\it right}) images of the Canis Major (CMa) OB1 association \citep{ruprecht1966,hetem2009}. All three images show the same area on the sky. Overlaid on the DSS R image in the left panel are  the members of the CMa OB1 \citep{claria1974b} and CMa R1 \citep{shevchenko1999} associations shown as blue and red `$\times$' symbols, respectively.  The blue box shows the boundaries of the CMa OB1 association \citep{ruprecht1966}. Several well-known open clusters (black circles, e.g., \citealt{dias2002}; \citealt{moitinho2006};  \citealt{kharchenko2013}) and stars (red `$\times$' symbols, \citealt{hetem2008} and references therein) are indicated in the left and middle panel, respectively. In the middle and right panels, the H\,{\sc ii} regions from the  \citet{sharpless1959} catalog are shown as blue circles. The brown circle indicates the approximate position of the H$\alpha$ ring.   The orange rectangle in the right panel shows the region we study in this paper, which we dubbed `CMa--$l224$'; it encloses the region rich in outflows (as traced by the 4.5 $\mu$m emission) and was selected based on the {\it Herschel} emission as described in Section~\ref{s:intro}.  \label{f:cmaob1}} 
\end{figure*}

The outflow area is located within the boundaries of the Canis Major (CMa) OB1 association (222$^{\circ}$ $<$ {\it l} $<$ 226$^{\circ}$, -3$\rlap.^{\circ}$4 $<$ {\it b} $<$  +0$\rlap.^{\circ}$7, see Fig.~\ref{f:cmaob1}; \citealt{ambartsumian1949}; \citealt{ruprecht1966}), $\sim$50$'$ north-east from the largest H\,{\sc ii} region in the area Sh 2-296 (hereafter S296; \citealt{sharpless1959}), at a distance of $\sim$1 kpc (\citealt{claria1974a}; \citealt{shevchenko1999}; \citealt{kaltcheva2000}). S296  is the most prominent feature in the images of the CMa OB1 association, forming a long arc of emission in the central part of the association. Another striking feature, at optical wavelengths in particular, is the H\,{\sc ii} region S292 (IC 2177) - a regularly-shaped nebulosity with a dark lane across its face. S292, as well as nearby H\,{\sc ii} regions S293, S295, and S297 are associated with reflection nebulae (RNe; \citealt{magakian2003}) and are members of the CMa R1 association. CMa R1 is defined by a group of stars embedded in RNe, located within the boundaries of the CMa OB1 association in the galactic latitude range from -3$\rlap.^{\circ}$4 to -2$\rlap.^{\circ}$0 \citep{vandenbergh1966}. The CMa OB1 and CMa R1 associations are physically related \citep{claria1974b}. 

Nineteen confirmed open clusters are located within the boundaries of CMa OB1 (e.g., \citealt{dias2002}; \citealt{moitinho2006}; \citealt{kharchenko2013}), three of them have been studied in detail:  NGC\,2335, NGC\,2343, and NGC\,2353 (see Fig.~\ref{f:cmaob1}). These three open clusters with distances similar to that of CMa OB1 association were suspected to be physically related to the OB1/R1 complex in early studies (e.g., \citealt{ruprecht1966}); however, the cluster ages estimated in later studies are inconsistent with this hypothesis ($\sim$15 to over 100 Myr vs. $\sim$3 Myr; e.g., \citealt{claria1974b}; \citealt{fitzgerald1990}; \citealt{lim2011}; \citealt{kharchenko2013}). 

 The optical and radio images of CMa OB1 association reveal a large scale ring of emission nebulosity with a diameter of $\sim$3$^{\circ}$ centered at $(l,b) \sim (225\rlap.^{\circ}5, -1\rlap.^{\circ}5)$, defined by S\,296 and fainter nebulae to the north and east (see Fig.~\ref{f:cmaob1}; \citealt{herbst1977}). The studies on kinematics of the neutral (H\,{\sc i}; \citealt{herbst1977} and references therein) and ionized (H$\alpha$ and [N\,{\sc ii}]; \citealt{reynolds1978}) gas revealed that the shell is expanding. The origin of the large-scale expanding shell of ionized gas in the CMa OB1 association is still being debated. One of the hypotheses is that the shell was produced by a supernova explosion that triggered star formation in the region \citep{herbst1977,nakano1984,comeron1998}; however, models involving strong stellar winds or an evolving H\,{\sc ii} region are not ruled out \citep{reynolds1978,nakano1984}.

CMa OB1 was covered by the $^{13}$CO($J$=1--0) survey of the Monoceros and Canis Major region ($\sim$560 square degrees; 208$^{\circ} \leq l \leq 230^{\circ}$, $-20^{\circ} \leq b \leq 10^{\circ}$) with the NANTEN millimeter-wave telescope at 2$\rlap.{'}$6 resolution \citep{kim2004}. \citet{kim2004} detected a giant molecular cloud complex associated with CMa OB1 association and comparable in size to S296. \citet{kim2004} identified 115 CO clouds in the survey area; 11 are within the boundaries of the CMa OB1 association.  The higher-resolution (43$''$) $^{12}$CO($J$=1--0) survey with the SEST telescope was conducted toward IRAS sources with colors of star-forming regions in the outer Galaxy ($b \leq 10^{\circ}$ and $85^{\circ} < l < 280^{\circ}$) by \citet{wouterloot1989}, detecting 18 CO clouds in the CMa OB1 association. 

\begin{figure*}
\centering
\includegraphics[width=0.8\textwidth]{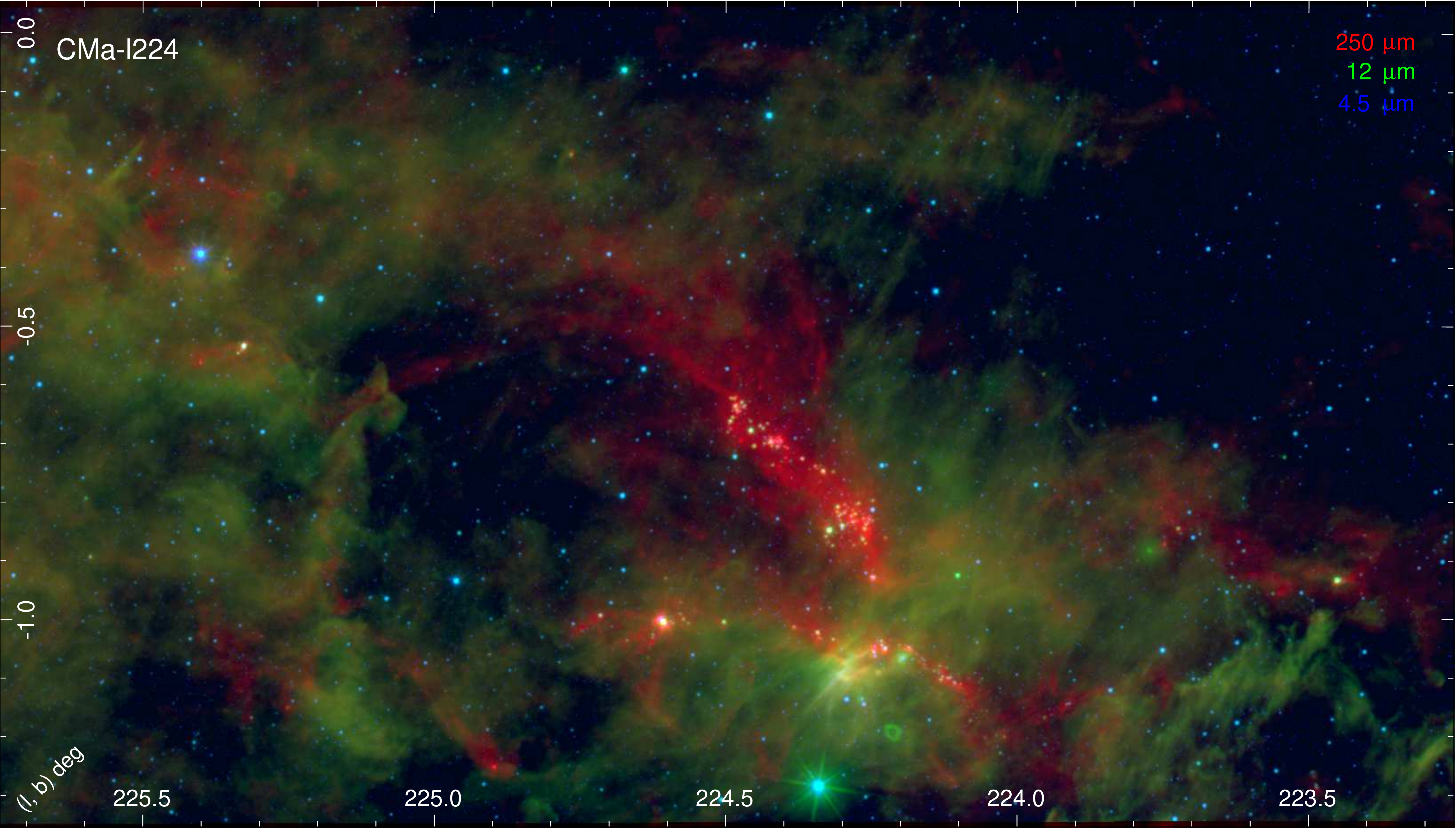}
\caption{The three-color composite image combining the {\it Herschel}/Hi-GAL SPIRE 250 $\mu$m ({\it red}), WISE 12 $\mu$m ({\it green}), and {\it Spitzer}/GLIMIPSE360 IRAC 4.5 $\mu$m ({\it blue}) mosaics. The image shows a region indicated with the orange rectangle in the right panel of Figure~\ref{f:cmaob1}. \label{f:herwisegl}}
\end{figure*}

The CMa--$l224$ outflows region overlaps with the first outer Galaxy region ($216\rlap.^{\circ}5 \lesssim l \lesssim 225\rlap.^{\circ}5, -2^{\circ} \lesssim b \lesssim  0^{\circ}$) studied using the {\it Herschel Space Observatory} ({\it Herschel}; \citealt{pilbratt2010}) data from the Open Time Key Project ``Herschel infrared Galactic Plane Survey'' (Hi-GAL; \citealt{molinari2010}). This Hi-GAL study by \citet{elia2013} based on the PACS 70 and 160 $\mu$m \citep{poglitsch2010} and SPIRE 250, 350, and 500 $\mu$m data \citep{griffin2010}, combined with the WISE 22 $\mu$m photometry, constructed a catalog of {\it Herschel} compact sources (cores and clumps) that they classify as proto-stellar (with a detection at 70 $\mu$m and/or at 22 $\mu$m and 160 $\mu$m; 255 sources) and starless (688).  They found that most of the proto-stellar sources are in the early accretion phase, while for pre-stellar sources the accretion has not started yet.  The temperature and H$_2$ column density maps derived using the 160-500 $\mu$m data reveal a variety of conditions with star formation preferentially distributed along filamentary structures. The filaments were identified in follow-up work by \citet{schisano2014} using the H$_2$ column density map. The outflows region includes areas characterized by the lowest temperatures and the highest H$_2$ column densities in the entire region studied by \citet{elia2013} and \citet{schisano2014}.  

\begin{figure*}
\centering
\includegraphics[width=\textwidth]{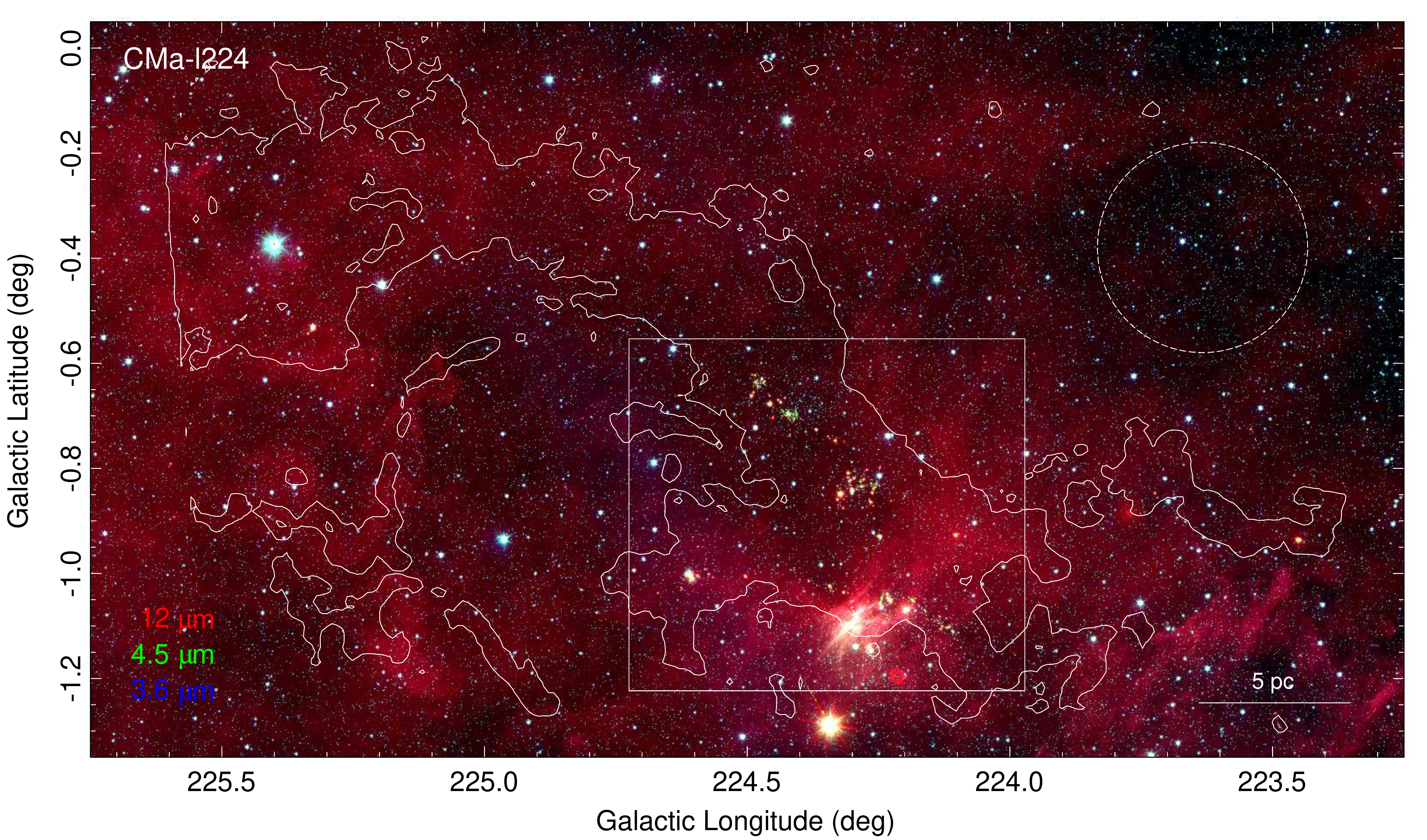}
\caption{The three-color composite image of CMa--$l224$ combining the WISE 12 $\mu$m ({\it red}), GLIMPSE360 4.5 $\mu$m ({\it green}), and GLIMPSE360 3.6 $\mu$m ({\it blue}) images.  The region rich in outflows (traced by 4.5 $\mu$m emission) is located in the central part of the image. The white box indicates the area shown in Fig.~\ref{f:outflows}; it is very bright at {\it Herschel} wavelengths, indicating the earliest stages of star formation. The white contour corresponds to the H$_{2}$ column density of 4 $\times$ 10$^{21}$ cm$^{-2}$.  The region within the white circle represents the background emission in our analysis.  The scale bar at the lower right corresponds to 5 pc assuming a distance of 1 kpc. \label{f:3color}}
\end{figure*}

The area studied in this work was selected based on the H$_2$ column density map from \citet{elia2013}. It encloses a high column density ring-like structure (as traced by the $4 \times 10^{21}$ cm$^{-2}$ contour; see Fig.~\ref{f:3color}); the western side of this structure is formed by the two brightest {\it Herschel} filaments in the region that coincide with the outflows. The size of the selected region is 2$\rlap.^{\circ}$5 $\times$ 1$\rlap.^{\circ}$4 with the center at ($l$, $b$) = (224$\rlap.^{\circ}$5, -0$\rlap.^{\circ}$65); see Figs.~\ref{f:cmaob1} and \ref{f:herwisegl}. We refer to this region as `CMa--$l224$' throughout the paper. 

The goal of this paper is to identify YSO candidates in CMa--$l224$ using the high-sensitivity GLIMPSE360 3.6 and 4.5 $\mu$m data combined with ancillary catalogs and images. We estimate the physical parameters of the YSO candidates and study their correlations with the dust and gas tracers. The methods developed for this relatively small region can be applied to the larger areas in the outer Galactic plane where {\it Spitzer} photometry at longer wavelengths is not available, making the methods developed for the nearby molecular clouds unsuitable (e.g., \citealt{gutermuth2009}; \citealt{dunham2015}). These new methods enable systematic studies of the outer Galaxy, covering a range of environments and star formation activities, to uncover and characterize a relatively unstudied population of intermediate- and low-mass YSOs and study the impact of the environment on the star formation process (e.g., outer vs. inner Galaxy, arm vs. inter-arm regions, position in the Galaxy thus chemical abundances, etc.).

In Sections 2 and 3, we describe the catalogs and images used in the analysis, and the catalog matching, respectively. The method of distance determination is described in Section 4. The initial YSO selection criteria are provided in Section 5. Sections 6--9 provide details on the subsequent analysis that selected the most reliable YSO candidates. We compare our list of YSO candidates to the catalog of previously known WISE YSO candidates in Section 10. In Sections 11 and 12, we compare the spatial distribution of YSO candidates to the dust and gas tracers. We discuss possible star formation scenarios in the region in Section 13. The summary and conclusions are provided in Section 14.

\begin{figure*}
\centering
\includegraphics[width=\textwidth]{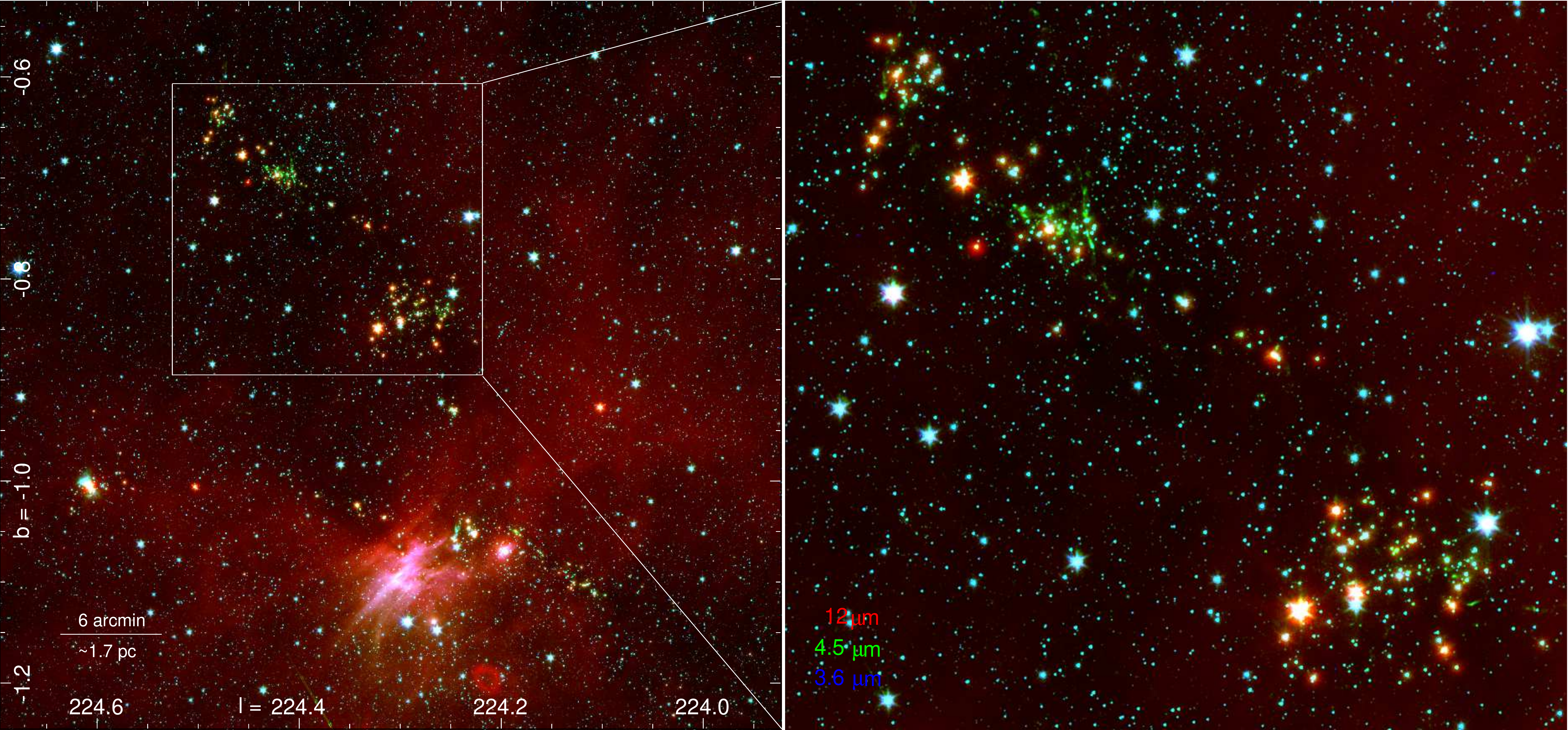}
\caption{{\it Left:} The image shows the region indicated with white box in Fig.~\ref{f:3color} in the same combination of colors.  This area harbors sources with outflows (traced by enhanced 4.5 $\mu$m emission). The source at ({\it l}, {\it b}) $\sim$ (224$\rlap.^{\circ}$6,-1$^{\circ}$) is the reflection nebula Gy 3--7  associated with the embedded cluster and the high-luminosity IRAS source IRAS 07069--1045 (Gy 3--7 \#6; \citealt{tapia1997}). {\it Right:} A zoom on the area enclosed in the white box in the left panel.  \label{f:outflows}}.
\end{figure*}

\begin{deluxetable*}{clcc}[ht!]
\tablecaption{Photometric Data Used in this Study and the Angular Resolutions of the Surveys \label{t:data}}
\tabletypesize{\small}
\tablewidth{0pt}

\tablehead{
\colhead{Wavelength} &
\colhead{Instrument} &
\colhead{Survey} &
\colhead{FWHM} \\
\colhead{($\mu$m)} &
\colhead{} &
\colhead{} &
\colhead{($''$)}}
\startdata
1.2 & {\it 2MASS} & All-Sky 2MASS & 2--3\tablenotemark{a} \\
1.6 & {\it 2MASS} & All-Sky 2MASS & 2--3 \\ 
2.2 & {\it 2MASS} & All-Sky 2MASS & 2--3 \\
3.6 & {\it Spitzer} IRAC & GLIMPSE360 &  1.7\\
4.5 & {\it Spitzer} IRAC & GLIMPSE360 &  1.7\\
8.3 & {\it MSX} SPIRIT III & MSX  & $\sim$20 \\
  9 & {\it AKARI} IRC & All-Sky Survey & 5.5 \\
 12 & {\it WISE}         & All-Sky WISE/ NEOWISE &  6.5 \\
12.1& {\it MSX} SPIRIT III & MSX  & $\sim$20 \\
14.7& {\it MSX} SPIRIT III & MSX  & $\sim$20 \\
 18 & {\it AKARI} IRC & All-Sky Survey & 5.7 \\
21.3&{\it MSX} SPIRIT III & MSX  & $\sim$20 \\
 22 & {\it WISE}         & All-Sky WISE/ NEOWISE &  12 \\ 
70  & {\it Herschel} PACS & Hi-GAL\tablenotemark{b}  & $\sim$10\tablenotemark{c} \\
160 & {\it Herschel} PACS & Hi-GAL  & 12\tablenotemark{c} \\
250 & {\it Herschel} SPIRE & Hi-GAL & 18\tablenotemark{c} \\
350 & {\it Herschel} SPIRE & Hi-GAL & 25\tablenotemark{c} \\
500 & {\it Herschel} SPIRE & Hi-GAL & 36\tablenotemark{c} \\
\enddata
\tablenotetext{a}{Depending on the atmospheric seeing.}
\tablenotetext{b}{The Hi-GAL source list from \citet{elia2013}.}
\tablenotetext{c}{\citet{molinari2016}}
\end{deluxetable*}

\begin{figure*}[ht!]
\includegraphics[width=0.23\textwidth]{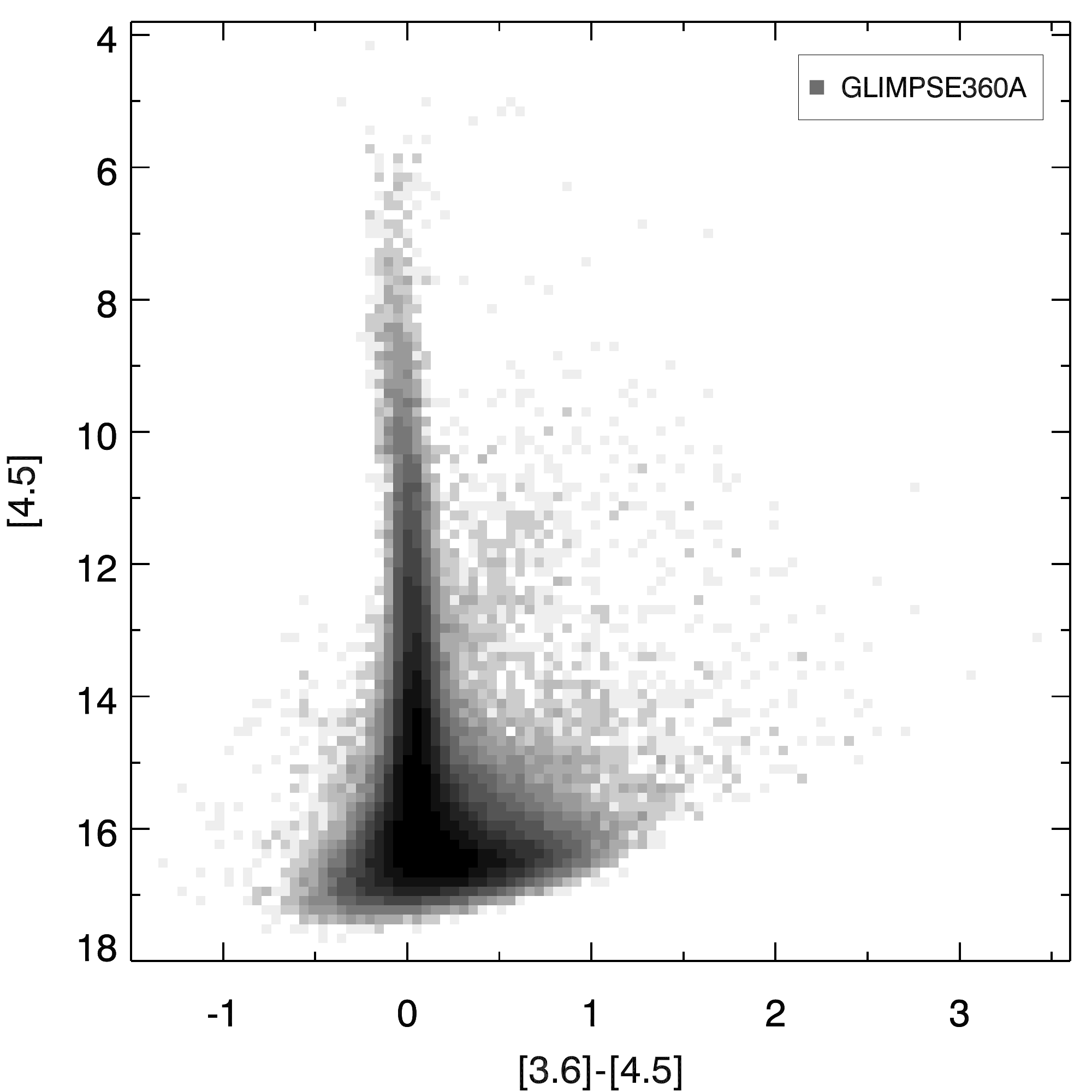}
\includegraphics[width=0.23\textwidth]{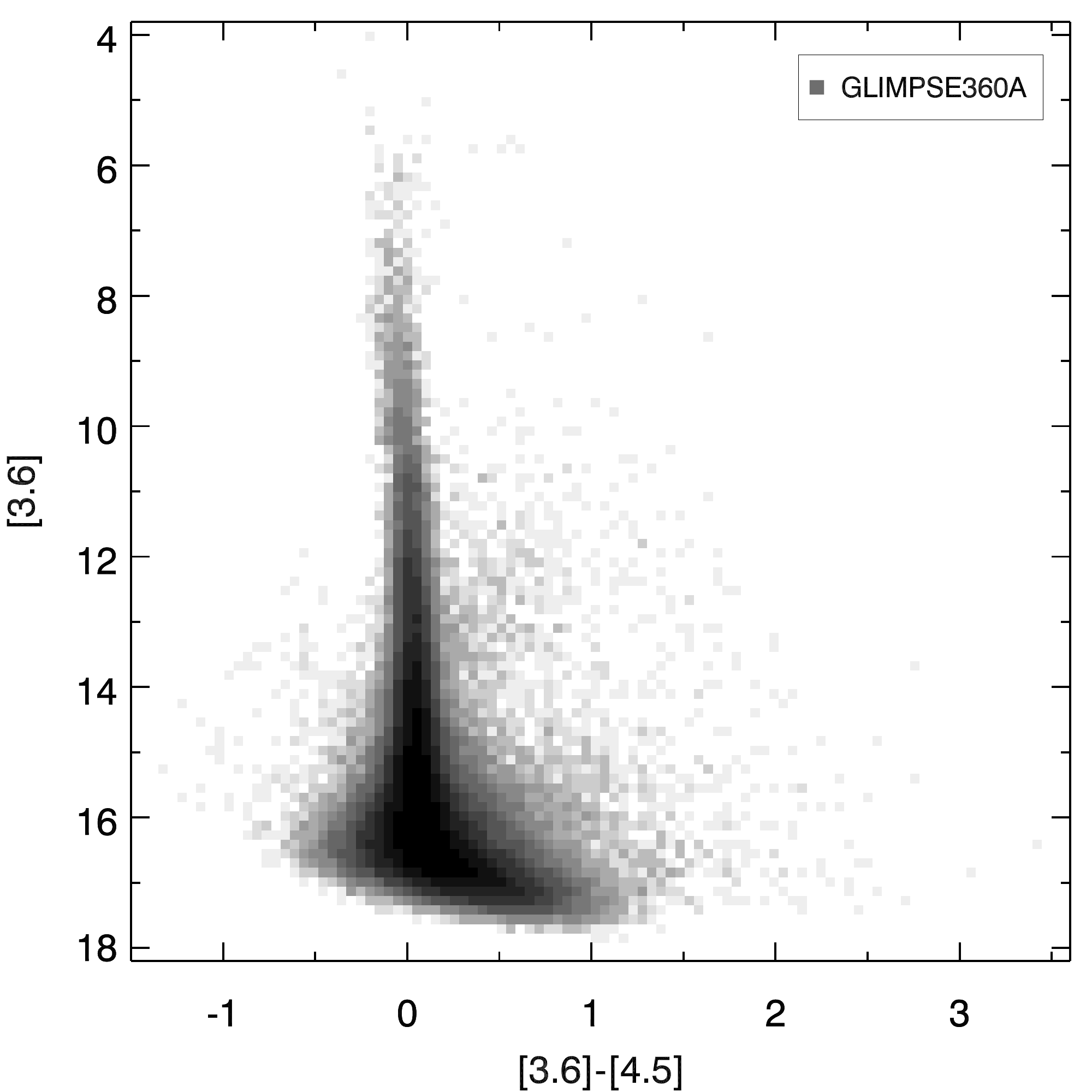}
\includegraphics[width=0.23\textwidth]{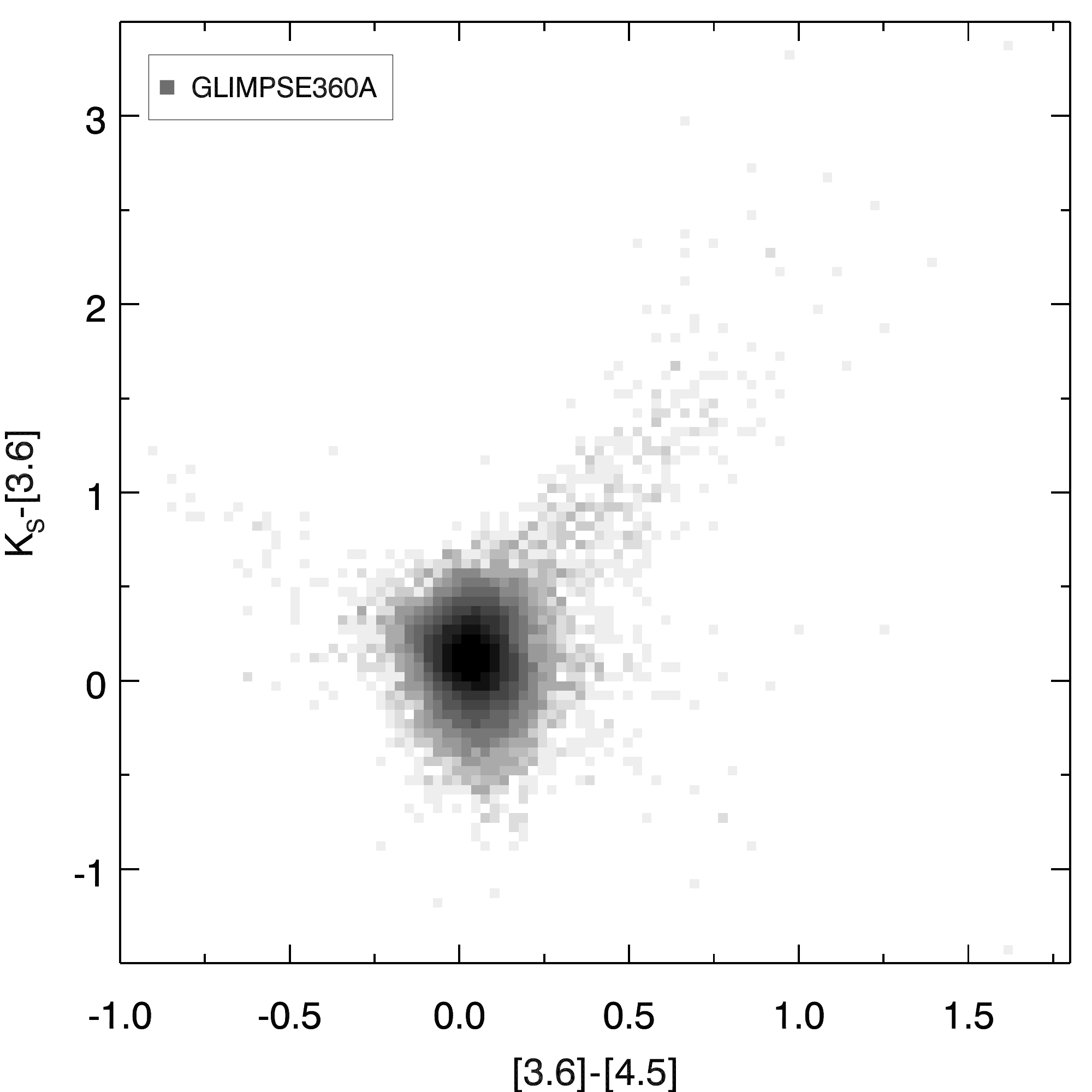}
\includegraphics[width=0.23\textwidth]{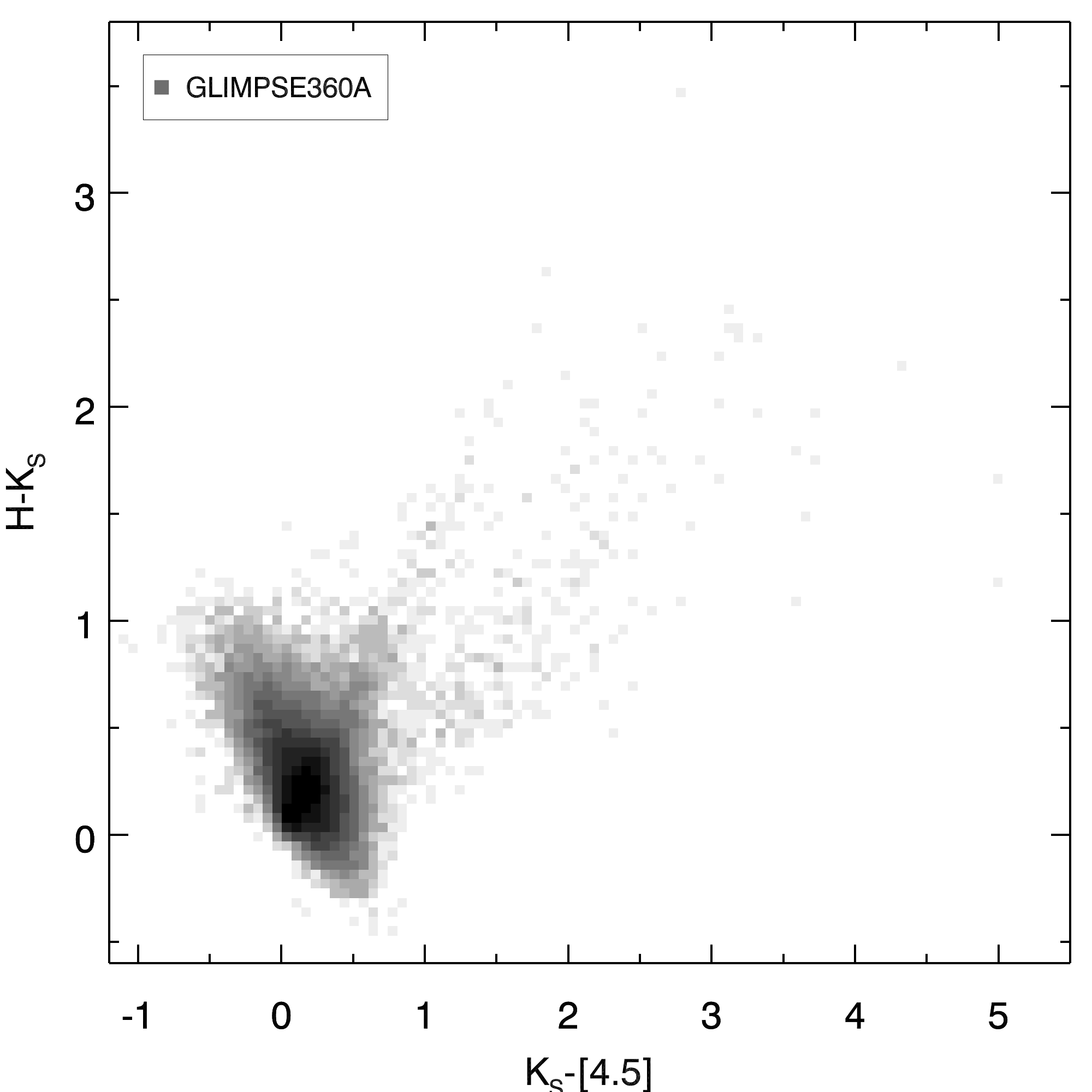}
\hfill
\includegraphics[width=0.23\textwidth]{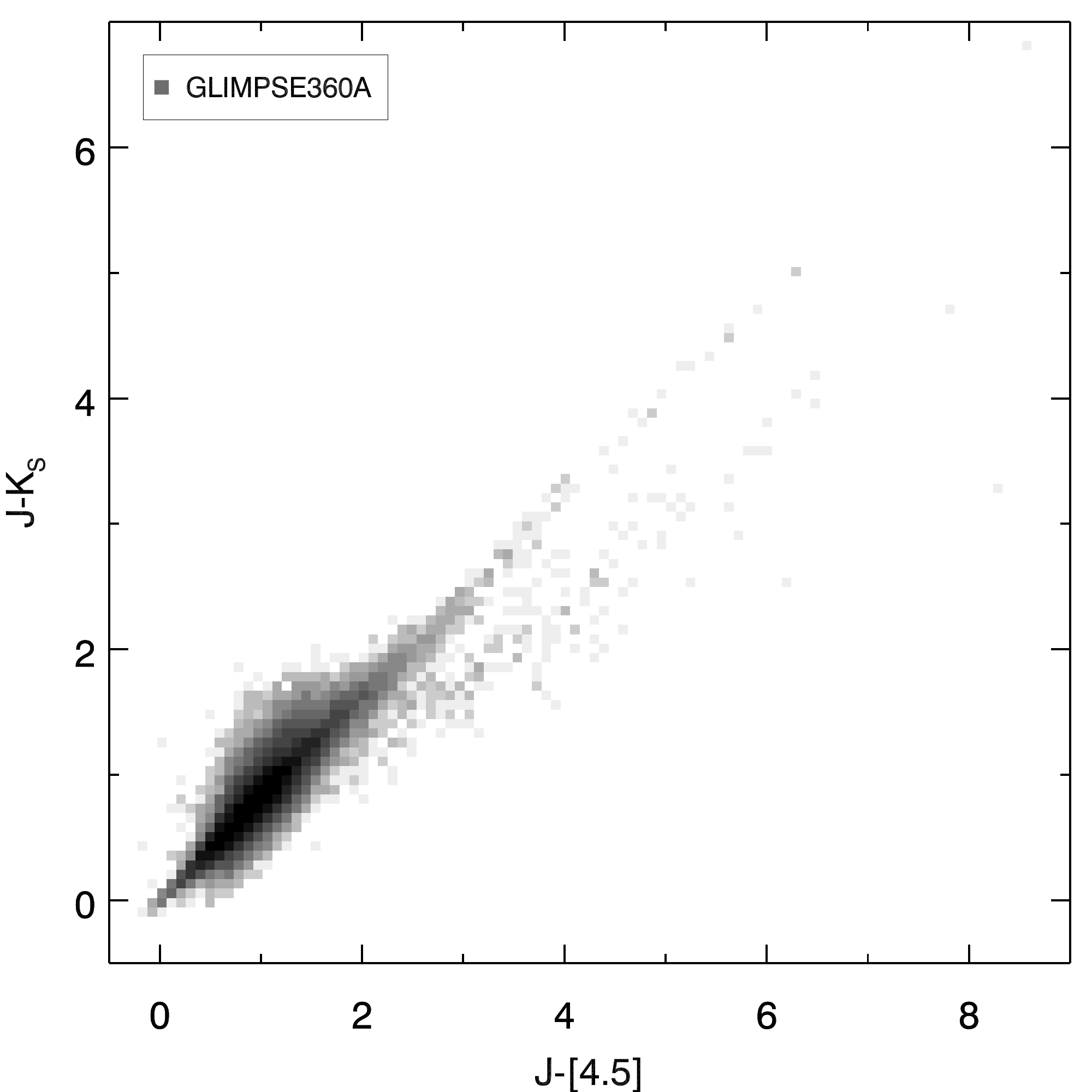}
\hfill
\includegraphics[width=0.23\textwidth]{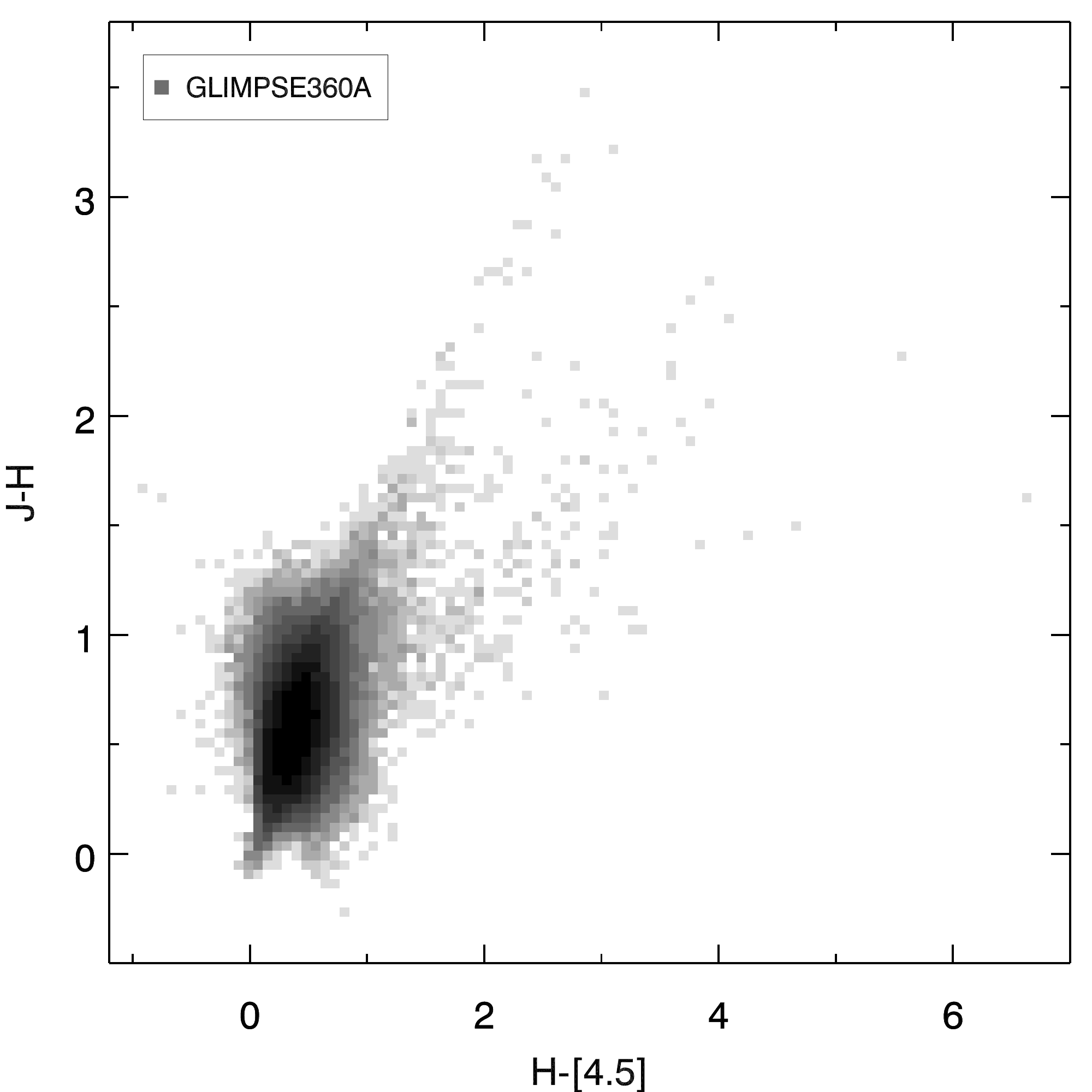}
\hfill
\includegraphics[width=0.23\textwidth]{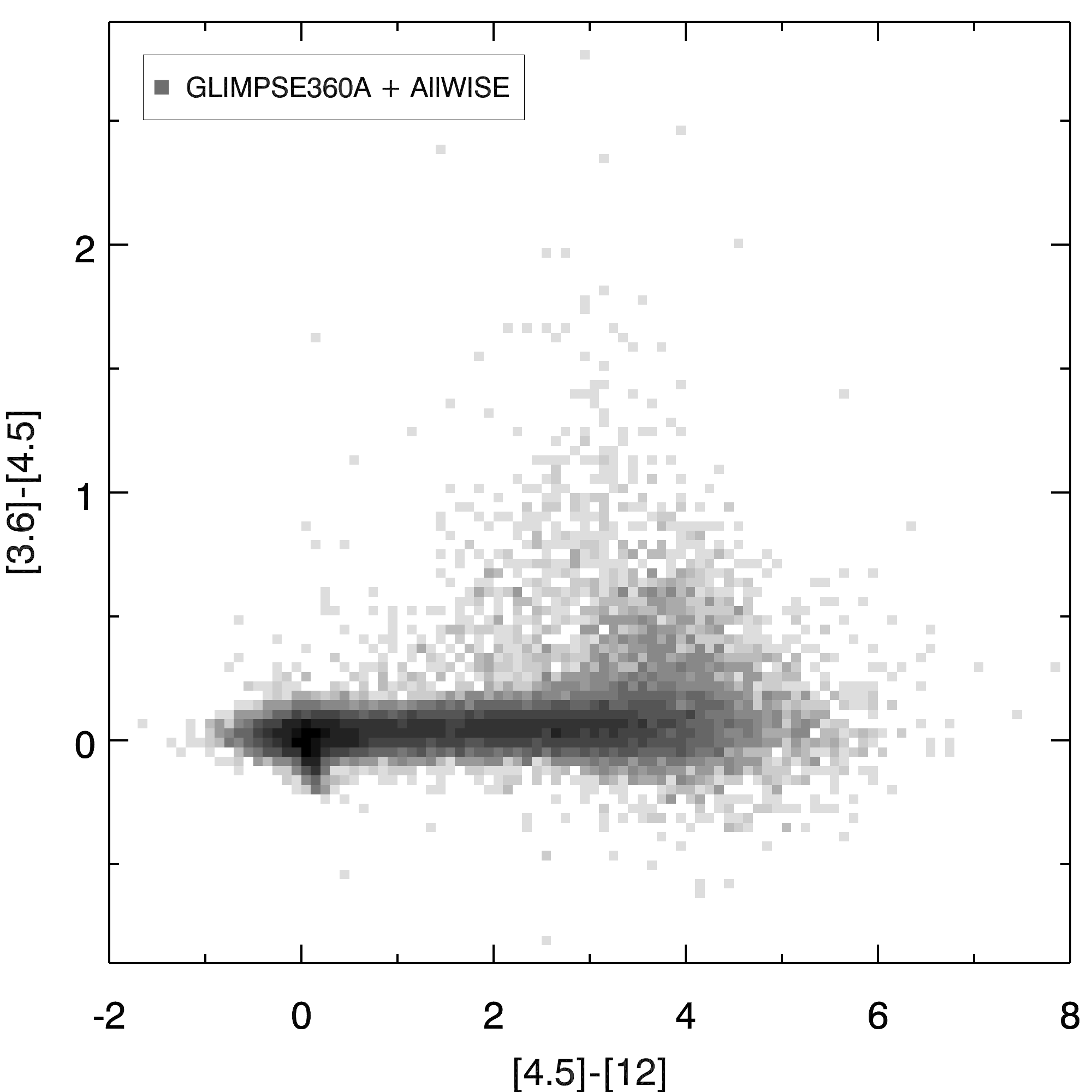}
\hfill
\includegraphics[width=0.23\textwidth]{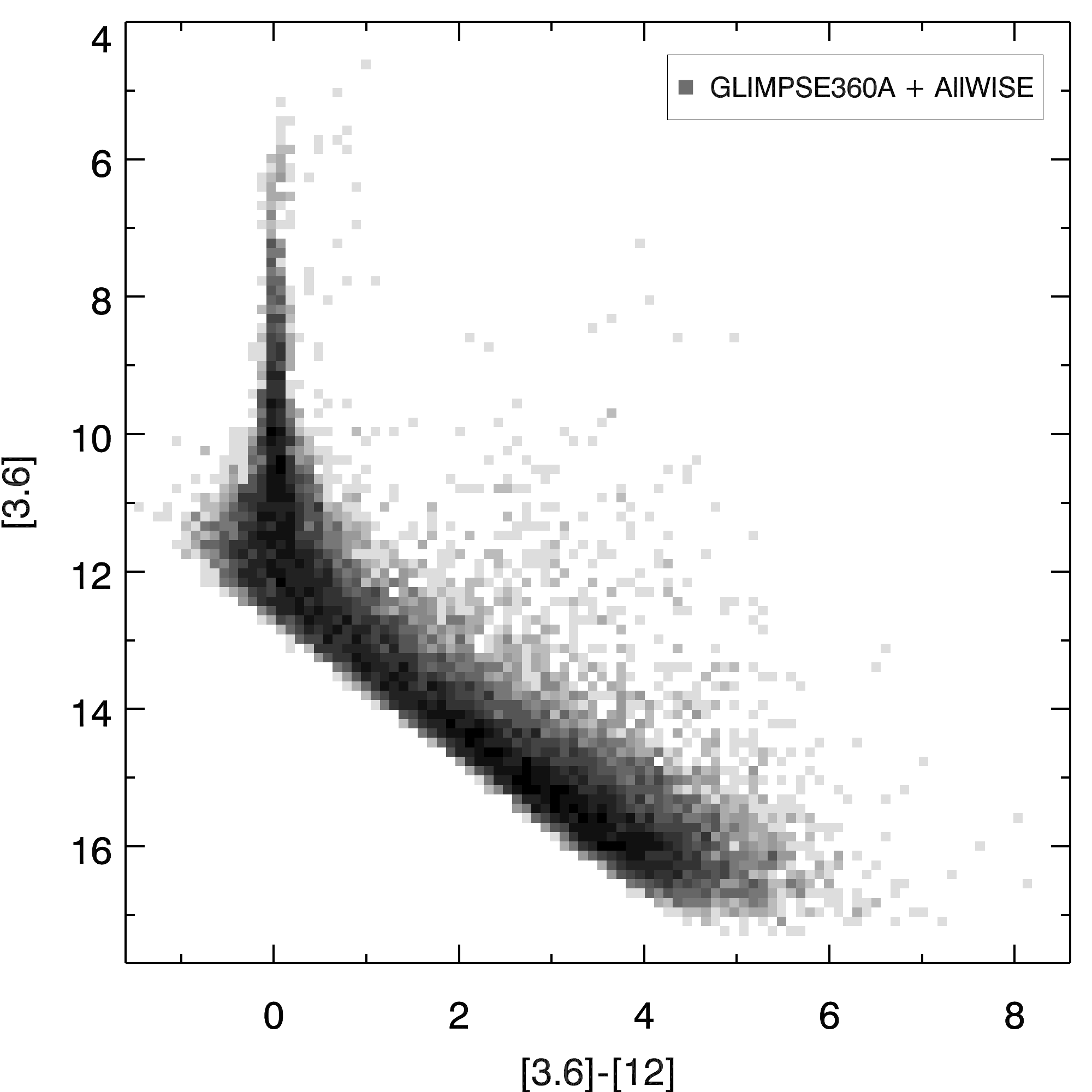}
\caption{Example CMDs and CCDs showing the distribution of all the GLIMPSE360 IRAC Archive (GLIMPSE360A) sources with flux measurements at both 3.6 and 4.5 $\mu$m ($[3.6]$ vs. $[3.6]-[4.5]$ and $[4.5]$ vs. $[3.6]-[4.5]$), and the distribution of the GLIMPSE360 sources with valid photometry in both {\it Spitzer} bands with 2MASS and/or AllWISE 12 $\mu$m matches (the remaining plots).  The data are displayed as Hess diagrams: the grey scale indicates the source density in logarithmic scale (the source density range is different in each plot).  The distribution of the combined catalog sources shown in these plots are compared in Section~\ref{s:2massspitzer} to the distribution of the YSO and non-YSO sources from literature and YSO models to develop the YSO selection criteria for CMa--$l224$ (see also Figs.~\ref{f:HKK45}--\ref{f:36454512}).  \label{f:data}}
\end{figure*} 

\section{The Data}
\label{s:data}

Our study is mainly based on the catalogs and images from the GLIMPSE360 survey (PI B. Whitney). GLIMPSE360 observed the Outer Galactic plane at 3.6 $\mu$m and 4.5 $\mu$m with the Infrared Array Camera (IRAC; \citealt{fazio2004}), covering a longitude range from 65$^{\circ}$ to 265$^{\circ}$, excluding regions covered by smaller surveys (SMOG: $l$ = 102$^{\circ}$ -- 109$^{\circ}$, PI S. Carey;  Cygnus-X:  $l$ = 76$^{\circ}$  -- 82$^{\circ}$, \citealt{beerer2010}). The GLIMPSE360 data products are described in the data delivery document available at the NASA/IPAC Infrared Science Archive \citep{gldoc}.  We complemented the GLIMPSE360 data with the ancillary data from previous near- to far-IR surveys (see Section~\ref{s:complementary}). Table~\ref{t:data} includes a list of photometric catalogs used in our analysis together with the angular resolutions of the observations.  

\subsection{Primary Dataset: {\it Spitzer} GLIMPSE360}
\label{s:glimpse360}

The GLIMPSE360 data products include a highly reliable IRAC Point Source Catalog and a more complete IRAC Point Source Archive (see \citealt{gldoc} for details). In our analysis, we use the IRAC Archive for completeness both in the number of sources and flux measurements at each wavelength. The sources included in the Archive fulfill less stringent criteria than those included in the Catalog, however, these criteria were developed to ensure that each source is a legitimate astronomical source and that the fluxes reported for the IRAC bands are of high quality. The {\it Spitzer} GLIMPSE360 data are matched to the near-IR ({\it JHK$_{\rm s}$}) data from the Two Micron All Sky Survey \citep[2MASS;][]{skrutskie2006}; 2MASS photometry is an integral part of the GLIMPSE360 Catalog and Archive.

About 212,000 GLIMPSE360 sources are located in CMa--$l224$ (see Table~\ref{t:stats}). The lower and upper sensitivity limits in the 3.6 $\mu$m band are 0.021 mJy (17.8 mag) and 1100 mJy (6.0 mag), respectively.  In the 4.5 $\mu$m band, the lower and upper sensitivity limits are 0.022 mJy (17.3 mag) and 1100 mJy (5.5 mag), respectively \citep{gldoc}. The minimum signal-to-noise ratio is 5 in both bands. Figure~\ref{f:data} shows example color-color  and color-magnitude diagrams (CCDs and CMDs) using the {\it Spitzer} and 2MASS data.

\begin{figure*}
\includegraphics[width=\textwidth]{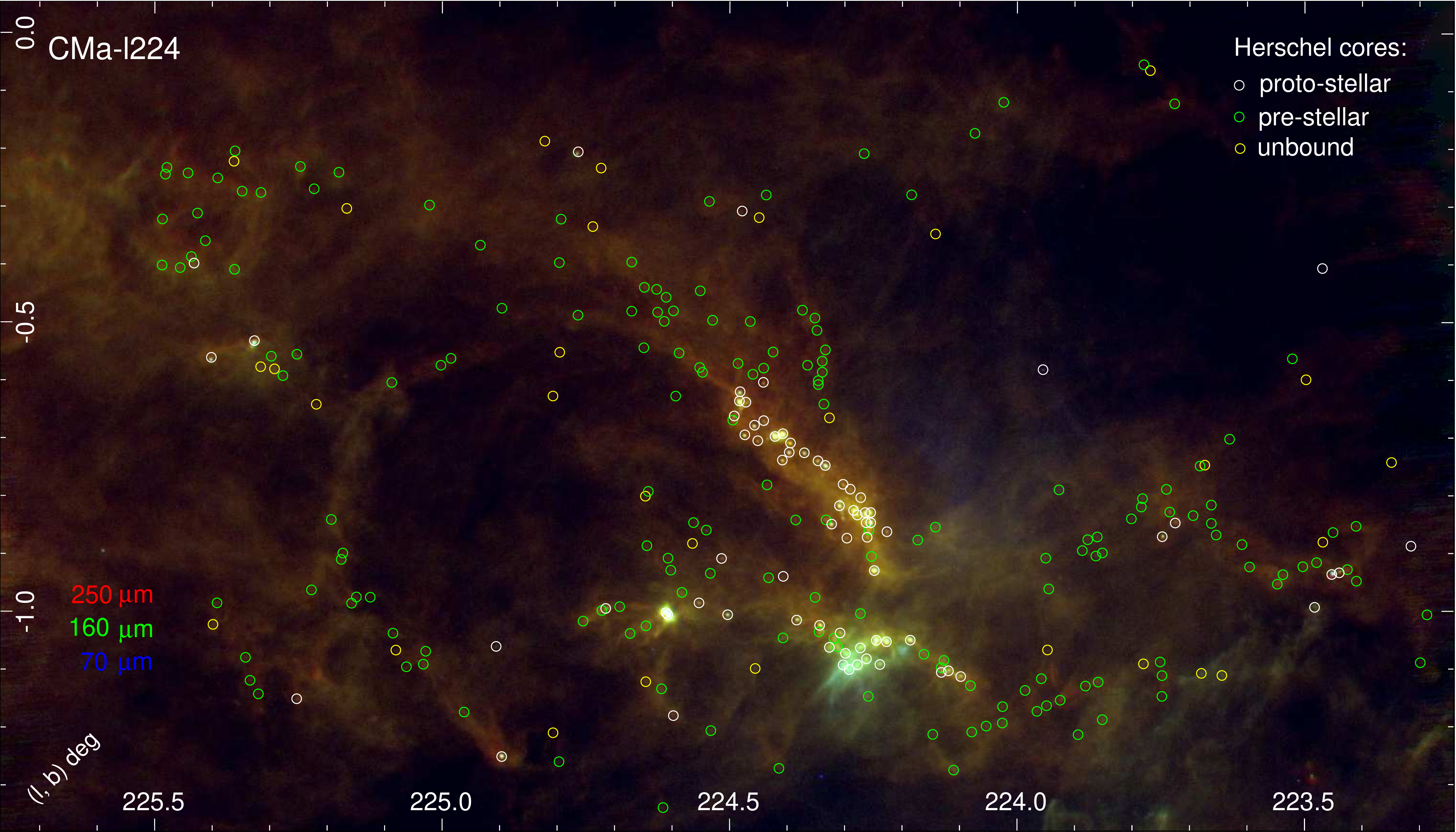}
\caption{The three-color image combining the {\it Herschel} SPIRE 250 $\mu$m ({\it red}), PACS 160 $\mu$m ({\it green}), and PACS 70 $\mu$m ({\it blue}) mosaics.  The white, green, and yellow circles indicate {\it Herschel} proto-stellar, starless, and unbound cores from \citet{elia2013}, respectively. \label{f:hercores}}
\end{figure*}

\subsection{Complementary Mid-Infrared to Submillimeter Data}
\label{s:complementary}

We supplement the {\it Spitzer} GLIMPSE360 data with the longer-wavelength WISE/NEOWISE (Section~\ref{s:allwise}), AKARI (Section~\ref{s:akari}), MSX (Section~\ref{s:msx}), and {\it Herschel} (Section~\ref{s:higal}) data (see Table~\ref{t:data}). The longer wavelength data probe the embedded population of YSOs more deeply.

\subsubsection{AllWISE}
\label{s:allwise}

We supplement the GLIMPSE360 data with the 12 $\mu$m and 22 $\mu$m catalogs and images from the AllWISE program. A longer wavelength photometry improves the YSO identification, allowing better separation between protostars and more evolved YSOs. The AllWISE program combines the data from ``The Wide-field Infrared Survey Explorer'' (WISE; \citealt{wright2010}) cryogenic and NEOWISE \citep{neowise} post-cryogenic survey phases. The AllWISE products have improved photometric sensitivity and accuracy, as well as astrometric precision compared to the original WISE All-Sky Data Release.   

The sensitivity of the WISE/NEOWISE survey is much lower than that of the GLIMPSE360 survey, and thus the combination of GLIMPSE360 and AllWISE bands will miss sources that were only detected at {\it Spitzer} bands.  We do not use WISE 3.4 and 4.6 $\mu$m filters which are similar to GLIMPSE360/IRAC 3.6 and 4.5 $\mu$m filters since the sensitivity limits of the GLIMPSE360 survey are much deeper and the saturation limits are higher. GLIMPSE360 also provides three times better spatial resolution (2$''$ vs. WISE 6$''$).  

\citet{koenig2014} investigate the performance of the AllWISE catalog in the Galactic Plane by analyzing the data for the W3 and W5 giant molecular clouds and part of the W4 region.  They found that in WISE Band 3 (12 $\mu$m) and Band 4 (22 $\mu$m), the AllWISE source extraction pipeline retrieves only 50\%-60\% and 75\% of sources seen in the WISE images, respectively. They also found that the vast majority of the AllWISE 12 $\mu$m and 22 $\mu$m catalog sources with a signal-to-noise ratio larger than 5 are likely spurious detections, especially in the high-sky background regions; they likely represent upper limit measurements of the nebular background. 

In this work, we use the profile fitting photometry -- columns $w?mpro$ and $w?msigpro$ (magnitude and magnitude uncertainty, respectively) in the online AllWISE catalog. The AllWISE catalog includes the photometric quality flag ({\it ph\_qual} column), which provides information on the quality of the profile-fit photometry measurement in each band based on the signal-to-noise ratio of the measurement.  The photometric quality flag `U' indicates that the quoted magnitude is computed from a 2$\sigma$ brightness upper limit. For valid flux measurements, {\it ph\_qual} is `A', `B', or `C', depending on the signal-to-noise ratio ($w?snr$ column):  $w?snr$ $\geq$ 10, 3 $<$ $w?snr$ $<$ 10,  2 $<$ $w?snr$ $<$ 3, respectively. Out of 80,171 AllWISE sources (74,726 point sources: extended source flag {\it ext\_flg} = 0) in CMa--$l224$, $\sim$74\% ($\sim$75\%) have 12 $\mu$m photometry and $\sim$90\% ($\sim$91\%) have 22 $\mu$m photometry flagged as an upper limit. Only 6,310 AllWISE sources (5,352 point sources) have valid both 12 $\mu$m and 22 $\mu$m photometry.

We use the zero magnitude flux of 31.674 Jy and 8.363 Jy \citep{jarrett2011} to calculate, respectively, the 12 $\mu$m and 22 $\mu$m fluxes from magnitudes.

\subsubsection{AKARI}
\label{s:akari}

To provide additional longer-wavelength photometry for GLIMPSE360 sources in CMa--$l224$, we use the data from the Infrared Astronomical Satellite AKARI Infrared Camera (IRS) All-Sky Survey \citep{murakami2007,onaka2007,ishihara2010}. The AKARI IRS catalog (version 1.0) provides photometry in two mid-IR bands: MIR-S (9 $\mu$m) and MIR-L (18 $\mu$m). The spatial resolutions (FWHM) in the 9 $\mu$m and 18 $\mu$m bands are $\sim$5$\rlap.{''}$5 and $\sim$5$\rlap.{''}$7, respectively. For a signal-to-noise ratio of $\sim$5, the detection limits in the MIR-S and MIR-L bands are about 0.05 and 0.09 Jy, respectively \citep{akaridoc}. Since the GLIMPSE360 observations are much deeper, we expect to find AKARI counterparts only for the brightest GLIMPSE360 sources.  There are 102 AKARI IRS point sources in the CMa--$l224$ region. 

The zero magnitude fluxes are 56.262 Jy and 12.001 Jy for the 9 $\mu$m and 18 $\mu$m bands, respectively \citep{tanabe2008}.

\subsubsection{MSX}
\label{s:msx}

Our analysis includes the data from The Midcourse Space Experiment (MSX) Point Source Catalog (\citealt{egan2003}; $\sim$20$''$ resolution). We use the data from the broad spectral bands: A (8.28 $\mu$m), C (12.13 $\mu$m), D (14.65 $\mu$m), and E (21.3 $\mu$m).  The survey sensitivity in Band (A, C, D, E) is (0.1--0.2, 1.1--3.1, 0.9--2, 2--6) Jy (early--late in the mission; \citealt{egan2003}). As for the AKARI data, due to much lower sensitivity of the MSX survey than the GLIMPSE360 survey, only a small fraction of {\it Spitzer} sources have MSX counterparts.  There are 78 MSX sources in CMa--$l224$. 

The zero magnitude fluxes are (58.49, 26.51, 18.29, 8.80) Jy for (A, C, D, E) bands \citep{egan2003}. 

\subsubsection{{\it Herschel} Hi-GAL }
\label{s:higal}

We use the catalog of {\it Herschel} Hi-GAL sources provided by \citet{elia2013} (see Section~\ref{s:intro}). Their full catalog includes the PACS 70 and 160 $\mu$m and SPIRE 250, 350, and 500 $\mu$m photometry for high-reliability {\it Herschel} compact sources eligible for grey-body fit from the $\sim$18 square degrees area in the third Galactic quadrant. Each source has valid fluxes in at least three consecutive bands and spectral energy distributions (SEDs) not showing dips and not peaking at 500 $\mu$m.  The \citet{elia2013} catalog provides not only the 70--500 $\mu$m photometry, but also kinematic distances (estimated based on the CO data), physical properties of the sources (source sizes, masses, temperatures, and bolometric luminosities), and their classification. Two hundred and seventy three {\it Herschel} sources from the \citet{elia2013} catalog are located in CMa--$l224$: 29 unbound, 171 pre-stellar, and 73 proto-stellar cores (see Fig.~\ref{f:hercores}). The {\it Herschel} cores/clumps are classified as proto-stellar if they are detected at the PACS 70 $\mu$m and/or WISE 22 $\mu$m, and at the PACS 160 $\mu$m bands.

\begin{deluxetable*}{lr}
\tablecaption{Source Statistics for CMa--$l224$ \label{t:stats}}
\tabletypesize{\small}
\tablewidth{0pt}

\tablehead{
\colhead{Source Type} &
\colhead{\# of sources} 
}
\startdata
{\it Spitzer}/GLIMPSE360 sources \dotfill & 211,900 \\
AllWISE sources \dotfill & 80,171 (74,726)\tablenotemark{a} \\
 AKARI/IRS sources \dotfill & 102 \\
 MSX Point Source Catalog sources \dotfill & 78 \\
{\it Herschel}/Hi-GAL sources (Elia et al. 2013)\tablenotemark{b} \dotfill  & 273 \\
{\it Herschel}/Hi-GAL sources (Full)\tablenotemark{b}  \dotfill  &  787 \\
 &\\
 GLIMPSE360 sources with valid 3.6 $\mu$m and 4.5 $\mu$m photometry \dotfill & 202,575 \\
 AllWISE sources with valid 12 $\mu$m photometry\tablenotemark{c} \dotfill & 21,255 (18,919)\tablenotemark{a} \\ 
 AllWISE sources with valid 22 $\mu$m photometry\tablenotemark{c} \dotfill & 8,200 (7,064)\tablenotemark{a} \\
 AllWISE sources with valid both 12 $\mu$m and 22 $\mu$m photometry\tablenotemark{c} \dotfill & 6,310 (5,352)\tablenotemark{a} \\
 {\it Herschel}/Hi-GAL sources (Elia et al. 2013) with valid 70 $\mu$m flux \dotfill & 64\\
 {\it Herschel}/Hi-GAL sources (Full) with valid 70 $\mu$m flux \dotfill & 113\\
 &\\
\hline
\multicolumn{2}{c}{Matching Results\tablenotemark{d}}\\
\hline \\
GLIMPSE360 (3.6 \& 4.5 $\mu$m) with AllWISE 12 $\mu$m matches \dotfill & 10,557/27,646\tablenotemark{e} \\
GLIMPSE360 (3.6 \& 4.5 $\mu$m) with AllWISE 22 $\mu$m matches \dotfill & 2,335/37,862\tablenotemark{e} \\
GLIMPSE360 (3.6 \& 4.5 $\mu$m) with AllWISE 12 \& 22 $\mu$m matches \dotfill & 1,878/26,371\tablenotemark{e} \\
GLIMPSE360 (3.6 \& 4.5 $\mu$m) with AKARI matches \dotfill & 58\\  
GLIMPSE360 (3.6 \& 4.5 $\mu$m) with MSX matches \dotfill & 26\\ 
GLIMPSE360 (3.6 \& 4.5 $\mu$m) with {\it Herschel} matches (Elia et al. 2013)\tablenotemark{b} \dotfill &  72 (40)\tablenotemark{f}\\ 
GLIMPSE360 (3.6 \& 4.5 $\mu$m) with {\it Herschel} matches (Full) \dotfill &  206 (75)\tablenotemark{g} \\
\enddata
\tablenotetext{a}{The number in the parenthesis corresponds to point sources ({\it ext\_flg}=0).}
\tablenotetext{b}{Based on the Hi-GAL catalog from \citet{elia2013}. In order to be included in the catalog, a source had to be detected in at least three consecutive bands.}
\tablenotetext{c}{AllWISE catalog sources with the {\it ph\_qual} of `A', `B', or `C' (i.e., no upper limits) in a corresponding band.}
\tablenotetext{d}{The matching radii between the GLIMPSE360 catalog and AllWISE, AKARI, MSX, {\it Herschel}/E10, and {\it Herschel}/Full are 1$''$, 1$''$, 1$\rlap.{''}$3, 5$''$, and 5$''$, respectively. All the constraints described in Section~\ref{s:matching} are applied.} 
\tablenotetext{e}{The numbers represent the number of sources with valid data points ({\it left}) and upper limits ({\it right}) at a corresponding wavelength(s).}
\tablenotetext{f}{The number in the parenthesis corresponds to sources with valid 70 $\mu$m flux. Out of 72 sources, 26 are SPIRE-only and are not used in the analysis.}
\tablenotetext{g}{The number in the parenthesis corresponds to sources with valid 70 $\mu$m flux. Out of 206 sources, 111 are SPIRE-only and are not used in the analysis.}
\end{deluxetable*}

In addition to the \citet{elia2013} catalog (their Table 3), we also use the more complete catalog that was used to select these sources (unpublished; D. Elia, private communication) and refer to it as the 'Full {\it Herschel} catalog'.  Although the \citet{elia2013} catalog is a subset of the Full {\it Herschel} catalog, we consider them separately since the sources from the \citet{elia2013} catalog were analyzed in detail and classified as described above, while the Full {\it Herschel} catalog includes all the sources extracted from the images that fulfill a set of the quality criteria (all five-band photometry is included regardless of the completeness/regularity of the SED). We are able to identify unreliable photometry in the subsequent analysis. There are 787 sources from the Full {\it Herschel} catalog in CMa--$l224$. 

In our analysis, we use the H$_{2}$ column density and temperature images derived by \citet{elia2013} through a pixel-to-pixel graybody fit using the Hi-GAL 160--500 $\mu$m images; they represent the properties of the cold component of dust. 

The {\it Herschel} PACS and SPIRE images from the Hi-GAL survey have been downloaded from the {\it Herschel Science Archive} (Proposal ID: OT1\_smolinar\_5).

\subsection{The CO Data}
\label{s:co}

In our analysis, we determine kinematic distances to the GLIMPSE360 sources using the NANTEN $^{12}$CO ({\it J}=1--0) Galactic Plane Survey data \citep{mizuno2004} presented by \citet{elia2013} for their survey area (see Section~\ref{s:intro}). The spatial resolution of the survey is 2$\rlap.{'}$6, significantly lower than the GLIMPSE360 resolution ($\sim$2$''$). The spectra for $|b|$ $\le$ 0$\rlap.^{\circ}$5 were obtained with a grid spacing of 4$'$. The CMa--$l224$ is covered in full by this survey.

\begin{figure*}[ht!]
\centering
\includegraphics[width=0.9\textwidth]{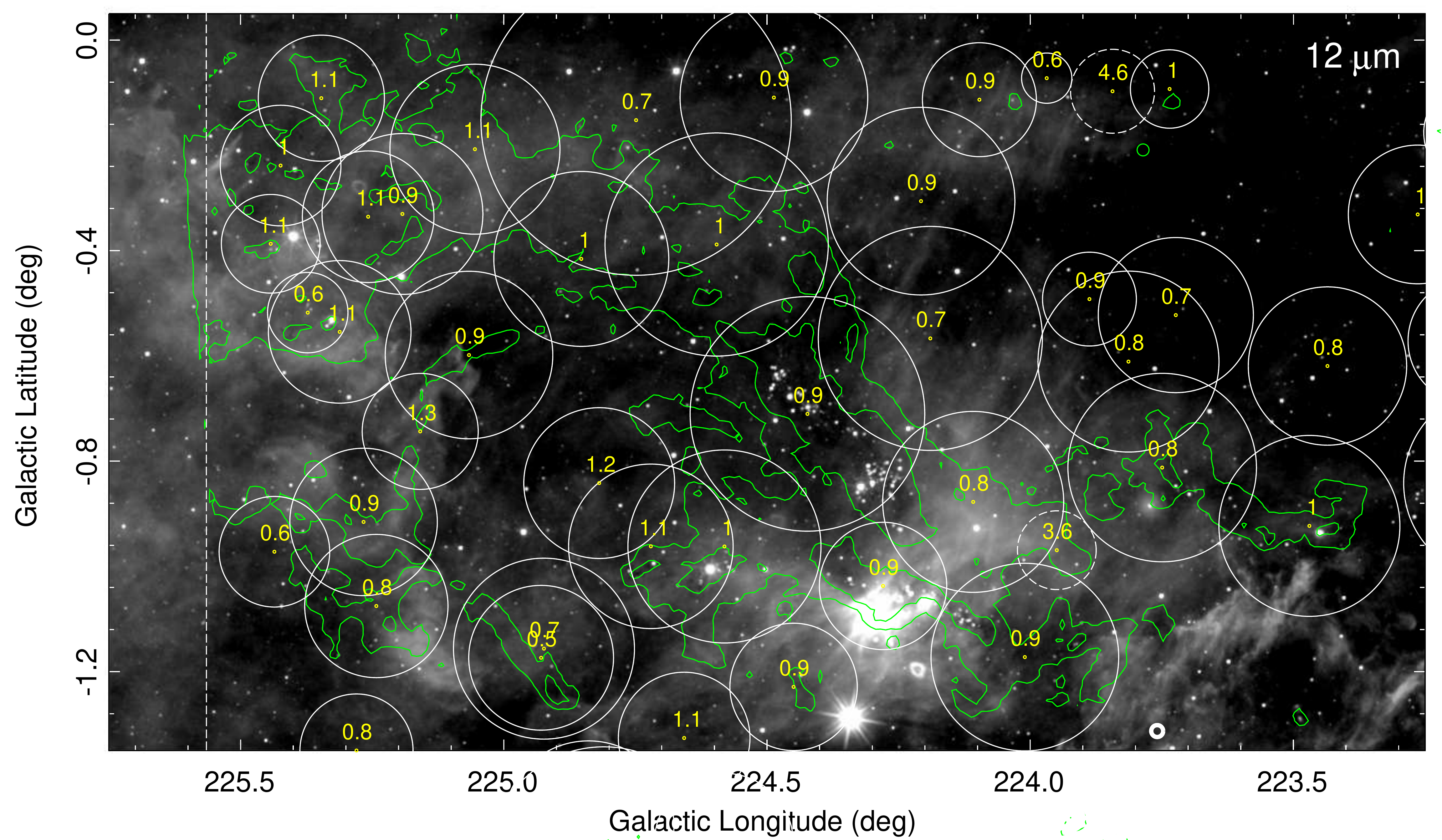}
\caption{The WISE 12 $\mu$m image with the positions and sizes of the CO clumps from the \citet{elia2013} catalog indicated with white circles. The yellow labels inside the circles provide the clumps' kinematic distances in kpc. Two clumps with distances larger than 3 kpc are shown as white dashed circles. The H$_{2}$ column density contour levels are (4, 8) $\cdot$ 10$^{21}$ cm$^{-2}$. The white dashed vertical line shows the extent of the CO map.  \label{f:clumps}}
\end{figure*}


\section{The Catalog Matching}
\label{s:matching}

We positionally cross-matched the {\it Spitzer}/GLIMPSE360 catalog (already matched to 2MASS) with the AllWISE, AKARI, MSX, and Hi-GAL catalogs to provide longer wavelength data that, in combination with near- and mid-infrared (mid-IR) data, would improve the YSO identification process, including the color-color/color-magnitude selection and the SED fitting. As a result of the much higher sensitivity of the GLIMPSE360 survey in comparison to other currently available surveys, only a small fraction of GLIMPSE360 sources have matches in other catalogs (see Table~\ref{t:stats}).

\noindent {\bf AllWISE:} To match the GLIMPSE360 catalog with the AllWISE catalog, we adopted a matching radius of 1$''$ and selected the closest matches. We found 80,171 matches; we used the quality flags provided in the AllWISE catalog to exclude unreliable sources. In our color-color/color-magnitude YSO selection process, we use the 12 and 22 $\mu$m ($w3$ and $w4$) photometry with a flag {\it ph\_qual} of `A' or  `B', excluding low signal-to-noise photometry and upper limits.  The 12 $\mu$m and 22 $\mu$m flux upper limits will be used for the SED fitting if both the $w1$ and $w2$ fluxes have a signal-to-noise ratio larger or equal 10 ({\it ph\_qual} = `A').  We remove fluxes flagged as likely spurious detections: Diffraction spikes ($cc\_flag$=`D'), Persistences (`P'), Halos (`H'), and Optical ghosts (`O'), and those flagged as extended sources with $ext\_flg$ $>$ 1. 

\citet{koenig2014} found that the signal-to-noise ($w?snr$) and profile-fit reduced chi-squared ($w?rchi2$) parameters have the strongest discriminatory effect in suppressing spurious source contamination in WISE Bands 3 and 4.  However, the cuts in the {\it w?snr} vs. $w?rchi2$ space designed to remove spurious sources from the catalog, also remove a large number of real sources from Bands 3 and 4. To increase the rate of the real source retrieval in Band 3 (up to $\sim$60\%), \citet{koenig2014} introduced additional criteria to recover the low signal-to-noise sources. For Bands 3, \citet{koenig2014} proposed the following set of criteria to remove the vast majority of spurious 12 $\mu$m sources from the AllWISE catalog: $w3snr \geq 5$ and $[(w3rchi2 < (w3snr - 8)/8)$ or $(0.45 < w3rchi2 < 1.15$)].  For WISE Band 4, the criteria are: non-null $w4sigmpro$ (i.e., {\it ph\_qual} of `A', `B', or `C') and $w4rchi2 < (2 \times w4snr - 20)/10.$
Only 8\% of all AllWISE sources with valid fluxes in CMa--$l224$ fulfill these criteria in Band 3 and $\sim$0.2\% in Band 4. We do not apply these criteria since they remove a large fraction of real sources. We will identify spurious AllWISE detections during the visual inspection. 

\noindent {\bf AKARI:} To search for counterparts to GLIMPSE360 sources in the AKARI catalog, we used a matching radius of 1$''$, which roughly corresponds to the mean AKARI survey's positional accuracy \citep{akaridoc}. We found 60 matches; 58 out of 60 GLIMPSE360 sources with AKARI matches have both 3.6 and 4.5 $\mu$m valid fluxes.  Out of these 58 sources, 91\% have valid 9 $\mu$m flux, 22\% have a valid 18 $\mu$m flux, and only 14\% have a valid both 9 and 18 $\mu$m fluxes.

\noindent {\bf MSX:} We inspected the flux quality flags ($q$) and selected fluxes with $q$ = 4 (excellent),  3 (good), and 2 (fair/poor, $\sim$15\%; SNR$>$5). The positional uncertainty estimated for sources with $q$ $\geq$ 2 in A band ($q\_a$) is $\sim$1$\rlap.{''}$3~\footnote{http://irsa.ipac.caltech.edu/data/MSX/docs/msxpsc2.3\_explguide.pdf}. Therefore, we used the matching radius of 1$\rlap.{''}$3 to search for counterparts of the GLIMPSE360 sources in the MSX catalog. For sources with fluxes at both 3.6 and 4.5 $\mu$m bands, we found (26, 3, 1, 1) matches in (A, C, D, E) band.

\noindent {\bf Hi-GAL (Elia et al. 2013):} We matched the GLIMPSE360 catalog to the \citet{elia2013} catalog and the Full {\it Herschel} catalog (see Section~\ref{s:higal}) using a matching radius of 5$''$.  There is a significant difference in angular resolutions between the {\it Spitzer} and {\it Herschel} bands, the SPIRE bands (250--500 $\mu$m) in particular (see Table~\ref{t:data}), thus the matching radius we selected is very conservative. However, we do not expect sources detected with {\it Spitzer} to be detected with SPIRE, but not with PACS.  Thus, we selected the matching radius based on the highest resolution {\it Herschel} PACS band, i.e. 70 $\mu$m ($\sim$5$''$).  

We found 72 (206) matches for the GLIMPSE360 sources in the \citet{elia2013} (Full {\it Herschel}) catalog with 40 out of 72 (75 out of 206) having a valid 70 $\mu$m flux. There are 26 out of 72 (111 out of out 206) SPIRE-only sources in the  \citet{elia2013} (Full {\it Herschel}) catalog, which will not be used in our analysis.

The matching results are summarized in Table~\ref{t:stats}. We verify the validity of the matches during the visual inspection of the images of pre-selected YSO candidates.


\section{Distance Determination}
\label{s:distance}

As the volume density in the outer Galaxy is lower, there is less line of sight confusion and it is easier to associate infrared emission with molecular clouds identified in spectral line surveys than in the inner Galaxy. The kinematic distances to these objects are more reliable as there is no distance ambiguity for lines of sight in quadrants II and III. The distance to YSOs can be estimated by association either with CO emission or H\,{\sc i} self-absorption from the cold H\,{\sc i} envelopes of molecular clouds. Since the CO data at relatively high resolution are available for the region we study, we will use the CO emission for distance determination. 

We determine kinematic distances to {\it Spitzer} sources by associating them with CO clumps using the technique described in detail in \citet{elia2013}. They used the $^{12}$CO ({\it J}=1--0) data from the NANTEN CO Galactic Plane Survey (\citealt{mizuno2004}; improved with respect to the original NANTEN survey mentioned earlier, \citealt{kim2004}) to search for spatially and kinematically coherent overdense structures that they call ``CO clumps''.  For the cloud decomposition, they used the CLUMPFIND \citep{williams1994} decomposition algorithm in the CLOUDPROPS package \citep{rosolowsky2006,rosolowsky2011}. This analysis provides morphological and physical properties of the clumps, e.g., velocity dispersion, distance, radius, mass, virial mass, and luminosity. \citet{elia2013} estimated heliocentric distances to CO clumps using the rotation curve of \citet[$R_{0}$ = 8.3 kpc, $\Theta_{0}$ = 239 km s$^{-1}$]{brunthaler2011}. 

\begin{figure}
\centering
\includegraphics[width=0.45\textwidth]{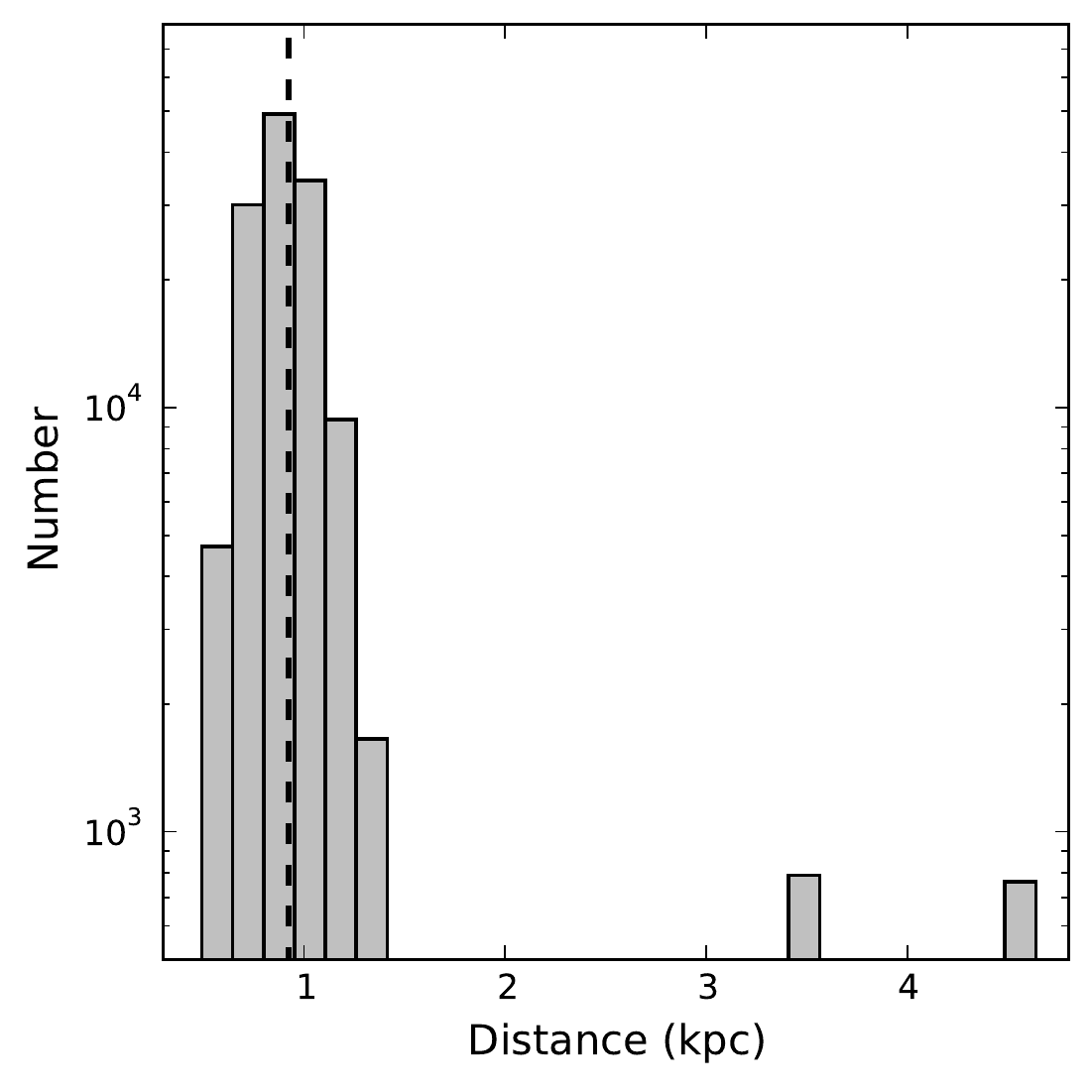}
\caption{Distribution of kinematic distances estimated using the CO data as described in Section~\ref{s:distance}. The dashed vertical line shows a median distance of 0.92 kpc calculated using distances $<$2~kpc. \label{f:dist}} 
\end{figure}

In the CO line of sight containing the {\it Spitzer} source, we consider the spectral channels assigned by CLOUDPROPS/CLUMPFIND to one or more clumps, and associate with the source the distance of the clump (if any) having the closest centroid (see Fig.~\ref{f:clumps}). The advantages of such strategy are discussed in \citet{elia2013}.

We assigned distances to $\sim$62\% of all {\it Spitzer} sources in CMa--$l224$; the distribution of distances is shown in Fig.~\ref{f:dist}. About 99\% of sources have distances between 0.5 and 1.3 kpc with the median distance of 0.92 kpc. This result is consistent with the previous distance determinations based on the photometric and spectroscopic data for OB stars in CMa OB1 region (e.g., \citealt{claria1974b}; \citealt{kaltcheva2000}, the early type stars in CMa R1 \citep{shevchenko1999}, or by comparing the observed density of foreground stars detected by 2MASS with the prediction of galactic models \citep{lombardi2011}.

A small fraction of sources at distances larger than 3 kpc are likely associated with the Perseus spiral arm. These sources are associated with two small CO molecular clumps at ({\it l}, {\it b}) $\sim$ (223$\rlap.^{\circ}$9, -0$\rlap.^{\circ}$9) and (223$\rlap.^{\circ}$8, -0$\rlap.^{\circ}$1), see Fig.~\ref{f:clumps}.


\section{Initial Selection of YSO Candidates}
\label{s:2massspitzer}

\begin{figure*}
\centering
\includegraphics[width=0.45\textwidth]{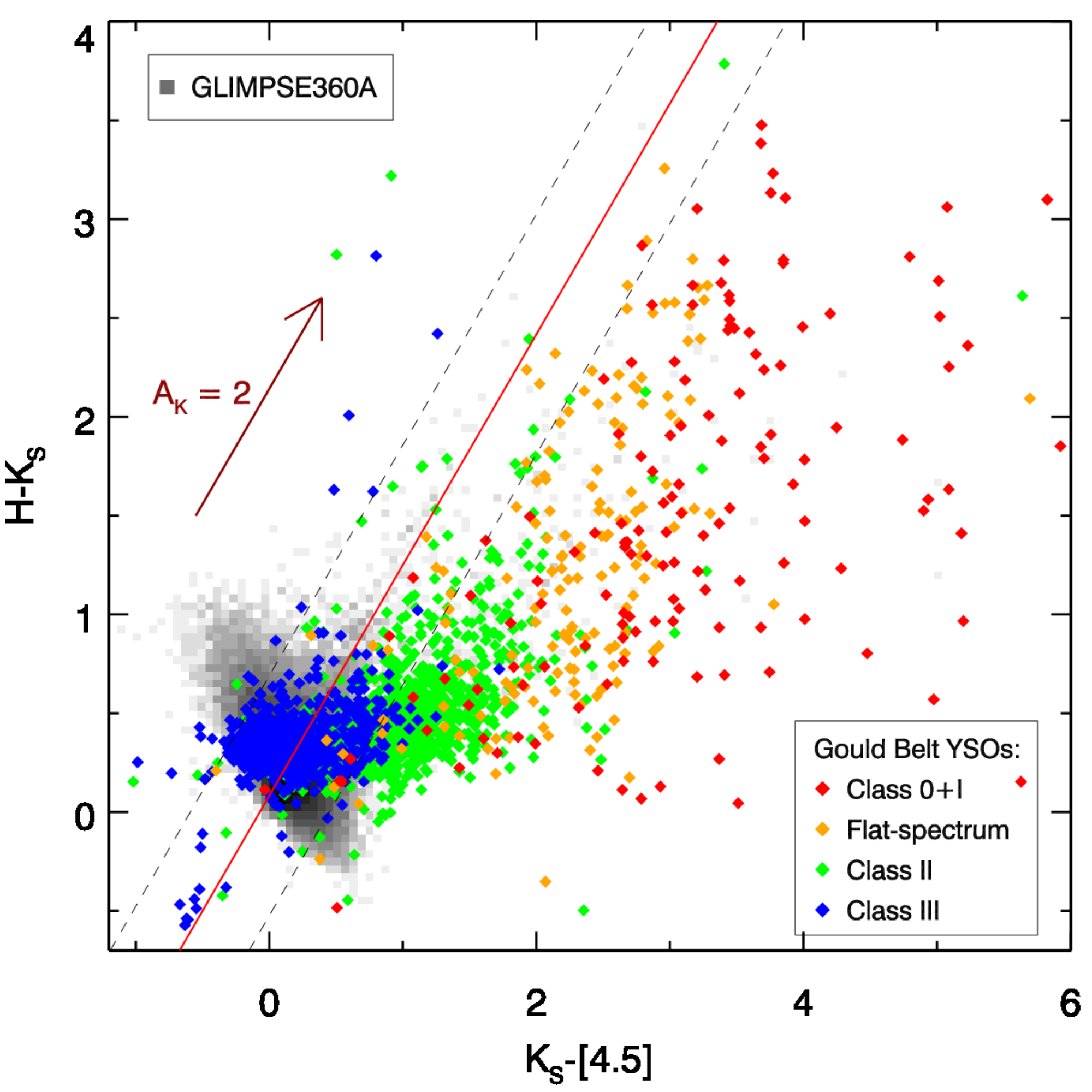}
\includegraphics[width=0.45\textwidth]{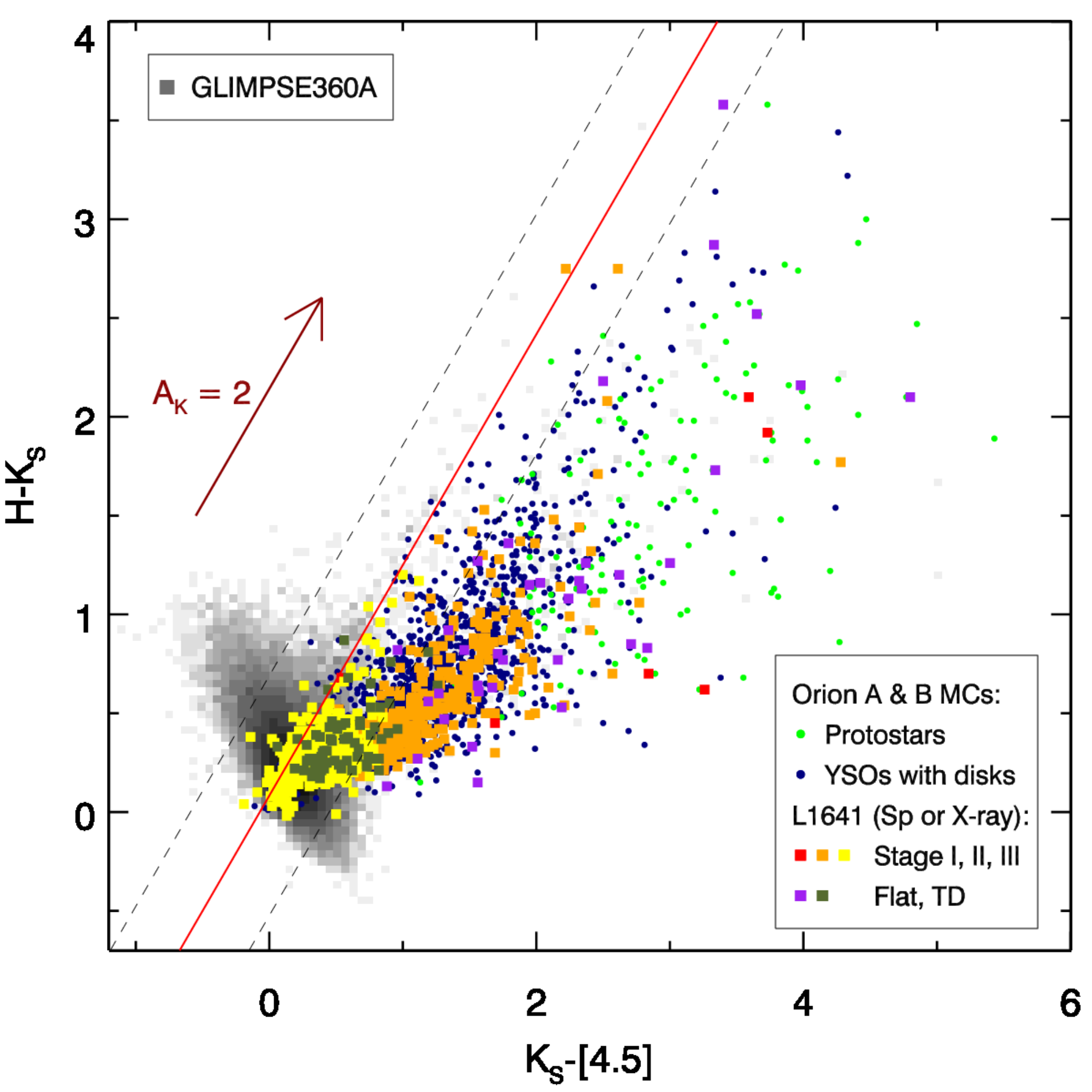}
\includegraphics[width=0.45\textwidth]{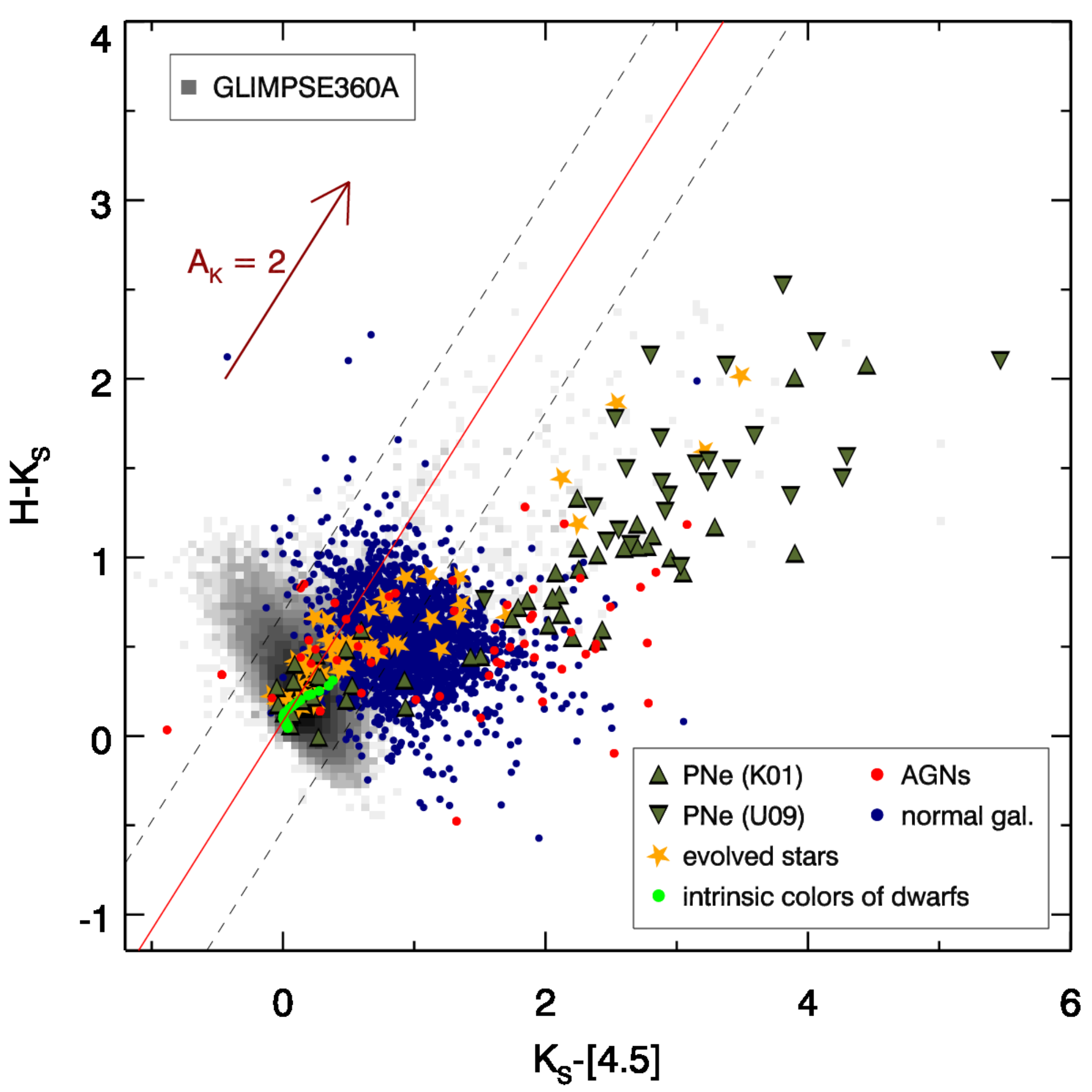}
\includegraphics[width=0.45\textwidth]{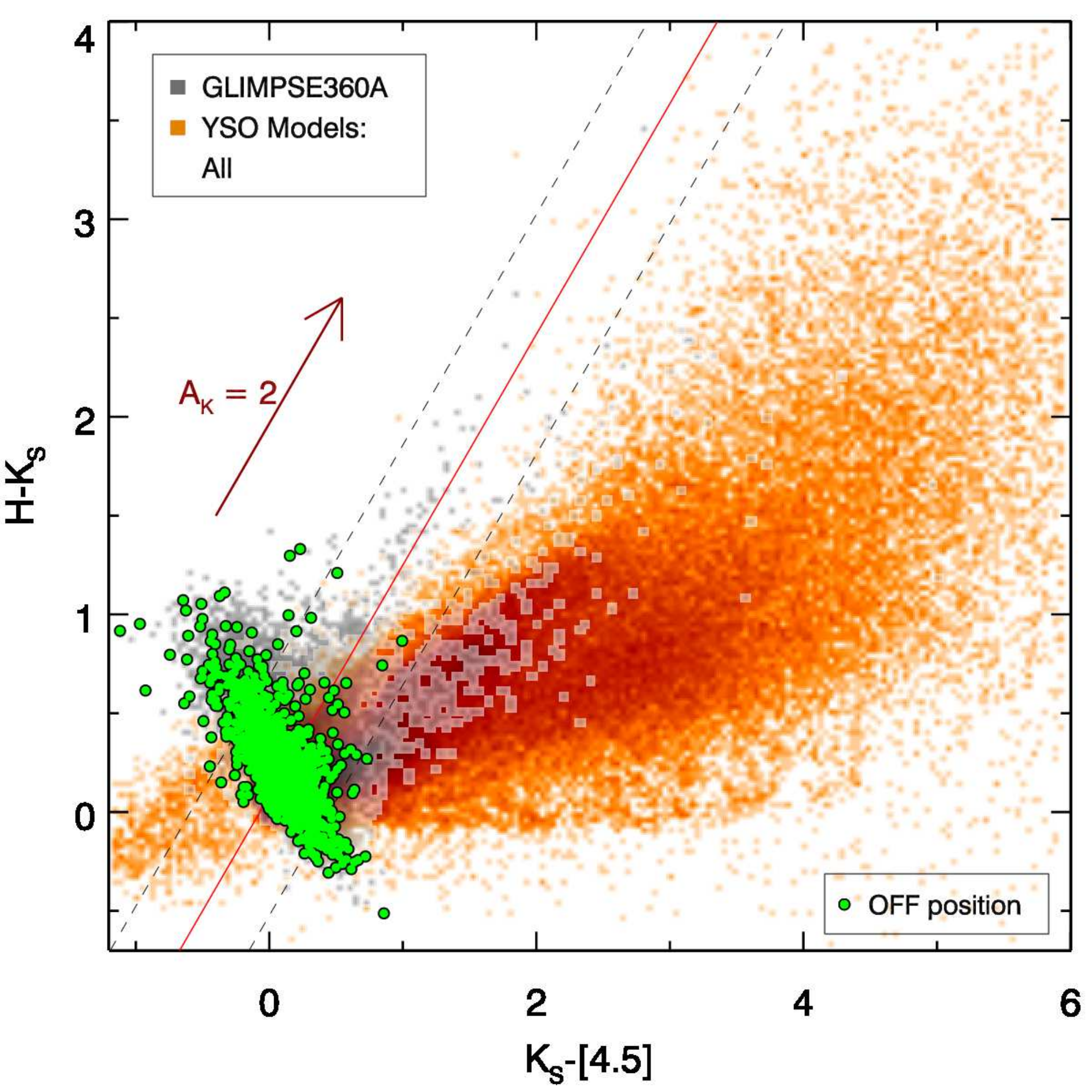}
\caption{Distribution of the GLIMPSE360 catalog sources (shown in grey as a Hess diagram) in the $H - K_{\rm s}$ vs. $K_{\rm s} - [4.5]$ CCD with respect to known Galactic YSOs (\citealt{fang2013}: L1641) and YSO candidates (\citealt{megeath2012}: Orion A and B molecular clouds; \citealt{dunham2015}: 18 nearby molecular clouds from the Gould Belt; {\it top panel}), evolved stars and extragalactic sources (\citealt{reiter2015,kochanek2012}; {\it bottom left}), and YSO 2D radiative transfer models (\citealt{robitaille2006}; {\it bottom right}). In the lower left panel, we also plot the locus of the intrinsic colors of dwarf stars from \citet{pecaut2013}; we use the $K_{\rm s}-w2$ color as a proxy of the $K_{\rm s}-[4.5]$ color. The catalogs from literature are described in Appendix A.  The green circles in the bottom right panel indicate sources detected in the area selected as a representation of the background (``off-region''; see Fig.~\ref{f:3color}). The red solid line is parallel to the reddening vector and originates at the position of the K0 dwarf.  Dashed lines represent 1$\sigma$ uncertainty in color. \label{f:HKK45}} 
\end{figure*}

Our general strategy on selecting YSO candidates based on the 2MASS and {\it Spitzer} data is to first identify IR-excess sources, and then apply a series of filters to remove various populations of contaminating sources. These include unresolved extragalactic objects detected in the outer Galactic plane in the deep GLIMPSE360 survey, unresolved planetary nebulae (PNe), and evolved stars. The subsequent analysis includes the SED fitting with YSO models and the visual inspection of multiwavelength images.

We identify regions in the CCD and CMD space that should be occupied predominantly by YSOs by comparing the distribution of the GLIMPSE360 catalog sources in the CCDs and CMDs to the distribution of non-YSOs from literature (evolved stars, AGNs, and normal galaxies: \citealt{reiter2015}; \citealt{kochanek2012}; \citealt{kimeswenger2001}; \citealt{urquhart2009}), previously identified Galactic YSOs \citep{fang2013} and YSO candidates \citep{megeath2012,dunham2015}, as well as 2D YSO radiative transfer models \citep{robitaille2006}. All the catalogs from literature are described in Appendix~A.  We assume the \citet{flaherty2007} interstellar extinction law for the {\it Spitzer} IRAC bands based on the data for five nearby star-forming regions.  The estimated extinction law for WISE bands comes from \citet{koenig2014}.

Out of 211,900 {\it Spitzer} GLIMPSE360 catalog sources in CMa--$l224$, 128,317 ($\sim$60.6\%) do not have counterparts in any other catalog used in this paper within the selected matching radii. With only one or two valid fluxes, these sources will not fulfill any of the YSO selection criteria described below and thus are removed from the analysis. A subset of these sources with both 3.6 and 4.5 $\mu$m catalog data (119,352) and additional {\it Spitzer} sources that have AllWISE matches but no valid 12 and 22 $\mu$m photometry (not in the catalog or of poor quality and removed) and no matches in other catalogs, are discussed separately in Section~\ref{s:gl360only}. Here, we analyze all the sources with valid fluxes in at least three shorter wavelength bands, i.e. excluding {\it Herschel} data.

\subsection{Interstellar Extinction toward CMa--$l224$}
\label{s:extinction}

In general, the interstellar extinction in the outer Galaxy is lower than within the solar circle; however, it is higher than average toward the CMa star formation region \citep{lombardi2011}. For this reason, we investigate the amount of extinction in more detail toward CMa--$l224$ using the N(H$_{2}$) map from \citet{elia2013}.

Based on the investigation of the extinction law in the Perseus molecular cloud, \citet{foster2013} found that the extinction law changes from the ``diffuse'' value of R$_{V}$ $\sim$ 3 to the ``dense cloud'' R$_{\rm V}$ $\sim$ 5. We adopt the measured N(H$_{2}$)/E(B-V) ratio of  2.9 $\times$ 10$^{21}$ cm$^{-2}$ \citep{bohlin1978} and assume R$_{\rm V}$ = A$_{\rm V}$/E(B-V) of 4.0. We calculate A$_{\rm V}$  (N(H$_{2}$)/(7.25 $\times$ 10$^{20}$) mag) for each pixel in the image;  each {\it Spitzer} source covered by the N(H$_{2}$) image is assigned A$_{\rm V}$ of the pixel it is associated with. For the remaining sources in the small region at the eastern boundary of CMa--$l224$, we use the maximum value of A$_{\rm V}$ in the ``off-region'' (see below). To calculate A$_{\rm K}$ from A$_{\rm V}$, we use a relation A$_{\rm K}$/A$_{\rm V}$ = 0.114 (\citealt{weingartner2001}, R$_{\rm V}$=4.0). We use the \citet{flaherty2007} and \citet{koenig2014} empirical extinction laws to determine extinction corrections for {\it Spitzer} and WISE bands, respectively. The resulting A$_{\rm V}$ ranges from $\sim$1.6 mag to $\sim$105 mag; $\sim$87\% sources are associated with A$_{\rm V}$ $<$ 5 mag and 0.2\% with A$_{\rm V}$ $>$ 40 mag. 

\begin{figure*}
\centering
\includegraphics[width=0.45\textwidth]{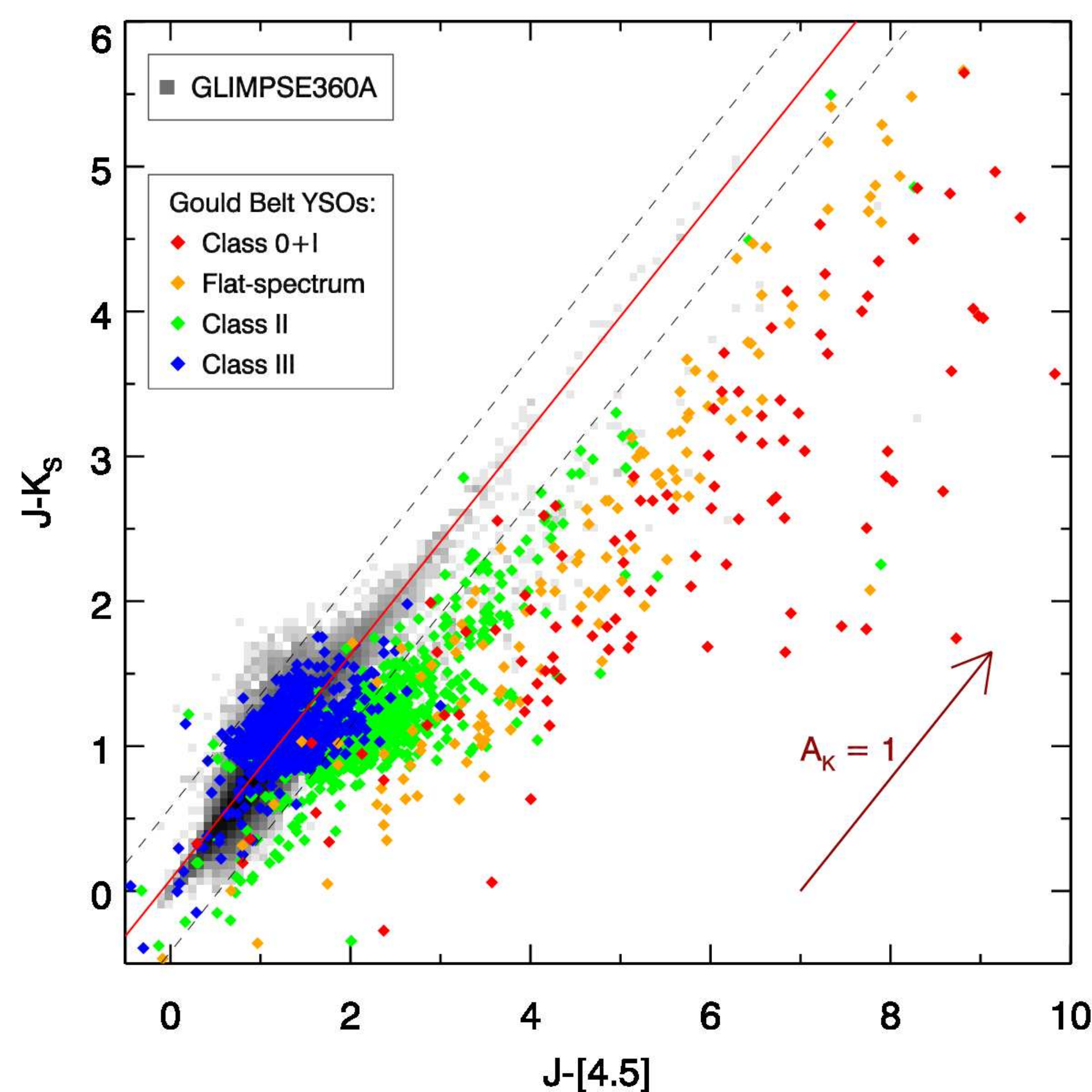}
\includegraphics[width=0.45\textwidth]{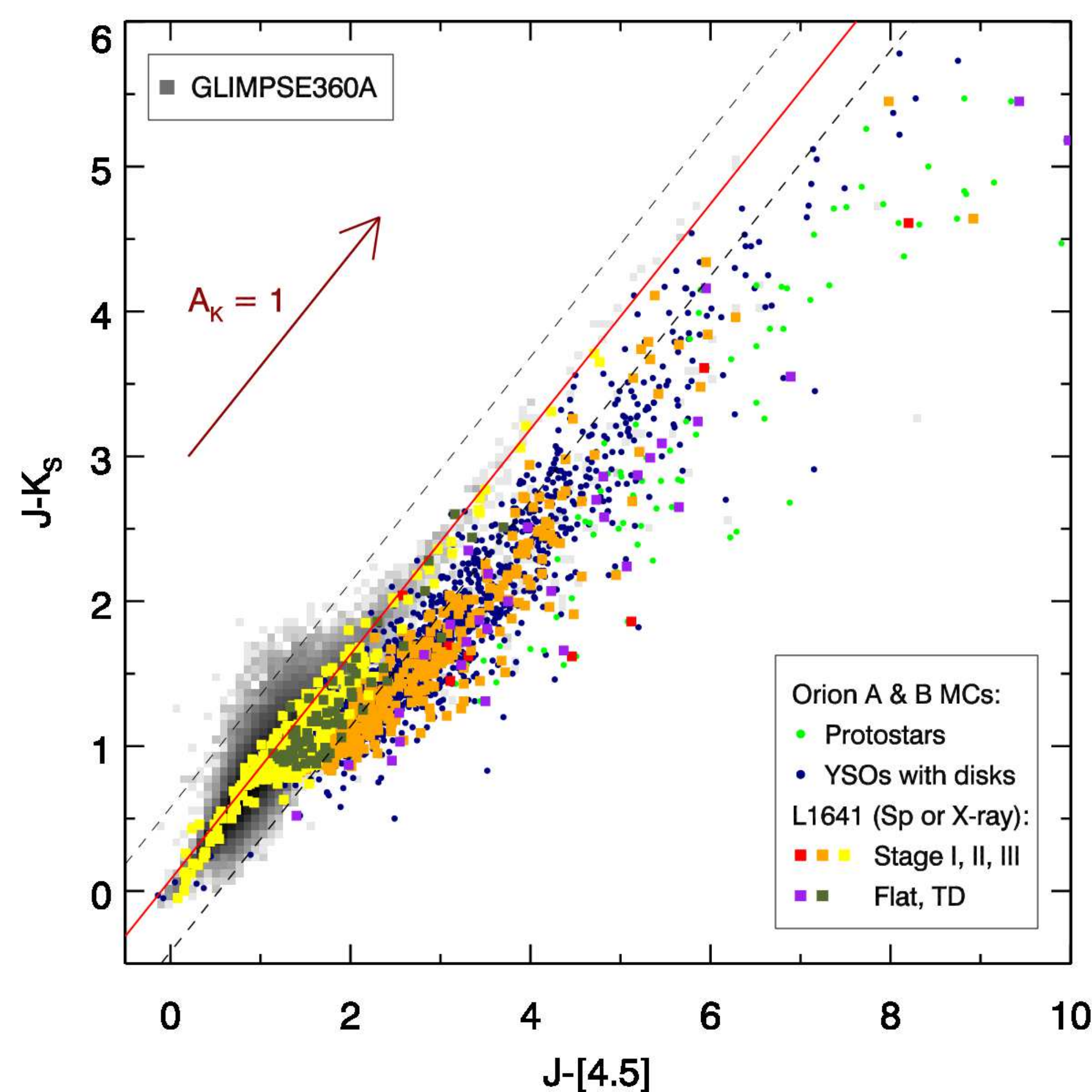}
\includegraphics[width=0.45\textwidth]{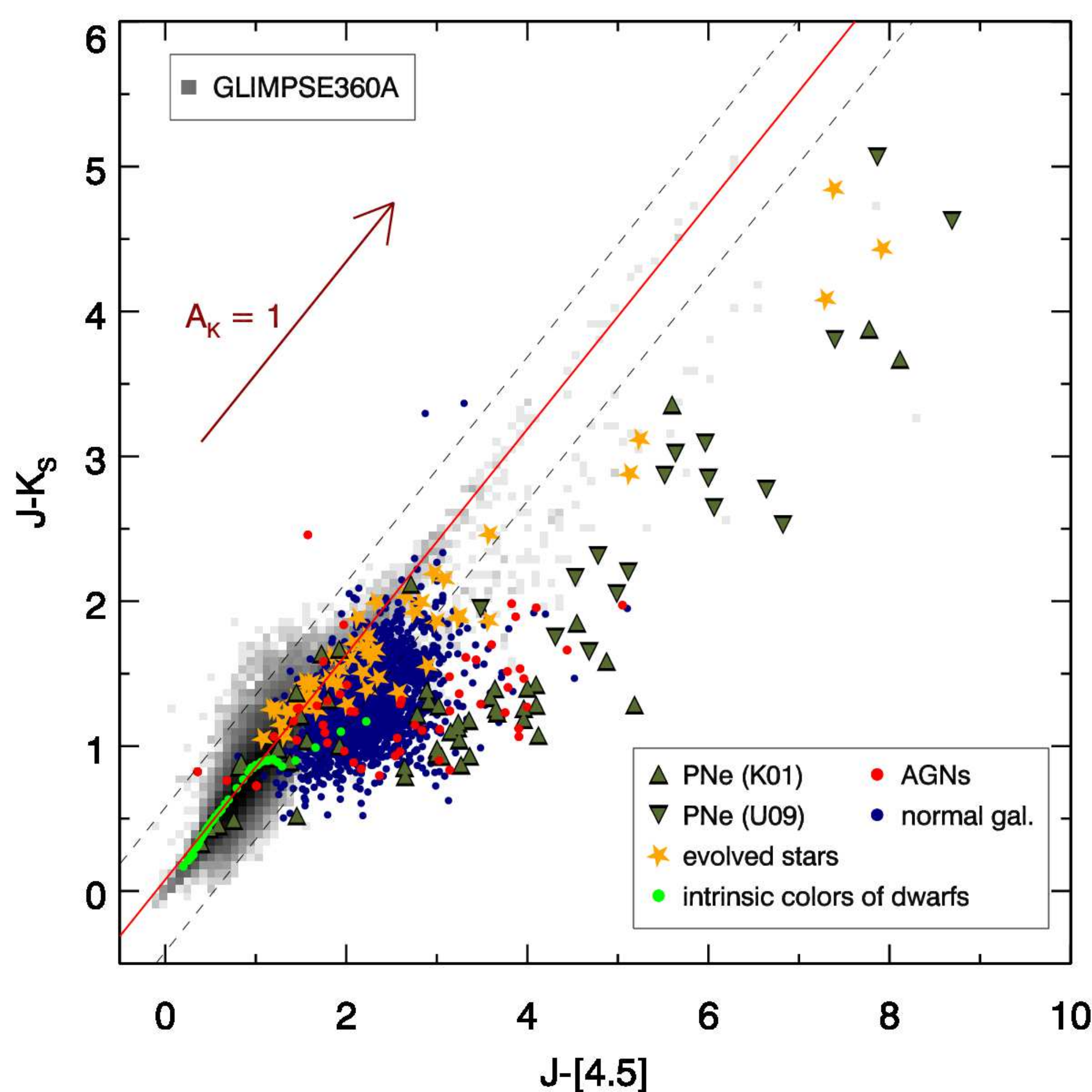}
\includegraphics[width=0.45\textwidth]{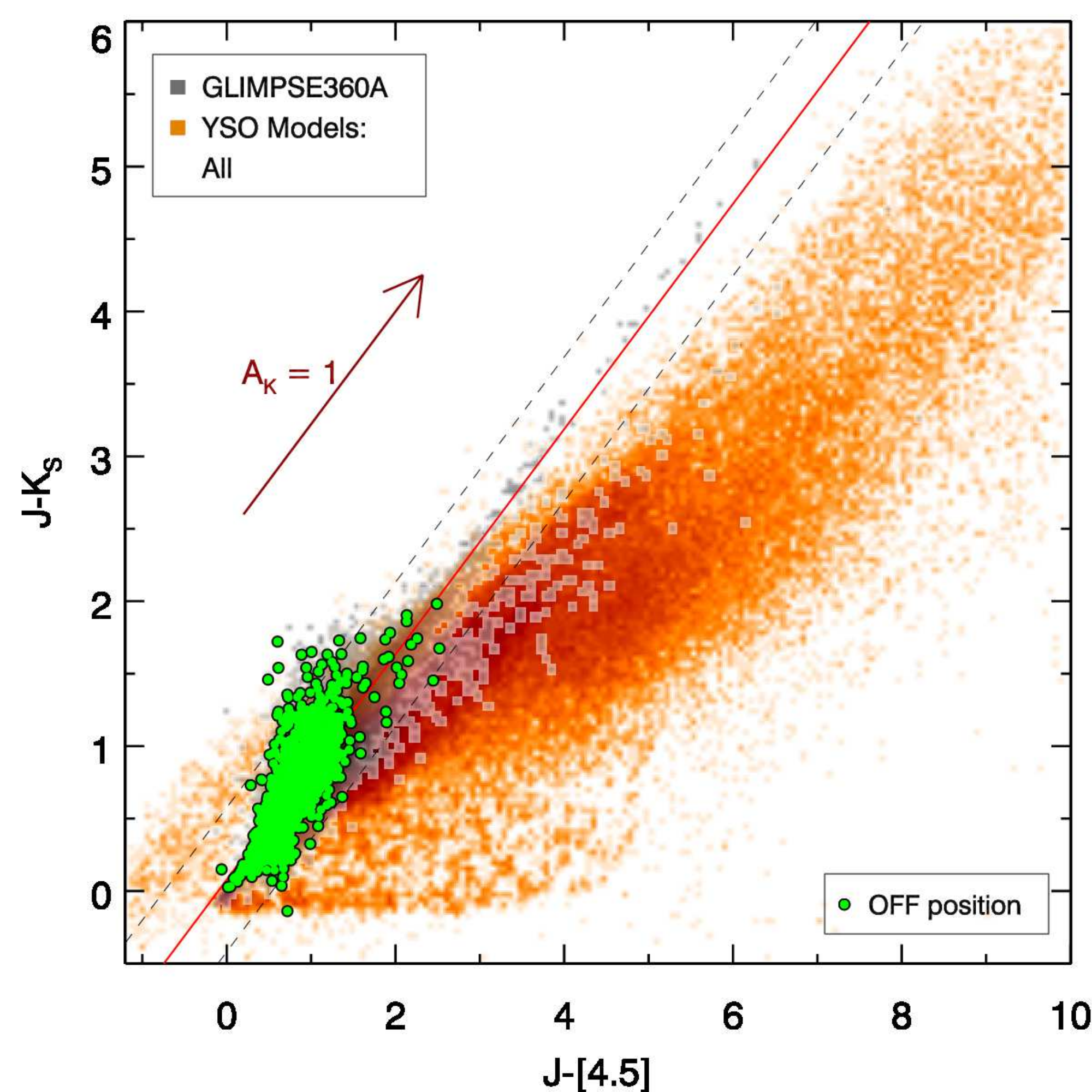}
\caption{The same set of plots as in Fig.~\ref{f:HKK45}, but for the  $J-K_{\rm s}$ vs. $J-[4.5]$ CCD. \label{f:JKJ45}} 
\end{figure*}

These calculations are useful, however, they only provide upper limits for A$_{\rm V}$. This is due to the fact that the {\it Herschel} emission used to construct the N(H$_{2}$) map can include envelope emission from the protostars, which is not part of the interstellar A$_{\rm V}$. Also, we do not know if the sources are on the near or far side of the star-forming region; some of the {\it Herschel} emission may originate from behind the cloud.

We estimated the maximum A$_{\rm V}$ of 4.1 mag in the ``off-region'' indicated with a circle in Fig.~\ref{f:3color} to represent the interstellar extinction up to the near side of the star forming cloud. This value is consistent with the high end of the A$_{\rm K}$ extinction range estimated by \citet{lombardi2011} toward the CMa star formation region based on the 2MASS data. 

\begin{figure*}
\centering
\includegraphics[width=0.45\textwidth]{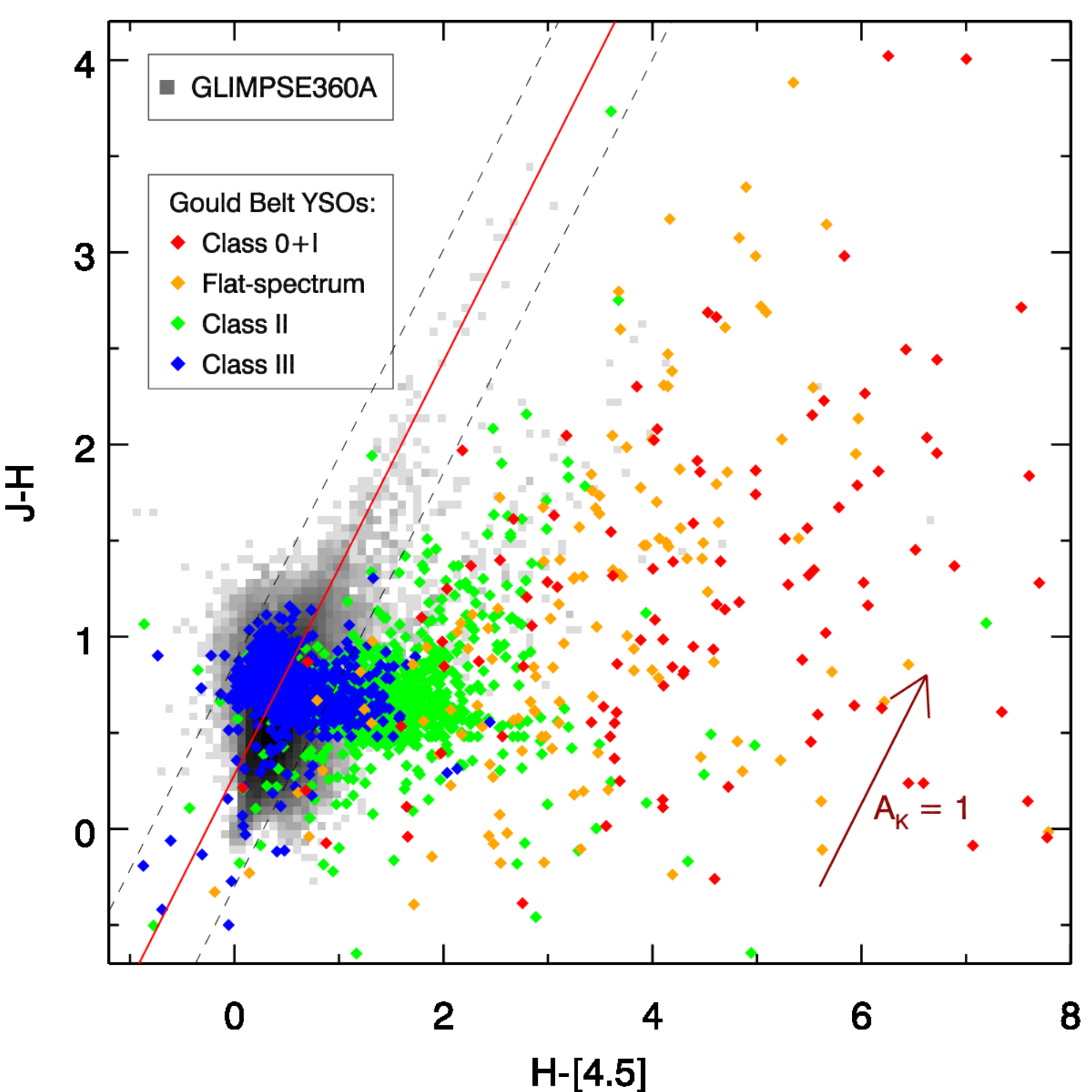}
\includegraphics[width=0.45\textwidth]{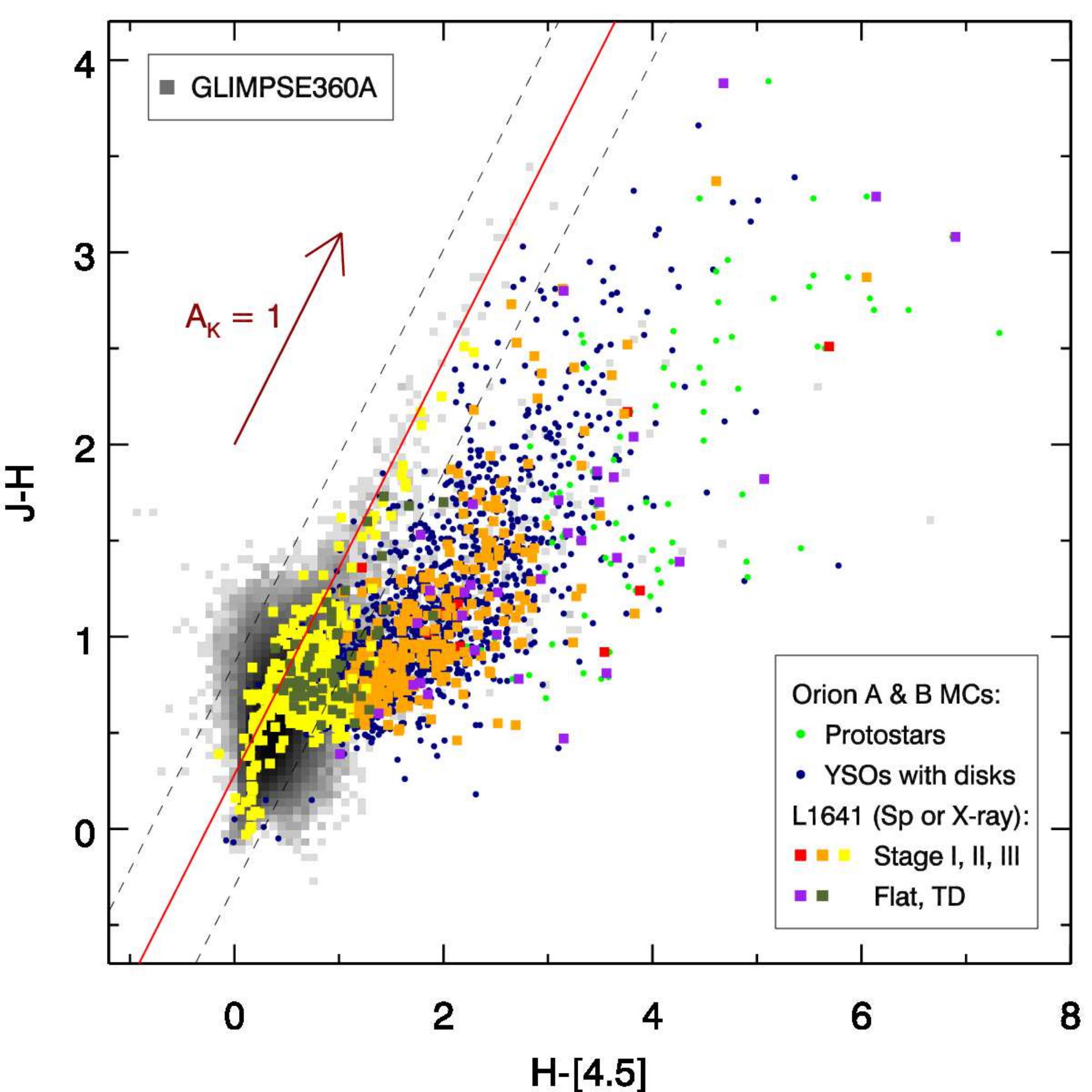}
\includegraphics[width=0.45\textwidth]{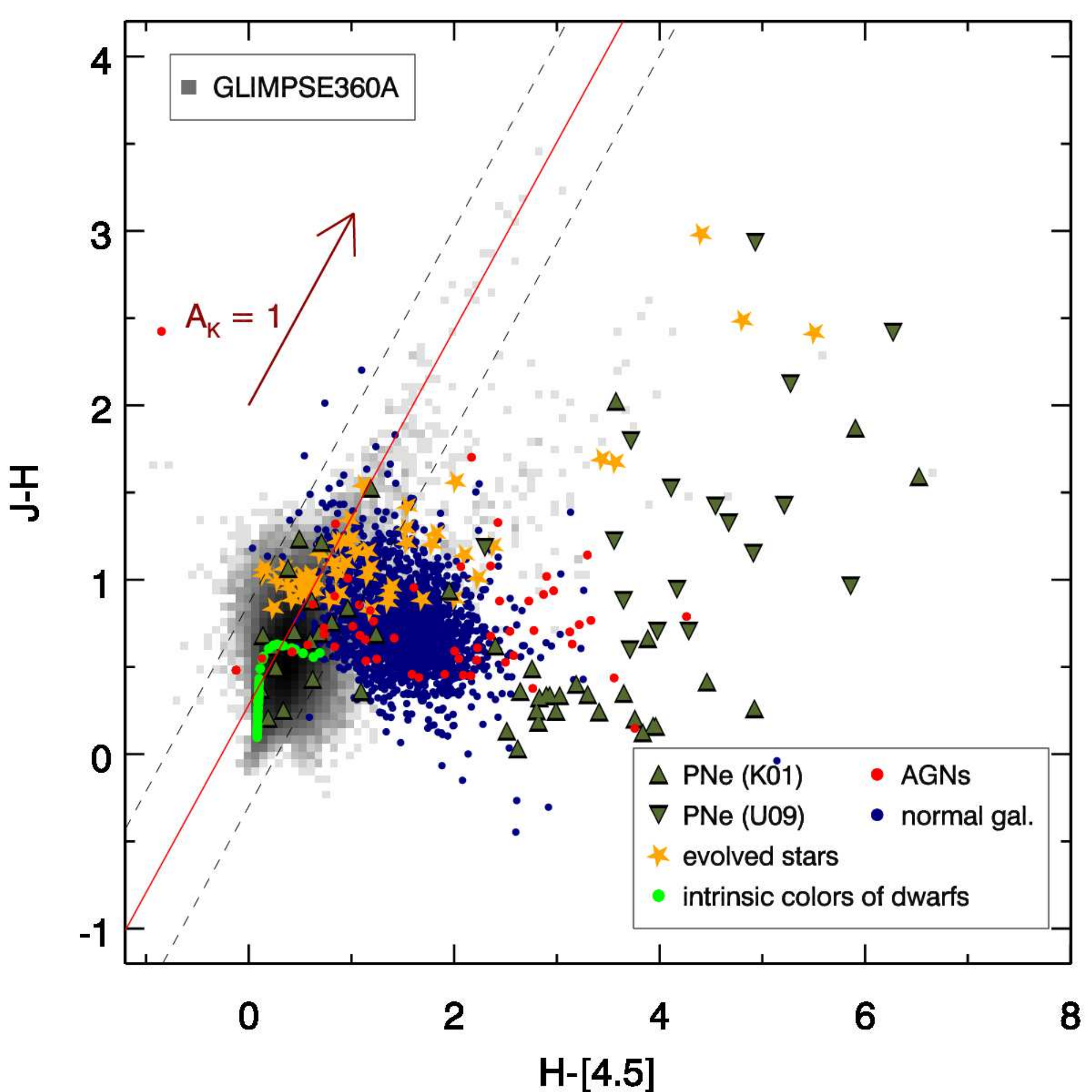}
\includegraphics[width=0.45\textwidth]{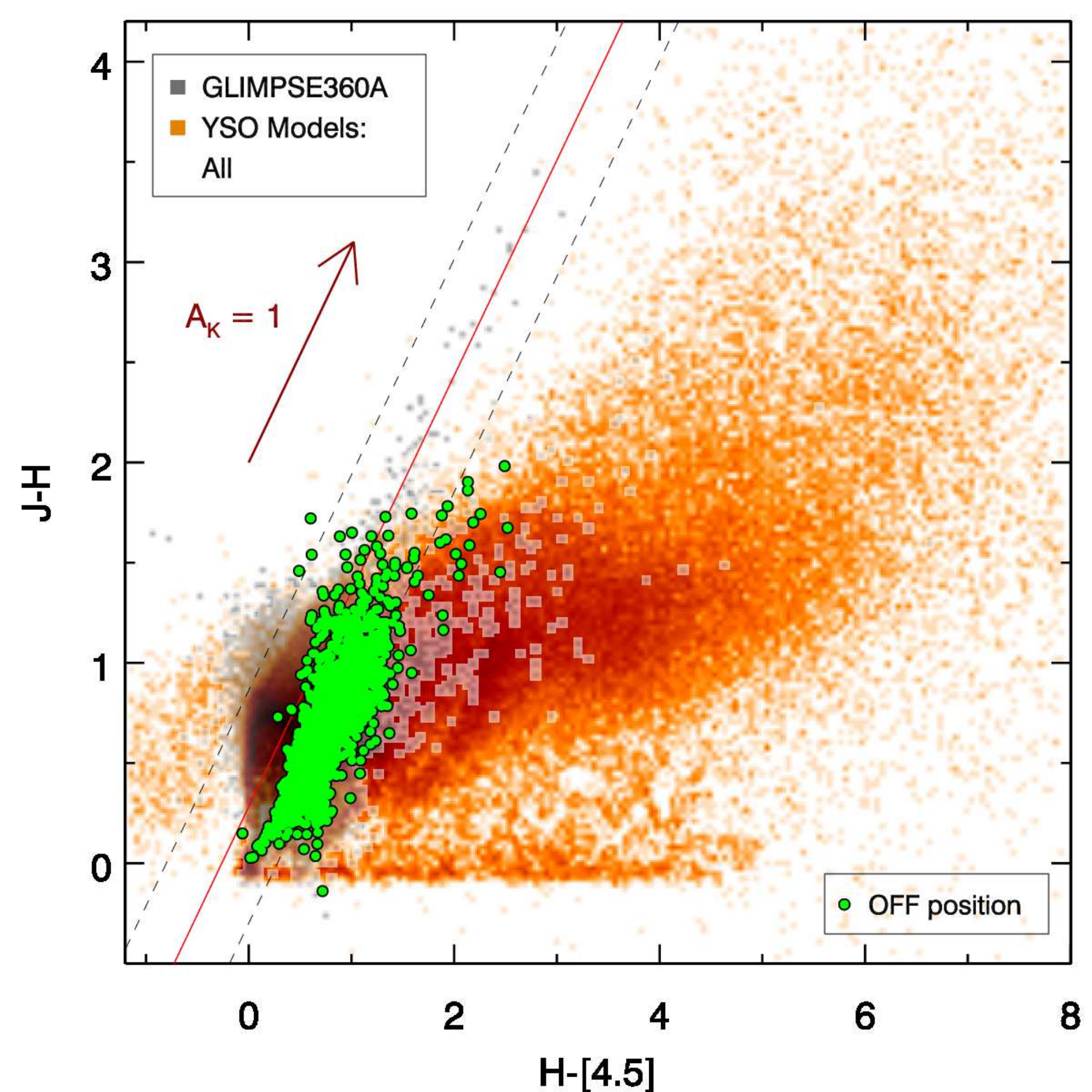}
\caption{The same set of plots as in Fig.~\ref{f:HKK45}, but for the $J-H$ vs. $H-[4.5]$ CCD. \label{f:JHH45}} 
\end{figure*}

\subsection{Color-Color Cuts: 2MASS and {\it Spitzer} Photometry}

We select the IR-excess sources based on a set of three CCDs that combine two 2MASS near-IR and the GLIMPSE360 4.5 $\mu$m  data: $H-K_{\rm s}$ vs. $K_{\rm s}-[4.5]$, $J-K_{\rm s}$ vs. $J-[4.5]$, and $J-H$ vs. $H-[4.5]$ (Fig.~\ref{f:HKK45}--\ref{f:JHH45}). A combination of these CCDs will allow us to include all {\it Spitzer} sources with IR-excess that have at least two detections in the 2MASS bands. 

Some of the sources in the $H-K_{\rm s}$ vs. $K_{\rm s}-[4.5]$ CCD have diagonal distribution in a direction roughly orthogonal to the reddening vector. We found that this feature corresponds to faint sources with large uncertainties. Unusually large photometric errors (especially at {\it K$_{\rm s}$} band), randomly shift the true color of the sources along the anti-diagonal in the plot. These  sources could be either faint background galaxies or faint halo stars, but they also could be low luminosity YSOs.  For our subsequent analysis, we use 2MASS photometry with uncertainties smaller than 0.2 mag to remove the least reliable photometry.

The red solid line parallel to the reddening vector in the 2MASS+{\it Spitzer} CCDs shown in Fig.~\ref{f:HKK45}-\ref{f:JHH45} originates at the position of the K0 dwarf \citep{pecaut2013}. The dashed line on either side of the red line in Fig.~\ref{f:HKK45}-\ref{f:JHH45} represents a 1$\sigma$ uncertainty in color; we adopt the maximum uncertainty in the colors displayed in each CCD. The sources located between the two dashed lines are expected to be reddened stellar photospheres (see e.g., \citealt{winston2007}; \citealt{willis2013}). For further analysis, we select sources rightward from the dashed line on the right in at least one of the three 2MASS+{\it Spitzer} CCDs:

\normalsize
\renewcommand{\arraystretch}{0.85}
\begin{eqnarray}
\label{e:spit2mass}
\left \lbrace
\begin{array}{r}
H-K_{\rm s} <  1.1674 \cdot (K_{\rm s}-[4.5]) - 0.5240 \; \;{\rm or} \\
 \\
  J-K_{\rm s} < 0.7774  \cdot (J-[4.5]) - 0.4196 \;\;{\rm or}   \\
  \\
  J-H < 1.0758 \cdot (H-[4.5]) - 0.2997 \; \;\; \;\;\\
\end{array}     \right \rbrace 
\end{eqnarray}
\normalsize
\renewcommand{\arraystretch}{1}

Out of $\sim$68,000 sources, 3144 fulfill the above criteria. We investigate how this number changes when we apply the selection criteria to the dereddened photometry. We deredden the photometry in two ways: (1) for each source, we apply the extinction corrections derived for the associated N(H$_{2}$) pixel;  (2) we calculate the extinction corrections using the maximum value of N(H$_{2}$)  in the ``off-region'' and apply it to all the sources. As discussed in Section~\ref{s:extinction}, the extinction corrections estimated based on N(H$_{2}$) are upper limits. For the  
$H-K_{\rm s}$ vs. $K_{\rm s}-[4.5]$ color-color cut, only six sources are removed when the dereddened photometry is used (both methods). The results remain the same for the  $J-K_{\rm s}$ vs. $J-[4.5]$ color-color cut. For the $J-H$ vs. $H-[4.5]$ cut, three more sources are selected for the photometry dereddened using the first method, and two using the second method with two sources overlapping. All but two of the sources removed/added to the list have [3.6] $>$ 14 mag and will be removed from the list in the further analysis. Since the difference between the results obtained using the observed and dereddened (with the extinction corrections being upper limits) photometry is negligible, we use the former in our analysis.  

This result is not unexpected since Eq.~\ref{e:spit2mass} is based on limits that follow the reddening path and the two dereddening methods should provide the same results. The difference of a few sources is likely due to computational round-off errors. 

\begin{figure*}
\includegraphics[width=0.33\textwidth]{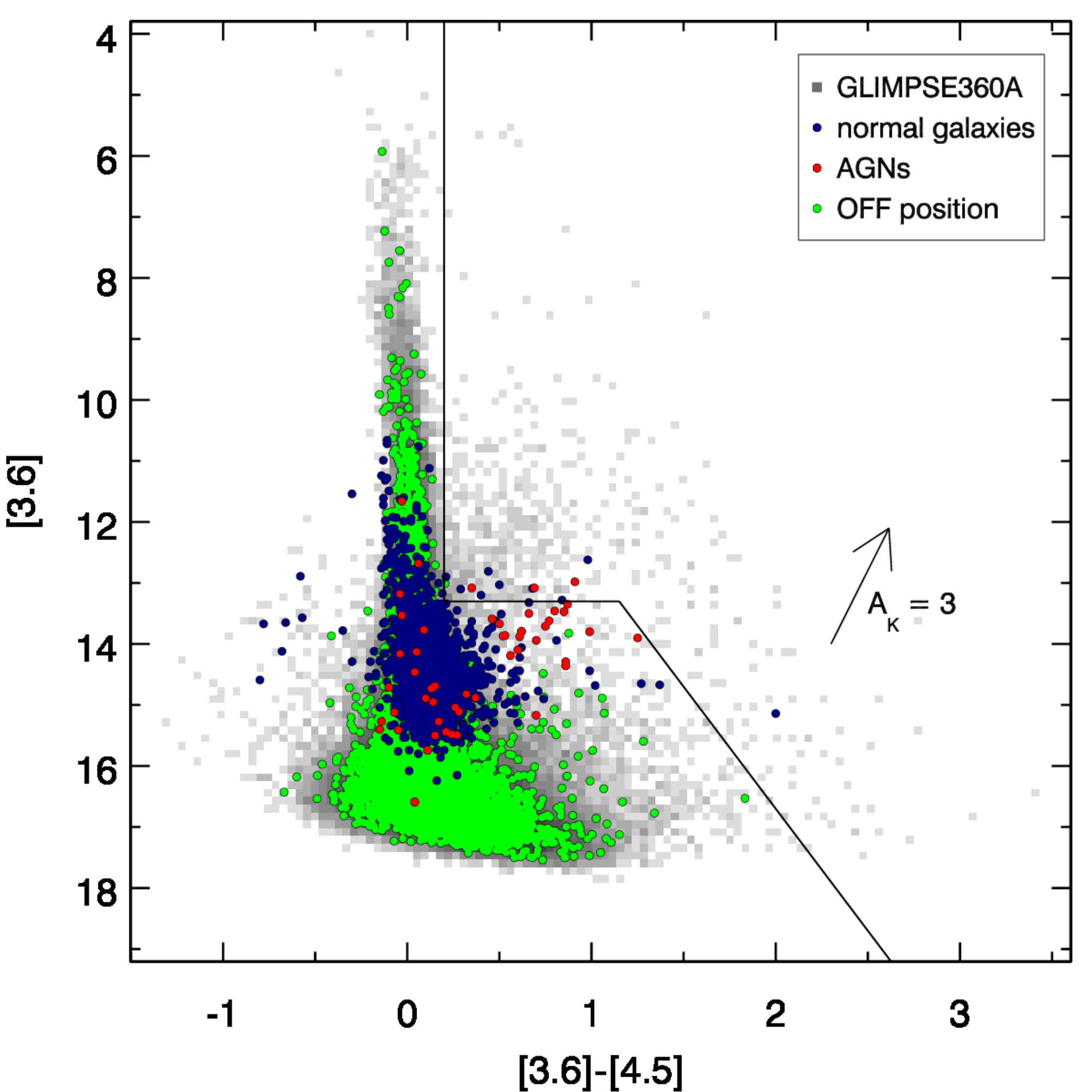}
\includegraphics[width=0.33\textwidth]{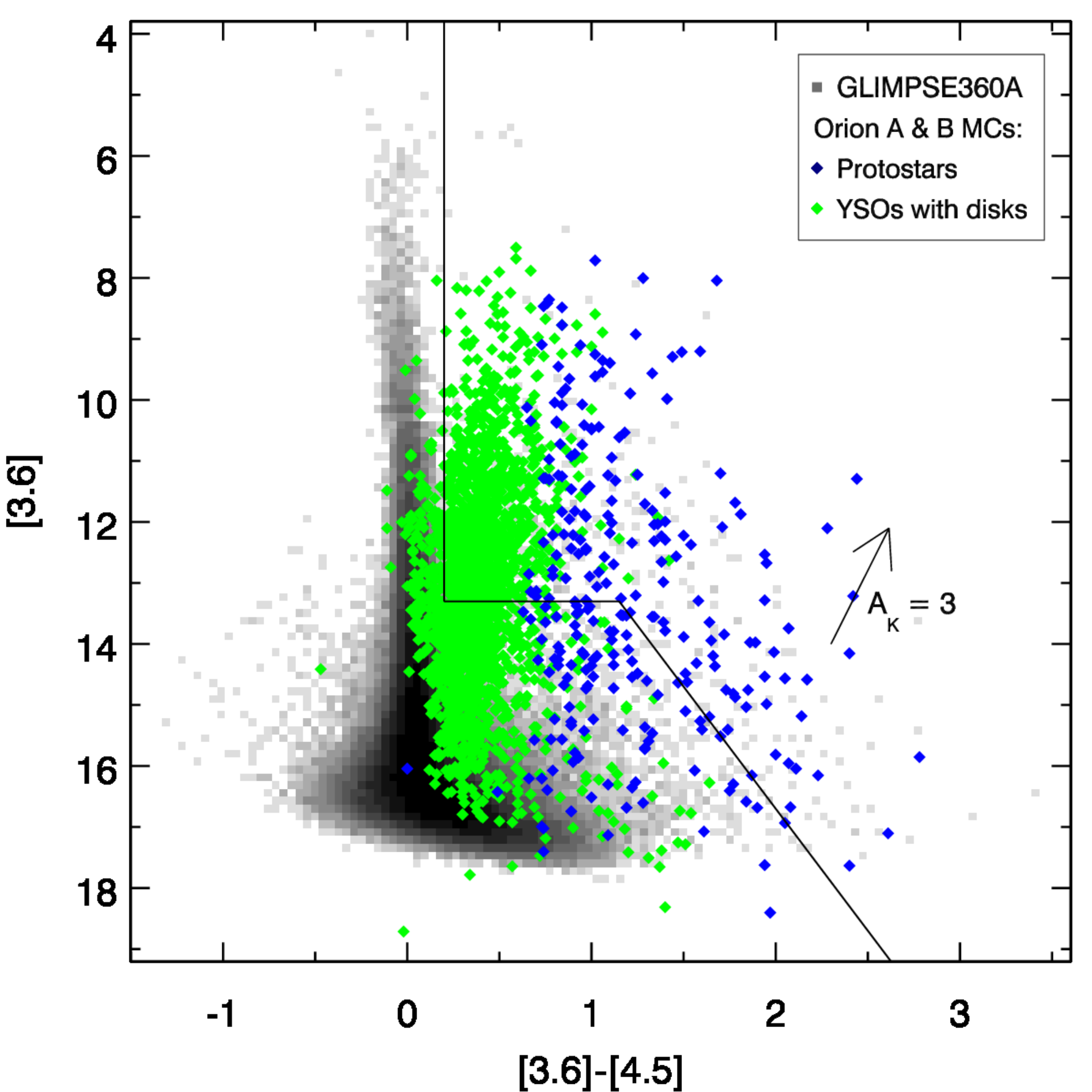}
\includegraphics[width=0.33\textwidth]{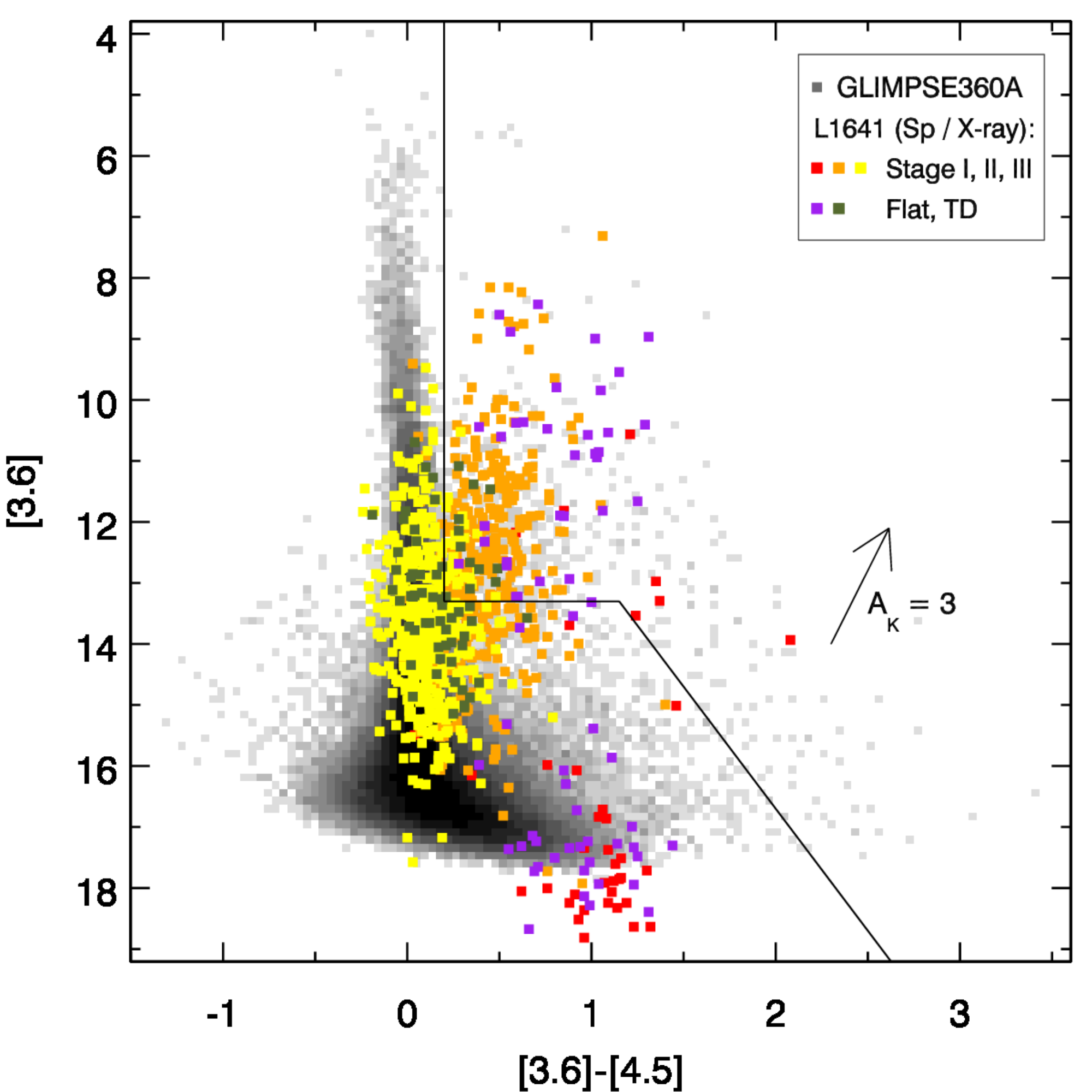}
\caption{The distribution of the GLIMPSE360 IRAC Archive sources in the $[3.6]$ vs. $[3.6]-[4.5]$ CMD compared to the distribution of  normal galaxies, AGNs, and sources from the ``off-position'' ({\it left}; \citealt{kochanek2012}), and confirmed and candidate YSOs ({\it middle} and {\it right} panels;  \citealt{kochanek2012}; \citealt{fang2013}) as indicated in the legends. To minimize a contamination from extragalactic sources and normal stars, we remove objects located to the left from the black lines from our color-selected source list as discussed in Section~\ref{s:removecontam}. \label{f:363645}} 
\end{figure*}

\subsubsection{Removing Contaminating Sources}
\label{s:removecontam}

As Figures~\ref{f:HKK45}-\ref{f:JHH45} show, the initial list of 3144 YSO candidates is still contaminated by background galaxies; there may also be some contamination from reddened stellar photospheres. We argue however that the contamination from evolved stars is negligible.

Thermally-pulsing (TP-)Asymptotic Giant Branch (AGB) stars can overlap significantly with YSOs in the diagrams using near-IR and mid-IR colors. To estimate the potential contamination from AGB stars in our sample, we use the TRILEGAL stellar population synthesis code \citep{girardi2005} updated with the latest COLIBRI TP-AGB models, including circumstellar dust \citep{marigo2013}. TRILEGAL predicts 69 TP-AGB stars in the direction of our target field (3.5 deg$^2$). Of these, 9 fall within all our 2MASS+{\it Spitzer} selection criteria (Eqs.~1 and 2), and they are also likely to fall within the WISE selection criteria based on WISE colors of similar TP-AGB stars in the Magellanic Clouds. However, all 9 stars are also predicted to be brighter than the survey's saturation limit (7~mag at 3.6~\micron), so TP-AGB stars are a negligible source of contamination among our YSO sample.

It is not possible to reliably separate low-luminosity YSO candidates from background galaxies with the photometric data alone. Therefore, we use the $[3.6]$ vs. $[3.6]-[4.5]$ CMD to identify regions in the color-magnitude space where the YSOs and background galaxies overlap, and remove the latter from our list. This cut makes our YSO candidates list more reliable (less contaminated), but incomplete as it removes low luminosity YSOs together with the most likely background galaxies. We also exclude sources overlapping with normal and evolved stars. From the list of the color-color selected YSO candidates, we keep 234 sources that fulfill the following criteria (see Figure~\ref{f:363645}): 

\normalsize
\renewcommand{\arraystretch}{0.85}
\begin{eqnarray}
\label{e:spitcmd}
\left \lbrace
\begin{array}{r}
[3.6]-[4.5] \geqslant 0.2   \; \;{\rm and} \\
  \\
{\rm [3.6]} \leqslant 13.3 \;\; {\rm or}\;\; {\rm [3.6]} \leqslant 4.0 \cdot ([3.6]-[4.5]) + 8.7\\
\end{array}     \right \rbrace 
\end{eqnarray}
\normalsize
\renewcommand{\arraystretch}{1}

To remove residual contamination from normally reddened stellar photospheres, we fit the sources' SEDs with the \citet{castelli2004} stellar photosphere models using the \citet{robitaille2007} fitting tool and remove well-fit sources. For the fitting, we use the \citet{weingartner2001} extinction law\footnote{http://www.astro.princeton.edu/$\sim$draine/dust/dustmix.html} for R$_{\rm V}$=4.0 and assume A$_{\rm V}$ ranging from 0 to 5 mag. We only consider sources with at least four valid data points (not including upper limits; 3143/3144). We quantify how well a source is fit by a given stellar photosphere model SED by considering the value of normalized $\chi^{2}$ for the best-fitting model per data point ($\chi^{2}_{\rm best}/{\rm pt}$). We determined the threshold $\chi^{2}_{\rm best}/{\rm pt}$ of 2 for sources to be well-fit, and thus removed 11 sources with $\chi^{2}_{\rm best}/{\rm pt}$ $\leq$ 2 from the sample, leaving 223 sources.

\subsection{Color-Color Cuts: {\it Spitzer} and AllWISE Photometry}
\label{s:spitzerwise}

\begin{figure*}
\centering
\includegraphics[width=0.45\textwidth]{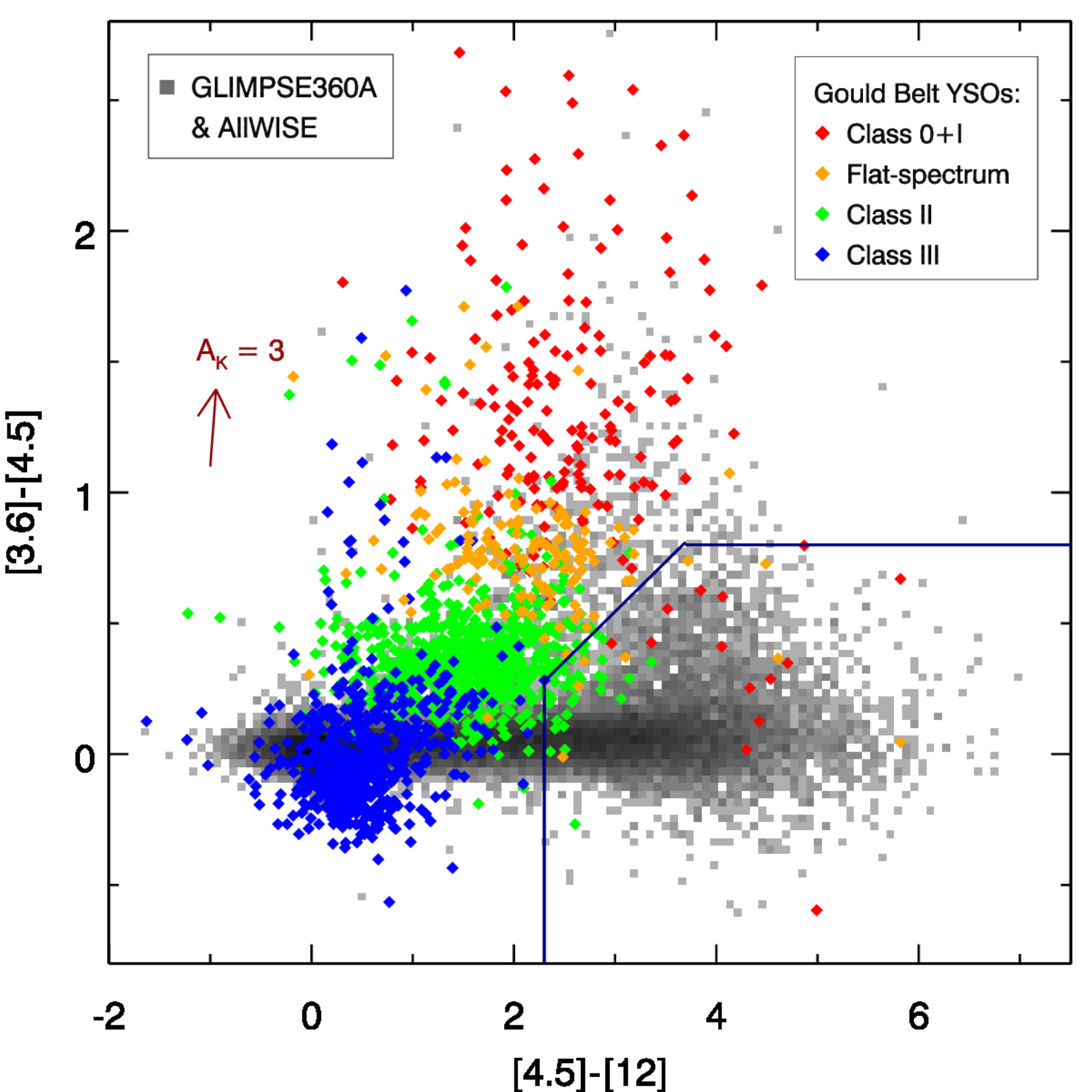}
\includegraphics[width=0.45\textwidth]{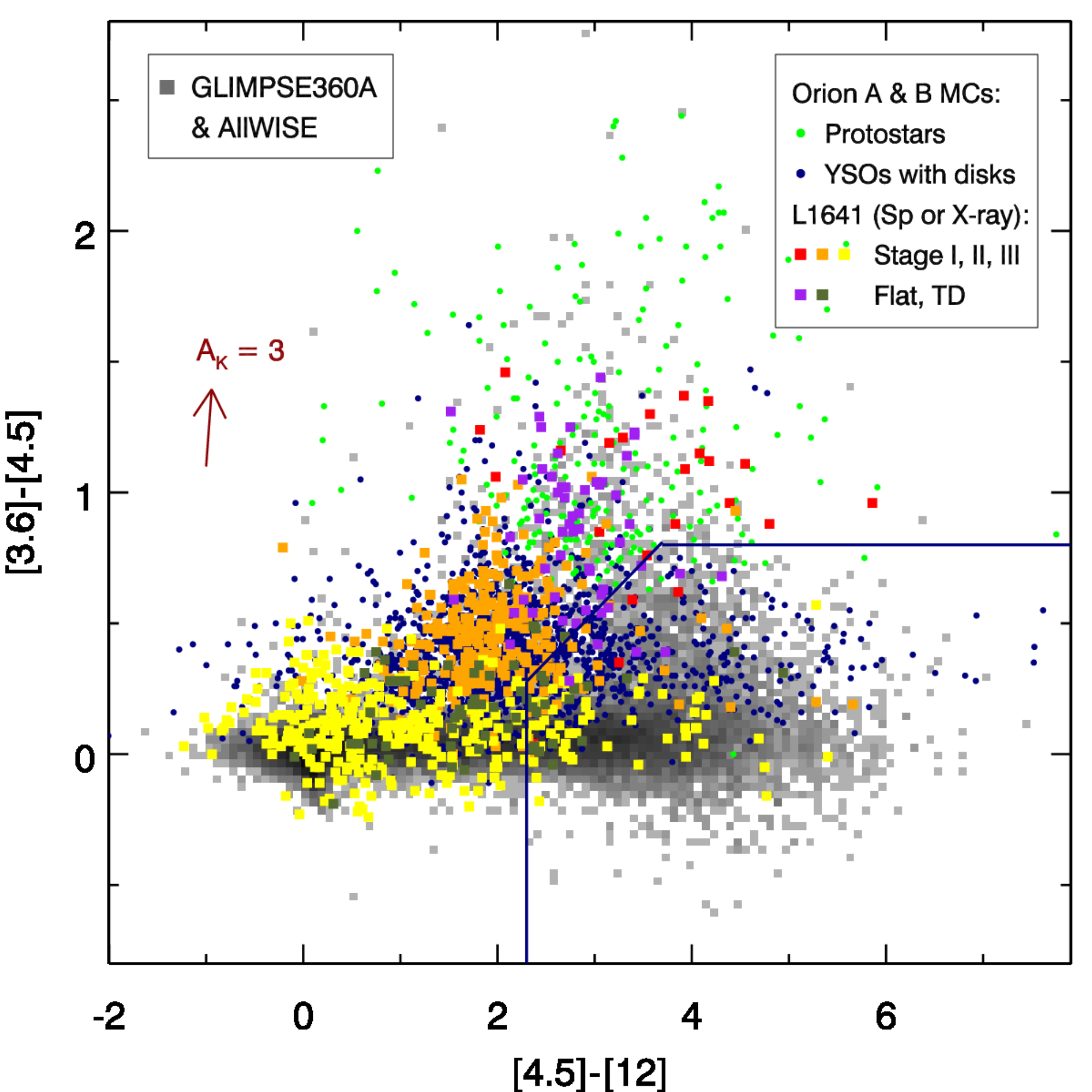}
\includegraphics[width=0.45\textwidth]{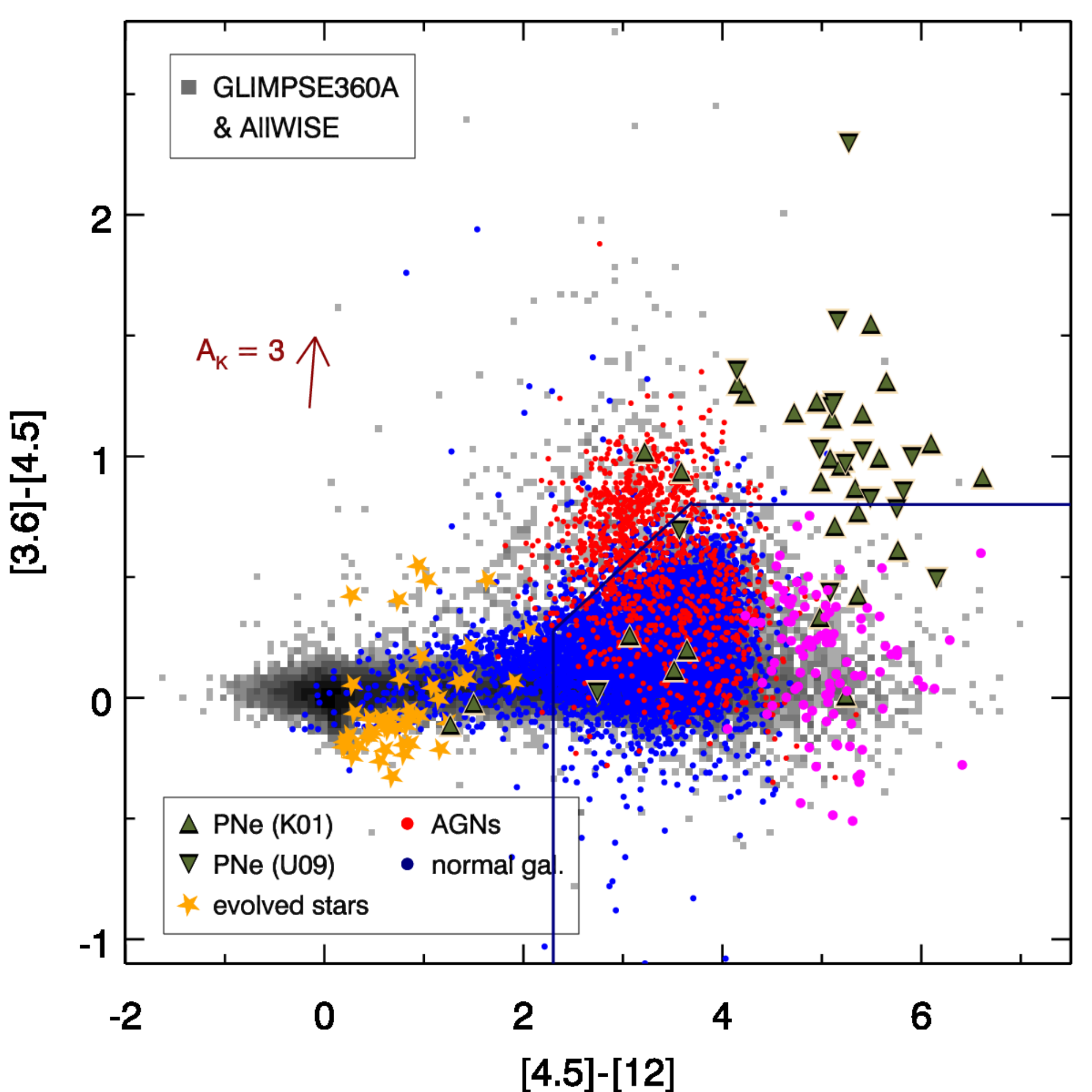}
\includegraphics[width=0.45\textwidth]{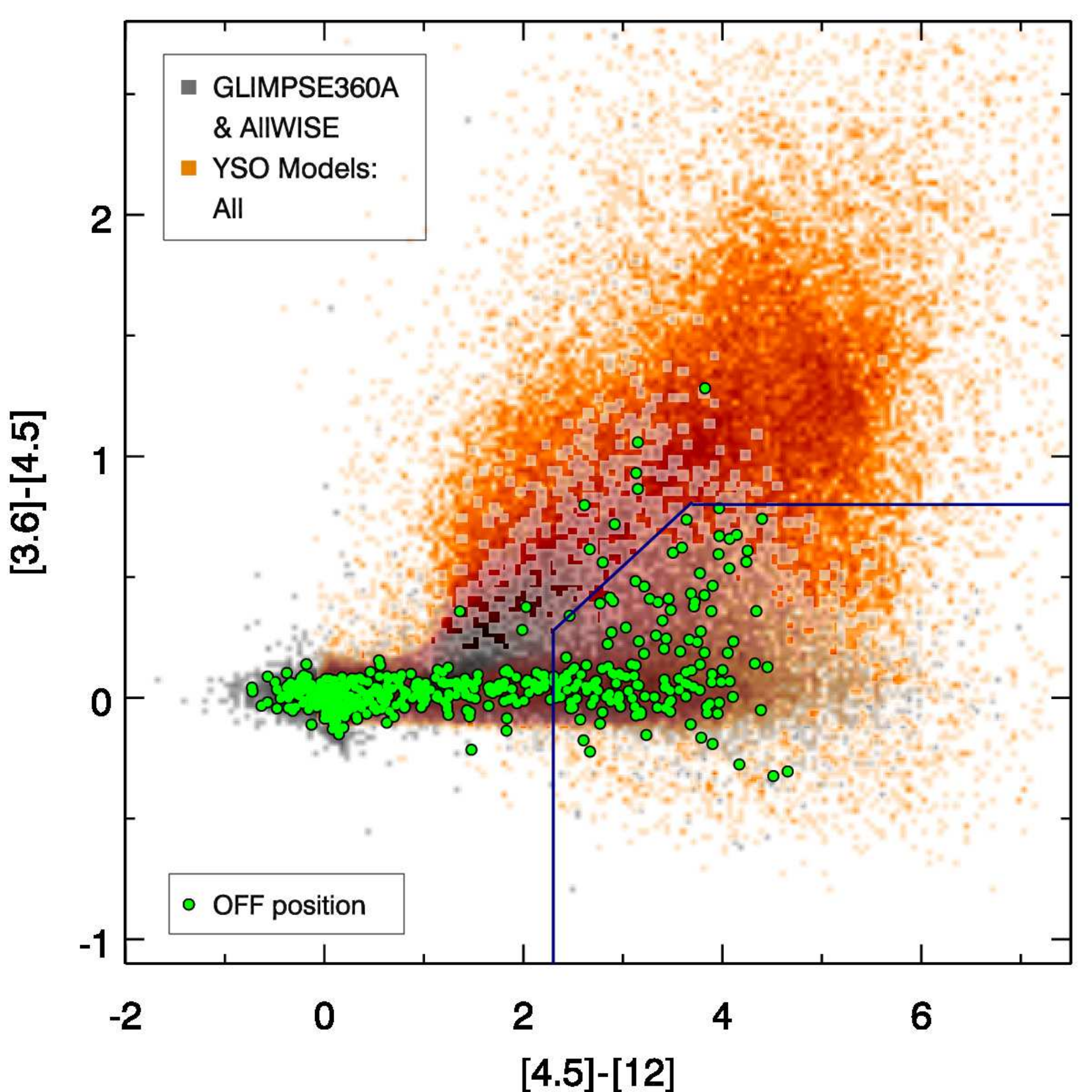}
\caption{The $[3.6]-[4.5]$ vs. $[4.5]-[12]$ CCD showing the YSO ({\it top panel}) and non-YSO ({\it lower left panel}) sources from literature (as indicated in the legends), and the sources from the ``off-position'' with the {\it Spitzer}--AllWISE selection criteria overlaid (Section~\ref{s:spitzerwise}). See the Fig.~\ref{f:HKK45} caption for references. \label{f:36454512}} 
\end{figure*}

We apply additional YSO selection criteria for sources with AllWISE 12 $\mu$m data that will allow us to identify younger YSOs without 2MASS detection or sources with low-quality 2MASS photometry. The AllWISE data have been used for the YSO selection in literature using the $[3.4]-[4.6]$ vs. $[12]-[22]$ (or $w1-w2$ vs. $w3-w4$) CCD, combined with additional criteria aimed to eliminate the background galaxy contamination  (e.g., \citealt{koenig2014}; \citealt{fischer2016}). 

We develop a similar set of the YSO selection criteria based on the GLIMPSE360 3.6 and 4.5 $\mu$m and AllWISE 12 $\mu$m photometry:

\normalsize
\renewcommand{\arraystretch}{0.85}
\begin{eqnarray}
\label{e:spit2wise}
\left \lbrace
\begin{array}{r}
 [3.6] - [4.5] \geqslant 0.46 \cdot ([4.5] - [12]) - 0.78 \; \;{\rm or} \\
 \\ \relax
 [3.6] - [4.5] \geqslant  0.8\;\; {\rm or}   \\
  \\ \relax
 [4.5] - [12] \leqslant 2.3\; \; \; \;\; \\
\end{array}     \right \rbrace 
\end{eqnarray}
\normalsize
\renewcommand{\arraystretch}{1}

These CCD cuts are shown graphically in Fig.~\ref{f:36454512}. As for the objects selected based on the {\it Spitzer} and 2MASS data, we fit SEDs of 6454 sources that fulfill these criteria and have at least 4 data points with the stellar photosphere models and remove well-fit sources. We also apply the criteria listed in Eq.~\ref{e:spitcmd} to remove background galaxies and normal stars. The {\it Spitzer}-AllWISE criteria identified 233 YSO candidates; 93 of those are new.

\subsection{Detailed Source Inspection}

We perform a detailed inspection of the images to asses the environment of the sources selected as YSO candidates and the quality of the photometry (identify artifacts, unresolved/blended sources in the lower-resolution, longer wavelength bands, etc.). We did not find any technical issues or clear evidence that the sources are not YSOs. The list of YSO candidates for further analysis contains 316 sources. 

\begin{deluxetable*}{cclc}
\tablecaption{Results of the SED fitting \label{t:modelstats}}
\tabletypesize{\small}
\tablewidth{0pt}
\tablehead{
\colhead{Model Set\tablenotemark{a}} &
\colhead{\# of sources} &
\colhead{Components} &
\colhead{Group}
}
\startdata
s$---$s$-$i	& 21 & star & \nodata \\
sp$--$s$-$i	& 72 & star $+$ passive disk; $R_{\rm inner}$ = $R_{\rm sub}$\tablenotemark{b} & $d$\\
sp$--$h$-$i	& 28 & star $+$ passive disk; variable $R_{\rm inner}$ & $d$ \\
s$---$smi	& 1  & star $+$ medium; $R_{\rm inner}$ = $R_{\rm sub}$ & \nodata \\
sp$--$smi	& 47 & star $+$ passive disk $+$ medium; $R_{\rm inner}$ = $R_{\rm sub}$ & $d$ \\
sp$--$hmi	& 26  & star $+$ passive disk $+$ medium; variable $R_{\rm inner}$ & $d$ \\
s$-$p$-$smi	&  1 & star $+$ power-law envelope $+$ medium; $R_{\rm inner}$ = $R_{\rm sub}$ & $e$\\
s$-$p$-$hmi & 0  & star $+$ power-law envelope $+$ medium; variable $R_{\rm inner}$ & $e$\\
s$-$pbsmi	& 2  & star $+$ power-law envelope $+$ cavity $+$ medium; $R_{\rm inner}$ = $R_{\rm sub}$ & $e$\\
s$-$pbhmi	& 5  & star $+$ power-law envelope $+$ cavity $+$ medium; variable $R_{\rm inner}$ & $e$\\
s$-$u$-$smi	&  2 & star $+$ Ulrich envelope $+$ medium; $R_{\rm inner}$ = $R_{\rm sub}$ & $e$\\
s$-$u$-$hmi	&  2 & star $+$ Ulrich envelope $+$ medium; variable $R_{\rm inner}$ & $e$\\
s$-$ubsmi & 0  &  star $+$ Ulrich envelope $+$ cavity $+$ medium; $R_{\rm inner}$ = $R_{\rm sub}$ & $e$\\
s$-$ubhmi	& 4  & star $+$ Ulrich envelope $+$ cavity $+$ medium; variable $R_{\rm inner}$ & $e$\\
spu$-$smi	& 11 & star $+$ passive disk $+$ Ulrich envelope $+$ medium; $R_{\rm inner}$ = $R_{\rm sub}$ & $d+e$\\
spu$-$hmi	&  4 & star $+$ passive disk $+$ Ulrich envelope $+$ medium; variable $R_{\rm inner}$ & $d+e$\\
spubsmi	&  3 & star $+$ passive disk $+$ Ulrich envelope $+$ cavity $+$ medium; $R_{\rm inner}$ = $R_{\rm sub}$ & $d+e$\\
spubhmi	& 3  & star $+$ passive disk $+$ Ulrich envelope $+$ cavity $+$ medium; variable $R_{\rm inner}$ & $d+e$\\
\enddata
\tablenotetext{a}{The model set names from \citet{robitaille2017}. Seven characters in the model set names indicate which component is present; they are (in order): s: star; p: passive disk, p or u: power-law or Ulrich envelope; b: bipolar cavities; h: inner hole; m: ambient medium; and i: interstellar dust. A dash ($-$) is used when a component is absent.}
\tablenotetext{b}{$R_{\rm inner}$ is the inner radius for the disk, envelope, and the ambient medium - when one or more of these components are present. $R_{\rm sub}$ is the dust sublimation radius.}
\end{deluxetable*}


\section{SED Fitting}
\label{s:sedfit}

We fit the SEDs of the YSO candidates with a new set of radiative transfer model SEDs for YSOs developed by \citet{robitaille2017} using the Robitaille et al. (2007) SED fitting tool.  As the \citet{robitaille2006} set of models that has been widely used to characterize thousands of YSOs and YSO candidates in both the Galaxy and the Magellanic Clouds, the new models span a wide range of evolutionary stages from the youngest, most embedded YSOs to the pre-main-sequence stars with little or no disk. The new models, however, include significant improvements. They do not depend on highly model-dependent parameters (such as the stellar age and mass) which depend on stellar evolutionary tracks, but they are defined using parameters that have a direct impact on the radiative transfer. The new models cover a much wider and more uniform region of parameter space and do not introduce correlations between the model parameters. The envelope outer radius for envelope models are now large enough to include 10-20 K dust that is essential for modeling far-IR and submm observations.    

Some limitations of the old set of models still remain. The new models still assume a single source of emission (the central star) and do not take into account the effects of the interstellar radiation fields, emission from transiently heated very small grains and polycyclic aromatic hydrocarbons (PAHs).  The models also assume a single dust model throughout the density structures and do not include accretion explicitly (i.e., only passive disks are included and thus the UV and optical fluxes for non-embedded objects cannot be reproduced by the models; \citealt{robitaille2017}). 

The \citet{robitaille2017} models consist of a combination of several components:  a star, disk, infalling envelope, bipolar cavities, and an ambient medium. They were computed as several sets of models with increasing complexity. The simple model has 2 parameters, while the most complex consisting of all of the components has 12.

For the fitting, we allow the distance to vary within 10\% from a median value of 0.92 kpc (see Section~\ref{s:distance}). We consider models with interstellar extinction $A_{\rm V}$ ranging from 0 to 40 mag in addition to the extinction caused by the YSO's circumstellar environment which is intrinsic to the models. We use the \citet{weingartner2001} extinction law for $R_{\rm V}$=4.0. We only select sources with a 12 $\mu$m detection, so the fitting can be better constrained. We require at least five valid flux measurements for sources to be included in the analysis. There are 232 sources ($\sim$73\% of the color-selected sample) that fulfill both criteria. 

\begin{figure*}[ht!]
\includegraphics[width=0.32\textwidth]{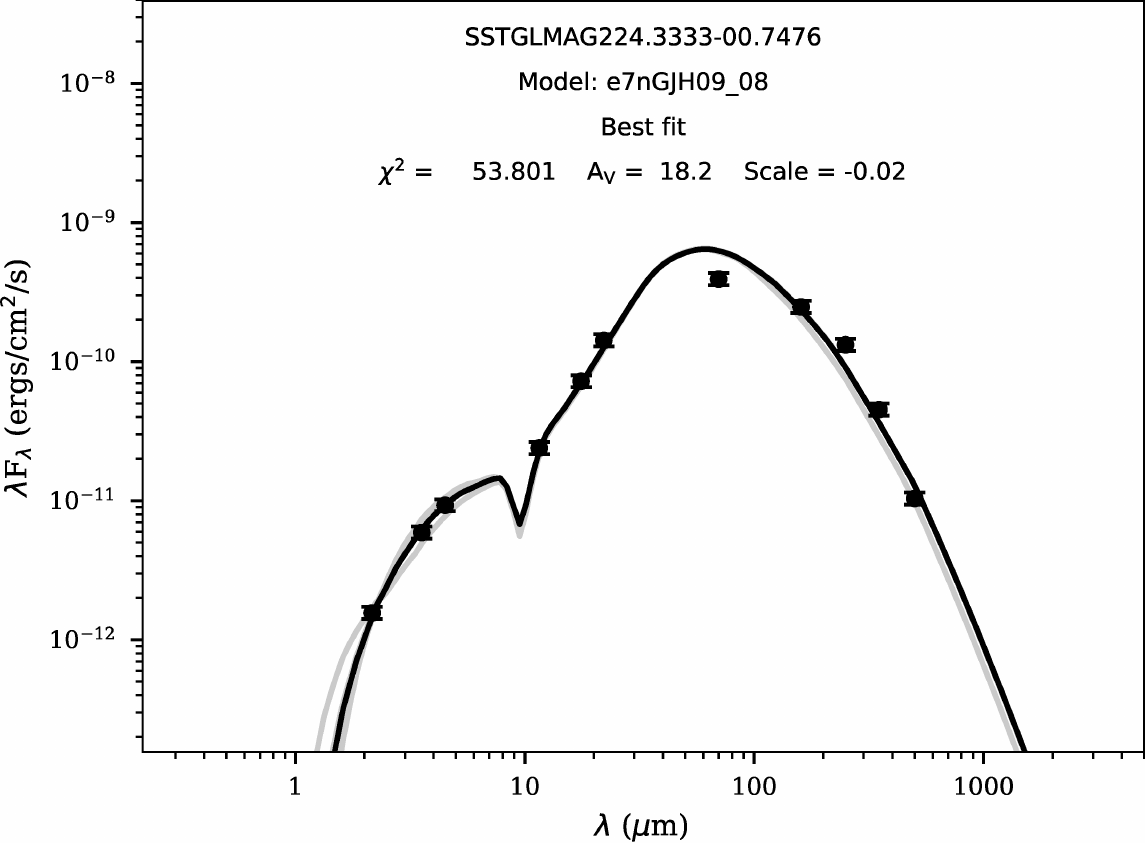} 
\hfill
\includegraphics[width=0.32\textwidth]{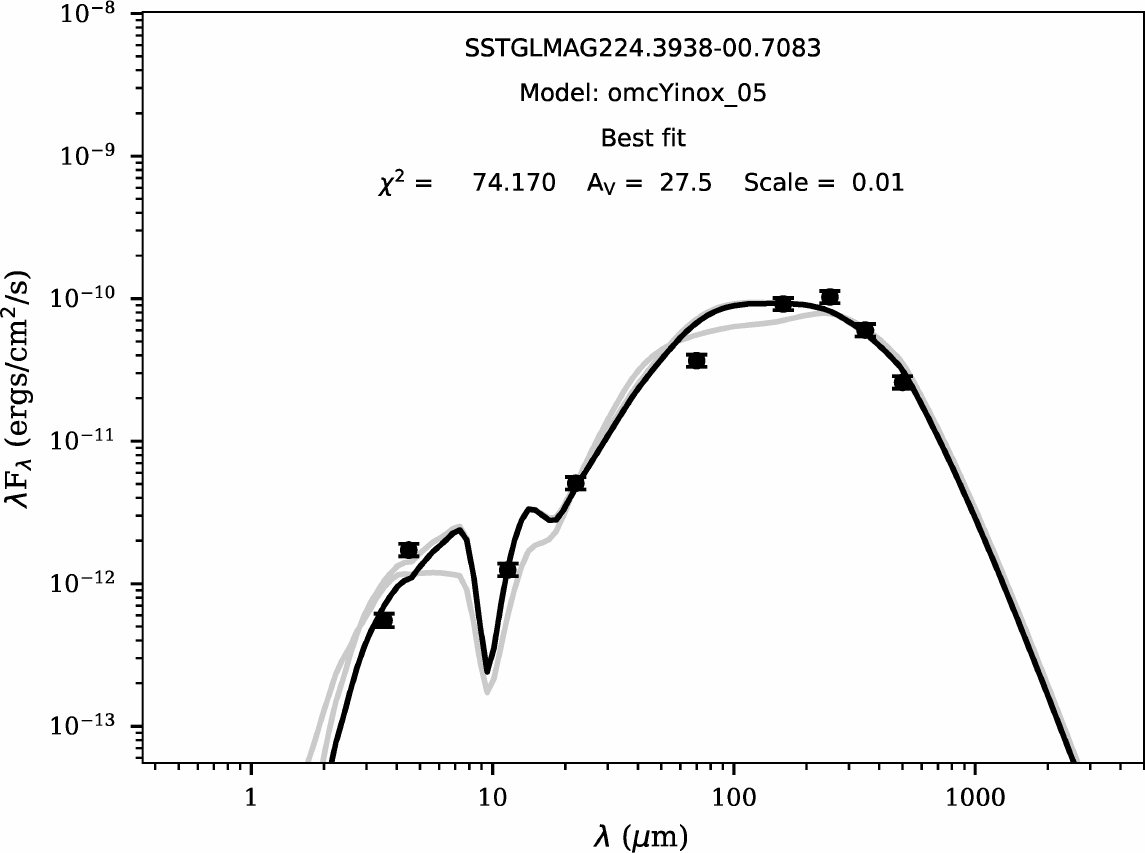} 
\hfill
\includegraphics[width=0.32\textwidth]{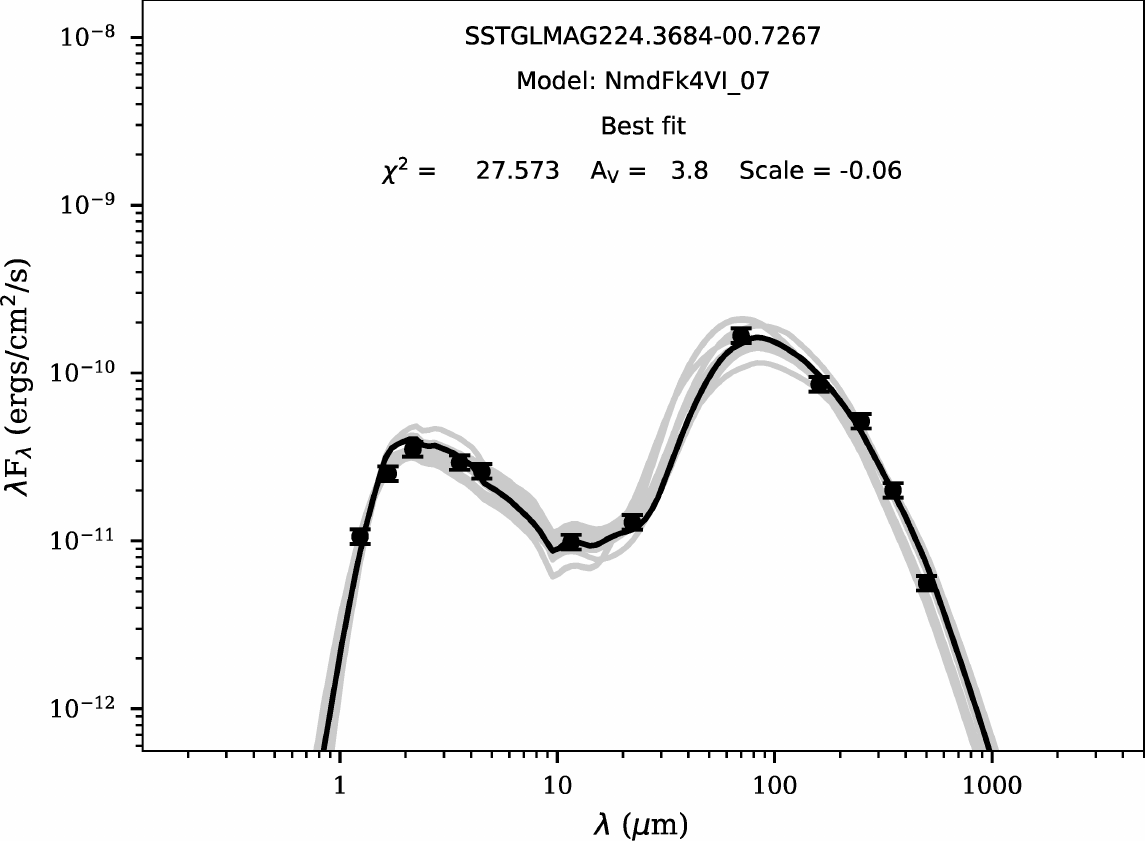} \par
\vspace{2mm}
\includegraphics[width=0.32\textwidth]{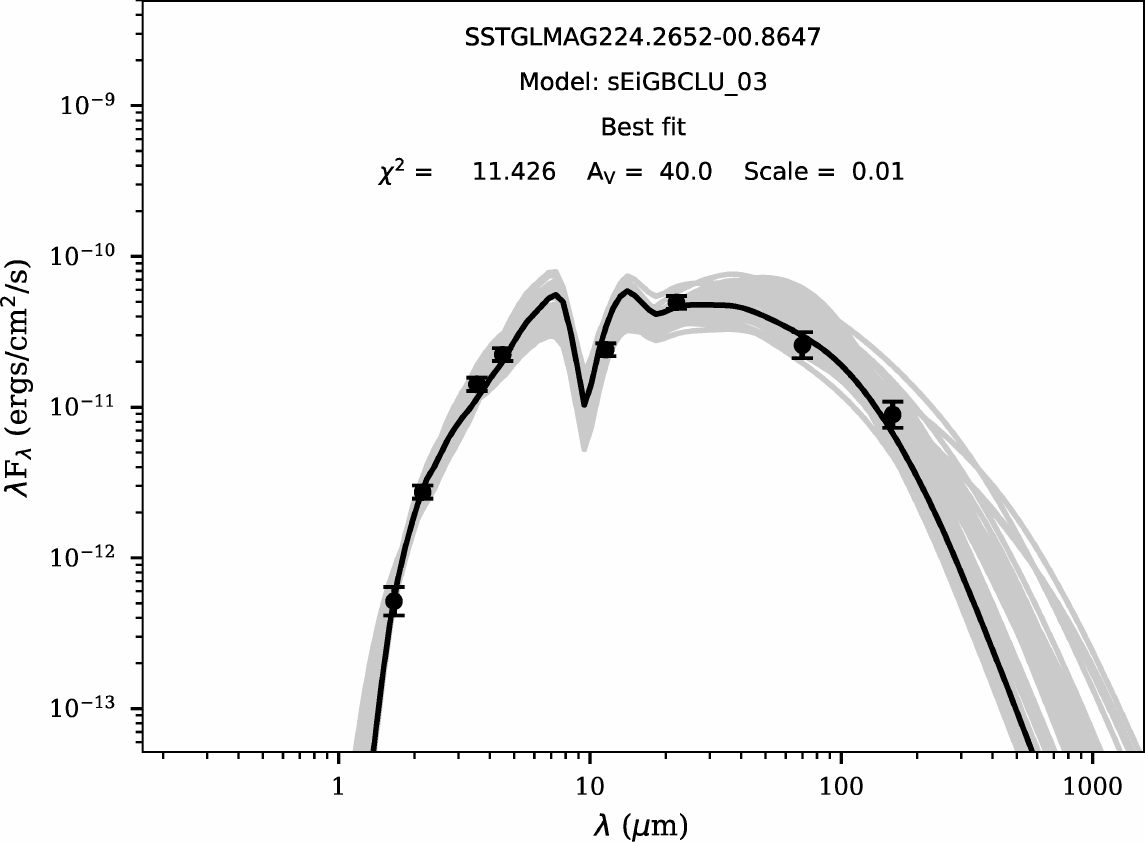} 
\hfill
\includegraphics[width=0.32\textwidth]{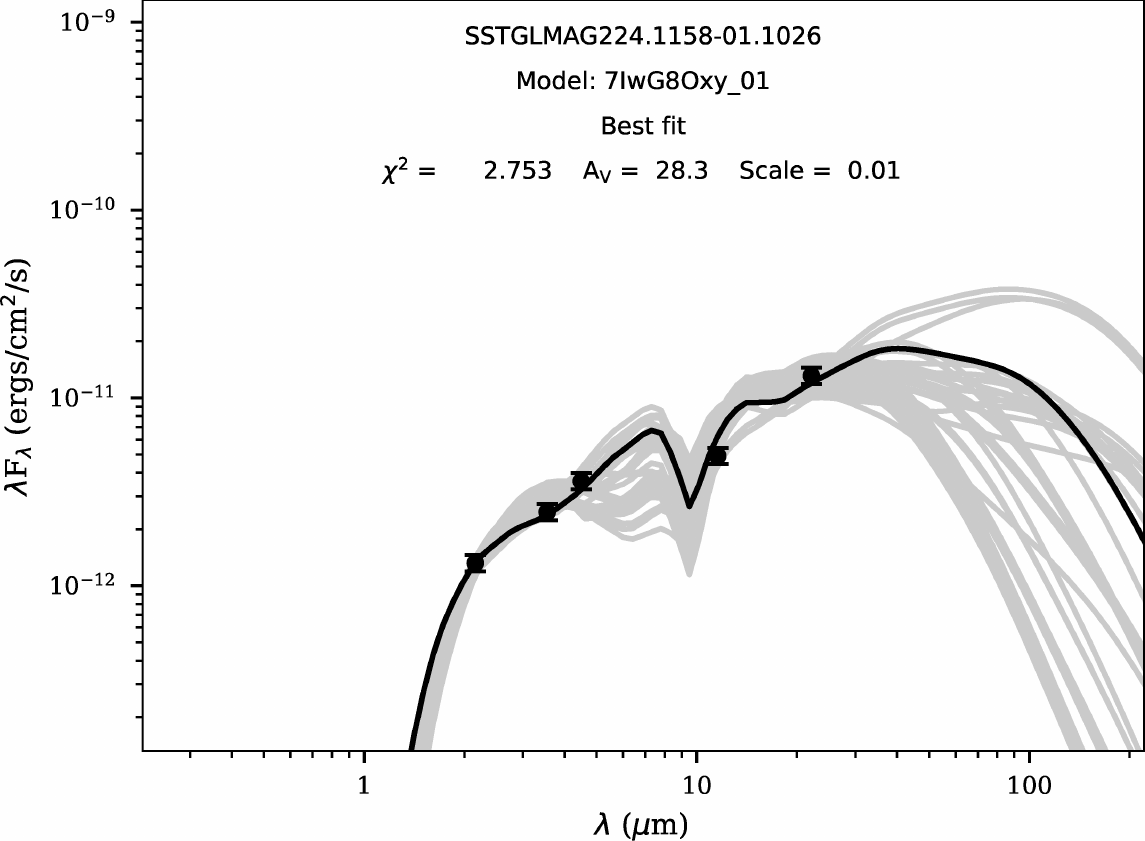} 
\hfill
\includegraphics[width=0.32\textwidth]{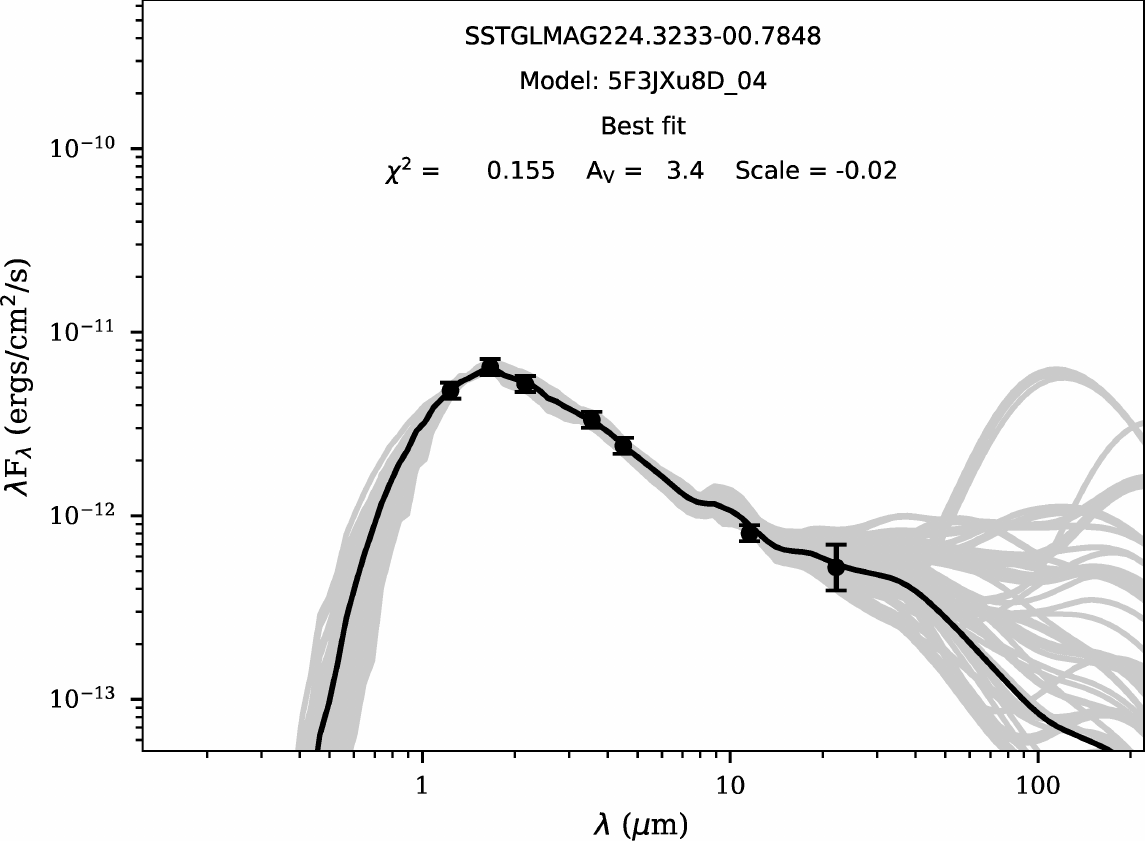} 
\caption{Example SEDs for sources at different evolutionary stages well-fit with \citet{robitaille2017} YSO models. The embedded objects with envelopes, but no disks (Group $e$) are shown in the left and middle panel in the upper row. The best-fit model for the source in the rightmost plot in the upper row contains both an envelope and disk (Group $d+e$). The lower row shows SEDs of sources best-fit with models with disks, but not envelopes (Group $d$). The solid black line in each plot shows the best fit YSO model. Filled circles and triangles are valid flux values and flux upper limits, respectively. The values of a reduced $\chi^{2}$, interstellar visual extinction, and a distance scale for the best-fit model are indicated in the plots. The flux error bars are plotted if larger than the data points. See Section~\ref{s:sedfit} for details. \label{f:exsed}}
\end{figure*}

The new set of models aims at covering the parameters space as uniformly as possible, but the trade off is that some of the models are not physical. We select for our fitting procedure only the physical models by comparing their positions on the Hertzsprung-Russell (HR) diagram with the PARSEC evolutionary tracks \citep{bressan2012,ychen2014,chen2015,tang2014}.   The PARSEC evolutionary tracks are produced by the revised Padova code and include the pre-main sequence (PMS) stage; the initial stellar masses range from 0.1 to 350 M$_{\odot}$  \citep{bressan2012,chen2015}.  The YSO models located outside the coverage of the PARSEC PMS tracks are excluded. 

The SED fitting results are highly degenerate. To find the best fitting model, we use a Bayesian approach to assign relative probabilities to the various models of increasing complexity; the technical details and the philosophy behind are described in \citet{robitaille2017}. In short, for each source after fitting every model from all the model sets, we first find the smallest $\chi^{2}$, namely $\chi^{2}_{\rm best}$. We define ``good'' fits are those with $\chi^{2} < (\chi^{2}_{\rm best} + F)$, where $F$ is a threshold parameter that is determined empirically. In our case, it is found that $F=3$ is a reasonable choice. We count the number of good fits provided by each model set, and the one with the largest number of good fits are selected as the best model set. The single best model is the one with the smallest $\chi^{2}$ from the best model set. Note that this single best model may not be the one with $\chi^{2}_{\rm best}$. That means, instead of directly using $\chi^{2}_{\rm best}$ to determine the best model, we first find the best model set, and from that set we select the best model. The reason is that each model set represents a combination of components (e.g, with envelope, disk) and  the most probable model set should be able to give the largest number of reasonably good fits, even if the single $\chi^{2}_{\rm best}$ is not found from there. On the contrary, the single $\chi^{2}_{\rm best}$ could be just a coincidence from a model set where most of the other fits are actually bad. In addition, we note that we at most have 10 photometric data points for each source, but some of the model sets have up to 13 parameters. Overfitting might occur especially for the sources with fewer data points.

Table~\ref{t:modelstats} lists the \citet{robitaille2017} model sets and their components, along with the number of sources well-fit with models from a given model set. For the subsequent analysis, we divide sources into three groups depending on the presence of the envelope and/or disk: `envelope-only' ($e$ in Table~\ref{t:modelstats}; 16 sources), `envelope and disk' ($d+e$; 21 sources), and `disk-only' ($d$; 173 sources). Sources in the `$e$' group might have a disk, but one was not needed to fit their SEDs. Twenty two sources are well-fit with models that contain only a star (21) or a star and an ambient medium (1). We remove these sources from the YSO candidates list; 294 sources remain (210 YSO candidates well-fit with YSO models and 84 YSO candidates for which the SED fitting was not performed due to the lack of photometry at 12 $\mu$m or too few data points). Example SEDs for YSO candidates at different evolutionary stages are shown in Fig.~\ref{f:exsed}. 

\subsection{Luminosity, Mass and Age Estimation}
\label{s:otherphyspar}

The new set of models does not provide the stellar luminosity, mass, and age. 
First we find the luminosity for each model with the Stephan-Boltzmann law using the stellar radius and effective temperature which are given. Then, on the HR diagram the closest PMS track to each model is found. The mass is taken as that of the closest track. The age is interpolated from the closest point on the track to the model on the HR diagram.

We compare the ages estimated from the evolutionary tracks to the results of the SED fitting, i.e., the presence of an envelope and/or disk, and adopt masses only for sources for which the ages are consistent with the evolutionary stage of the YSO candidates. We adopt YSO lifetimes determined by \citet{dunham2015} based on a sample of $\sim$3000 YSO candidates from nearby star-forming regions. They calculated the lifetimes of 0.46--0.72 Myr for Class 0+I sources and 0.30--0.46 Myr for Flat-spectrum sources, adopting the Class II + III as a reference class with the duration of 3 Myr. The duration of the Class 0 and Class I phases are 0.335 and 0.665 $\times$ the Class 0 + I lifetime. For simplicity, we assume that all the `envelope-only' sources are Class 0 YSOs, `envelope and disk' sources are Class I or Flat-spectrum YSOs, and `disk-only' sources are Class II and Class III YSOs. 

The physical parameters estimated from the SED fitting and the HR diagram analysis for 210 YSO candidates are listed in Table~\ref{t:physpar}.

\begin{figure*}[ht!]
\centering
\includegraphics[width=1.\textwidth]{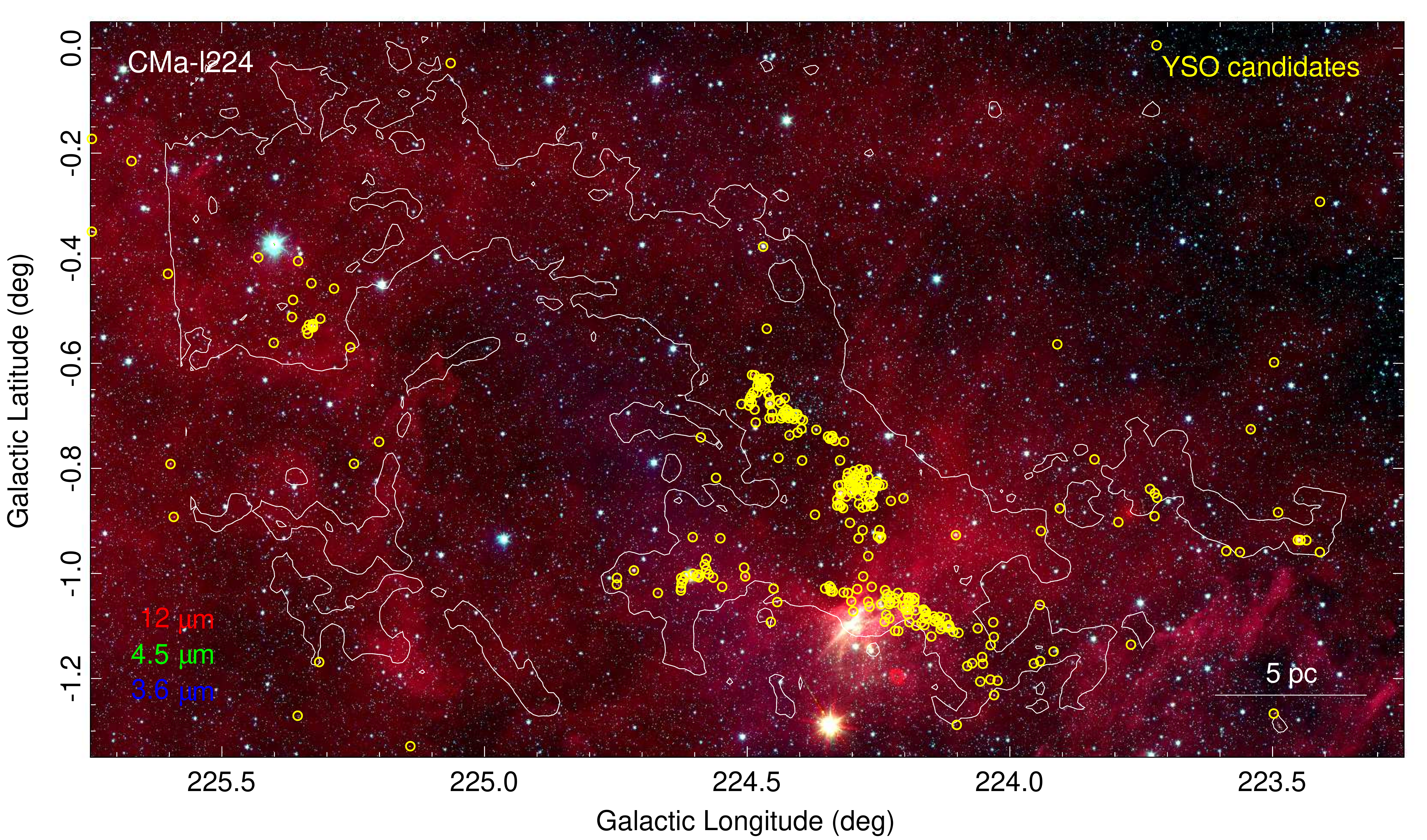}
\caption{The three-color composite image of CMa--$l224$ combining the WISE 12 $\mu$m ({\it red}), GLIMPSE360 4.5 $\mu$m ({\it green}), and GLIMPSE360 3.6 $\mu$m ({\it blue}) images; the same as in Fig.~\ref{f:3color}. The positions of YSO candidates identified in this paper are indicated with yellow circles. The white contour corresponds to the H$_{2}$ column density of 4 $\times$ 10$^{21}$ cm$^{-2}$. \label{f:distrall}}
\end{figure*}

\begin{deluxetable*}{ccccccccccc}
\tablecaption{Physical Parameters for a Subset of YSO Candidates in CMa--$l224$ \label{t:physpar}}
\tabletypesize{\small}
\tablewidth{0pt}
\tablehead{
\colhead{IRAC Designation} &
\colhead{$n_{\rm pt}$\tablenotemark{a}} &
\colhead{Model} &
\colhead{Model} &
\colhead{($\chi^{2}_{\rm min}$/} &
\colhead{($n_{\rm fits}$)$_{set}$\tablenotemark{d}} &
\colhead{$R_{\rm \star}$} &
\colhead{$T_{\rm \star}$} &
\colhead{$L_{\rm \star}$} &
\colhead{$M_{\rm \star}$\tablenotemark{e}} &
\colhead{$Age$\tablenotemark{e}} 
\\
\colhead{`SSTGLMC'} &
\colhead{} &
\colhead{Set\tablenotemark{c}} &
\colhead{Name} &
\colhead{$pt$)$_{\rm set}$\tablenotemark{b}} &
\colhead{} &
\colhead{($R_{\odot}$)} &
\colhead{($T_{\odot}$)} &
\colhead{($L_{\odot}$)} &
\colhead{($M_{\odot}$)} &
\colhead{(Myr)} 
}
\startdata
G223.4100-00.9594 & 5 & spu$-$smi & O9yTeubb\_01 & 0.363 & 132 & 1.63 & 3690 & 0.443 & \nodata & \nodata \\
G223.4356-00.9369 & 6 & sp$--$s$-$i & bRpK1icY\_01 & 0.0262 & 461 & 0.939 & 4180 & 0.243 & \nodata & \nodata \\
G223.4467-00.9365 & 6 & spu$-$hmi & lcceyFSO\_03 & 3.69 & 30 & 7.99 & 6880 & 129 & 4 & 0.766 \\
G223.4519-00.9364 & 18 & spubhmi & UYPUxRp7\_07 & 2.93 & 13 & 12.3 & 6270 & 210 & 5.2 & 0.275 \\
G223.4894-00.8837 & 7 & sp$--$hmi & szXsXcnT\_09 & 0.743 & 68 & 1.4 & 5740 & 1.92 & \nodata & \nodata \\
G223.5416-00.7255 & 7 & sp$--$h$-$i & ddJqkFr2\_03 & 1.41 & 42 & 2.03 & 5520 & 3.43 & 1.7 & 3.93 \\
G223.5622-00.9594 & 7 & sp$--$s$-$i & r9oRQITx\_03 & 0.135 & 328 & 2.61 & 9610 & 52.2 & 2.4 & 3.85 \\
G223.5881-00.9575 & 6 & sp$--$smi & 5FMrwNtB\_08 & 0.0817 & 664 & 1.78 & 5900 & 3.46 & \nodata & \nodata \\
G223.7198-00.8560 & 7 & sp$--$s$-$i & J29C1Wcg\_04 & 0.581 & 353 & 1.24 & 6300 & 2.16 & \nodata & \nodata \\
G223.7210+00.0052 & 7 & sp$--$hmi & 74RteWgE\_07 & 0.841 & 32 & 0.975 & 5020 & 0.544 & \nodata & \nodata \\
\enddata 
\tablenotetext{a}{$n_{\rm pt}$ is the number of valid data points used for the fitting.}
\tablenotetext{b}{See Table~\ref{t:modelstats} footnotes for the meaning of characters in the model set names.}
\tablenotetext{c}{($\chi^{2}_{\rm min}/pt$)$_{\rm set}$ is $\chi^{2}$ per data point for the best-fit model from the most likely model set.}
\tablenotetext{d}{($n_{\rm fits}$)$_{set}$ is the number of fits with $\chi^2/pt$ between ($\chi^{2}_{\rm min}/pt$)$_{\rm set}$ and ($\chi^{2}_{\rm min}/pt$)$_{\rm set}$$+$2 in the most likely model set.}
\tablenotetext{e}{We provide stellar masses ($M_{\rm \star}$) and ages ($Age$) only for sources with the most reliable estimates of these parameters (see Section~\ref{s:otherphyspar} for details).}
(This table is available in its entirety in a machine-readable form in the online journal. A portion is shown here for guidance regarding its form and content.)
\end{deluxetable*}


\section{SIMBAD Search}

We searched the SIMBAD database for matches within 5$''$ from the {\it Spitzer}/GLIMPSE360 positions of 294 YSO candidates from our list to check whether the sources have been classified in literature. We have found four matches with known sources that were not from the \citet{elia2013} or \citet{fischer2016} catalogs.

Source SSTGLMAG\,224.4696-00.3780 is classified in literature as a Be star (HD\,55135; e.g., \citealt{fremat2006}); we removed it from the list of YSO candidates.

Source SSTGLMAG\,225.3263-00.5314 has been identified in literature as a galaxy (2MASX\,J07122445-1115336; e.g., \citealt{paturel2003}), a massive YSO with an extended bipolar H$_2$ emission (MSX6C\,G225.3266-00.5318, \citealt{navarete2015}), and a {\it Herschel}/Hi-GAL proto-stellar core (G225.32629-00.53244, \citealt{elia2013}). Since the extragalactic nature of this source is not confirmed, we keep the source on the list of YSO candidates.

Source SSTGLMAG\,224.3005-00.8417 was identified as a member of the EX Lupi-like class of YSO outbursts by \citet{miller2015}: iPTF 15afq. These outbursts with a duration from a few weeks to several months, are interpreted as the accretion events driven by disk instabilities (e.g., \citealt{audard2014}). The EX Lupi objects have circumstellar disks and no envelopes; this is consistent with the results of the SED fitting. 

Source SSTGLMAG\,225.1412-01.3289 coincides with a radio source PMN\,J0709-1128 (\citealt{wright1996}; \citealt{vollmer2010} and references therein) with a spectral index $\alpha$ between 1.4 GHz and 4.9 GHz of 0.32 ($S_{\nu}\propto\nu^{\alpha}$, \citealt{vollmer2010}). The extragalactic nature of this source is not ruled out \citep{petrov2011}, but we keep it on the list of YSO candidates with a warning that the classification is uncertain.  

In summary, based on the SIMBAD search, we removed one source from the list of YSO candidates; 293 sources remain on the list. Their spatial distribution is shown in Fig.~\ref{f:distrall}. All the photometry used in the analysis is provided in Table~\ref{t:catdata}.


\section{Classification of YSO Candidates in Color-Color Space}

In Table~\ref{t:catdata}, we provide the Class I and Class II classification for YSO candidates from our list based on the classification criteria from literature. For sources with valid 12 $\mu$m photometry, we use the \citet{fischer2016} criteria, otherwise the classification is based on the \citet{gutermuth2009} criteria. We include a flag indicating which criteria were used. Four sources do not fulfill either \citet{fischer2016} or \citet{gutermuth2009} criteria; they are classified as YSO candidates with disks based on the SED fitting. 

\begin{figure*}
\includegraphics[width=0.32\textwidth]{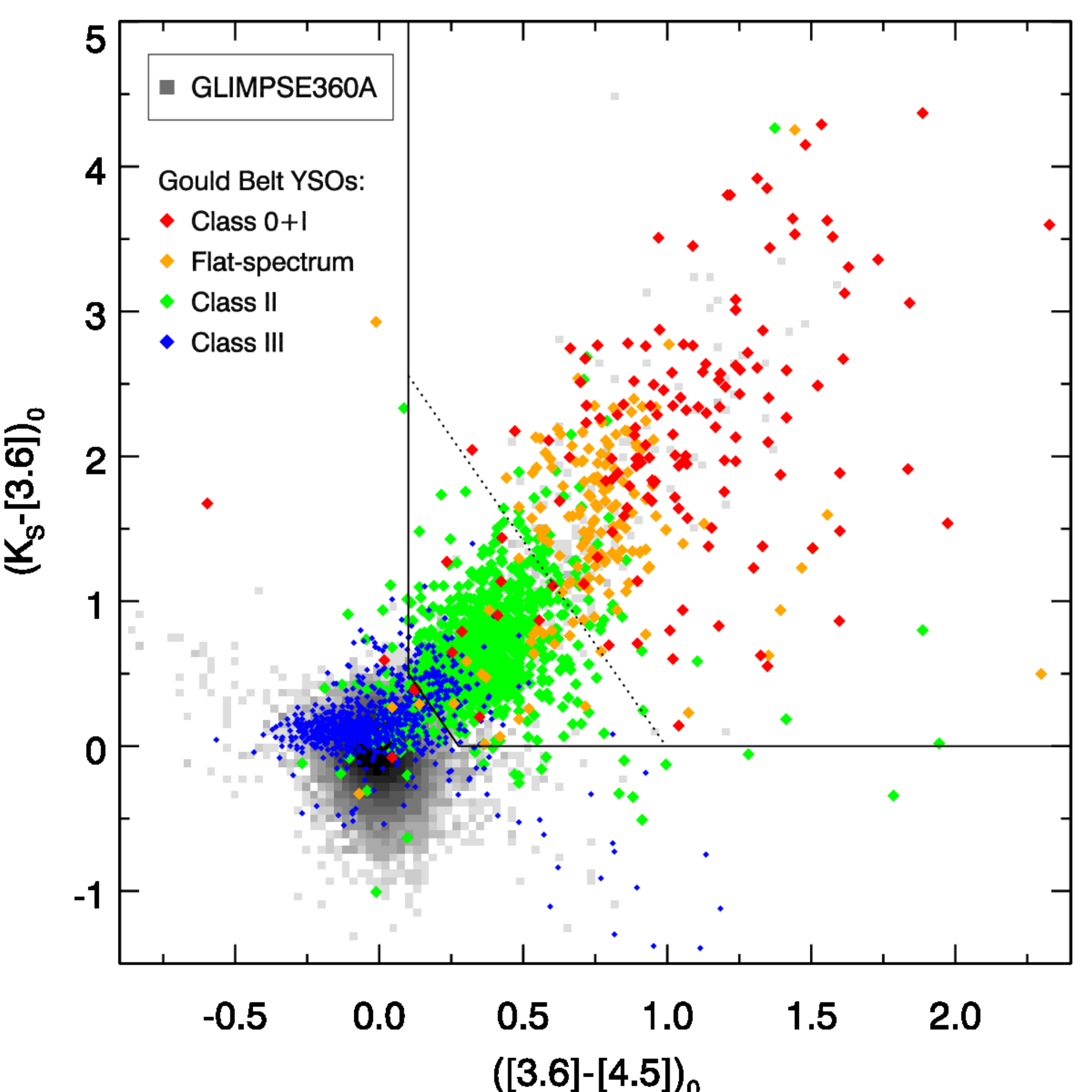}
\includegraphics[width=0.32\textwidth]{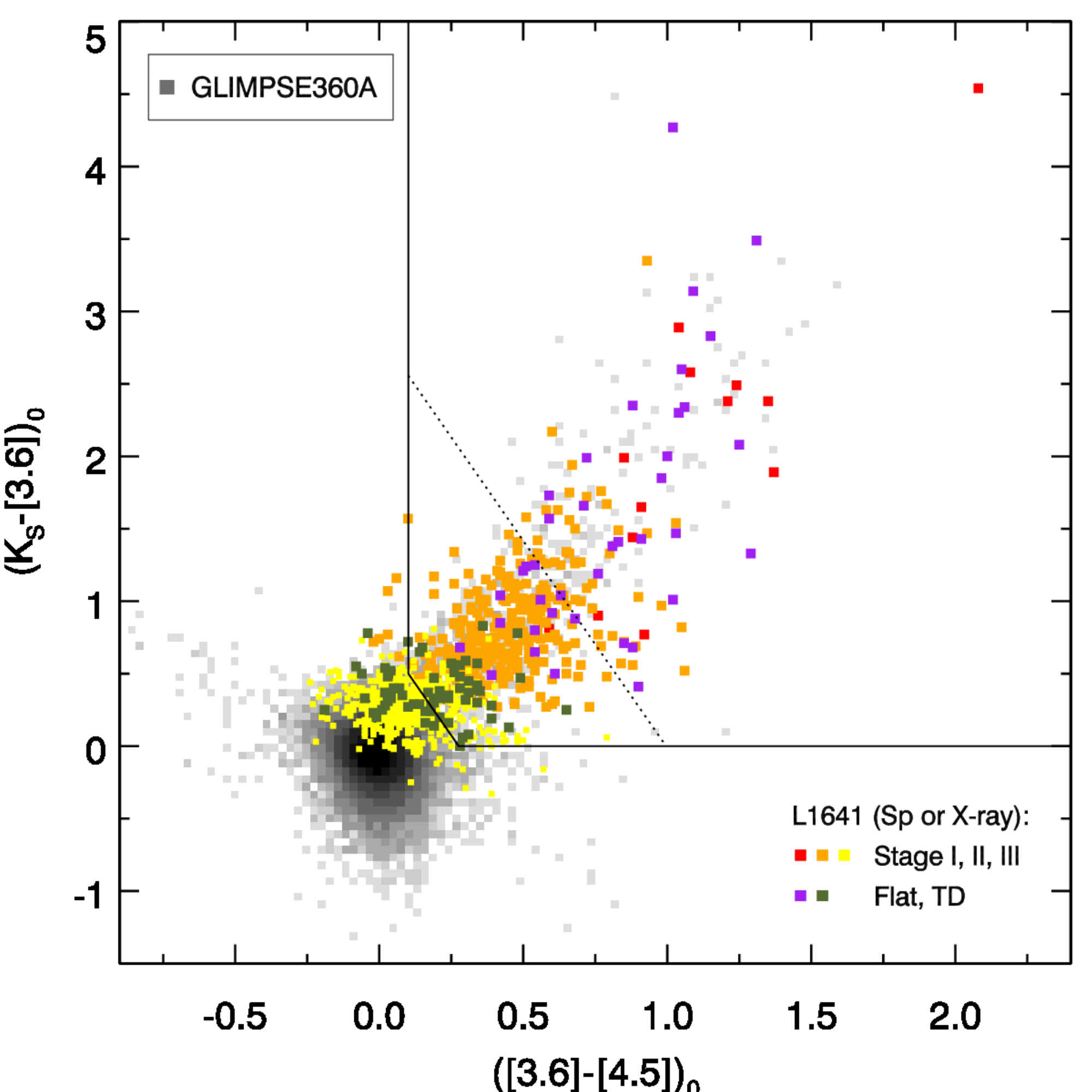}
\includegraphics[width=0.32\textwidth]{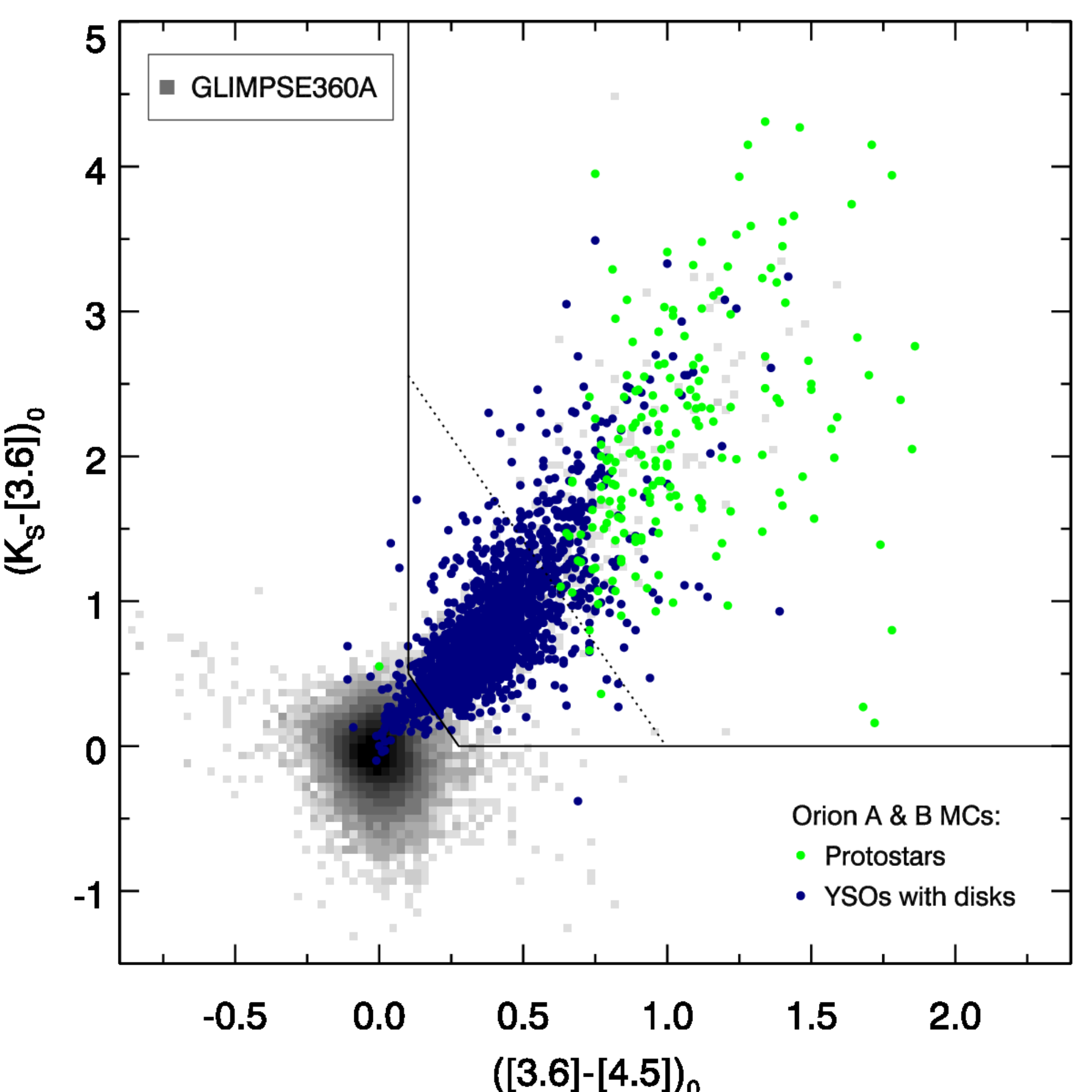}
\includegraphics[width=0.32\textwidth]{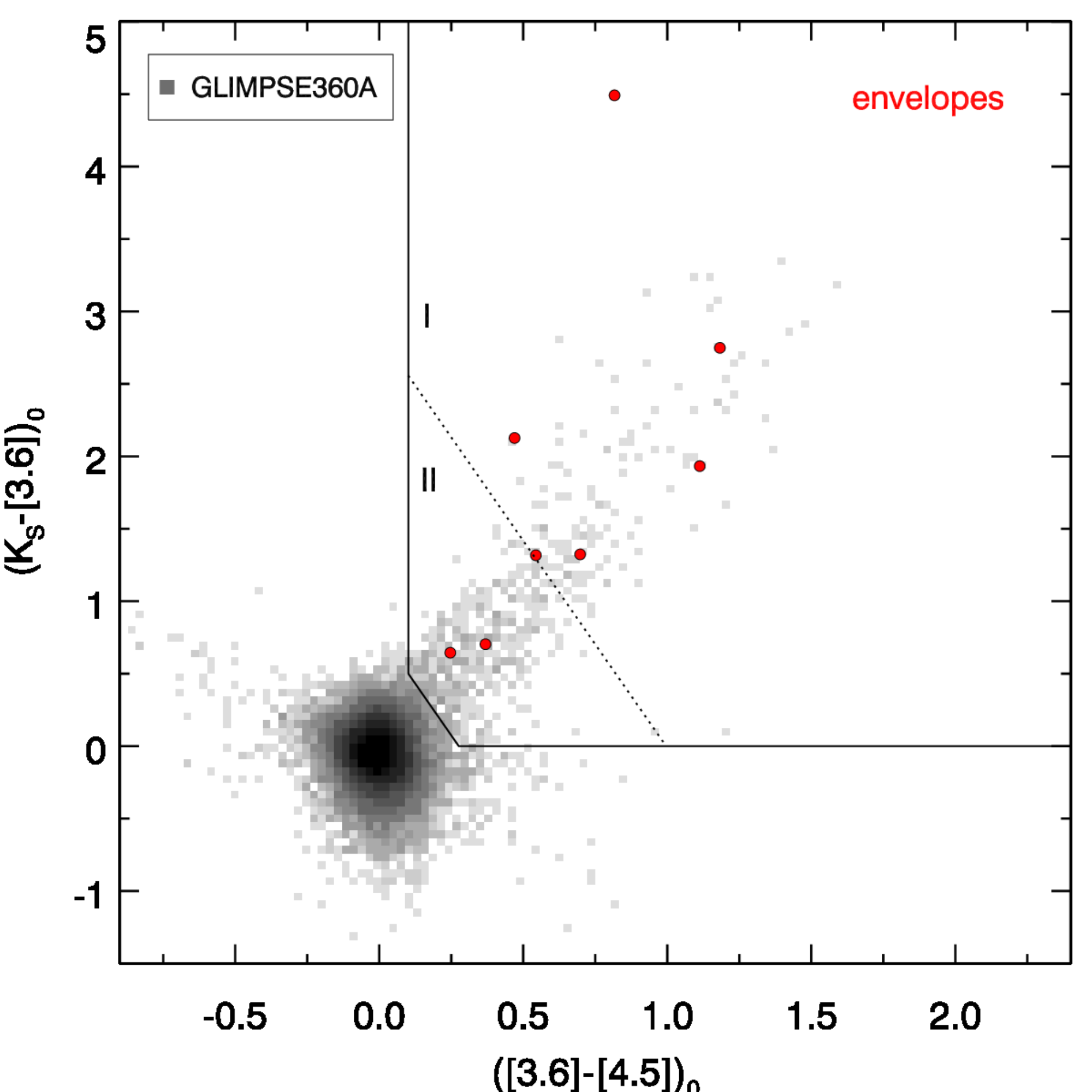}
\hfill
\includegraphics[width=0.32\textwidth]{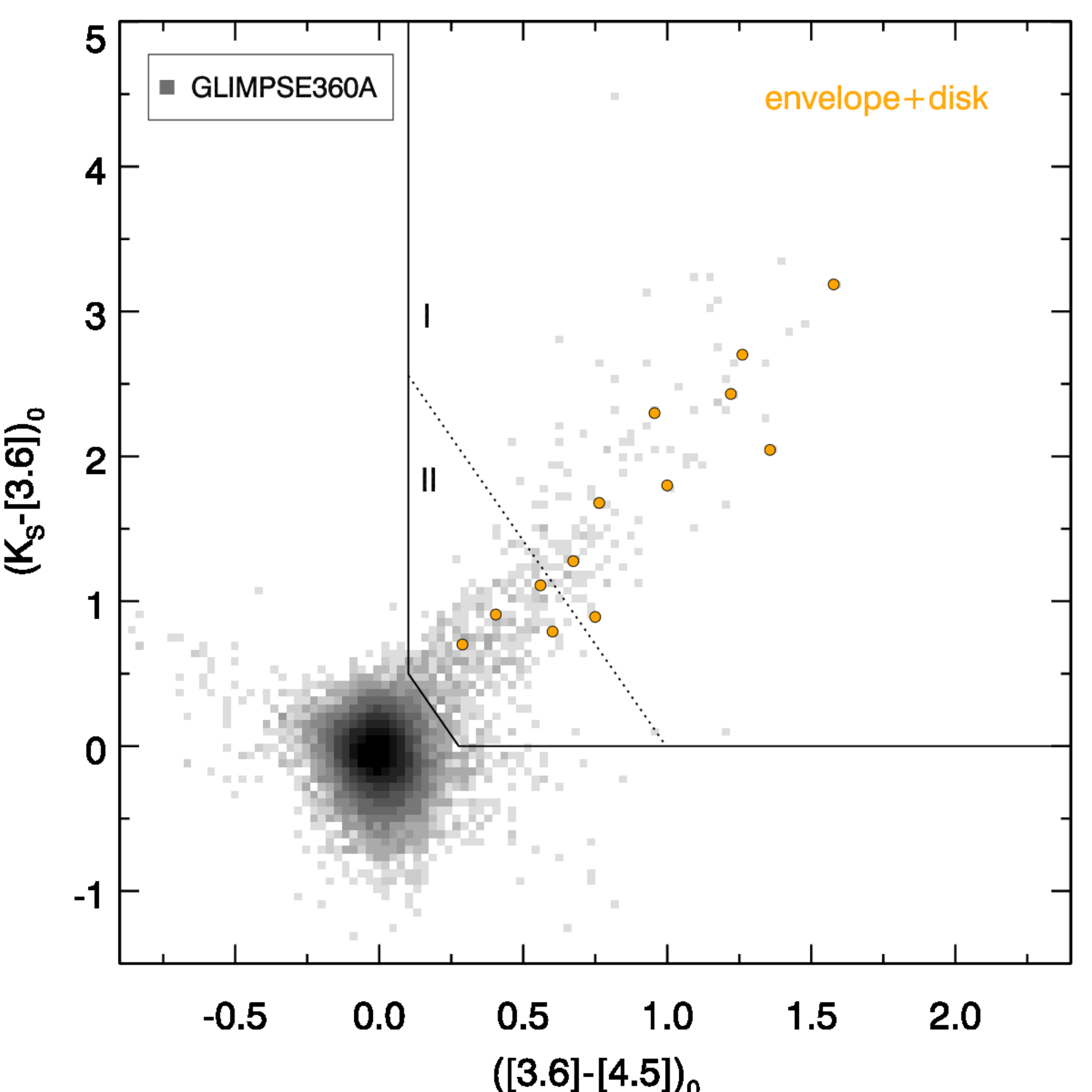}
\hfill
\includegraphics[width=0.32\textwidth]{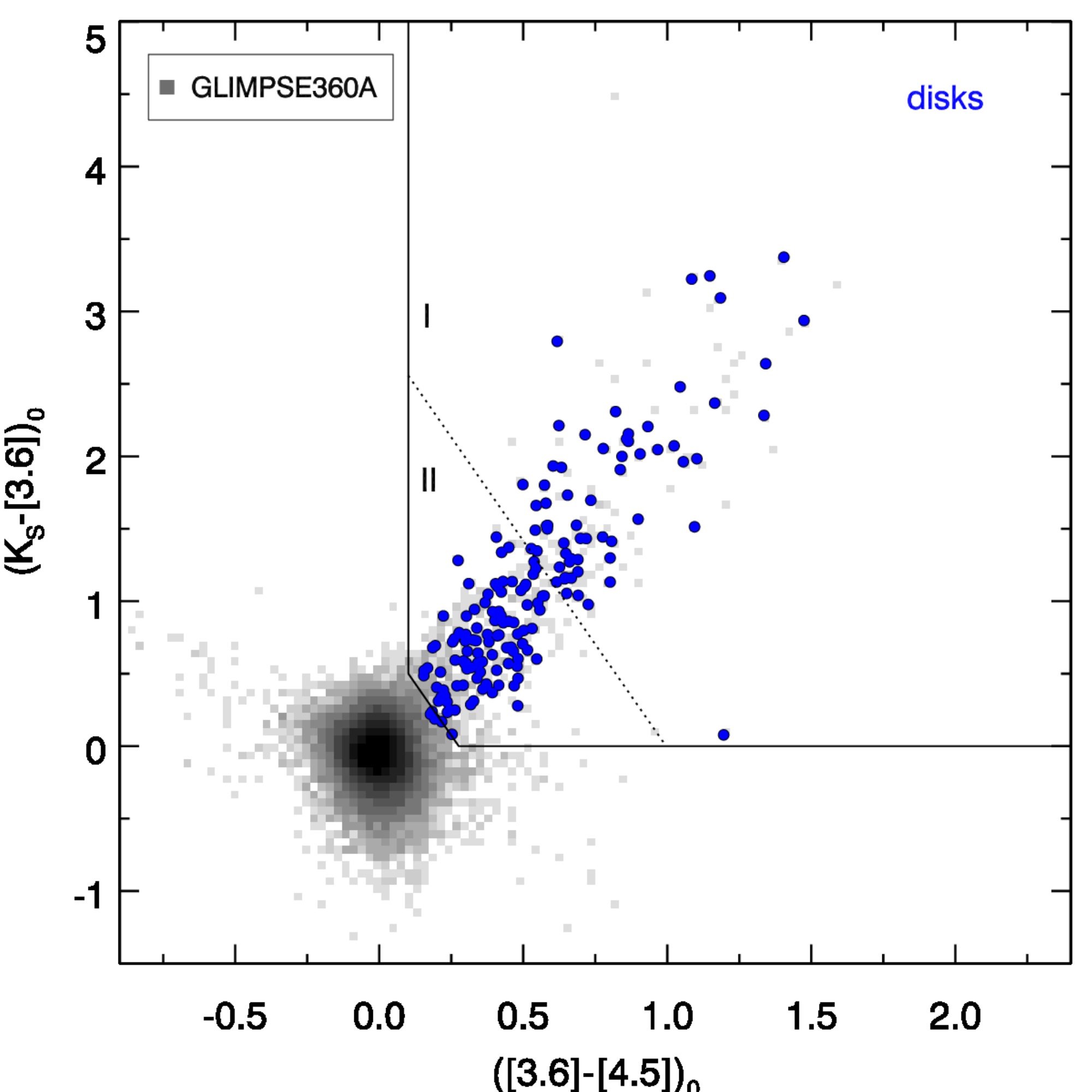}
\caption{{\it All panels:} The distribution of the GLIMPSE360 IRAC Archive sources  in the $K_{\rm s}-[3.6]$ vs. $[3.6]-[4.5]$ CCD; the photometry has been corrected for extinction using `method 2' (see Section~\ref{s:2massspitzer}).  Black solid lines show the YSO classification criteria developed by \citet{gutermuth2009} to classify sources with no longer wavelength IRAC photometry available. The black dashed line separates Class I and Class II YSO candidates. The top panel shows CCDs with YSO candidates identified in \citet[{\it upper left}]{dunham2015}, \citet[{\it upper middle}]{fang2013}, and \citet[{\it upper right}]{megeath2012}.  The YSO candidates identified in CMa--$l224$ (see Sections~\ref{s:2massspitzer}--\ref{s:sedfit}) are plotted in the CCDs in the lower panel: `$e$' ({\it left}), `$d+e$' ({\it middle}), and `$d$' ({\it right}). See text for details. \label{f:K363645}} 
\end{figure*}

We use the \citet{gutermuth2009} criteria for Class I and Class II based on the ($K_{\rm s}-[3.6]$)$_{0}$ vs. ($[3.6]-[4.5]$)$_{0}$ CCD (see Fig.~\ref{f:K363645}), where the subscript `0' indicates dereddened colors. The extinction is estimated based on the 2MASS colors (they require at least $H$ and $K_{\rm s}$ photometry and magnitude uncertainties $<$0.1 mag). To minimize the contamination from extragalactic sources, they apply the brightness limit using dereddened 3.6 $\mu$m photometry: [3.6]$_{0}$ $<$ 15.0 and [3.6]$_{0}$ $<$ 14.5 for Class I and Class II YSO candidates, respectively.

\begin{figure*}
\includegraphics[width=0.32\textwidth]{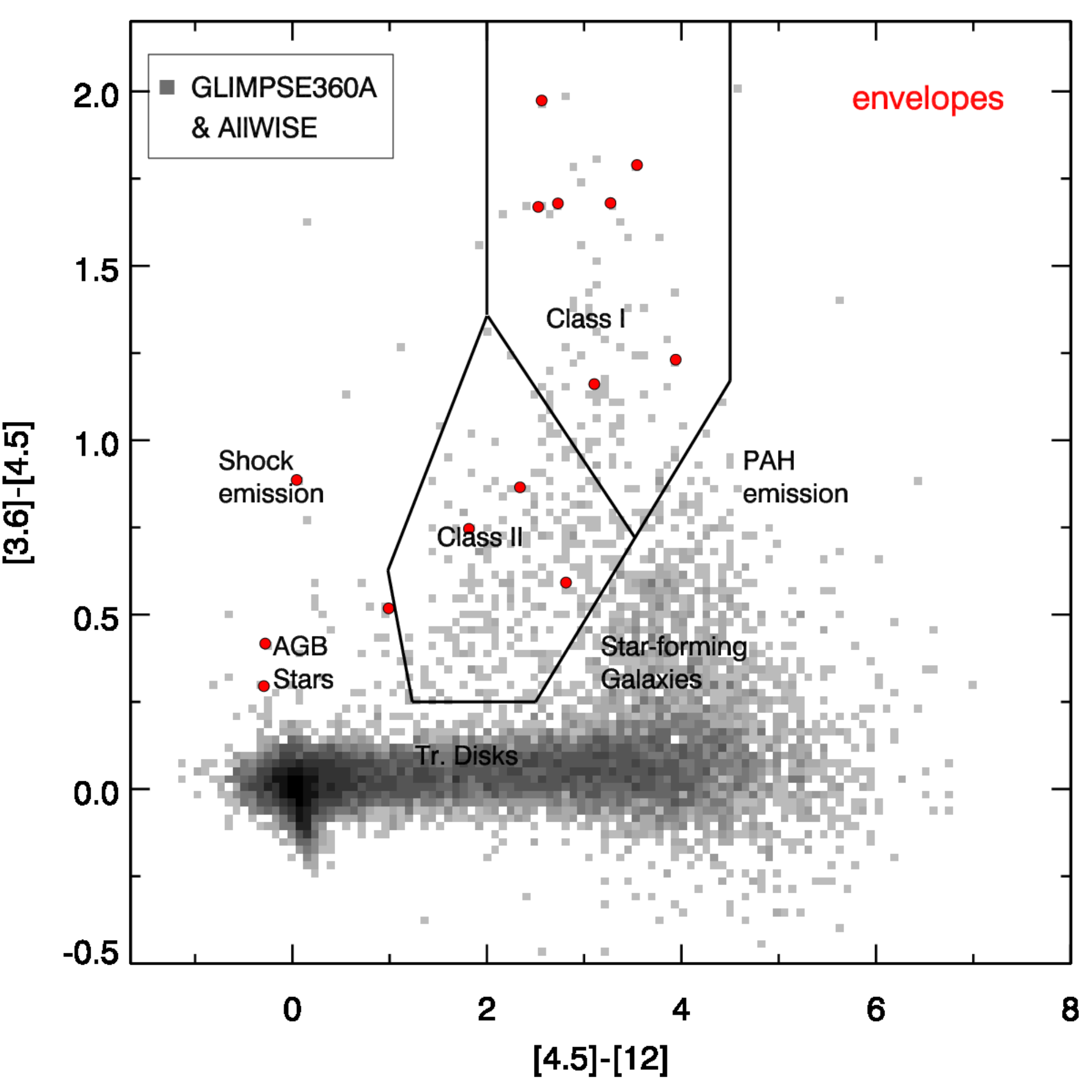}
\includegraphics[width=0.32\textwidth]{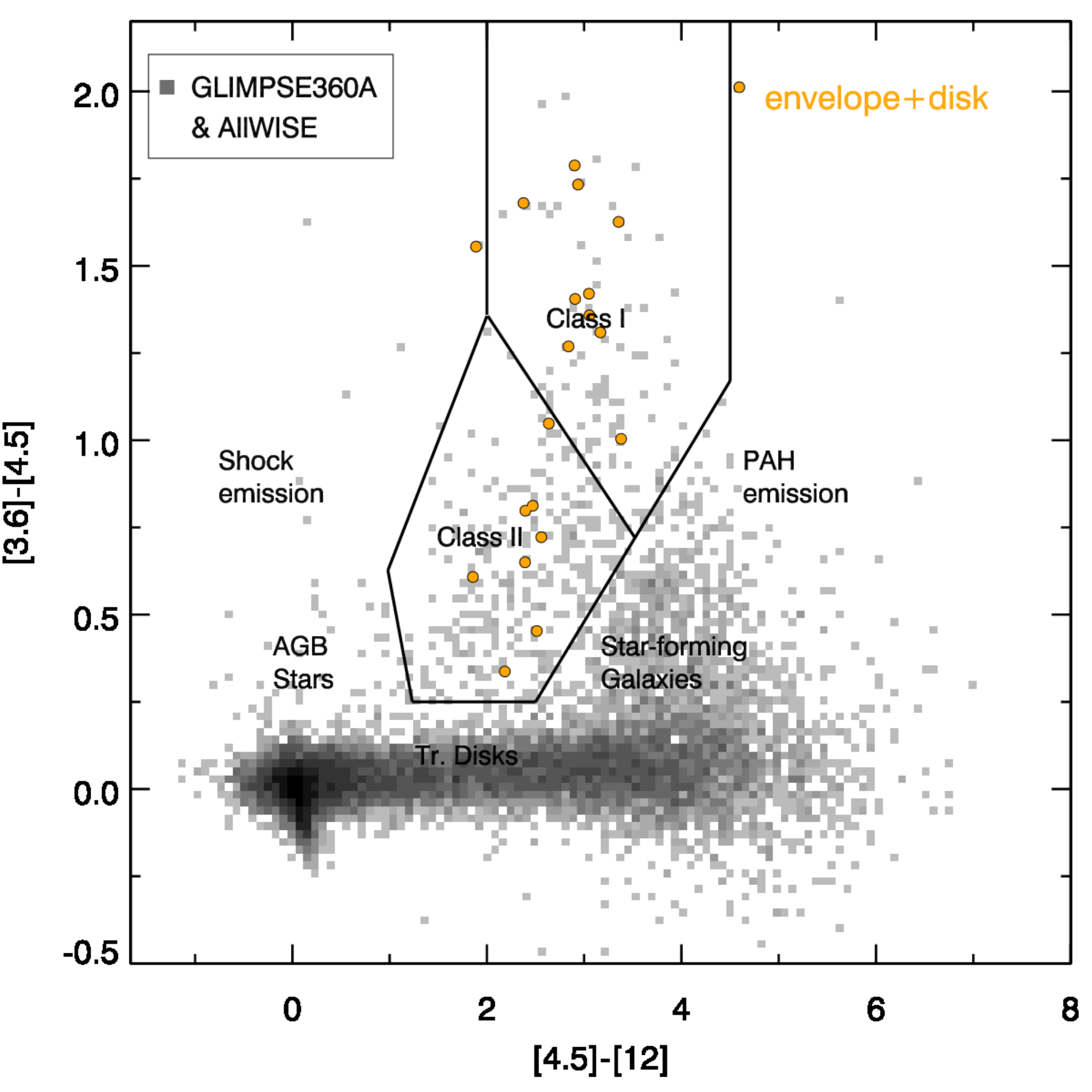}
\includegraphics[width=0.32\textwidth]{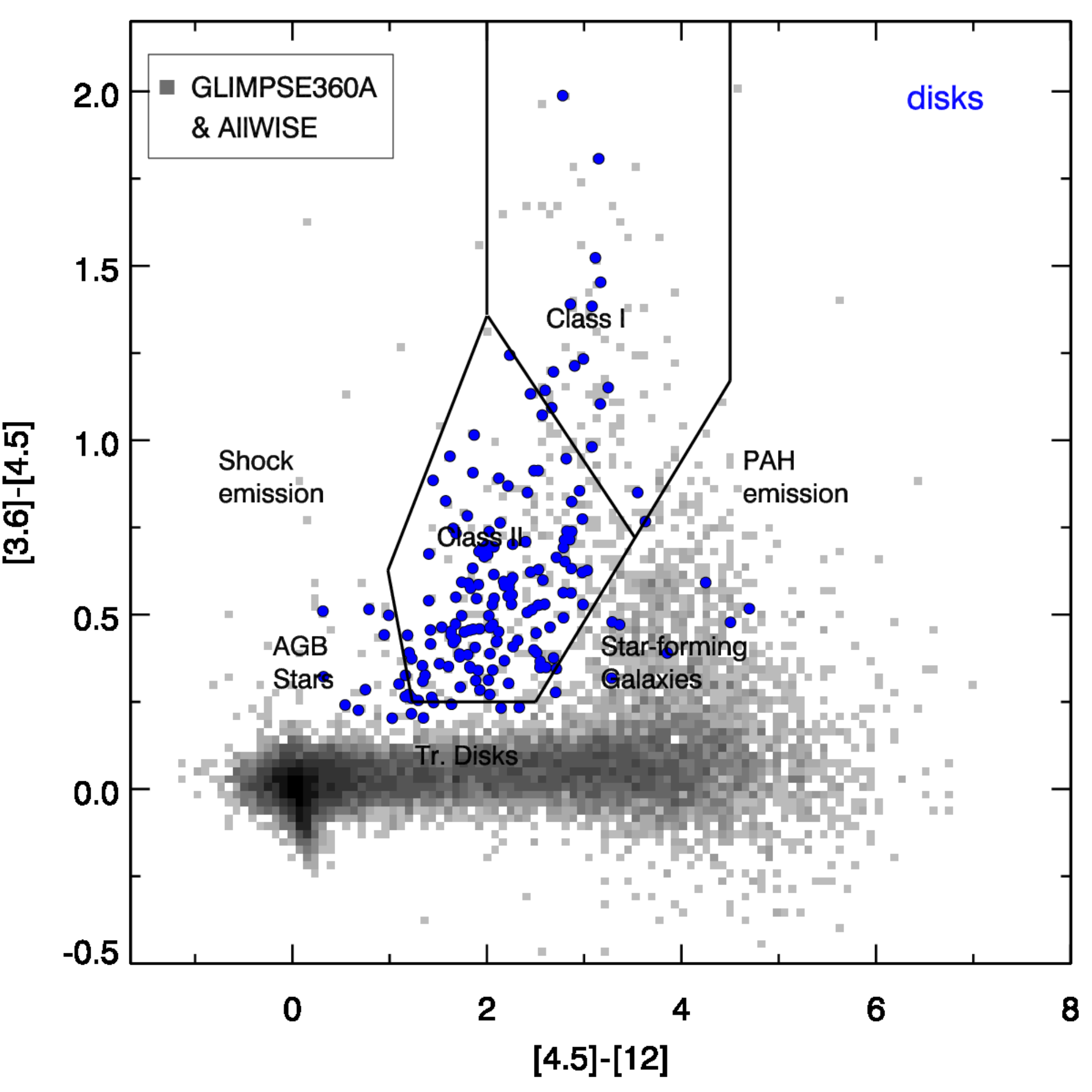}
\caption{The $[3.6]-[4.5]$ vs. $[4.5]-[12]$ CCDs showing the distribution of all GLIMIPSE360 sources with AllWISE matches with the AllWISE YSO selection criteria from \citet{fischer2016} overlaid. We use the GLIMIPSE360 3.6 and 4.5 $\mu$m bands rather than AllWISE $w1$ and $w2$ bands. The YSO candidates classified as `$e$' ({\it left}), `$d+e$' ({\it middle}), and `$d$' ({\it right}) based on the SED fitting are overlaid (see Section~\ref{s:sedfit}). The AGB stars can be found between $[3.6]-[4.5]$ = 0 and 2 and they can also be red in $[4.5]-[12]$, but we only expect a small contamination from AGB stars. \label{f:wise}}
\end{figure*}

The \citet{fischer2016} criteria, a slightly modified version of the \citet{koenig2014} criteria, use the WISE $w1-w2$ vs. $w2-w3$ CCD (see Fig.~\ref{f:wise}). We replace the WISE bands $w1$ and $w2$ with GLIMIPSE360 3.6 $\mu$m and 4.5 $\mu$m bands, respectively.

The YSO candidates from our list are over-plotted on the CCDs used for the classification: ($K_{\rm s}-[3.6]$)$_{0}$ vs. ($[3.6]-[4.5]$)$_{0}$ in Fig.~\ref{f:K363645} \citep{gutermuth2009} and $[3.6]-[4.5]$ vs. $[4.5]-[12]$ in Fig.~\ref{f:wise} \citep{fischer2016}. 

The top panel shows CCDs with YSO candidates identified in \citet[{\it upper left}]{dunham2015}, \citet[{\it upper middle}]{fang2013}, and \citet[{\it upper right}]{megeath2012}.


\section{[3.6][4.5]-only Sources}
\label{s:gl360only}

About 56\% of the GLIMPSE360 Archive sources with both 3.6 and 4.5 $\mu$m photometry do not have matches in any of the other catalogs, thus we cannot apply to them the YSO selection criteria described above. These sources may be too faint in the near-IR to be detected by 2MASS, which is not a good match for the high-sensitivity GLIMPSE360 survey. However, a non-detection at shorter wavelengths may also indicate the sources' youth, making them interesting targets for future studies. Although we cannot provide strong evidence that any of these sources are YSOs, we make an effort to identify the best candidates.   
 
In this analysis, we also include {\it Spitzer} sources that have AllWISE matches but no valid 12 and 22 $\mu$m photometry (not in the catalog or of poor quality and removed) and no matches in other catalogs. 

\begin{figure*}[ht!]
\centering
\includegraphics[width=0.32\textwidth]{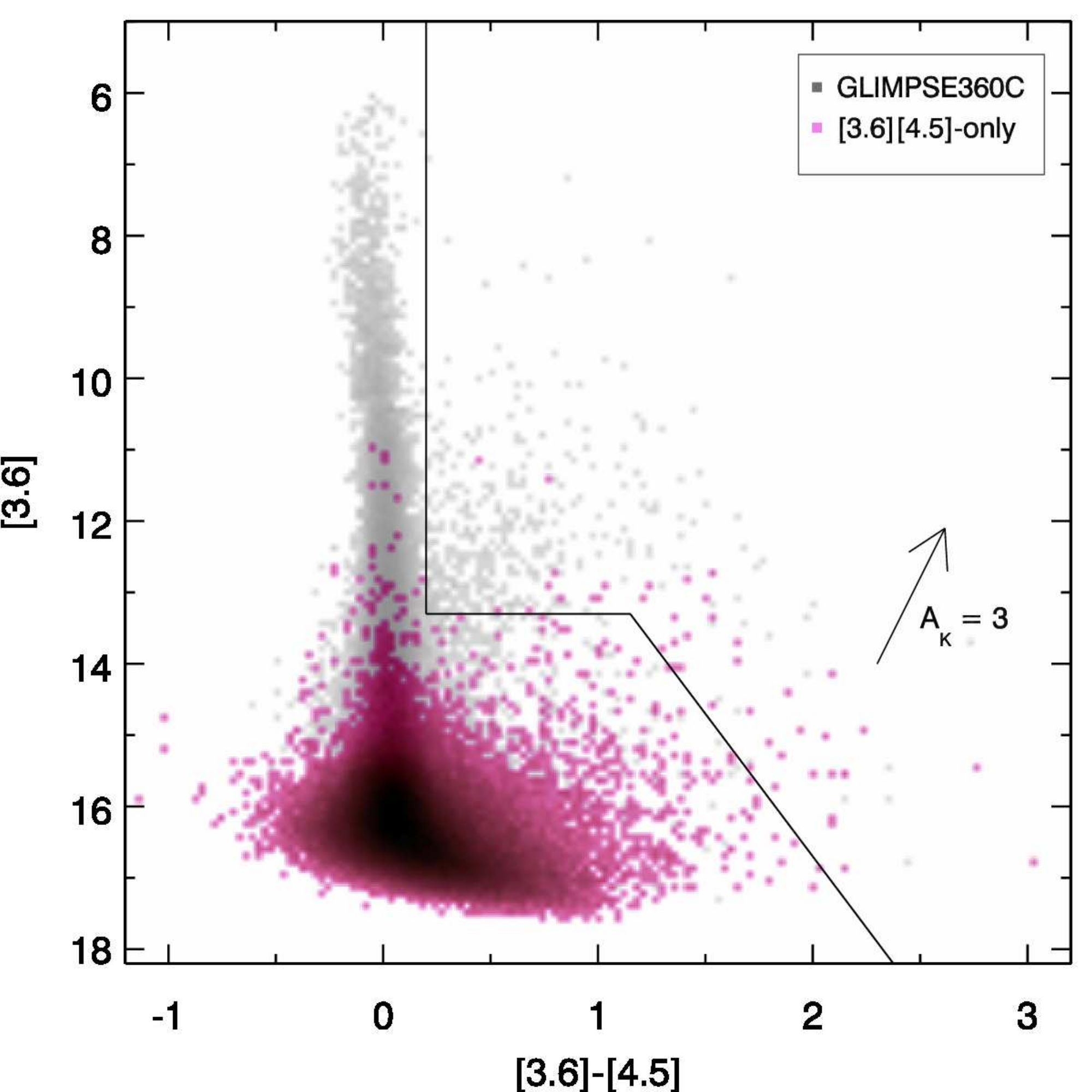}
\includegraphics[width=0.32\textwidth]{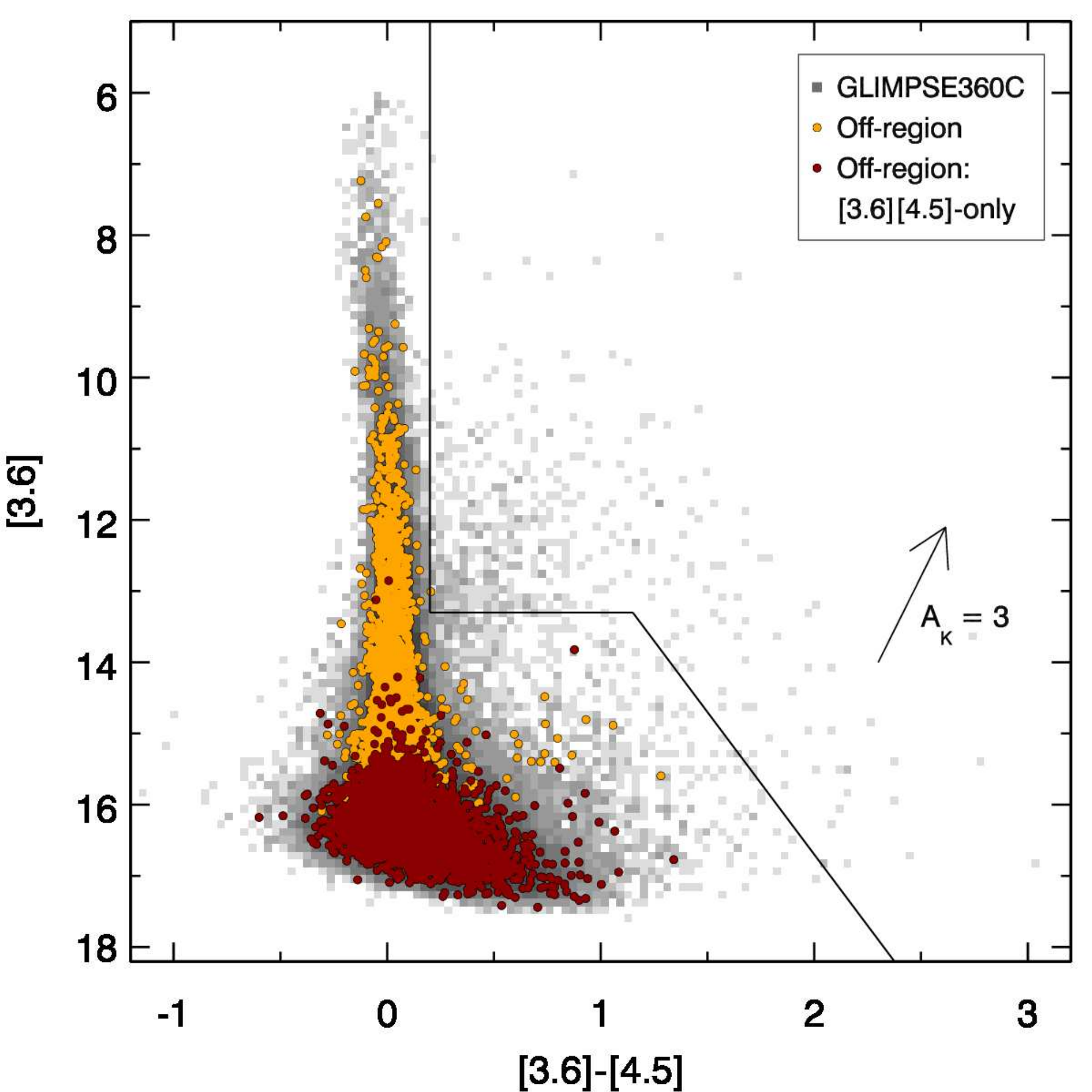}
\includegraphics[width=0.32\textwidth]{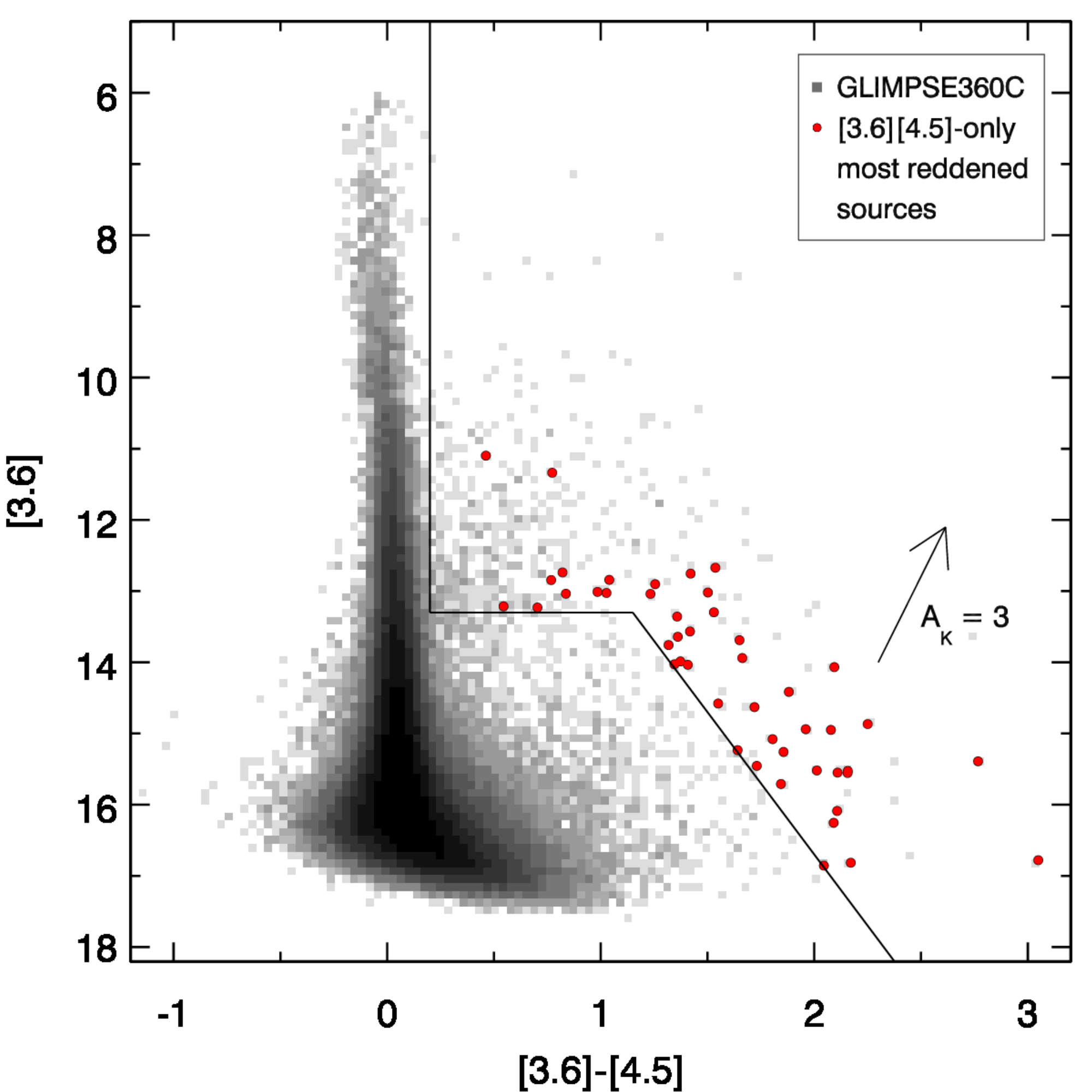}
\caption{The [3.6] vs. [3.6]-[4.5] CMDs showing the distribution of the GLIMPSE360 Catalog (rather than the less reliable, but more complete Archive) sources in CMa--$l224$ ({\it grey in both panels}). The [3.6][4.5]-only Catalog sources are shown in pink as a Hess diagram in the {\it left panel}. The {\it middle panel} shows all the GLIMPSE360 Catalog sources in the ``off-position'' (see Fig.~\ref{f:3color}) that represent the background sources in yellow and those with [3.6][4.5]-only photometry in red.  We select sources located to the right from the black solid line indicated in all panels (Eq. 2; see also Section~\ref{s:removecontam}) as potential YSO candidates (plotted in red in the {\it right panel}); more firm classification requires further investigation. See Section~\ref{s:gl360only} for details. \label{f:gl360only}}
\end{figure*}

\begin{figure*}[ht!]
\centering
\includegraphics[width=0.8\textwidth]{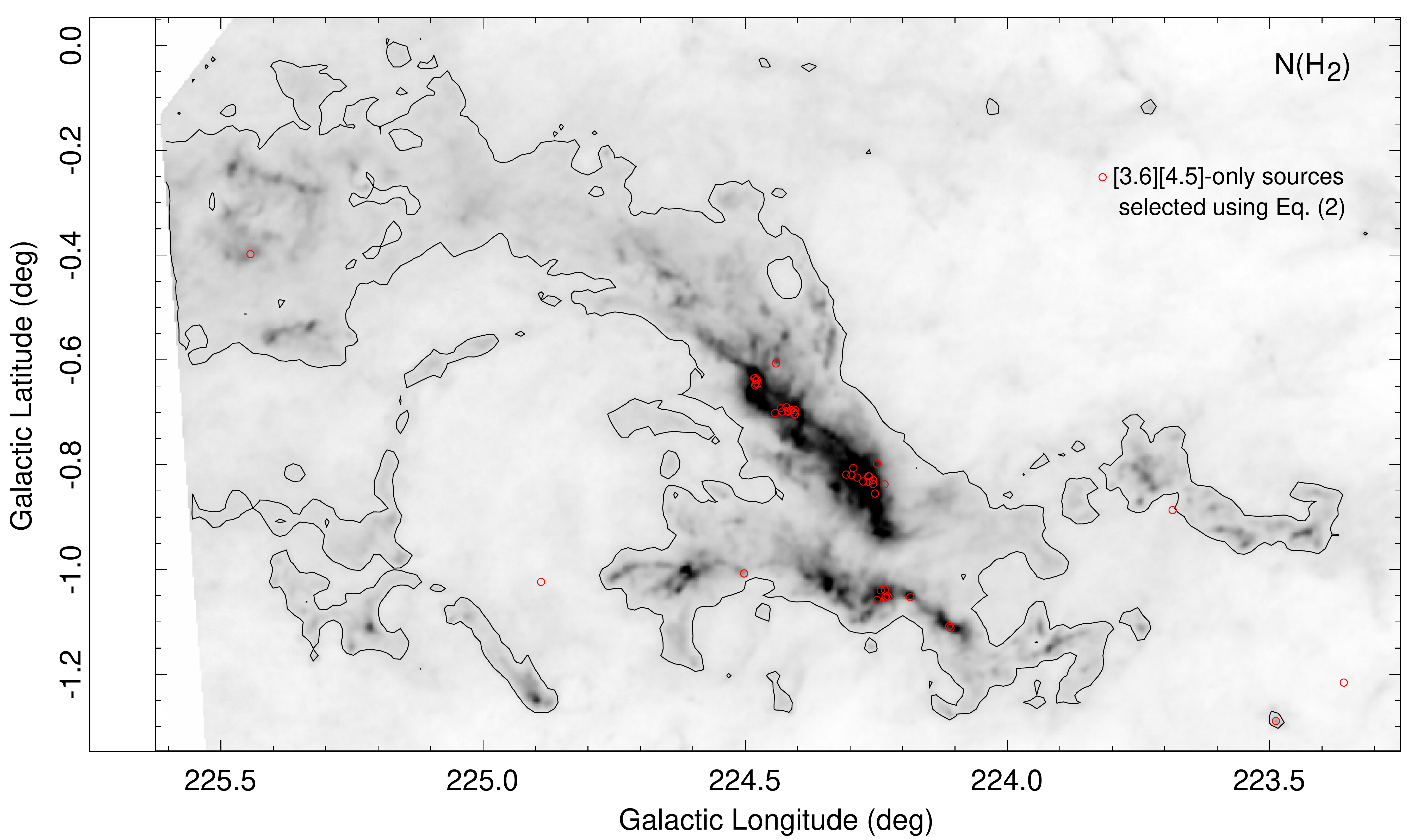}
\caption{The distribution of [3.6][4.5]-only sources selected as `possible' YSO candidates using Eq.~\ref{e:spitcmd} ({\it red circles}; see also  Fig.~\ref{f:gl360only}) overlaid on the H$_{2}$ column density image. The black contour corresponds to N(H$_{2}$) of 4.0 $\times$ 10$^{21}$ cm$^{-2}$. \label{f:gl360onlydist}}.  
\end{figure*}

We only consider GLIMPSE360 Archive sources that are also present in the highly reliable GLIMPSE360 Catalog (see Section~\ref{s:glimpse360}) to make sure all the [3.6][4.5]-only sources are legitimate astronomical objects with the highest quality photometry.  We found Catalog counterparts for 125,918 out of 133,152 (or $\sim$95\%) Archive sources; 74\% of these Catalog sources still have both 3.6 and 4.5 $\mu$m photometry. The Catalog has higher signal-to-noise thresholds and slightly more stringent acceptance criteria. Less reliable fluxes present in the Archive were removed from the Catalog. The subsequent analysis includes 92,962 sources and uses the Catalog photometry.  

The left panel of Fig.~\ref{f:gl360only} shows the distribution of all GLIMPSE360 Catalog sources in CMa--$l224$ and a subset with the 3.6 $\mu$m and 4.5 $\mu$m photometry only in the [3.6] vs. [3.6]-[4.5] CMD. For comparison, the middle panel shows the distribution of Catalog sources from the ``off-position'' in the same CMD. The ``off-position'' represents the background/foreground populations.  We apply color-magnitude cuts listed in Eq. 2 to remove background galaxies (see the right panel in Fig.~\ref{f:gl360only}). We consider the remaining 47 sources as ``possible YSO candidates''

The spatial distribution of the [3.6][4.5]-only possible YSO candidates is shown overlaid on the H$_{2}$ column density image in Fig.~\ref{f:gl360onlydist}. The vast majority of the sources are associated with the highest N(H$_{2}$) regions, coincident with clusters of YSO candidates; this location supports their classification as possible YSO candidates.

Table~\ref{t:photposs} provides the {\it Spitzer}/GLIMPSE360 IRAC Catalog photometry for [3.6][4.5]-only sources which we identified as ``possible YSO candidates''. For sources in regions covered by the H$_{2}$ column density map, the value of the N(H$_{2}$) pixel a given {\it Spitzer} source is associated with is also listed; it may be useful to identify the most interesting/reliable sources for follow-up studies. 

We searched the SIMBAD Astronomical Database to check if any of the [3.6][4.5]-only possible YSO candidates is associated with a known astronomical object. We used a search radius of 5$''$ and found only three matches. Two of the sources are associated with IRAS sources (IRAS 07067-1040 and IRAS 07077-1026) and one with the \citet{elia2013} {\it Herschel} core (G224.48274-00.63637), all at distances $>$4$''$ (see Table~\ref{t:photposs}). Therefore, we do not remove any sources based on the SIMBAD search.


\section{A Comparison to the WISE YSO Candidates List}

Ninety three WISE YSO candidates from \citet{fischer2016} are located within CMa--$l224$. We matched these sources to our final list of YSO candidates using the AllWISE designation. Twenty out of 93 WISE sources are not present (not matched to the GLIMPSE360 sources) in our list of YSO candidates. A detailed comparison between the two catalogs, including AllWISE designations, is provided in Appendix~\ref{s:F16comp}. 

Twelve of the AllWISE sources identified as YSO candidates by \citet{fischer2016} were matched to the GLIMPSE360 sources with distances larger than 1$''$; the matches were not considered real and AllWISE photometry was not used.  Two of these GLIMPSE360 sources were identified as YSO candidates based on {\it Spitzer} and 2MASS photometry and are on our final list, and four sources are on the [3.6][4.5]--only list of ``possible YSO candidates''. One [3.6][4.5]--only source does not have a GLIMPSE360 Catalog counterpart and was removed from the list of ``possible YSO candidates'' (see Section~\ref{s:gl360only} for the selection criteria). The remaining five sources (without the AllWISE photometry) do not fulfill one of our YSO selection criteria (see Appendix~\ref{s:F16comp} for details). 

Five AllWISE sources were matched to GLIMPSE360 sources within 1$''$ and both the {\it Spitzer} and AllWISE photometry was used in the analysis. Two of these sources were removed as background galaxy candidates. Three sources are on our final list of YSO candidates, but the AllWISE matches/photometry were removed during the analysis.  

Two out of 20 WISE sources do not have a {\it Spitzer} detection and one was resolved into two {\it Spitzer} sources. 

In summary, five of the GLIMPSE360 sources that were initially matched to the `missing' \citet{fischer2016} sources are on our list of YSO candidates and four are on the [3.6][4.5]-only list of  ``possible YSO candidates'' (indicated in Table~\ref{t:photposs}). One [3.6][4.5]--only source was removed due to the lack of the GLIMPSE360 Catalog counterpart. Seven GLIMPSE360 sources that were initially matched to AllWISE sources were removed from the list based on the 2MASS and GLIMPSE360 photometry. Three \citet{fischer2016} sources do not have a clear {\it Spitzer} counterpart.  

\begin{figure*}
\centering
\includegraphics[width=0.83\textwidth]{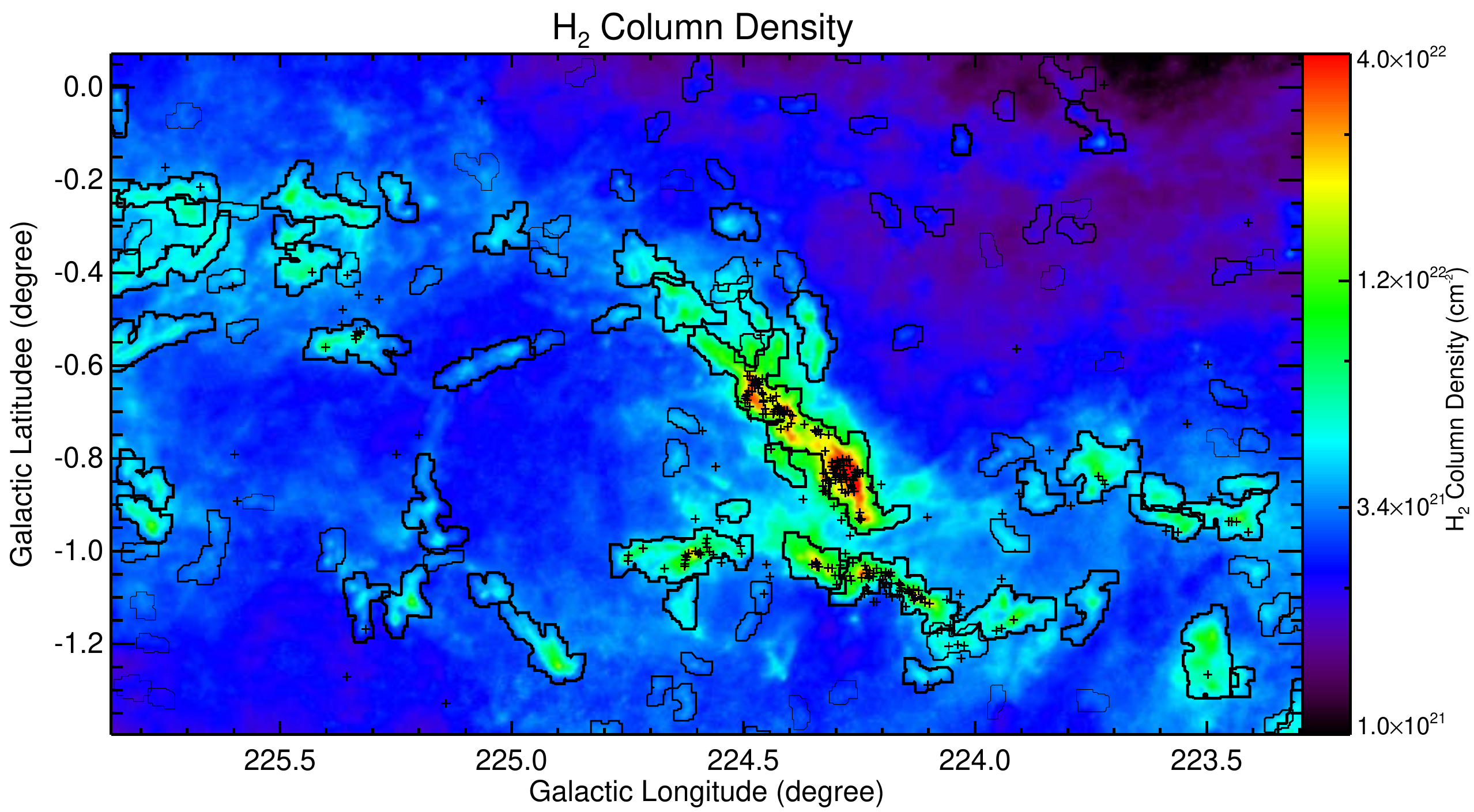}
\includegraphics[width=0.8\textwidth]{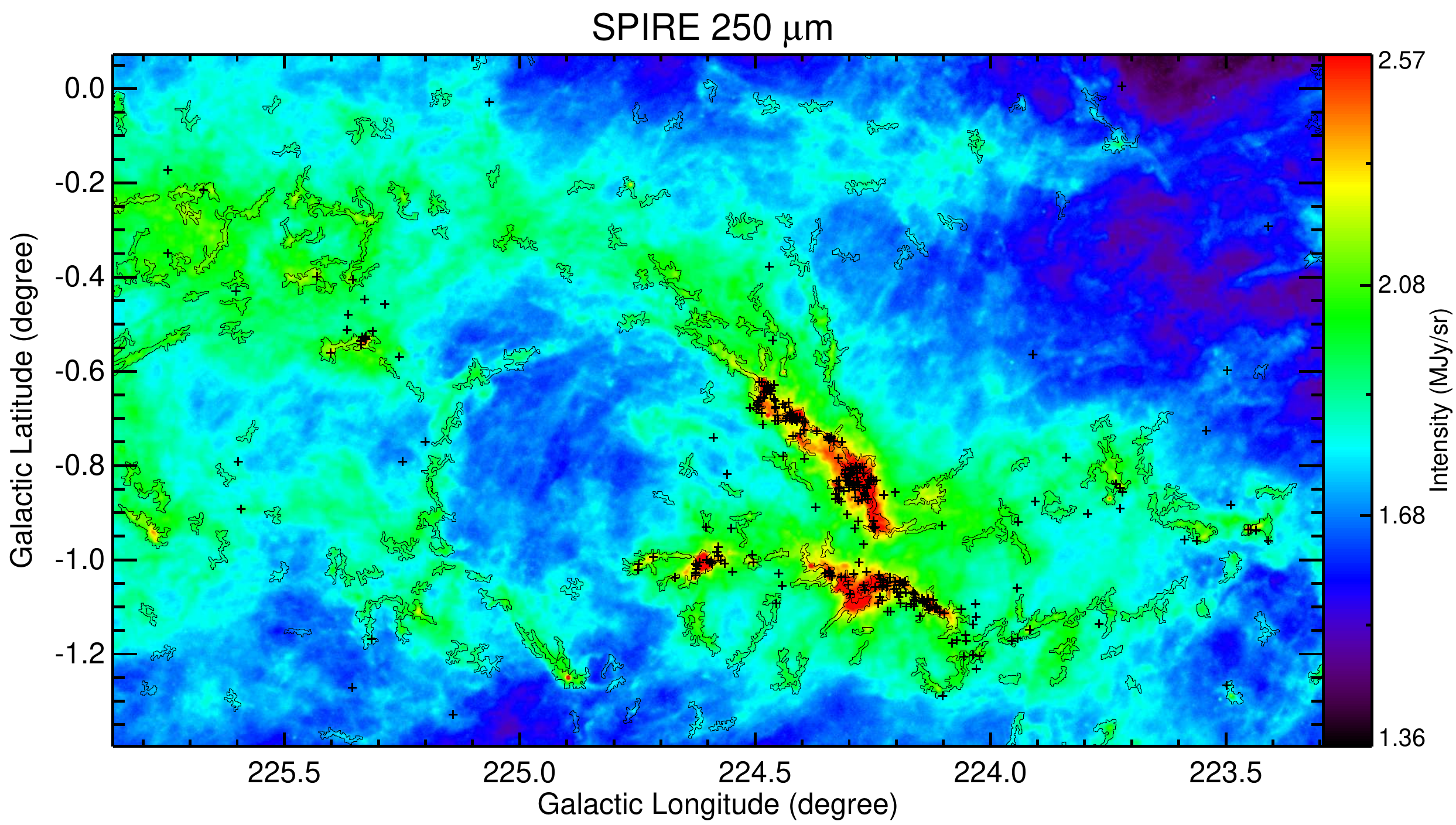} 
\caption{The distribution of YSO candidates with respect to the filaments identified based on the H$_2$ column density map ({\it top}; \citealt{schisano2014}) and the {\it Herschel} 250 $\mu$m image ({\it bottom}; this paper and E. Schisano, submitted). The boundaries of the filaments are indicated with black contours and the positions of YSO with black `$+$' symbols. The vast majority of the YSO candidates are associated with the three brightest filaments in the region. \label{f:fullfilam}}
\end{figure*}


\section{YSO Candidates vs. {\it Herschel} Filaments}

A wealth of filamentary structures are detected in CMa--$l224$ as in other regions in the Galaxy \citep{molinari2010}. Filaments can be identified in the {\it Herschel} images using algorithms designed to extract features with a filamentary shape (e.g., \citealt{schisano2014}, \citealt{koch2015}).  It is worth noting that a term `filament' is widely adopted in literature, but it still lacks a precise definition. 

\citet{schisano2014} identified filaments in the region of the Galactic plane studied by \citet{elia2013} that covers CMa-$l$224.  They define a `filament' as an extended 2D feature in the H$_2$ column density map with an elongated, cylinder-like shape and a relatively high contrast with respect to the background emission. They define filament `branches' as the group of segments tracing the skeleton of the filamentary region and as a `spine', a subset of branches forming the main axis of the region. They adopted the algorithm based on the Hessian matrix to identify filament extraction algorithm as regions of the map where the shape of the emission has a downward concavity, resembling the appearance of a cylinder. \citet{schisano2014} measured the physical properties of the filaments, finding that these structures cover a wide range of lengths ($\sim$1~pc to $\sim$30~pc), widths (0.1~pc to 2.5~pc), and average column densities ($\sim$10$^{20}$~cm$^{-2}$ to $\sim$10$^{22}$~cm$^{-2}$). This work has recently been revised by \citet{schisano2018}, who developed the catalog of the filamentary structure candidates extracted from the {\it Herschel} Hi-GAL images over the entire Galactic plane \citep{schisano2018}. 

\begin{figure*}
\centering
\includegraphics[width=0.48\textwidth]{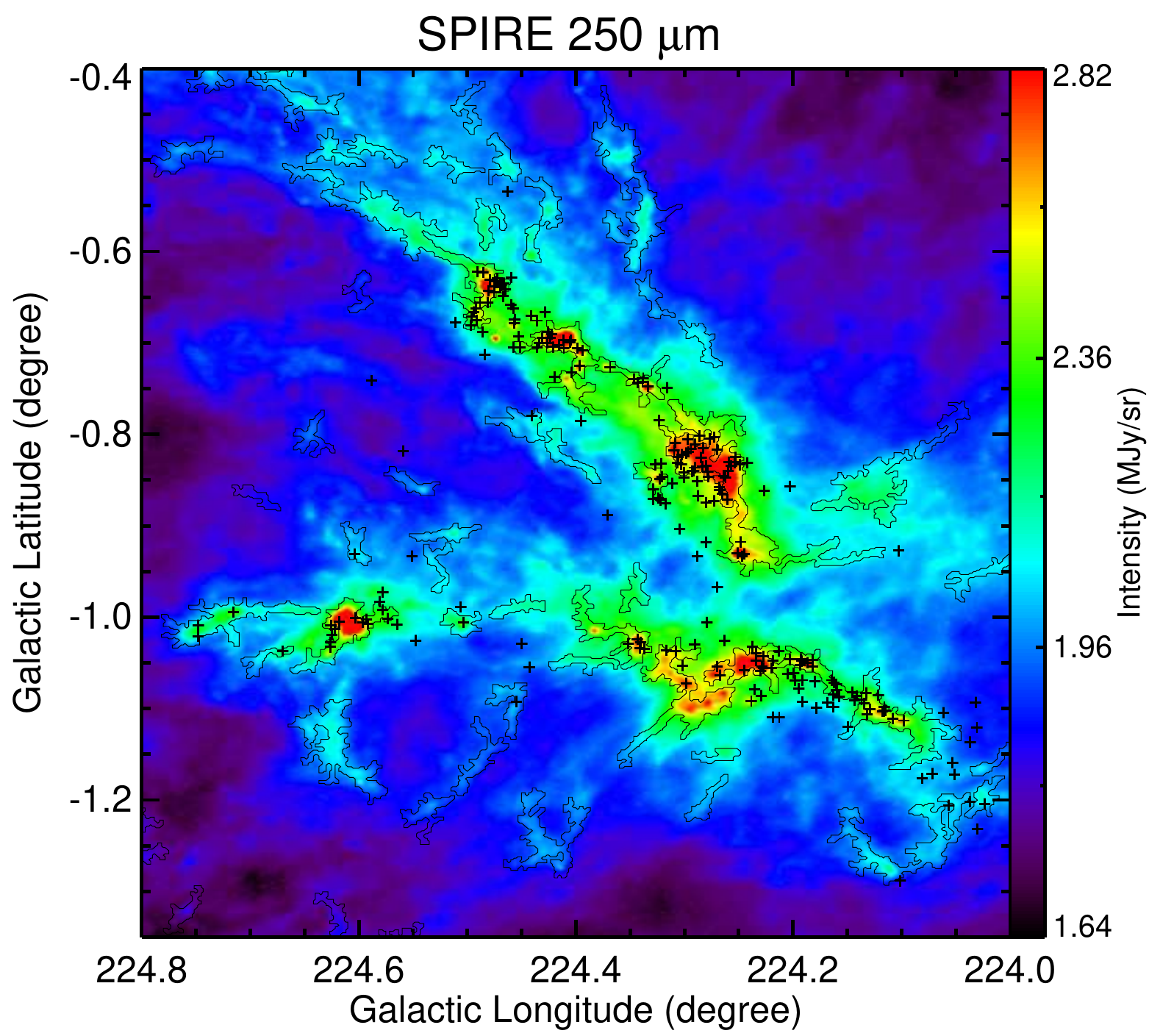}
\includegraphics[width=0.48\textwidth]{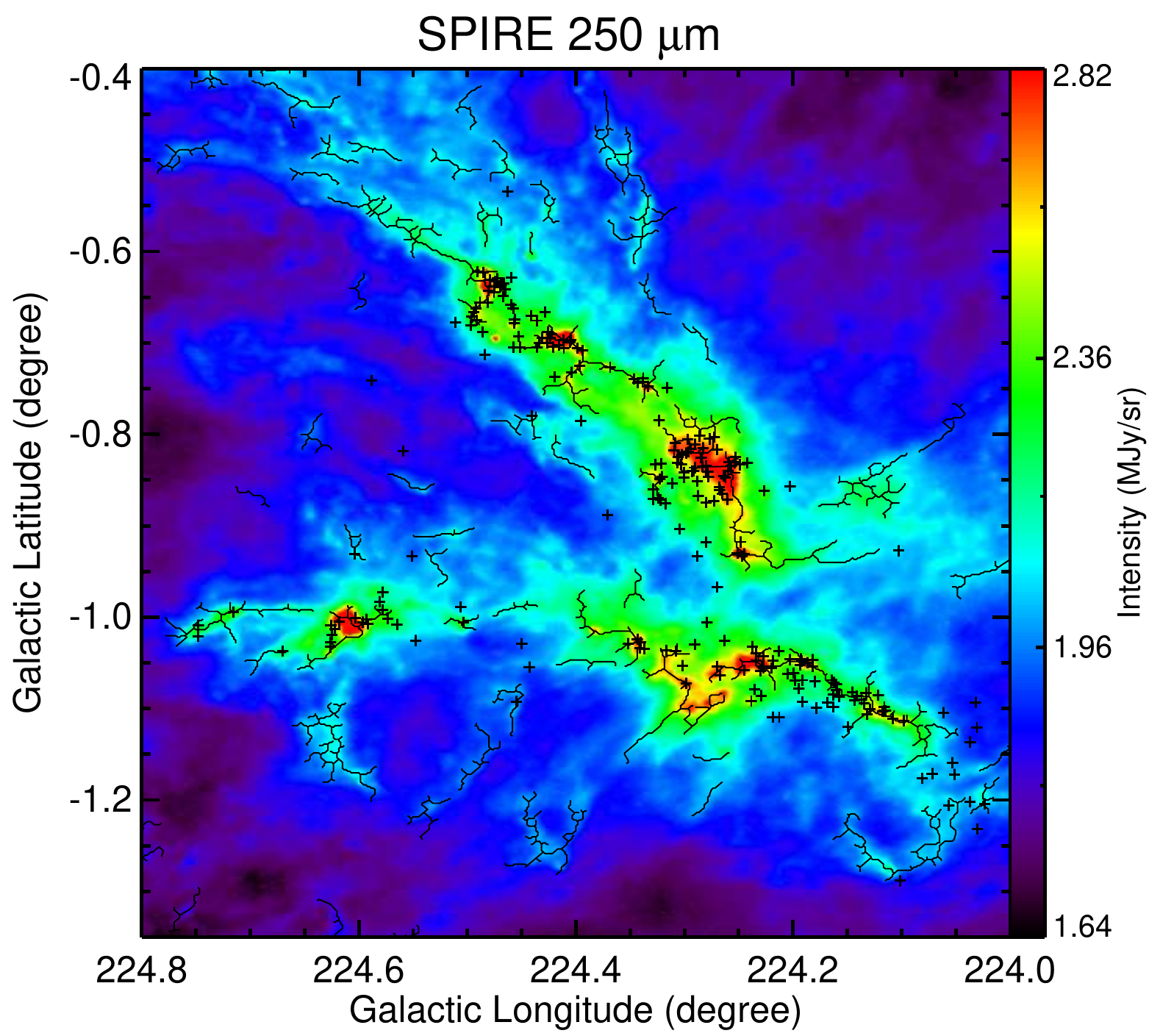}
\caption{{\it Left}: The same image as in the lower panel in Fig.~\ref{f:fullfilam}, zoomed-in on the three brightest filaments. {\it Right}: The 250 $\mu$m image with the filament ``skeletons'' ({\it black solid lines}) and YSO candidate positions (`$+$') indicated. \label{f:zoomfilam}}
\end{figure*}

A total of 212 filament candidates from \citet{schisano2018} are located in the region 223.3$^{\circ}\leq\,l\,\leq$ 225.8$^{\circ}$ corresponding to CMa-$l$224; they are shown in the top panel of Fig.~\ref{f:fullfilam}, overlaid on the H$_2$ column density map they were identified on. The contours represent the filament candidate borders that are extended with respect to the position traced by the algorithm in order to find the position where the filamentary emission merges with the background. The width of the contours in Fig.~\ref{f:fullfilam} is a function of contrast (`C'), i.e.,  the ratio between the intensity along the filament branches and its border \citep{schisano2018}.  The thick contours indicate features with $C>1.1$ (or contrast  $>10\%$ of the average background level), which are the most robust, while the thin contours indicate features with $C\,\leq\,1.1$.  In CMa-$l$224, we found 78, 94, and 40 features with $C\leq\,1.1$, $1.1\,\leq\,C\,\leq\,1.25$, and $C\geq\,1.25$, respectively.

The large number of structures in CMa--$l224$ indicates that the dense clouds in the region have a significant number of filamentary substructures, many of which might not be identified due to the coarse spatial resolution of the {\it Herschel} H$_2$ column density map (36$''$). The limited spatial resolution of this data set also makes the association between the filaments and {\it Spitzer} YSOs uncertain as the positions of the latter are generally known with the $\sim$1$''$ accuracy. To overcome these issues, we applied the filament identification algorithm to the {\it Herschel}/SPIRE 250 $\mu$m image, which is one of the images that clearly shows a filamentary network and has a two times higher spatial resolution than the H$_2$ column density image (18$''$). We adopt a similar set of parameters as in \citet{schisano2018}, where the extraction was done over the column density map of the entire Galactic plane.  

\begin{figure}
\centering
\includegraphics[width=0.495\textwidth]{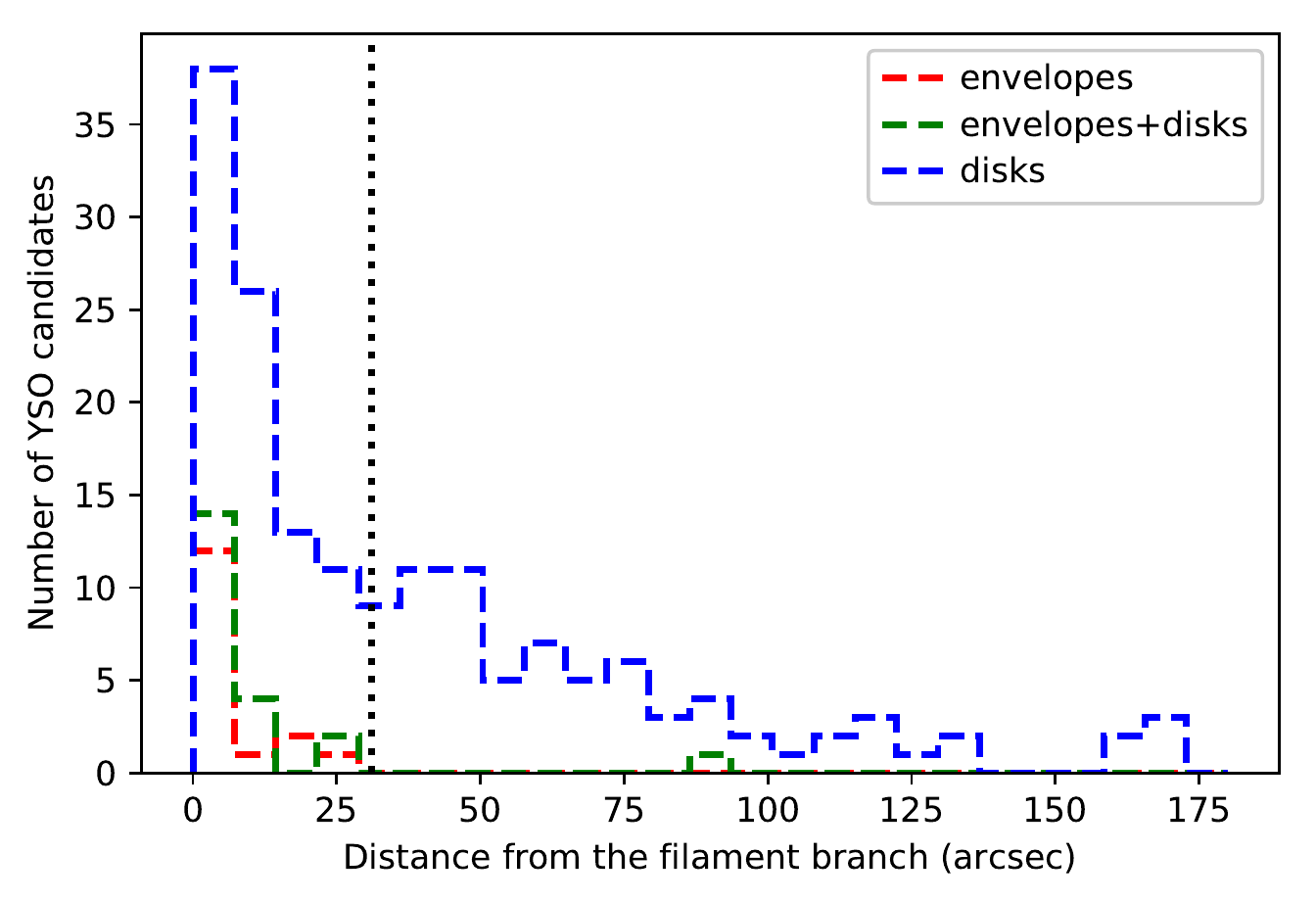}
\caption{Histograms of distances from the filament branches for YSO candidates well-fit with YSO models containing envelopes only ({\it red}), envelopes+disks ({\it green}), and disks only ({\it blue}). The vertical dotted line corresponds to 31$''$ or 0.15 pc at a distance of 1 kpc. \label{f:histfilam}}
\end{figure}

\begin{figure*}[ht!]
\centering
\includegraphics[width=1.\textwidth]{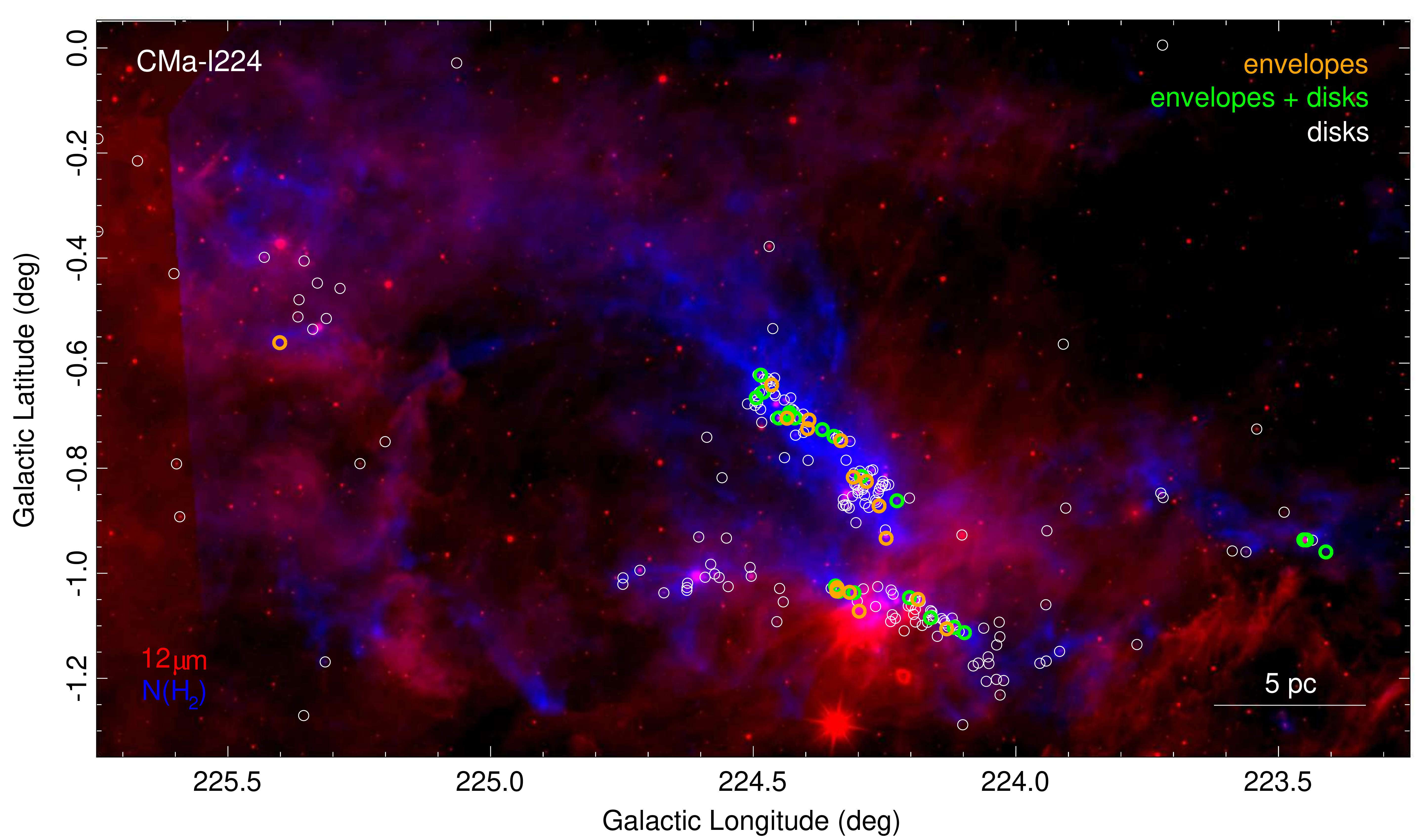}
\caption{The two color composite image combining the WISE 12 $\mu$m image ({\it red}) and the H$_2$ column density image ({\it blue}). The YSO candidates well-fit with YSO models with the envelopes-only (`envelopes'; {\it orange circles}), envelopes and disks (`envelopes + disks'; {\it green circles}), and disks-only (`disks'; {\it white circles}). The `envelopes-only' and `envelopes + disks' are concentrated in regions with the highest N(H$_2$). \label{f:distcomp}}
\end{figure*}

After visual inspection of the initial results, we decided to redo the analysis with a slightly lower threshold value.  The threshold level is given in terms of the local standard deviation of the eigenvalue of the Hessian Matrix.  We lowered the threshold from 3 times the standard deviation used by \citet{schisano2018} to 2.7. The lower threshold level adopted in this work slightly changes the number of identified structures, but more importantly merges adjacent regions that appeared to be connected to a visual inspection.  The goal of the filament extraction based on the 250 $\mu$m image is to improve the reliability of associating the {\it Spitzer} sources with {\it Herschel} filaments rather than determining the physical parameters of the latter, thus we do not extend the filament borders to encompass the entire filament emission, but we only trace the innermost regions of the filamentary structures where the emission has a downward concavity. This approach allows us to better estimate the position of the branches forming the skeleton of the region.

The bottom panel of Fig.~\ref{f:fullfilam} shows the {\it Herschel} 250 $\mu$m image with contours outlying the filamentary structures identified using the method described above. There is a good correspondence between the features extracted based on the H$_2$ column density map (the top panel of Fig.~\ref{f:fullfilam}) and those based on the 250 $\mu$m image, but the latter shows more substructures as a result of the better spatial resolution. Figure~\ref{f:zoomfilam} shows a zoom-in on the densest clouds in the center of CMa--$l224$ with the positions of the {\it Spitzer} YSOs and {\it Herschel} filaments extracted from the 250 $\mu$m image (left panel) and the filament branches (right panel) indicated. The vast majority of the YSO candidates are located in the proximity of the filament branches, indicating that stars likely form along the filaments as suggested by \citet{andre2014}. The dense cores, the precursors of the stars, are mostly found along or at the intersection of filaments (e.g., \citealt{konyves2015}).

To investigate whether our results are consistent with this star formation scenario, we measure the YSO distances from the filament branches. Figure~\ref{f:histfilam} shows the distance histograms color-coded by the YSO evolutionary stage as defined in Section~\ref{s:sedfit}: `envelope-only' (the youngest YSOs dominated by the emission from the circumstellar envelope), `envelope and disk' (more evolved YSOs with the contribution from both the envelope and the disk), and `disk-only' (the most evolved YSOs in our sample).  Our results indicate that younger YSOs, those with envelopes and envelopes$+$disks, are found within $\sim$30$''$ from filament branches. This angular distance corresponds to 0.15~pc at a distance of 1~kpc, close to the characteristic width of $\sim$0.1~pc of filaments in nearby clouds (e.g., \citealt{andre2017} and references therein). The older population of disk-only YSOs  is more spread out with a wide range of distances from the filament branches.  The spatial distribution of YSO candidates as a function of the evolutionary stage is shown in Fig.~\ref{f:distcomp}.

The fact that the younger population of YSOs is closely associated with the central regions of the filaments while the oldest YSOs are more widely distributed supports the idea that stars are formed in  the filaments and become more dispersed with time. Similar results are found in nearby star-forming regions like Taurus \citep{hartmann2002}, Orion \citep{salji2015}, Lupus \citep{benedettini2015}, and Corona Australis \citep{bresnahan2018}.


\section{Maser, Molecular Line, and Continuum Observations from Literature}
\label{s:surveys}

To put the results of our work into broader context, we collect information from literature on the maser, molecular line, and continuum observations overlapping with the CMa--$l224$ region.

\subsection{Water Masers and SiO Emission}
\label{s:waterSiO}

Two 22 GHz H$_{2}$O masers from the Arcetri Catalog are within the boundaries of CMa--$l224$ (\citealt{valdettaro2001}; 32-m Medicina telescope, HPBW$\sim$1$\rlap.^{'}$9). They were observed toward two IRAS sources: IRAS\,06040-1241 and IRAS\,06011-1445. IRAS\,06040-1241 is associated with the main filament, while IRAS\,06011-1445 is located close to the southern boundary of the region. The latter source is classified in SIMBAD as a star (BD-14 1318, TYC 5361-812-1); it is saturated in {\it Spitzer} and WISE images and coincides with a {\it Herschel} source. 

\citet{harju1998} detected SiO ({\it J}=2--1) and ({\it J}=3--2) emission toward the position of the H$_{2}$O maser in the main filament with the SEST telescope (HPBW$\sim$57$''$ and 40$''$, respectively). The detection of the SiO emission indicates the presence of shocks.

\subsection{CS Emission}
\label{s:CS}

The CS (2--1) emission was detected toward five IRAS sources with the SEST telescope (HPBW$\sim$50$''$) in the \citet{bronfman1996} survey of the IRAS point sources with the characteristics of UC H\,{\sc ii} regions. All of the IRAS sources detected in CS are associated with {\it Herschel} 
 filaments (see Fig.\ref{f:distrYSOsKnown}). The correlation between the CS emission and the most active star formation sites is expected since CS is a tracer of the high molecular gas column density.

\subsection{Planck Cold Cores}
\label{s:planck}

\begin{figure*}[ht!]
\centering
\includegraphics[width=0.45\textwidth]{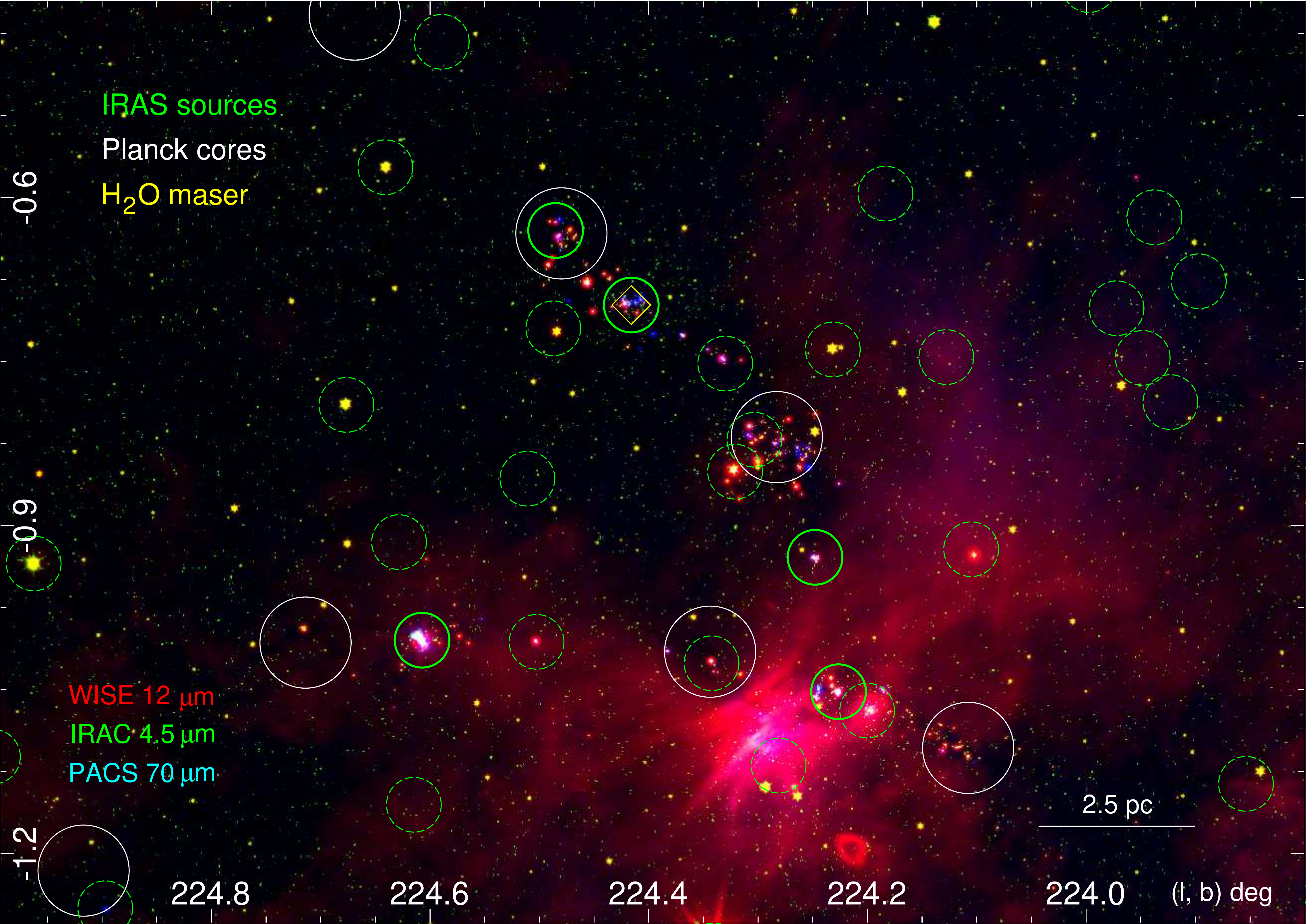}
\includegraphics[width=0.45\textwidth]{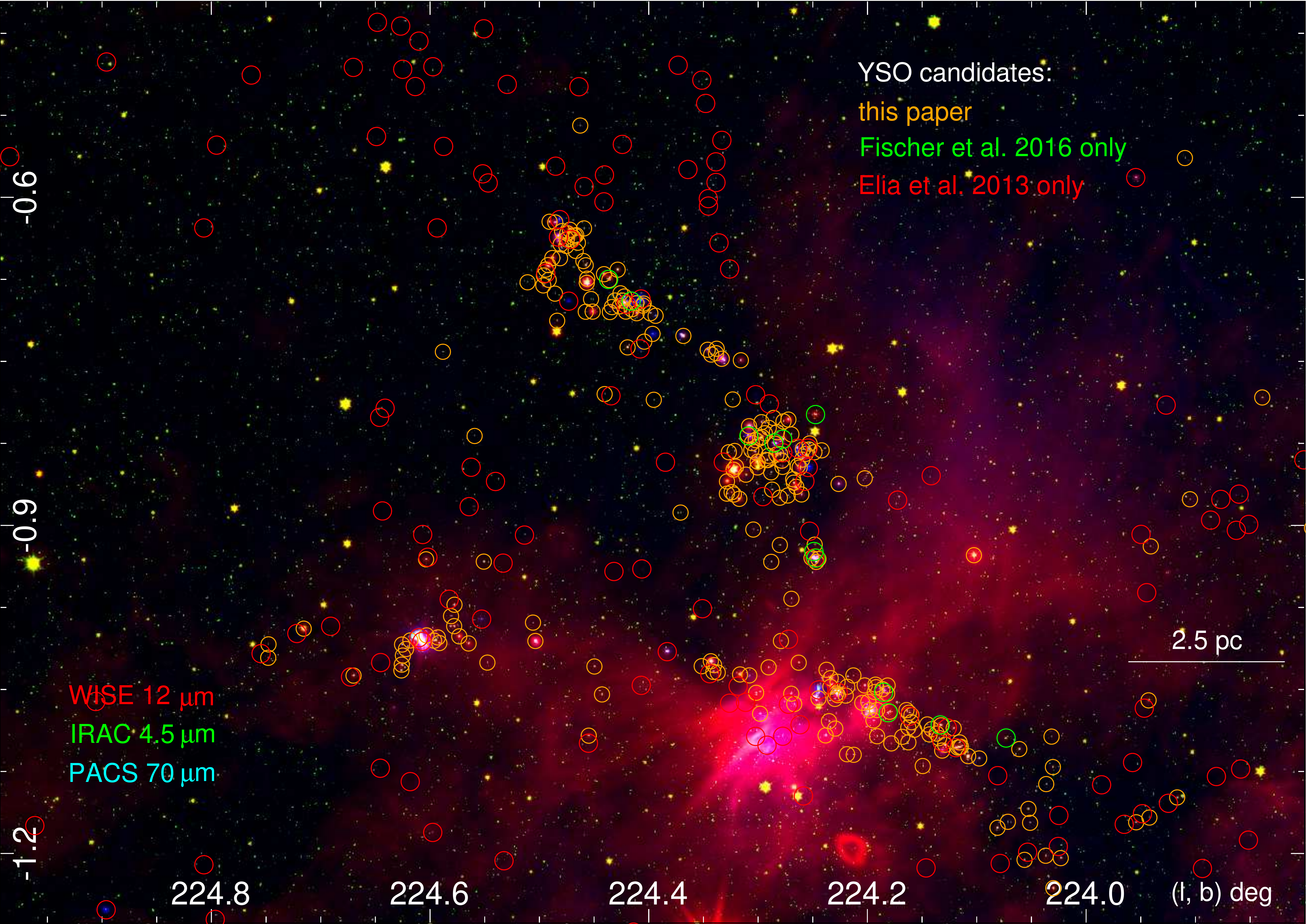}
\caption{{\it Left and right}: The three color composite image combining the WISE 12 $\mu$m ({\it red}), {\it Spitzer}/GLIMPSE360 4.5 $\mu$m ({\it green}), and {\it Herschel}/Hi-GAL PACS 70 $\mu$m ({\it blue}) images. The position of the IRAS sources, Planck cold cores, and a H$_2$O maser ({\it left}), and YSO candidates from this paper, those from \citet{fischer2016} only and \citet{elia2013} only are overlaid as indicated in the legends. The green solid circles in the left panel indicate IRAS sources associated with the CS emission. See Section~\ref{s:surveys} for details. \label{f:distrYSOsKnown}}
\end{figure*}

Sixteen cold cores from the {\it Planck} Catalogue of Galactic Cold Clumps are located in CMa--$l224$ (\citealt{planck2016}; 350, 550, and 850 $\mu$m, HPBW $\sim$ 4$\rlap.{'}$8, 4$\rlap.{'}$7, and 4$\rlap.{'}$3, respectively). As expected, {\it Planck} cores (extracted from the images convolved to a common 5$'$ resolution), tracing cold dust within dense regions in the molecular clouds, are well-correlated with the {\it Herschel} emission. Two of the clumps overlap with regions with a high concentration of outflows traced by the 4.5 $\mu$m emission in the main {\it Herschel} filament: PGCC\,G224.28-0.82 and PGCC\,G224.48-0.63 (see Fig.\ref{f:distrYSOsKnown}). \citet{zahorecz2016} identified PGCC\,G224.48-0.63 as a potential site of massive star and cluster formation based on the empirical mass-size threshold of massive star formation proposed by \citet{kauffmann2010}.

\subsection{CO Emission}
\label{s:CO}

\citet{olmi2016} selected the two most prominent {\it Herschel}/Hi-GAL filaments from \citet{elia2013} for follow-up CO observations. These filaments are located in CMa--$l224$ and they were mapped in $^{12}$CO, $^{13}$CO, C$^{17}$O, and C$^{18}$O ($J$=1--0) lines with the MOPRA radio telescope at a 38$''$  resolution.  \citet{olmi2016} studied the main filament associated with the largest number of the Hi-GAL protostellar clumps and the nearby, more quiescent (`secondary') filament associated with starless clumps (see Fig.~\ref{f:hercores} and \ref{f:CO}). Our results confirm the evolutionary status of these filaments; the main filament is associated with a large number of YSO candidates including the youngest sources in the region, while no YSO candidates are identified in the secondary filament.  

\citet{olmi2016} found that the {\it Herschel} protostellar clumps are more luminous and more turbulent than the starless clumps and they lie in regions where the ambient gas has broader lines. 

\citet{olmi2016} observations cover two out of three regions associated with the extended 4.5 $\mu$m emission (likely outflows from YSO candidates) and {\it Herschel} protostellar clumps/cores; all three regions are indicated in Fig.~\ref{f:CO} where the CO emission is compared to the 12 $\mu$m and 4.5 $\mu$m emission (see also Fig.~\ref{f:outflows}). \citet{olmi2016} detected the velocity gradients along the filaments; however, they did not find any strong evidence for  accretion flows feeding these two main YSO candidate/{\it Herschel} cores/clumps clusters located in the main filament.  

Based on the analysis of the gravitational stability of the filaments, \citet{olmi2016} concluded that the main, higher density filament (thermally supercritical and gravitationally bound) is at a later stage of evolution than the secondary filament (unbound or near to virial equilibrium and hosts less turbulent ambient gas), consistent with the evolutionary stages of the associated Hi-GAL clumps.  

\section{Star Formation in CM\MakeLowercase{a}--\MakeLowercase{{\it l}}\,{\rm 224}}
\label{s:starformation}

The presence of the H$\alpha$ expanding shell and the absence of luminous stars at the center of expansion, led \citet{herbst1977} to propose that star formation in the CMa OB1/R1 complex was triggered by a supernova event.  The main filament in CMa--$l224$ lies at the rim of the H$\alpha$ shell, hinting on a possible connection between the supernova event and star formation in the region.

The stars in CMa R1 association lie between the kinematic center of the expansion and the densest concentration of molecular gas, where star formation is on-going. \citet{herbst1977} identified a runaway star HD\,54662 (Fig.~\ref{f:cmaob1}) that could be associated with the event that produced the supernova remnant. They estimated an age of the supernova shell of 5 $\times$ 10$^{5}$ years by comparing the observed physical properties of the shell in CMa OB1/R1 complex to the models of supernova remnants evolving in a uniform medium. 

The scenario of the supernova-induced star formation was supported by a study by \citet{herbst1978} who derived similar ages for most of the members of the R1 association; the positions of stars with respect to the shell were also consistent with the model proposed by \citet{herbst1977}. However, the later study by \citet{shevchenko1999} showed that the distribution of CMa R1 candidate members in the Hertzsprung-Russell diagram indicates a large spread in stellar ages, which is inconsistent with a scenario of a single burst of star formation induced by a shock wave.  They found that only a small fraction of young stars in CMa R1 association could have been created by the supernova explosion, which indicates that the supernova-induced star formation is a minor process in this region.  \citet{shevchenko1999} further argue that even in the presence of external compression,  quiescent cloud collapse seems to be more efficient. 

An alternative scenario of the origin of the shell in CMa R1 was proposed by \citet{reynolds1978} who studied the kinematics, temperature, and ionization state of gas based on the H$\alpha$, [N\,{\sc ii}], and [O\,{\sc iii}] lines. They estimated a radius of the shell of $\sim$34 pc (assuming a distance of 1.1 kpc), a temperature of $\sim$8000 K, and an expansion velocity of $\sim$13~km~s$^{-1}$, ionized by two O stars in the CMa OB1 association. They argue that although their observations are consistent with the hypothesis that the shell was produced by a supernova explosion that triggered star formation in the region, models involving strong stellar winds or an evolving H\,{\sc ii} region are not ruled out. Another possible explanation of the nature of the shell was proposed by \citet{nakano1984}. Based on the radio continuum observations, they showed that OB stars in the area are sufficient to ionize the whole nebula. Similarly to the \citet{reynolds1978} study, \citet{nakano1984} results are still consistent with the supernova scenario. 

\begin{figure*}
\centering
\includegraphics[width=0.49\textwidth]{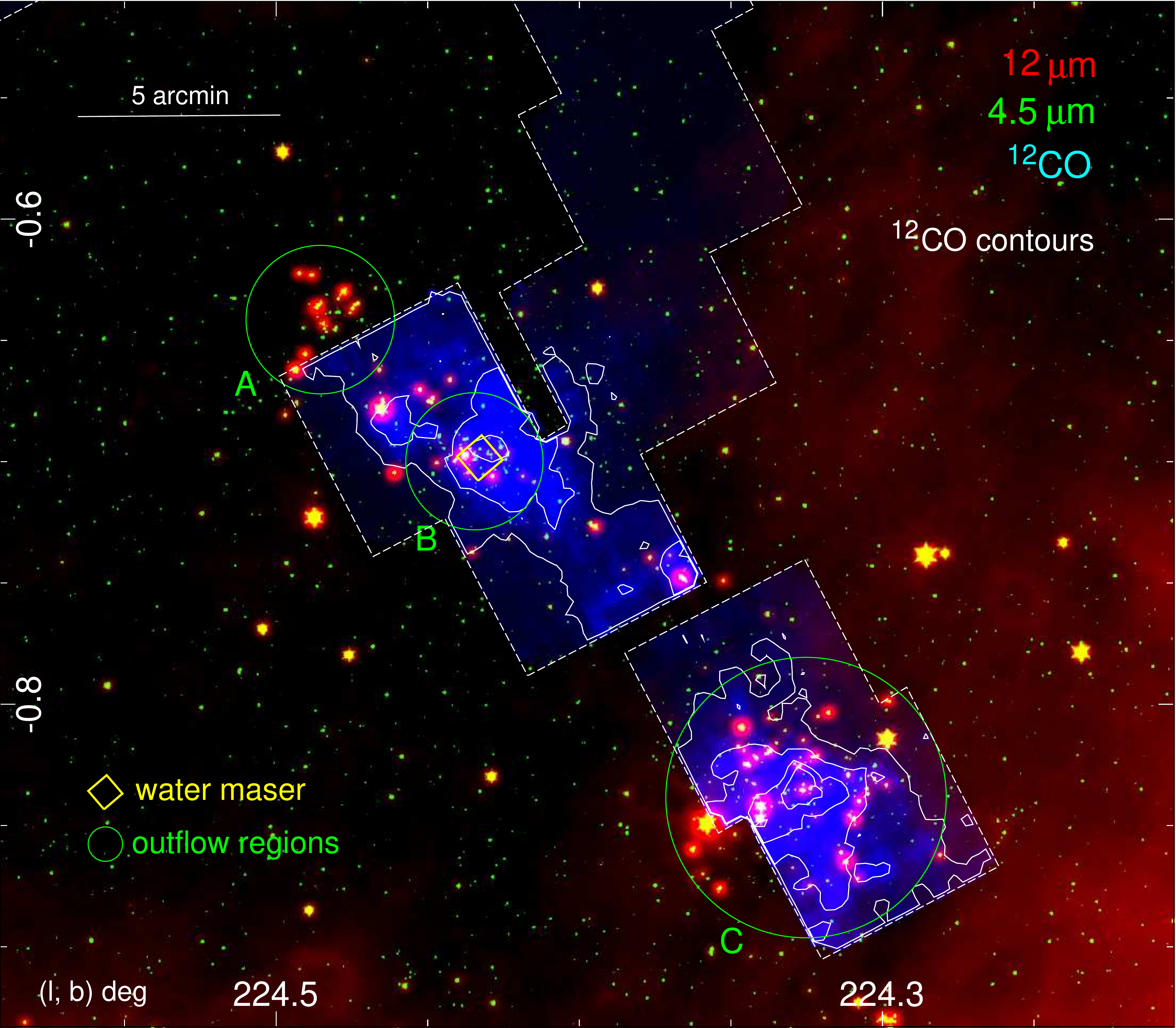}
\includegraphics[width=0.47\textwidth]{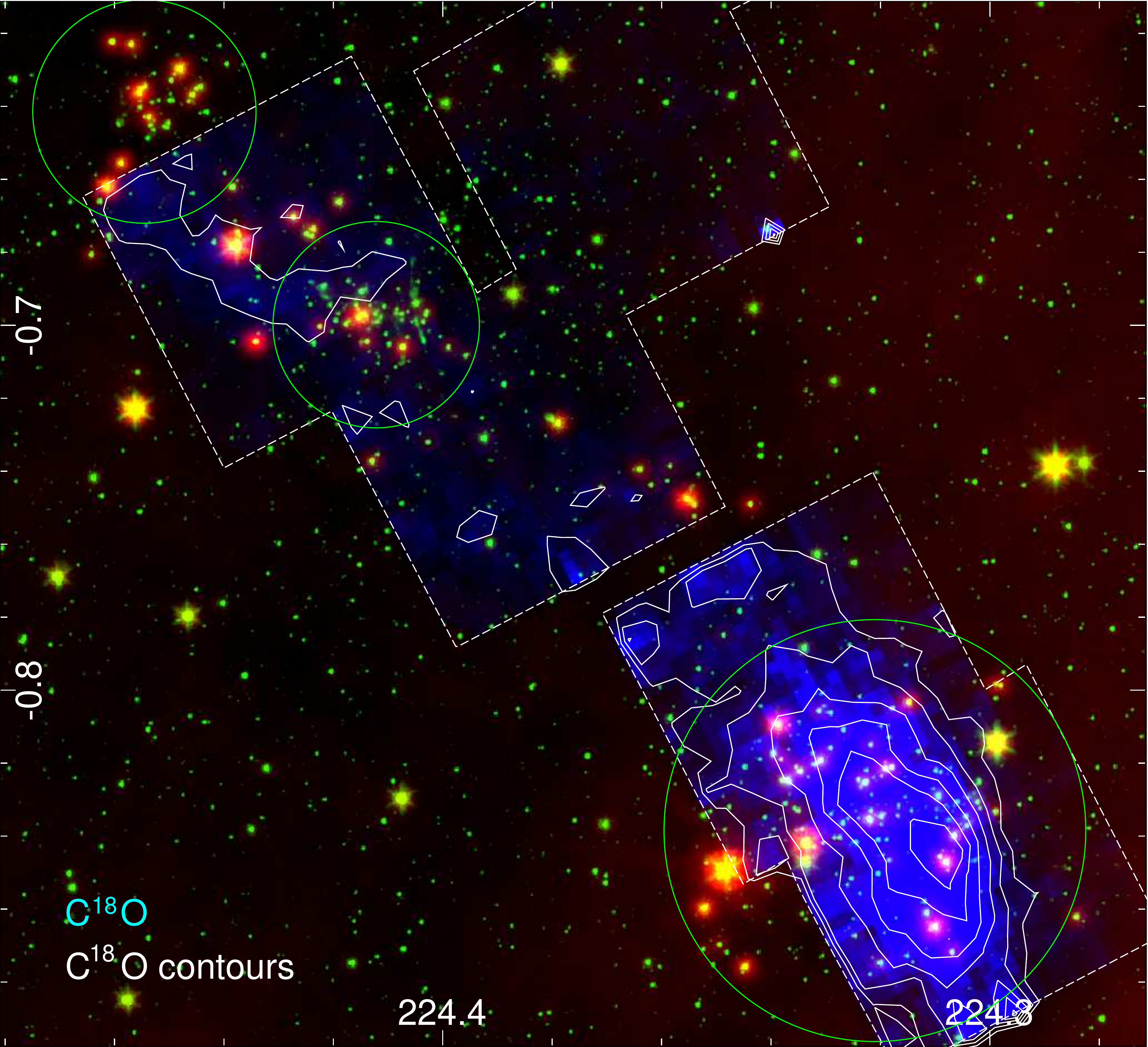} 
\caption{The three-color composite images combining the WISE 12 $\mu$m ({\it red}), {\it Spitzer}/IRAC 4.5 $\mu$m ({\it green}), and CO ({\it blue}; \citealt{olmi2016}) images: $^{12}$CO ({\it left}) and C$^{18}$O ({\it right}). The white dashed lines show the coverage of the \citet{olmi2016} CO observations.  The CO images are scaled to highlight the brightest regions. The white contours in the left panel correspond to the $^{12}$CO emission at the level of (70, 80, 90, 95)\% of the peak emission of 29.8 Jy beam$^{-1}$ km s$^{-1}$. The position of the H$_2$O maser is indicated with a yellow diamond. The image in the right panel is slightly zoomed in on the main filament. The white contours in the right panel correspond to the C$^{18}$O emission; the contour levels are (40, 50, 60, 70, 80, 90)\% of the C$^{18}$O peak of 2.9 Jy beam$^{-1}$ km s$^{-1}$. The approximate positions of the regions containing sources associated with the extended 4.5 $\mu$m emission (likely outflows) are indicated with the green circles. \label{f:CO}}.
\end{figure*}

Some further evidence supporting the supernova-induced star formation was provided by \citet{comeron1998} who studied the patterns of expanding motions in CMa OB1/R1 complex measured by the Hipparcos satellite. They found that the spatial arrangement and age of different structures in CMa OB1 association, the kinematic pattern, the existence of runaway stars, and the energetic requirements implied by the observed motions, are all compatible with the supernova explosion. \citet{comeron1998} found an expansion center of the shell at $(l,b) \sim (226\rlap.^{\circ}5, -1\rlap.^{\circ}6)$, an expansion velocity of $\sim$15~km~s$^{-1}$, and an approximate age of 1.5 Myr. They found the existence of a new runaway star revealed by its large proper motion directed away from the center of expansion, further supporting the supernova scenario.

\citet{fischer2016} studied the 10$^{\circ}$ $\times$ 10$^{\circ}$ region centered on the Canis Major star-forming region with WISE 3.4-22 $\mu$m data. They identified 144 Class I and 335 Class II YSO candidates in this 100 deg$^2$ region with 53\% of the sources organized in 16 groups of more than four members. Four out of 16 groups have more than 25 members. \citet{fischer2016} found that the distribution of the WISE YSO candidates are consistent with supernova-induced star formation; however, the Class II to Class I ratios in different regions are unexpected providing they trace age. They conclude that this quantity depends on the initial conditions if the supernova explanation is correct. 

The {\it Spitzer} YSO sources associated with extended 4.5 $\mu$m emission indicating outflows are members of the group \citet{fischer2016} identified as `Group\,00' with the largest number of members (41). Our analysis shows that Group\,00 is not a single group, but rather two (possibly three) different YSO candidate clusters.

We dub regions with the high concentration of YSO candidates in the upper, middle, and lower section of the main filament `cluster A', `cluster B', and 'cluster C', respectively (see Fig.~\ref{f:CO}, the sizes of the clusters indicated in the image are arbitrary). All the sources in these three clusters are the \citet{fischer2016}'s Group\,00 members. 

Each cluster is associated with an IRAS source (A: IRAS\,07081-1028, B: IRAS\,07077-1026. C: IRAS\,07070-1024) with the IRAS source positions offset from the cluster's center in Cluster A and C (see Fig.~\ref{f:distrYSOsKnown}). Cluster B is the most compact, while Cluster C is the most extended. Both Cluster A and C are associated with a {\it Planck} source (Section~\ref{s:planck}). Cluster B was not detected with {\it Planck}, but it is associated with the H$_2$O maser and the SiO emission (SiO observations were targeted toward the H$_2$O maser; see Section~\ref{s:waterSiO}) -- both considered as good protostellar outflow tracers, confirming the on-going outflow activity in the region. IRAS sources in Clusters A and B were observed in CS (2--1) and the CS emission was detected toward both sources (see Section~\ref{s:CS}). 

\citet{olmi2016} CO observations cover Clusters B and C, but not Cluster A (see Section~\ref{s:CO} and Fig.~\ref{f:CO}). The $^{12}$CO emission peaks coincide with Clusters B and C; the $^{12}$CO emission is brighter toward Cluster C.  The C$^{18}$O emission that traces the dense gas peaks toward this cluster, while significantly fainter C$^{18}$O emission is detected toward Cluster B (see Fig.~\ref{f:CO}). 

The properties of the YSO candidate clusters in the main filament indicate that Cluster B, with the reduced molecular gas reservoir and association with the H$_2$O maser, may be at a later evolutionary stage than Clusters A and C.

The \citet{olmi2016} CO observations indicate that Cluster B is located at the intersection of two filaments -- the main filament associated with {\it Herschel} protostellar cores and YSO candidates (including Clusters A, B, and C), and a more quiescent filament associated with pre-stellar cores, directed toward north-east from the main filament and connected with it near Cluster B's position (see \citealt{olmi2016}). \citet{olmi2016} detected a velocity gradient from northwest to southeast (from Cluster C toward Cluster B) along the main filament and along the secondary filament in the direction of Cluster B in the main filament. These velocity gradients hint at the possibility of the filamentary accretion. The star formation in the main filament might have started at the location of Cluster B where mass would be accreted from both the main and the secondary filaments.

The positional coincidence between the main filament in CMa--$l224$ and the H$\alpha$ shell indicates that there may be a connection between the supernova event in the CMa OB1/R1 association and star formation in CMa--$l224$. According to one of the theories of filament formation, the filaments are formed by the large-scale compression of the interstellar material in supersonic magneto-hydrodynamic flows, which can be turbulent or non-turbulent (e.g., \citealt{andre2017} and references therein). The waves from the supernova explosion are possible driving sources for the large-scale turbulence (e.g., \citealt{rybakin2018}), thus the filaments in CMa--$l224$ could have been formed as a result of the supernova event in the CMa OB1/R1 association. The formation of the filament is followed by the gravitational fragmentation and the formation of pre-stellar cores, the precursors of stars. Such a two-step star-formation scenario (i.e., the filament formation followed by the pre-stellar cores formation by gravitational instability) is supported by the results of the {\it Herschel} studies on filaments (e.g., \citealt{andre2017}).

One of the kinematic signatures of filament formation within shock-compressed layers generated by large-scale supersonic flows are velocity gradients in the direction perpendicular to the main axis of the filament in addition to those along the filaments (e.g., \citealt{cchen2014}; \citealt{inutsuka2015}). Such kinematic patterns were detected in the molecular line observations of the filaments (e.g., \citealt{kirk2013}; \citealt{palmeirim2013}). The transverse velocity gradients are often observed along the entire length of the filaments. The \citet{olmi2016} single-dish observations show some indication of the transverse velocity gradients in the C$^{18}$O data; however, higher resolution observations are needed to draw reliable conclusions. 

In an alternative scenario, star formation in CMa--$l224$ could have been induced by the supernova in the pre-existing filaments or, if no supernova event happened in Canis Major as some studies suggest, by the spontaneous  gravitational collapse of the filaments.  The latter scenario is supported by the ages of the YSO candidates; e.g., in the main filament, half of the YSO candidates have ages longer than the highest estimate of the SNR's age (i.e., 1.5 Myr; \citealt{comeron1998}).


\section{Summary}

Using {\it Spitzer}/GLIMPSE360 3.6 and 4.5 $\mu$m data, combined with  2MASS $JHK_s$, AllWSISE 12 and 22 $\mu$m, and {\it Herschel}/Hi-GAL 70--500 $\mu$m data, we identify YSO candidates in a 3.5 deg$^2$ region in the Canis Major OB1 association, which we dubbed CMa--$l224$. CMa--$l224$ contains the largest concentration of sources associated with extended 4.5 $\mu$m emission (an outflow tracer) in the entire GLIMPSE360 survey that coincides with bright {\it Herschel} filaments. 

Our YSO selection method is based on {\it Spitzer}--2MASS, and {\it Spitzer}--AllWISE color-color cuts, combined with the visual inspection of the images and a set of criteria to remove non-YSO contaminants. The final list of YSO candidates in CMa--$l224$ contains 293 sources. 

We identify an additional 47 sources with the GLIMPSE360 3.6 and 4.5 $\mu$m photometry only (with no AllWISE matches or with poor-quality AllWISE photometry) that we consider `possible YSO candidates'. The vast majority of these sources are associated with high H$_2$ column density regions and are good targets for follow-up studies. 

We determine kinematic distances to {\it Spitzer} sources in CMa--$l224$ using CO data and determine a median distance of 0.92 kpc, consistent with the distance measurement to the CMa OB1 association. We perform SED fitting with the \citet{robitaille2017} YSO models for sources with at least five valid flux measurements, including a 12 $\mu$m flux. Based on the SED fitting results and the PARSEC evolutionary tracks, we estimate physical parameters for 210 sources (e.g., stellar luminosities, masses, and ages). We divide sources into three groups depending on the presence of the envelope and/or disk: `envelope-only' (16), `envelope and disk' (21), and `disk-only' (173). 

The {\it Herschel}/SPIRE emission in CMa--$l224$ is highly filamentary.  We compare the distribution of the YSO candidates with respect to the filaments identified by \citet{schisano2014} based on the H$_2$ column density map and the filaments we identify based on the higher resolution 250 $\mu$m image. We find that the younger population of YSO candidates, i.e., sources with `envelope-only' and `envelope and disk',  is closely associated with central regions of the filaments, while the more evolved YSO candidates (`disk-only') are distributed more widely. This result supports the idea that stars are formed in the filaments and become more dispersed with time. 

The clumpy distribution of the YSO candidates along the filaments is a very good evidence that contamination from older objects like AGB stars is very low because those would be more evenly distributed.

We compare our list of YSO candidates to the list of YSO candidates from \citet{fischer2016} identified in Canis Major based on the WISE data. Eleven out of 93 \citet{fischer2016} WISE YSO candidates located within the boundaries of CMa--$l224$ are not on our final list of YSO candidates: two were not detected by {\it Spitzer}, one was resolved into two {\it Spitzer} sources, three were removed as non-YSOs, five did not fulfill the 2MASS--{\it Spitzer} (4) or [3.6][4.5]-only (1) criteria (the AllWISE photometry was not used since the distance between the AllWISE and {\it Spitzer} sources is larger than 1$''$).  We found that Group\,00 of WISE YSO candidates from \citet{fischer2016} corresponds to three {\it Spitzer} YSO candidate clusters distributed along the main filament (A, B, and C). The selection criteria based on the GLIMPSE360 data recovered more of the evolved YSO candidates (`disk-only').  

We discuss star formation in CMa--$l224$ in a context of the larger region -- the CMa OB1 association. The main filament in CMa--$l224$ lies at the rim of the H$\alpha$ shell, hinting on a possible connection between the supernova event in CMa OB1 association and star formation in the region. We discuss possible star formation scenarios: 1) the formation of the filaments in CMa--$l224$ as a result of the supernova explosion in the CMa OB1 association, followed by the gravitational fragmentation and the formation of pre-stellar cores; 2) supernova-induced star formation in the pre-existing filaments; 3) the spontaneous gravitational collapse of the filaments. All the star formation scenarios seem to be plausible; however, our results indicate that the spontaneous gravitational collapse of filaments is the most likely scenario.  More evidence is needed to draw stronger conclusions.

\acknowledgments

We thank the anonymous referee for reading the paper carefully and providing useful comments. We thank Leo Girardi for providing the TRILEGAL simulation.  The work of M.S. was supported by an appointment to the NASA Postdoctoral Program at the Goddard Space Flight Center, administered by Universities Space Research Association (USRA) under contract with NASA and by NASA under award number 80GSFC17M0002. This work was supported by the Polish National Science Center grant 2014/15/B/ST9/02111. This paper is based largely on observations made with the {\it Spitzer} Space Telescope, which is operated by the Jet Propulsion Laboratory, California Institute of Technology, under contract with NASA. This publication also makes use of data products from: (1) the Two Micron All Sky Survey, which is a joint project of the University of Massachusetts and the Infrared Processing and Analysis Center/California Institute of Technology, funded by the National Aeronautics and Space Administration and the National Science Foundation; (2) the Wide-field Infrared Survey Explorer, which is a joint project of the University of California, Los Angeles, and the Jet Propulsion Laboratory/California Institute of Technology, funded by the National Aeronautics and Space Administration; (3) the {\it Herschel} ESA space observatory with science instruments provided by European-led Principal Investigator consortia and with important participation from NASA;  (4) AKARI, a JAXA project with the participation of ESA; (5)  the Midcourse Space Experiment: Processing of the data was funded by the Ballistic Missile Defense Organization with additional support from NASA Office of Space Science. This research has made use of the SIMBAD database, operated at CDS, Strasbourg, France.


\bibliographystyle{aasjournal}
\bibliography{CMa_refs.bib}

\begin{deluxetable*}{clp{11cm}}
\tablecaption{Data from the 2MASS, {\it Spitzer}/GLIMPSE360, AllWISE, AKARI, MSX, and {\it Herschel}/Hi-GAL Point Source Catalogs and the Classification of YSO Candidates \label{t:catdata}}
\tabletypesize{\small}
\tablewidth{0pt}

\tablehead{
\colhead{Column} &
\colhead{Name} &
\colhead{Description} 
}
\startdata
1     &  IRACDesignation           & GLIMPSE360 IRAC Archive source name: 'SSTGLMA GLLL.llll$\pm$BB.bbbb', where LLL.llll and BB.bbbb are the galactic longitude and latitude, respectively\\
2     &  RA(J2000)                 & Right Ascension, J2000 (deg)  \\
3     &  Dec(J2000)                & Declination, J2000 (deg) \\
4     &  Objid\_2MASS              & identification number for a 2MASS source (cntr in the IRSA/GATOR) \\ 
5     &  Objid\_AllWISE            & `designation' from the AllWISE catalog \\ 
6     &  Objid\_AKARI              & `objname' from the AKARI catalog \\ 
7     &  Objid\_MSX                & `MSX6C\_ID' from the MSX catalog  \\
8    &  Objid\_HerschelE13          & Herschel source identification number from \citet{elia2013} \\ 
9    &  Dist\_GL360-AllWISE       & a distance between GLIMPSE360 and AlLWISE matching sources (arcsec); a null value is -9.999 \\ 
10    &  Dist\_GL360-AKARI         & a distance between GLIMPSE360 and AKARI matching sources (arcsec); a null value is -9.999 \\ 
11    &  Dist\_GL360-MSX         & a distance between GLIMPSE360 and MSX matching sources (arcsec); a null value is -9.999 \\ 
12    &  Dist\_GL360-HerschelE13      & a distance between GLIMPSE360 and \citet{elia2013} {\it Herschel} matching sources (arcsec); a null value is -9.999 \\ 
13    &  Dist\_GL360-HerschelFull      & a distance between GLIMPSE360 and {\it Herschel} Full matching sources (arcsec); a null value is -9.999 \\ 
14--27 &  mag{\it i}, dmag{\it i}   & 2MASS {\it JHK$_{\rm s}$} ({\it i}=1-3), GLIMPSE360 IRAC 3.6 and 4.5 $\mu$m ({\it i}=4-5), AllWISE 12 and 22 $\mu$m ({\it i}=6-7); A null value for mag{\it i} and dmag{\it i} is 99.999. For magnitudes corresponding to flux upper limits, the magnitude uncertainties are set to -999.9.\\
28--63 & flux{\it i} , dflux{\it i} & 2MASS {\it JHK$_{\rm s}$} ({\it i}=1-3), GLIMPSE360 IRAC 3.6 and 4.5 $\mu$m ({\it i}=4-5), AllWISE 12 and 22 $\mu$m ({\it i}=6-7), AKARI 9 and 18 $\mu$m ({\it i}=8-9), MSX ({\it i}=10-13), Hi-GAL PACS 70 and 160 $\mu$m ({\it i}=14-15) and SPIRE 250, 350, and 500 $\mu$m ({\it i}=16-18) fluxes and flux uncertainties (mJy); A null value for flux{\it i} and dflux{\it i} is -999.9. For flux upper limits, flux{\it i} $>$ 0 and dflux{\it i} = -999.9. \\
64 & Classification\_SED\_Fitting & YSO classification based on the SED fitting: envelope-only (`$e$'), disk and envelope (`$d+e$'), and disk-only (`$d$') \\
65 & Classification\_Phot & Class I (`I') and Class II (`II') classification of YSO candidates based on the criteria from literature; when the 12 $\mu$m data are available, the \citet{fischer2016} criteria are used, otherwise the sources are classified based on the \citet{gutermuth2009} criteria. \\
66 & Classification\_Phot\_Ref & a flag indicating what criteria were used to determine the YSO class provided in the `Classification\_Phot' column: F16 -- \citet{fischer2016}, G09 -- \citet{gutermuth2009}\\
\enddata
\end{deluxetable*}

\begin{deluxetable*}{cccccccccccc}
\tablecaption{The {\it Spitzer} GLIMPSE360 IRAC Catalog Photometry of ``Possible YSO Candidates'' in CMa--$l224$ \label{t:photposs}}
\tabletypesize{\small}
\tablewidth{0pt}
\tablehead{
\colhead{IRAC Designation} &
\colhead{R.A.} &
\colhead{Decl.} &
\colhead{$F_{3.6 \mu m}$} &
\colhead{$\sigma_{F_{3.6 \mu m}}$} &
\colhead{$F_{4.5 \mu m}$} &
\colhead{$\sigma_{F_{4.5 \mu m}}$} &
\colhead{[3.6]} &
\colhead{$\sigma$[3.6]} &
\colhead{[4.5]} &
\colhead{$\sigma$[4.5]} &
\colhead{$N(H_2)$} 
\\
\colhead{'SSTGLMC'} &
\colhead{($^{\circ}$, J2000)} &
\colhead{($^{\circ}$, J2000)} &
\colhead{(mJy)} &
\colhead{(mJy)} &
\colhead{(mJy)} &
\colhead{(mJy)} &
\colhead{(mag)} &
\colhead{(mag)} &
\colhead{(mag)} &
\colhead{(mag)} &
\colhead{(10$^{21}$ cm$^{-2}$)} 
}
\startdata
G223.3585-01.2156 & 106.56161 & -9.82876 & 0.169 & 0.027 & 0.755 & 0.054 & 15.55 & 0.17 & 13.44 & 0.08 & 2.5     \\
G223.4879-01.2890\tablenotemark{a} & 106.55575 & -9.97743 & 2.256 & 0.282 & 3.076 & 0.088 & 12.74 & 0.14 & 11.92 & 0.03 & 8.29    \\
G223.6849-00.8865 & 107.01062 & -9.96732 & 0.221 & 0.027 & 0.782 & 0.08 & 15.26 & 0.13 & 13.4 & 0.11 & 3.88      \\
G224.1086-01.1116 & 107.00566 & -10.44693 & 8.202 & 0.404 & 10.69 & 0.417 & 11.34 & 0.05 & 10.56 & 0.04 & 22.69  \\
G224.1110-01.1081 & 107.00999 & -10.44751 & 1.349 & 0.058 & 3.531 & 0.081 & 13.3 & 0.05 & 11.77 & 0.03 & 24.75   \\
G224.1854-01.0529\tablenotemark{a} & 107.09463 & -10.48809 & 2.224 & 0.089 & 5.264 & 0.127 & 12.75 & 0.04 & 11.33 & 0.03 & 25.07  \\
G224.2321-01.0525 & 107.11684 & -10.52937 & 1.731 & 0.104 & 2.851 & 0.131 & 13.03 & 0.07 & 12 & 0.05 & 31.25     \\
G224.2282-01.0508 & 107.11656 & -10.52519 & 0.94 & 0.123 & 2.748 & 0.155 & 13.69 & 0.14 & 12.04 & 0.06 & 41.42   \\
G224.2329-01.0535 & 107.11625 & -10.53055 & 0.294 & 0.017 & 1.276 & 0.079 & 14.95 & 0.06 & 12.87 & 0.07 & 31.25  \\
G224.2337-01.0385 & 107.13019 & -10.52436 & 0.688 & 0.08 & 1.518 & 0.102 & 14.03 & 0.13 & 12.68 & 0.07 & 27.78   \\
G224.2344-00.8375 & 107.31198 & -10.43238 & 0.169 & 0.02 & 0.785 & 0.07 & 15.55 & 0.13 & 13.4 & 0.1 & 12.06      \\
G224.2417-01.0397 & 107.13287 & -10.53207 & 0.226 & 0.028 & 0.657 & 0.059 & 15.23 & 0.13 & 13.59 & 0.1 & 36.41   \\
G224.2524-00.8550 & 107.30466 & -10.4564 & 0.103 & 0.008 & 0.459 & 0.013 & 16.09 & 0.09 & 13.98 & 0.03 & 74.17   \\
G224.2485-01.0577 & 107.11986 & -10.54636 & 1.455 & 0.077 & 1.538 & 0.1 & 13.21 & 0.06 & 12.67 & 0.07 & 28.44    \\
G224.2485-00.7983 & 107.35396 & -10.42686 & 1.754 & 0.074 & 2.78 & 0.072 & 13.01 & 0.05 & 12.03 & 0.03 & 22.41   \\
G224.2560-00.8293 & 107.32951 & -10.44776 & 0.414 & 0.039 & 1.104 & 0.059 & 14.58 & 0.1 & 13.03 & 0.06 & 71.58   \\
G224.2564-00.8370 & 107.32273 & -10.45165 & 2.398 & 0.086 & 6.317 & 0.177 & 12.67 & 0.04 & 11.13 & 0.03 & 68.13  \\
G224.2636-00.8223 & 107.33935 & -10.45124 & 0.713 & 0.033 & 1.617 & 0.05 & 13.99 & 0.05 & 12.62 & 0.03 & 69.69   \\
G224.2648-00.8325 & 107.33072 & -10.45707 & 0.055 & 0.006 & 0.578 & 0.03 & 16.78 & 0.13 & 13.73 & 0.06 & 83.2    \\
G224.2650-00.8219 & 107.34042 & -10.45236 & 1.937 & 0.061 & 3.94 & 0.105 & 12.9 & 0.03 & 11.65 & 0.03 & 60.77    \\
G224.2751-00.8320 & 107.33607 & -10.46596 & 0.185 & 0.018 & 0.582 & 0.036 & 15.46 & 0.11 & 13.72 & 0.07 & 94.93  \\
G224.2862-00.8244\tablenotemark{a} & 107.34805 & -10.47232 & 0.663 & 0.029 & 2.916 & 0.092 & 14.07 & 0.05 & 11.97 & 0.03 & 82.63  \\
G224.2938-00.8062 & 107.36805 & -10.47061 & 2.049 & 0.079 & 3.417 & 0.102 & 12.84 & 0.04 & 11.8 & 0.03 & 52.83   \\
G224.2973-00.8198 & 107.35741 & -10.48002 & 1.714 & 0.117 & 2.37 & 0.09 & 13.04 & 0.07 & 12.2 & 0.04 & 60.76     \\
G224.3074-00.8188 & 107.36304 & -10.48851 & 0.981 & 0.054 & 2.198 & 0.063 & 13.64 & 0.06 & 12.28 & 0.03 & 53.08  \\
G224.4056-00.7035 & 107.51322 & -10.52242 & 0.297 & 0.036 & 1.155 & 0.152 & 14.94 & 0.13 & 12.98 & 0.14 & 41.17  \\
G224.4071-00.7001 & 107.51694 & -10.52218 & 0.088 & 0.015 & 0.388 & 0.057 & 16.25 & 0.18 & 14.16 & 0.16 & 54.81  \\
G224.4045-00.6961 & 107.51929 & -10.51803 & 1.275 & 0.064 & 2.847 & 0.094 & 13.36 & 0.05 & 12 & 0.04 & 61.7      \\
G224.4053-00.7043 & 107.51229 & -10.52257 & 0.481 & 0.054 & 1.74 & 0.14 & 14.42 & 0.12 & 12.54 & 0.09 & 36.93    \\
G224.4127-00.6954 & 107.52379 & -10.52505 & 0.395 & 0.036 & 1.232 & 0.059 & 14.63 & 0.1 & 12.91 & 0.05 & 72.02   \\
G224.4182-00.6989 & 107.52321 & -10.53149 & 0.196 & 0.027 & 1.602 & 0.117 & 15.39 & 0.15 & 12.62 & 0.08 & 68.11  \\
G224.4186-00.6969 & 107.52527 & -10.53095 & 0.317 & 0.042 & 1.612 & 0.108 & 14.87 & 0.14 & 12.62 & 0.07 & 80.2   \\
G224.4213-00.6901 & 107.53262 & -10.53017 & 0.174 & 0.019 & 0.71 & 0.067 & 15.52 & 0.12 & 13.51 & 0.1 & 47.39    \\
G224.4295-00.6986 & 107.52878 & -10.54137 & 2.047 & 0.07 & 2.656 & 0.086 & 12.84 & 0.04 & 12.08 & 0.04 & 33.02   \\
G224.4322-00.6921 & 107.536 & -10.54081 & 0.173 & 0.024 & 0.806 & 0.113 & 15.53 & 0.15 & 13.37 & 0.15 & 33.15    \\
G224.4429-00.7014 & 107.53261 & -10.55457 & 0.146 & 0.009 & 0.511 & 0.03 & 15.71 & 0.06 & 13.87 & 0.06 & 27.22   \\
G224.4412-00.6068 & 107.61717 & -10.50934 & 0.051 & 0.004 & 0.214 & 0.013 & 16.86 & 0.08 & 14.81 & 0.07 & 15.23  \\
G224.4764-00.6455 & 107.59878 & -10.55849 & 0.683 & 0.065 & 1.599 & 0.065 & 14.04 & 0.1 & 12.63 & 0.04 & 49.1    \\
G224.4817-00.6345 & 107.61111 & -10.55808 & 0.882 & 0.041 & 1.9 & 0.078 & 13.76 & 0.05 & 12.44 & 0.05 & 59.15    \\
G224.4798-00.6389 & 107.60636 & -10.55844 & 0.261 & 0.015 & 0.88 & 0.038 & 15.08 & 0.06 & 13.28 & 0.05 & 62.66   \\
G224.4780-00.6384 & 107.60594 & -10.55656 & 1.433 & 0.047 & 1.752 & 0.046 & 13.23 & 0.04 & 12.53 & 0.03 & 62.66  \\
G224.4802-00.6451 & 107.60091 & -10.5617 & 1.051 & 0.039 & 2.484 & 0.069 & 13.57 & 0.04 & 12.15 & 0.03 & 52.52   \\
G224.4827-00.6351 & 107.61112 & -10.55922 & 1.74 & 0.067 & 4.438 & 0.16 & 13.02 & 0.04 & 11.52 & 0.04 & 59.15    \\
G224.4810-00.6486 & 107.5981 & -10.56394 & 0.745 & 0.077 & 2.207 & 0.143 & 13.94 & 0.11 & 12.28 & 0.07 & 48.45   \\
G224.5022-01.0075 & 107.28408 & -10.74833 & 1.711 & 0.056 & 3.406 & 0.154 & 13.04 & 0.04 & 11.81 & 0.05 & 9.37   \\
G224.8892-01.0235 & 107.45126 & -11.09911 & 0.053 & 0.006 & 0.249 & 0.013 & 16.82 & 0.13 & 14.64 & 0.06 & 3.18   \\
G225.4435-00.3982\tablenotemark{a} & 108.27774 & -11.30138 & 10.22 & 0.245 & 10.02 & 0.219 & 11.1 & 0.03 & 10.63 & 0.02 & 8.05 \\
\enddata
\tablenotetext{a}{Sources SSTGLMC\,G224.1854-01.0529, SSTGLMC\,G224.2862-00.8244, SSTGLMC\,G225.4435-00.3982, and SSTGLMC\,G223.4879-01.2890 correspond to WISE YSO candidates J070822.75-102916.3, J070923.42-102818.8, J071306.61-111803.8, and   J070613.51-095837.9, respectively, from \citet{fischer2016}.}
\end{deluxetable*}

\clearpage

\appendix

\section{Catalogs from Literature Used for the Initial Source Selection}

We used the catalogs of non-YSOs, confirmed YSOs, and YSO candidates from literature to identify regions in the color-color and color-magnitude space where YSOs can be confused with other categories of sources. This information allowed us to select the initial list YSO candidates that is as free of contaminants as possible.  The catalogs are described below. 

\subsection{Non-YSOs}

\begin{packed_enum}

\item[--] {\it Normal galaxies and AGNs}:  AGNs and galaxies with spectroscopic redshifts from The AGN and Galaxy Evolution Survey (AGES; \citealt{kochanek2012}). The AGES catalog provides spectroscopic redshifts for over 18,000 galaxies to {\it I} = 20 mag and $\sim$5,000 AGNs to {\it I} = 22.5 mag in the 7.7 deg$^{2}$ field. The catalog also includes photometric data for the optical ({\it B$_{W}$, R}, and {\it I}), near-IR ({\it J, K}, and {\it K$_{\rm s}$}), and mid-IR (IRAC 3.5-8.0 $\mu$m and MIPS 24 $\mu$m) bands. The AGES survey does not provide the {\it H}-band photometry, which we use in the YSO selection process. To obtain the {\it H}-band photometry, we matched the AGES catalog to the 2MASS catalog and selected the nearest neighbors within 1 arcsec.  

\item[--] {\it Planetary Nebulae (PNe)}:   
We use the \citet{kimeswenger2001} catalog of southern Galactic PNe and the list of PNe identified in \citet{urquhart2009}. The \citet{kimeswenger2001} catalog includes $\sim$1000 ($>$99\%) sources from The Strasbourg-ESO Catalogue of Galactic Planetary Nebulae (and the first supplement to this catalog), but it provides more accurate optical coordinates than the original catalog.   \citet{urquhart2009} identified $>$50 PNe in the northern hemisphere as part of The Red MSX Source (RMS) survey, which is a multiwavelength observational program that aims to identify a large sample of massive YSOs in the Galaxy. \citet{urquhart2009} identified PNe based on a combination of 6 cm radio, CO, near- to far-IR data and removed them from the YSOs list. We used the General Catalog Query Engine in the NASA/IRSA Infrared Science Archive to search for the mid-IR counterparts to the PNe from \citet{kimeswenger2001} and \citet{urquhart2009} in the {\it Spitzer} GLIMPSE surveys\footnote{A zoomable web browser showing the GLIMPSE surveys can be found at www.alienearths.org/glimpse, and at www.spitzer.caltech.edu/glimpse360 for the GLIMPSE360 survey.}: GLIMPSE I, GLIMPSE II, GLIMPSE\,3D, GLIMPSE360, and Deep\,GLIMPSE \citep{benjamin2003}. We selected the nearest matches in the IRAC Archive within 1$''$. The matching provided us with the 2MASS and IRAC photometry for $\sim$120 PNe. 

\item[--] {\it Evolved stars}: The sample of 54 Galactic evolved stars come from \citet{reiter2015} and includes 48 Asymptotic Giant Branch (AGB) stars (22 O-rich, 19 C-rich, and 7 S-type) and 6 supergiants. The \citet{reiter2015} catalog provides the 2MASS and IRAC photometry. As for PNe, we searched the AllWISE catalog for the closest matches to evolved stars within 1$''$. The resulting sample of evolved stars with AllWISE photometry contains 45 sources. For all of these sources, the $w1$ and $w2$ magnitudes are upper limits, thus this sample of evolved stars is not included in CCDs and CMDs that use these bands.   

\end{packed_enum}

\subsection{Known Galactic YSOs and YSO Candidates}

\begin{packed_enum}
\item[--] {\it Photometrically selected YSOs}:  We make use of the \citet{dunham2015} catalog of YSOs identified in 18 molecular clouds covered by the  {\it Spitzer} Legacy Surveys ``From Molecular Cores to Planet-Forming Disks" (c2d; \citealt{evans2003}) and ``Gould's Belt: Star Formation in the Solar Neighborhood'' \citep{allen2007}. Using the YSO selection techniques developed for these surveys combined with the visual inspection, \citet{dunham2015} identified 2966 YSO candidates. Out of 2966, 326 (11\%), 210 (7\%), 1248 (42\%), and 1182 (40\%) are classified as Class 0 + I, Flat-spectrum, Class II, and Class III, respectively. We only include sources that are not flagged as potential AGB star contaminants. 

\item[--] {\it YSOs observed spectroscopically}: We use the catalog of $\sim$1390 YSOs in L1641 from \citet{fang2013}. The YSOs were identified using the multi-wavelength photometry (including {\it Spitzer}, WISE, 2MASS, and XMM) and optical spectroscopy. We selected sources that were either observed spectroscopically or are associated with X-ray emission (an accretion tracer).  About 72\% of sources fulfill these criteria (62\% observed spectroscopically): 38, 323, 503, 66, and 66 YSOs are classified as Stage I, Stage II, Stage II, Flat, and Transition Disks (TD), respectively. 
\end{packed_enum}

\section{Spectral Energy Distribution and YSO Model Fits for YSO Candidates}

We present SEDs and the \citet{robitaille2017} YSO model fits for YSO candidates listed in Table~\ref{t:physpar}. The symbols and lines are as in Fig.~\ref{f:exsed}. The YSO fitting is described in Section~\ref{s:sedfit}.

\begin{figure*}
\includegraphics[width=0.32\textwidth]{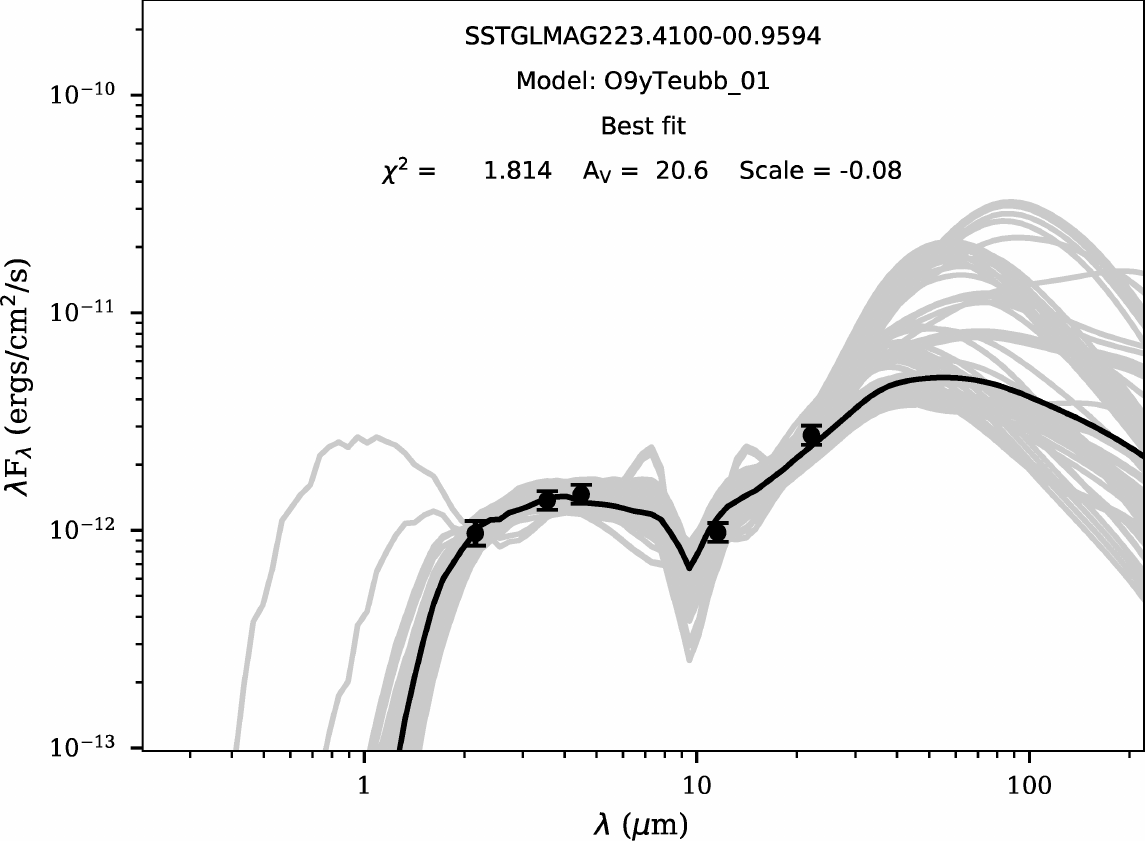}
\hfill
\includegraphics[width=0.32\textwidth]{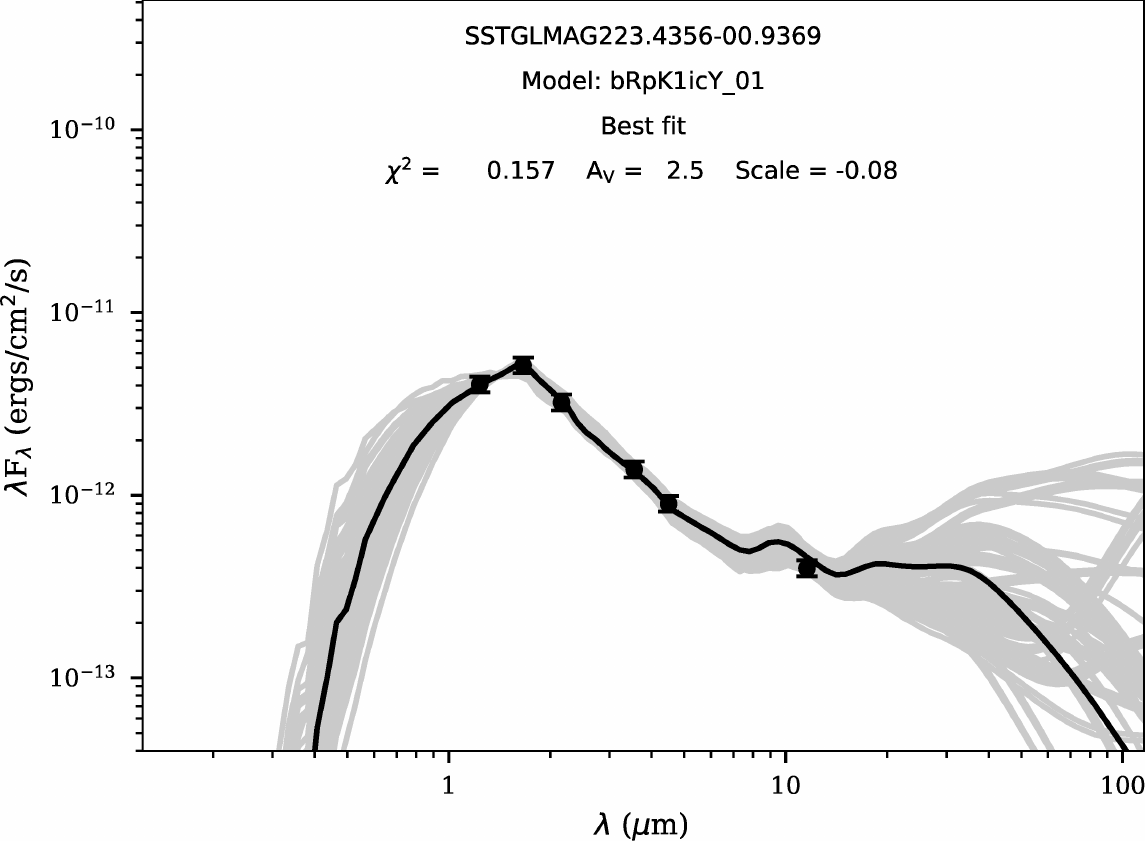}
\hfill
\includegraphics[width=0.32\textwidth]{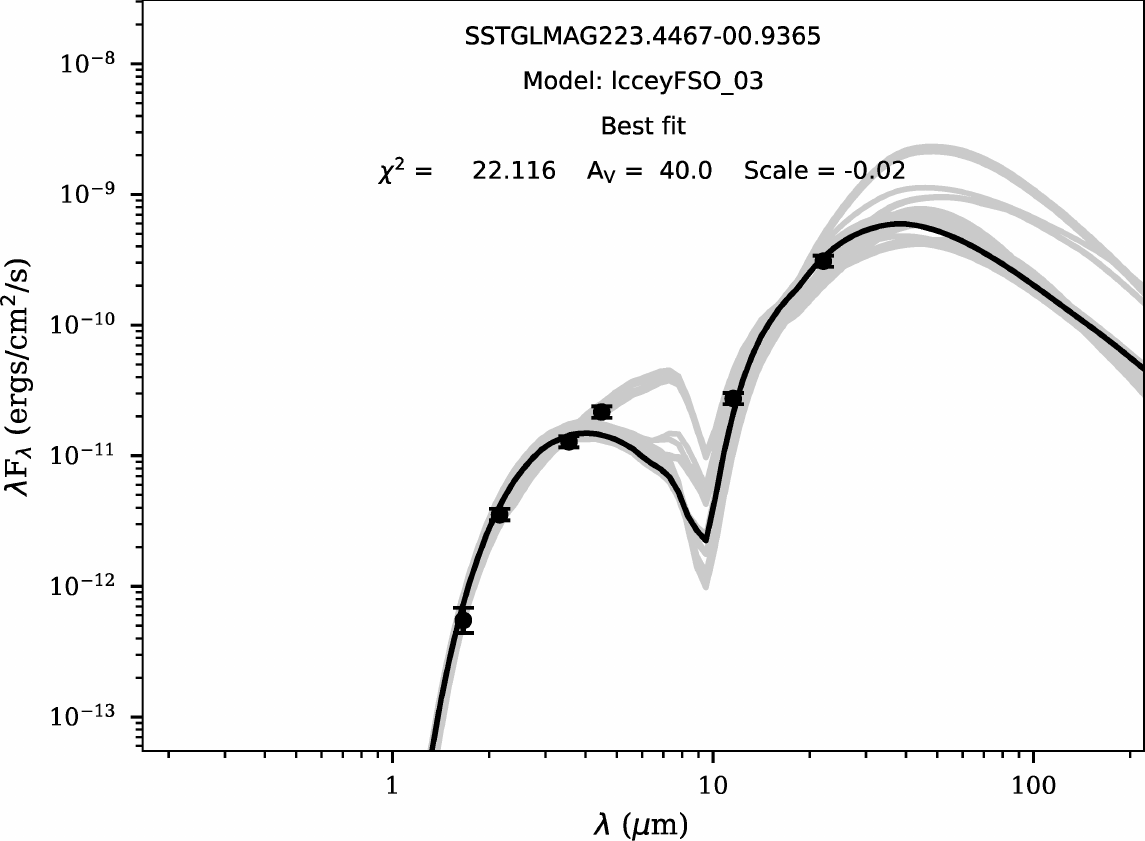} \par
\vspace{2mm}
\includegraphics[width=0.32\textwidth]{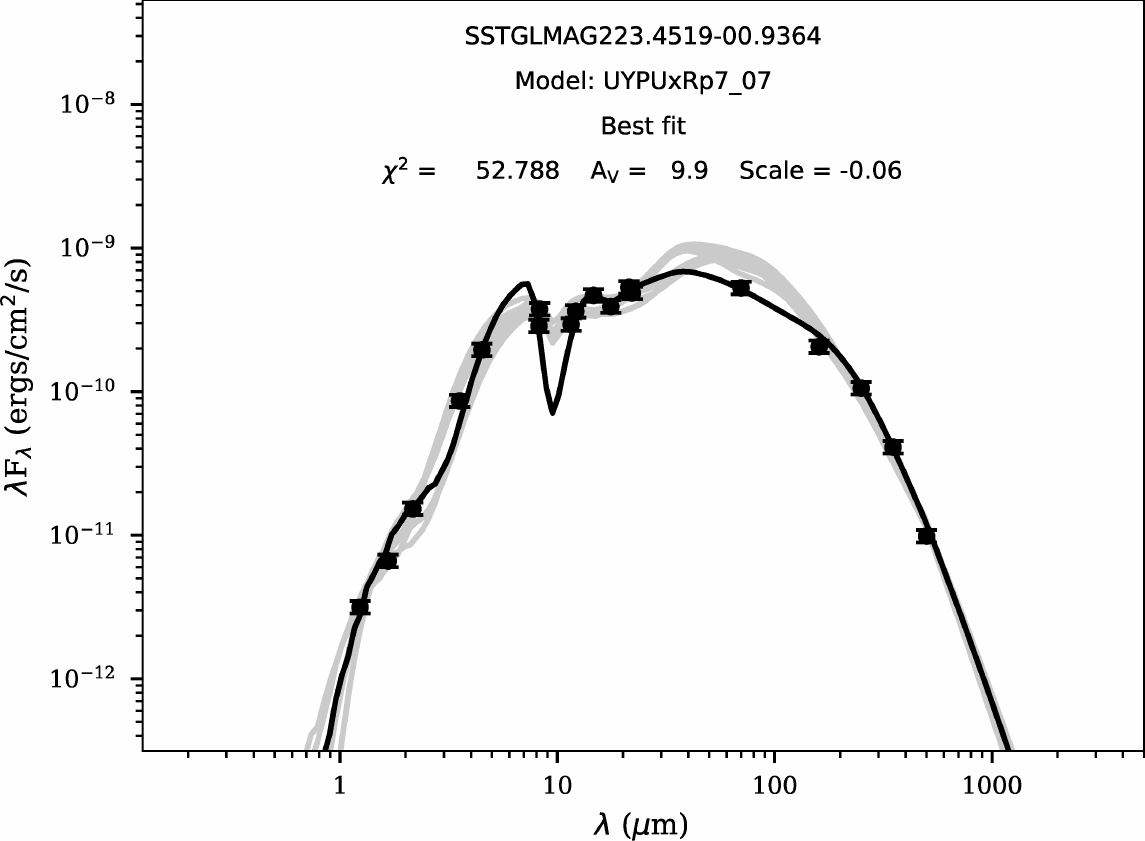}
\hfill
\includegraphics[width=0.32\textwidth]{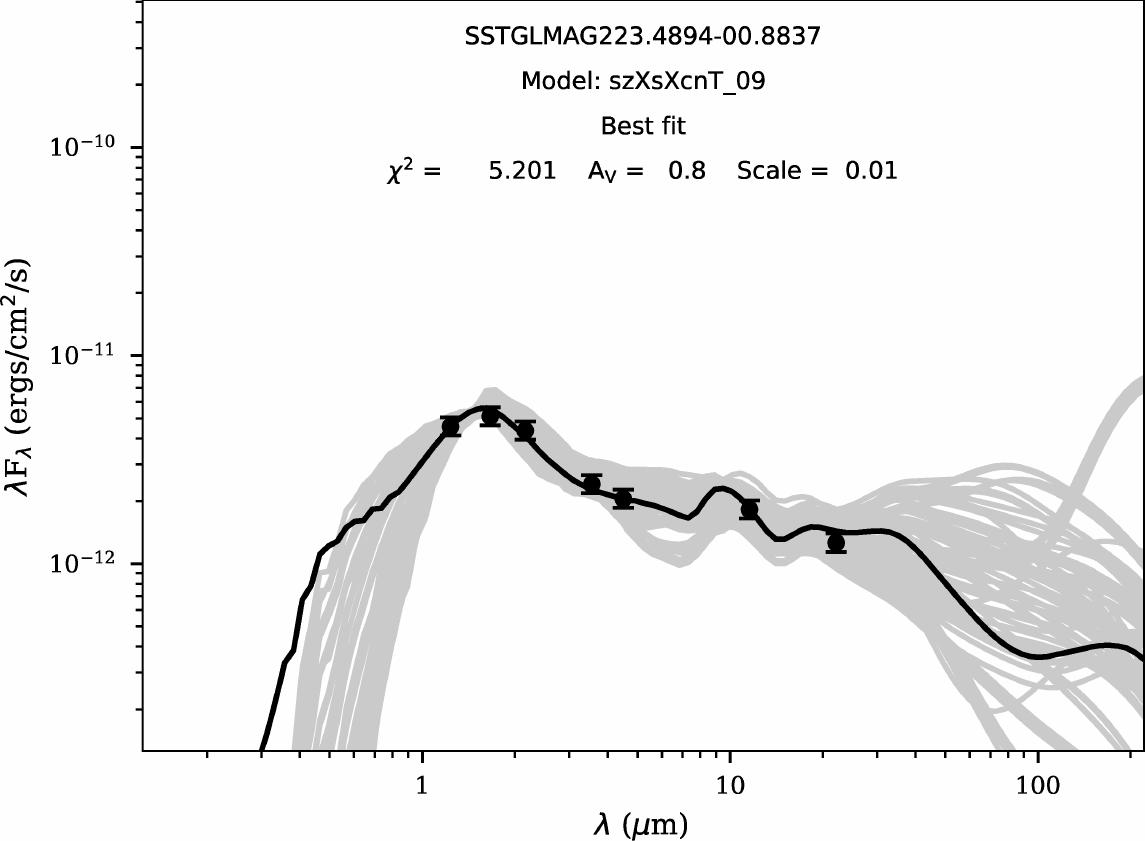}
\hfill
\includegraphics[width=0.32\textwidth]{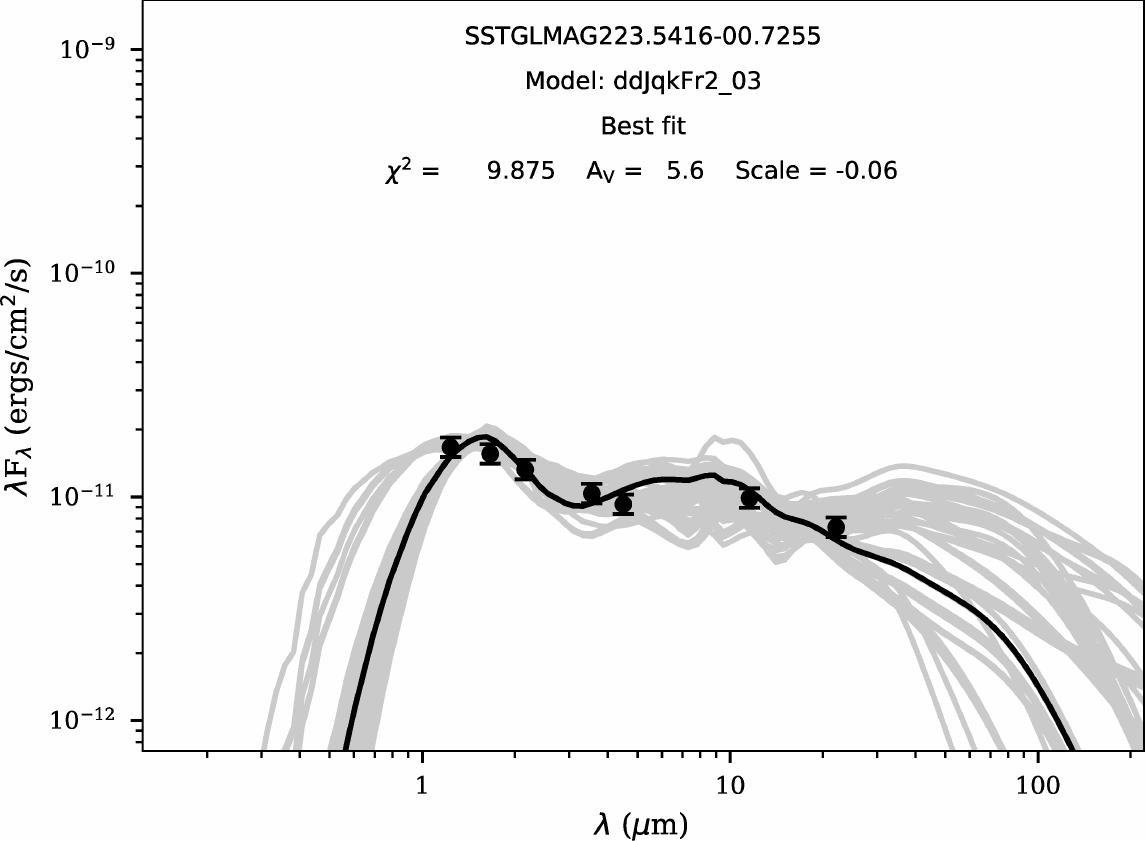} \par
\vspace{2mm}
\includegraphics[width=0.32\textwidth]{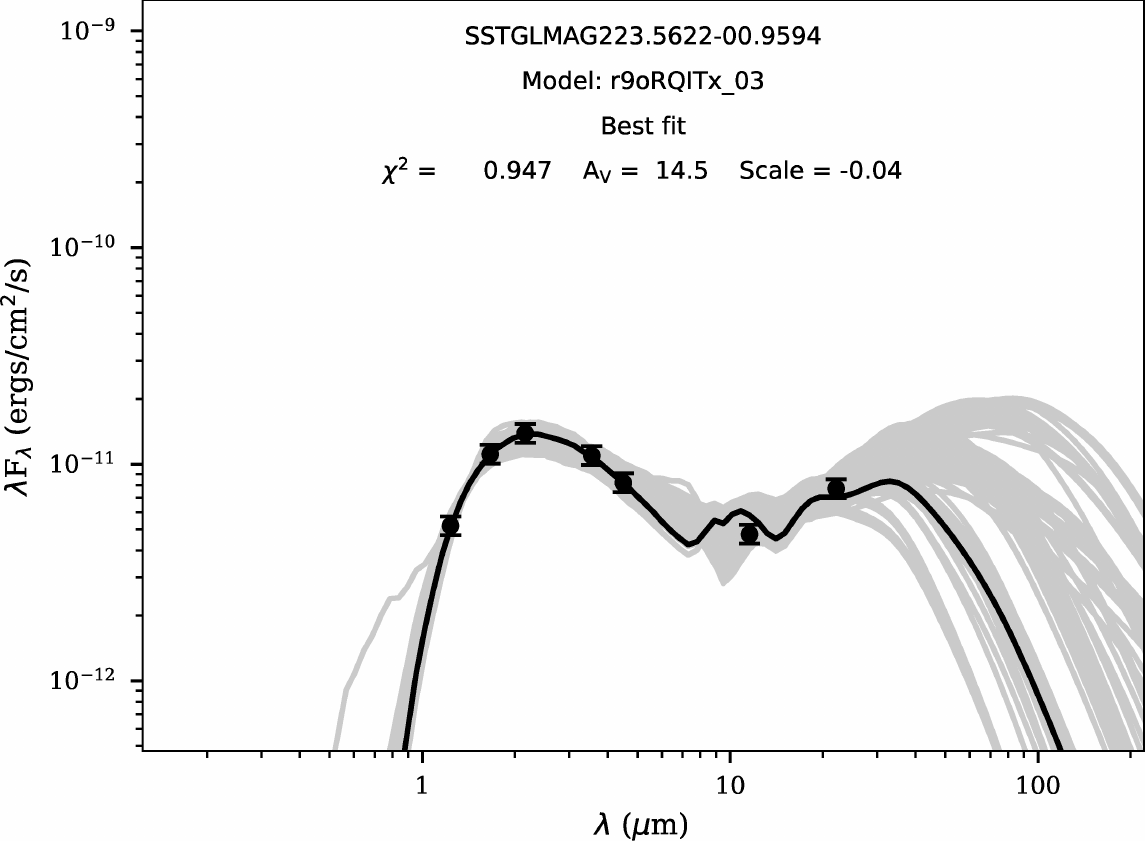}
\hfill
\includegraphics[width=0.32\textwidth]{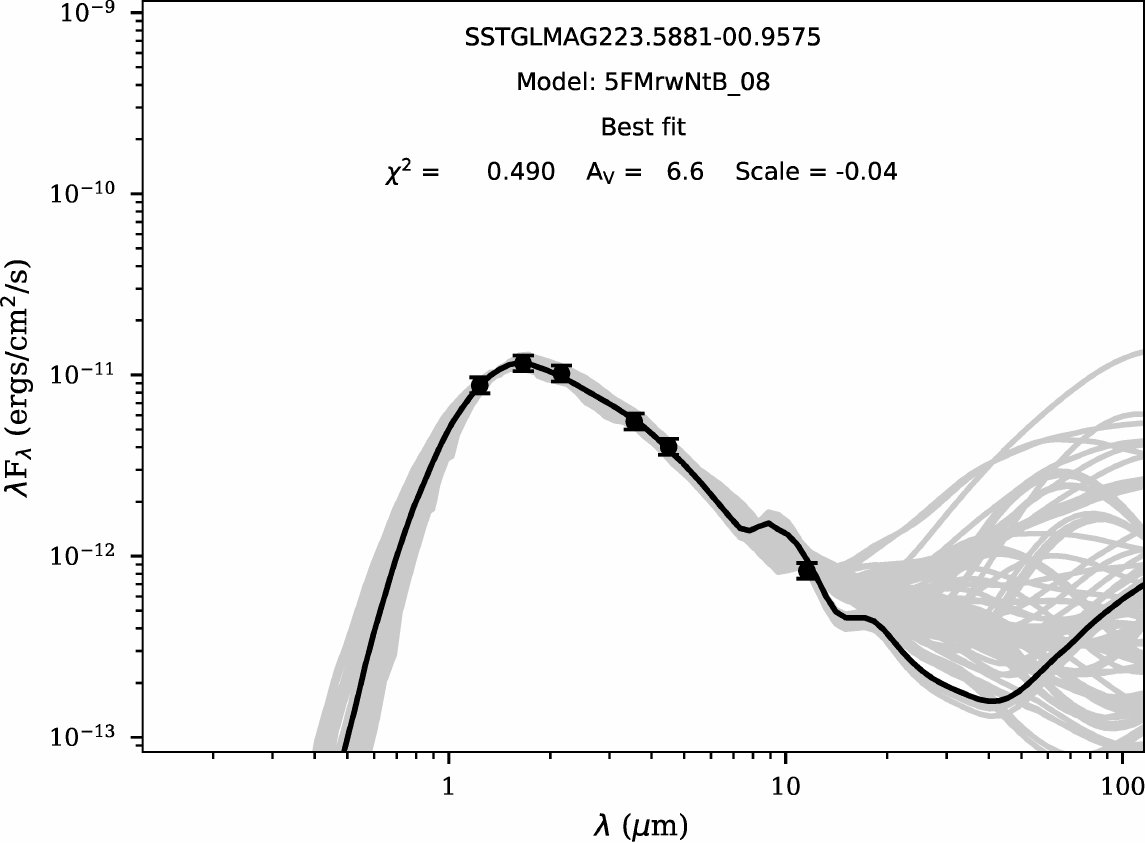}
\hfill
\includegraphics[width=0.32\textwidth]{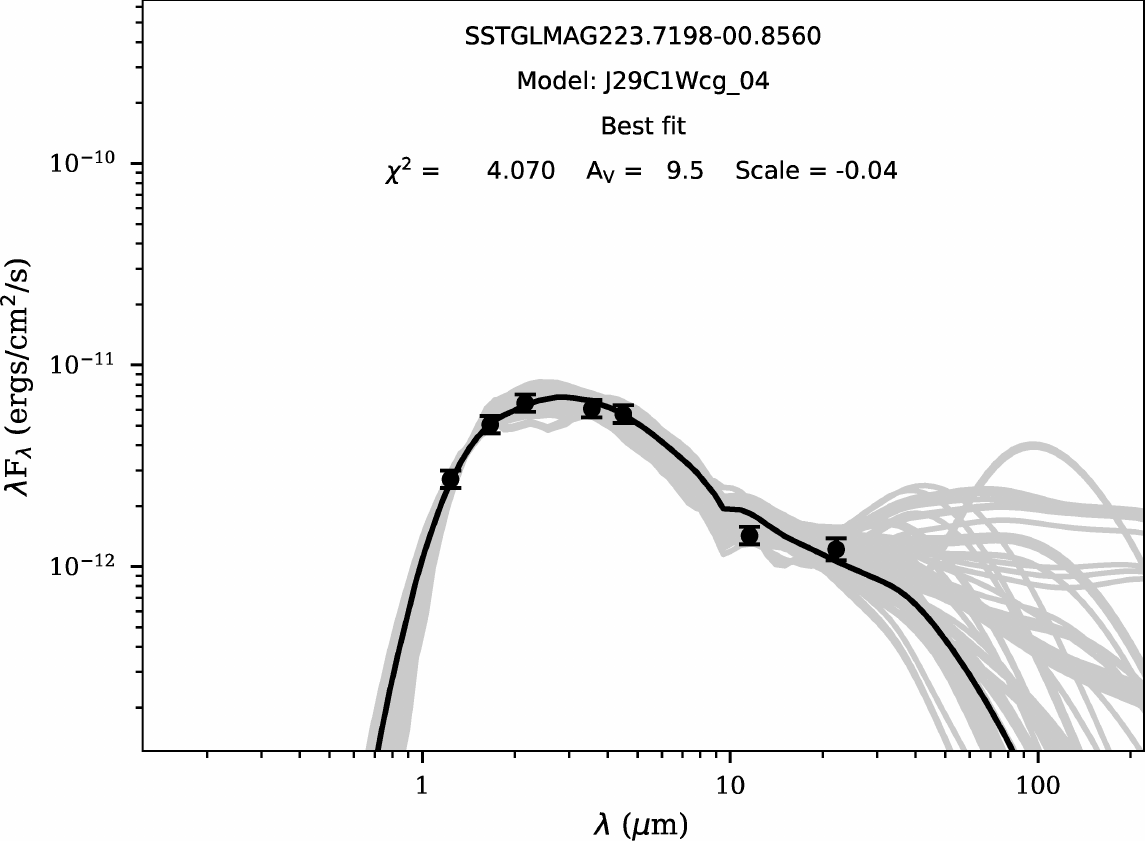} \par
\vspace{2mm}
\includegraphics[width=0.32\textwidth]{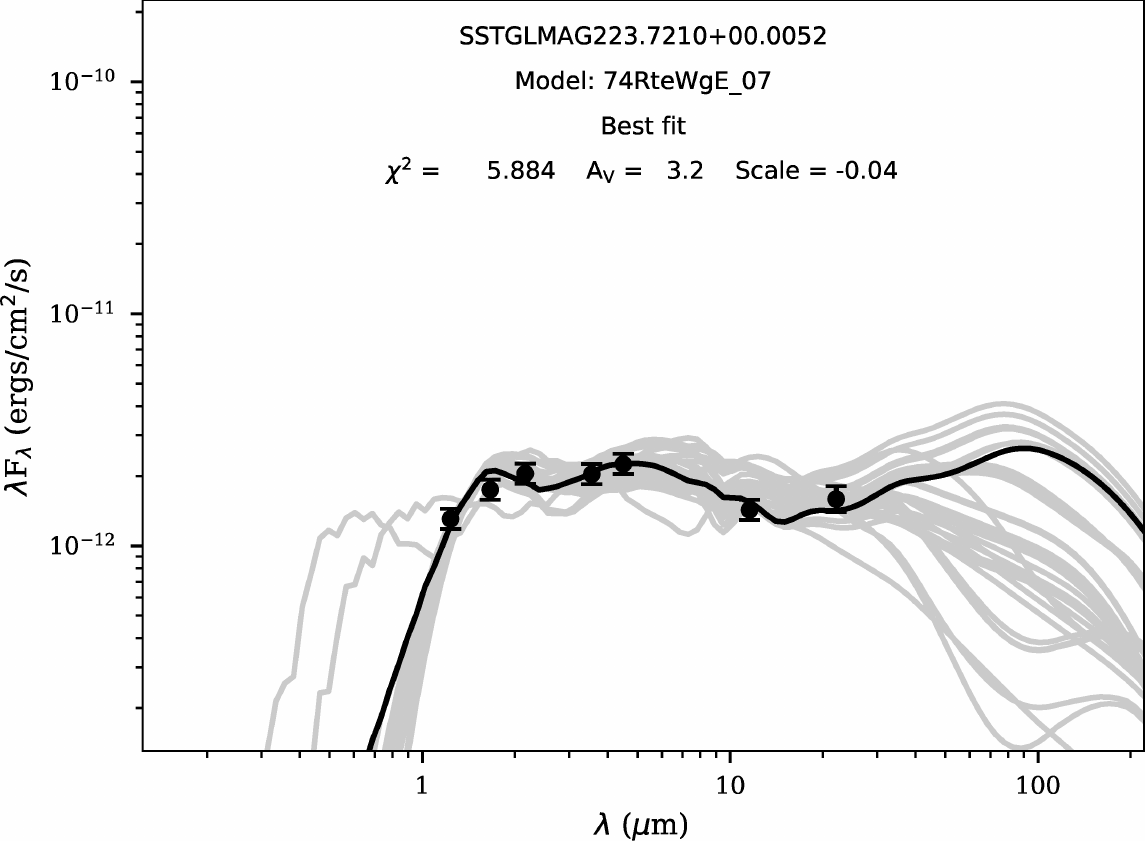}
\hfill
\includegraphics[width=0.32\textwidth]{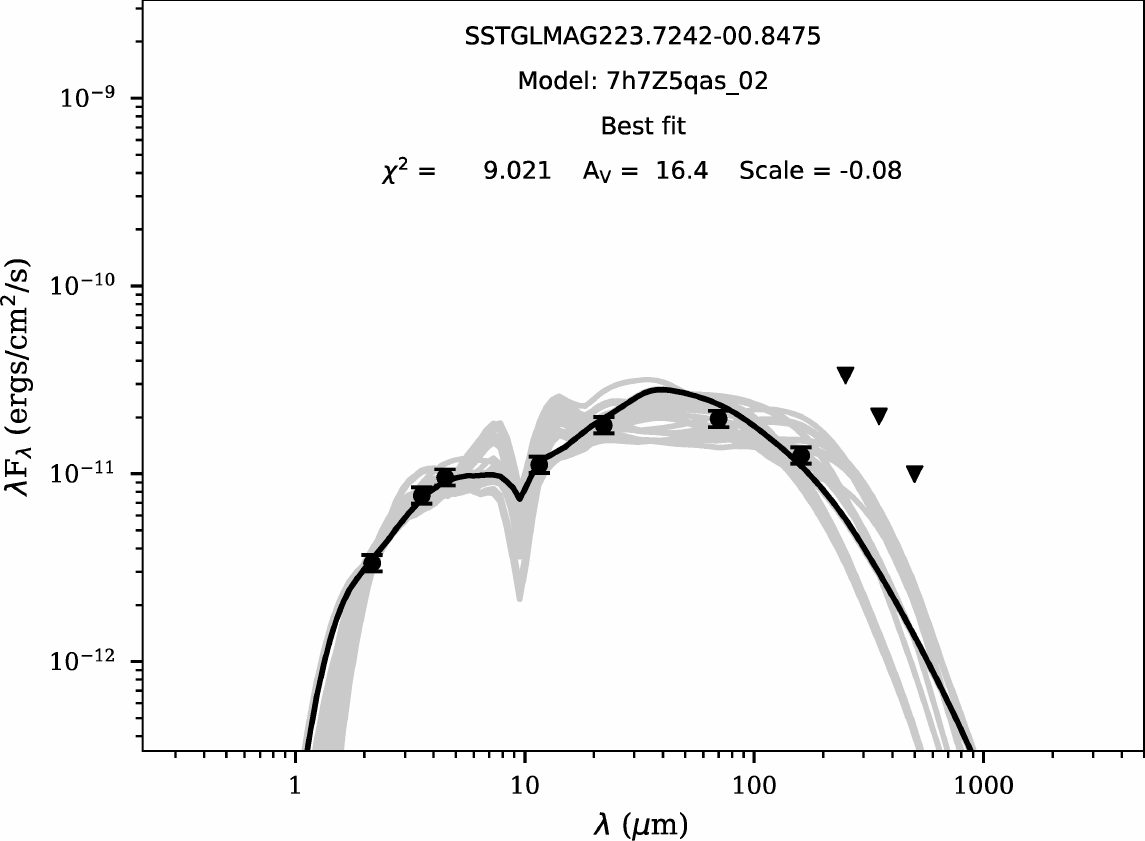}
\hfill
\includegraphics[width=0.32\textwidth]{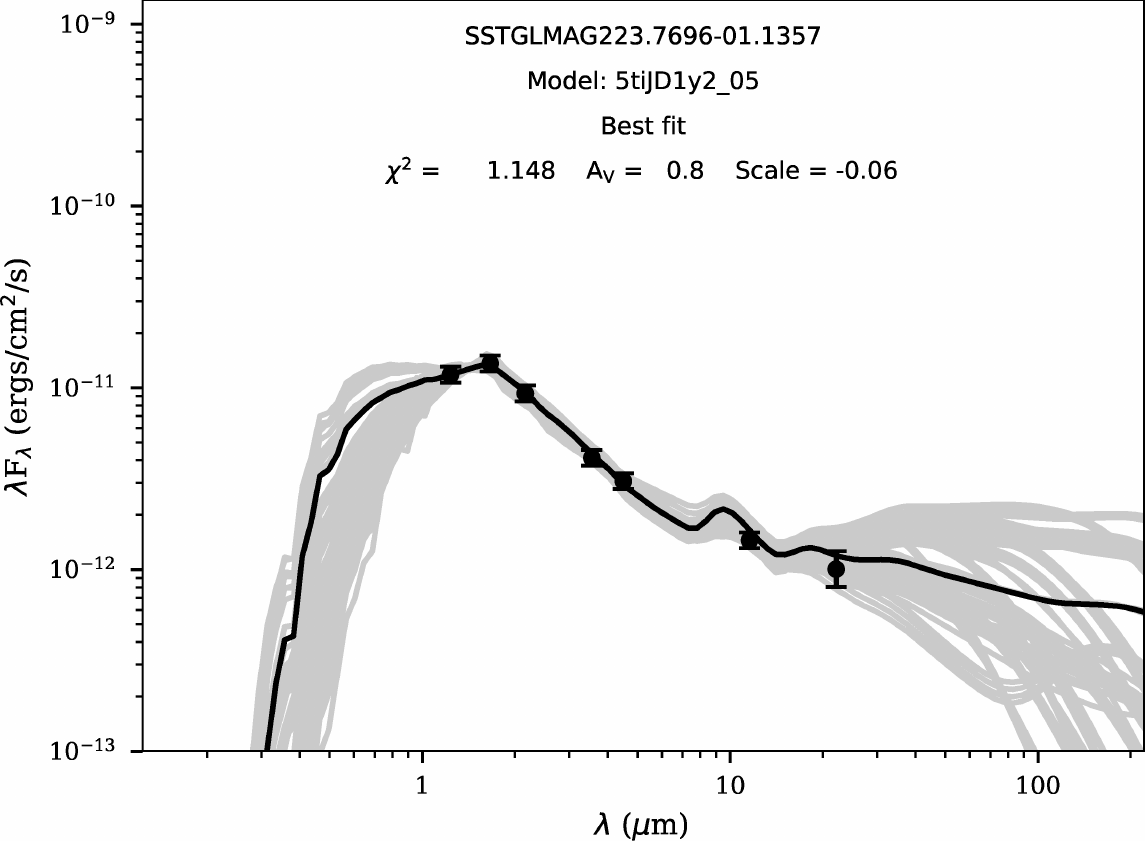} \par
\vspace{2mm}
\includegraphics[width=0.32\textwidth]{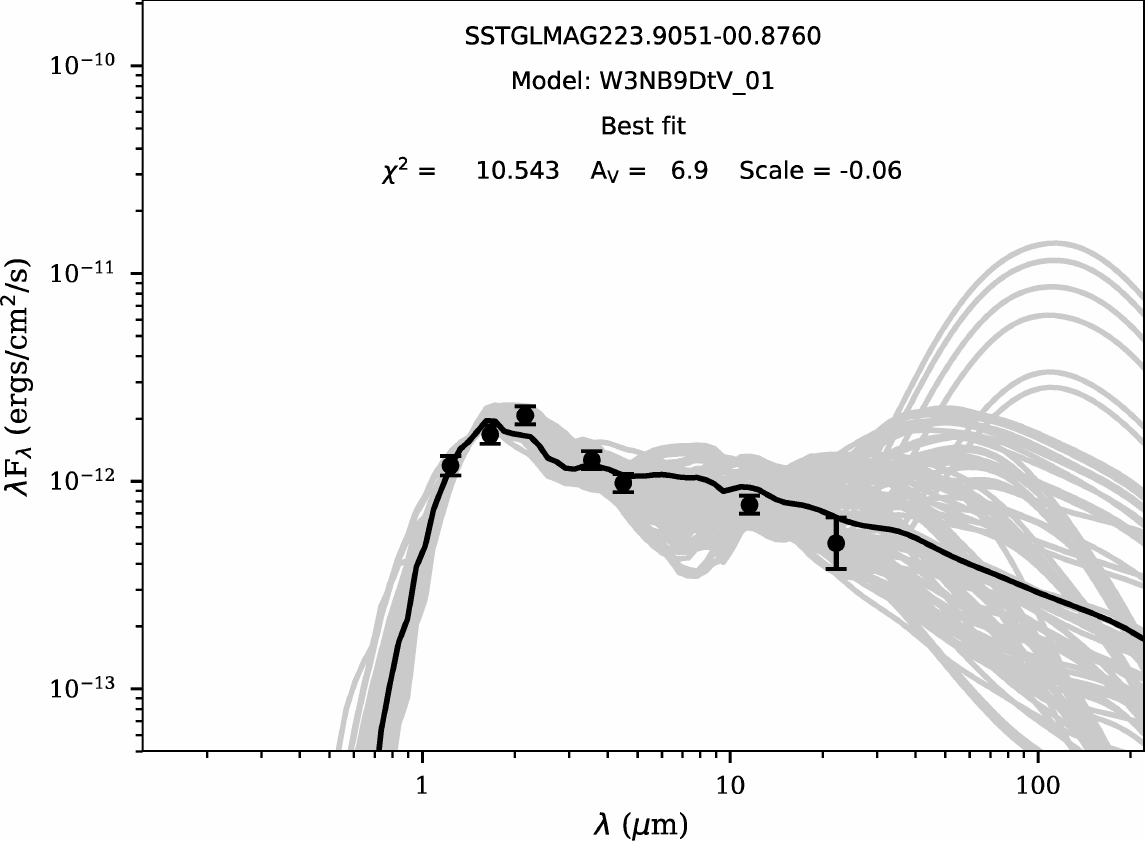}
\hfill
\includegraphics[width=0.32\textwidth]{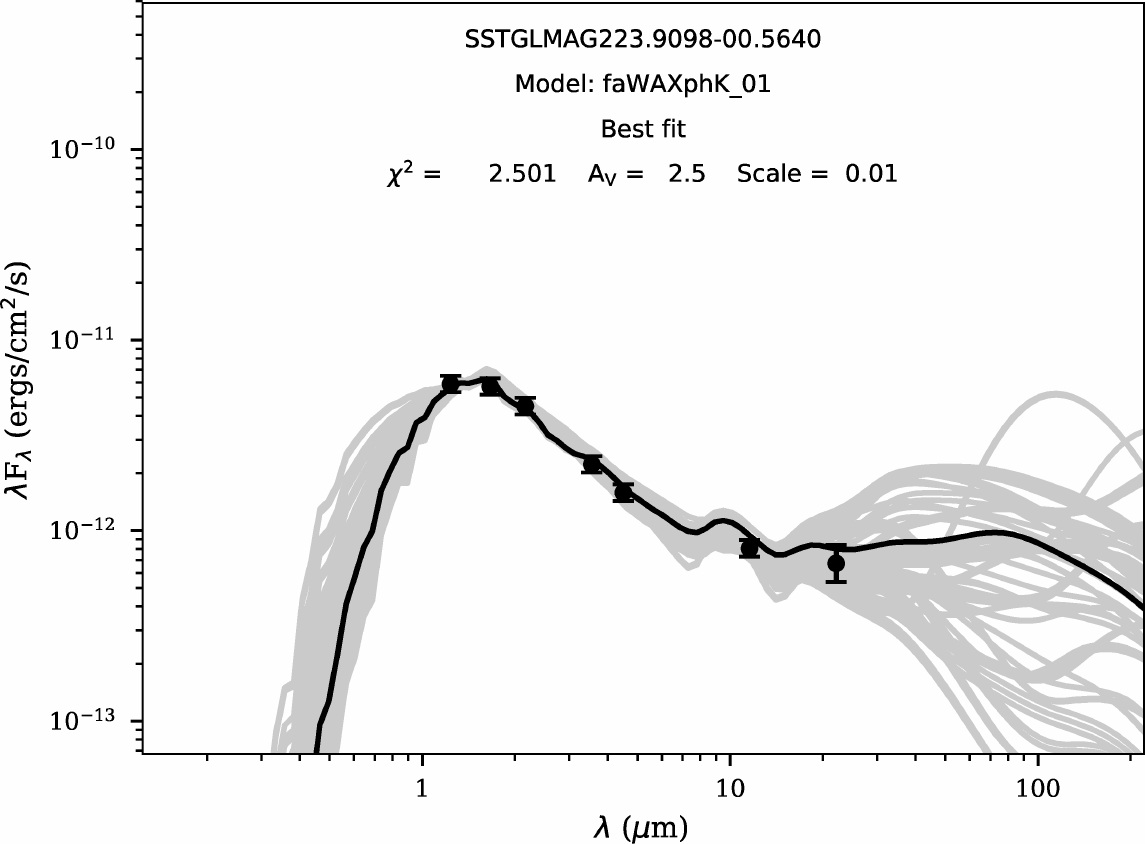}
\hfill
\includegraphics[width=0.32\textwidth]{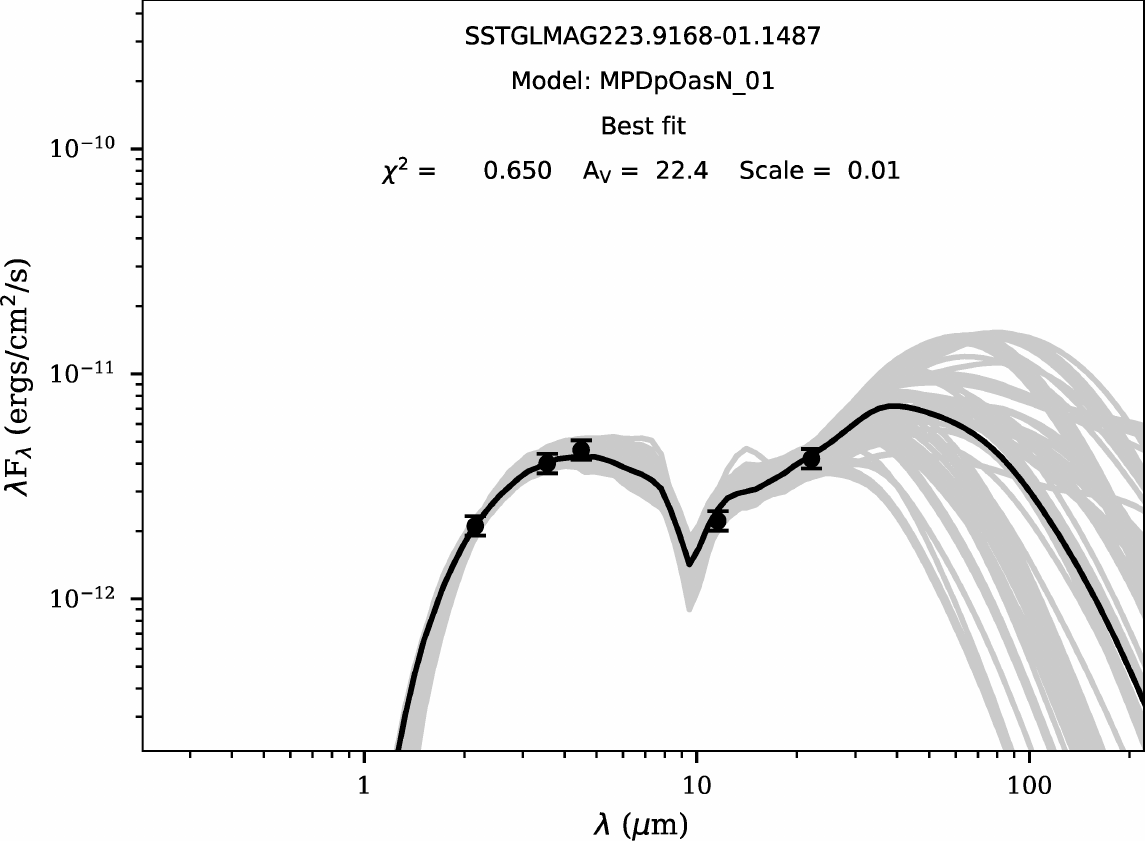}
\caption{SEDs and the \citet{robitaille2017} YSO model fits for YSO candidates listed in Table~\ref{t:physpar} excluding SEDs shown in Fig.~\ref{f:exsed}. The symbols and lines are as in Fig.~\ref{f:exsed}. \label{f:SEDs1}}
\end{figure*}

\begin{figure*}
\includegraphics[width=0.32\textwidth]{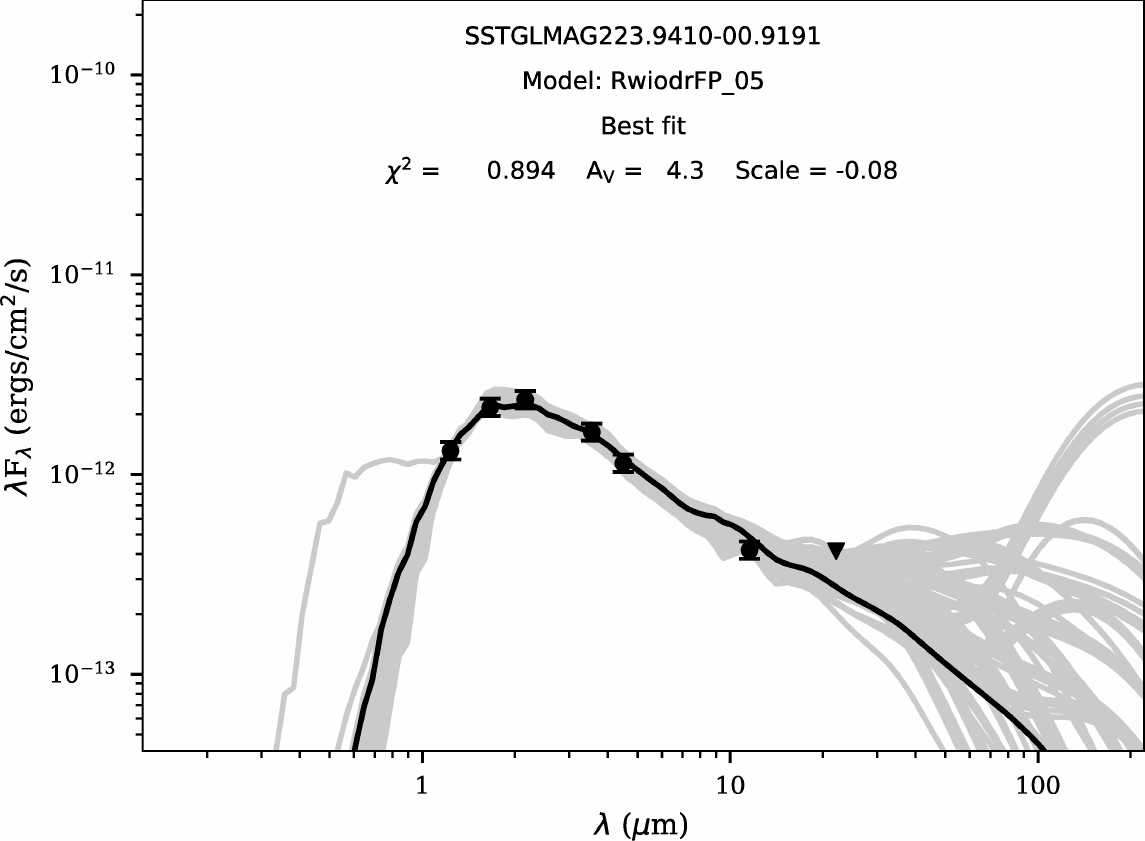}
\hfill
\includegraphics[width=0.32\textwidth]{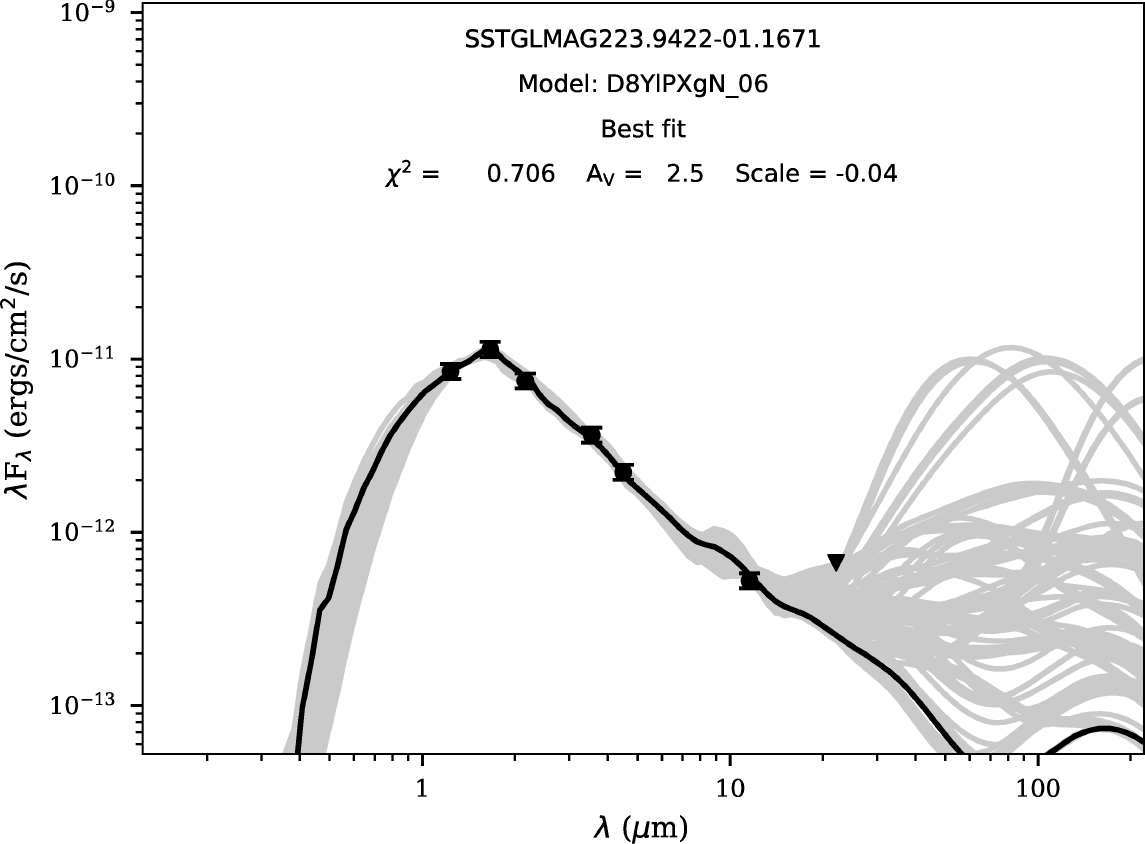}
\hfill
\includegraphics[width=0.32\textwidth]{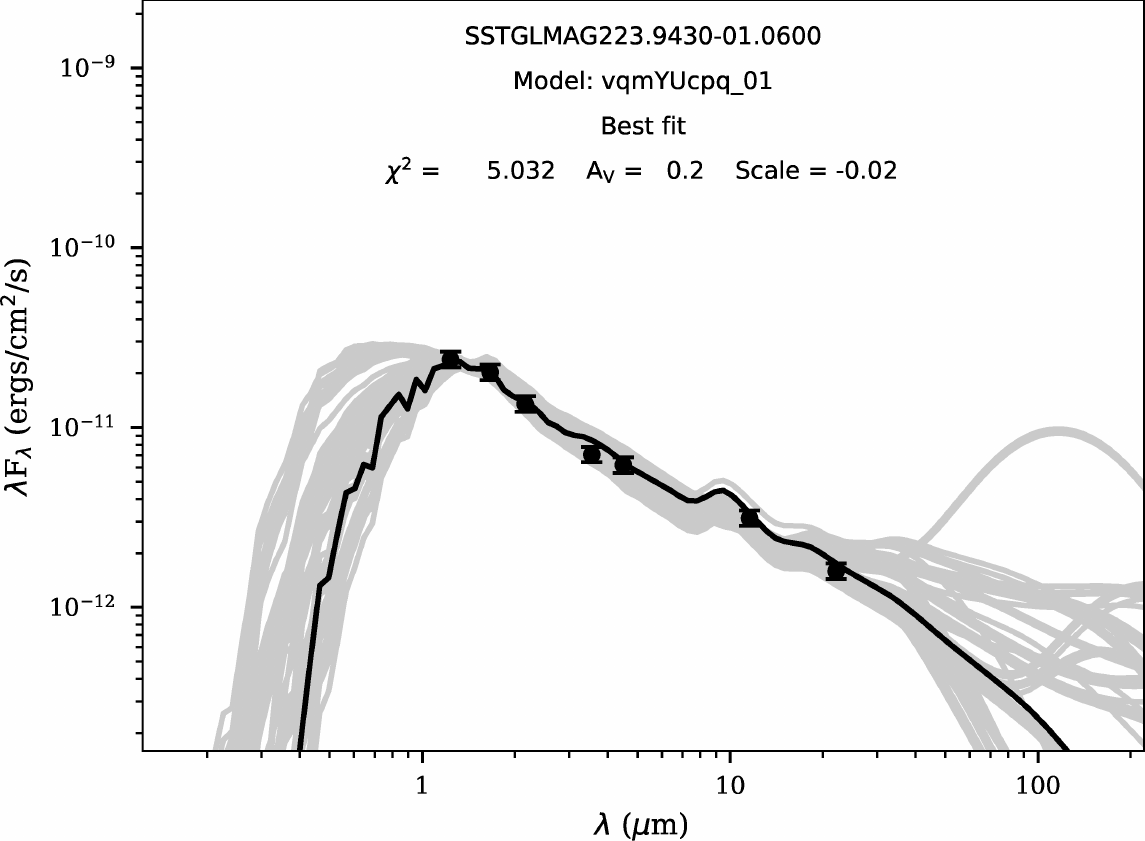} \par
\vspace{2mm}
\includegraphics[width=0.32\textwidth]{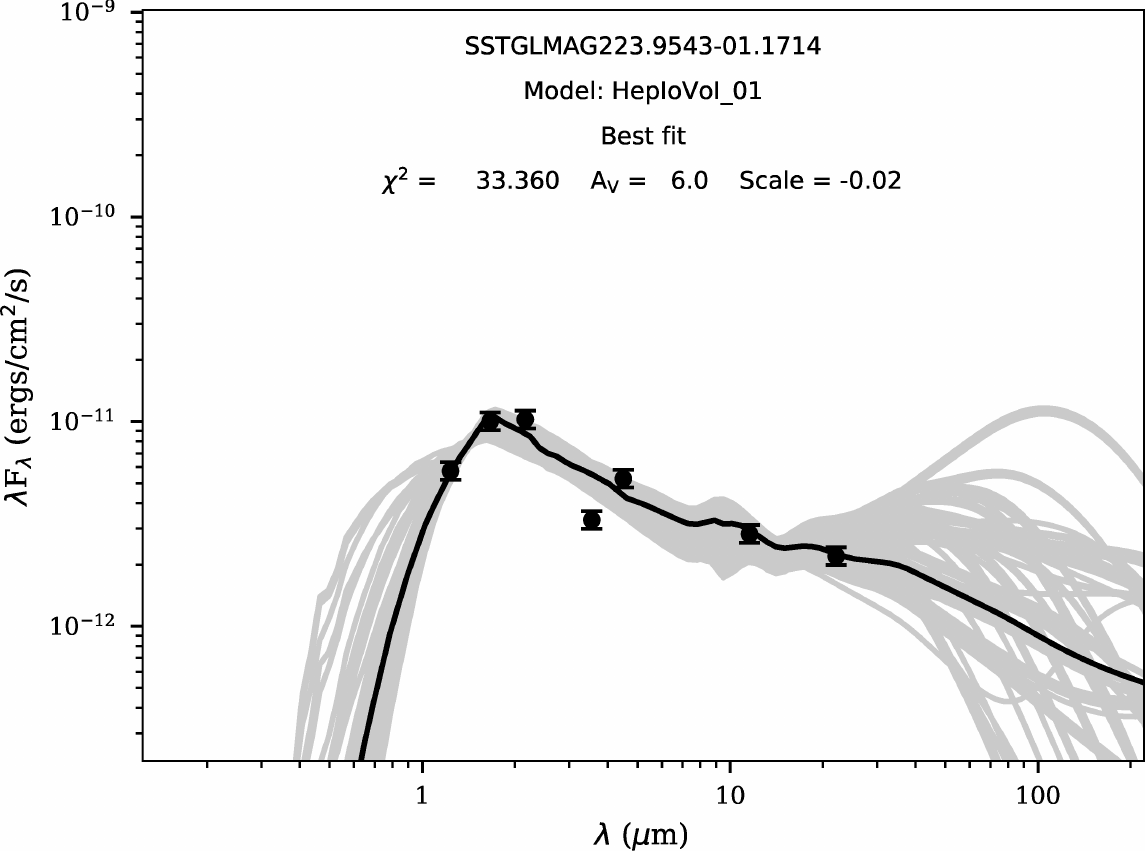}
\hfill
\includegraphics[width=0.32\textwidth]{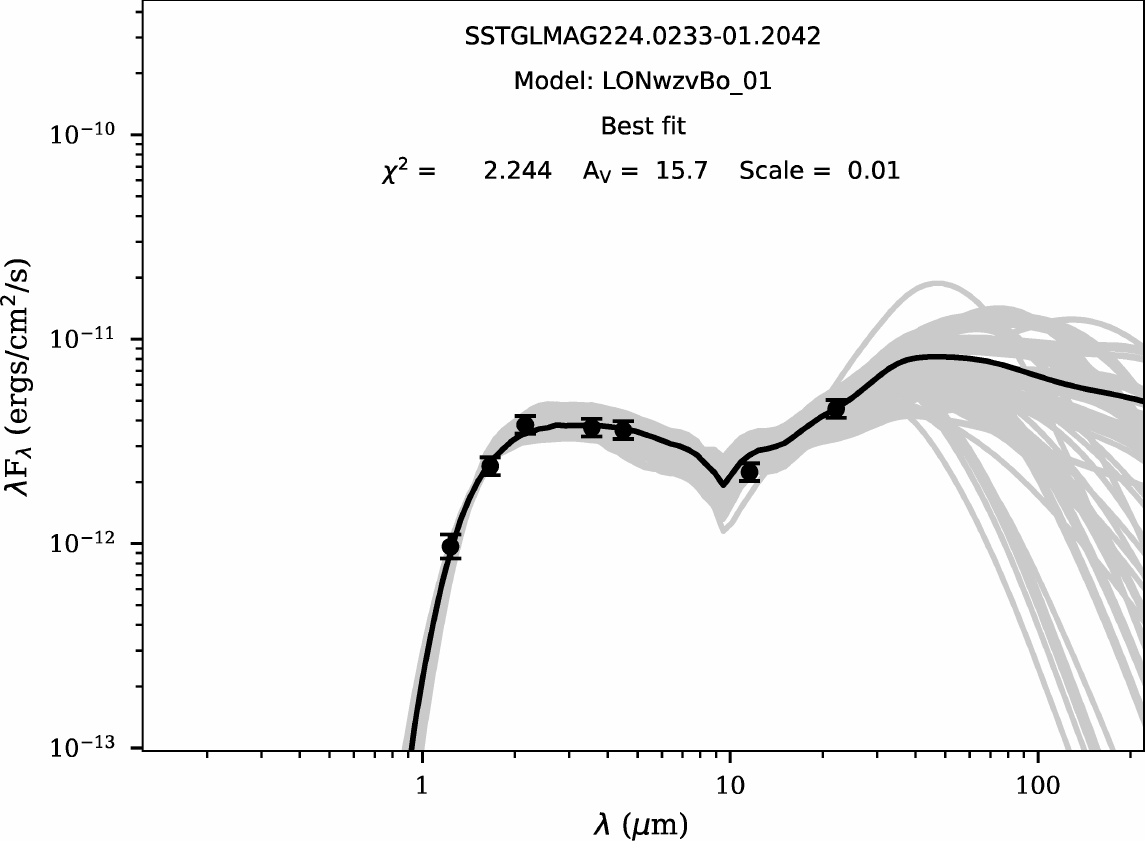}
\hfill
\includegraphics[width=0.32\textwidth]{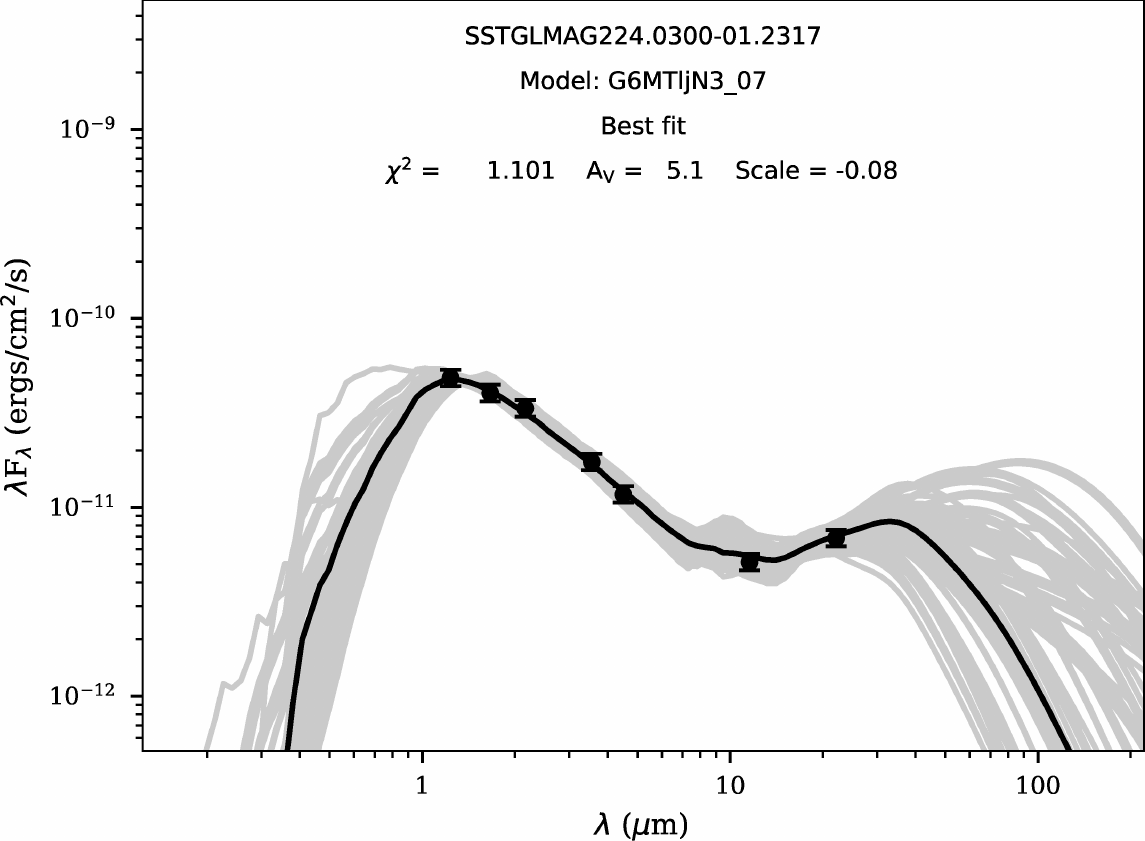} \par
\vspace{2mm}
\includegraphics[width=0.32\textwidth]{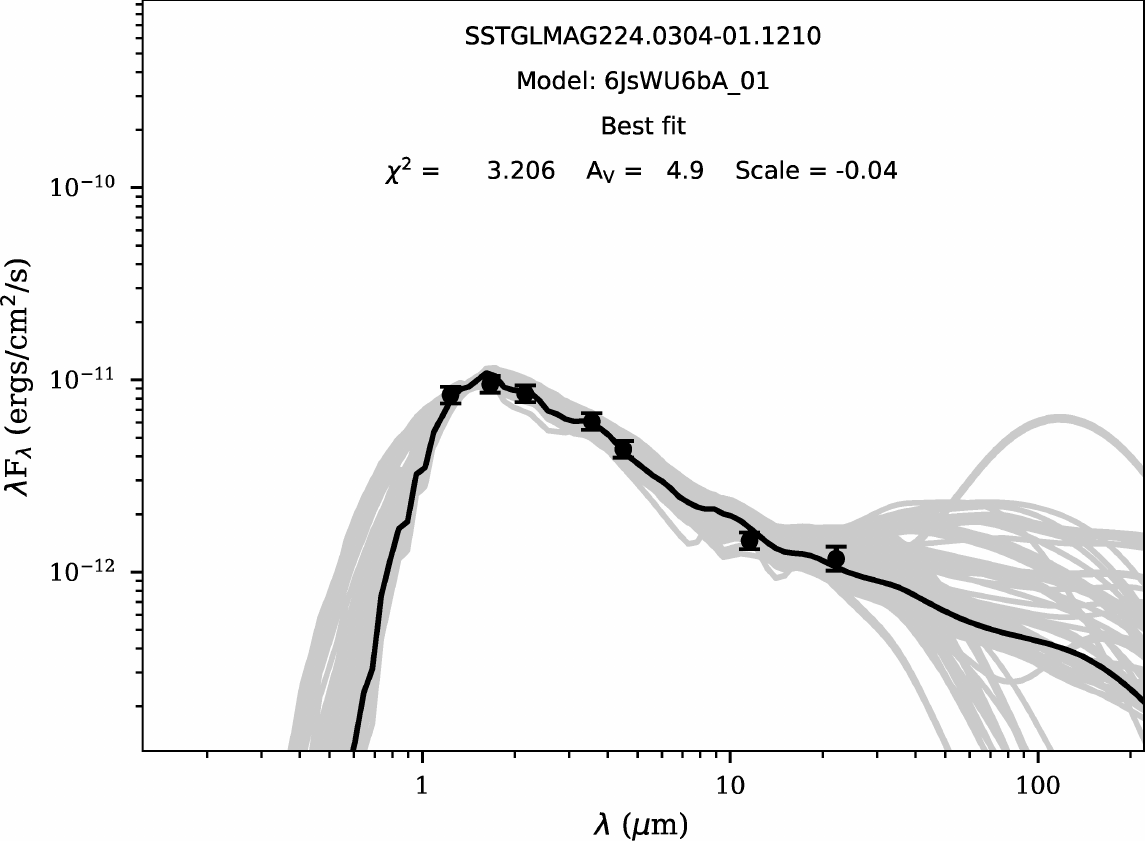}
\hfill
\includegraphics[width=0.32\textwidth]{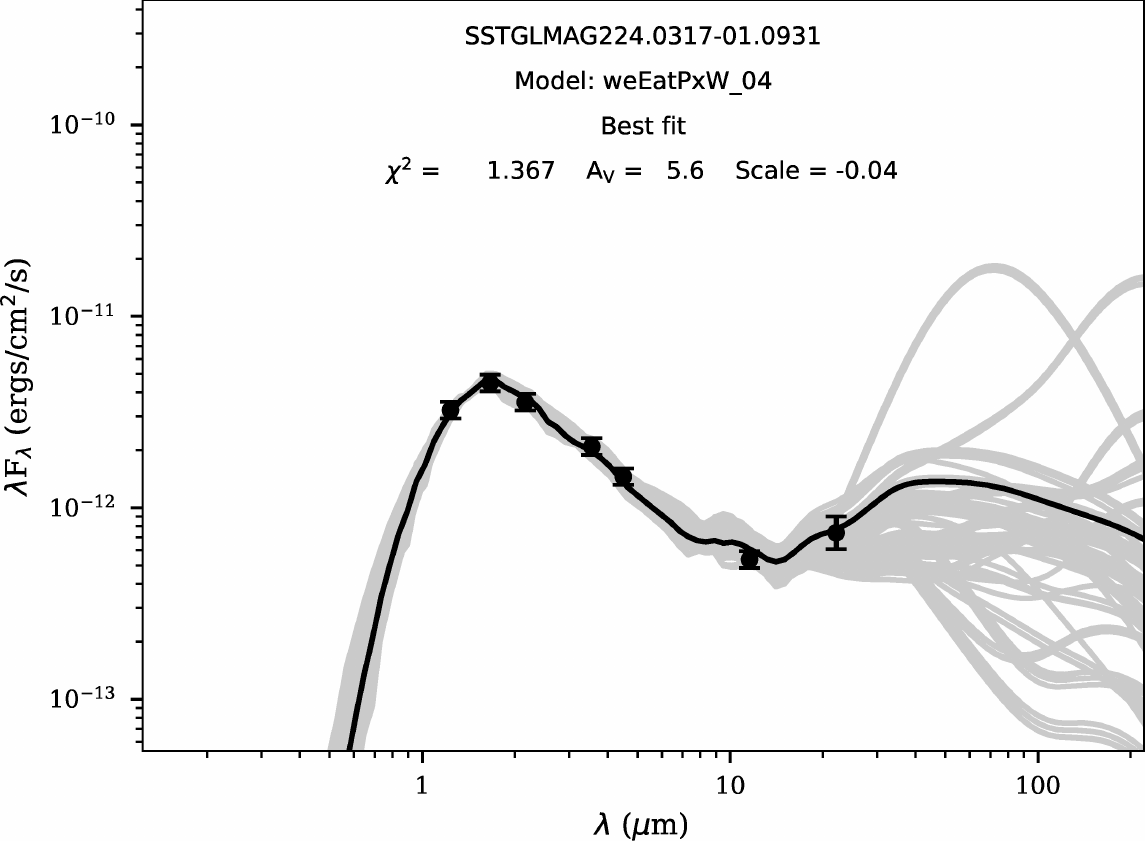}
\hfill
\includegraphics[width=0.32\textwidth]{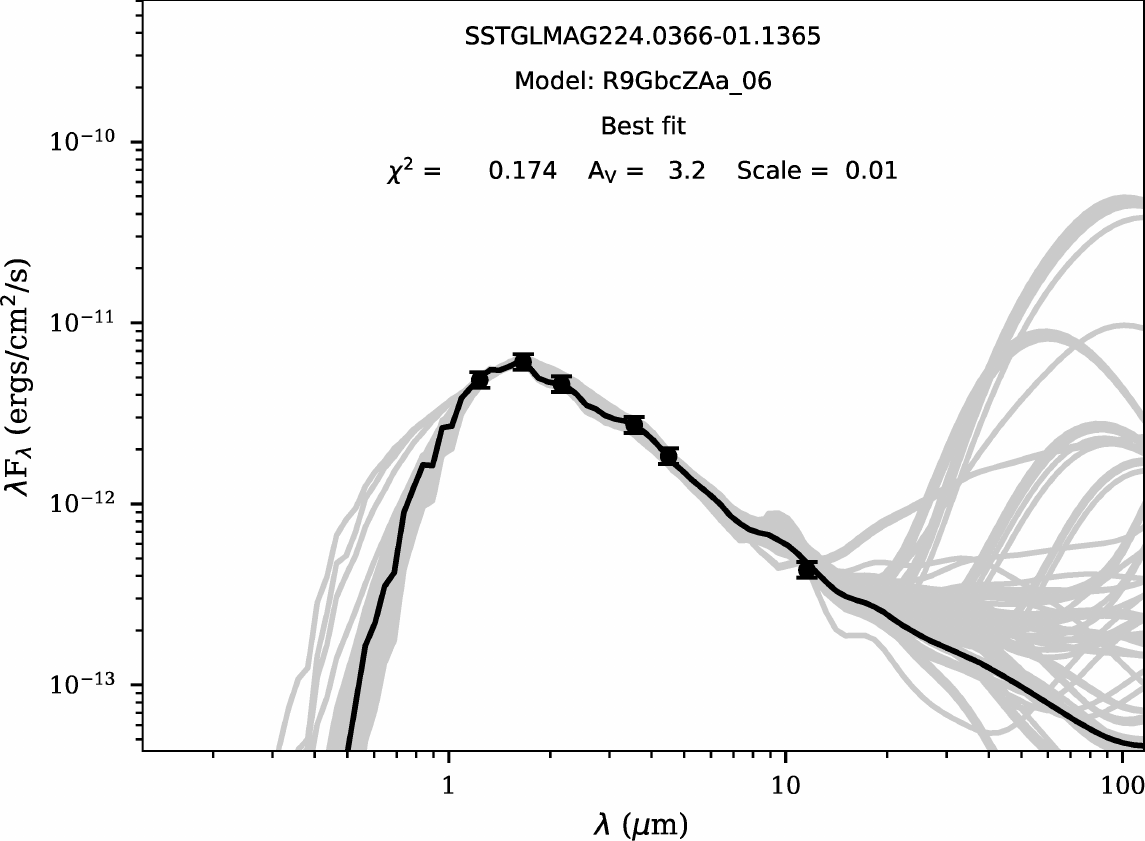} \par
\vspace{2mm}
\includegraphics[width=0.32\textwidth]{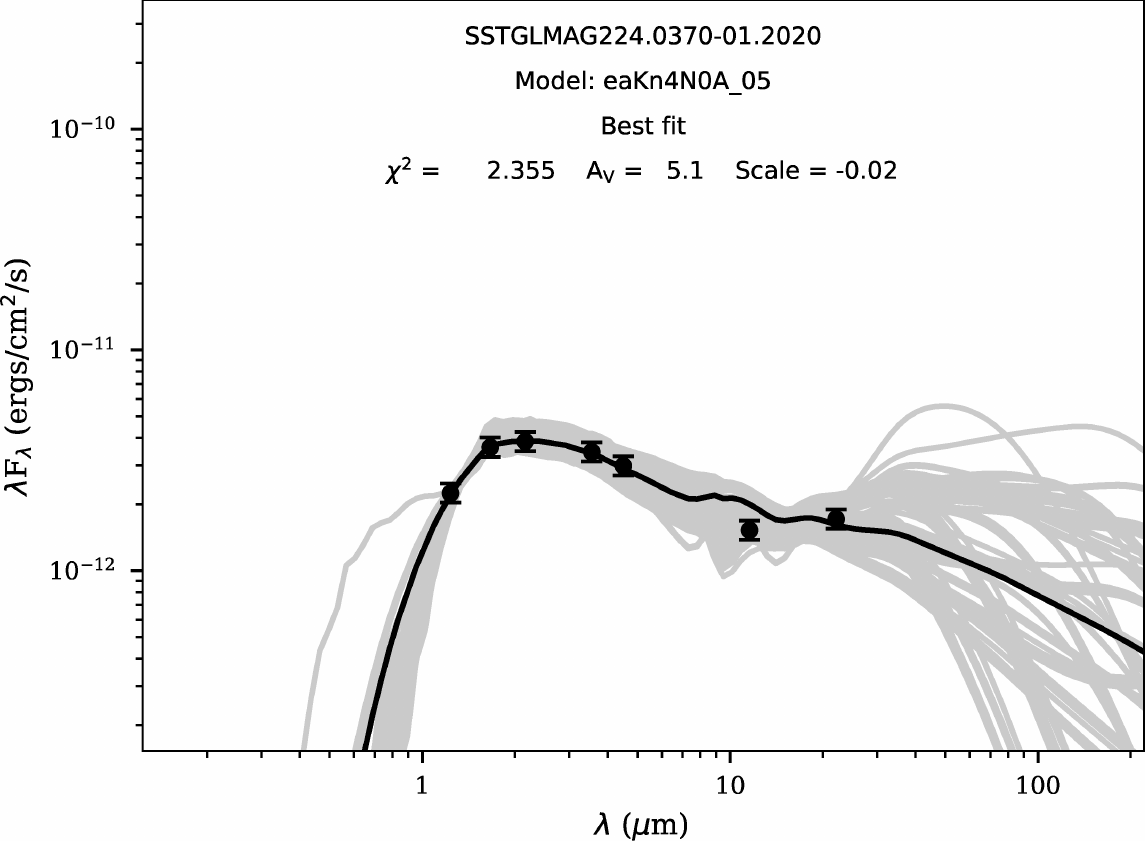}
\hfill
\includegraphics[width=0.32\textwidth]{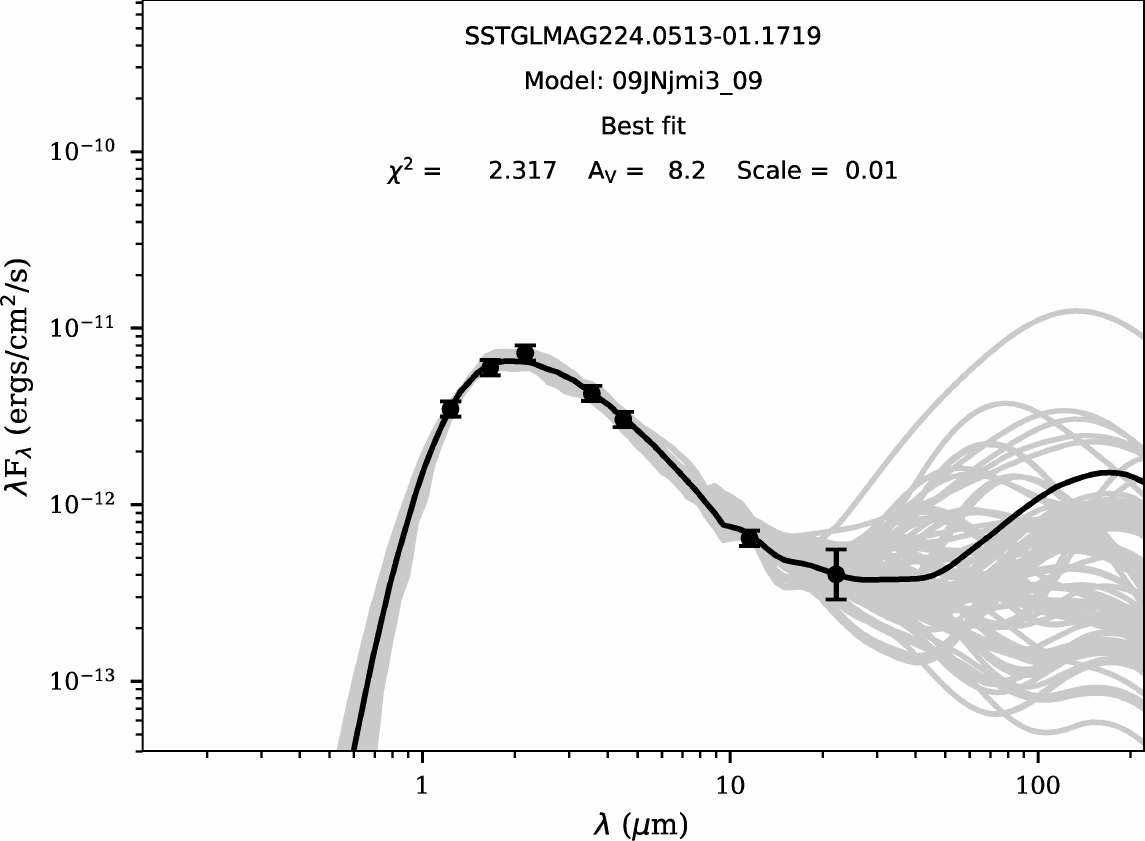}
\hfill
\includegraphics[width=0.32\textwidth]{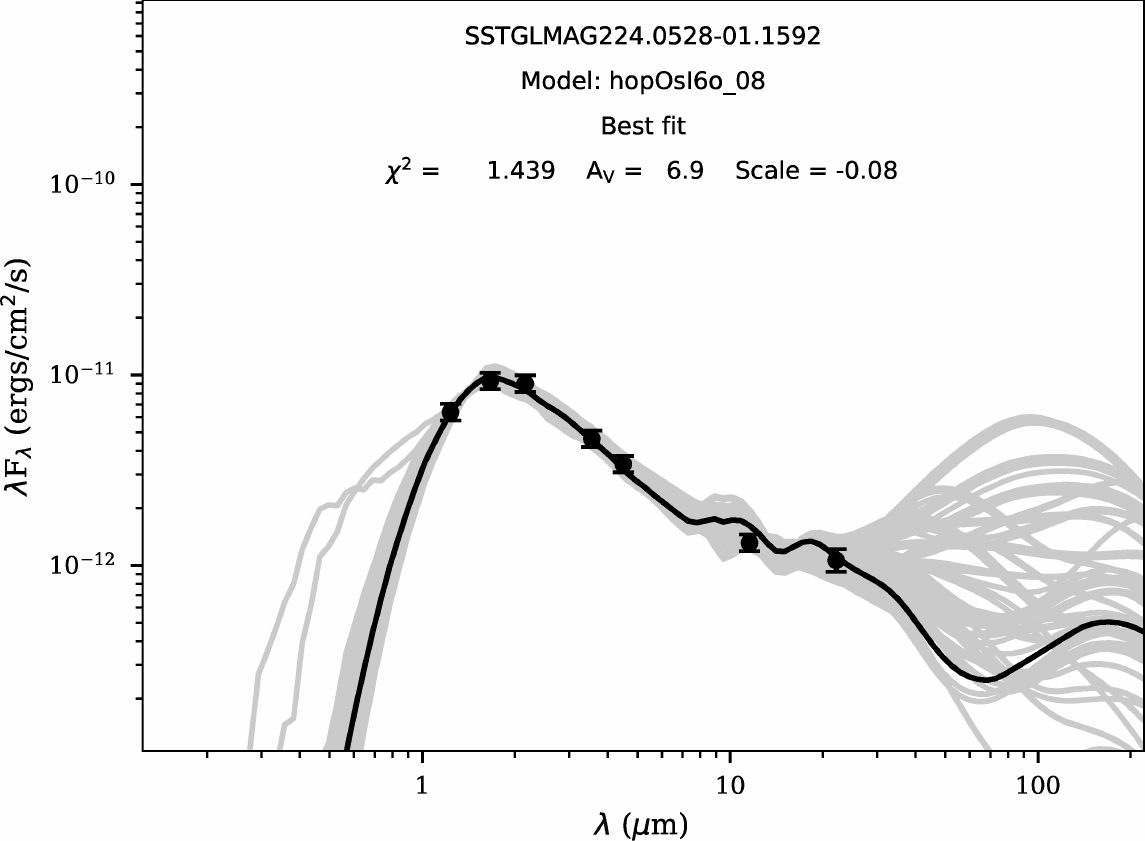} \par
\vspace{2mm}
\includegraphics[width=0.32\textwidth]{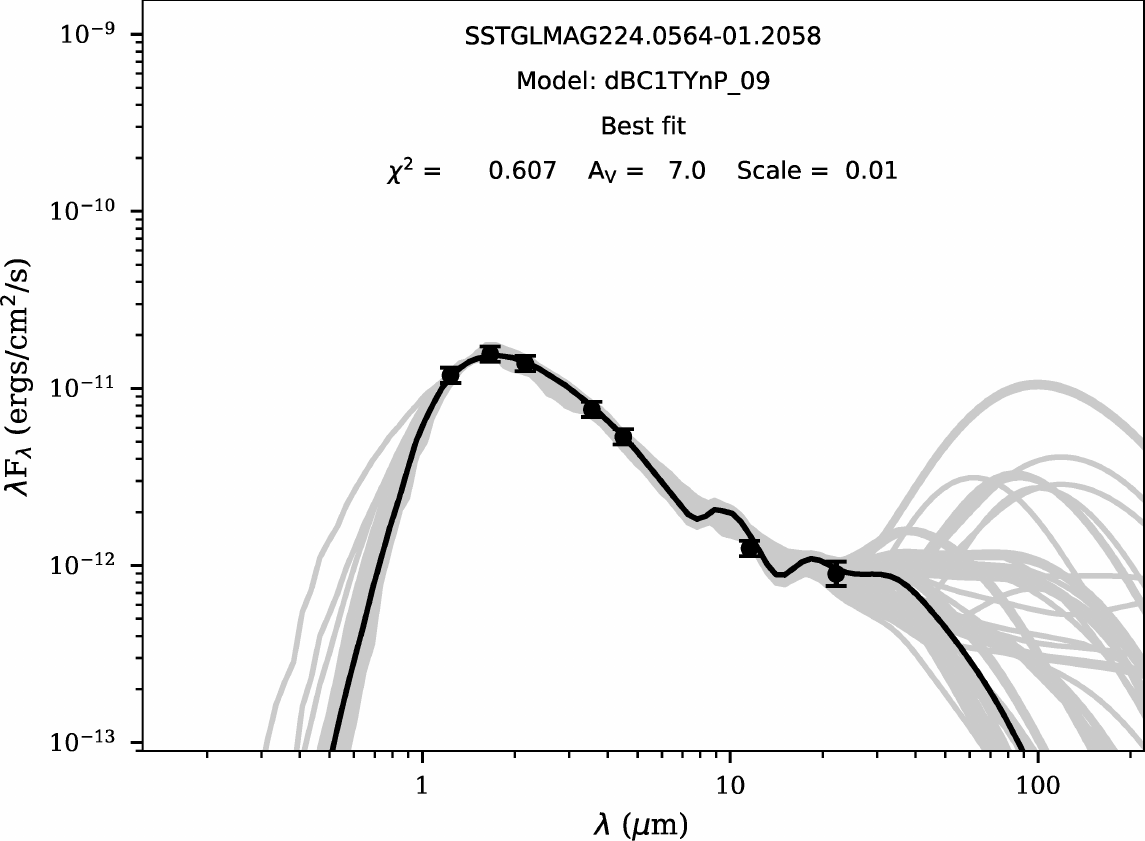}
\hfill
\includegraphics[width=0.32\textwidth]{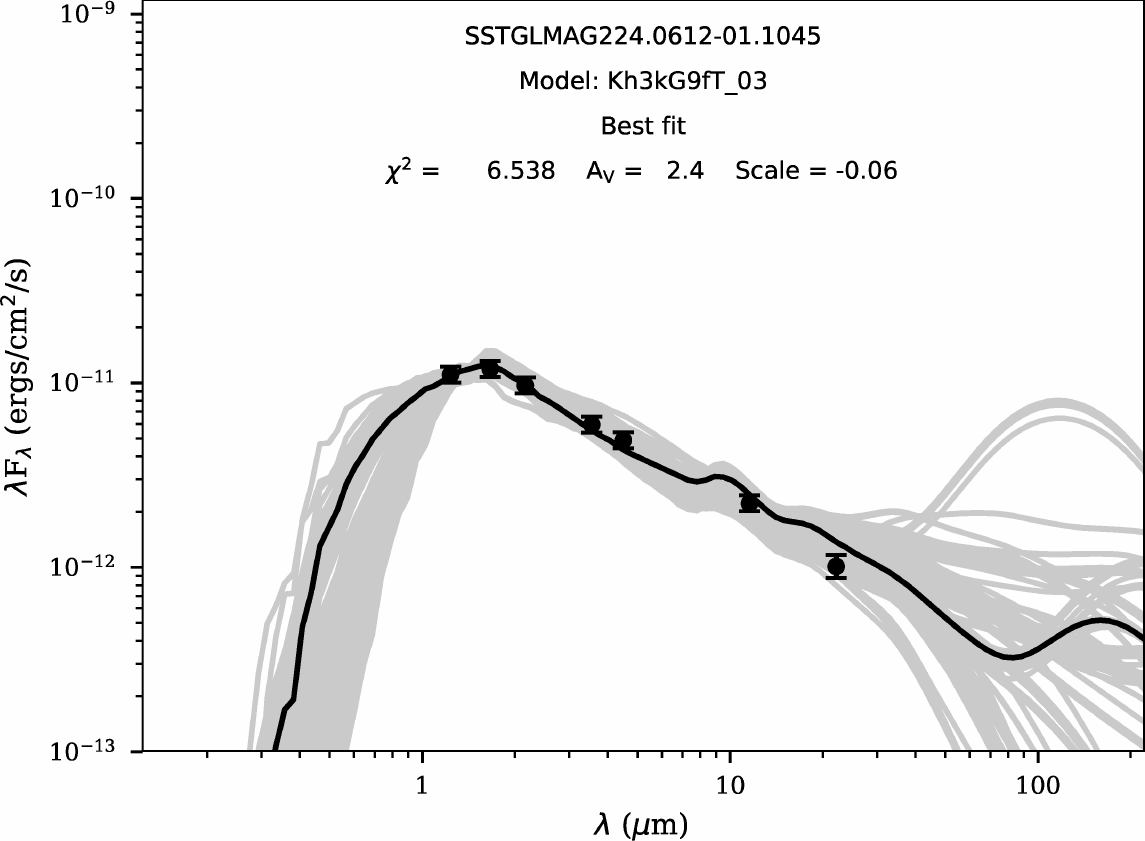}
\hfill
\includegraphics[width=0.32\textwidth]{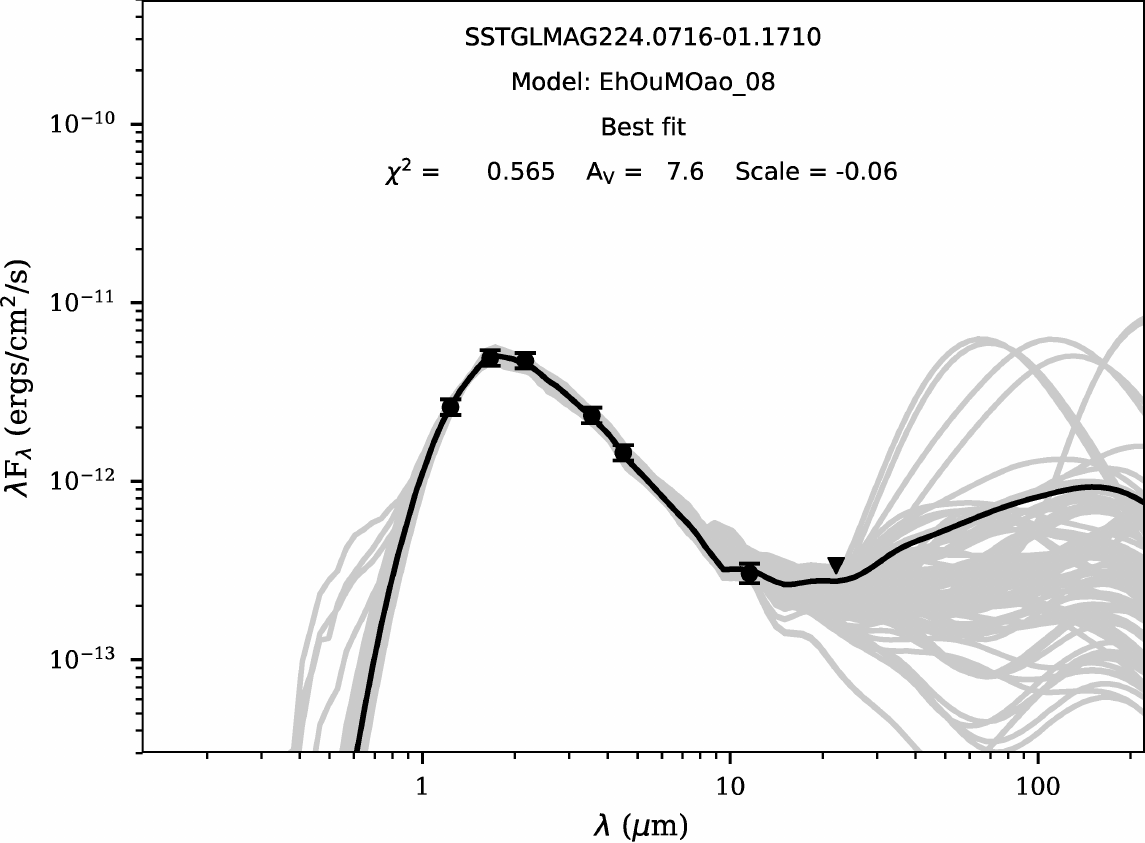}
\caption{Same as Fig.~\ref{f:SEDs1}  \label{f:SEDs2}}
\end{figure*}

\begin{figure*}
\includegraphics[width=0.32\textwidth]{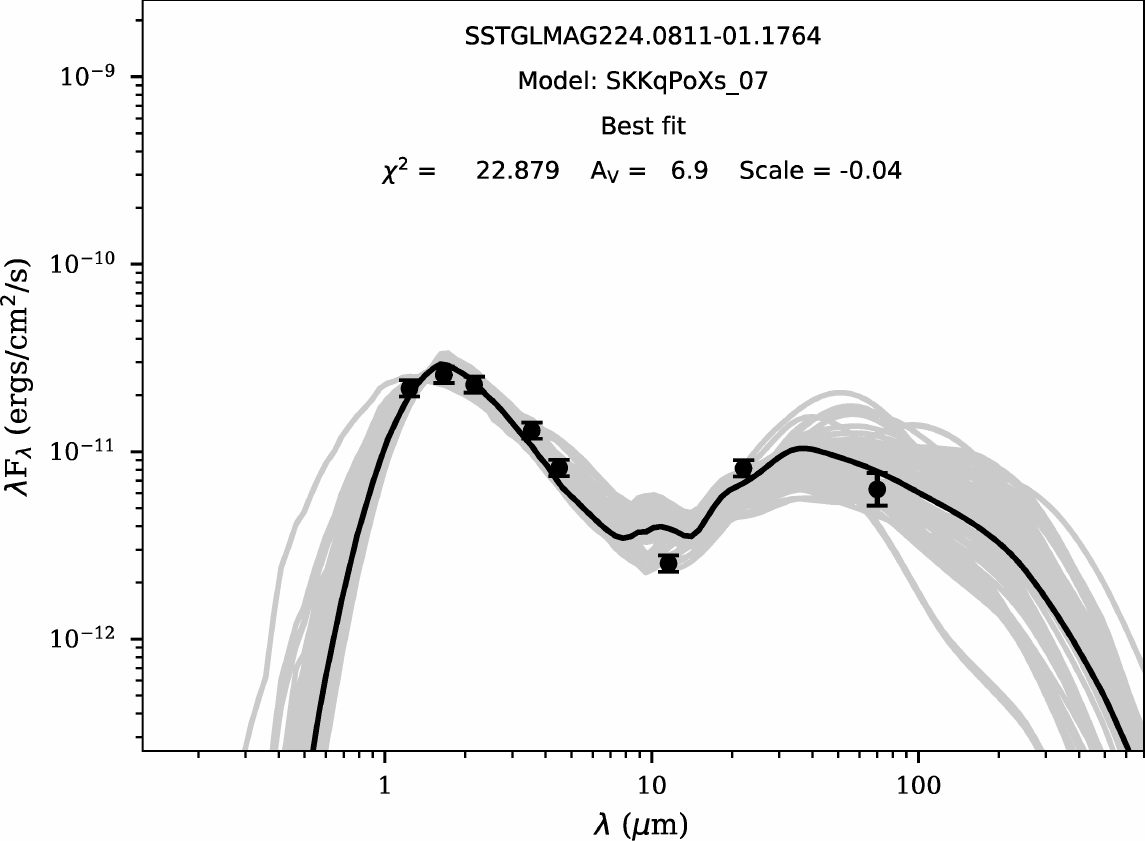}
\hfill
\includegraphics[width=0.32\textwidth]{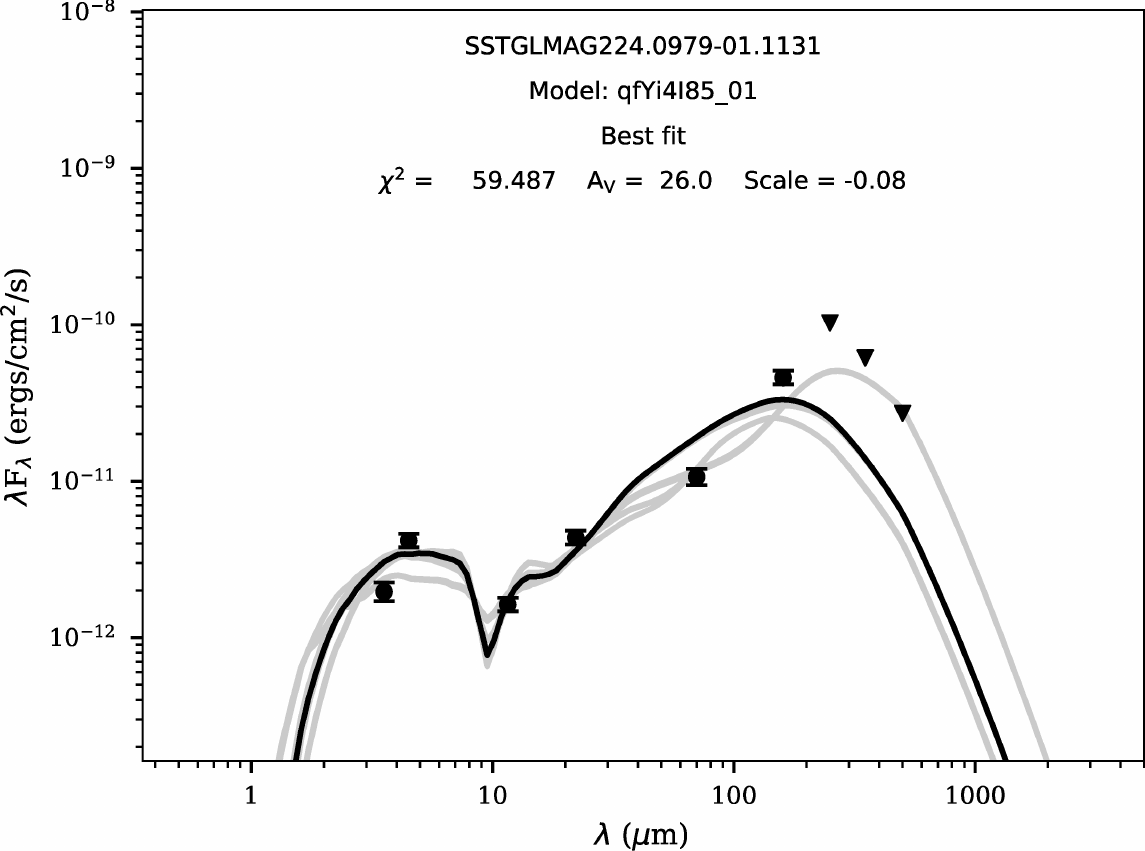}
\hfill
\includegraphics[width=0.32\textwidth]{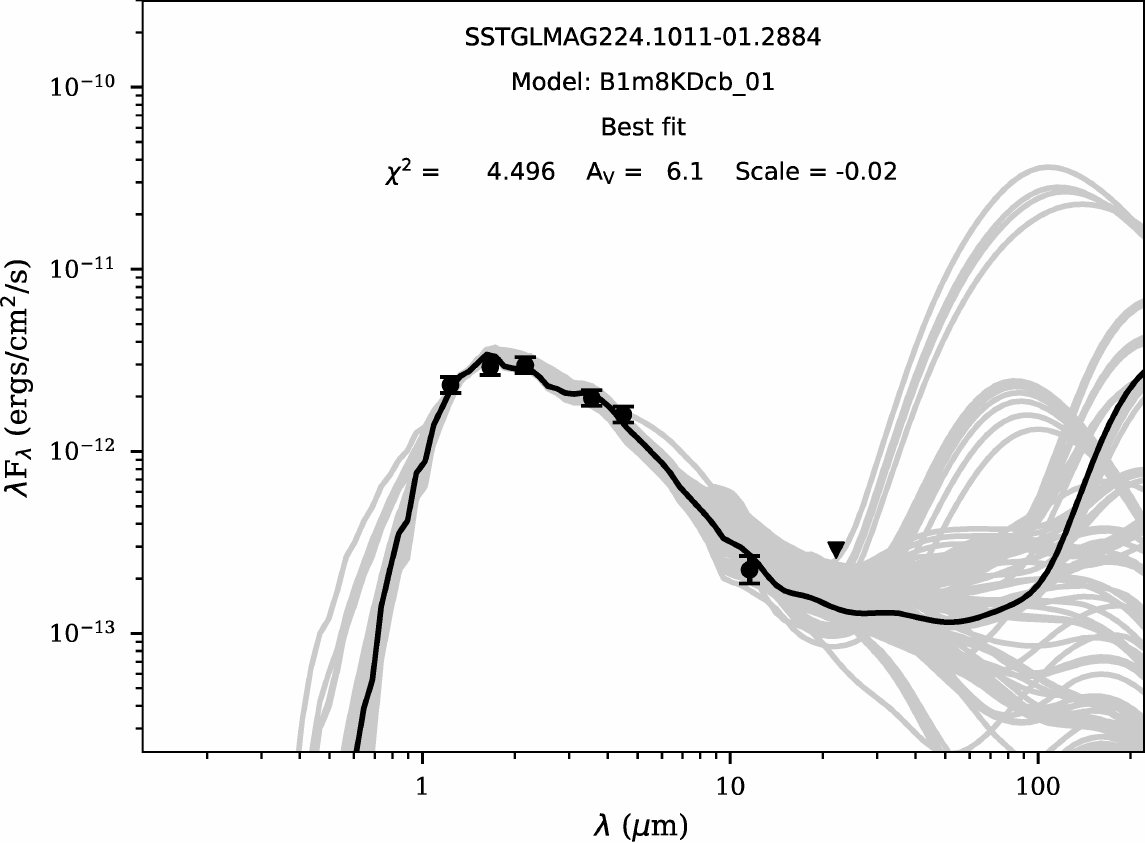} \par
\vspace{2mm}
\includegraphics[width=0.32\textwidth]{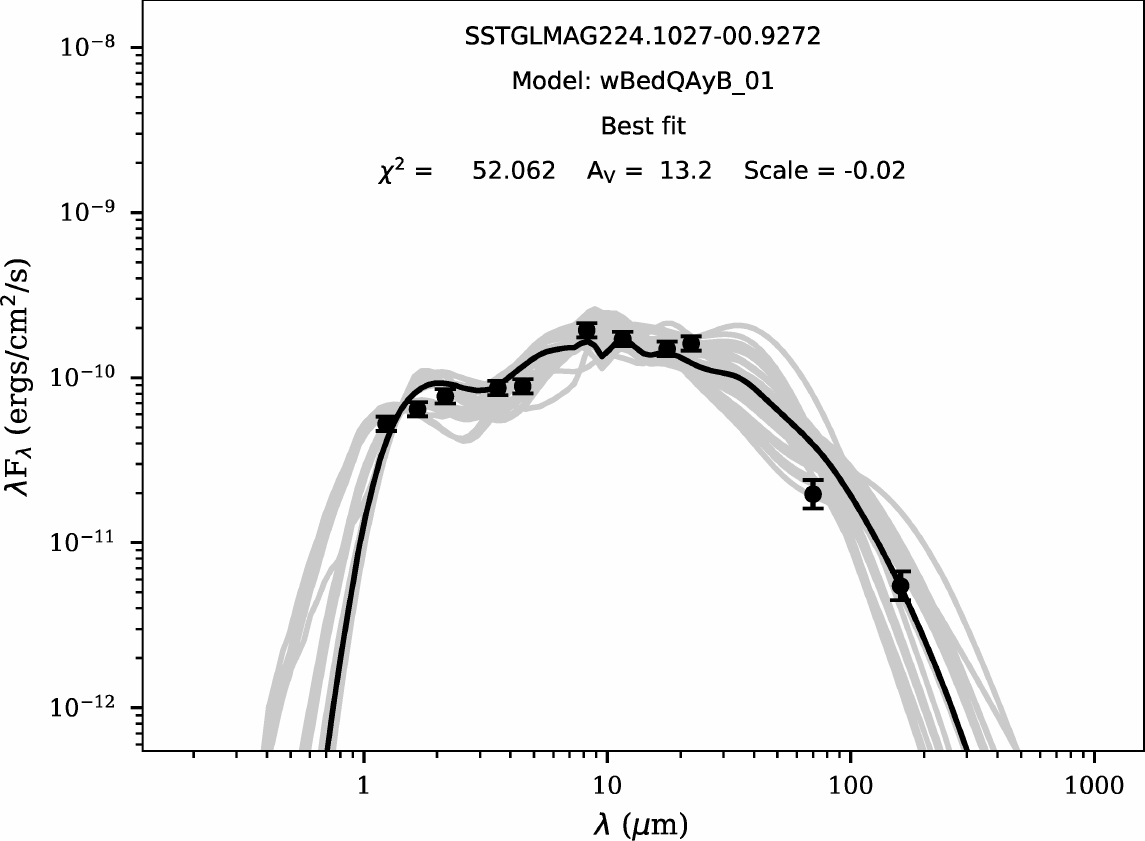}
\hfill
\includegraphics[width=0.32\textwidth]{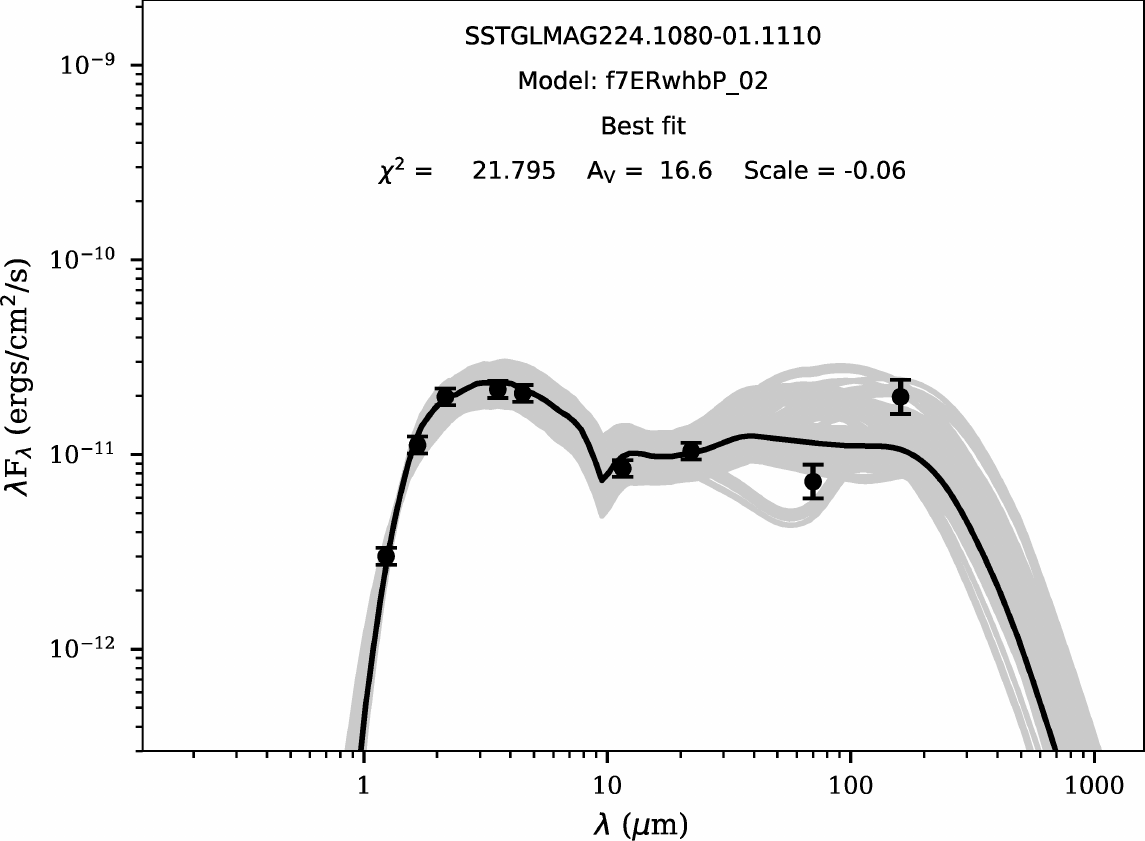}
\hfill
\includegraphics[width=0.32\textwidth]{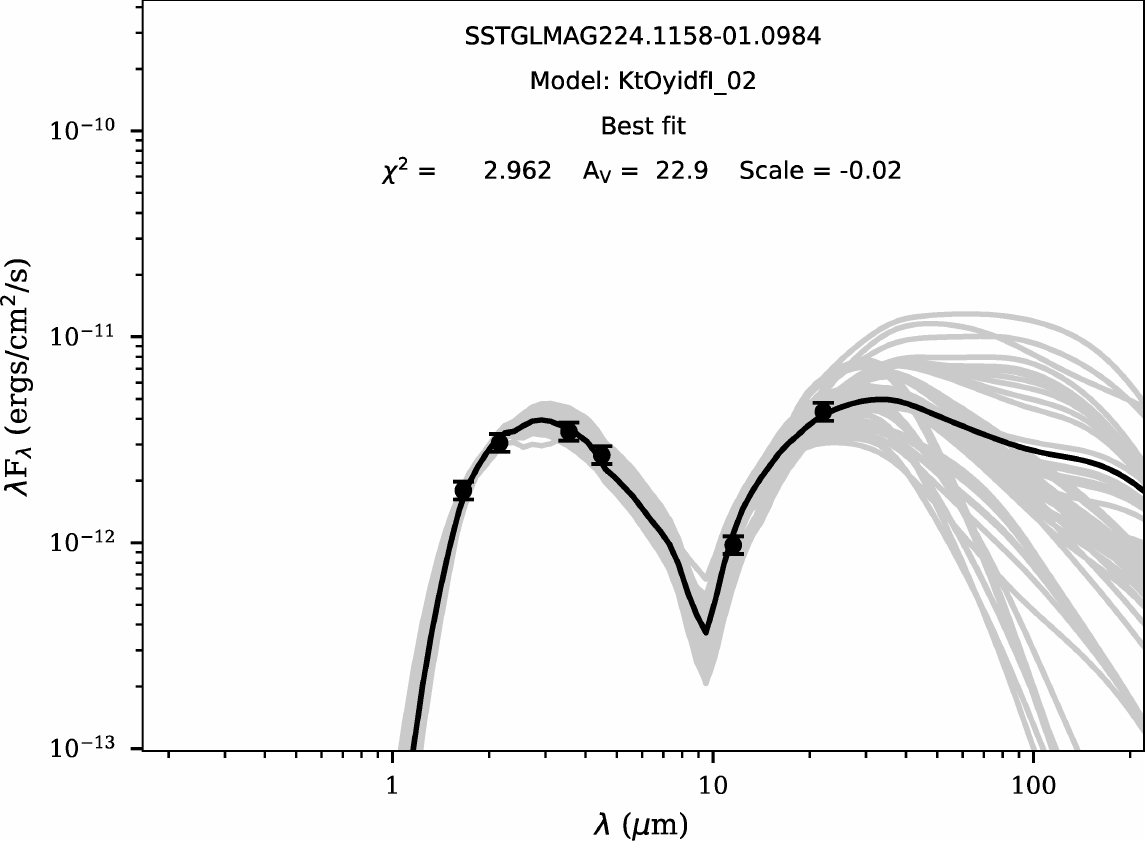} \par
\vspace{2mm}
\includegraphics[width=0.32\textwidth]{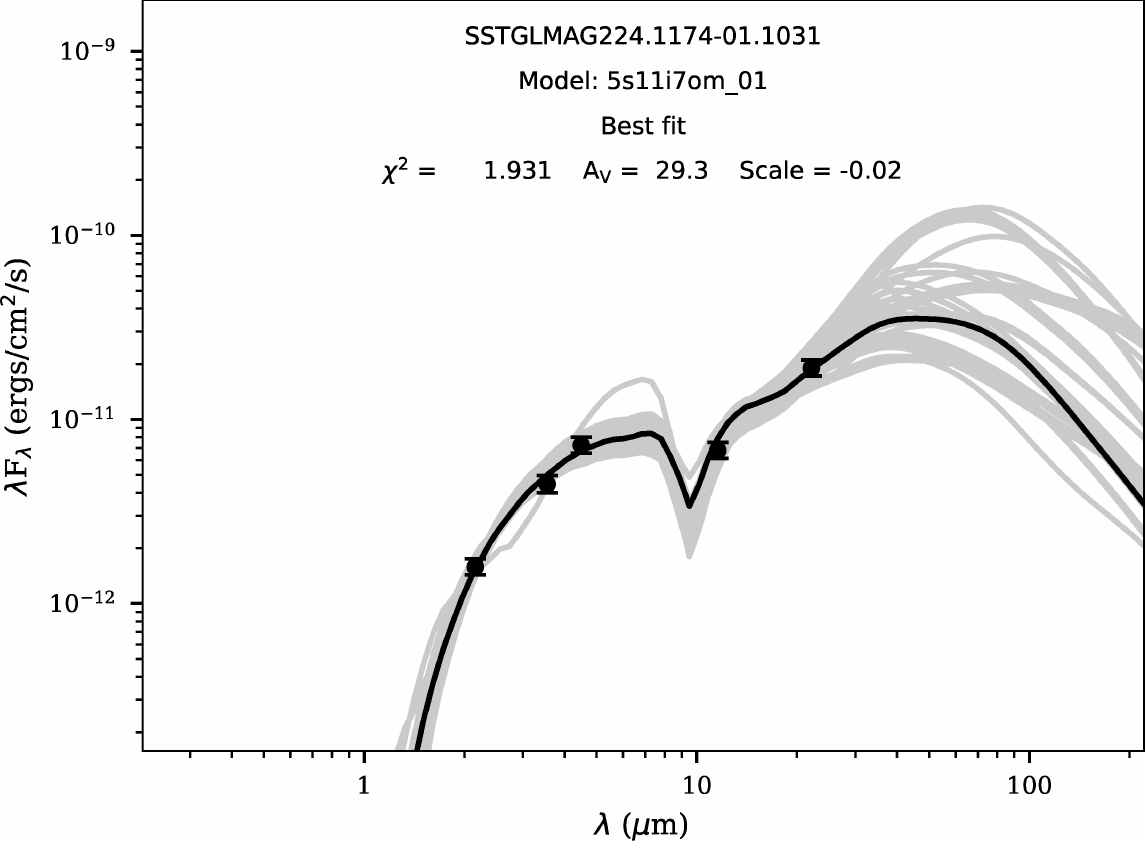}
\hfill
\includegraphics[width=0.32\textwidth]{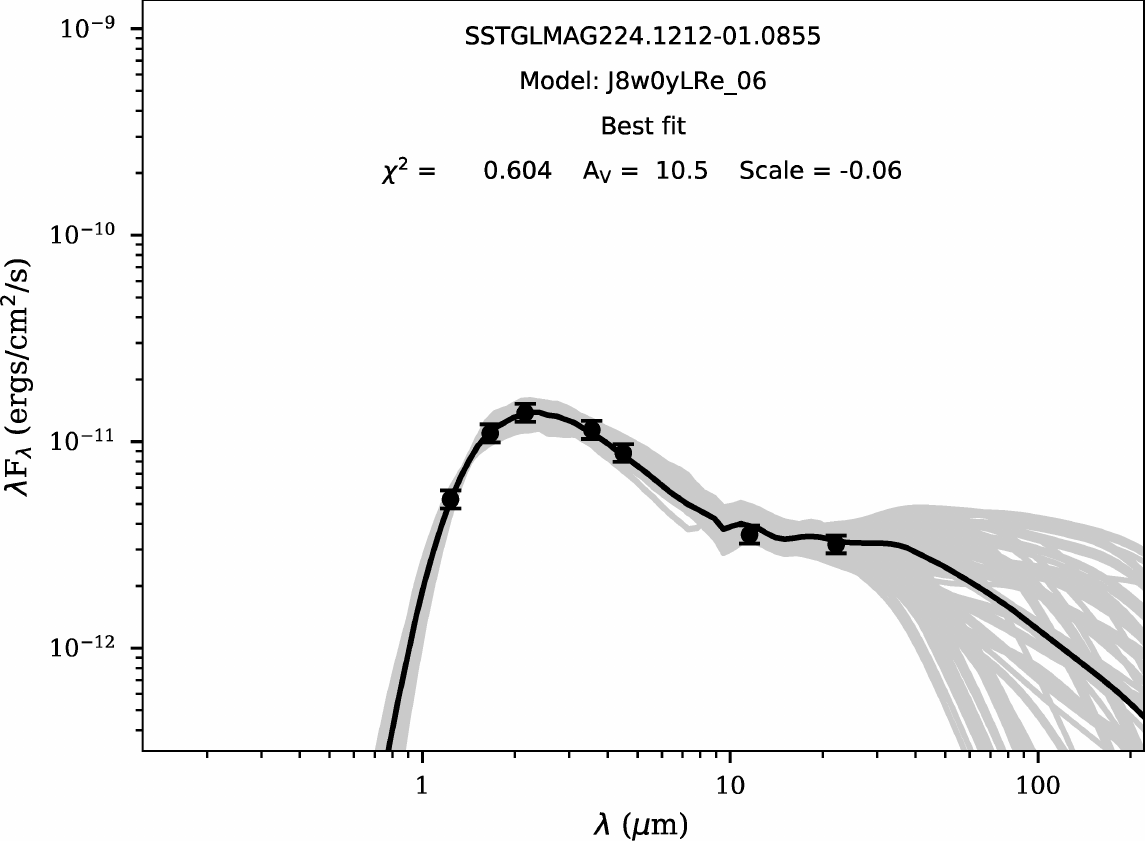}
\hfill
\includegraphics[width=0.32\textwidth]{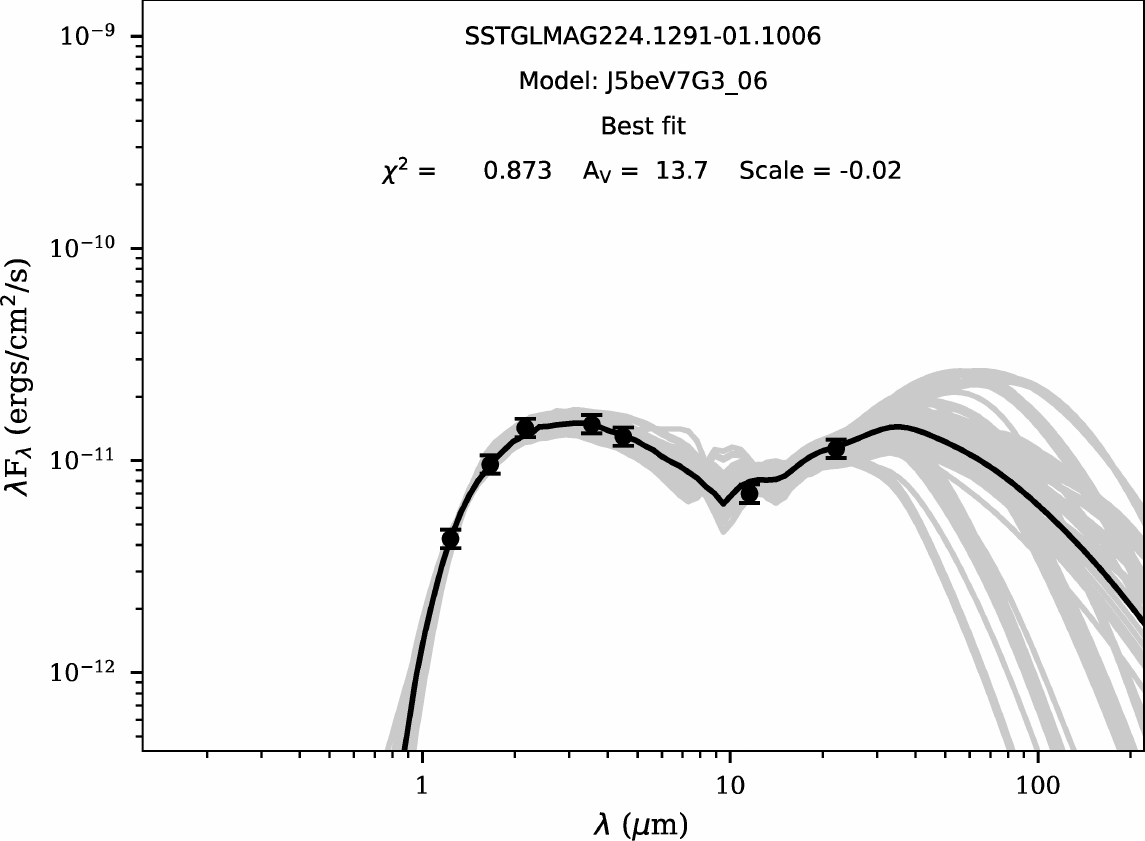} \par
\vspace{2mm}
\includegraphics[width=0.32\textwidth]{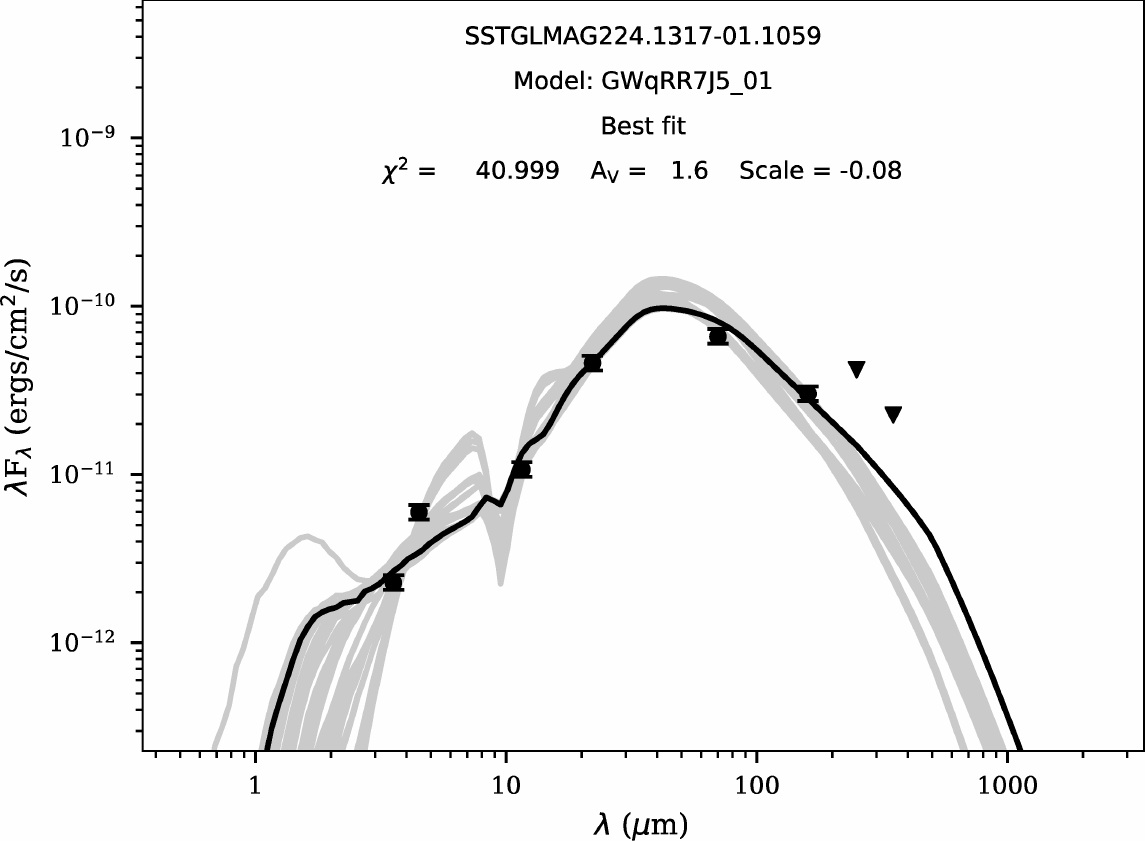}
\hfill
\includegraphics[width=0.32\textwidth]{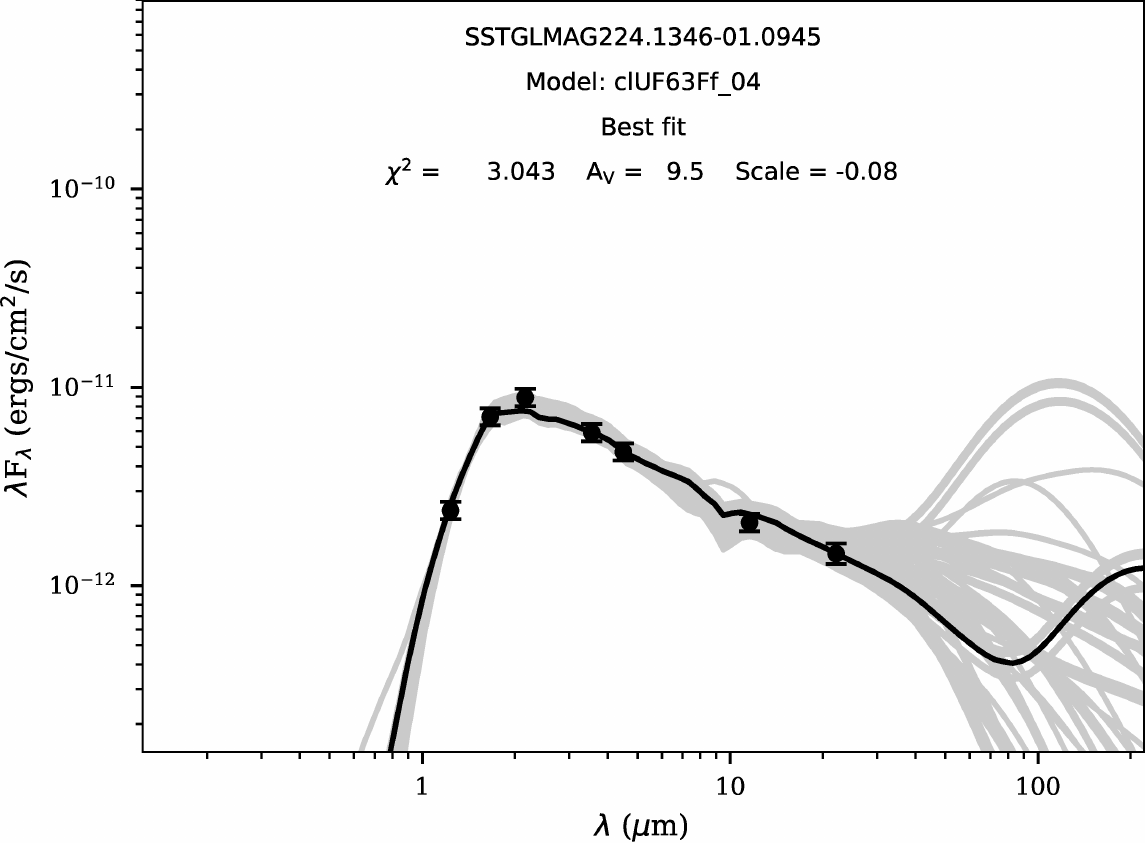}
\hfill
\includegraphics[width=0.32\textwidth]{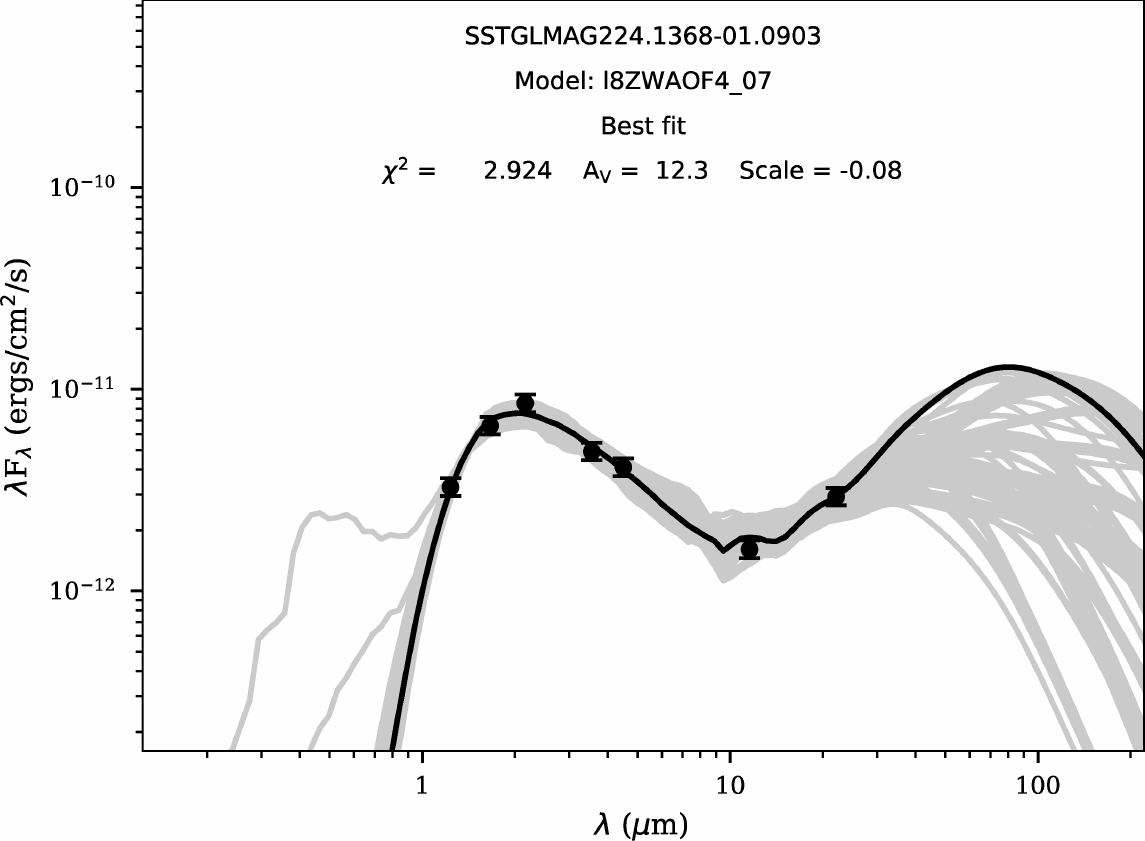} \par
\vspace{2mm}
\includegraphics[width=0.32\textwidth]{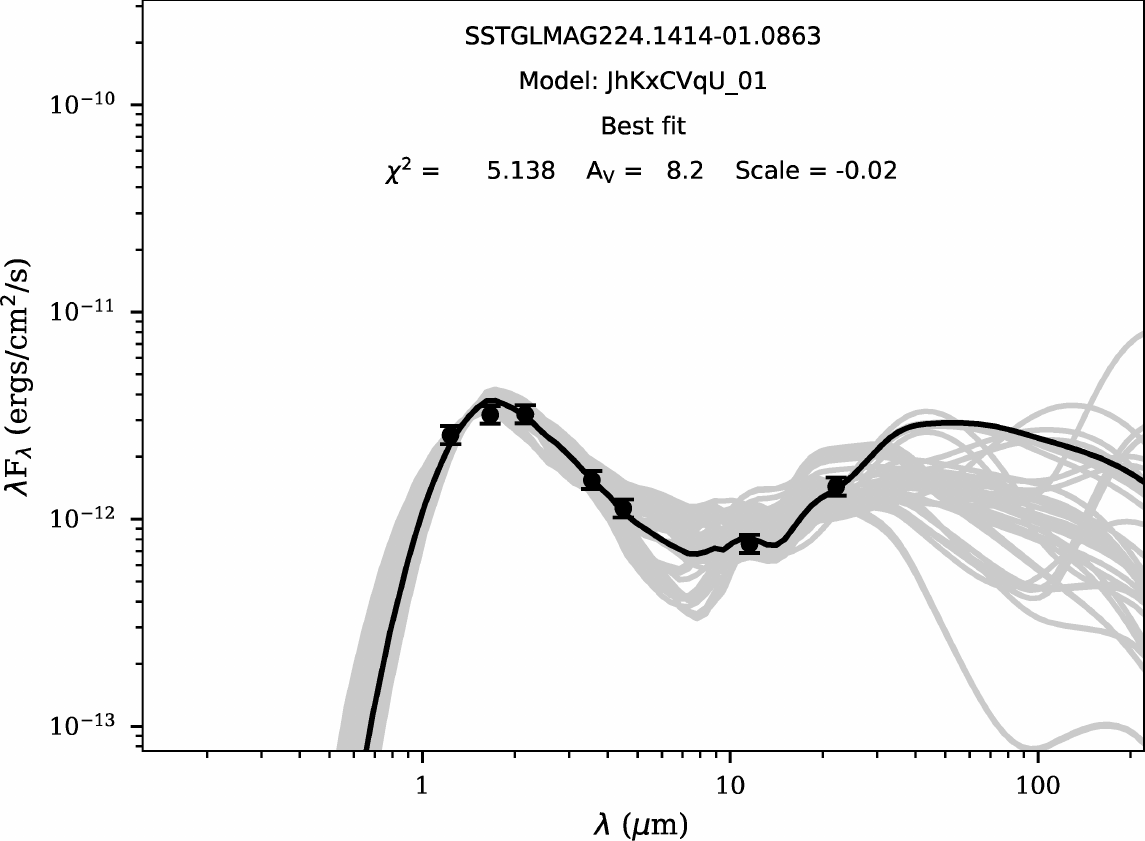}
\hfill
\includegraphics[width=0.32\textwidth]{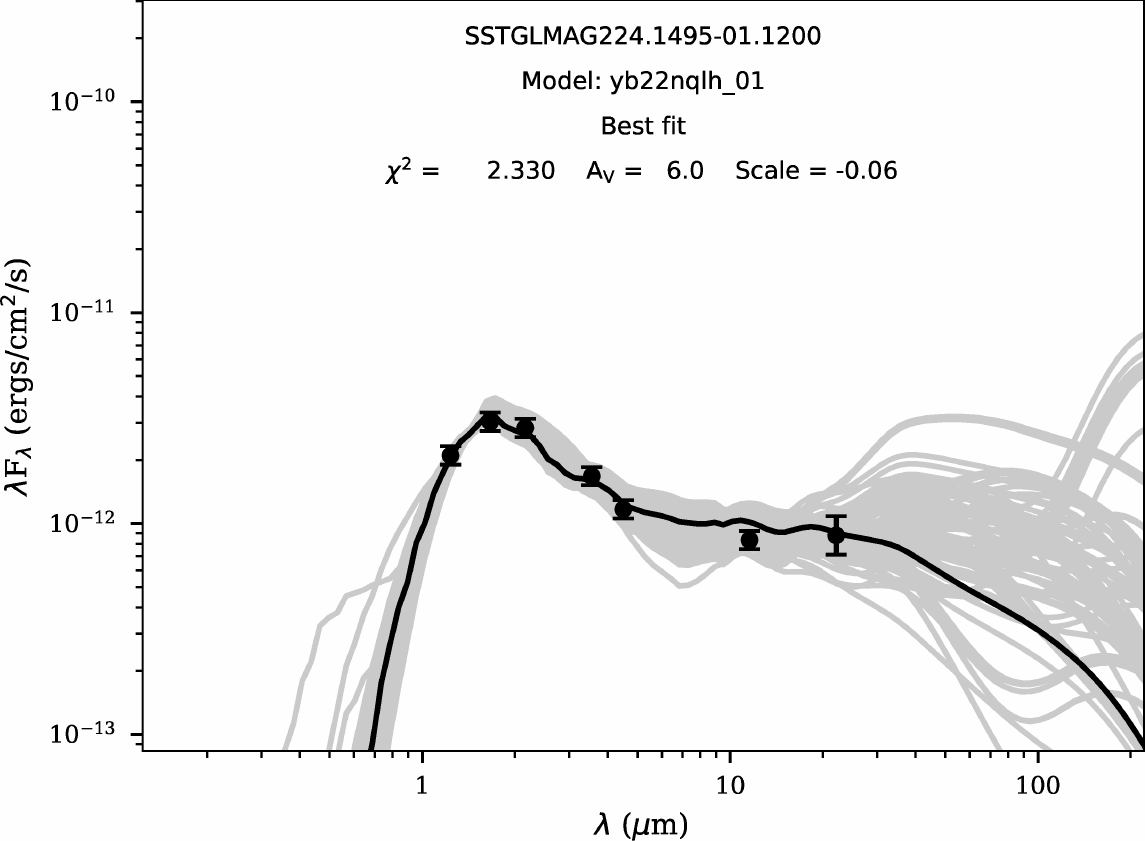}
\hfill
\includegraphics[width=0.32\textwidth]{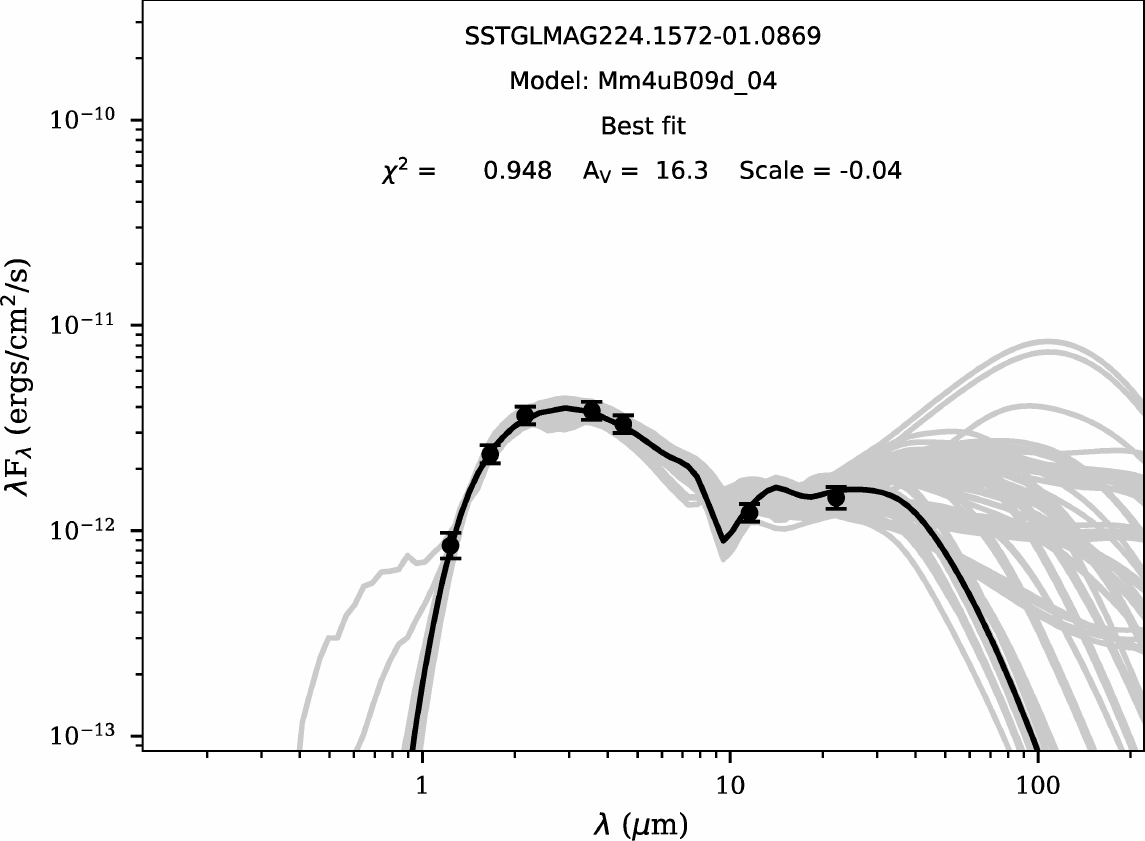}
\caption{Same as Fig.~\ref{f:SEDs1}  \label{f:SEDs3}}
\end{figure*}

\begin{figure*}
\includegraphics[width=0.32\textwidth]{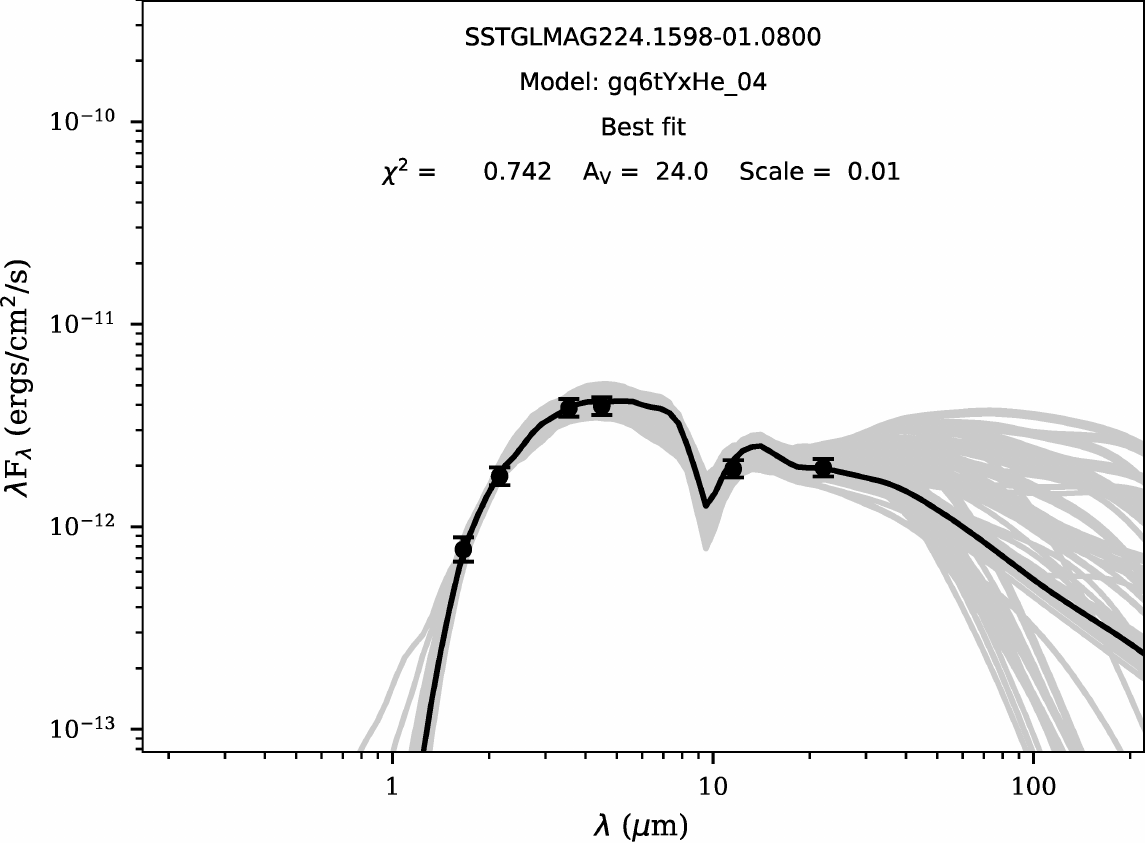}
\hfill
\includegraphics[width=0.32\textwidth]{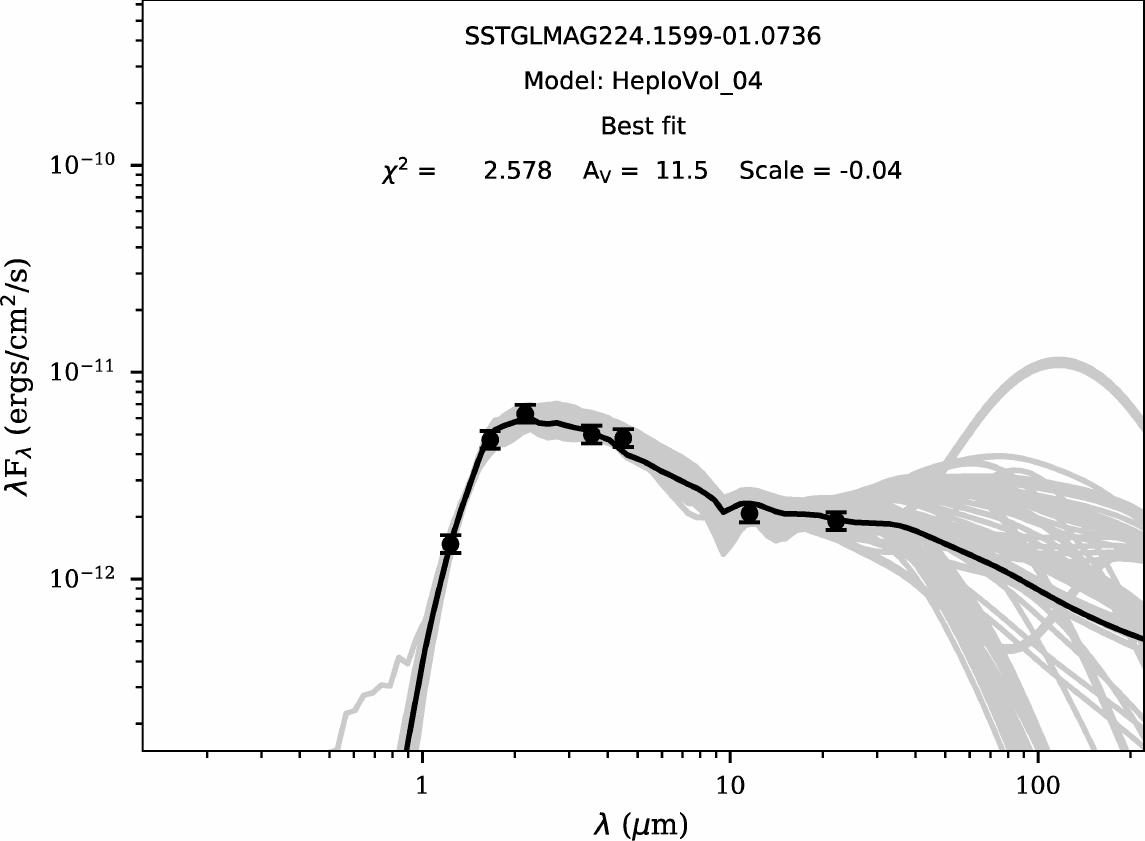}
\hfill
\includegraphics[width=0.32\textwidth]{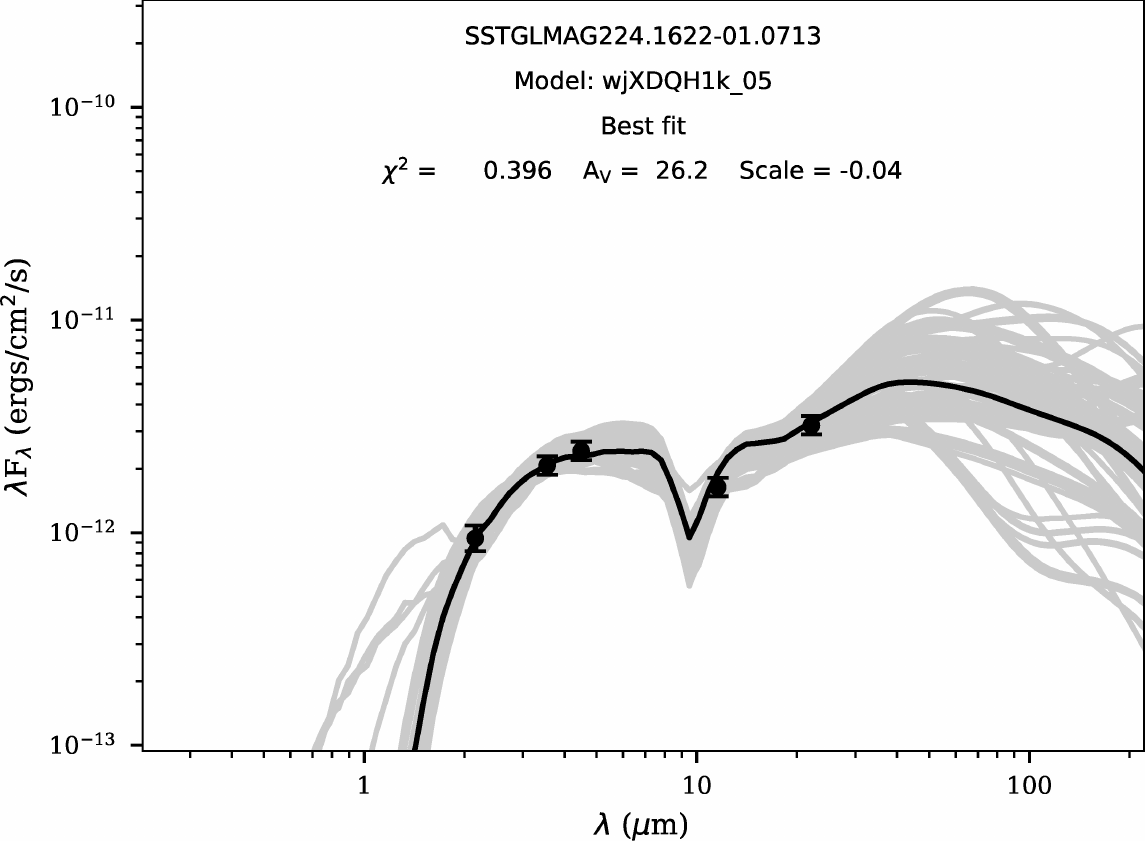} \par
\vspace{2mm}
\includegraphics[width=0.32\textwidth]{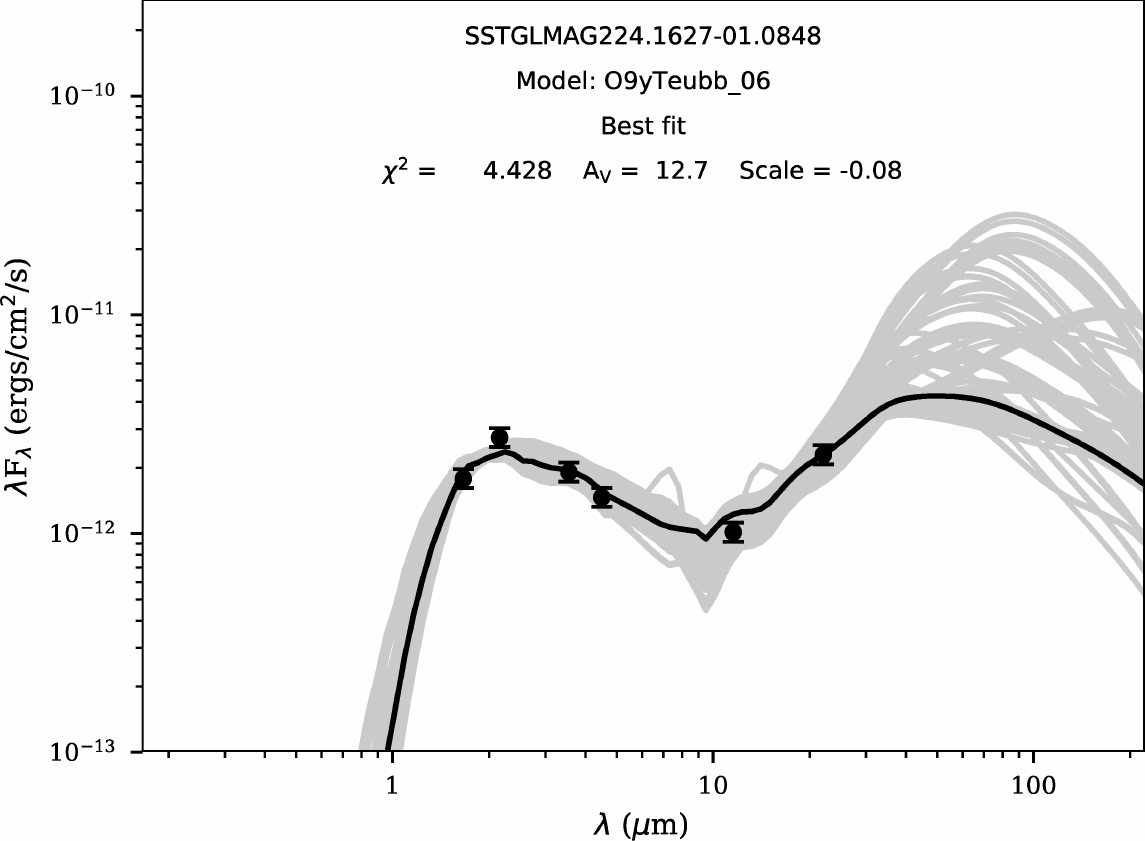}
\hfill
\includegraphics[width=0.32\textwidth]{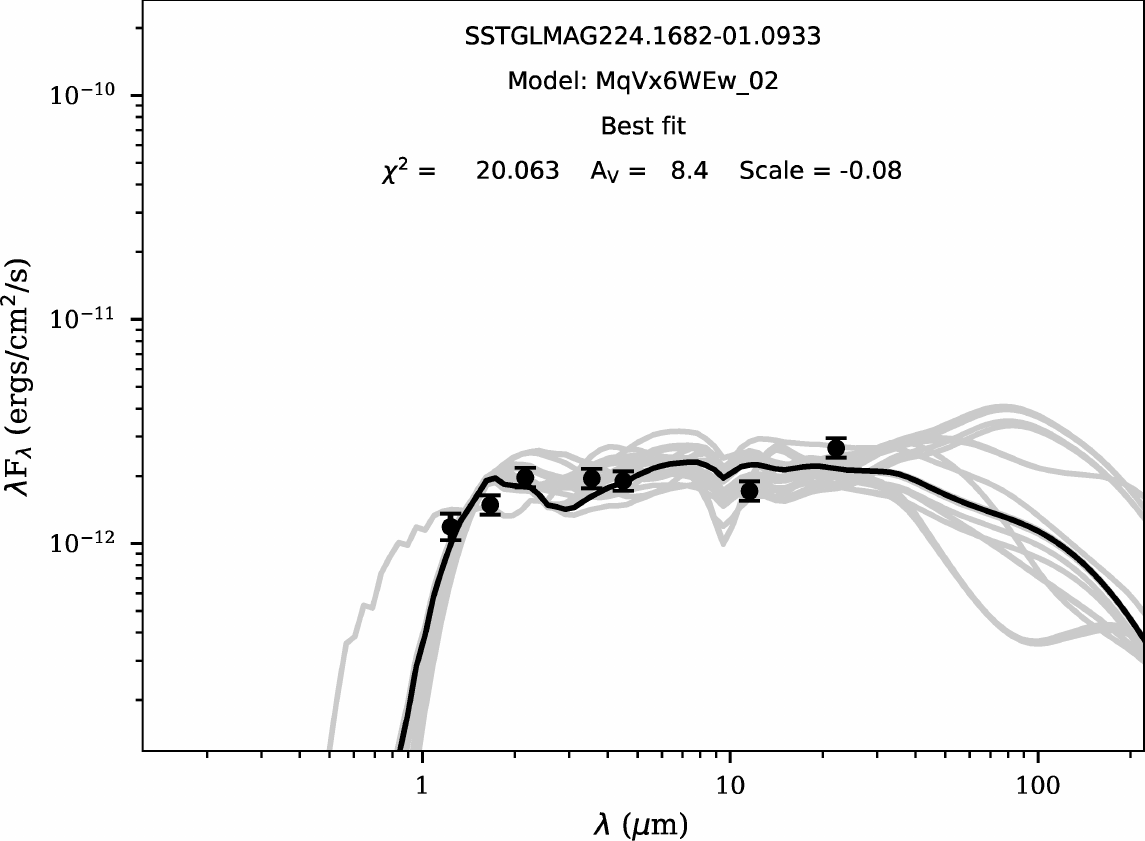}
\hfill
\includegraphics[width=0.32\textwidth]{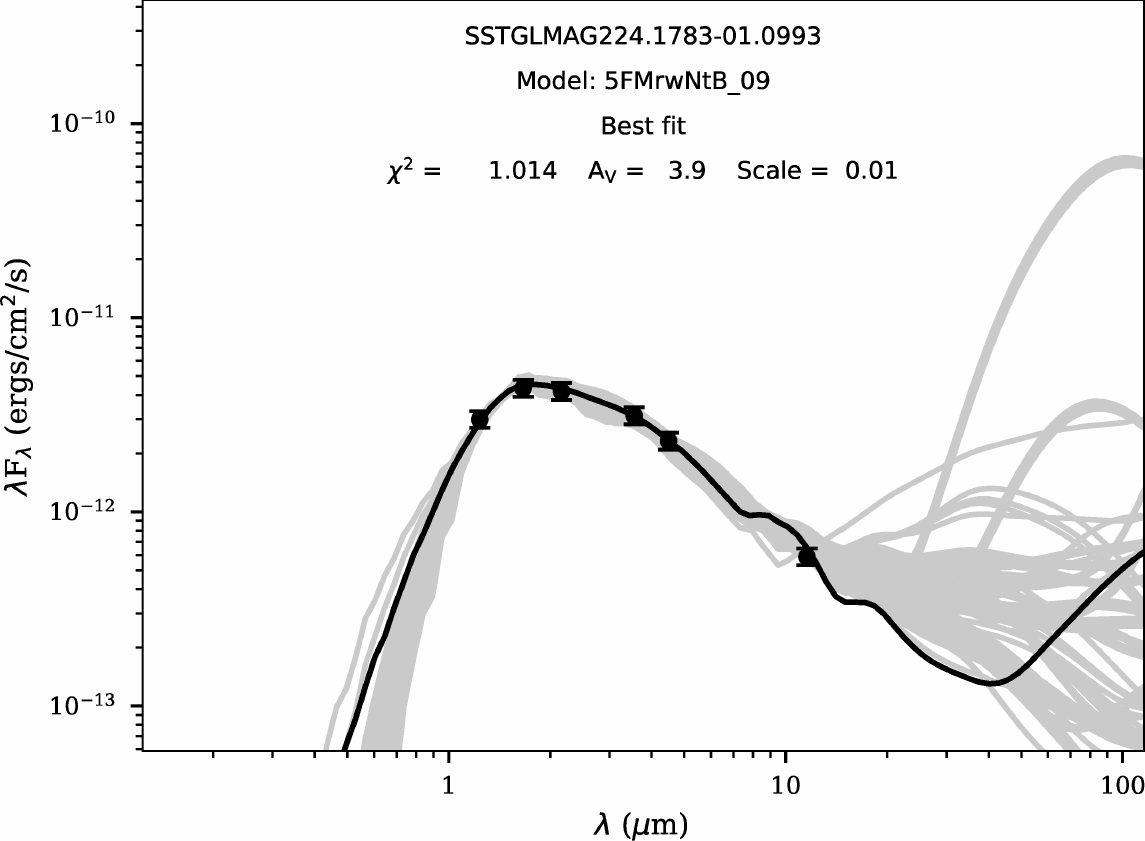} \par
\vspace{2mm}
\includegraphics[width=0.32\textwidth]{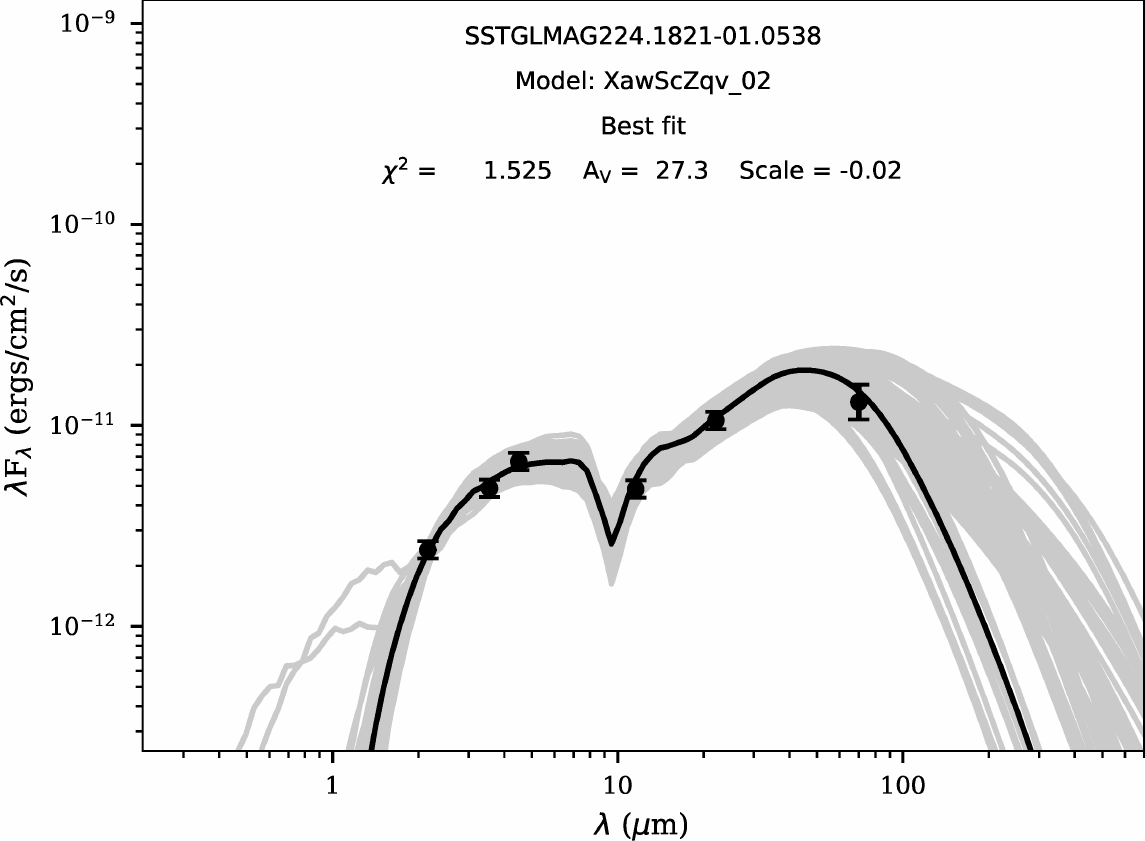}
\hfill
\includegraphics[width=0.32\textwidth]{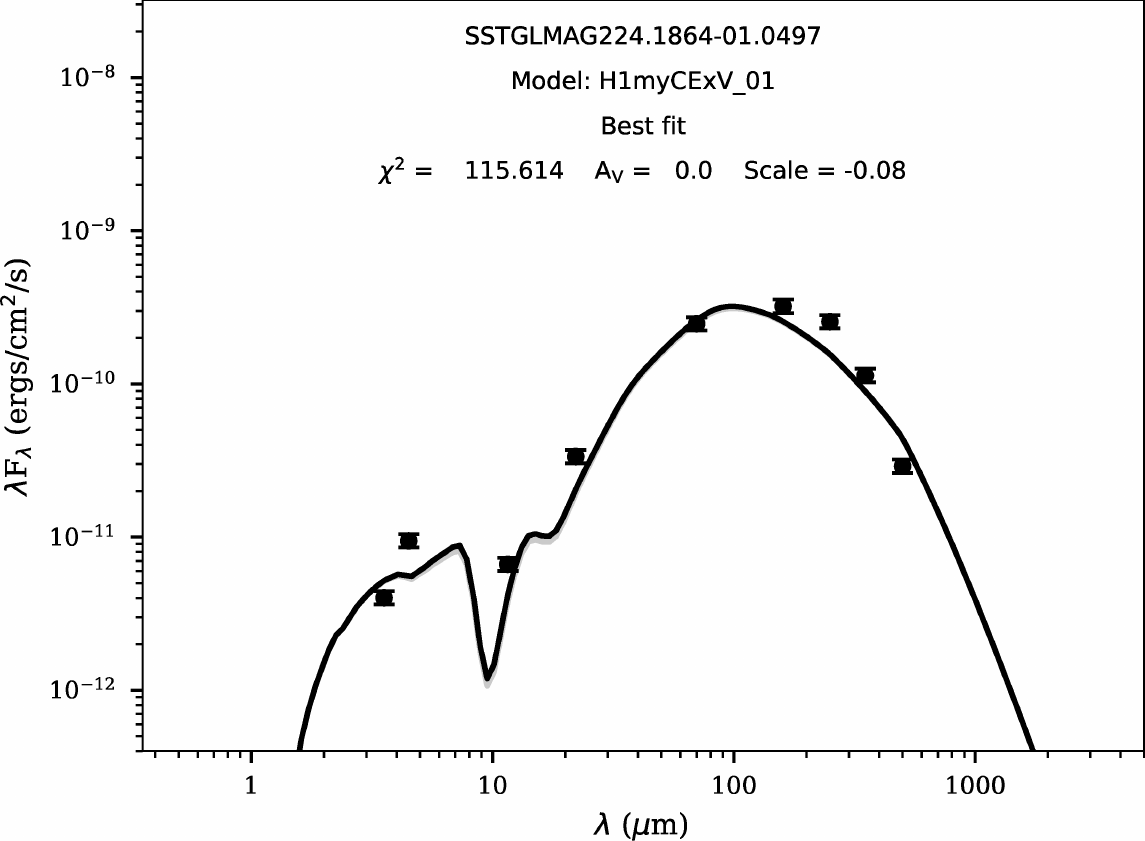}
\hfill
\includegraphics[width=0.32\textwidth]{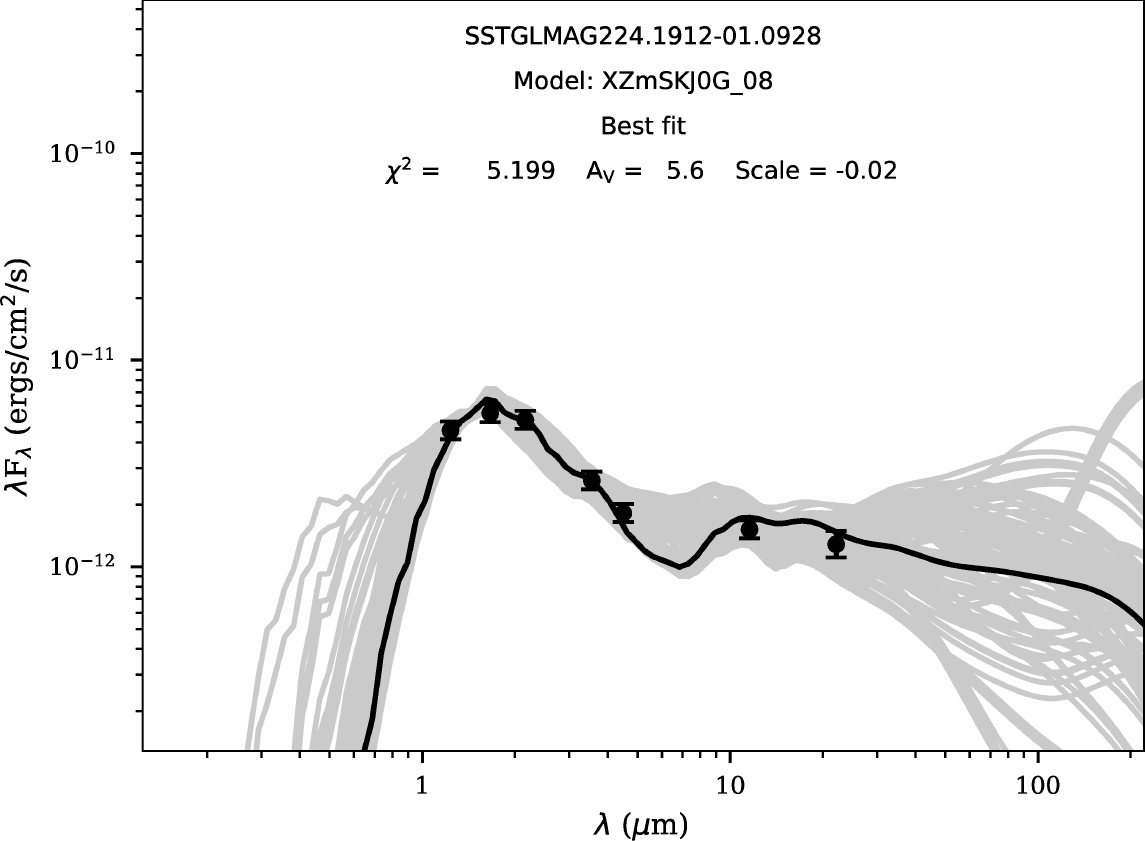} \par
\vspace{2mm}
\includegraphics[width=0.32\textwidth]{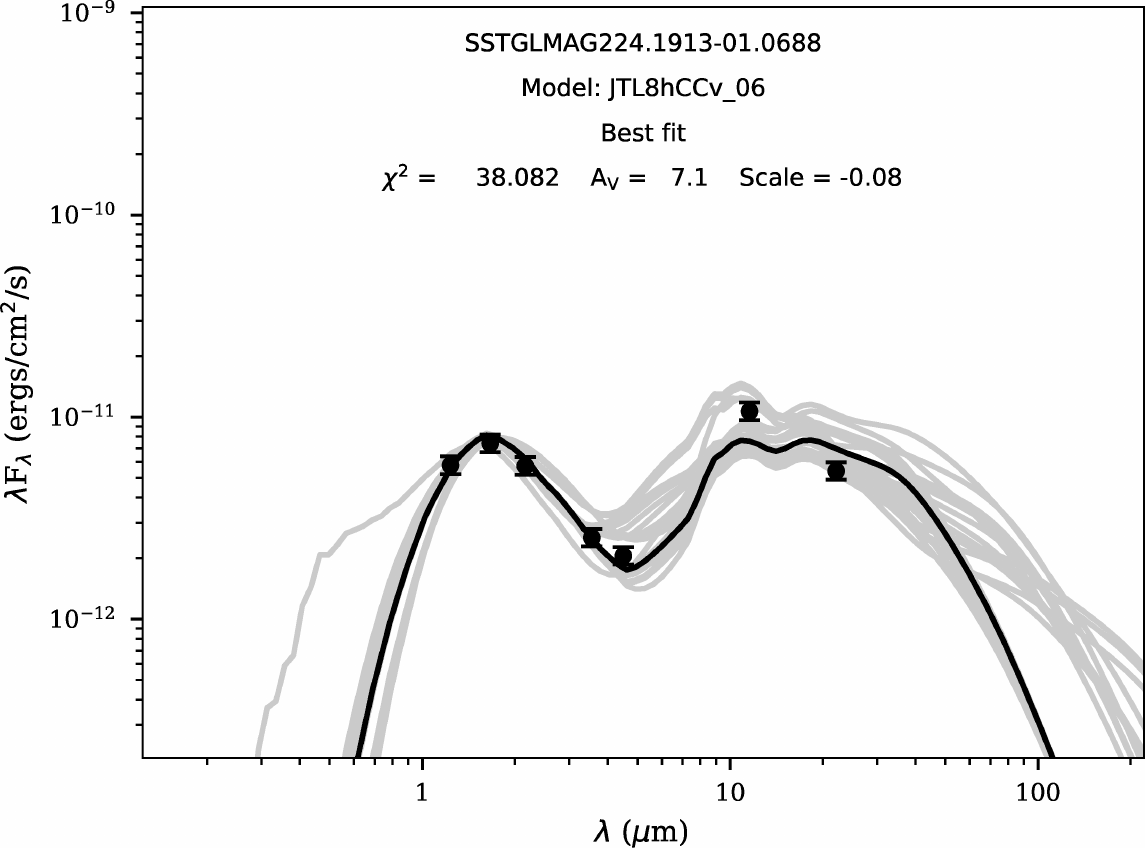}
\hfill
\includegraphics[width=0.32\textwidth]{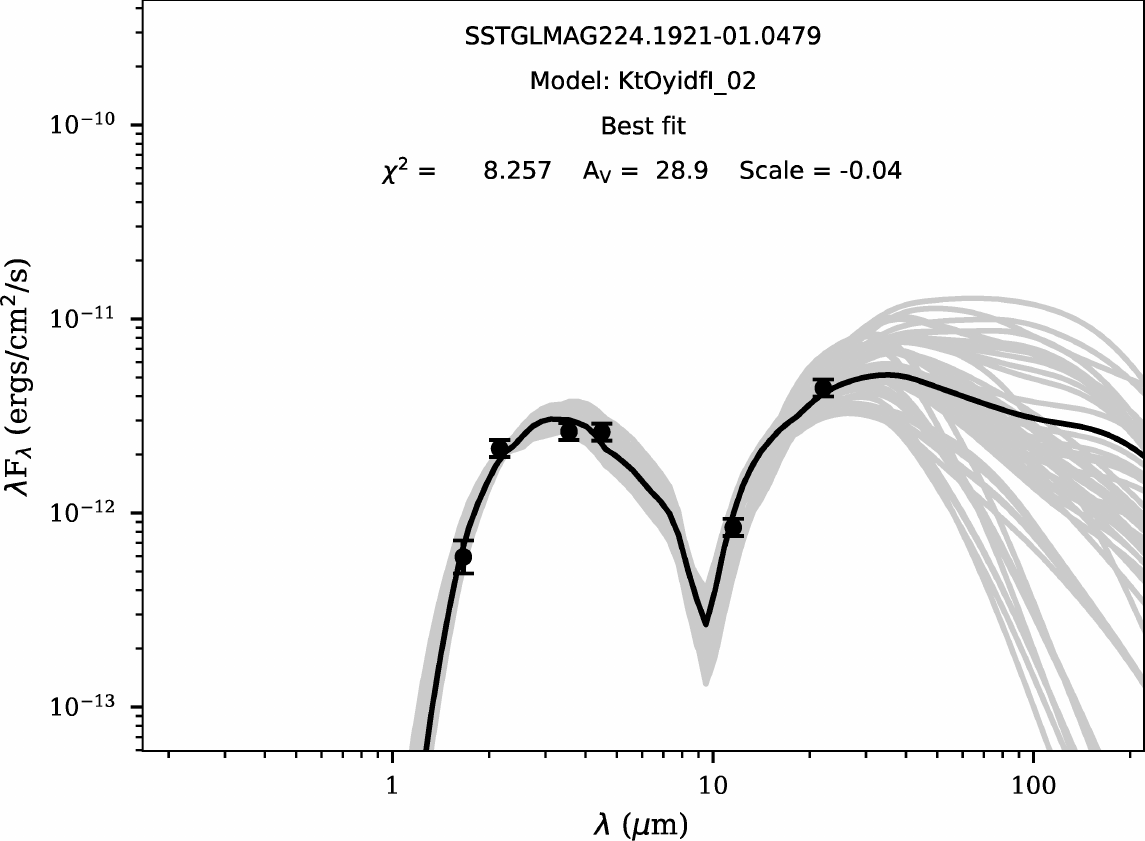}
\hfill
\includegraphics[width=0.32\textwidth]{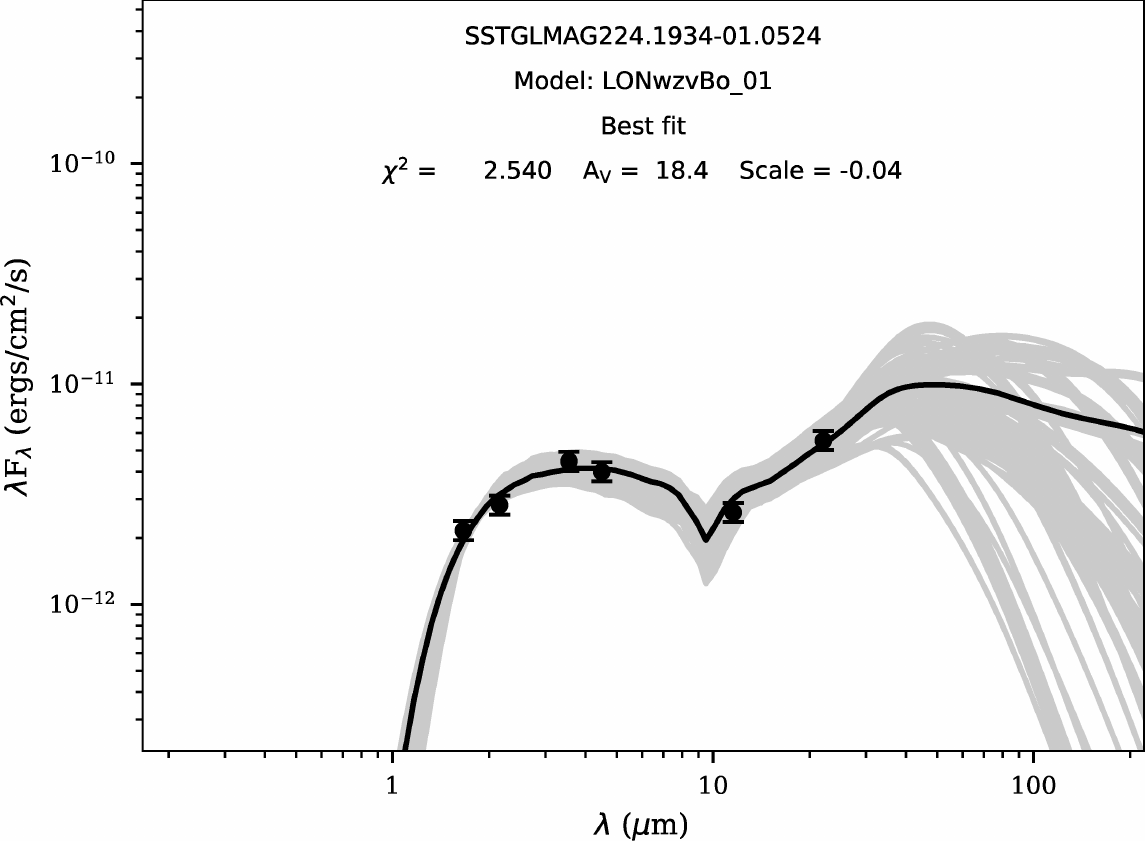} \par
\vspace{2mm}
\includegraphics[width=0.32\textwidth]{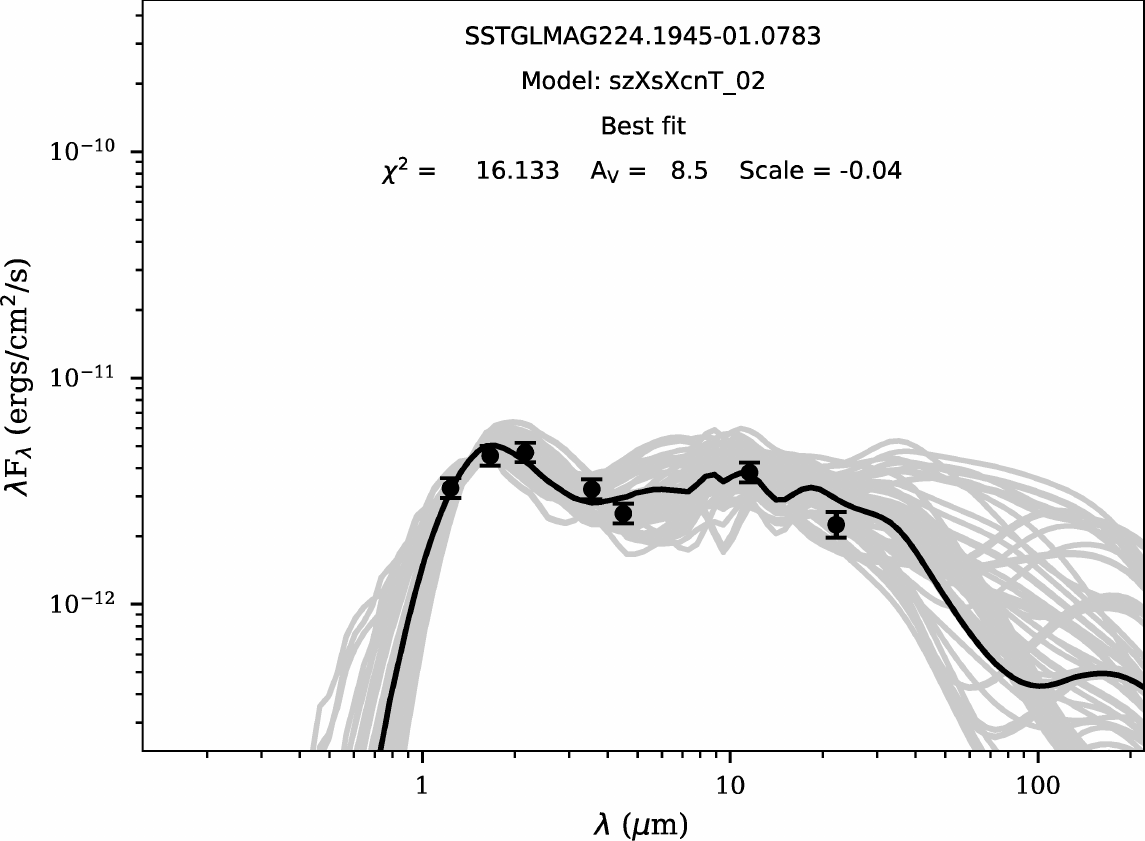}
\hfill
\includegraphics[width=0.32\textwidth]{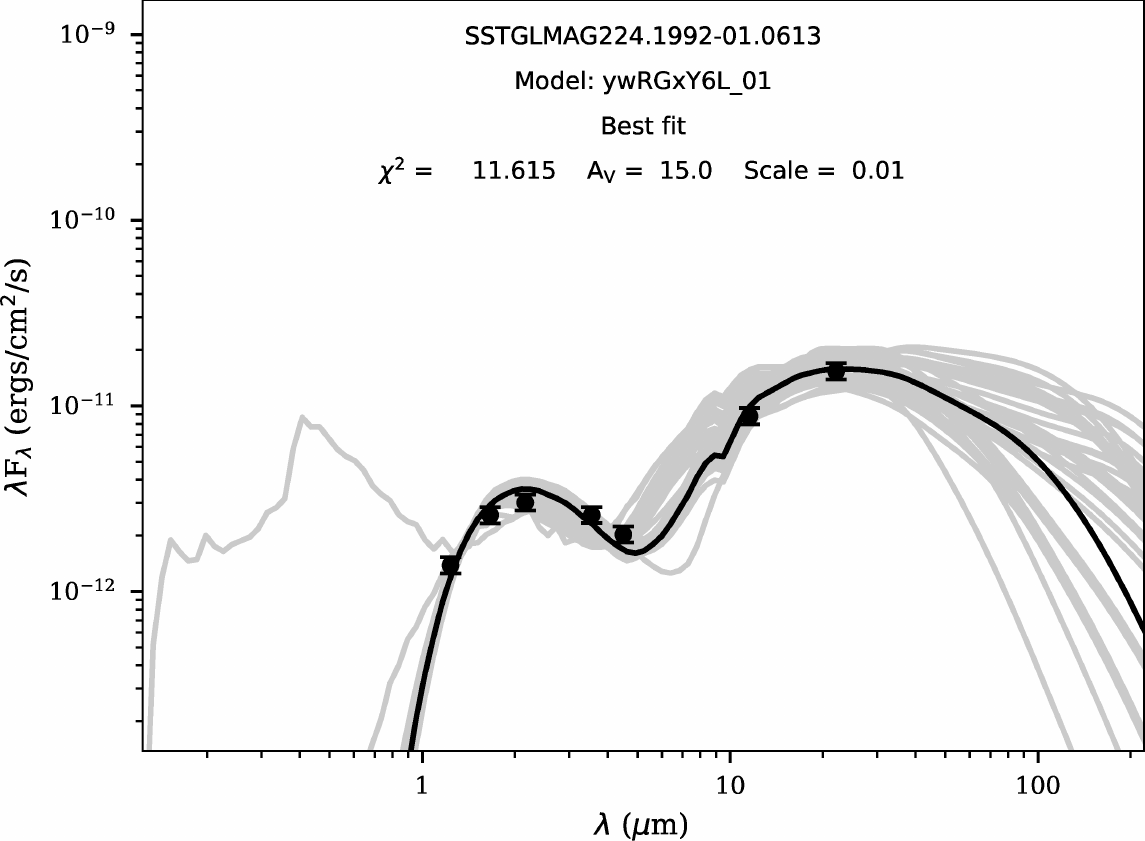}
\hfill
\includegraphics[width=0.32\textwidth]{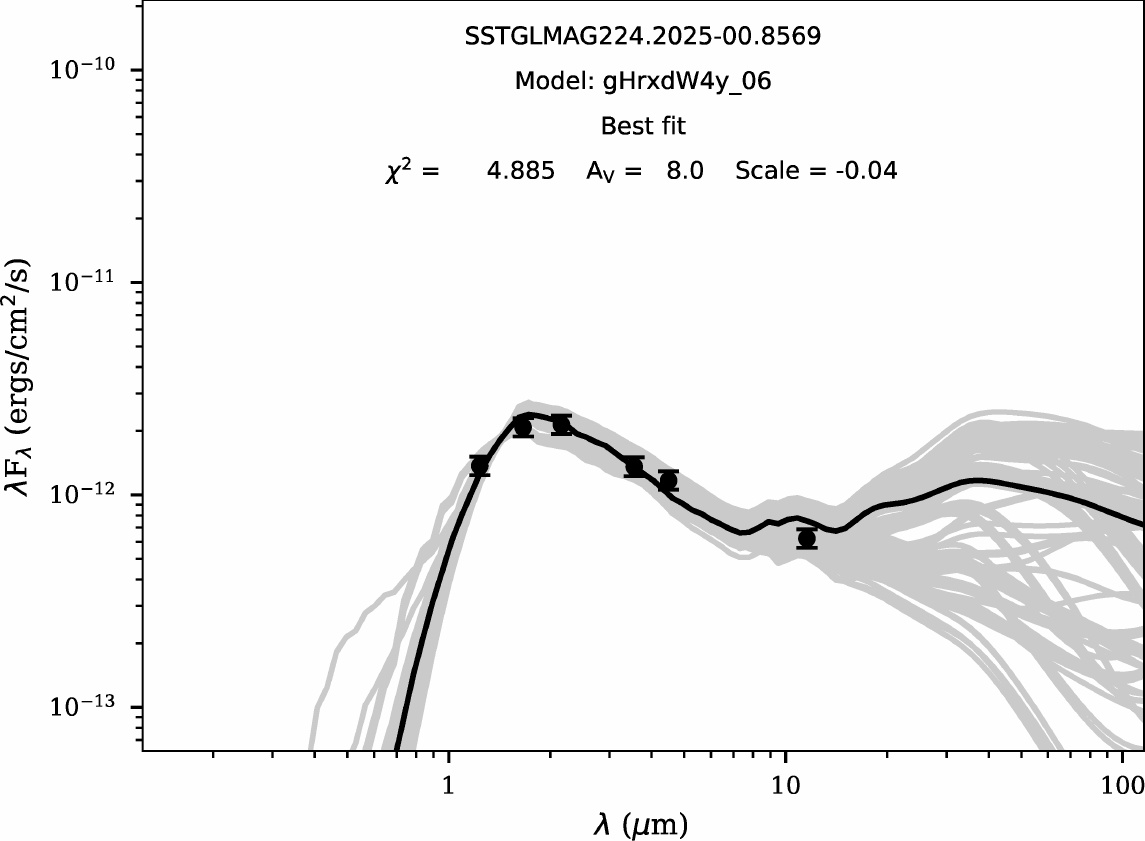}
\caption{Same as Fig.~\ref{f:SEDs1}  \label{f:SEDs4}}
\end{figure*}

\begin{figure*}
\includegraphics[width=0.32\textwidth]{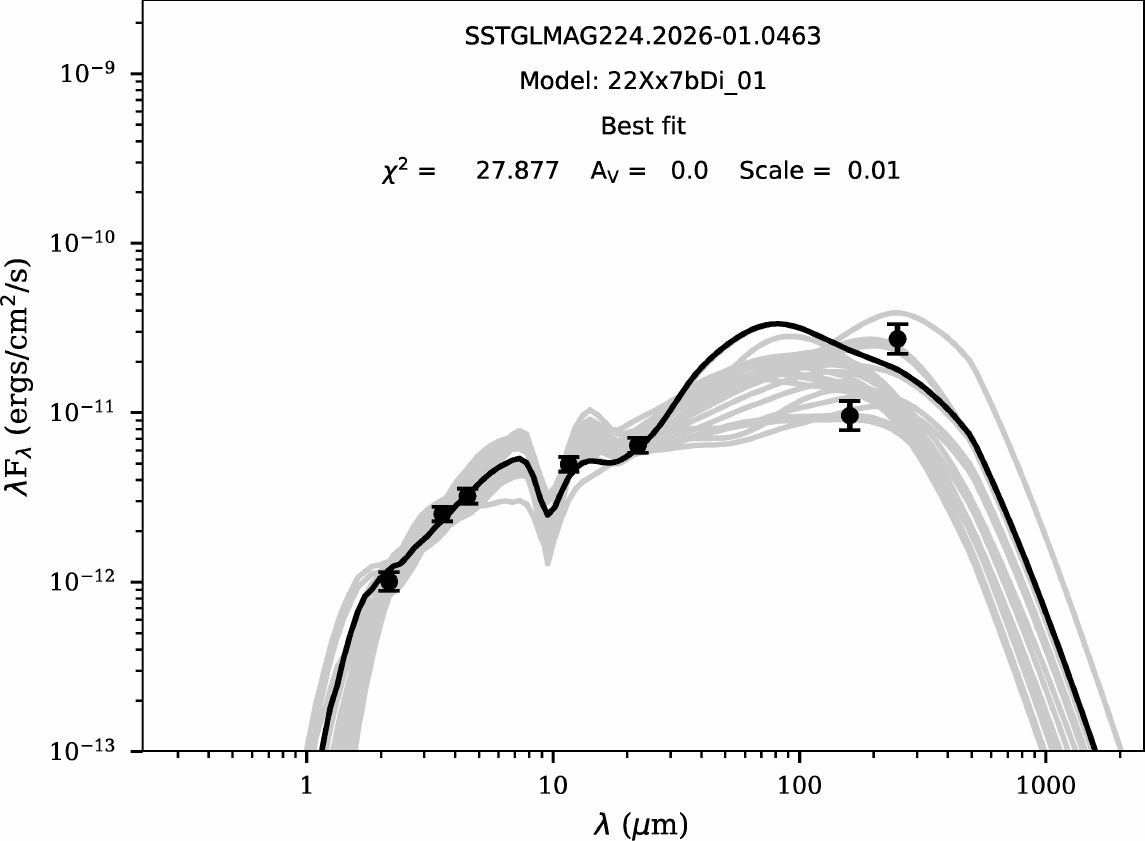}
\hfill
\includegraphics[width=0.32\textwidth]{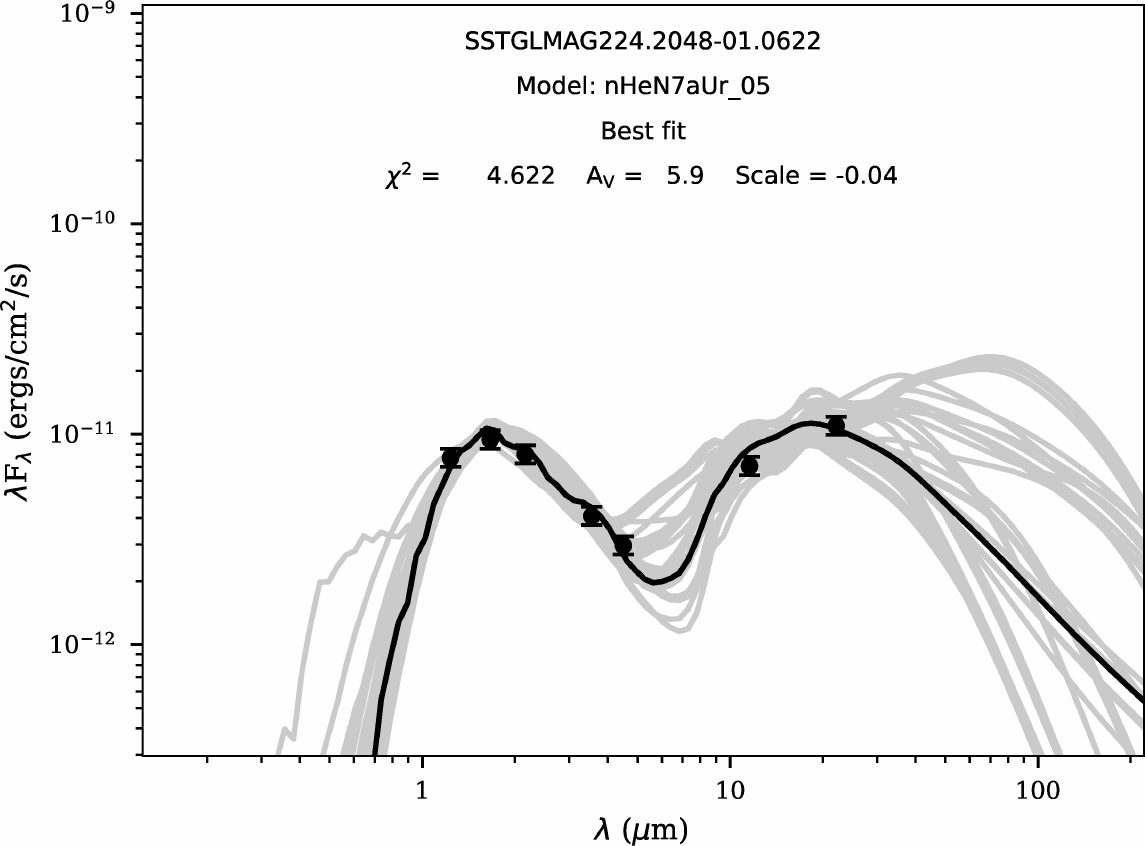}
\hfill
\includegraphics[width=0.32\textwidth]{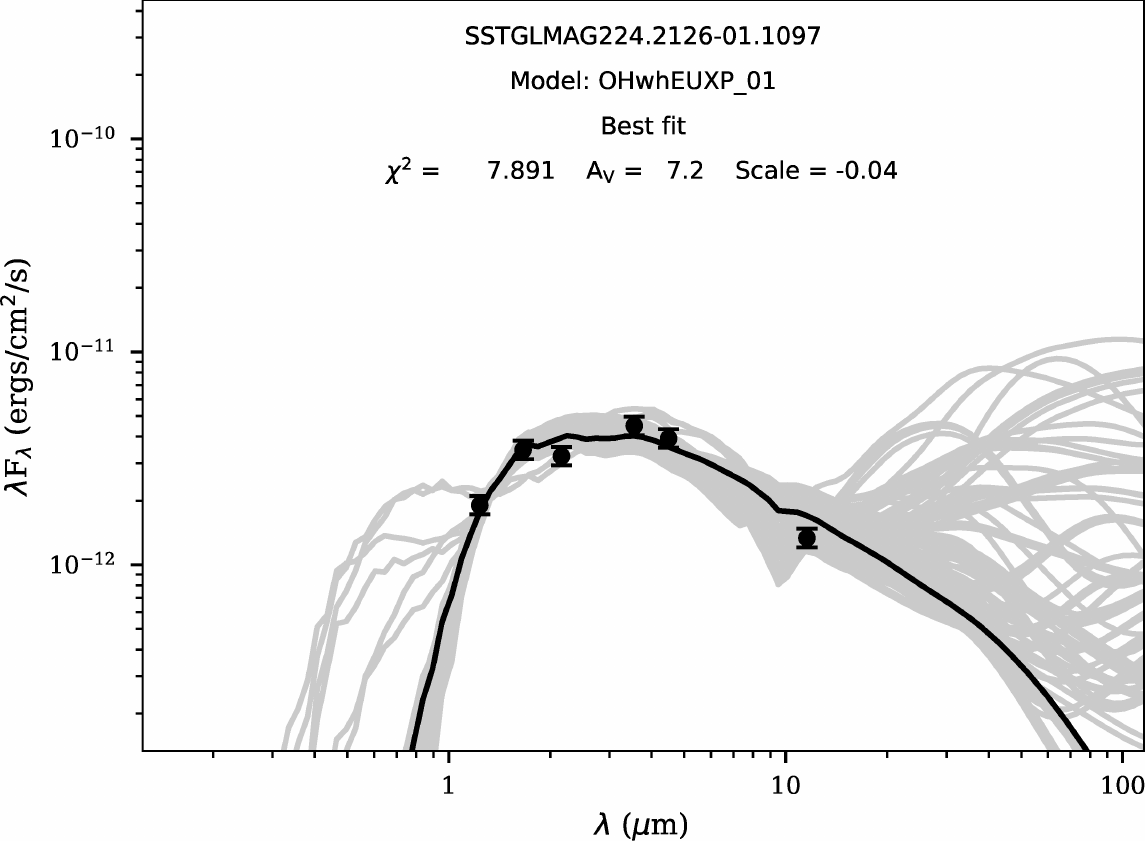} \par
\vspace{2mm}
\includegraphics[width=0.32\textwidth]{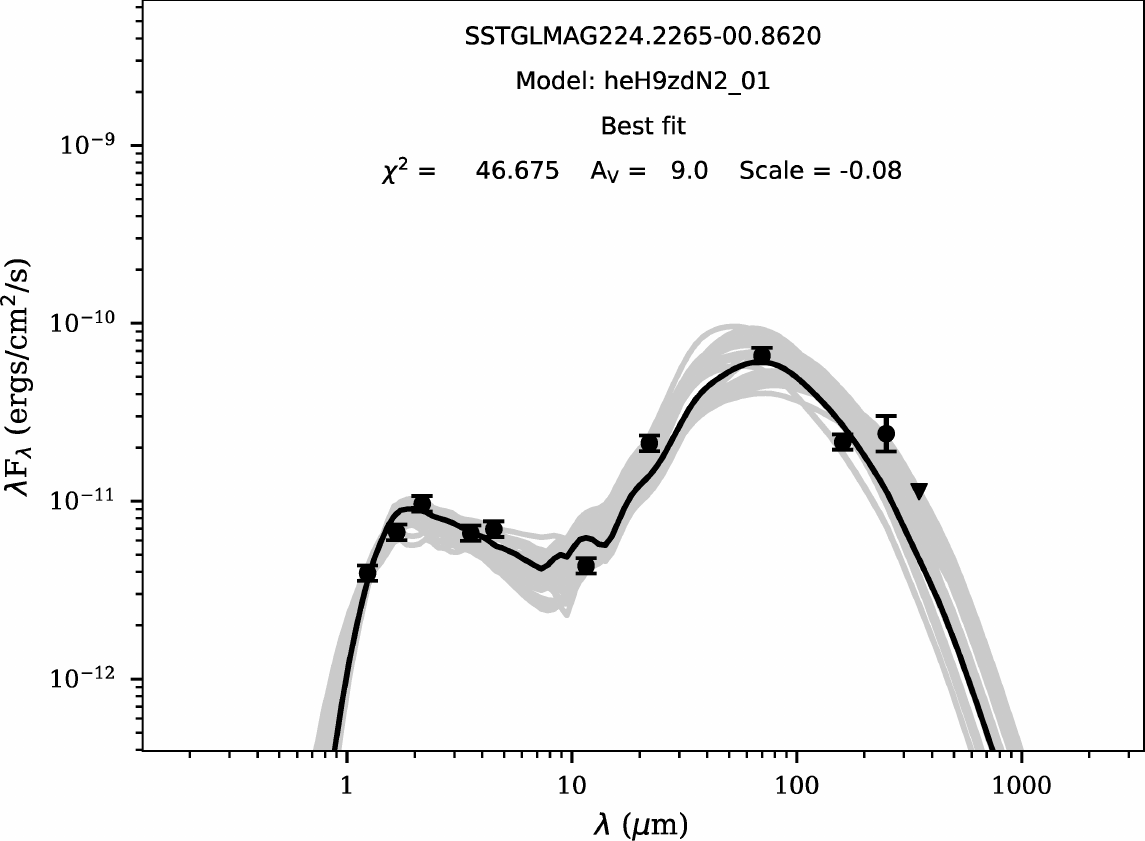}
\hfill
\includegraphics[width=0.32\textwidth]{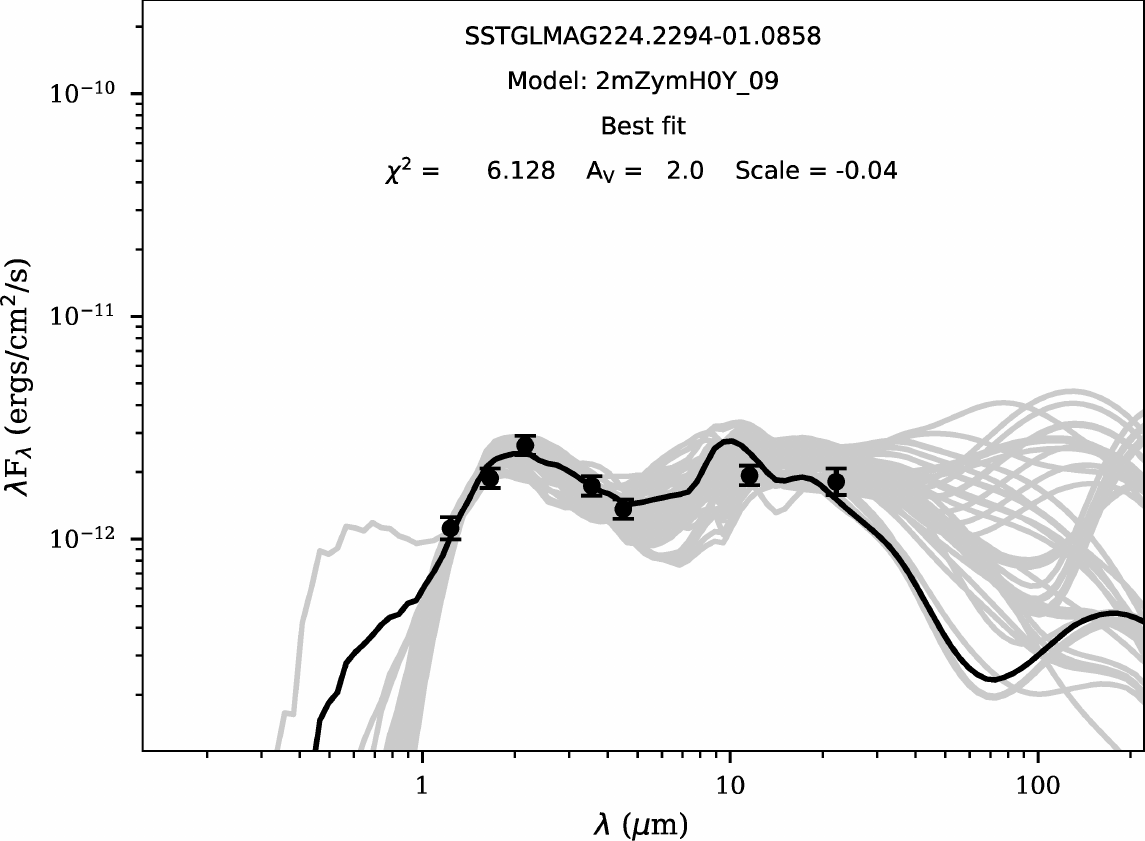}
\hfill
\includegraphics[width=0.32\textwidth]{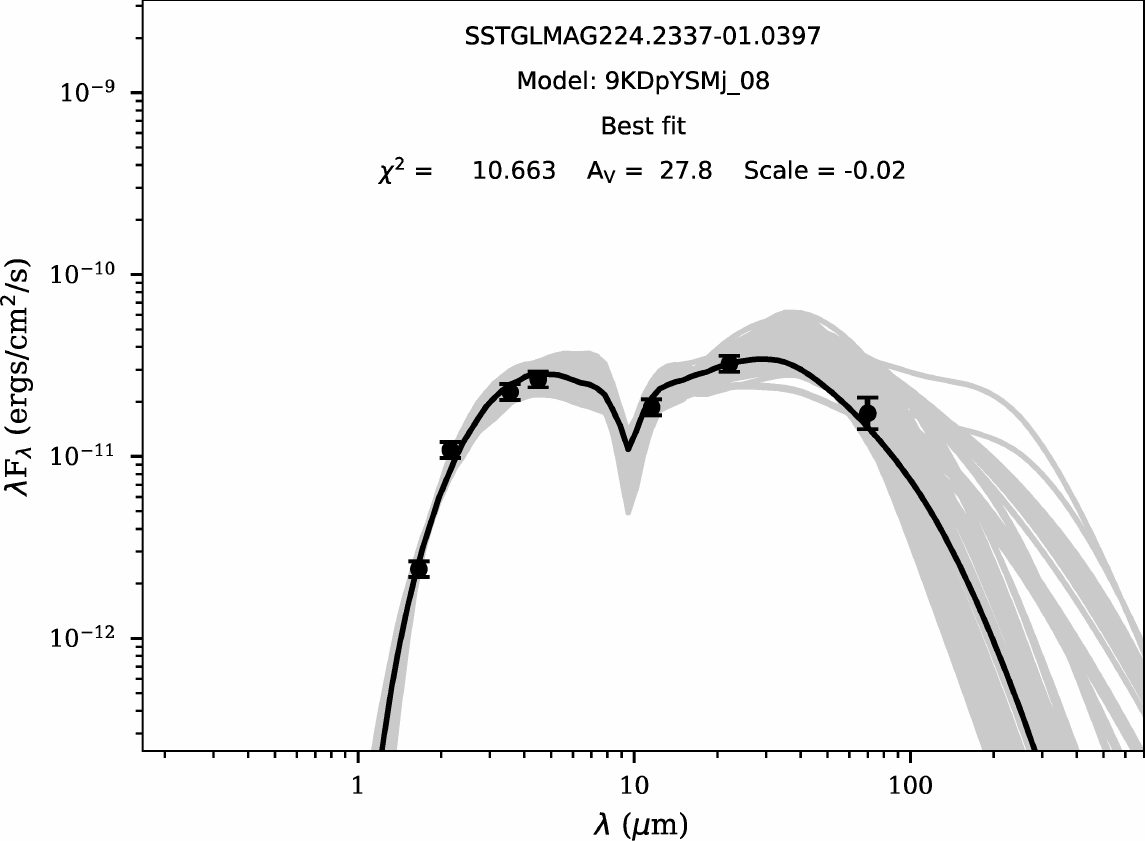} \par
\vspace{2mm}
\includegraphics[width=0.32\textwidth]{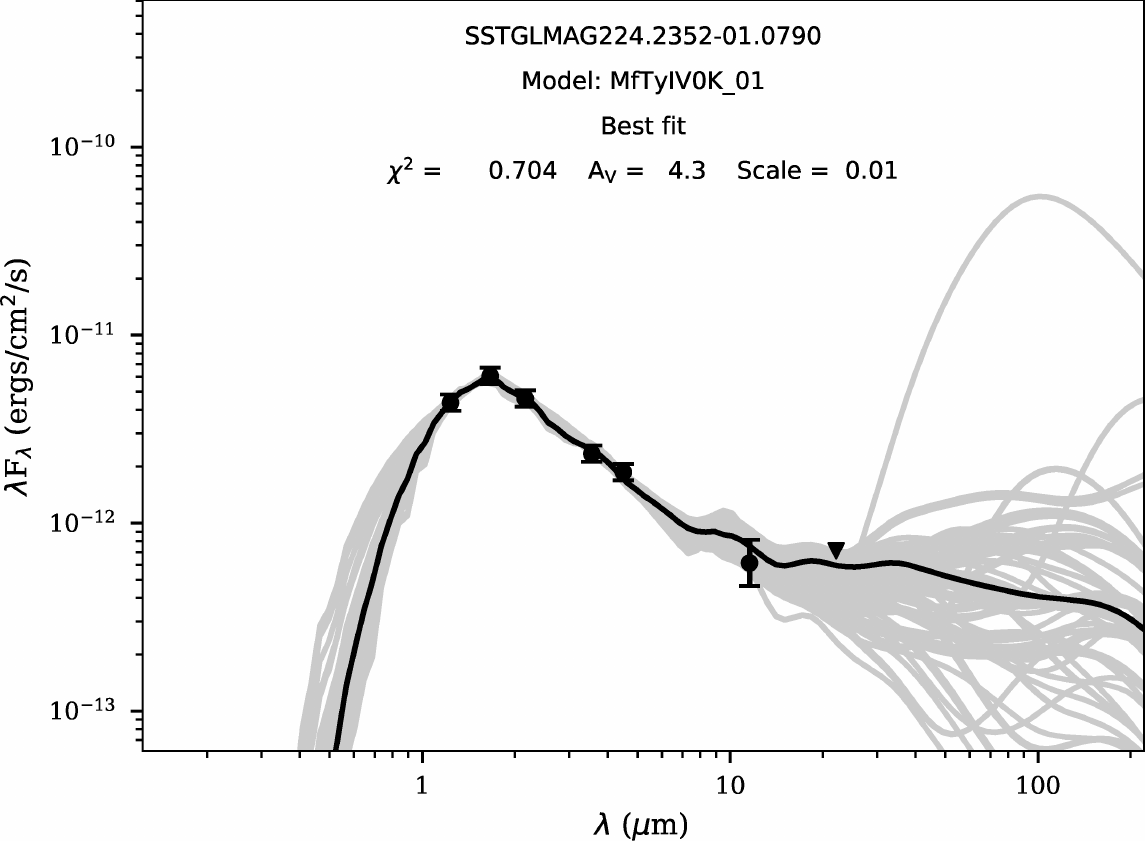}
\hfill
\includegraphics[width=0.32\textwidth]{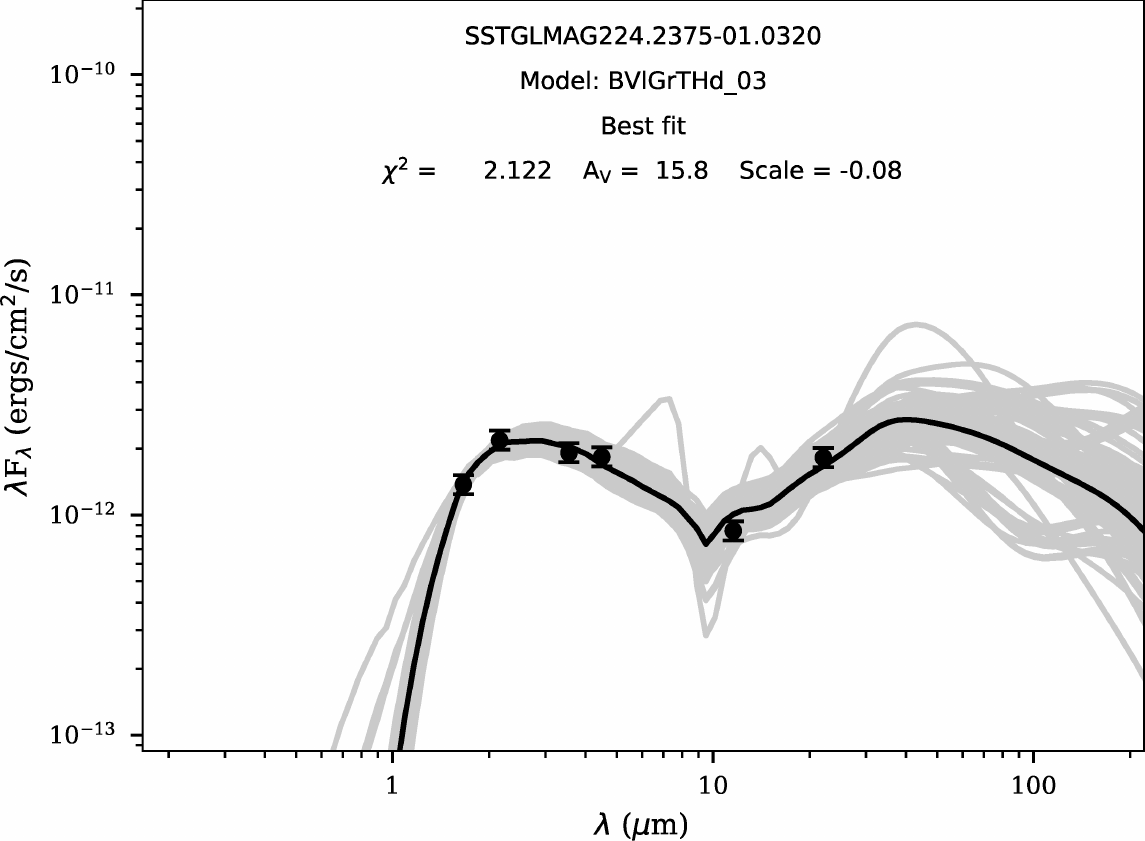}
\hfill
\includegraphics[width=0.32\textwidth]{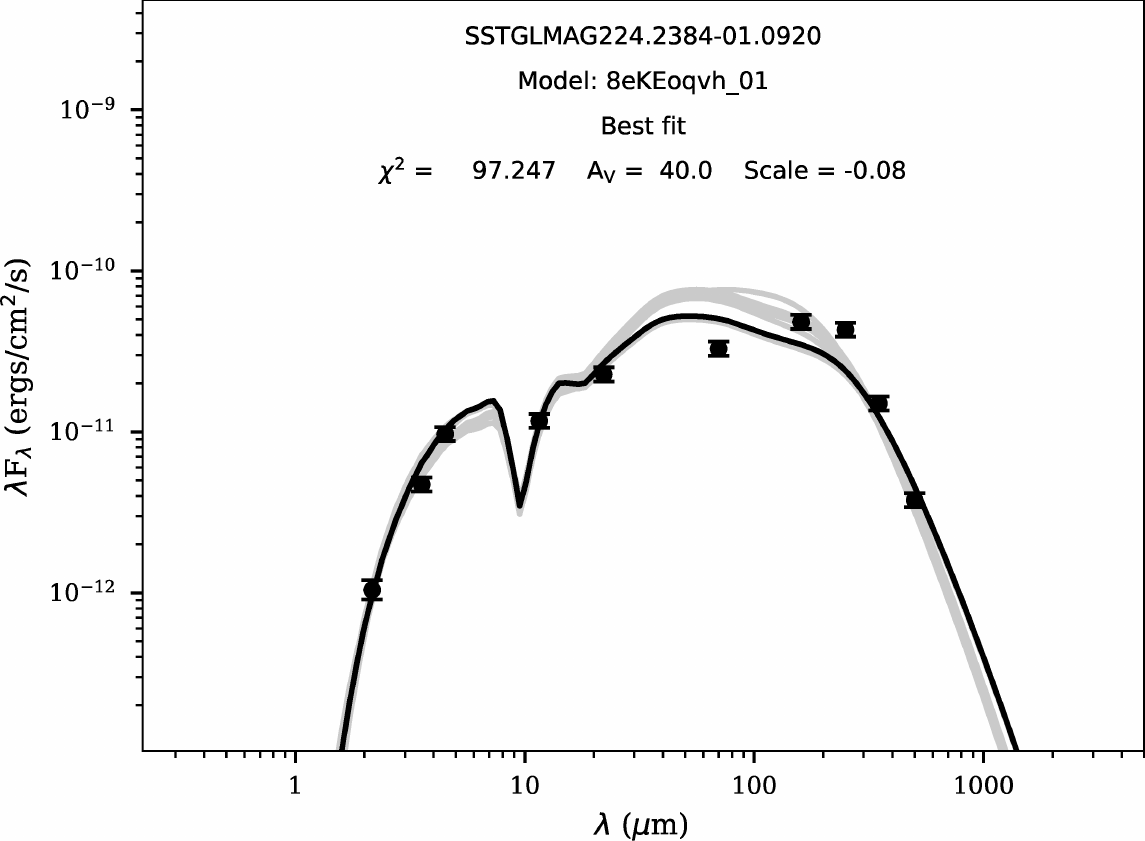} \par
\vspace{2mm}
\includegraphics[width=0.32\textwidth]{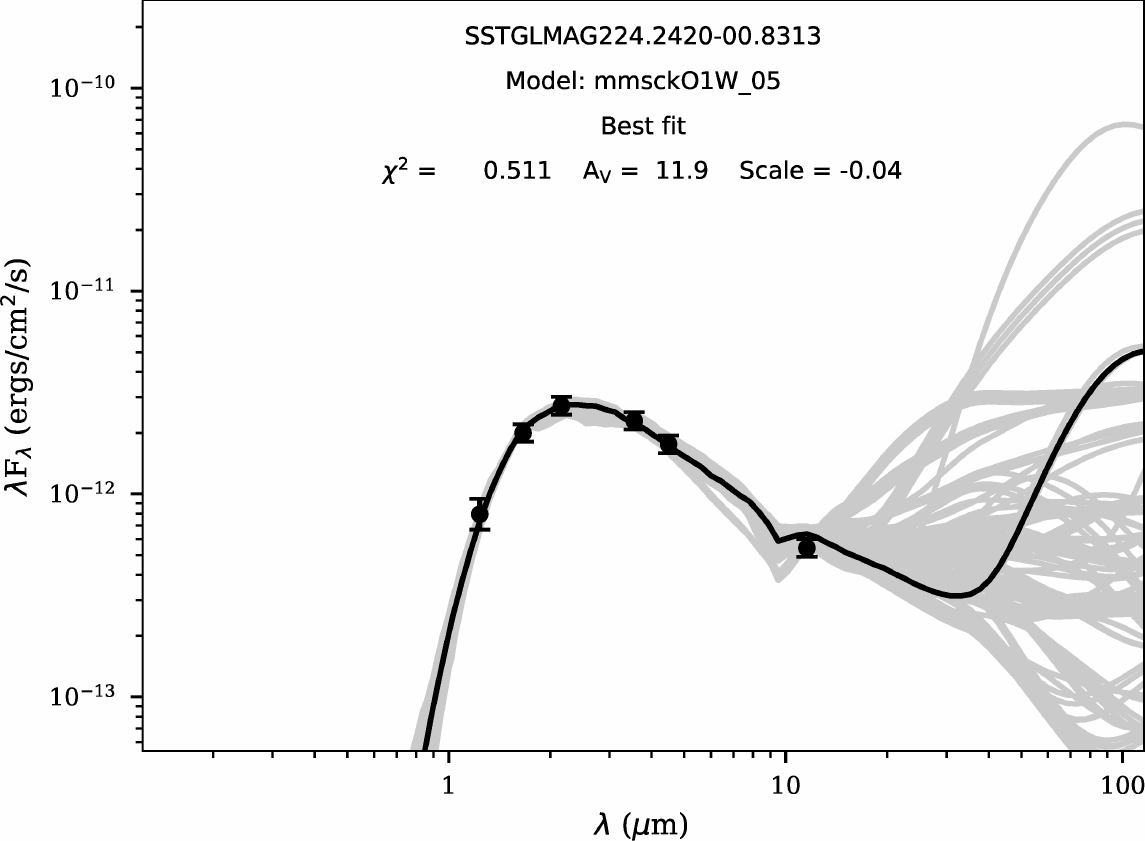}
\hfill
\includegraphics[width=0.32\textwidth]{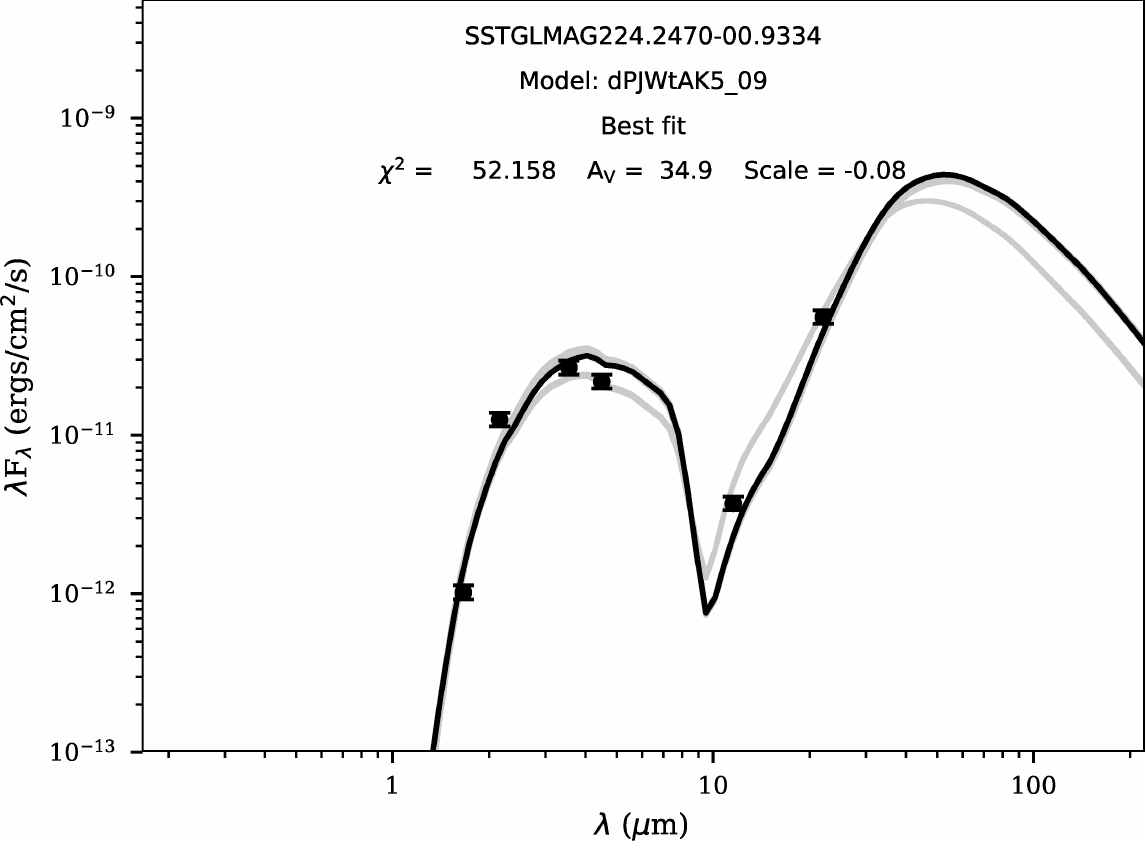}
\hfill
\includegraphics[width=0.32\textwidth]{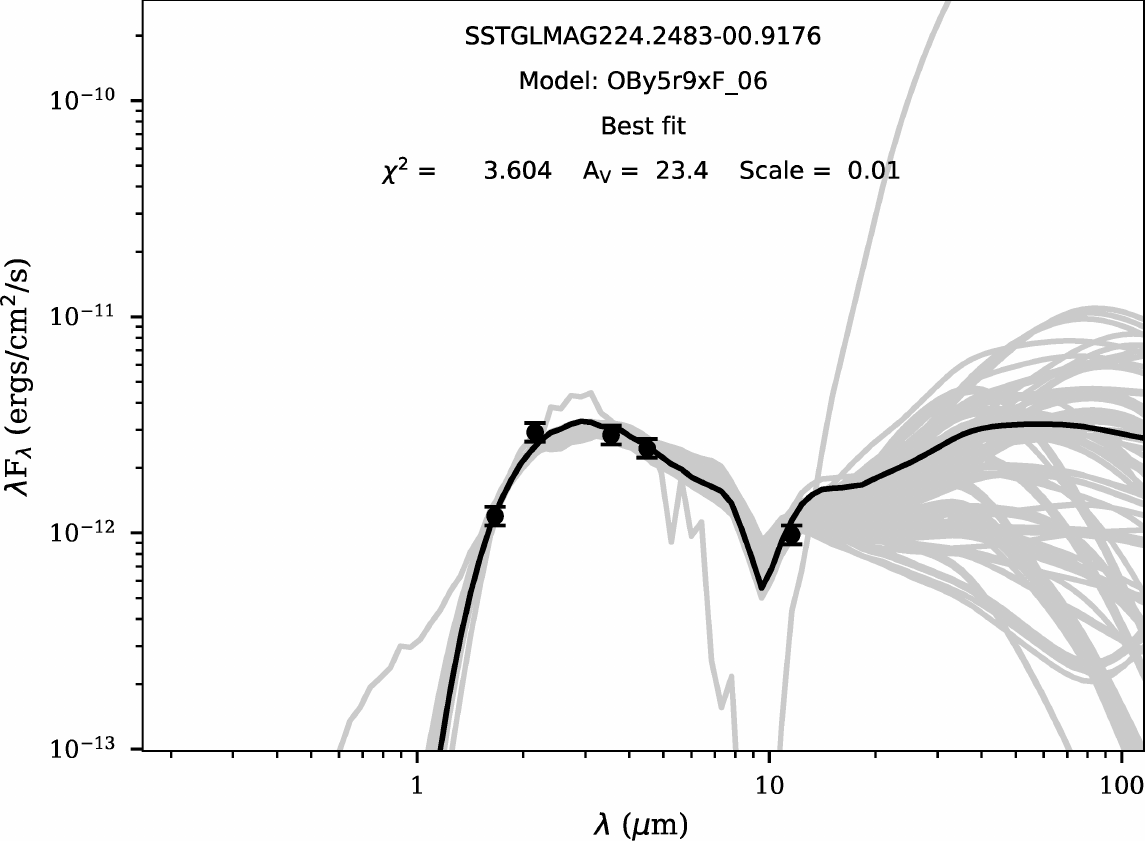} \par
\vspace{2mm}
\includegraphics[width=0.32\textwidth]{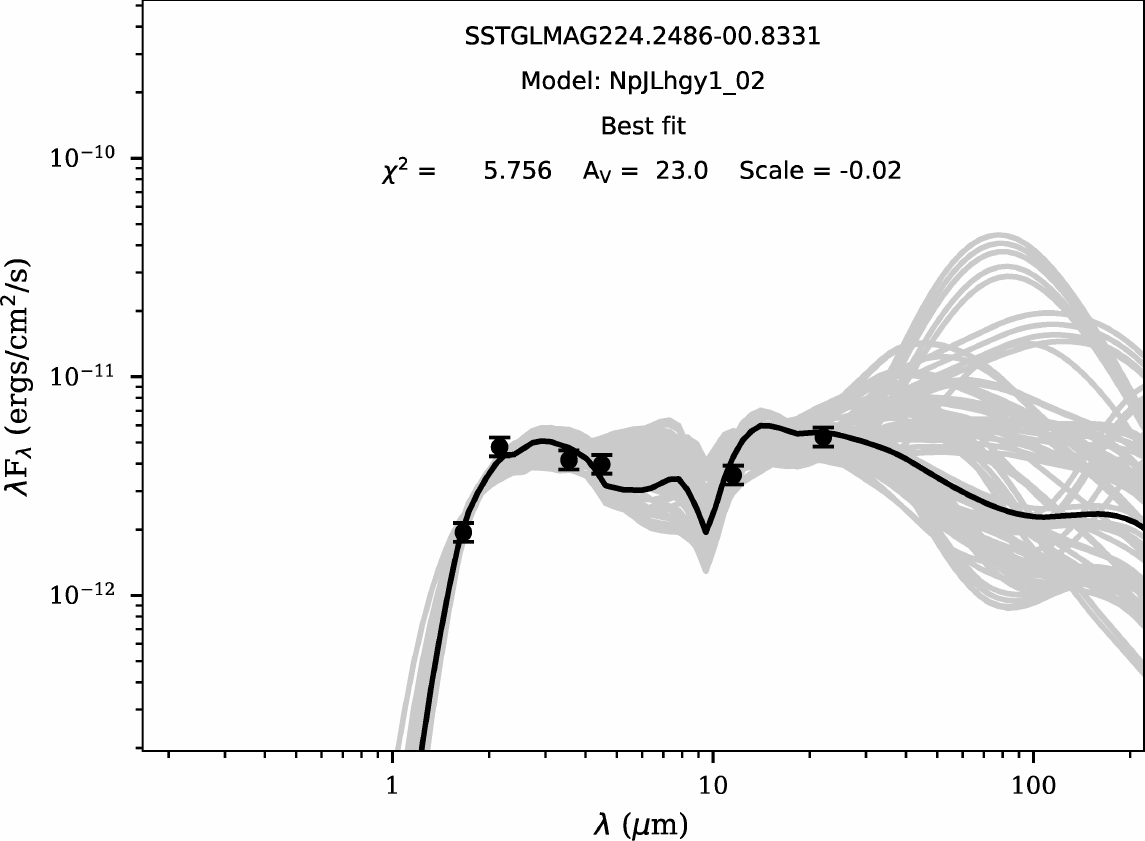}
\hfill
\includegraphics[width=0.32\textwidth]{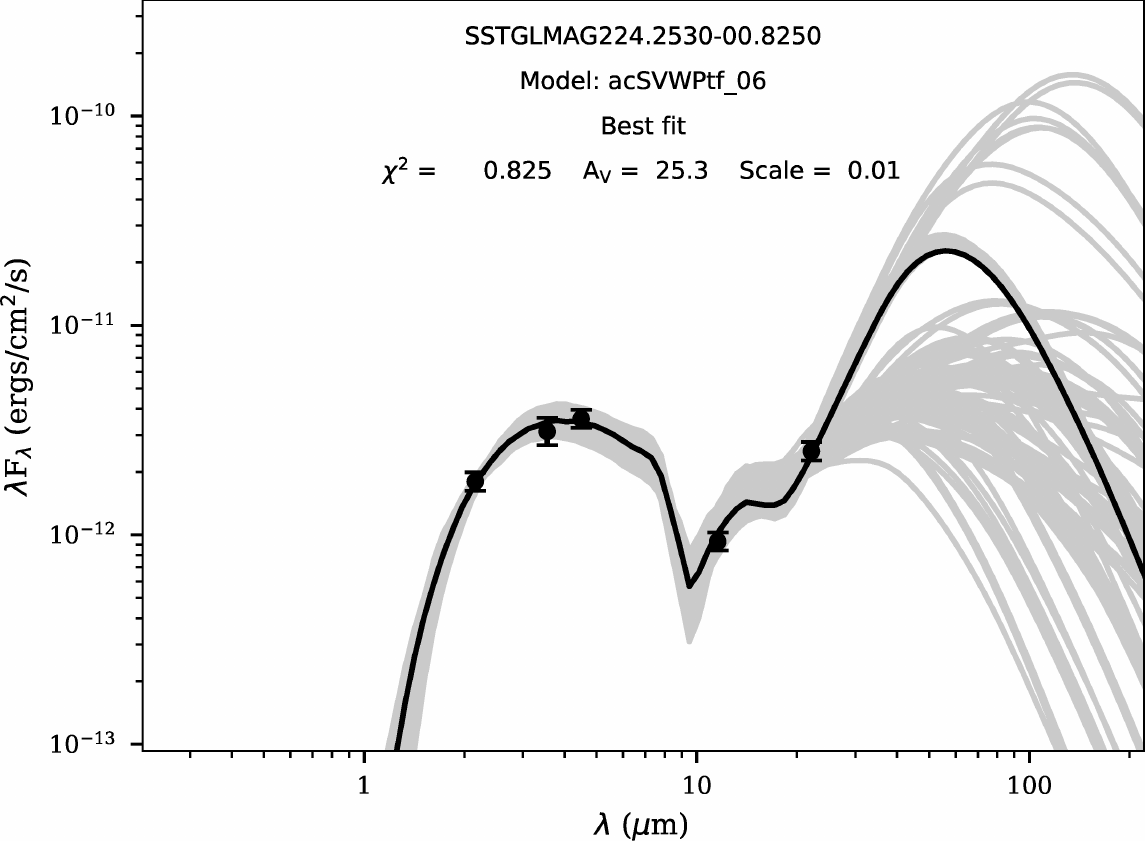}
\hfill
\includegraphics[width=0.32\textwidth]{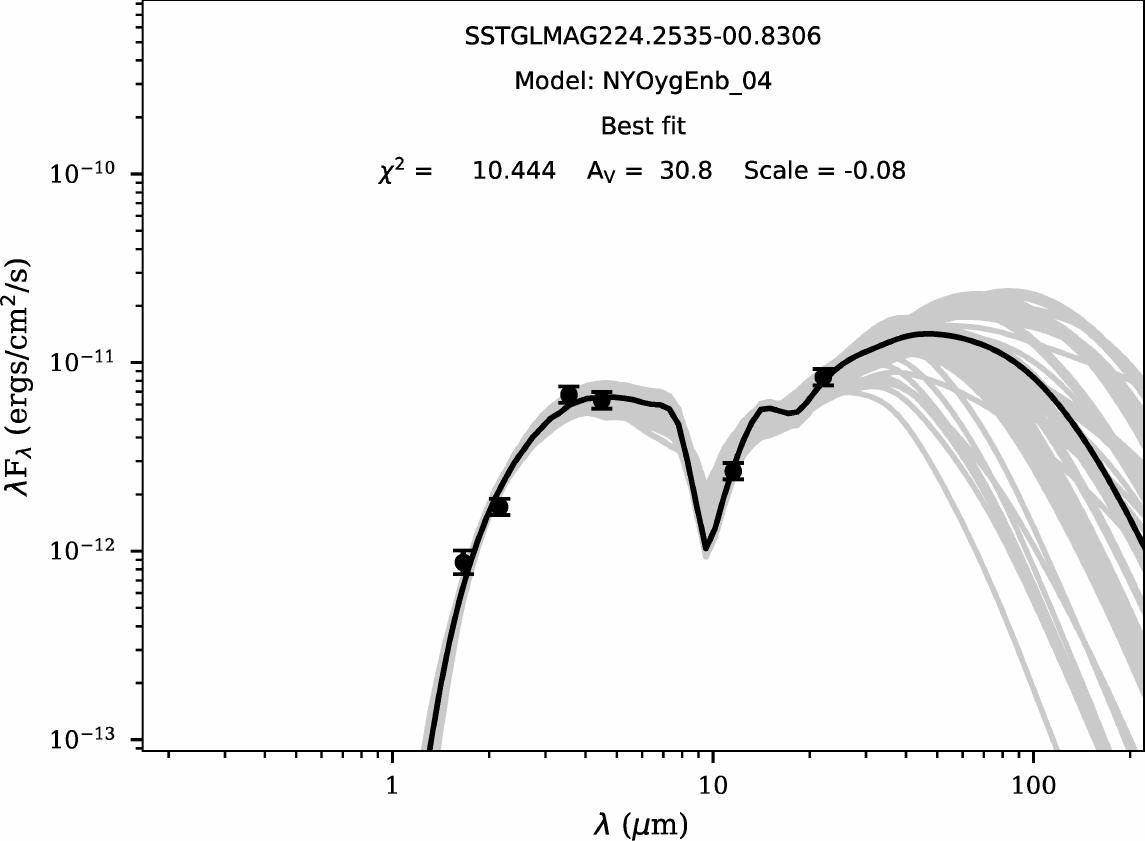}
\caption{Same as Fig.~\ref{f:SEDs1}  \label{f:SEDs5}}
\end{figure*}

\begin{figure*}
\includegraphics[width=0.32\textwidth]{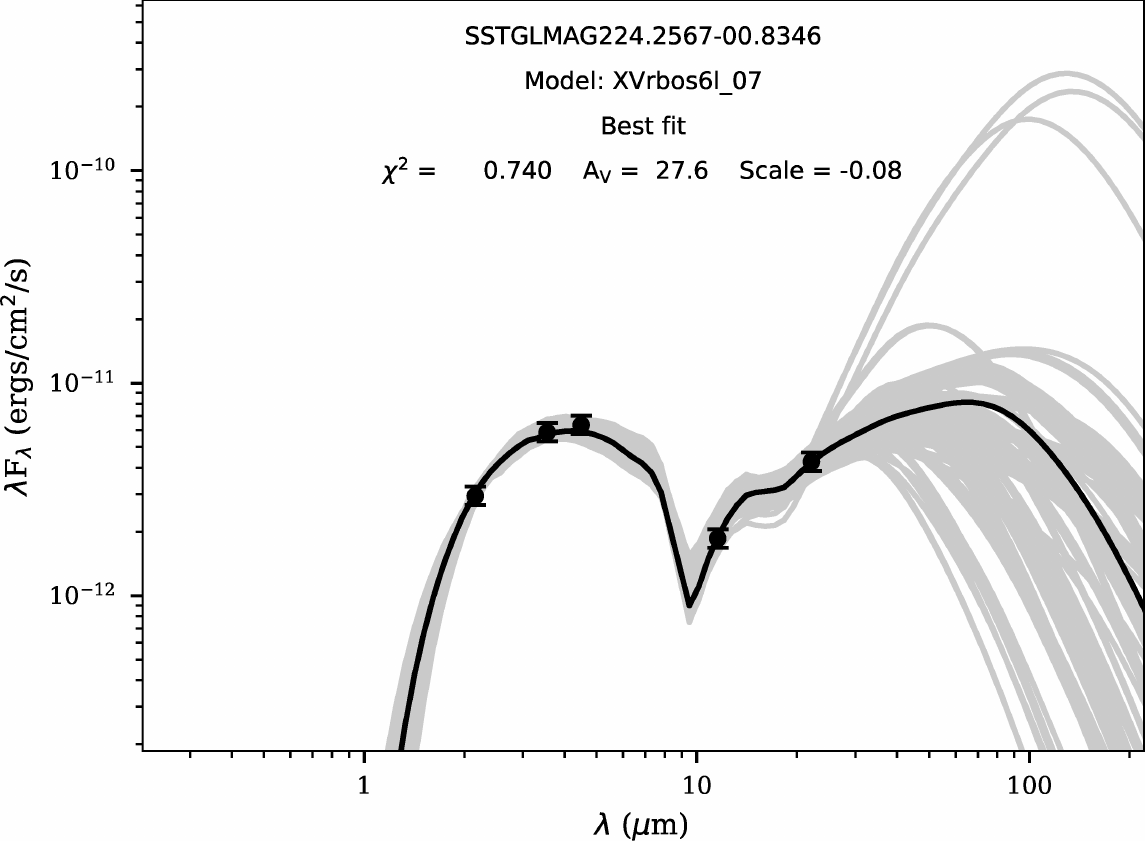}
\hfill
\includegraphics[width=0.32\textwidth]{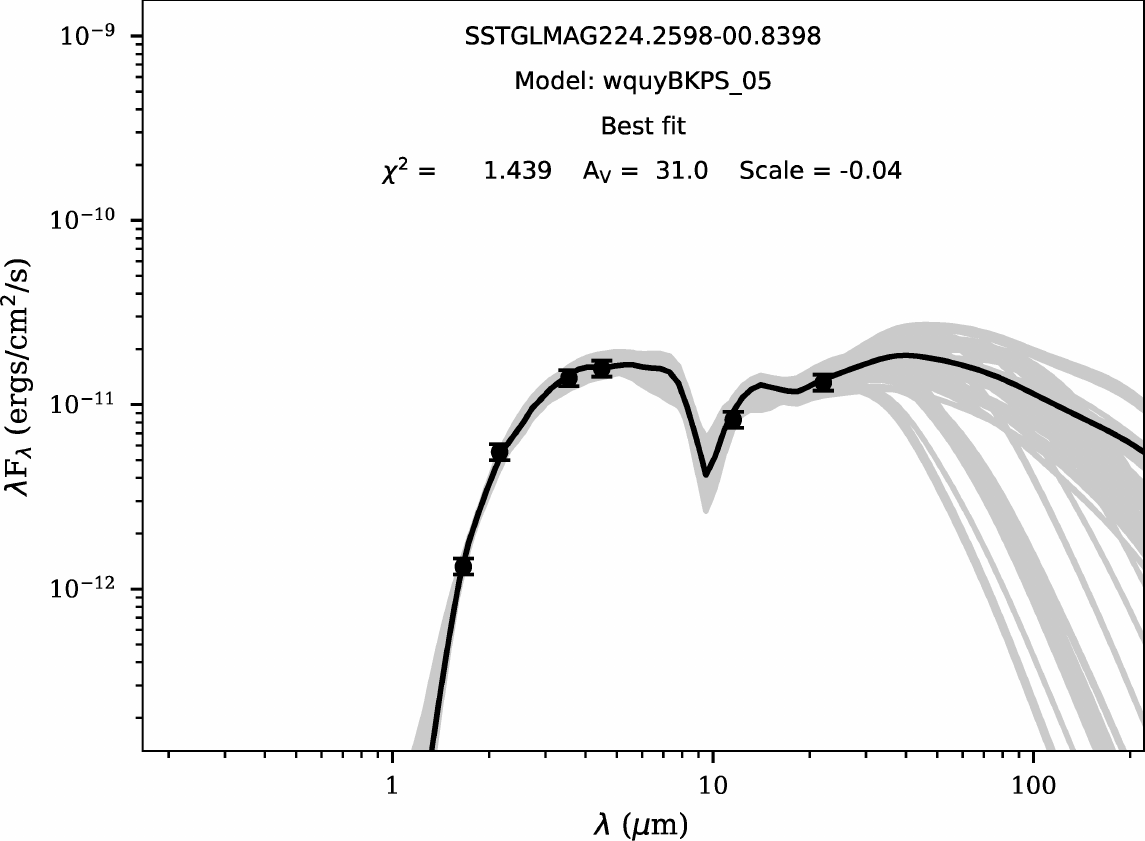}
\hfill
\includegraphics[width=0.32\textwidth]{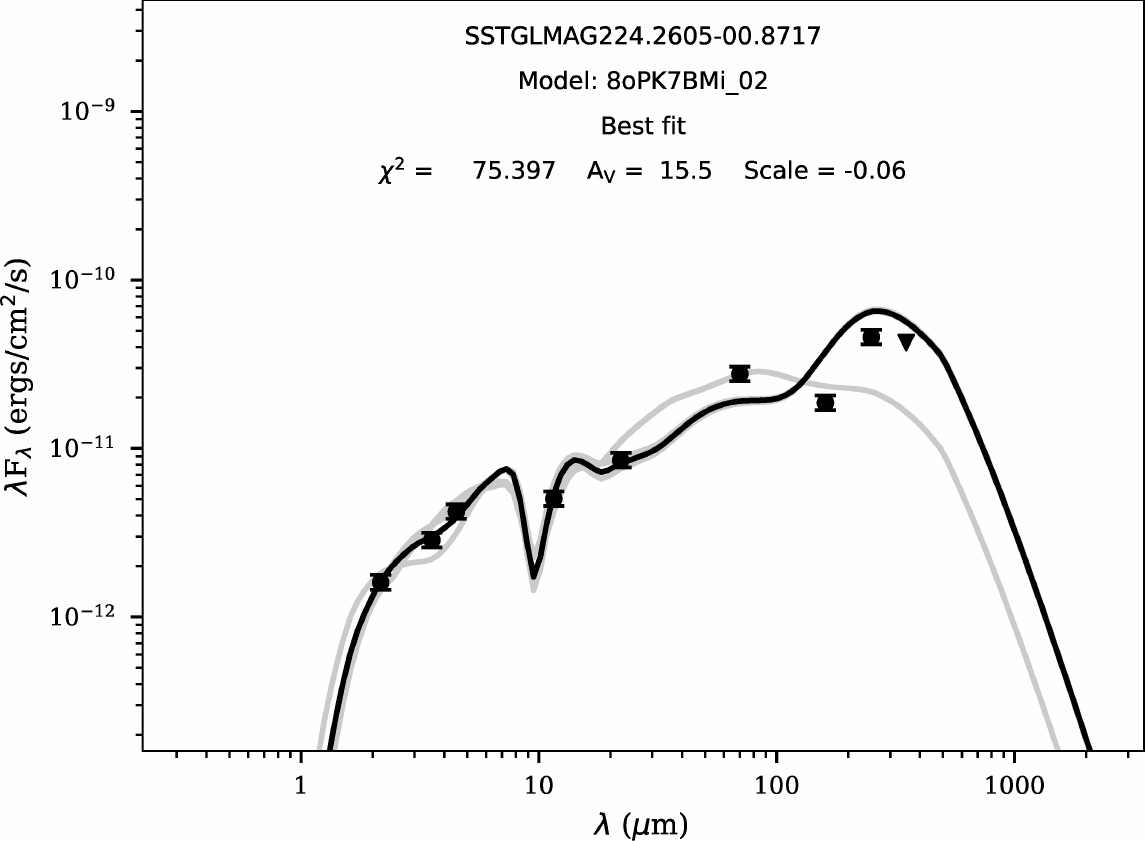} \par
\vspace{2mm}
\includegraphics[width=0.32\textwidth]{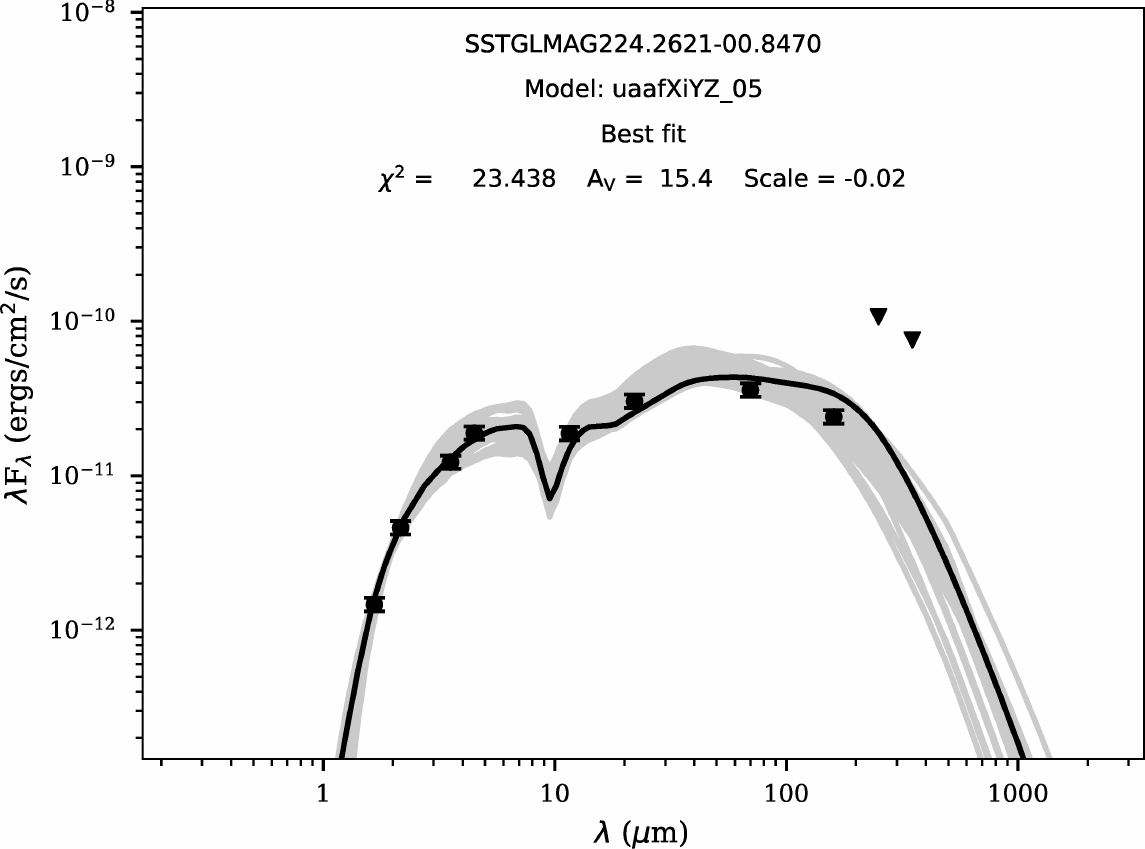}
\hfill
\includegraphics[width=0.32\textwidth]{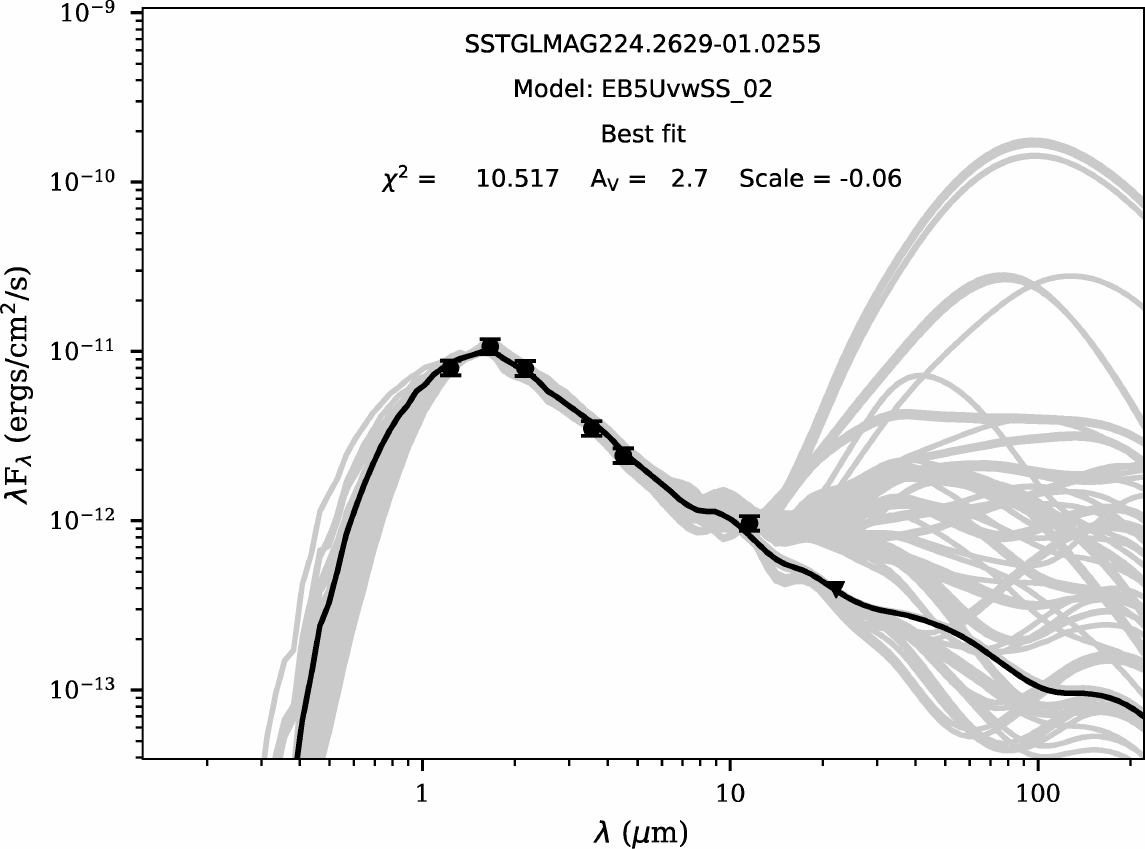}
\hfill
\includegraphics[width=0.32\textwidth]{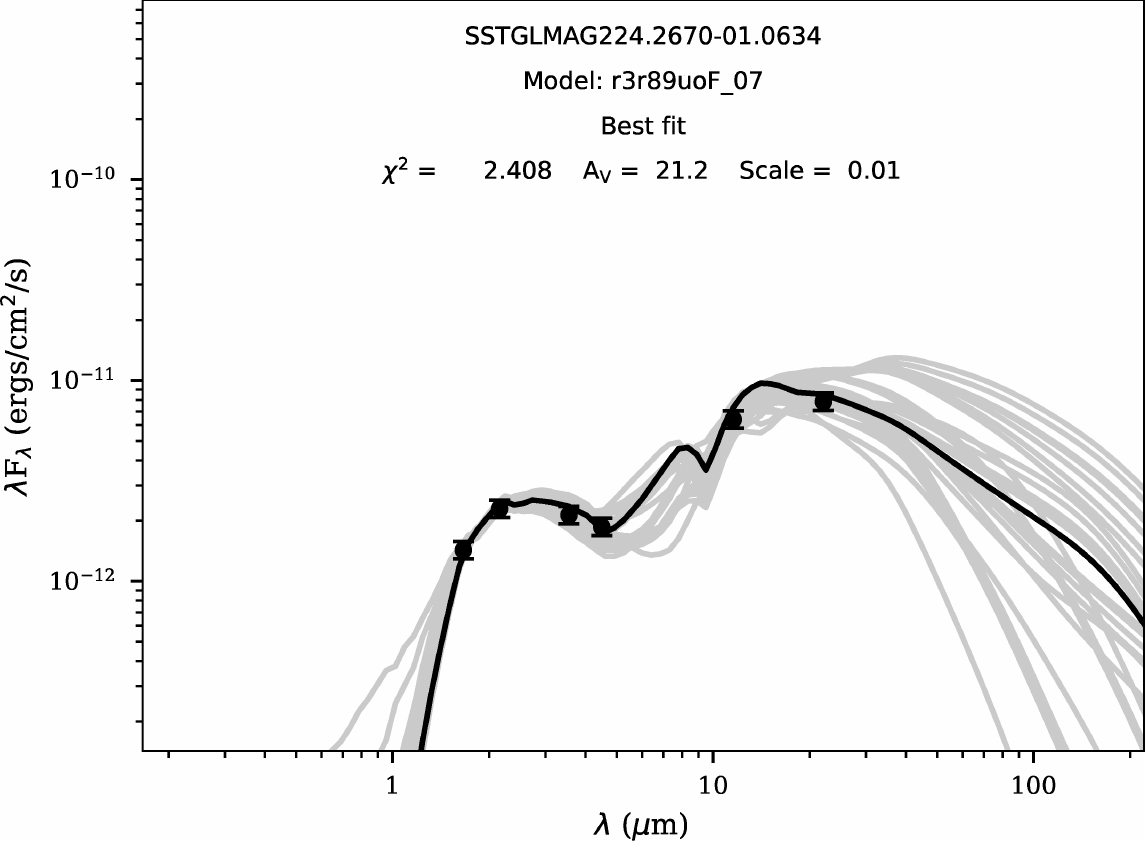} \par
\vspace{2mm}
\includegraphics[width=0.32\textwidth]{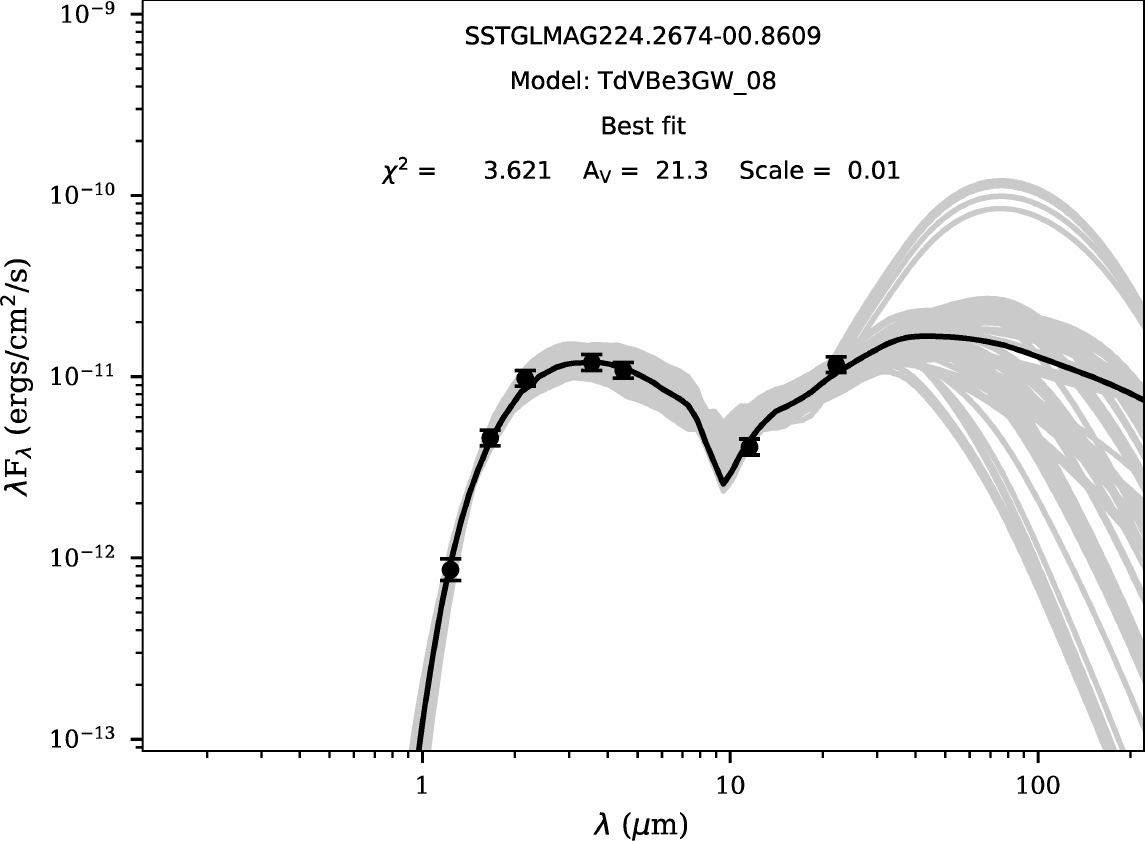}
\hfill
\includegraphics[width=0.32\textwidth]{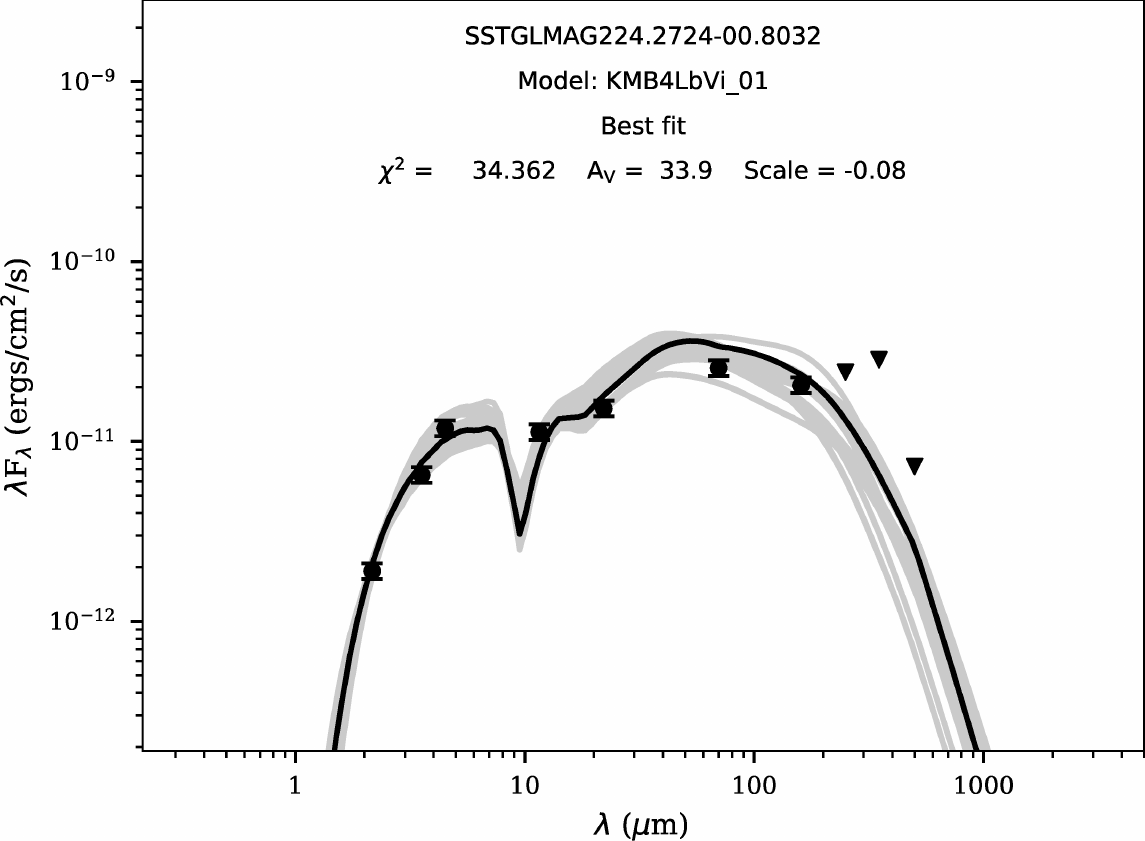}
\hfill
\includegraphics[width=0.32\textwidth]{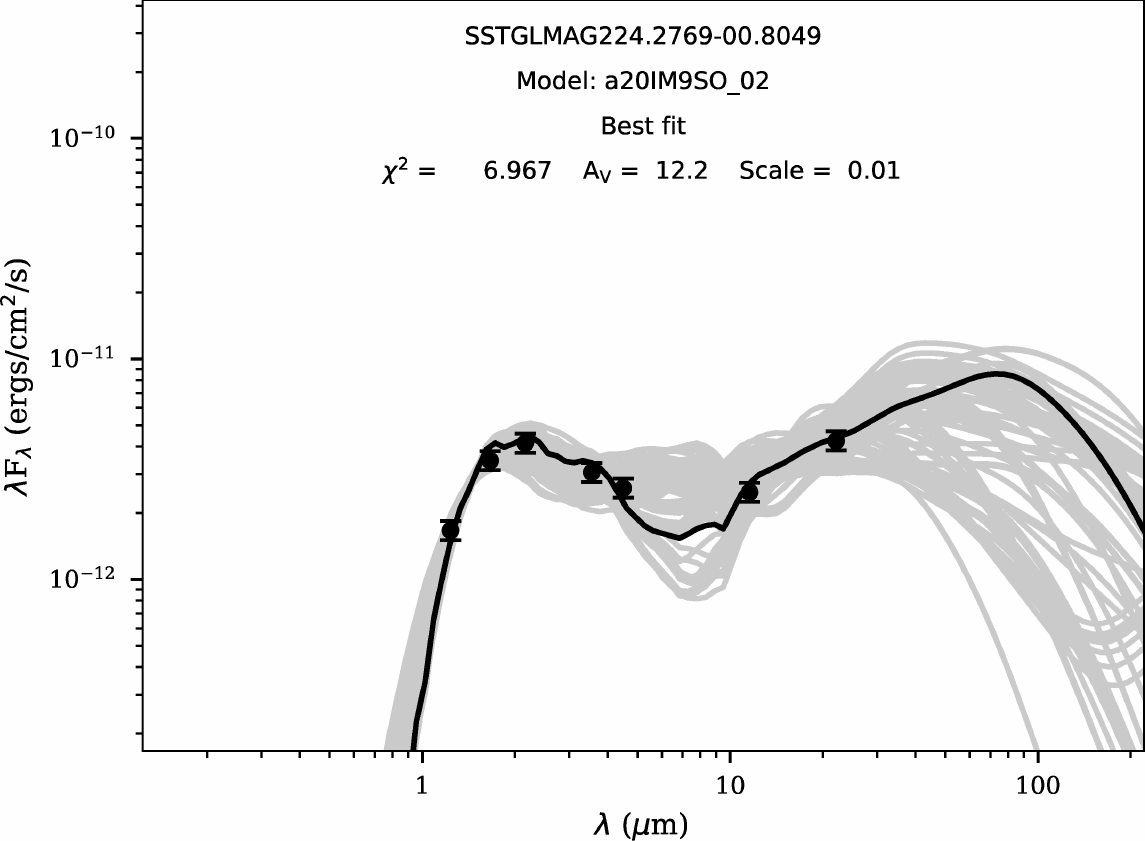} \par
\vspace{2mm}
\includegraphics[width=0.32\textwidth]{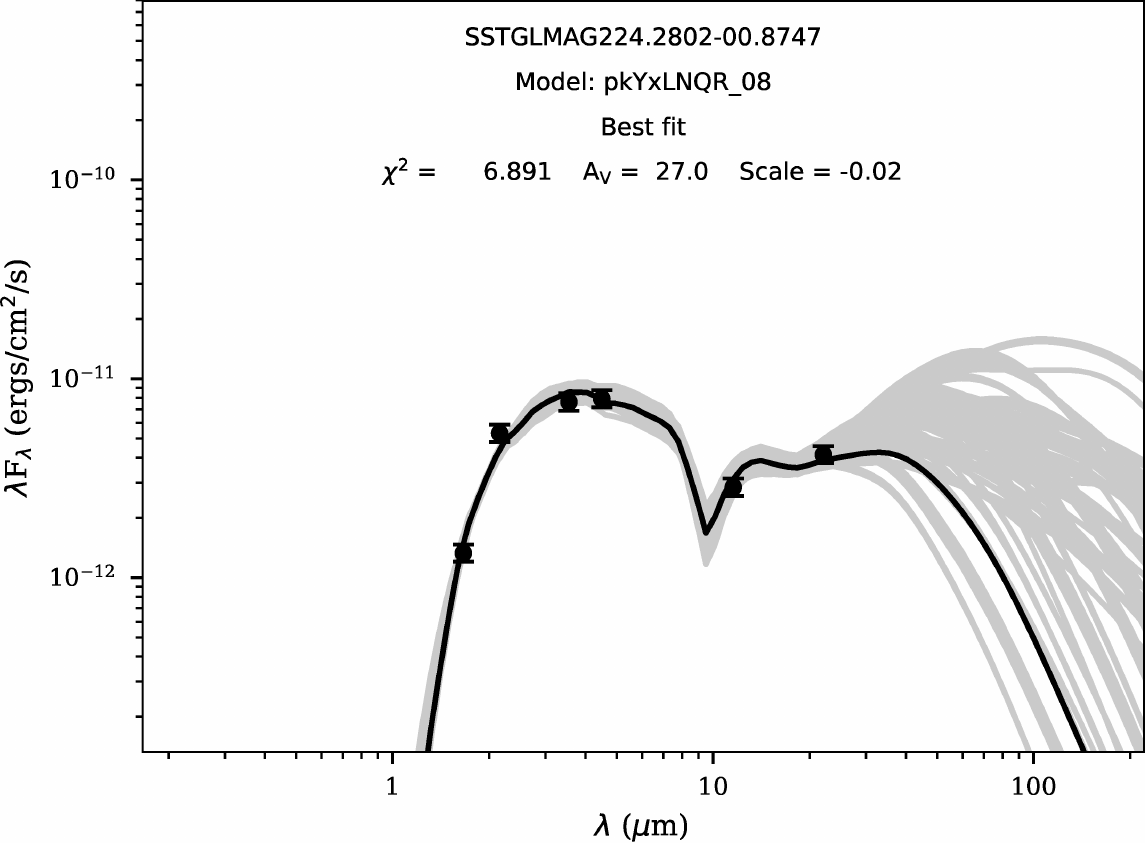}
\hfill
\includegraphics[width=0.32\textwidth]{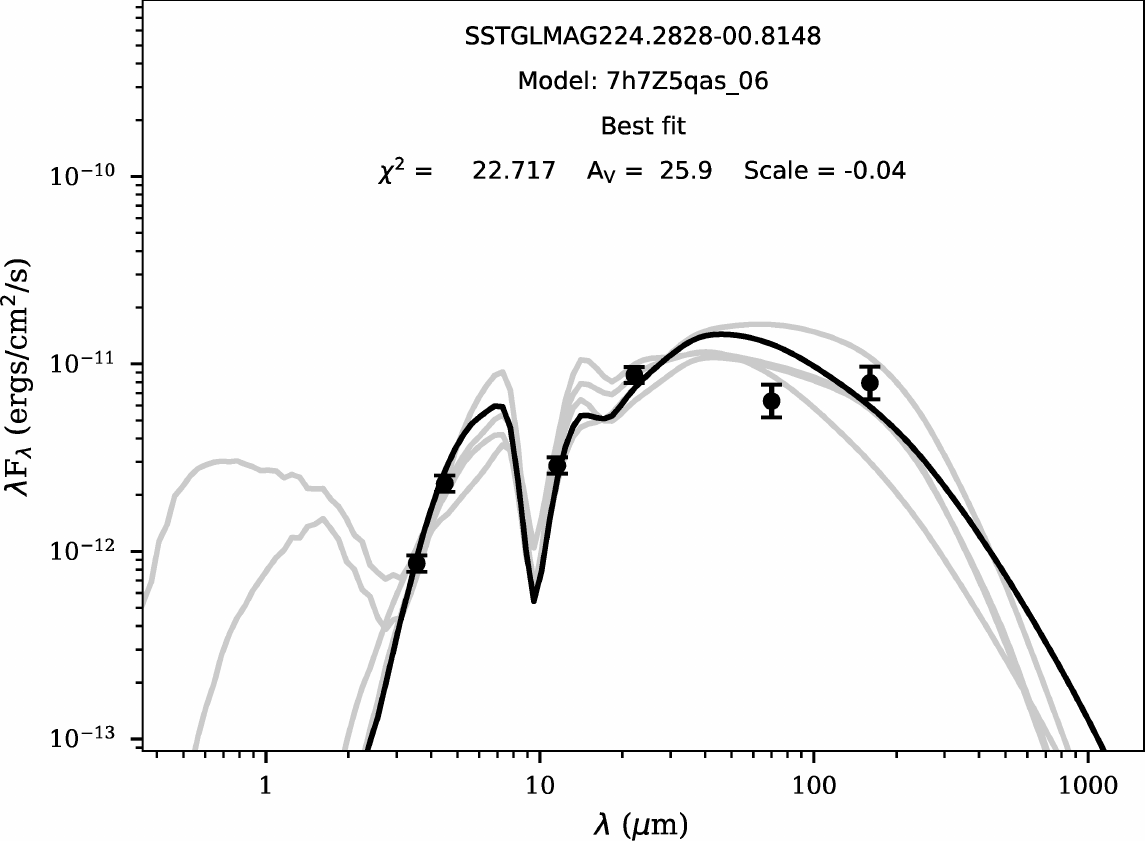}
\hfill
\includegraphics[width=0.32\textwidth]{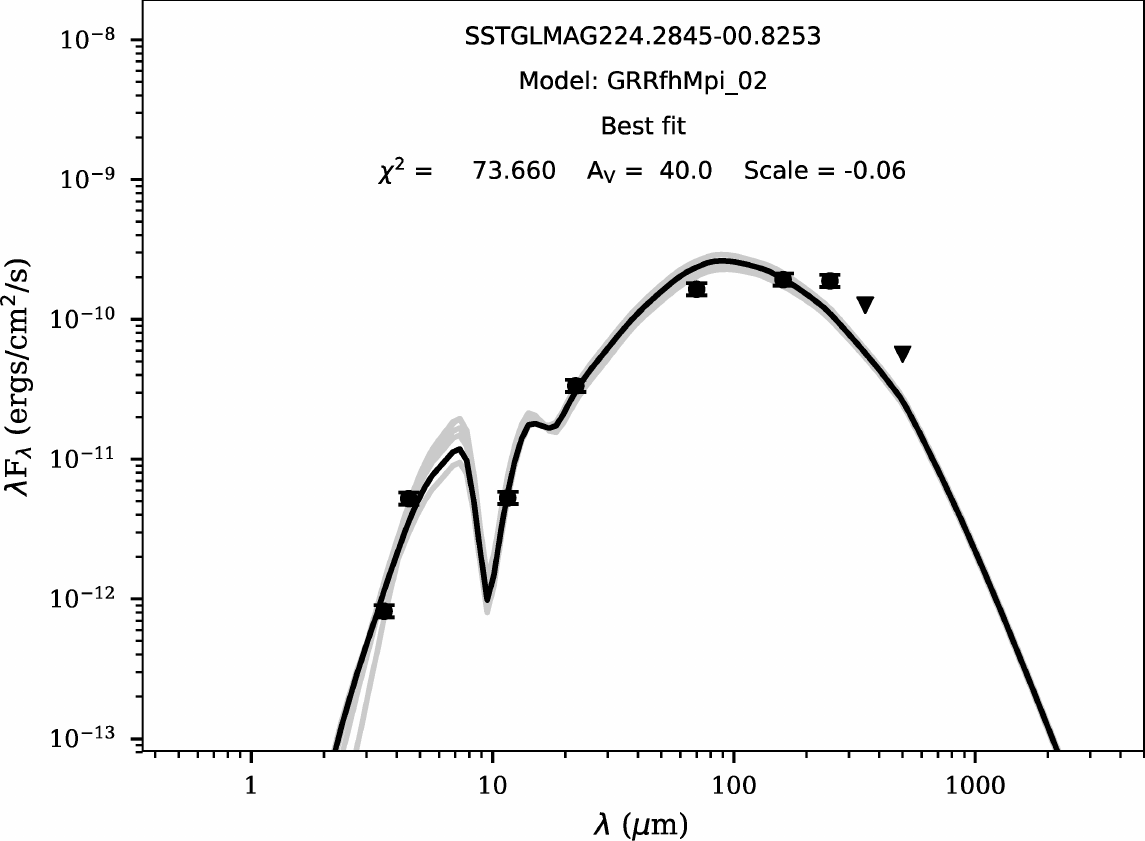} \par
\vspace{2mm}
\includegraphics[width=0.32\textwidth]{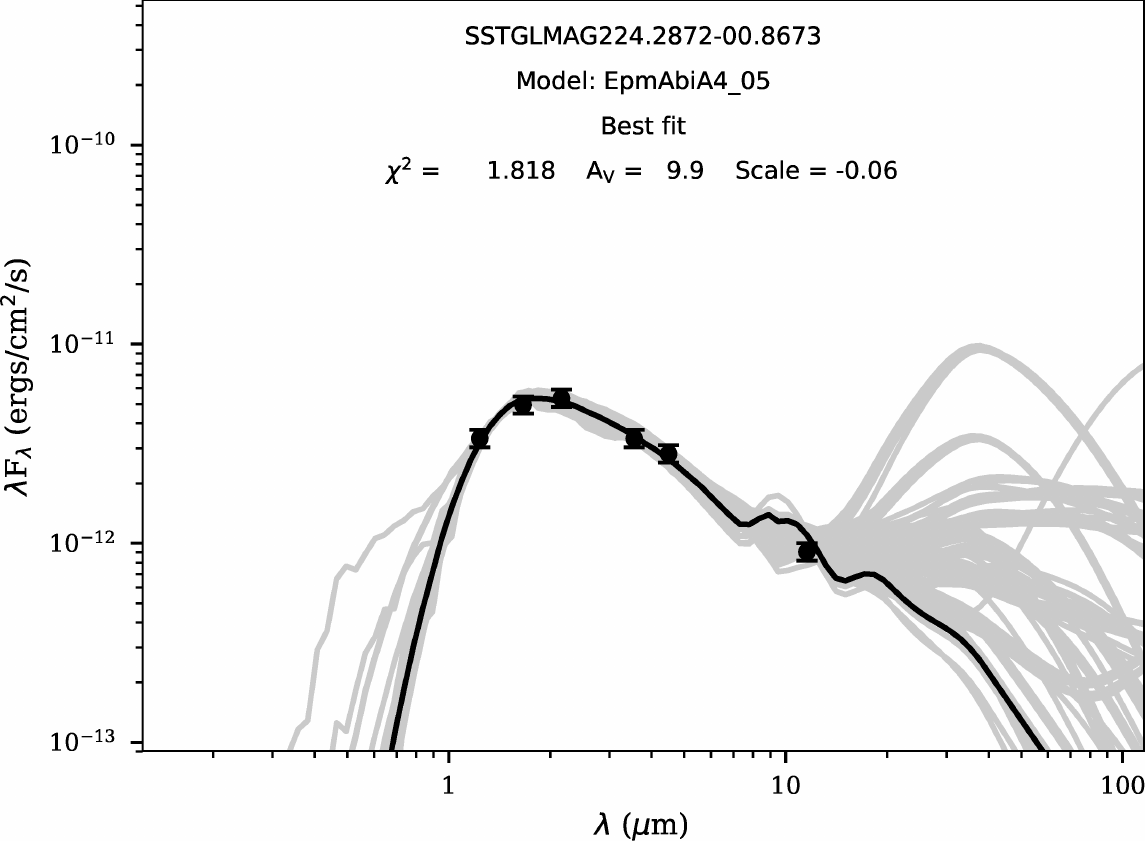}
\hfill
\includegraphics[width=0.32\textwidth]{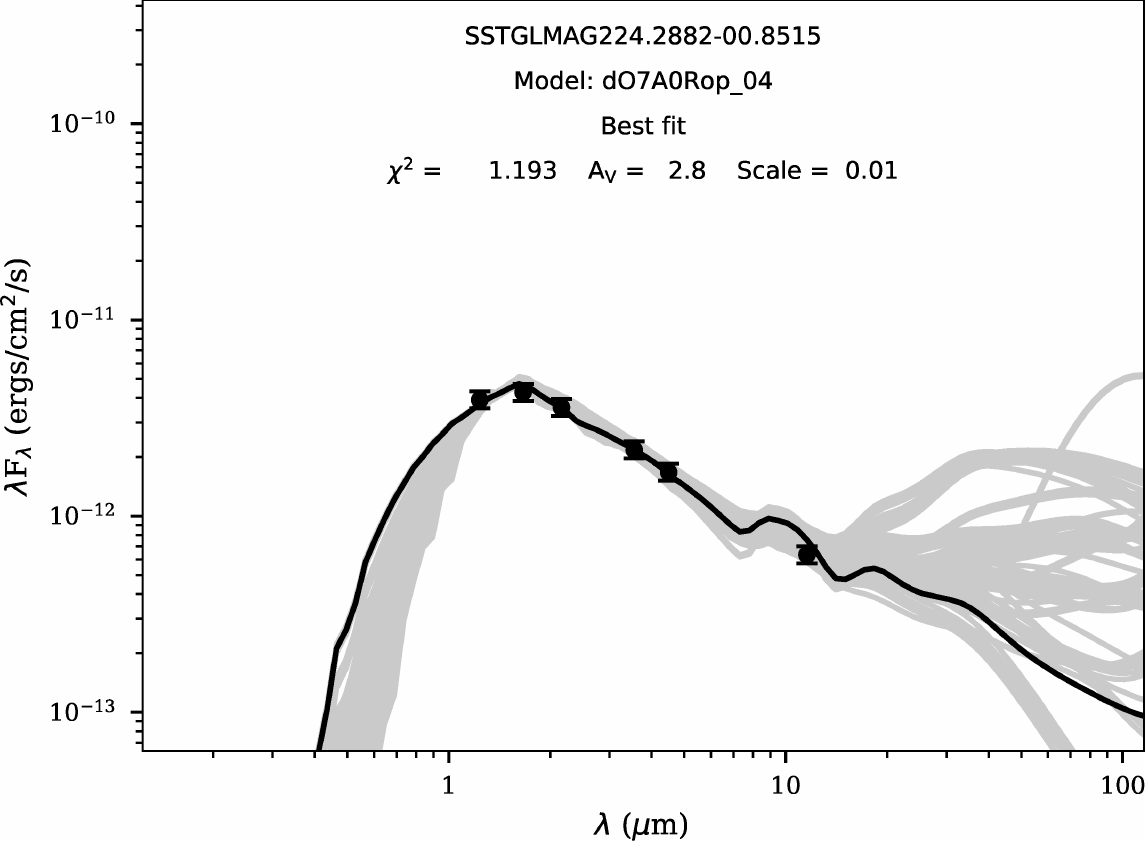}
\hfill
\includegraphics[width=0.32\textwidth]{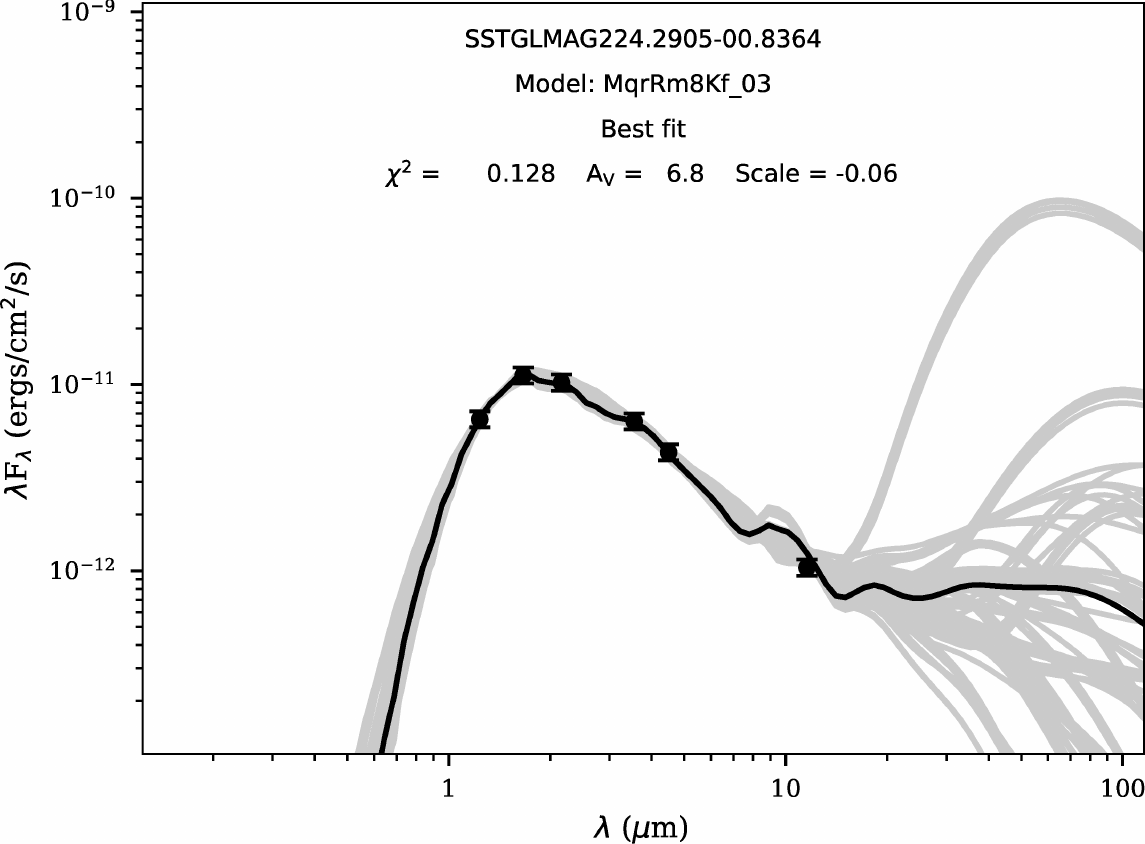}
\caption{Same as Fig.~\ref{f:SEDs1}  \label{f:SEDs6}}
\end{figure*}

\begin{figure*}
\includegraphics[width=0.32\textwidth]{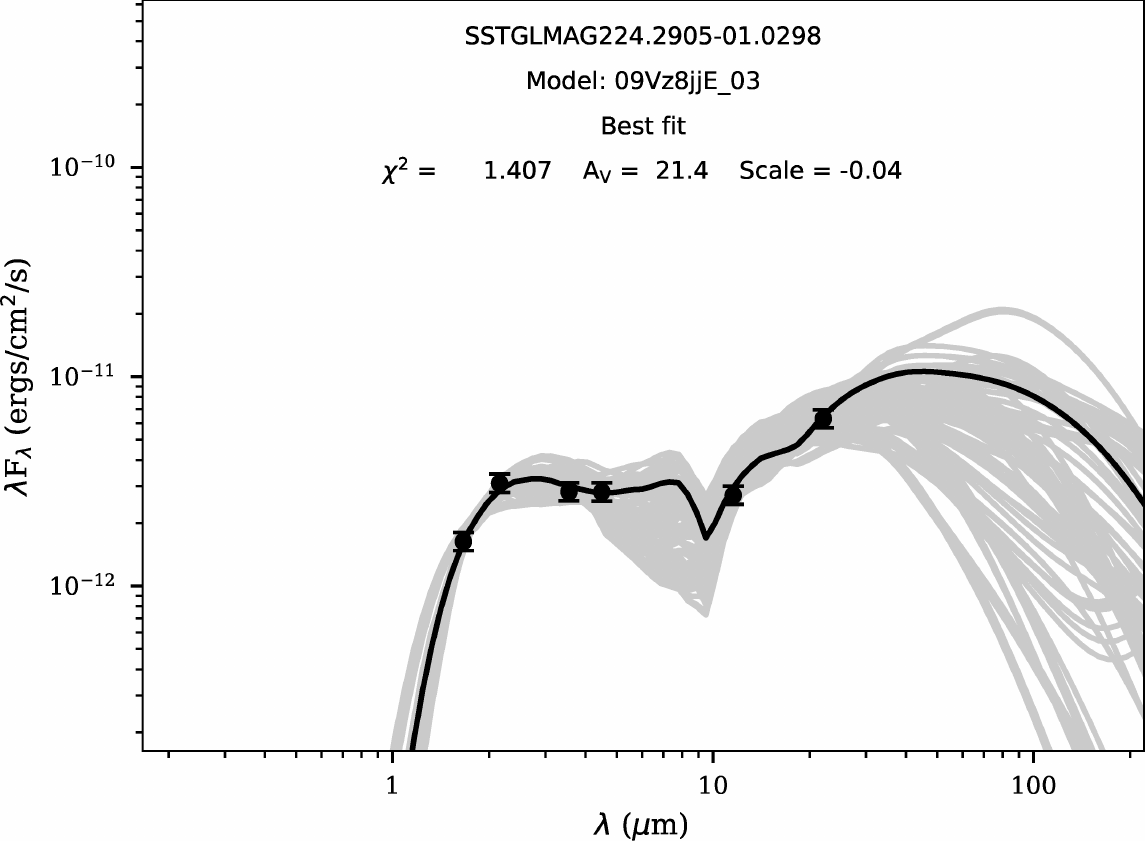}
\hfill
\includegraphics[width=0.32\textwidth]{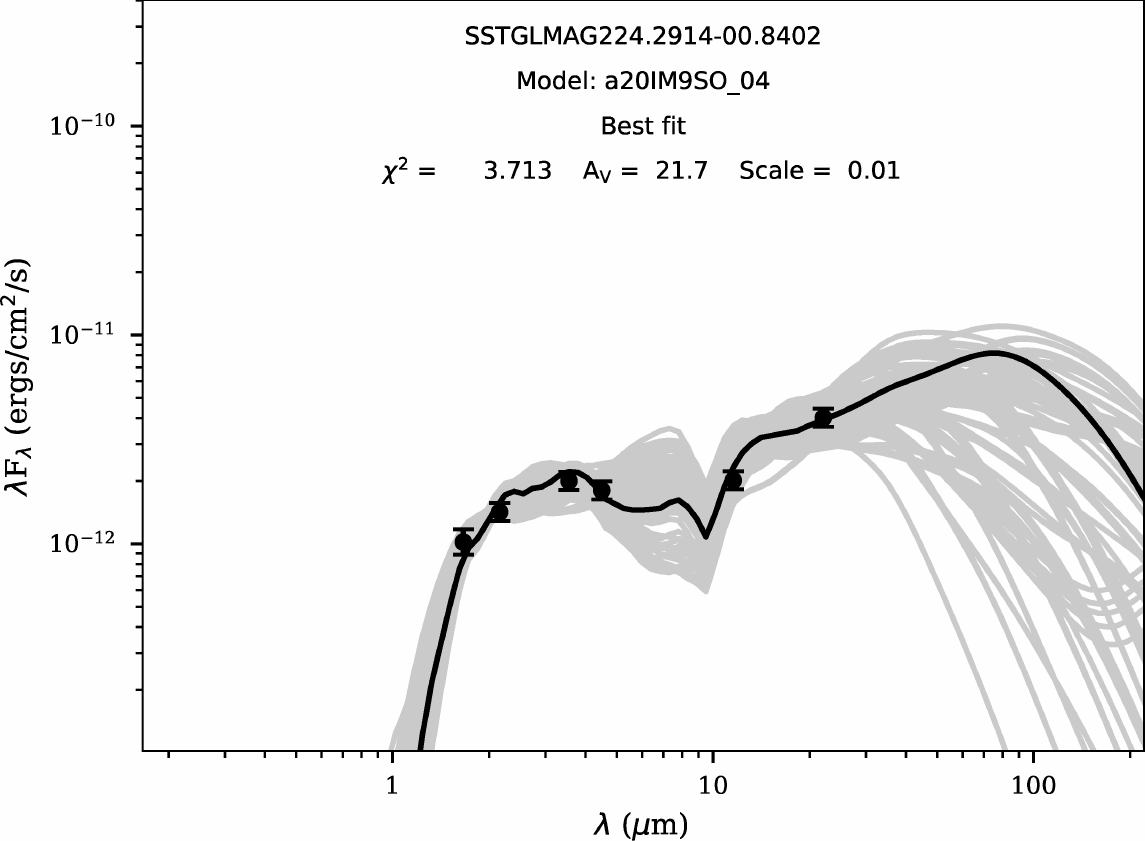}
\hfill
\includegraphics[width=0.32\textwidth]{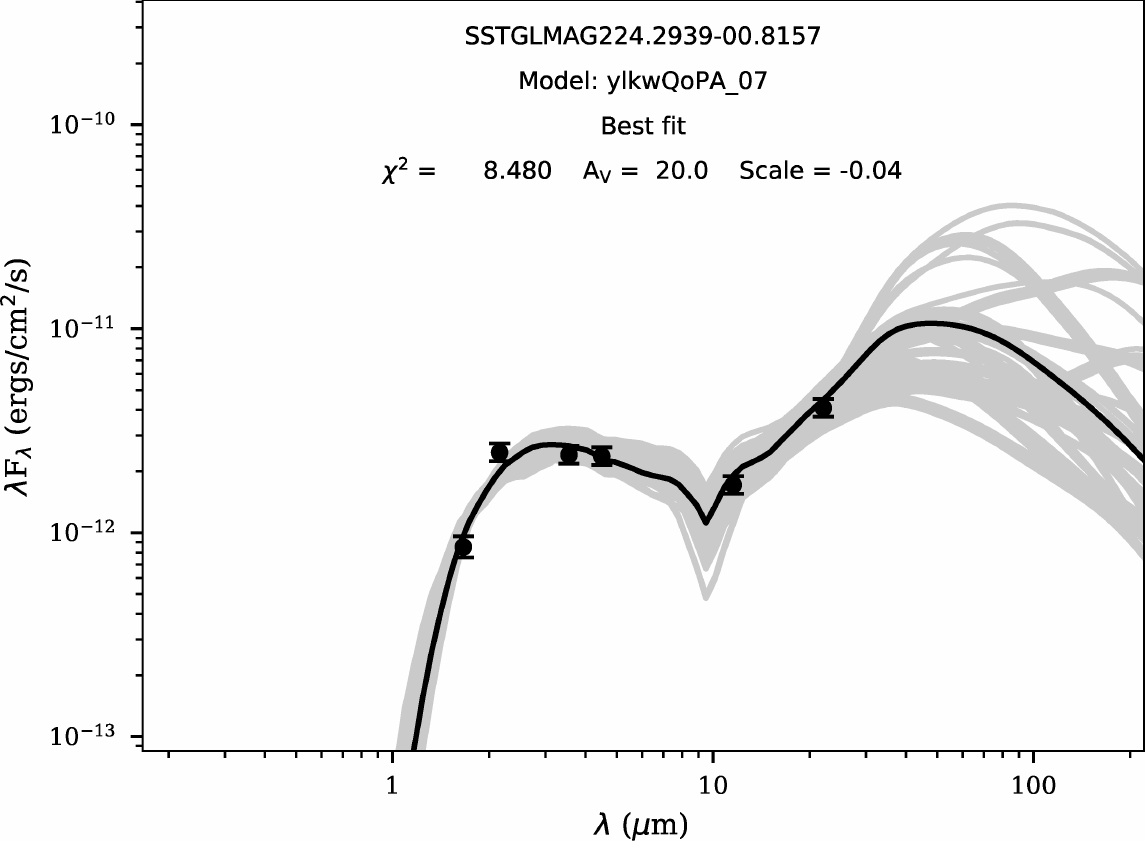} \par
\vspace{2mm}
\includegraphics[width=0.32\textwidth]{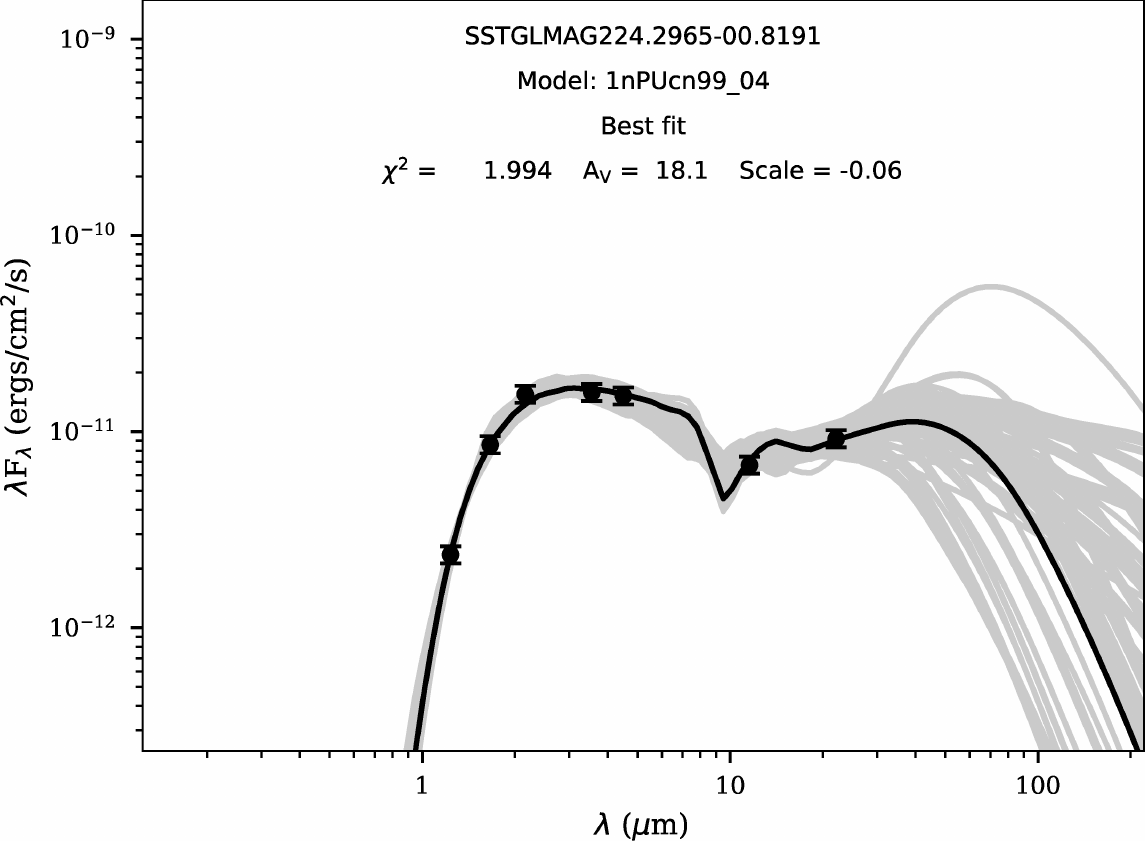}
\hfill
\includegraphics[width=0.32\textwidth]{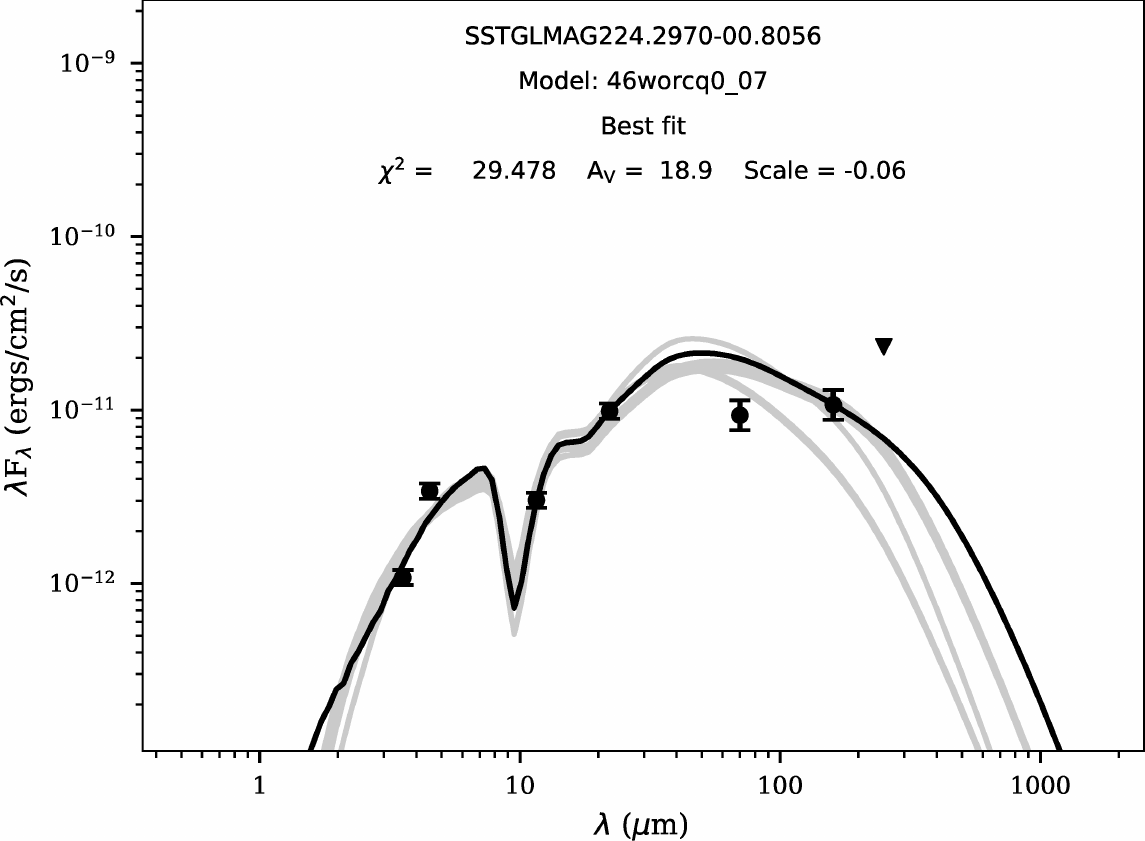}
\hfill
\includegraphics[width=0.32\textwidth]{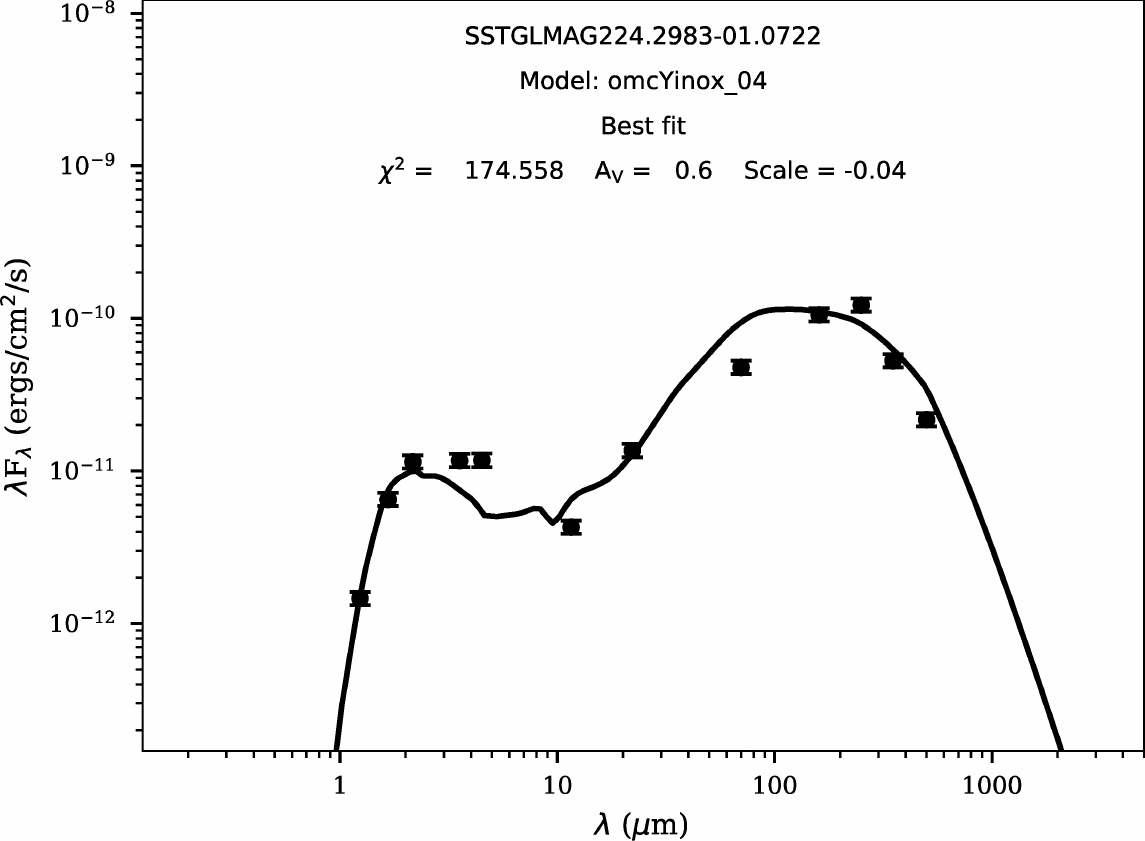} \par
\vspace{2mm}
\includegraphics[width=0.32\textwidth]{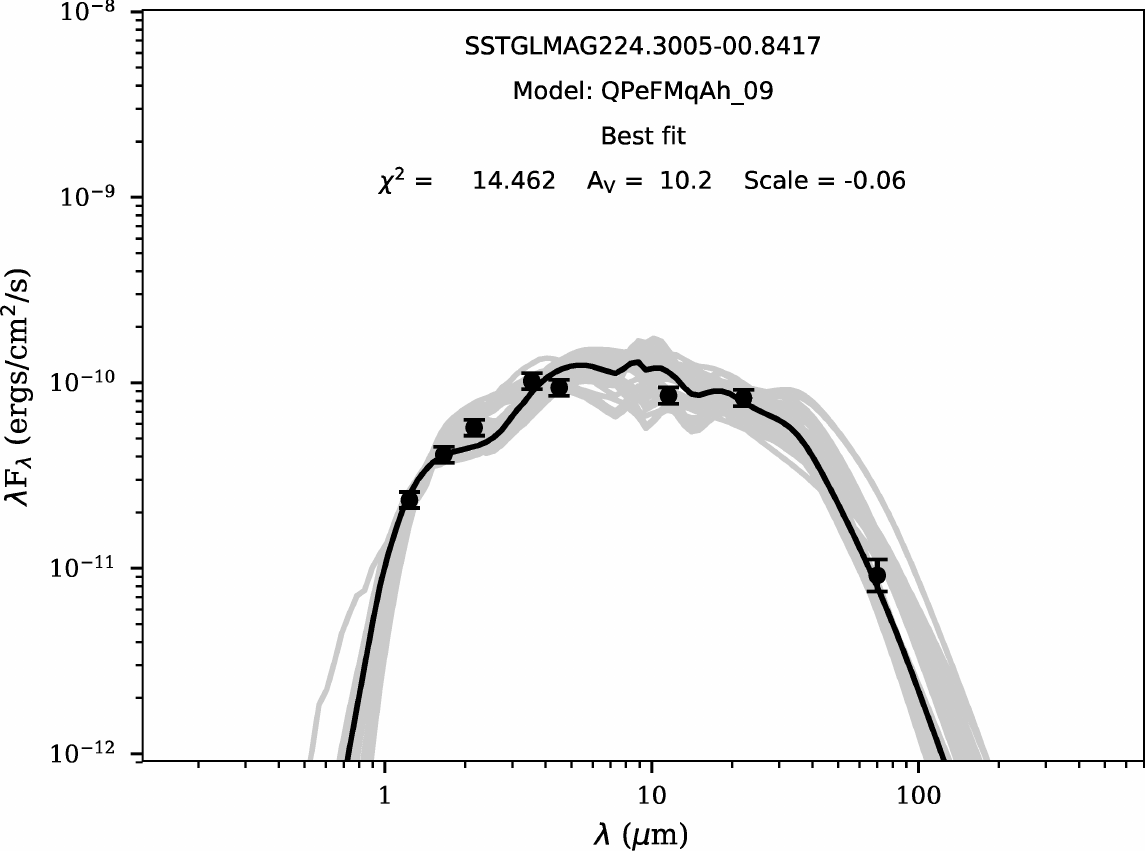}
\hfill
\includegraphics[width=0.32\textwidth]{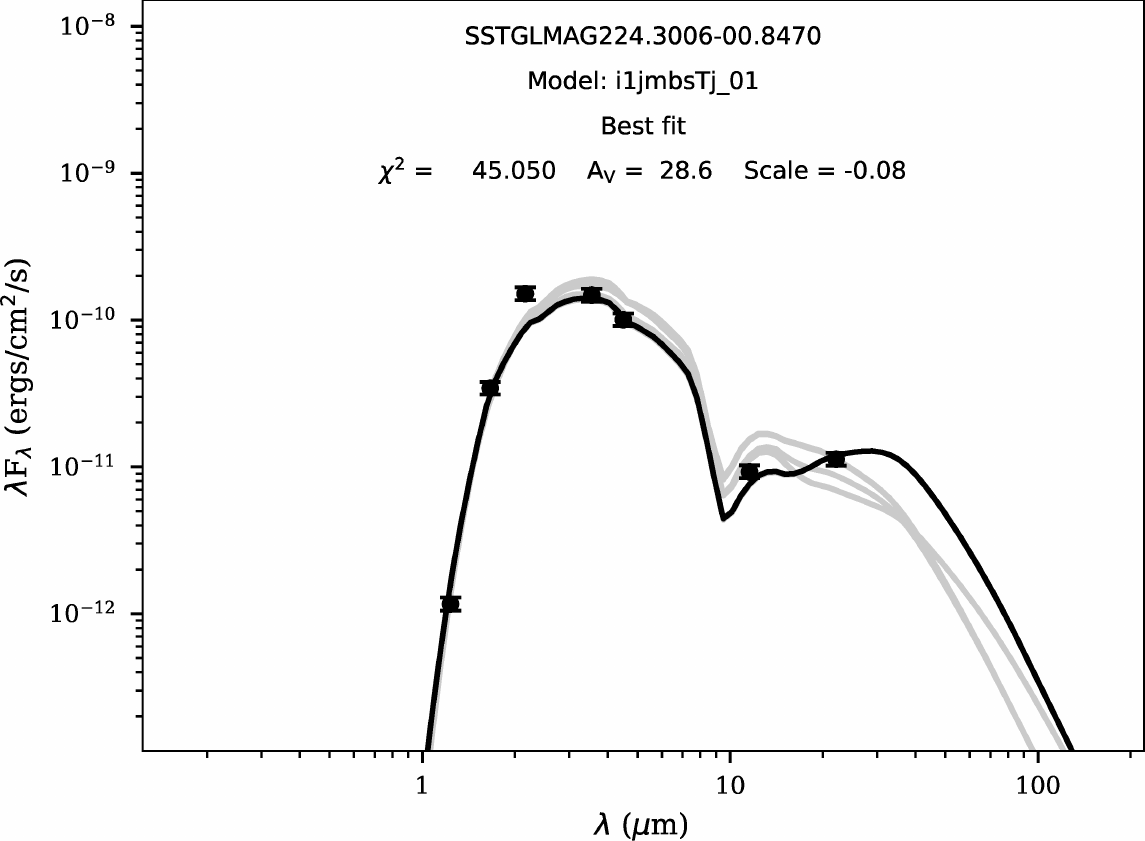}
\hfill
\includegraphics[width=0.32\textwidth]{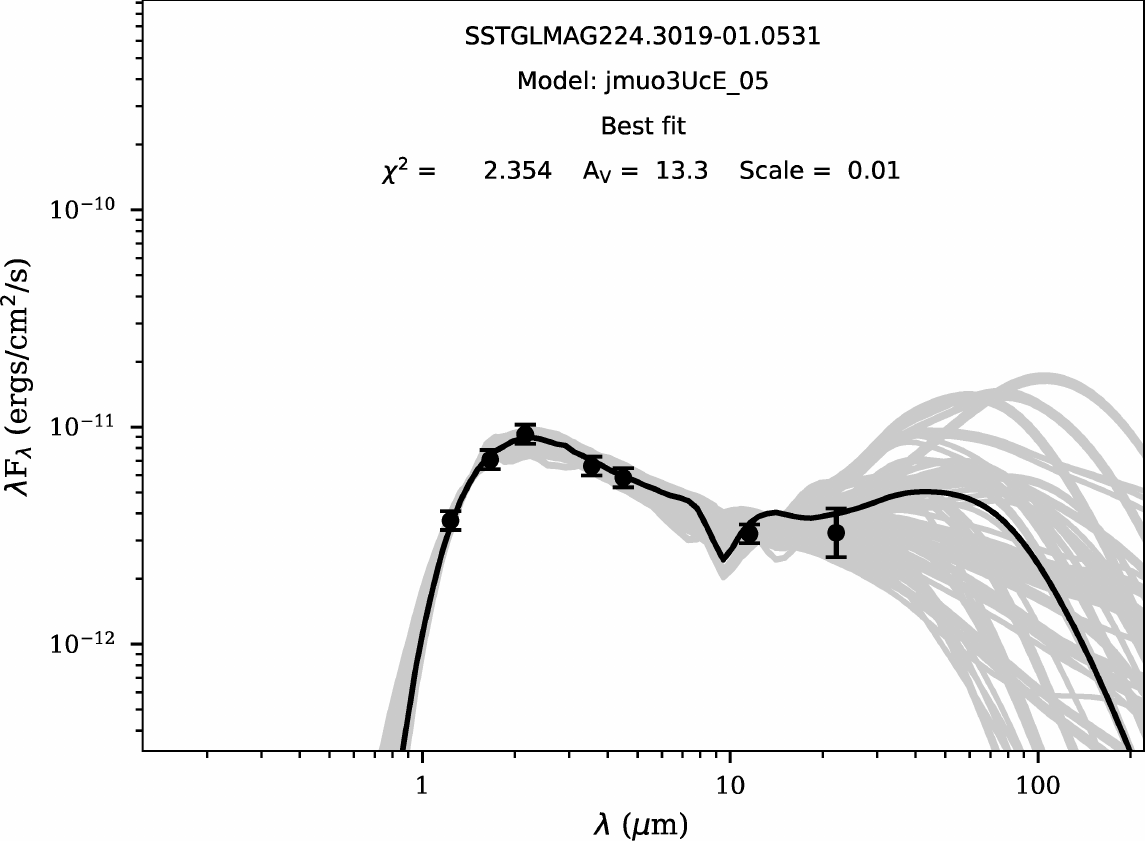} \par
\vspace{2mm}
\includegraphics[width=0.32\textwidth]{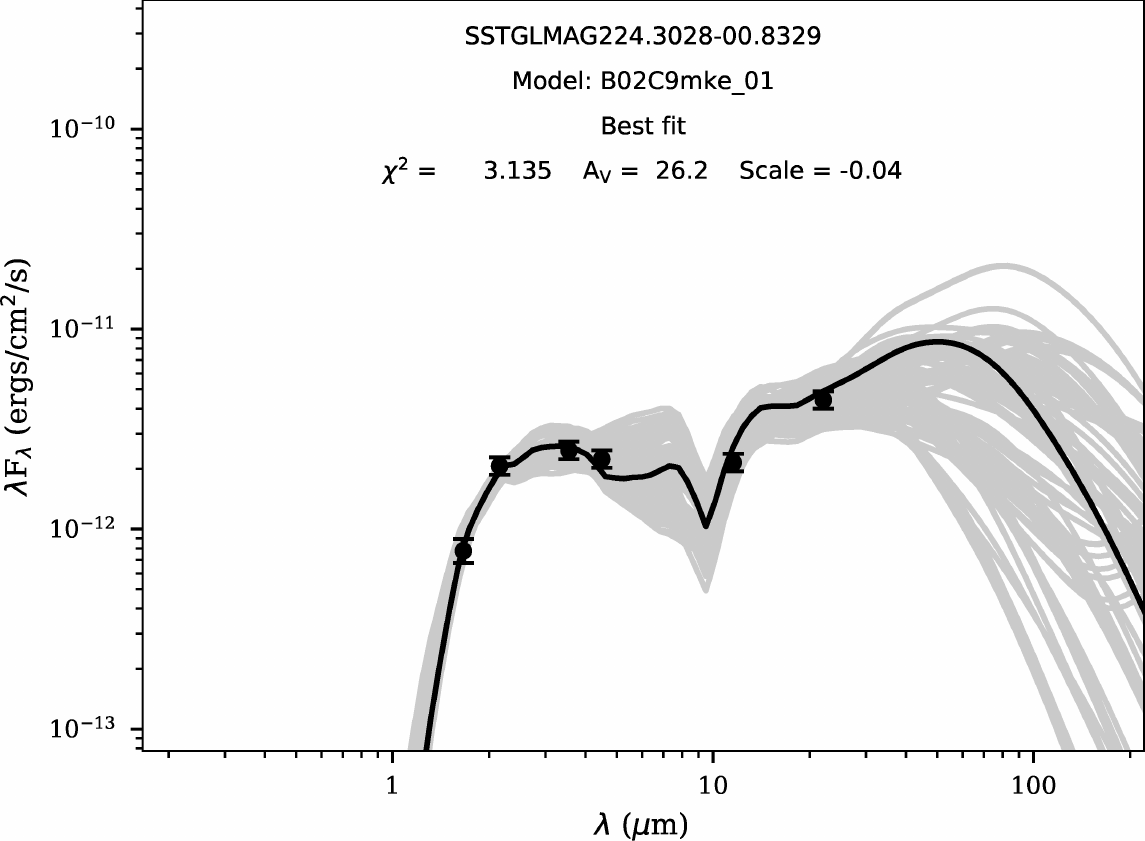}
\hfill
\includegraphics[width=0.32\textwidth]{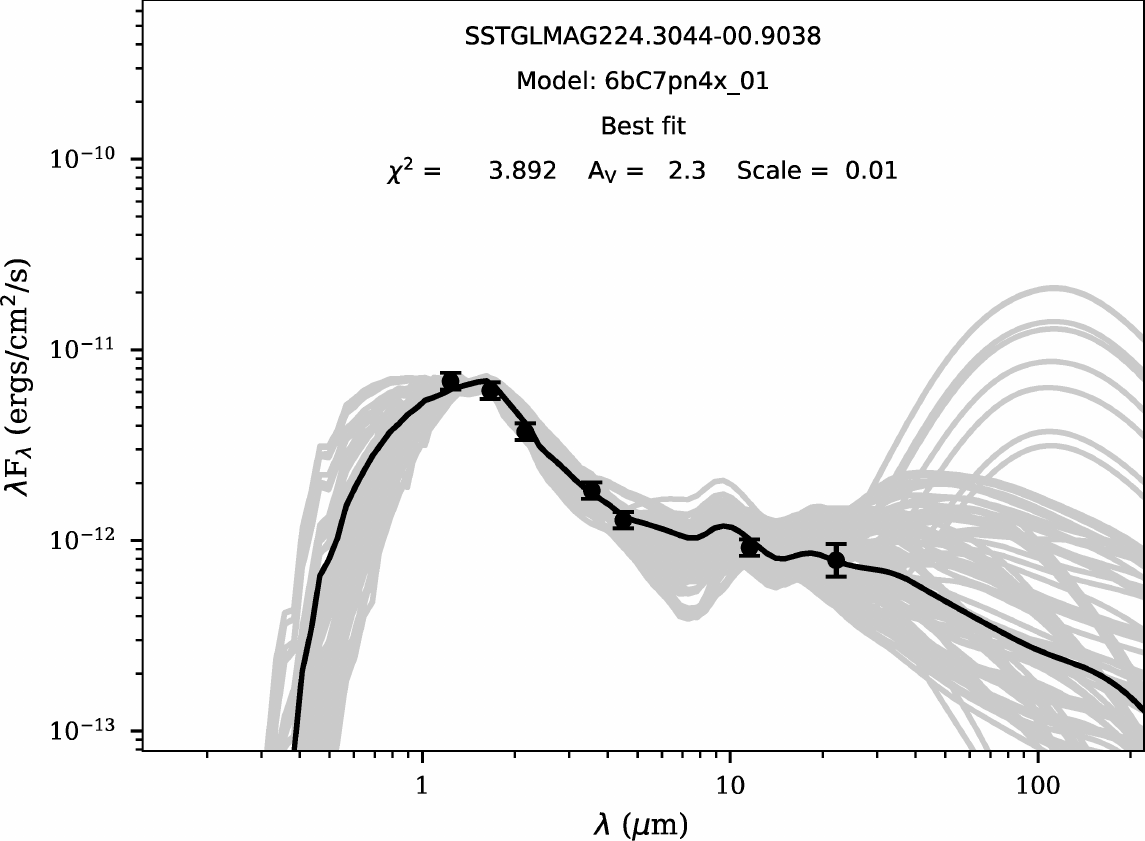}
\hfill
\includegraphics[width=0.32\textwidth]{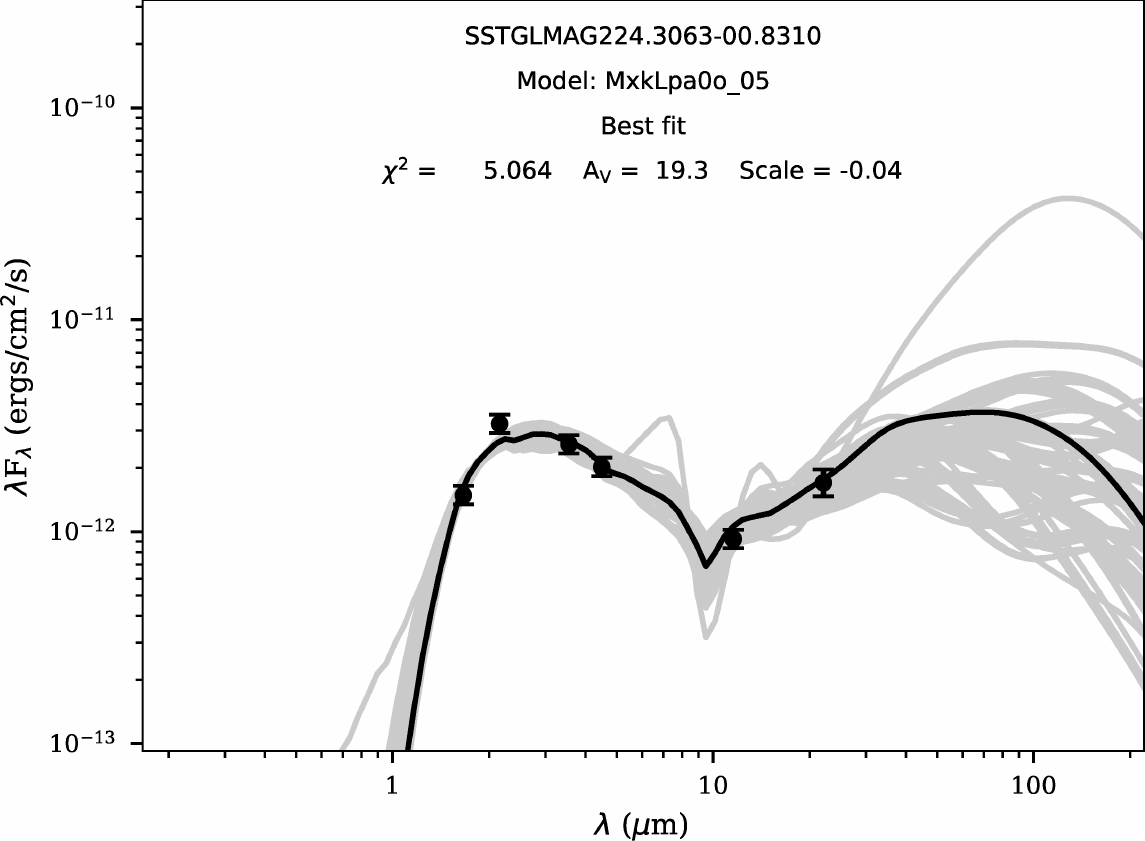} \par
\vspace{2mm}
\includegraphics[width=0.32\textwidth]{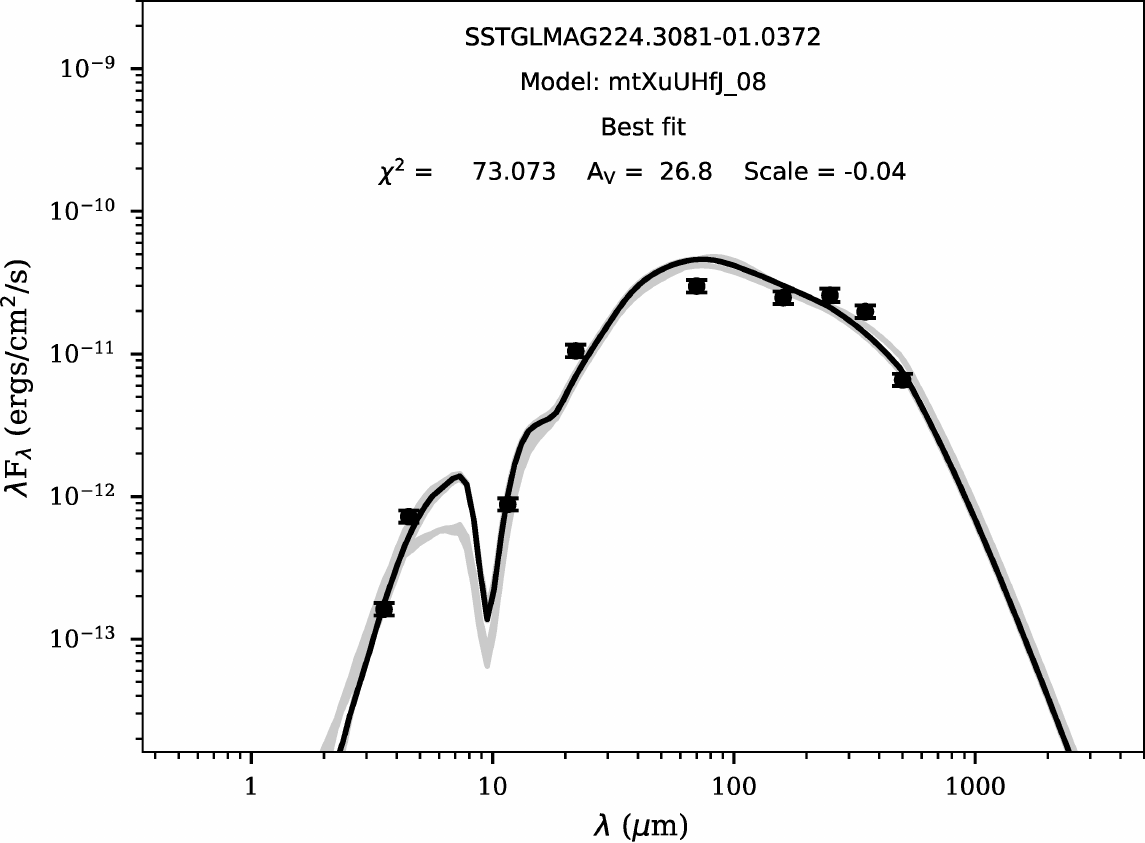}
\hfill
\includegraphics[width=0.32\textwidth]{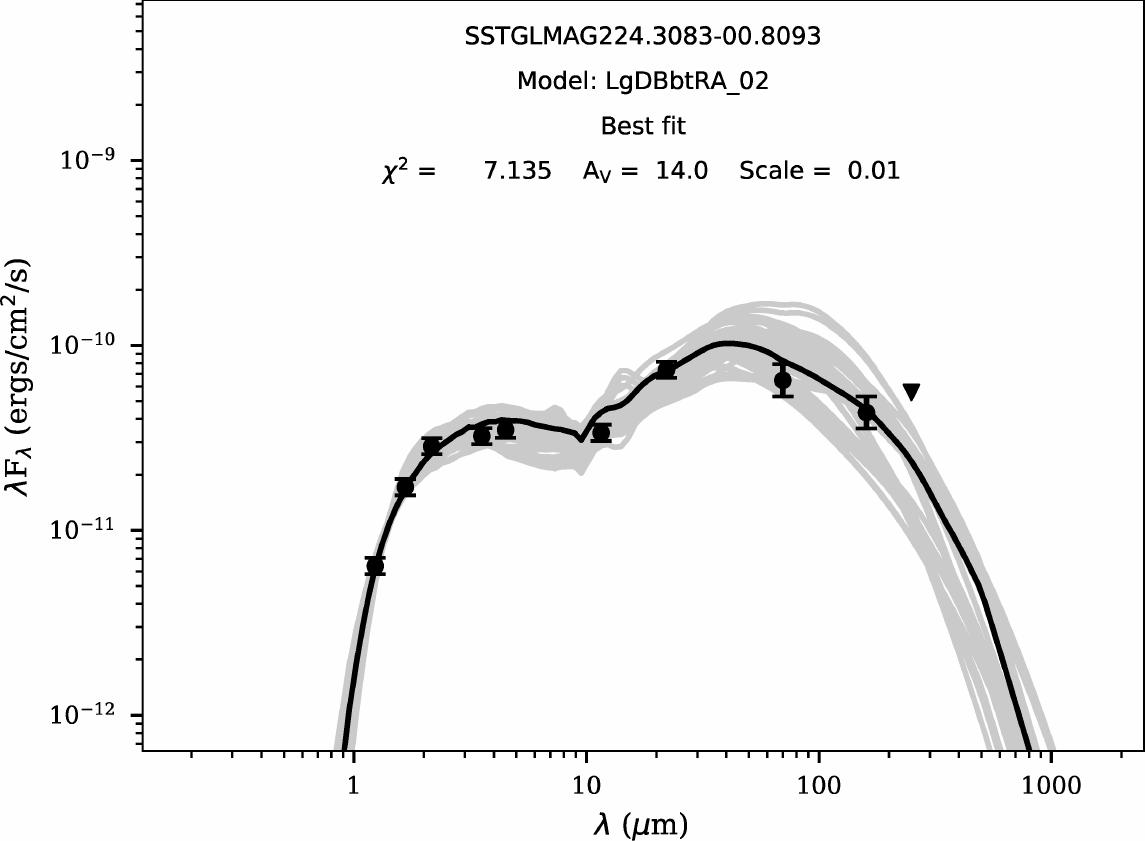}
\hfill
\includegraphics[width=0.32\textwidth]{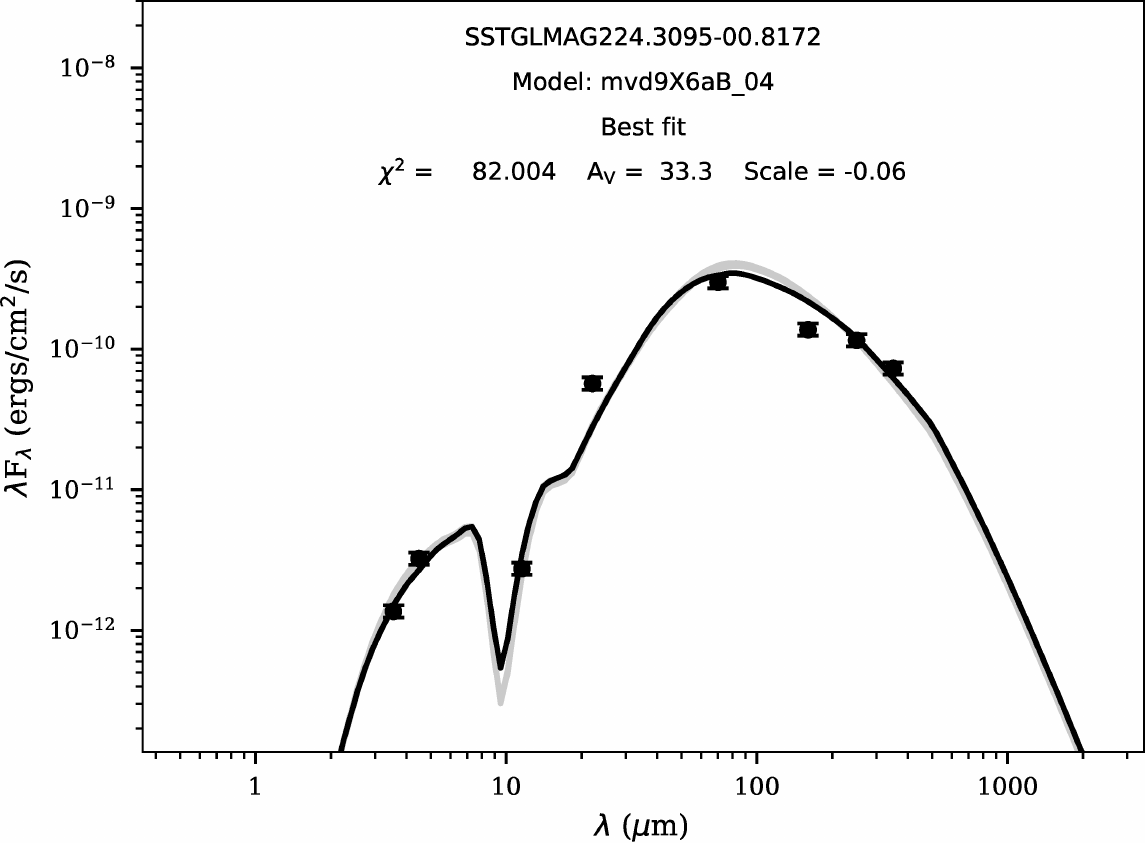}
\caption{Same as Fig.~\ref{f:SEDs1}  \label{f:SEDs7}}
\end{figure*}

\begin{figure*}
\includegraphics[width=0.32\textwidth]{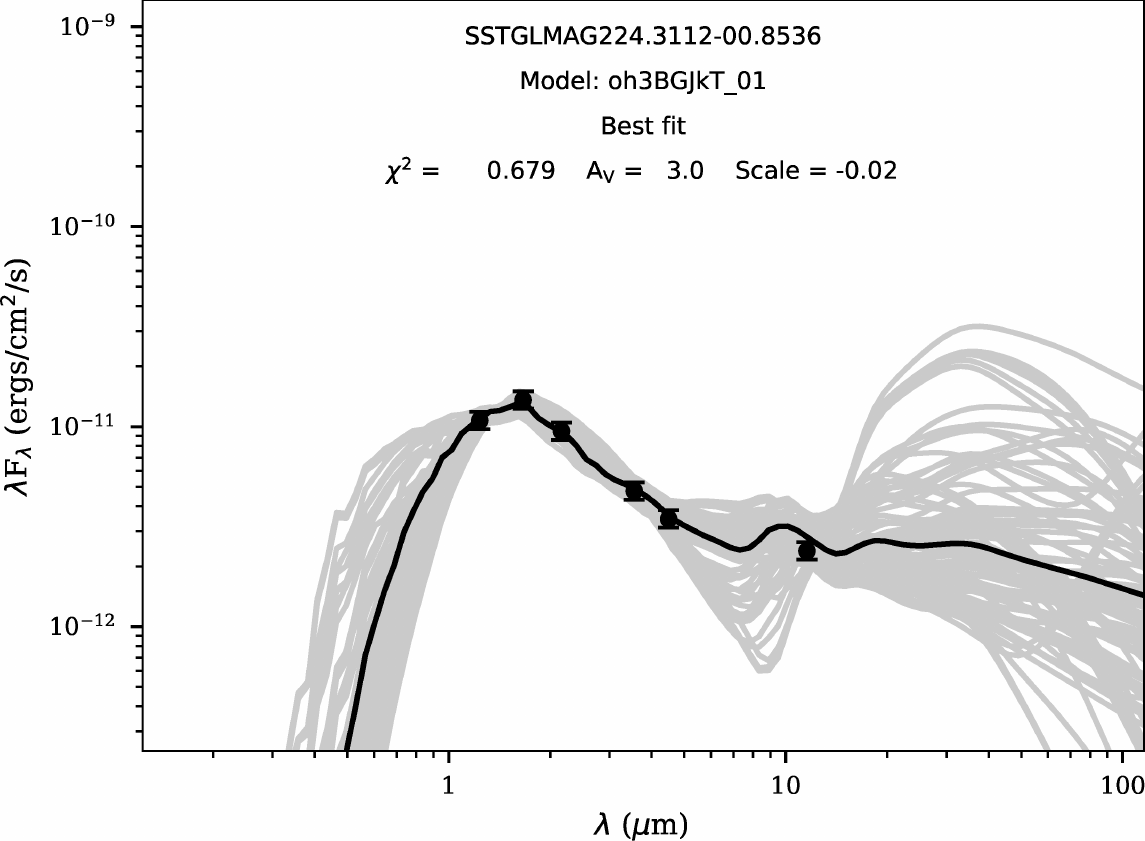}
\hfill
\includegraphics[width=0.32\textwidth]{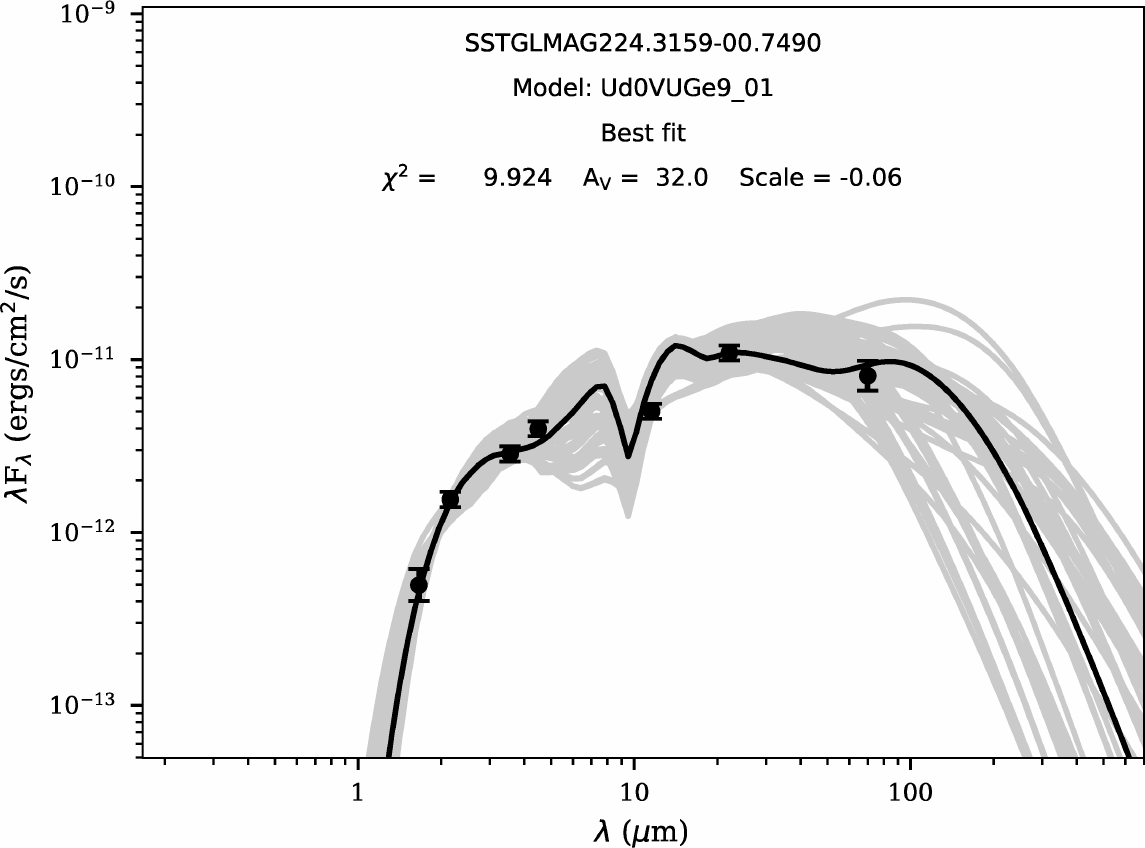}
\hfill
\includegraphics[width=0.32\textwidth]{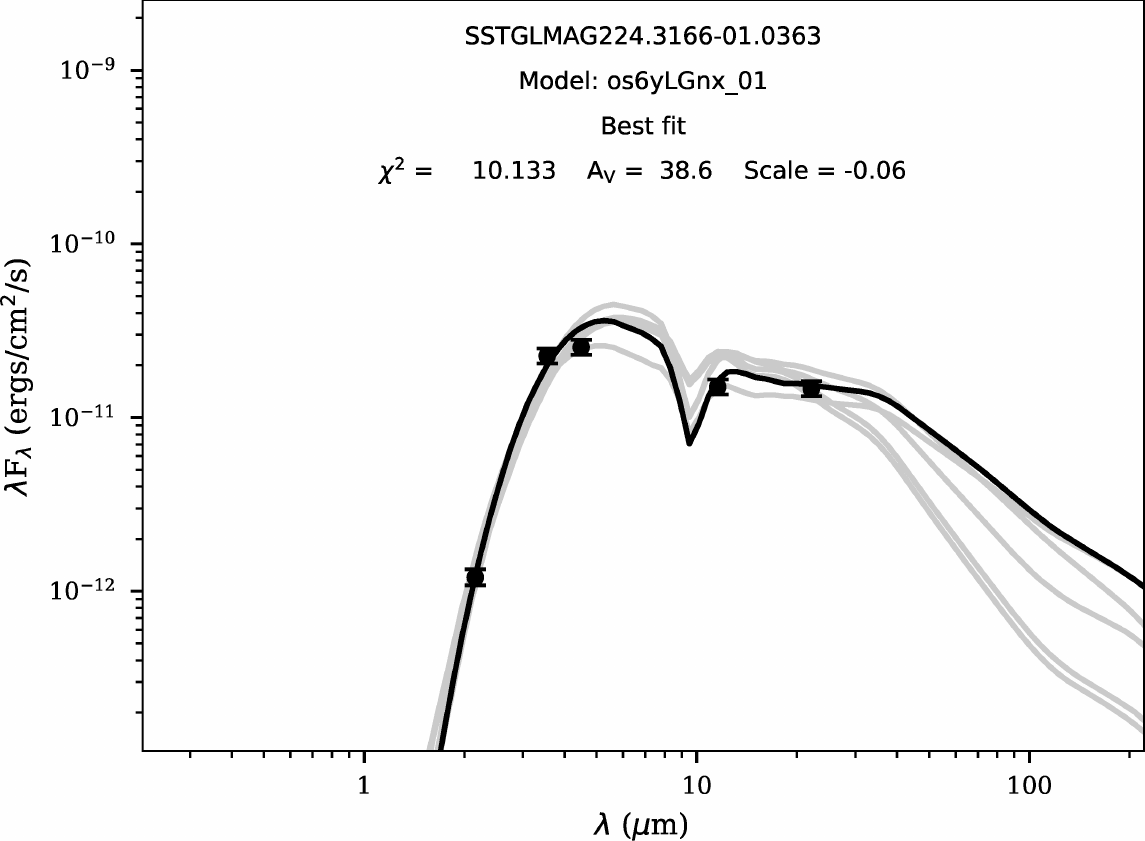} \par
\vspace{2mm}
\includegraphics[width=0.32\textwidth]{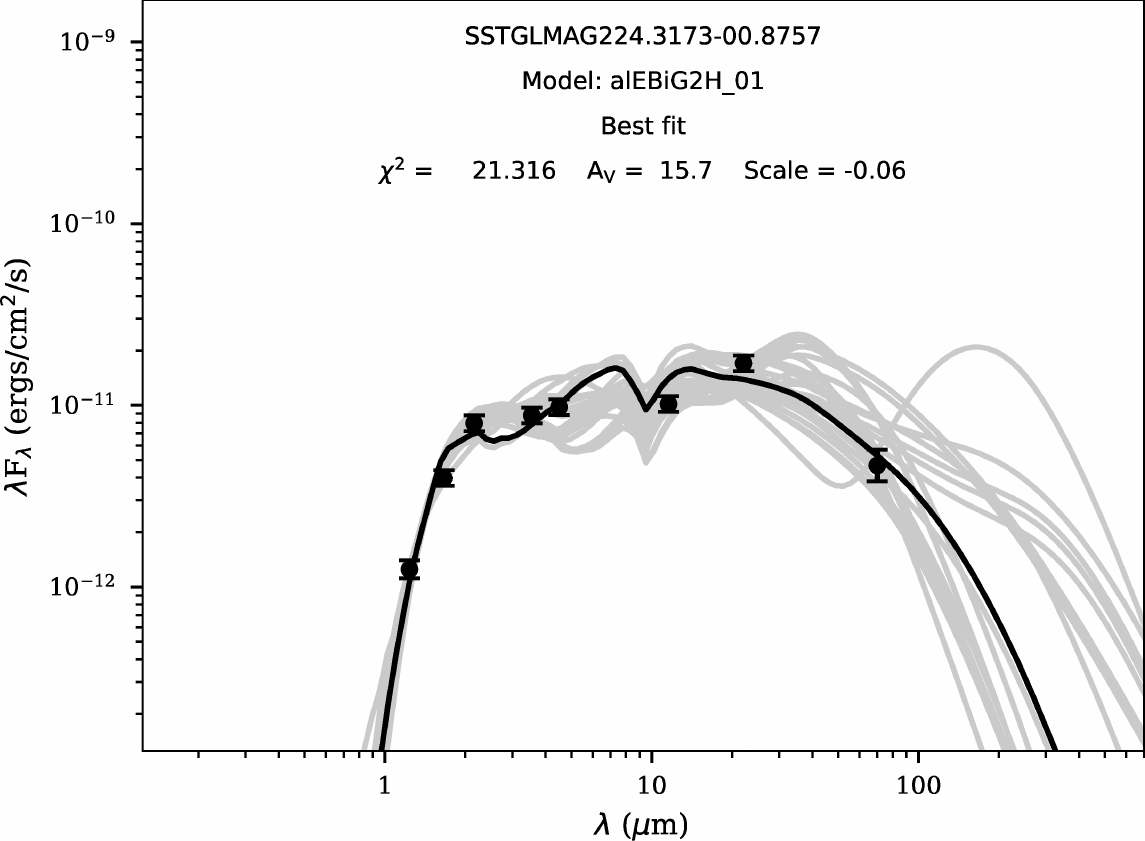}
\hfill
\includegraphics[width=0.32\textwidth]{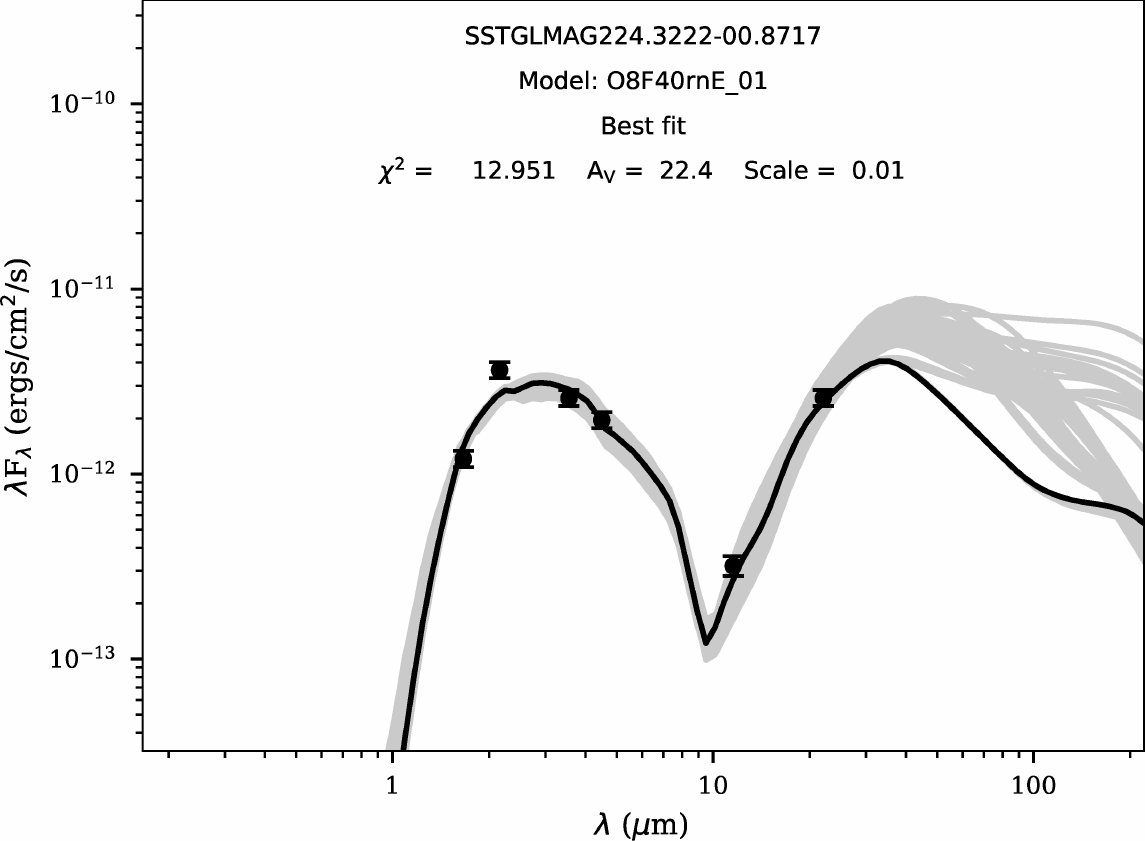}
\hfill
\includegraphics[width=0.32\textwidth]{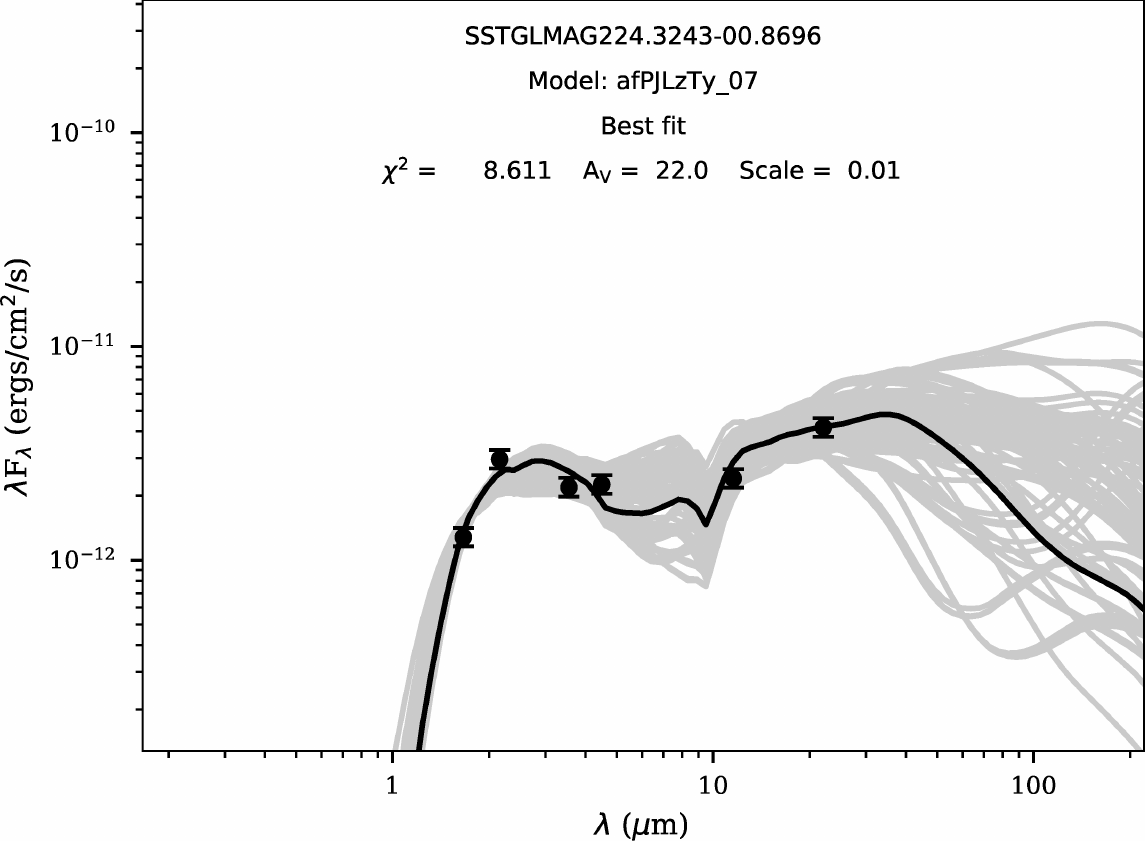} \par
\vspace{2mm}
\includegraphics[width=0.32\textwidth]{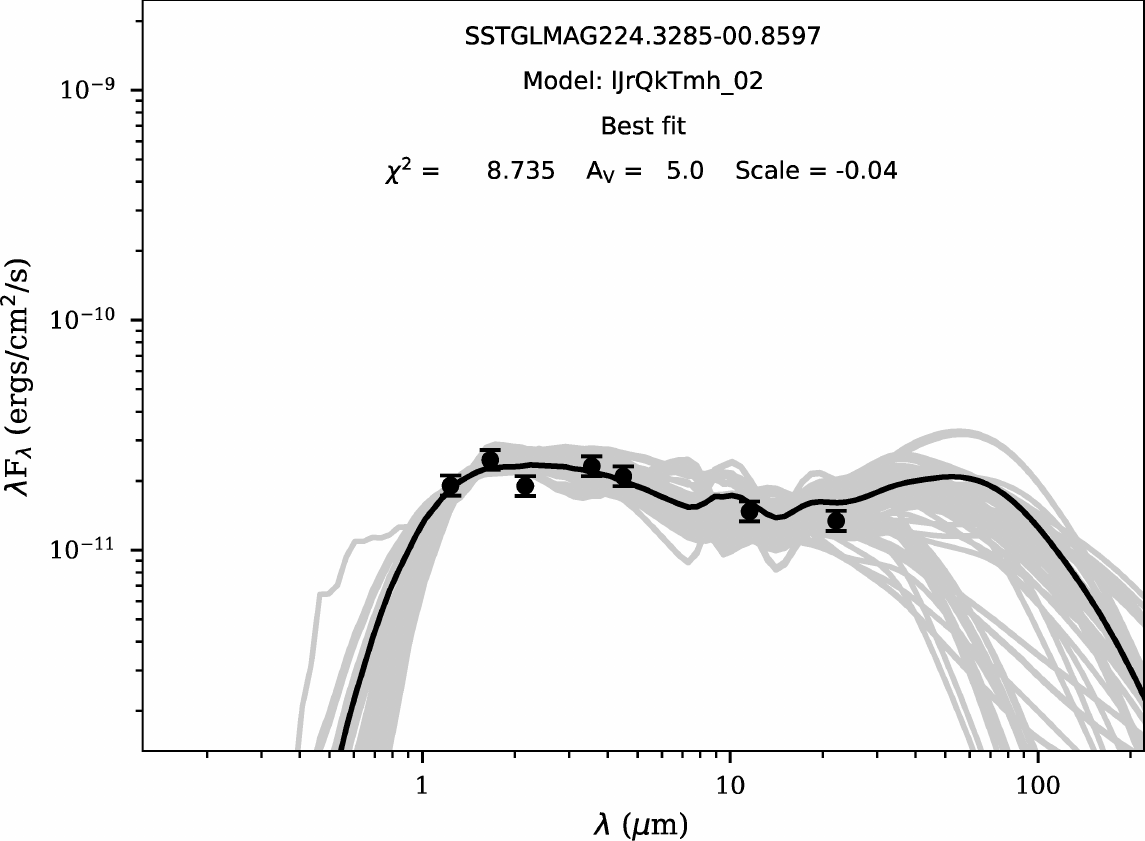}
\hfill
\includegraphics[width=0.32\textwidth]{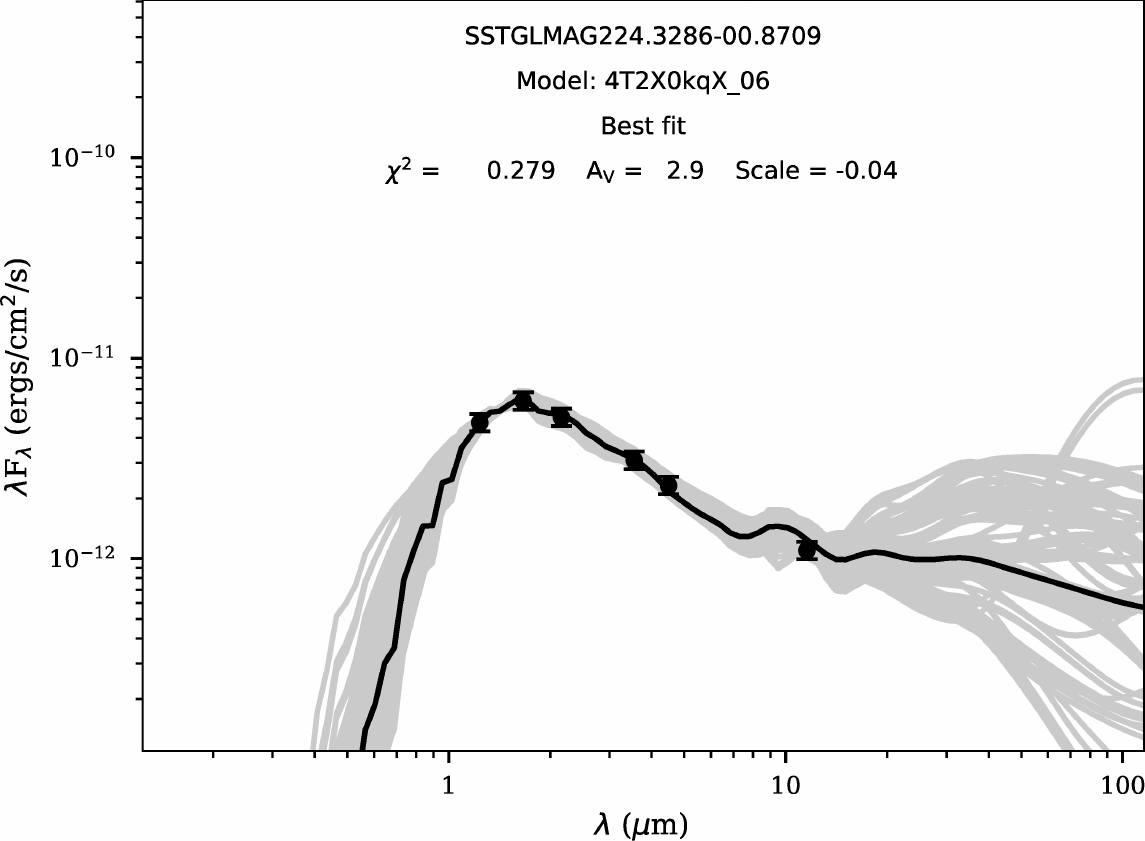}
\hfill
\includegraphics[width=0.32\textwidth]{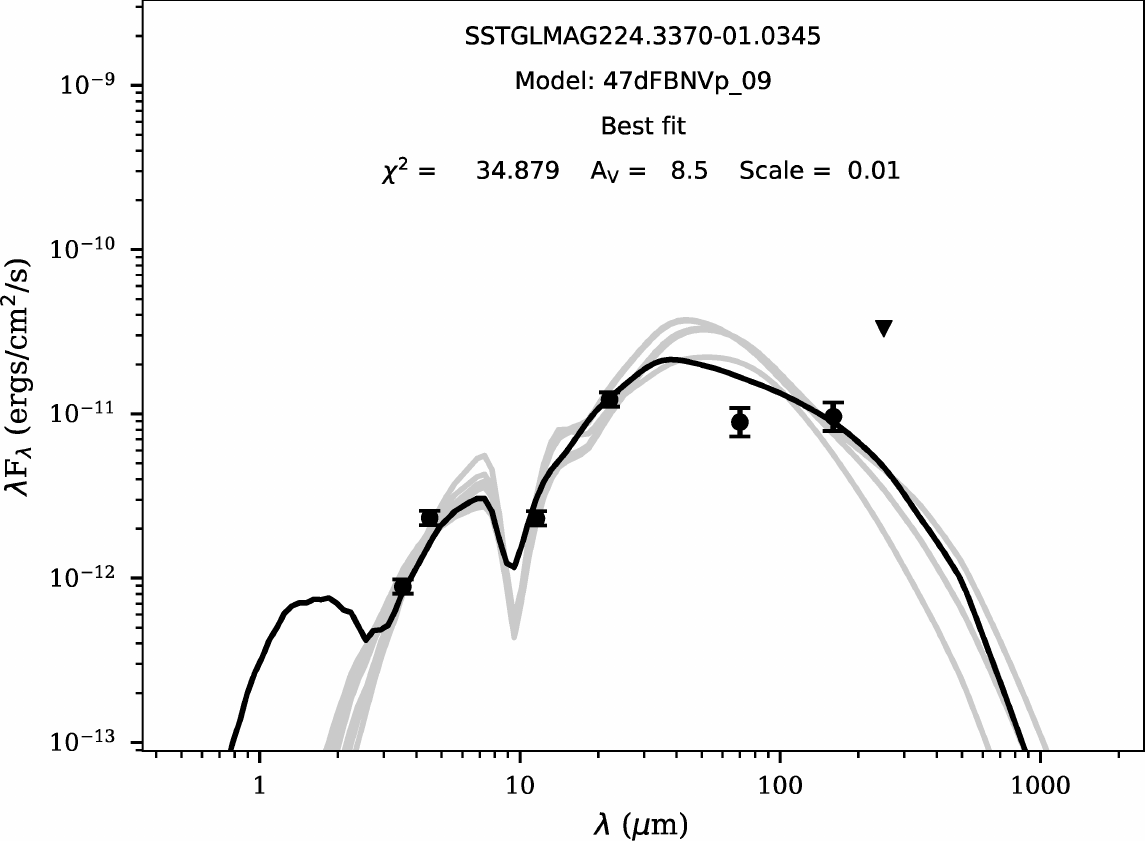} \par
\vspace{2mm}
\includegraphics[width=0.32\textwidth]{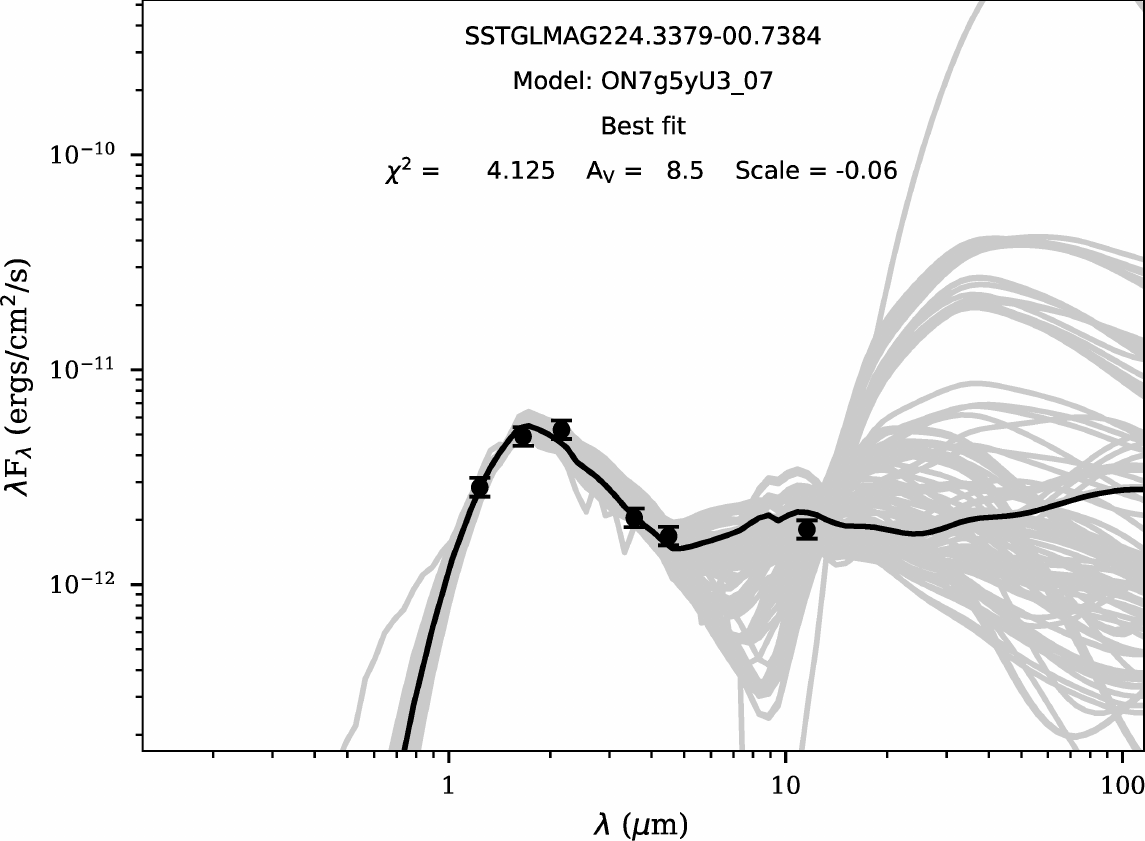}
\hfill
\includegraphics[width=0.32\textwidth]{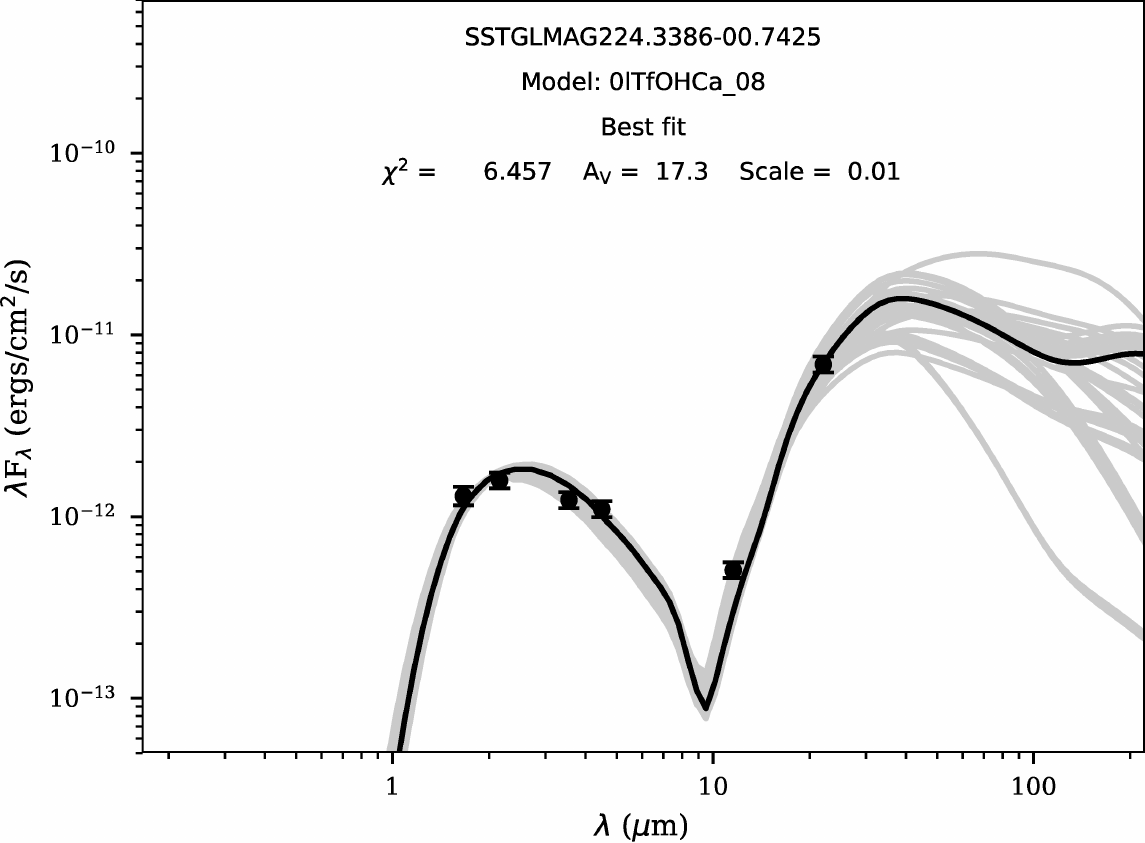}
\hfill
\includegraphics[width=0.32\textwidth]{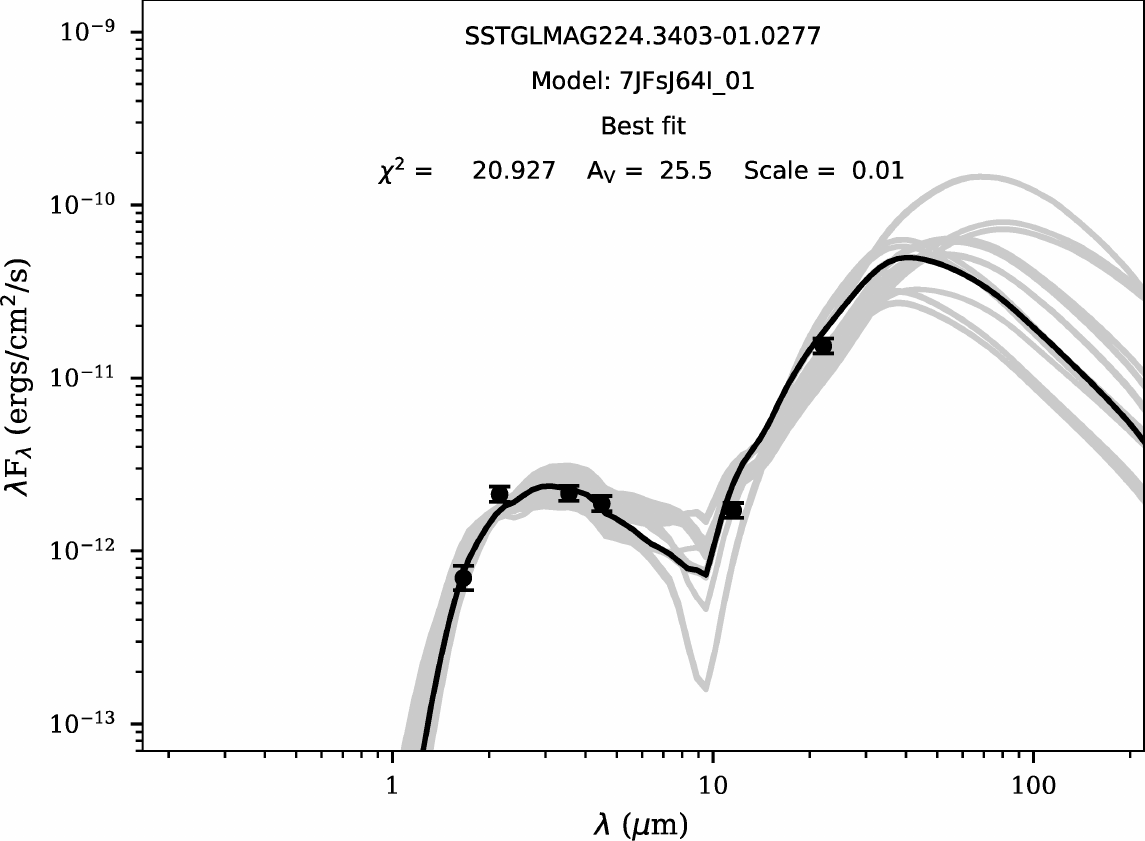} \par
\vspace{2mm}
\includegraphics[width=0.32\textwidth]{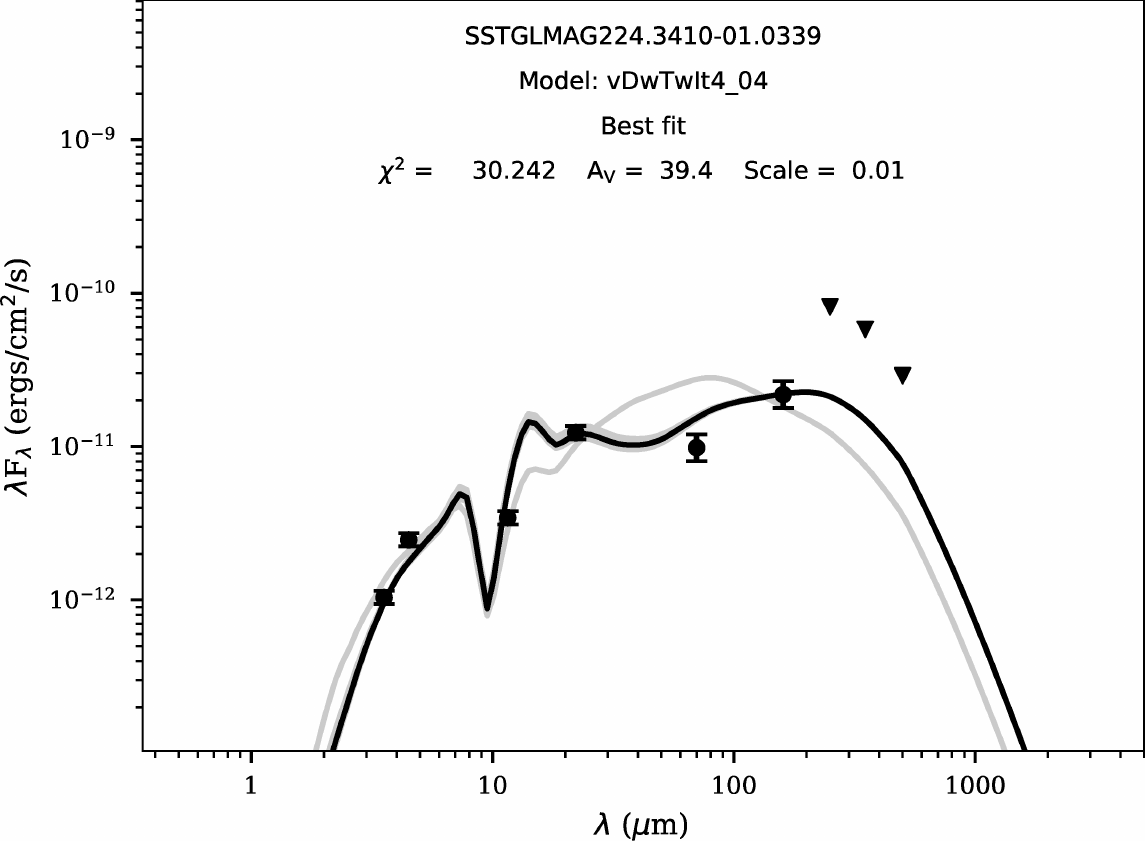}
\hfill
\includegraphics[width=0.32\textwidth]{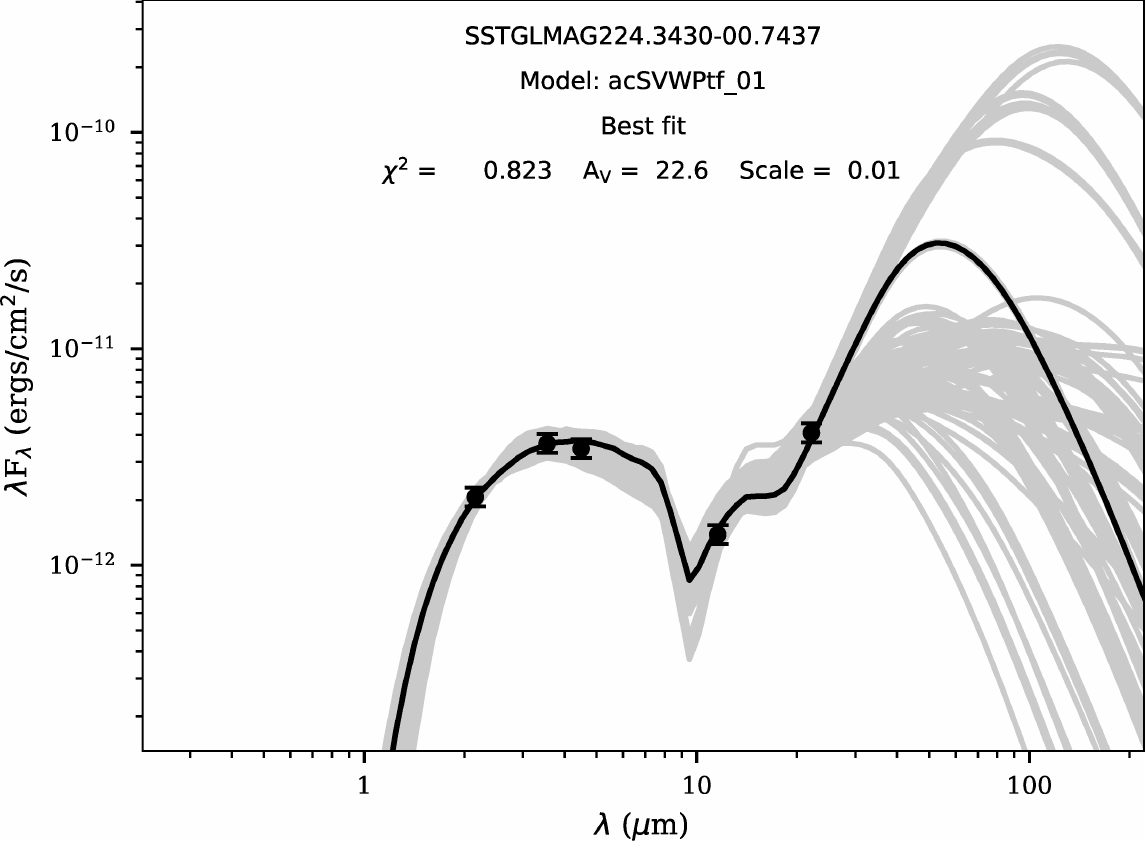}
\hfill
\includegraphics[width=0.32\textwidth]{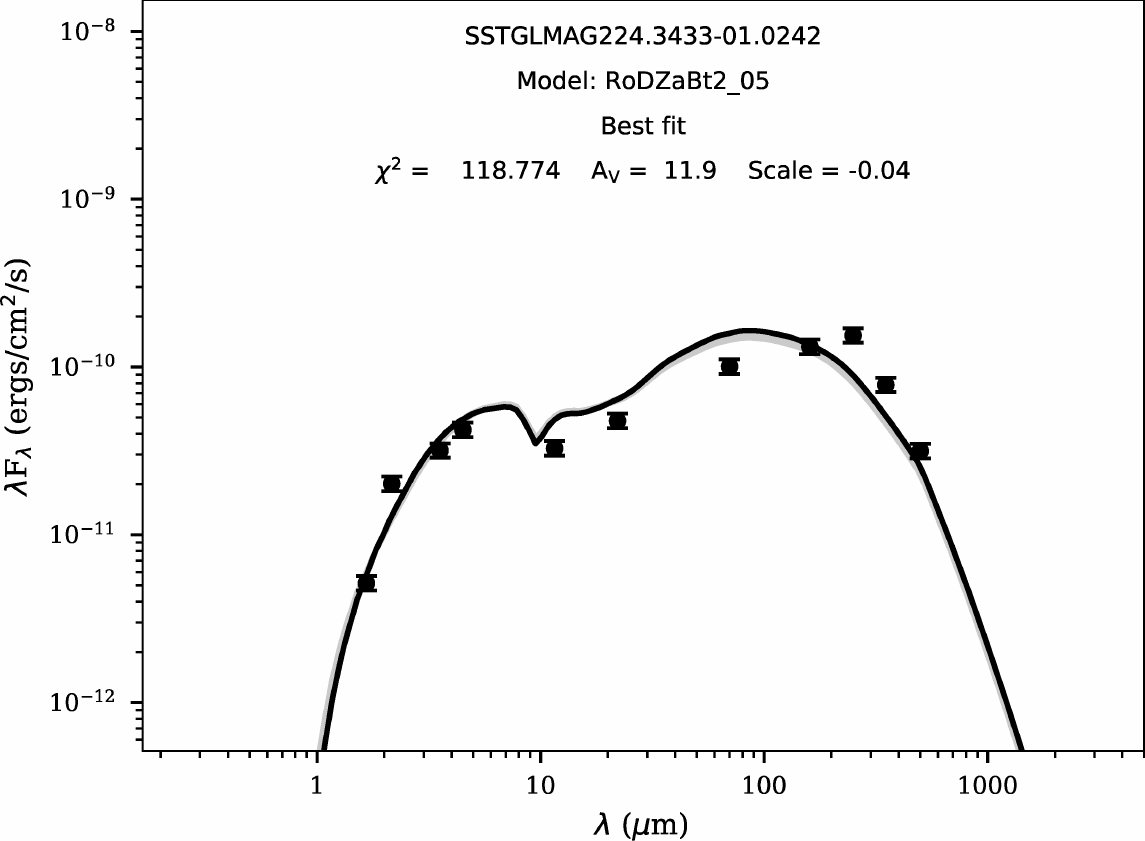}
\caption{Same as Fig.~\ref{f:SEDs1}  \label{f:SEDs8}}
\end{figure*}

\begin{figure*}
\includegraphics[width=0.32\textwidth]{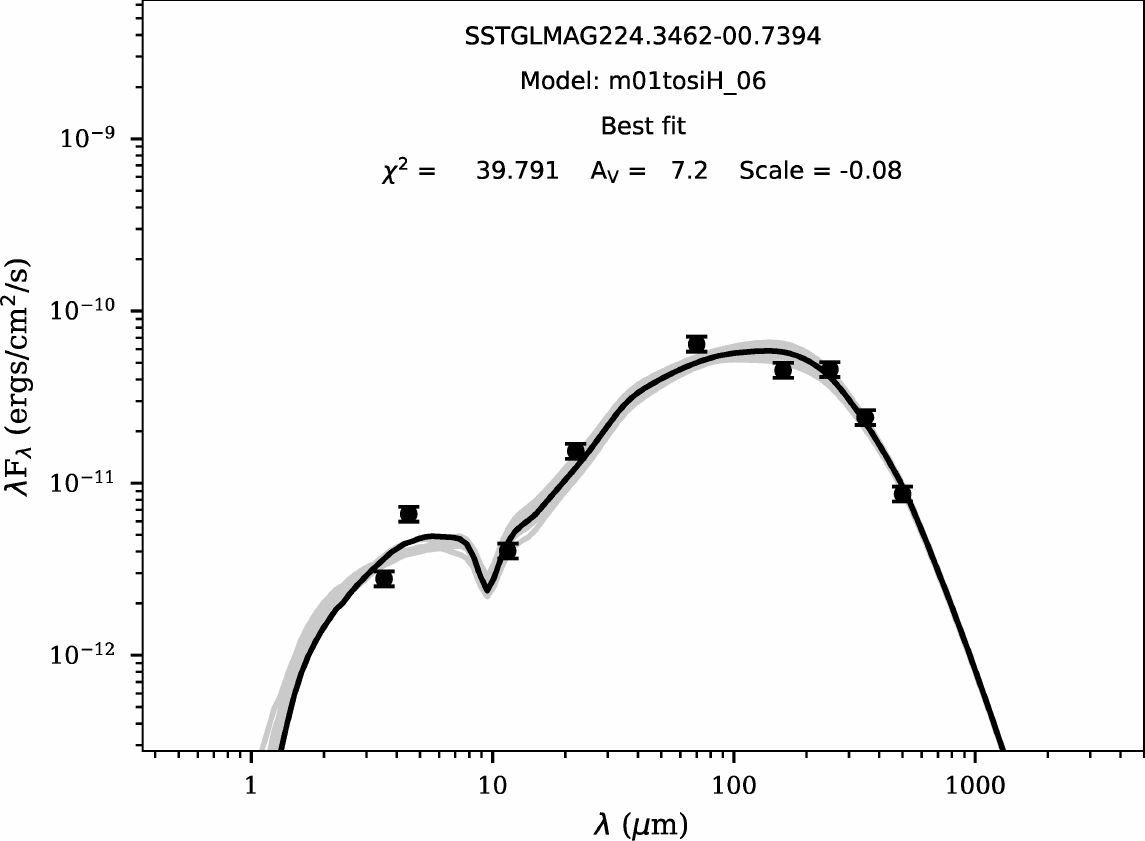}
\hfill
\includegraphics[width=0.32\textwidth]{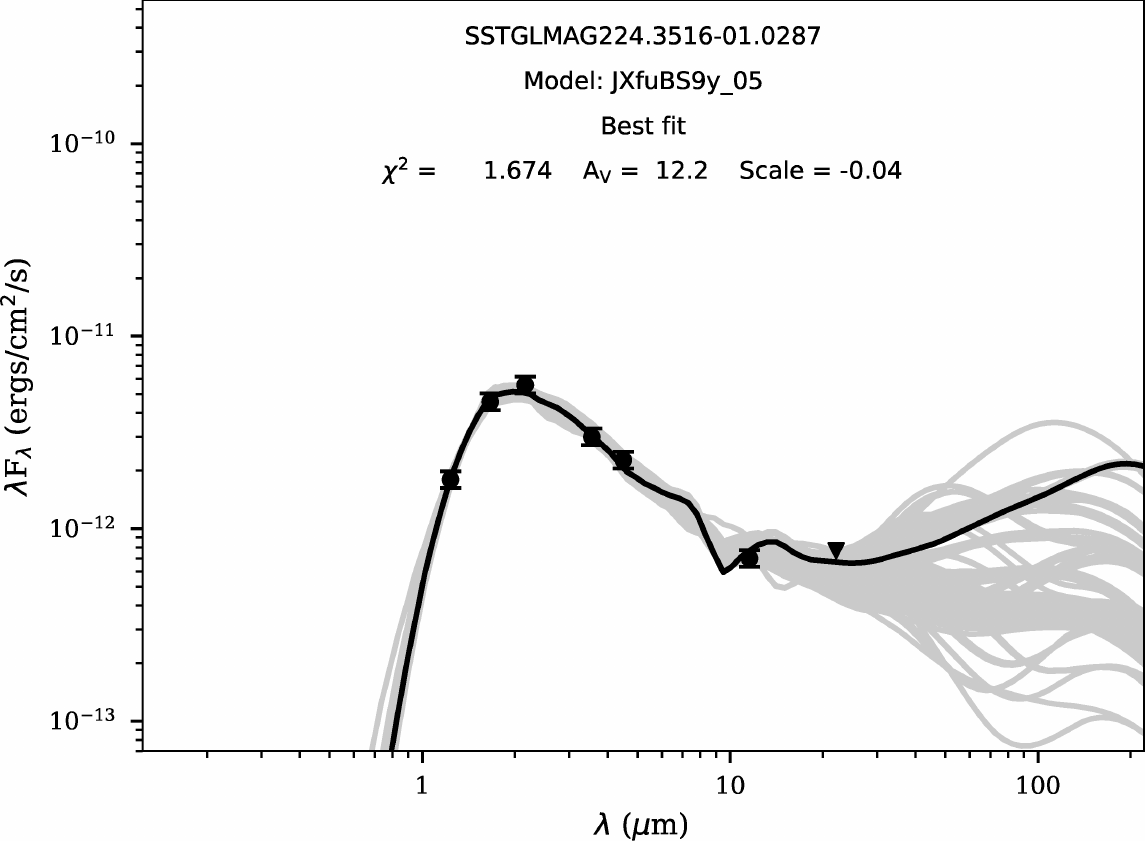}
\hfill
\includegraphics[width=0.32\textwidth]{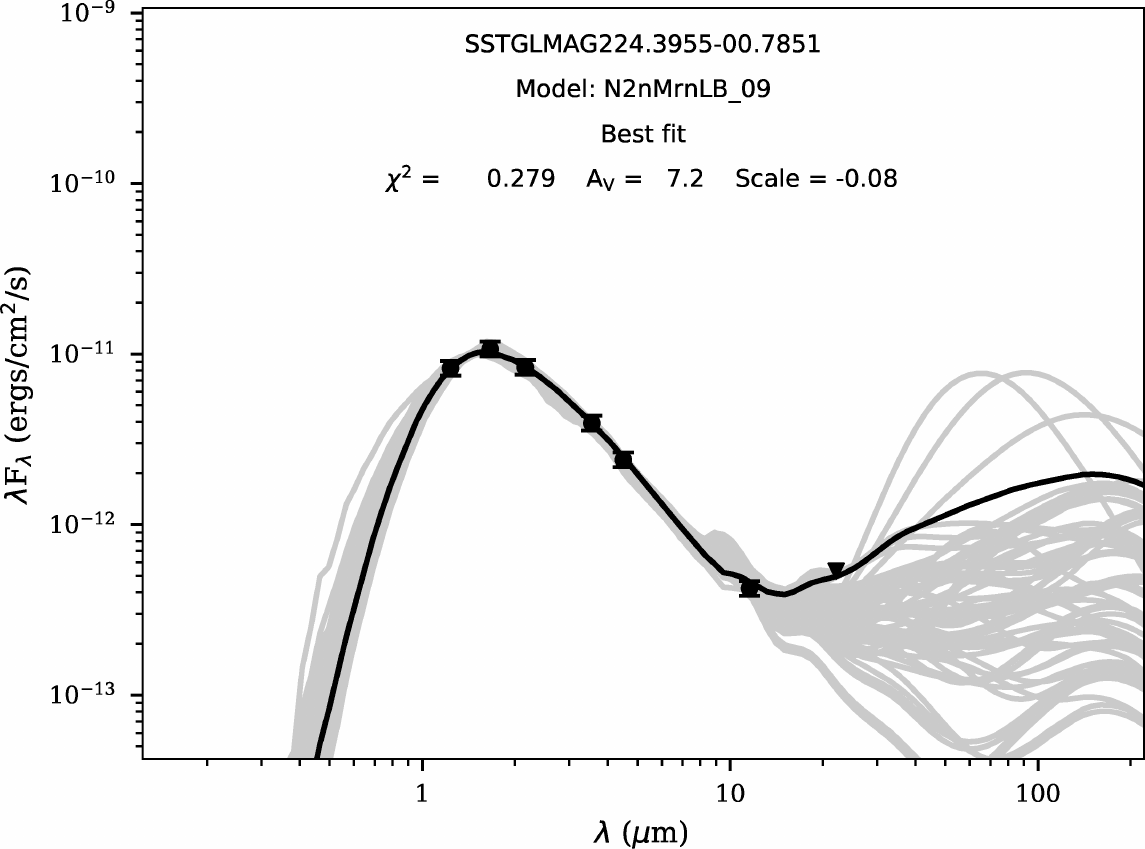} \par
\vspace{2mm}
\includegraphics[width=0.32\textwidth]{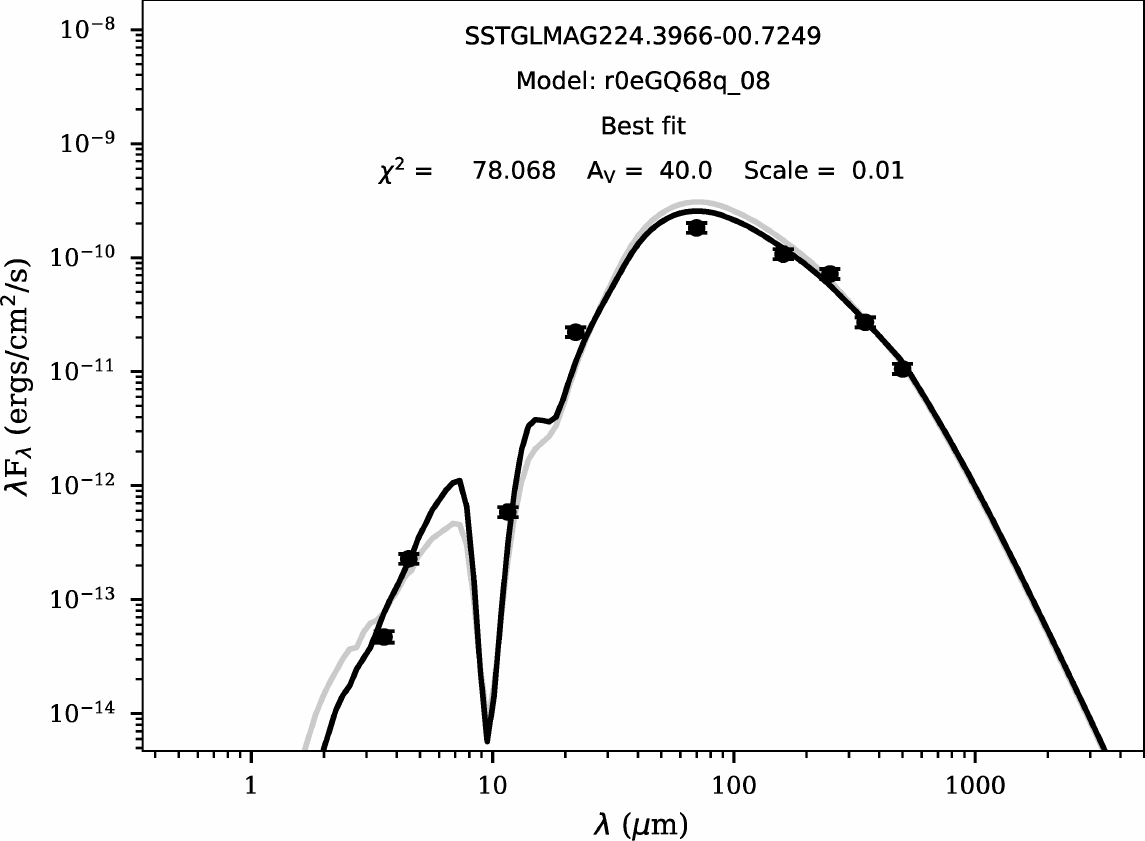}
\hfill
\includegraphics[width=0.32\textwidth]{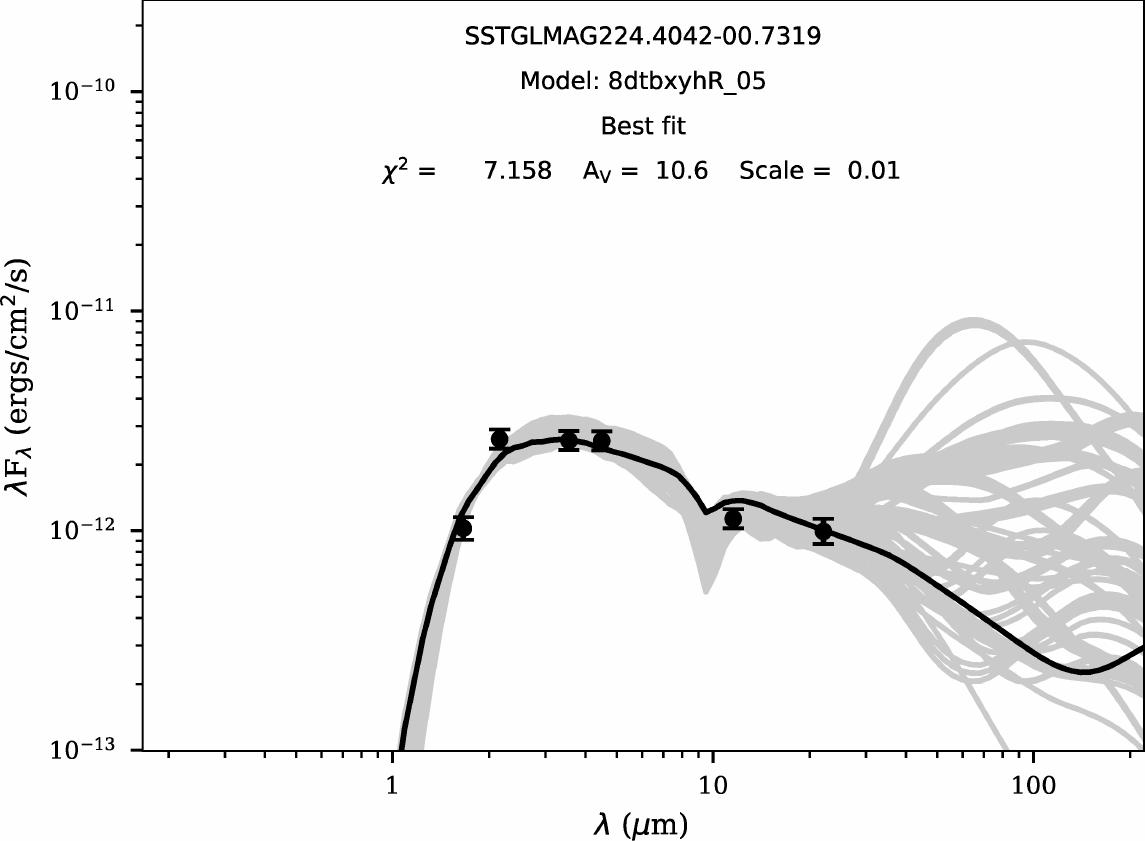}
\hfill
\includegraphics[width=0.32\textwidth]{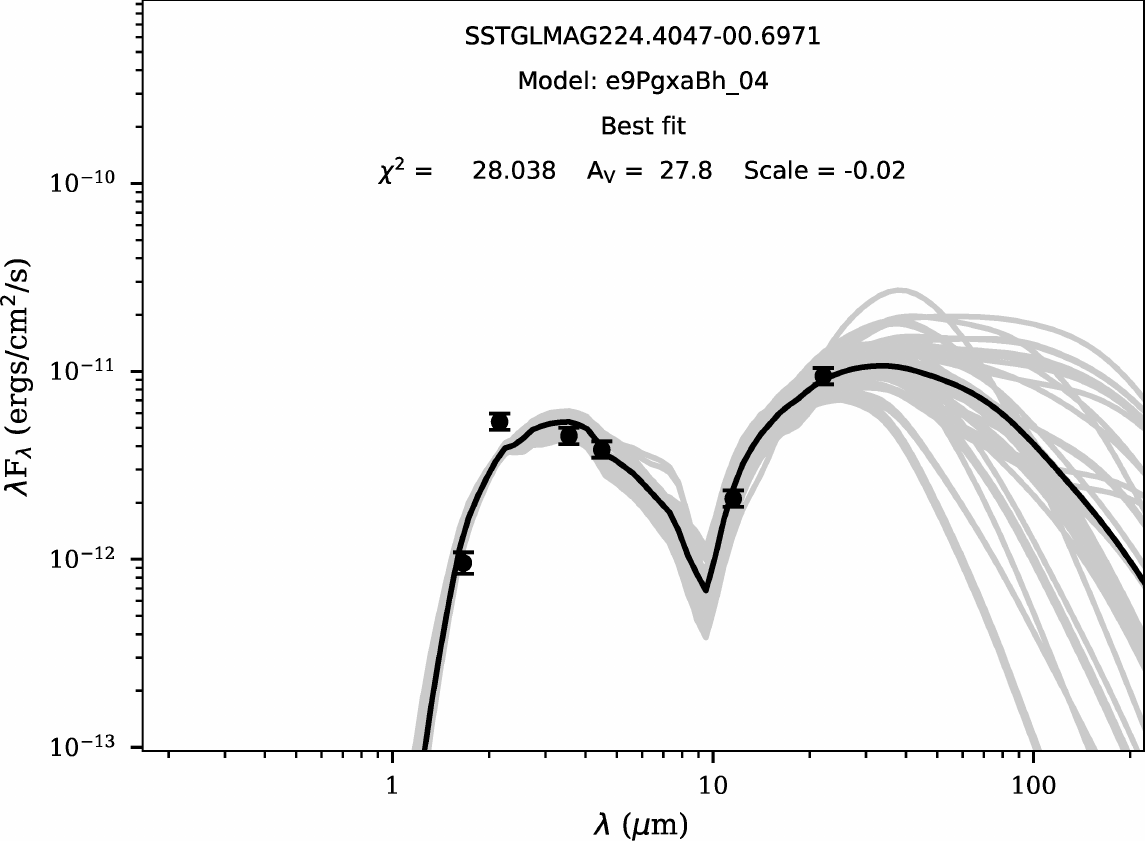} \par
\vspace{2mm}
\includegraphics[width=0.32\textwidth]{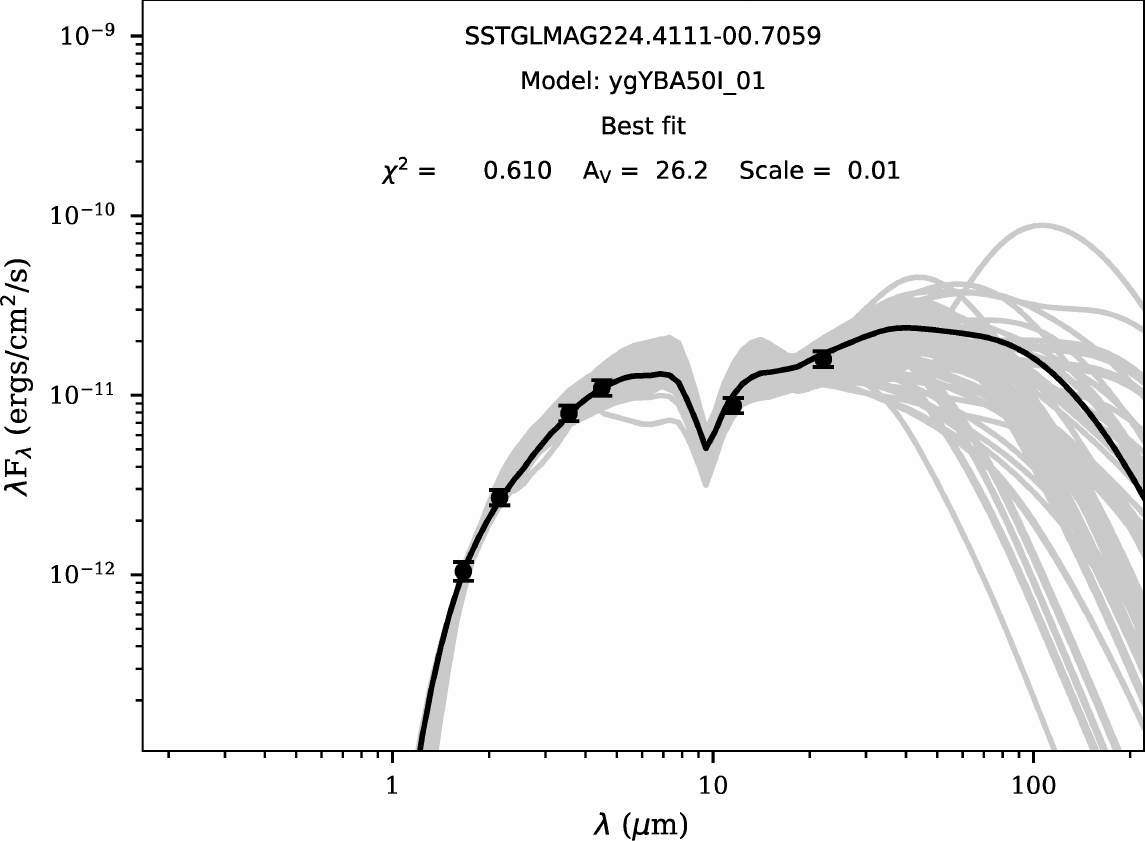}
\hfill
\includegraphics[width=0.32\textwidth]{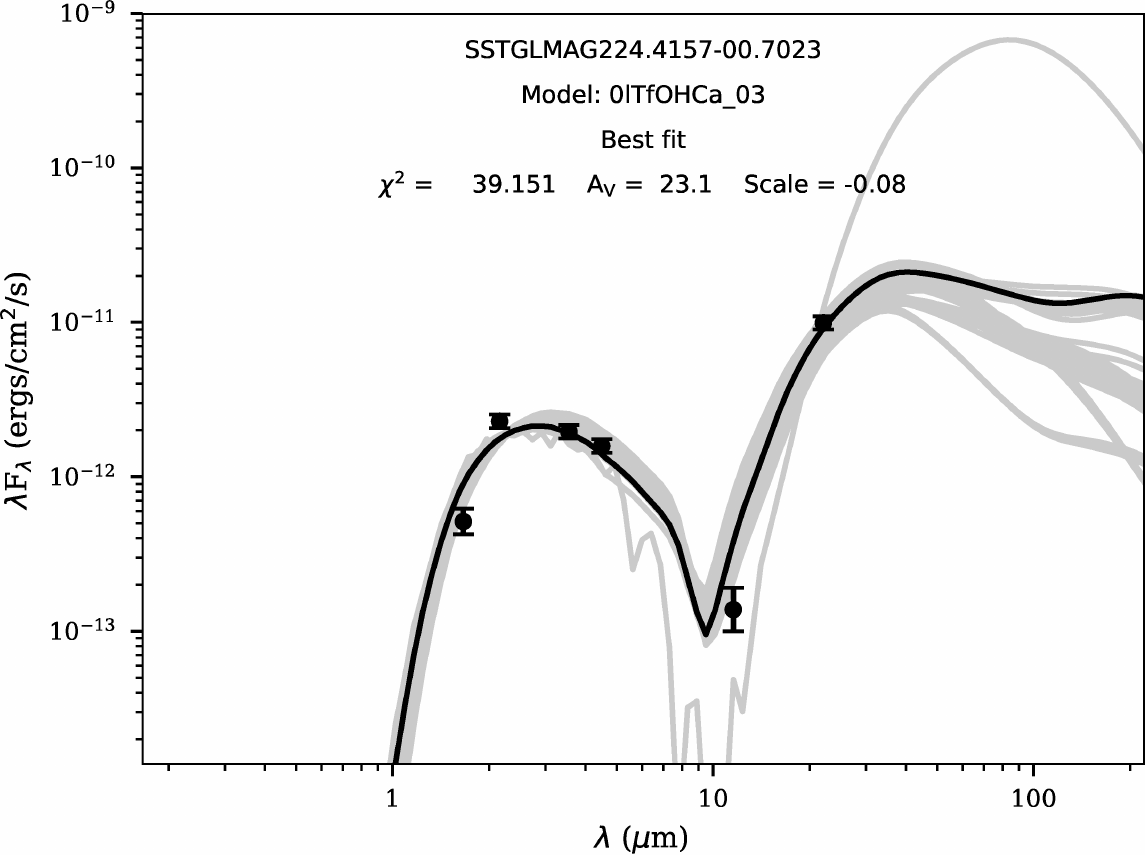}
\hfill
\includegraphics[width=0.32\textwidth]{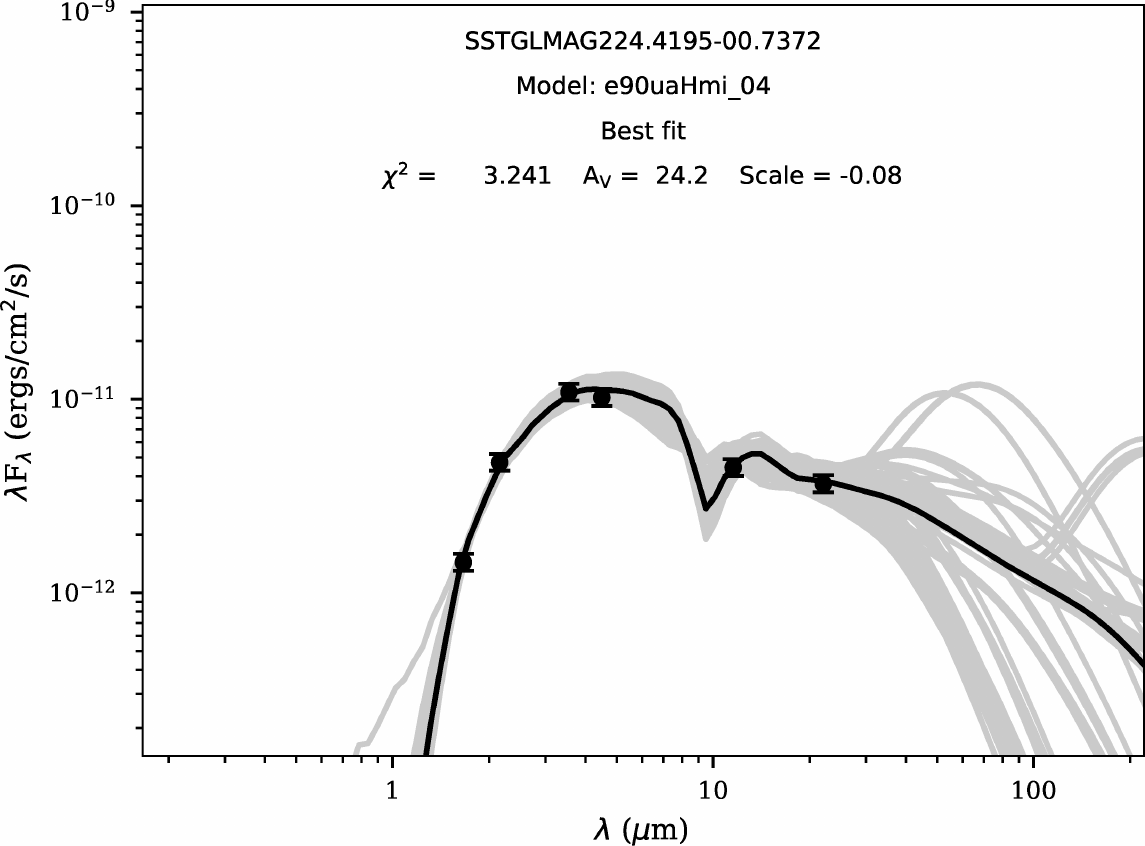} \par
\vspace{2mm}
\includegraphics[width=0.32\textwidth]{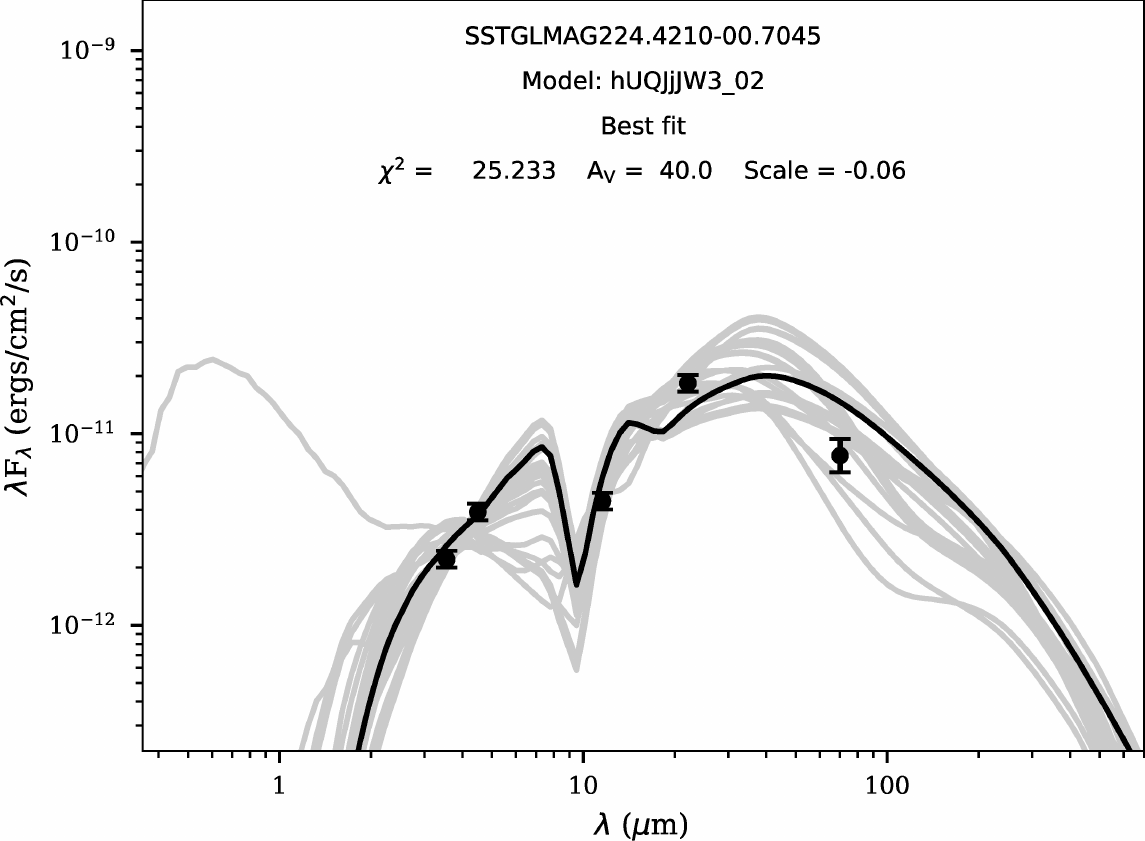}
\hfill
\includegraphics[width=0.32\textwidth]{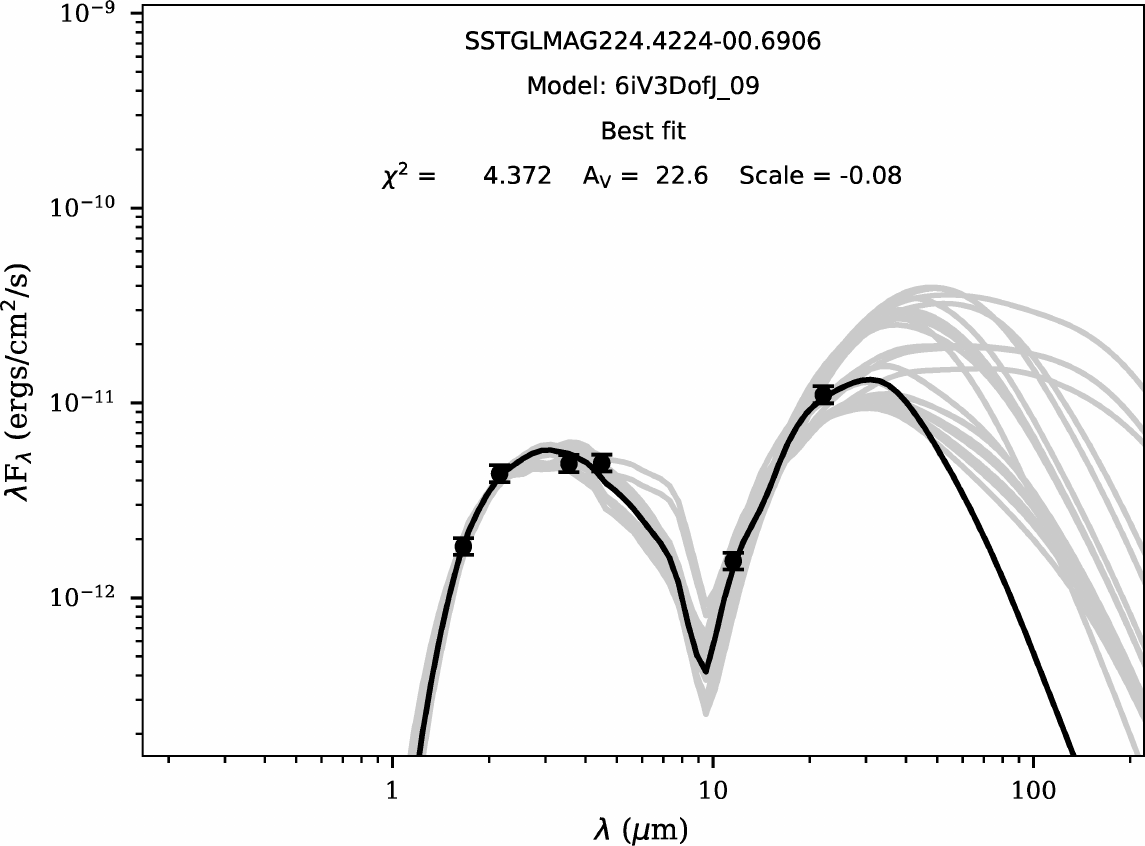}
\hfill
\includegraphics[width=0.32\textwidth]{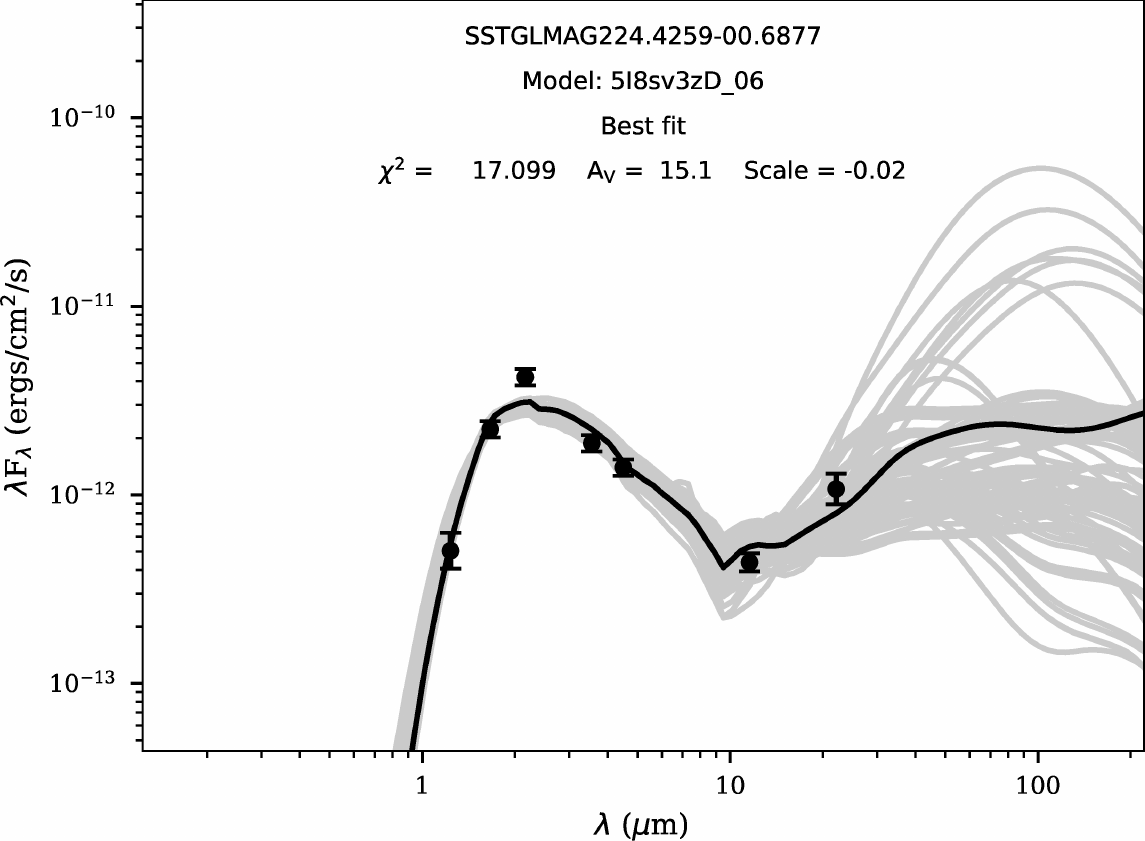} \par
\vspace{2mm}
\includegraphics[width=0.32\textwidth]{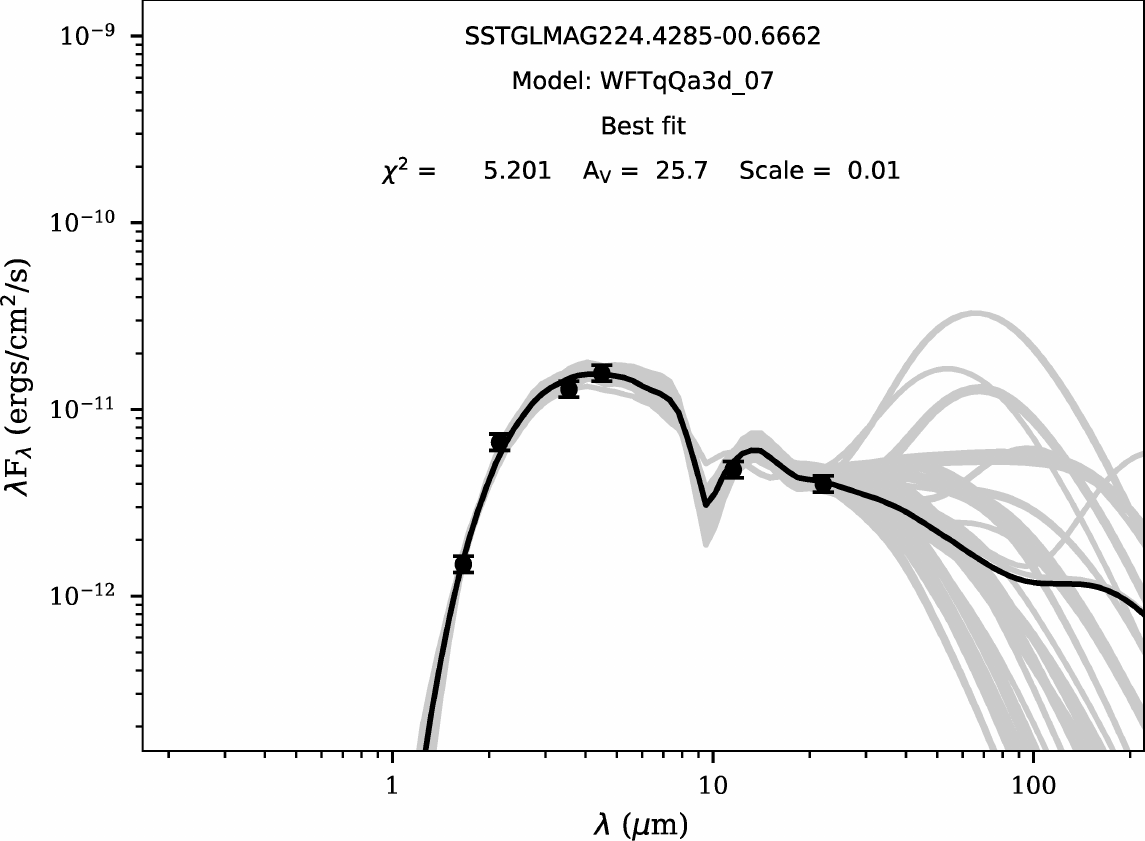}
\hfill
\includegraphics[width=0.32\textwidth]{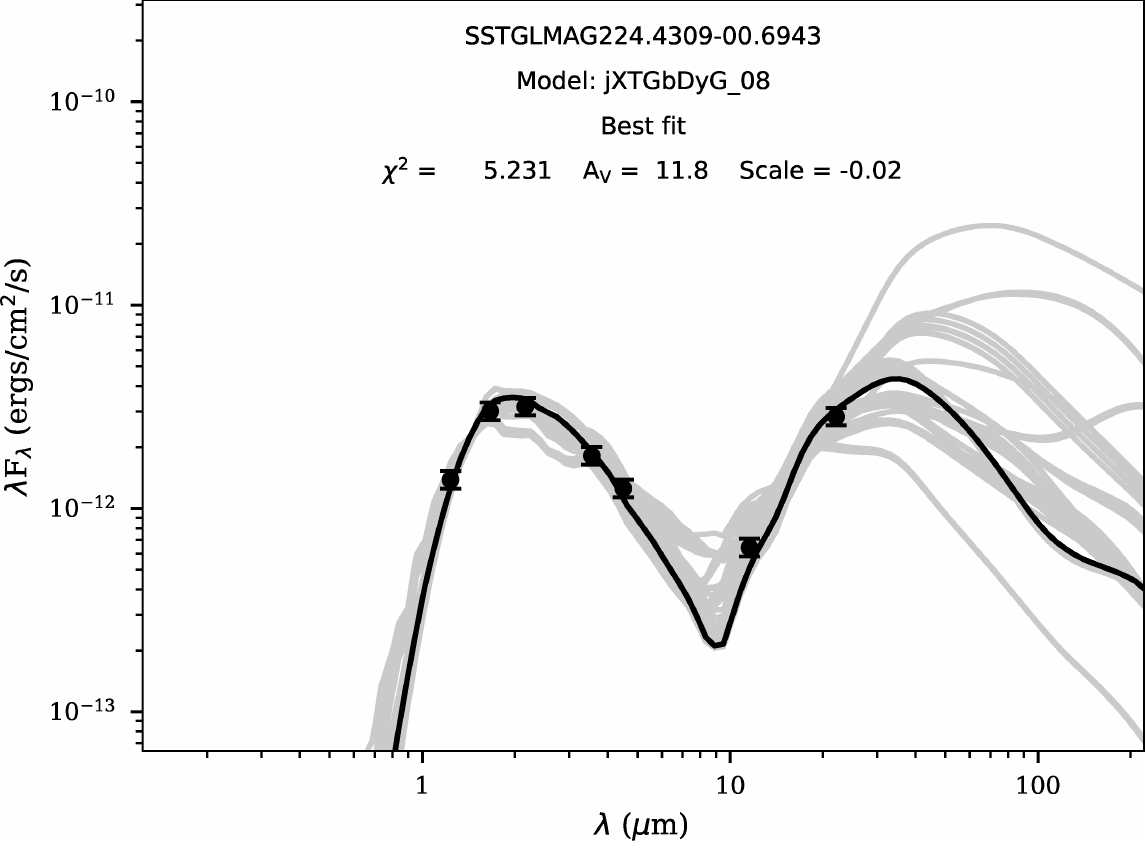}
\hfill
\includegraphics[width=0.32\textwidth]{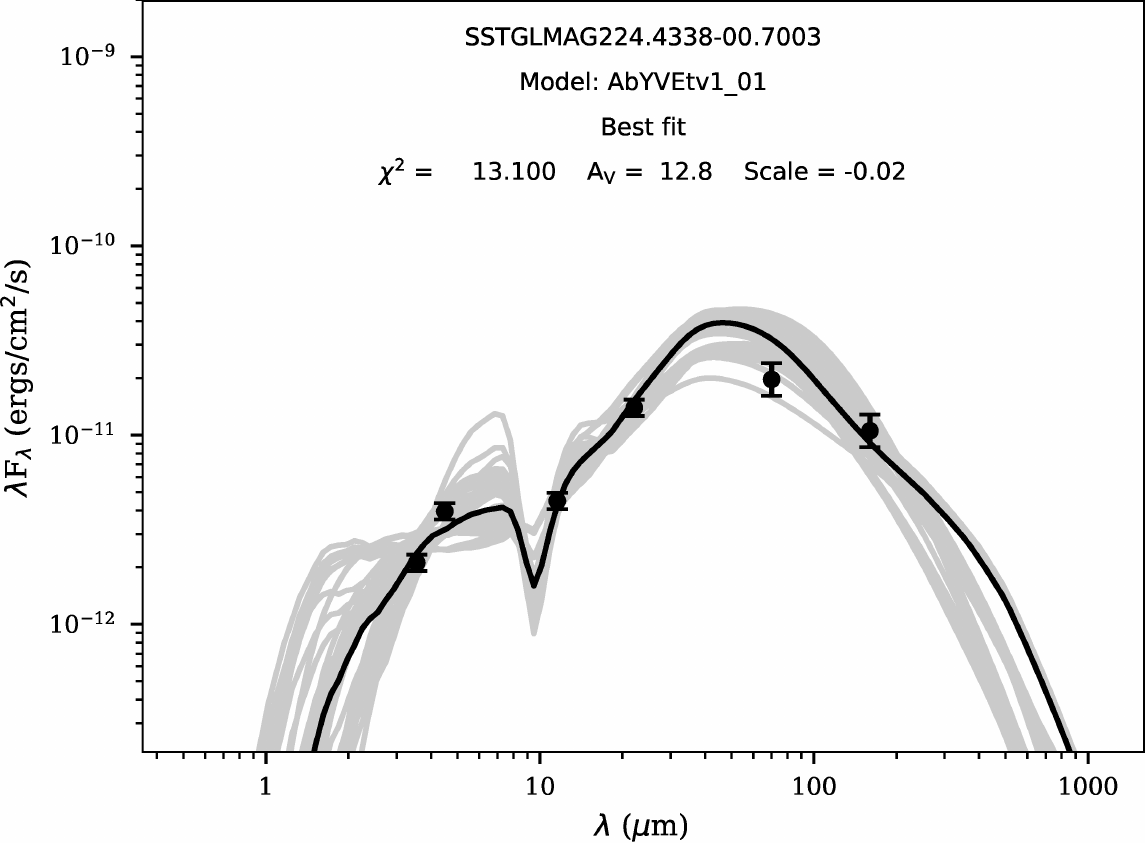}
\caption{Same as Fig.~\ref{f:SEDs1}  \label{f:SEDs9}}
\end{figure*}

\begin{figure*}
\includegraphics[width=0.32\textwidth]{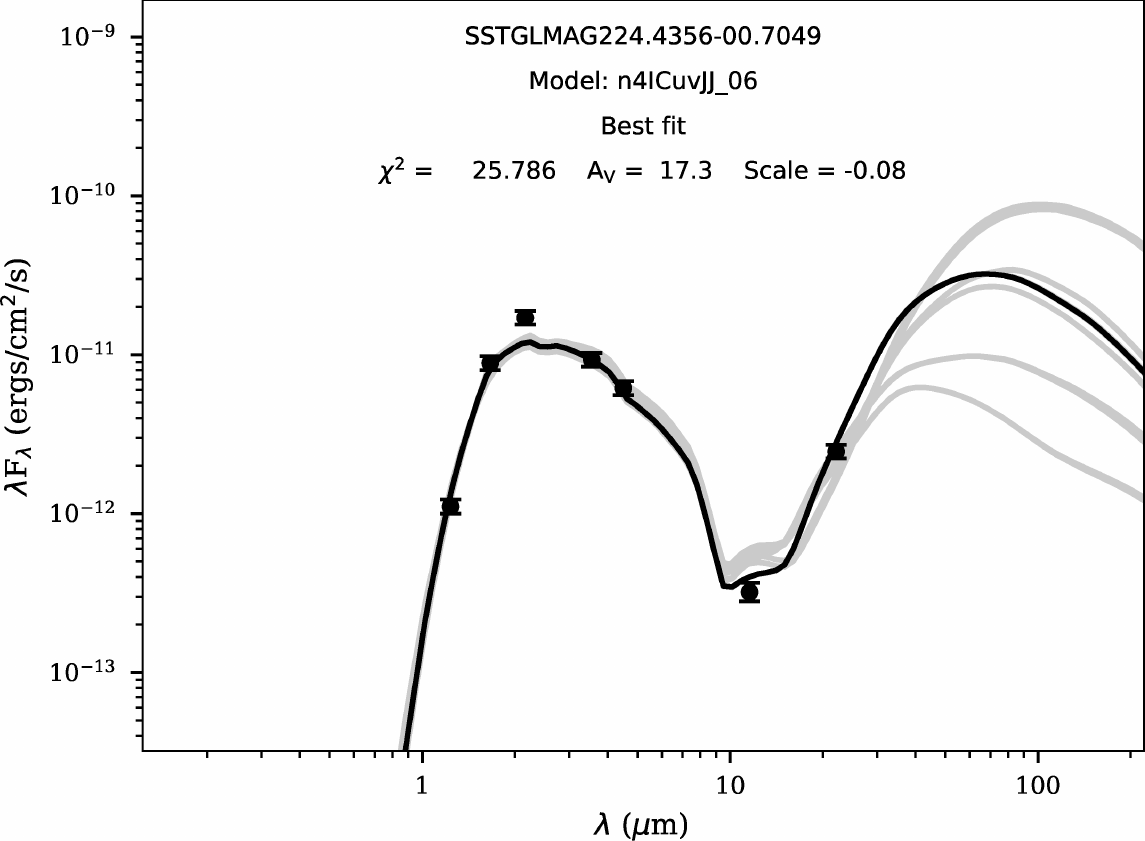}
\hfill
\includegraphics[width=0.32\textwidth]{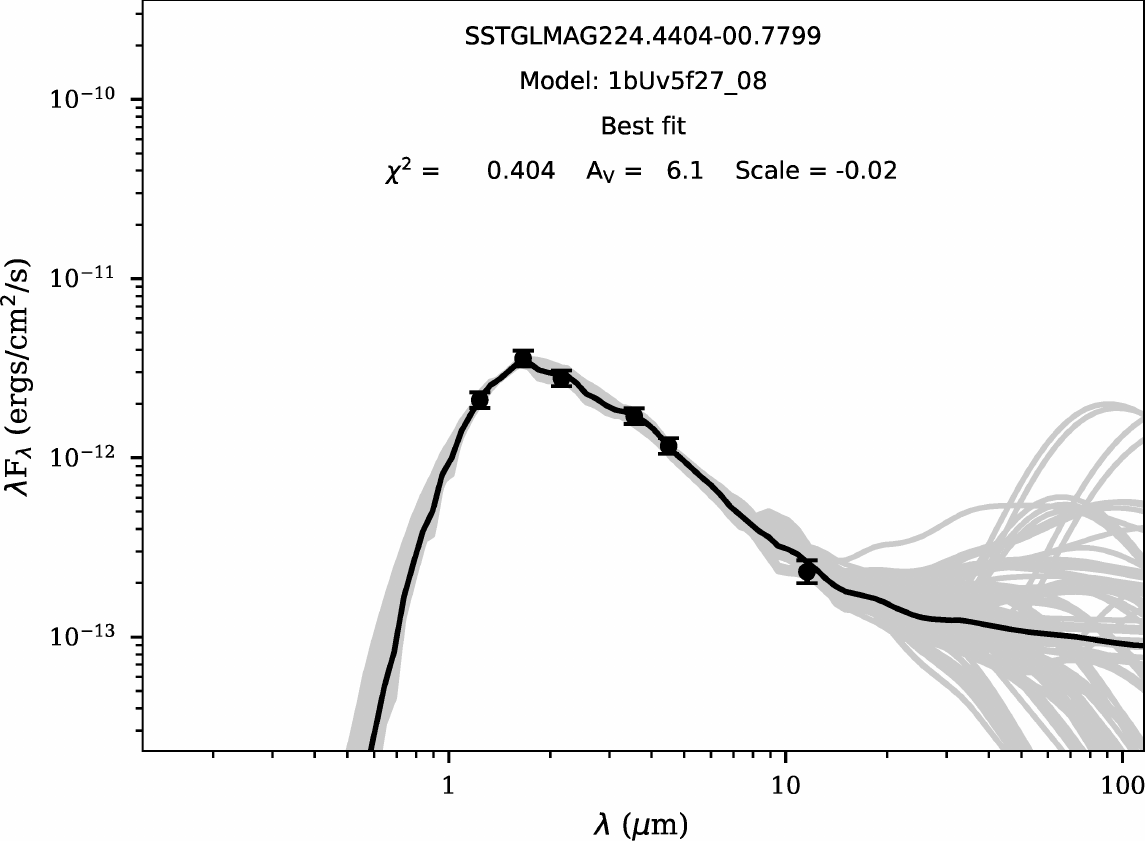}
\hfill
\includegraphics[width=0.32\textwidth]{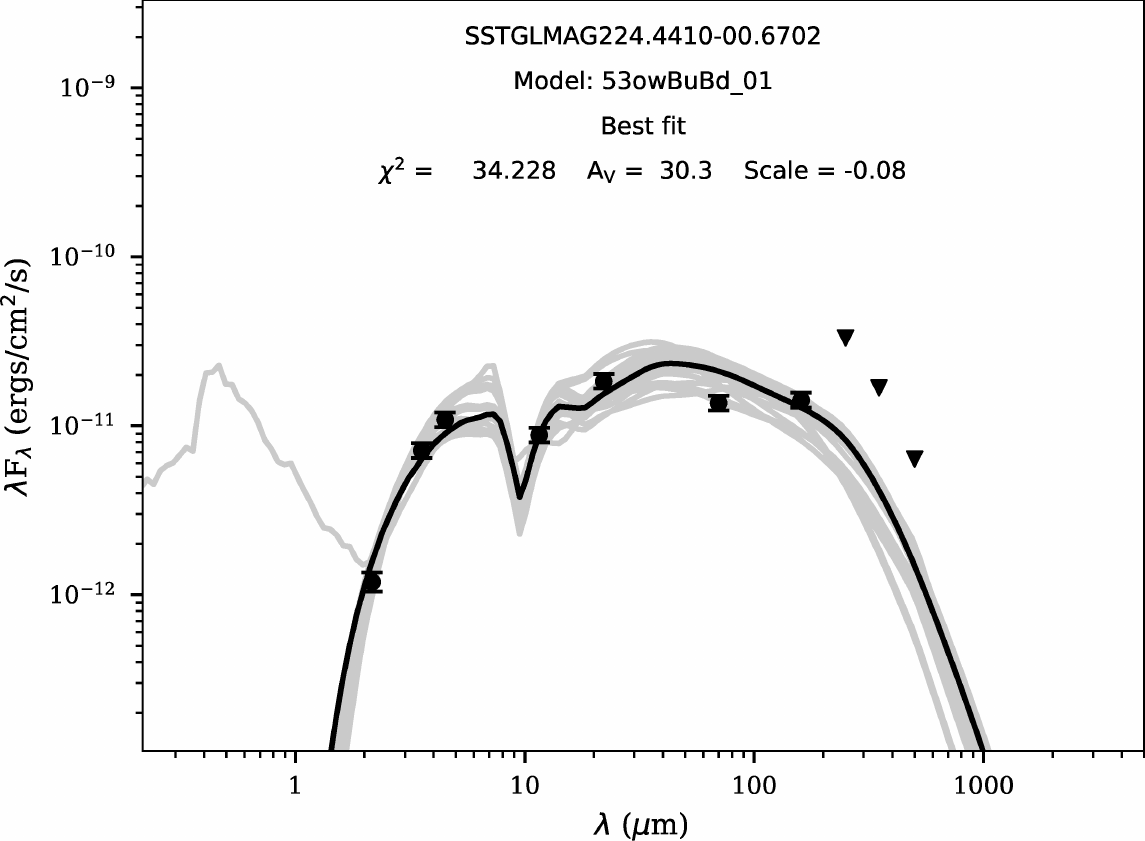} \par
\vspace{2mm}
\includegraphics[width=0.32\textwidth]{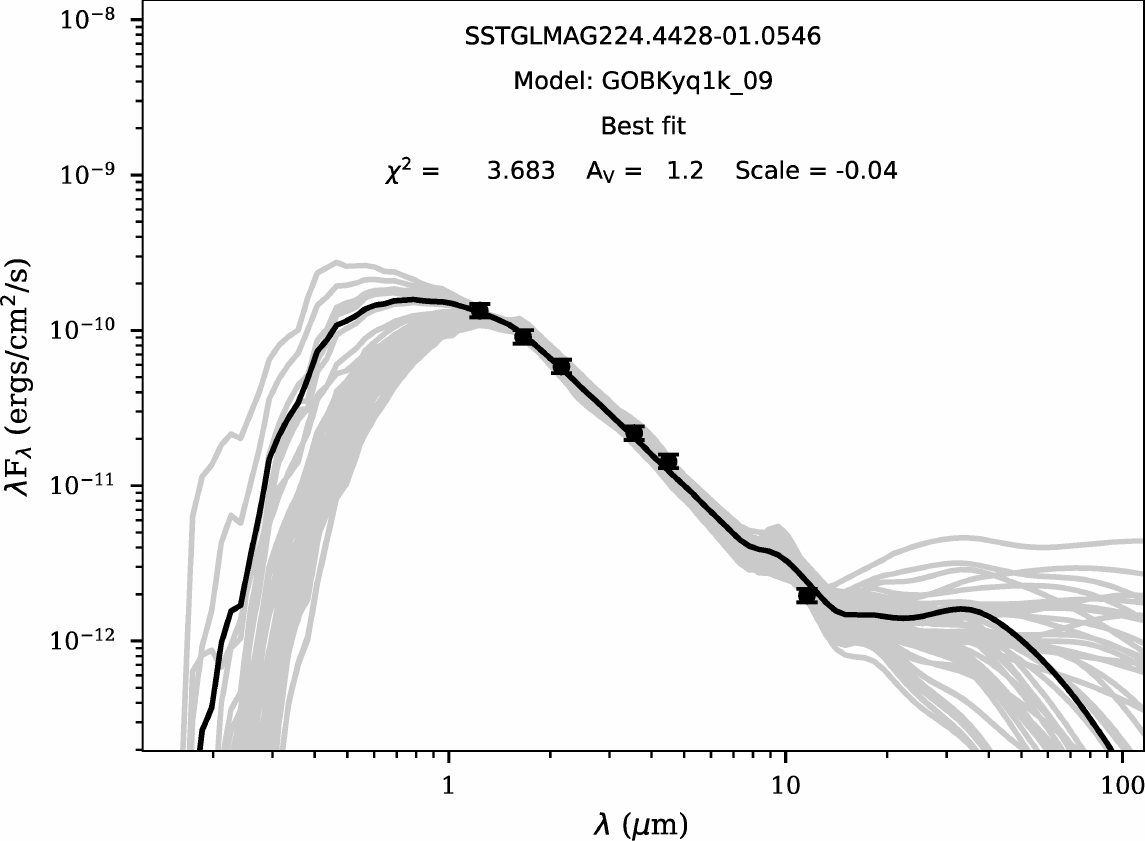}
\hfill
\includegraphics[width=0.32\textwidth]{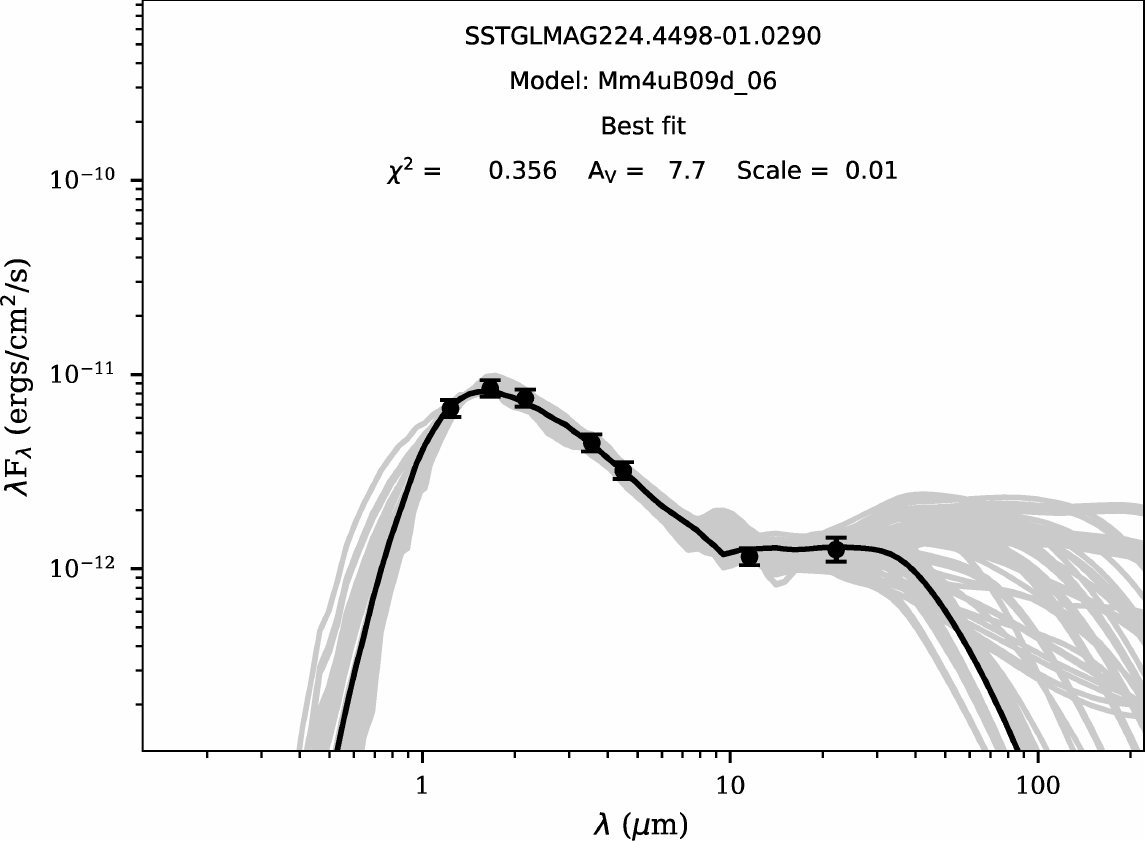}
\hfill
\includegraphics[width=0.32\textwidth]{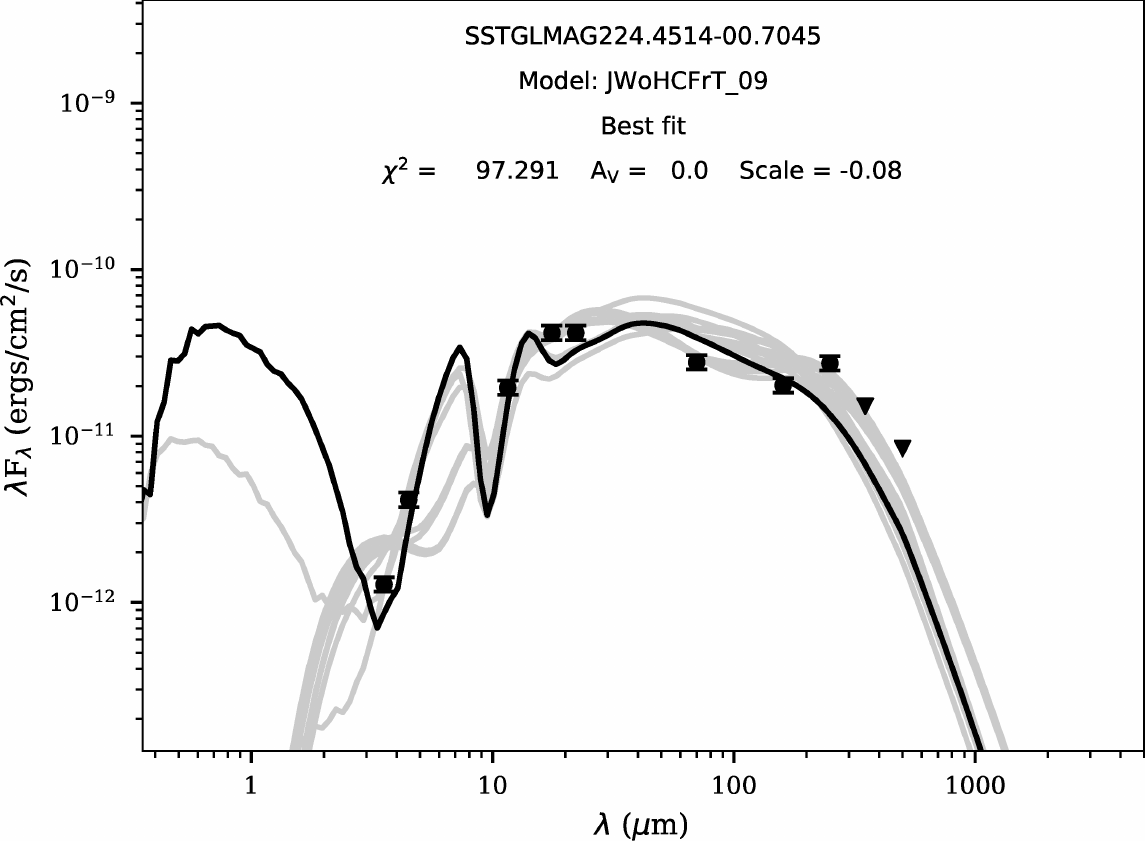} \par
\vspace{2mm}
\includegraphics[width=0.32\textwidth]{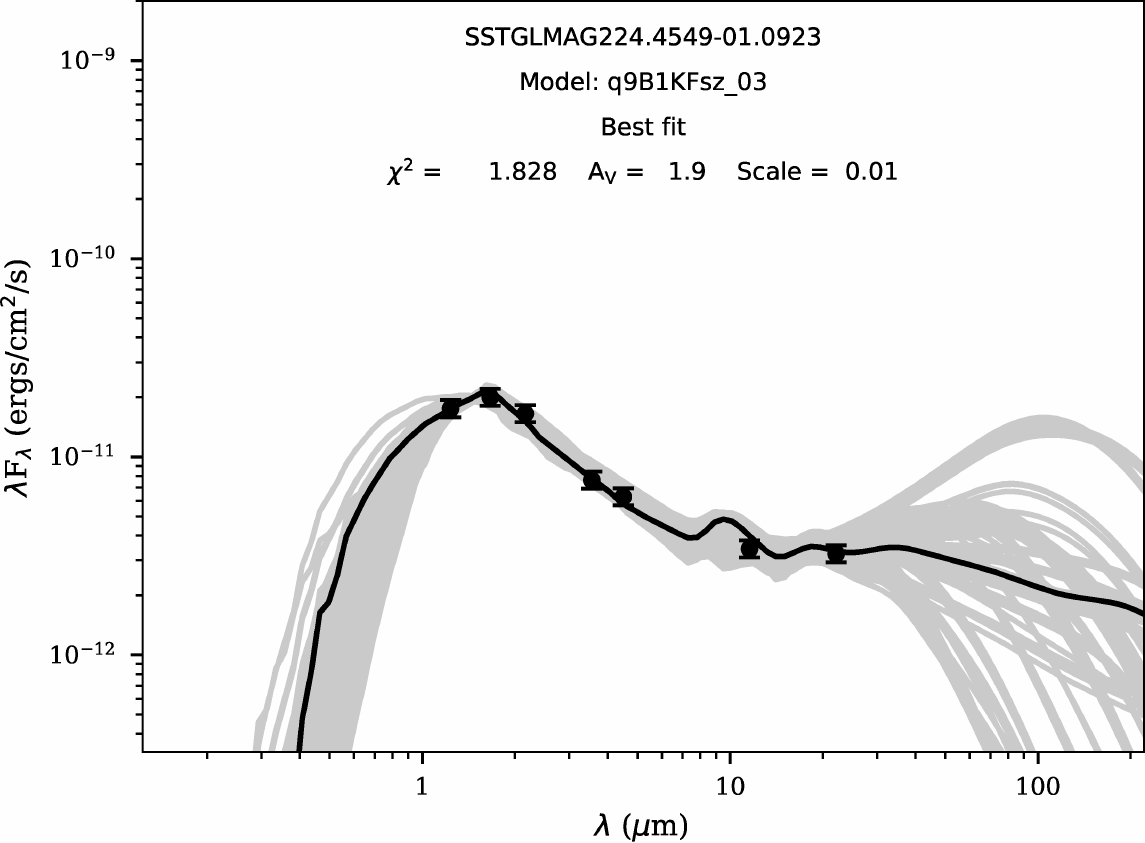}
\hfill
\includegraphics[width=0.32\textwidth]{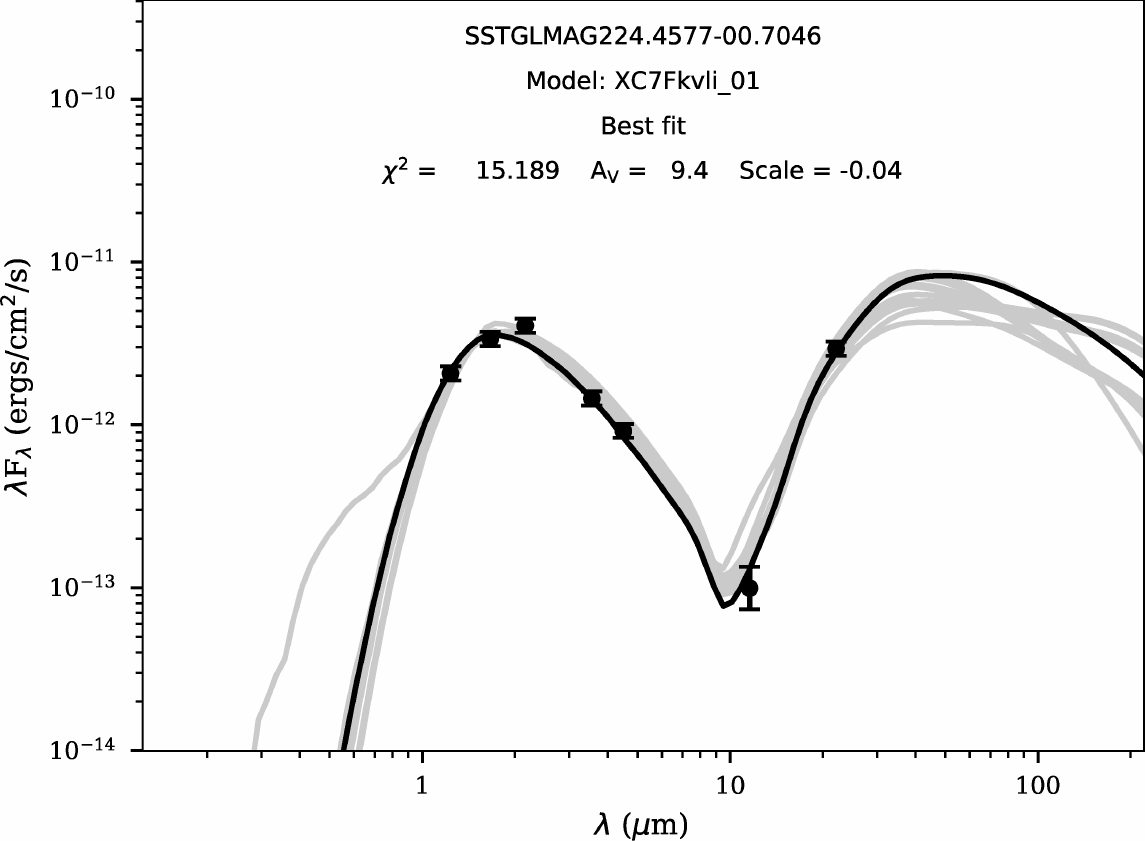}
\hfill
\includegraphics[width=0.32\textwidth]{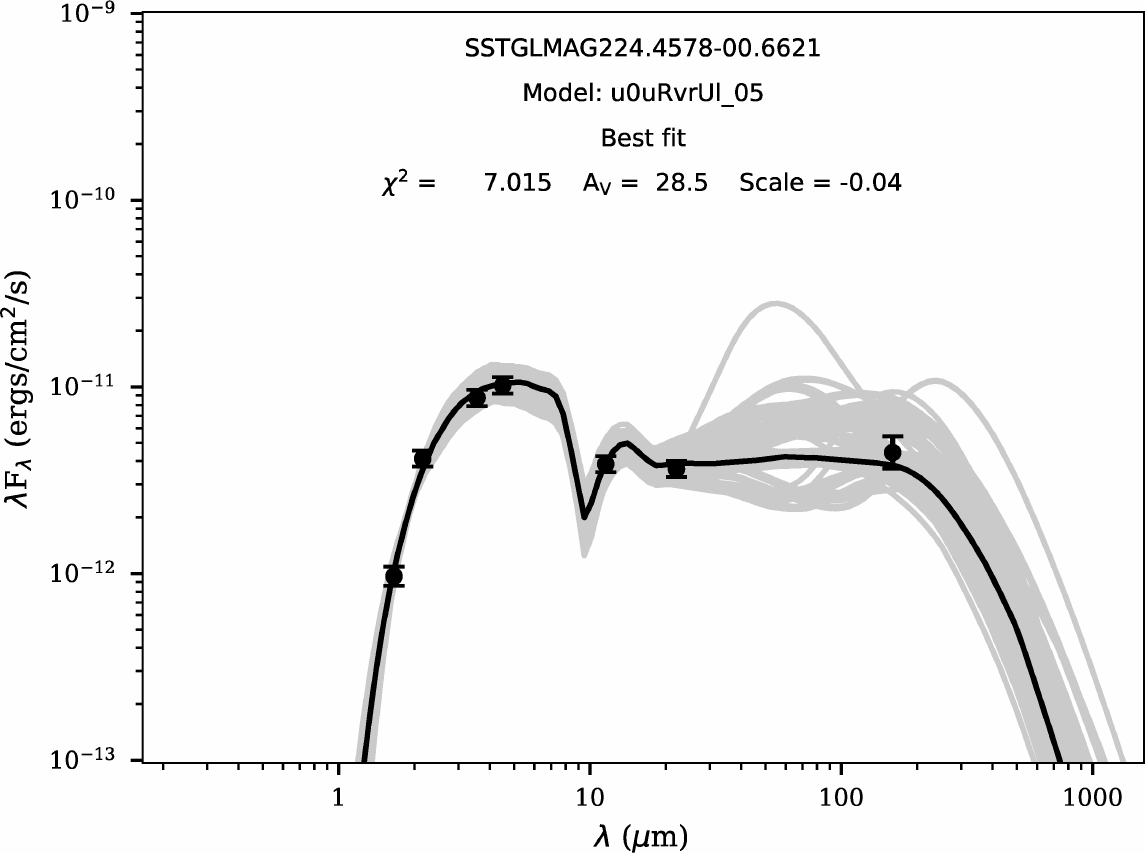} \par
\vspace{2mm}
\includegraphics[width=0.32\textwidth]{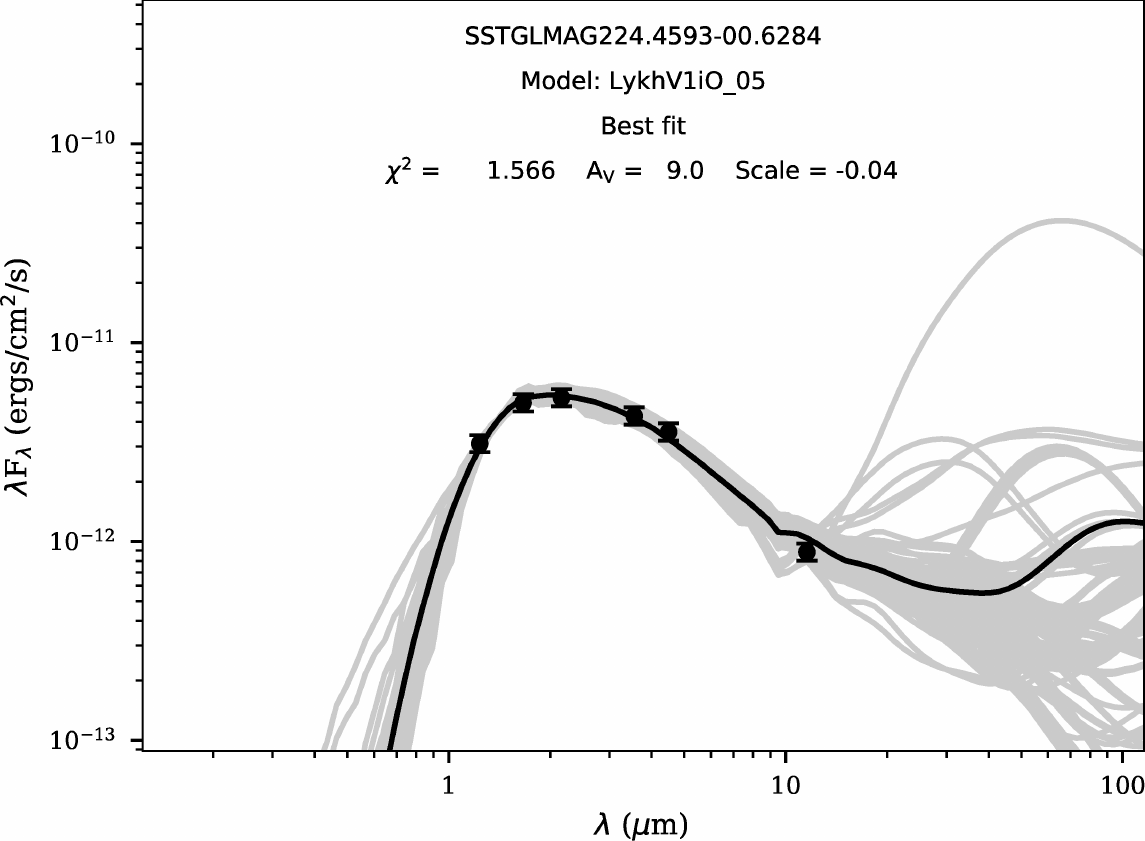}
\hfill
\includegraphics[width=0.32\textwidth]{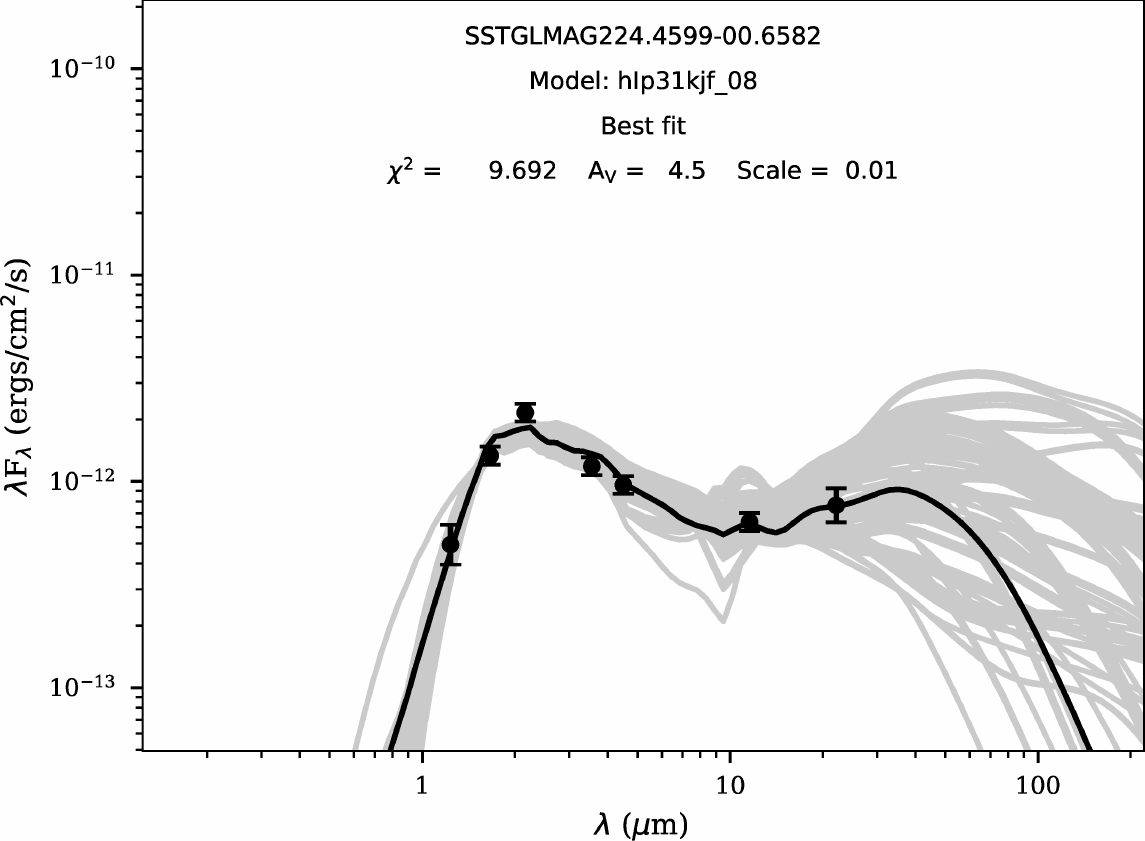}
\hfill
\includegraphics[width=0.32\textwidth]{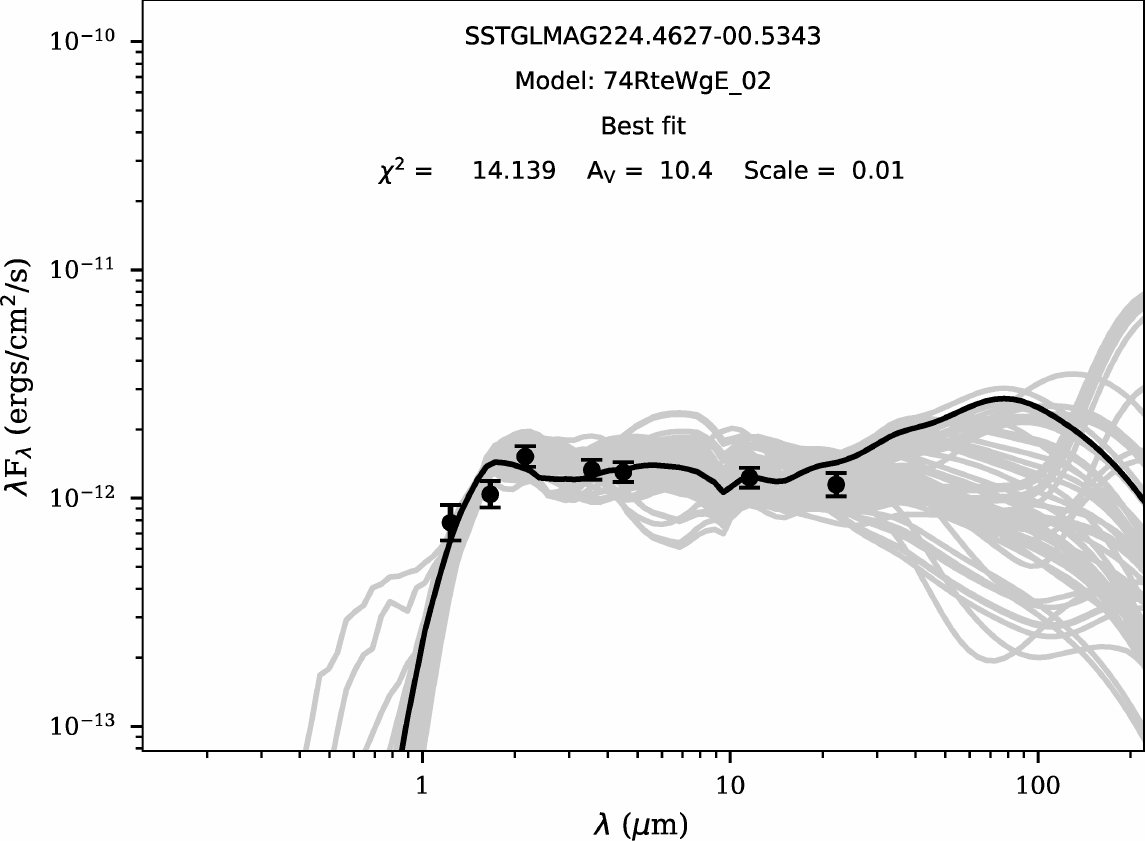} \par
\vspace{2mm}
\includegraphics[width=0.32\textwidth]{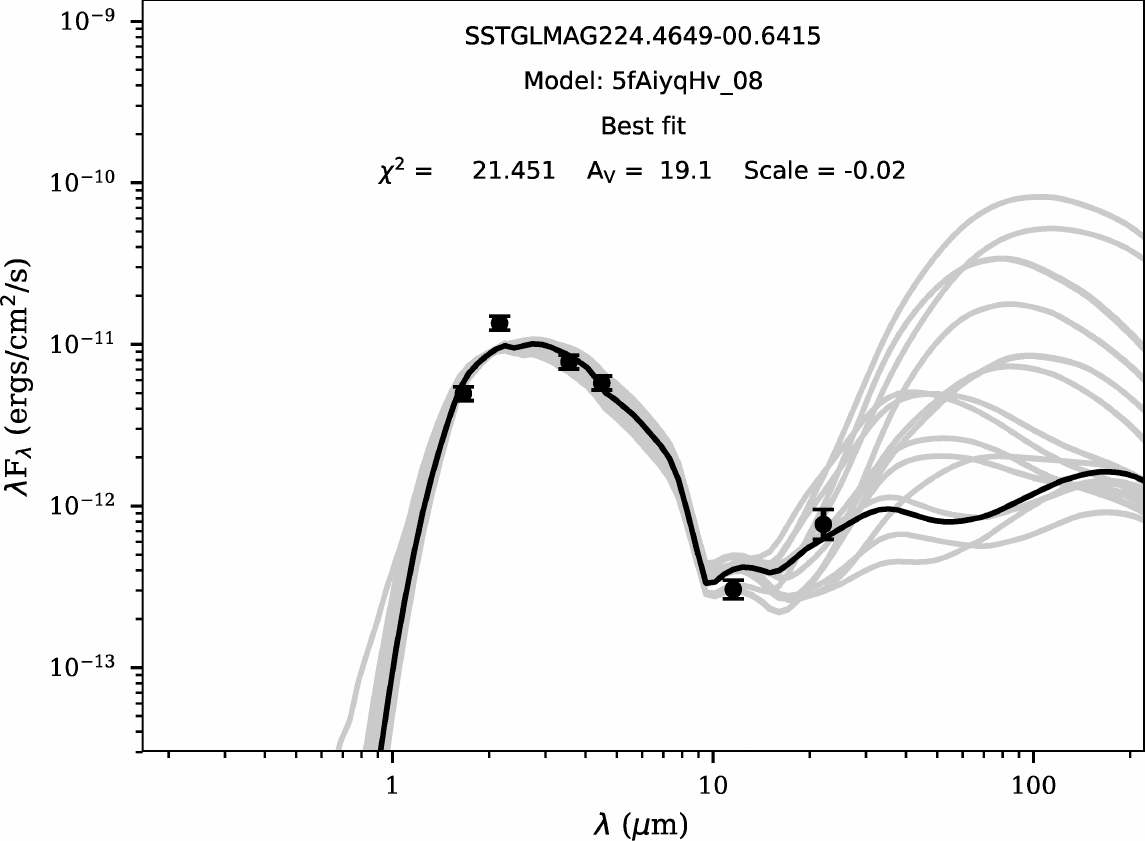}
\hfill
\includegraphics[width=0.32\textwidth]{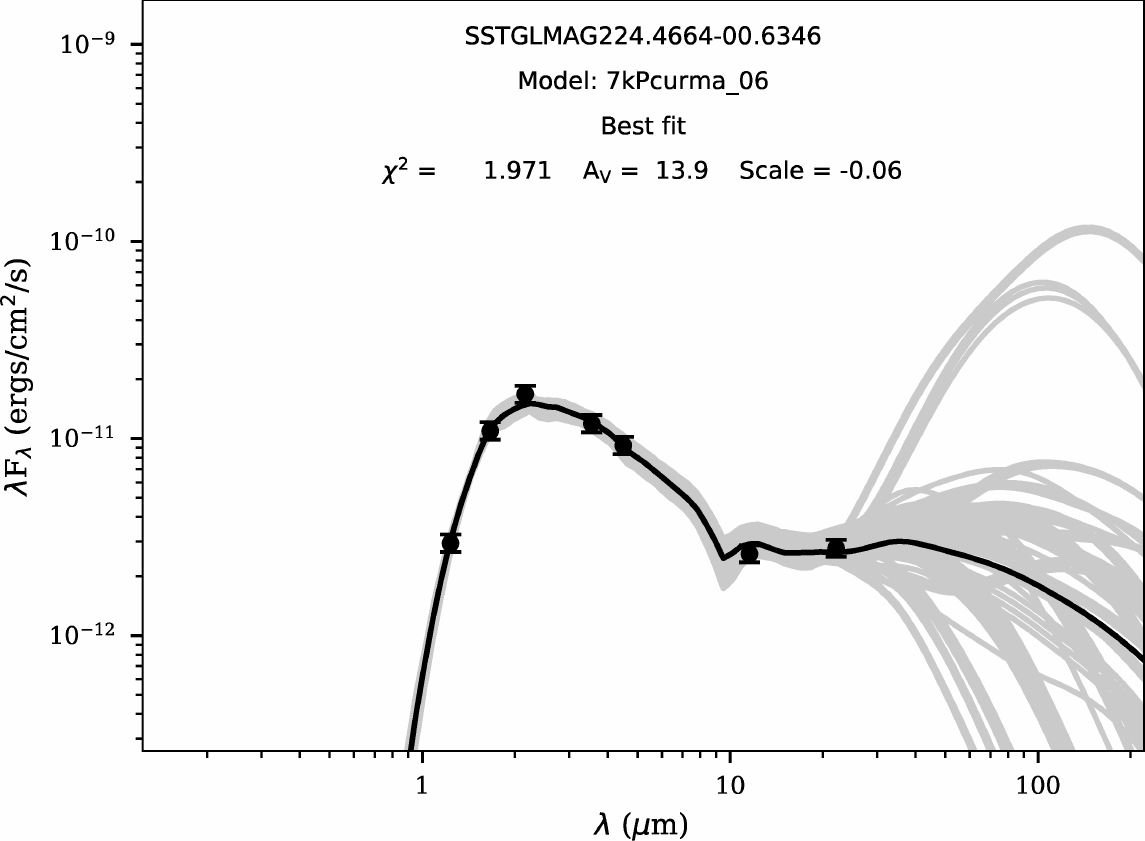}
\hfill
\includegraphics[width=0.32\textwidth]{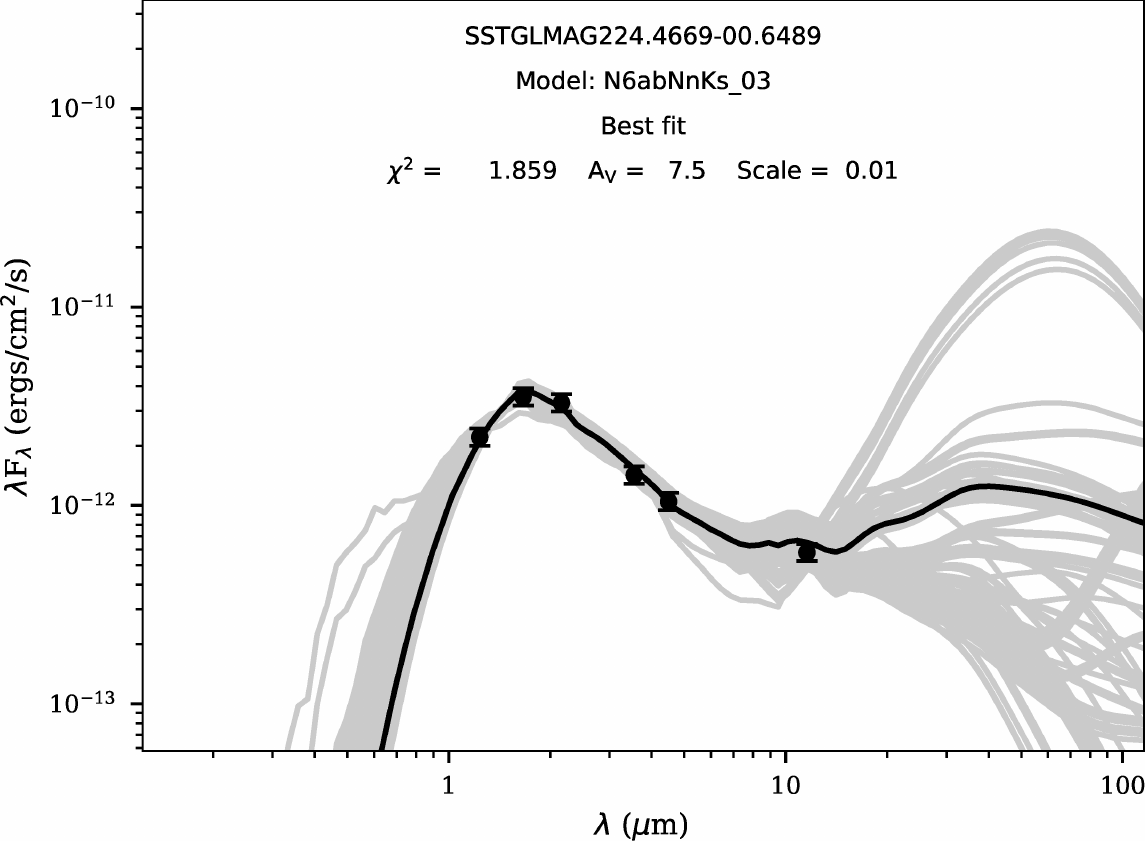}
\caption{Same as Fig.~\ref{f:SEDs1}  \label{f:SEDs10}}
\end{figure*}

\begin{figure*}
\includegraphics[width=0.32\textwidth]{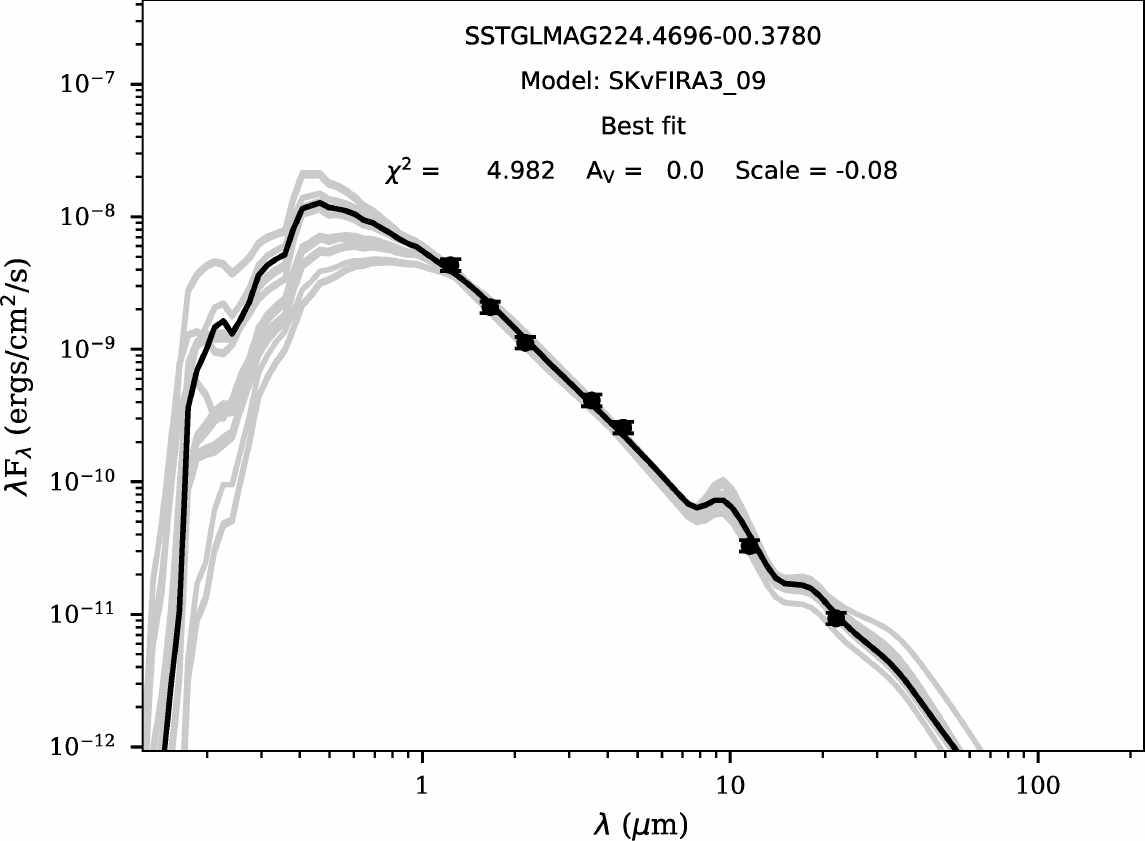}
\hfill
\includegraphics[width=0.32\textwidth]{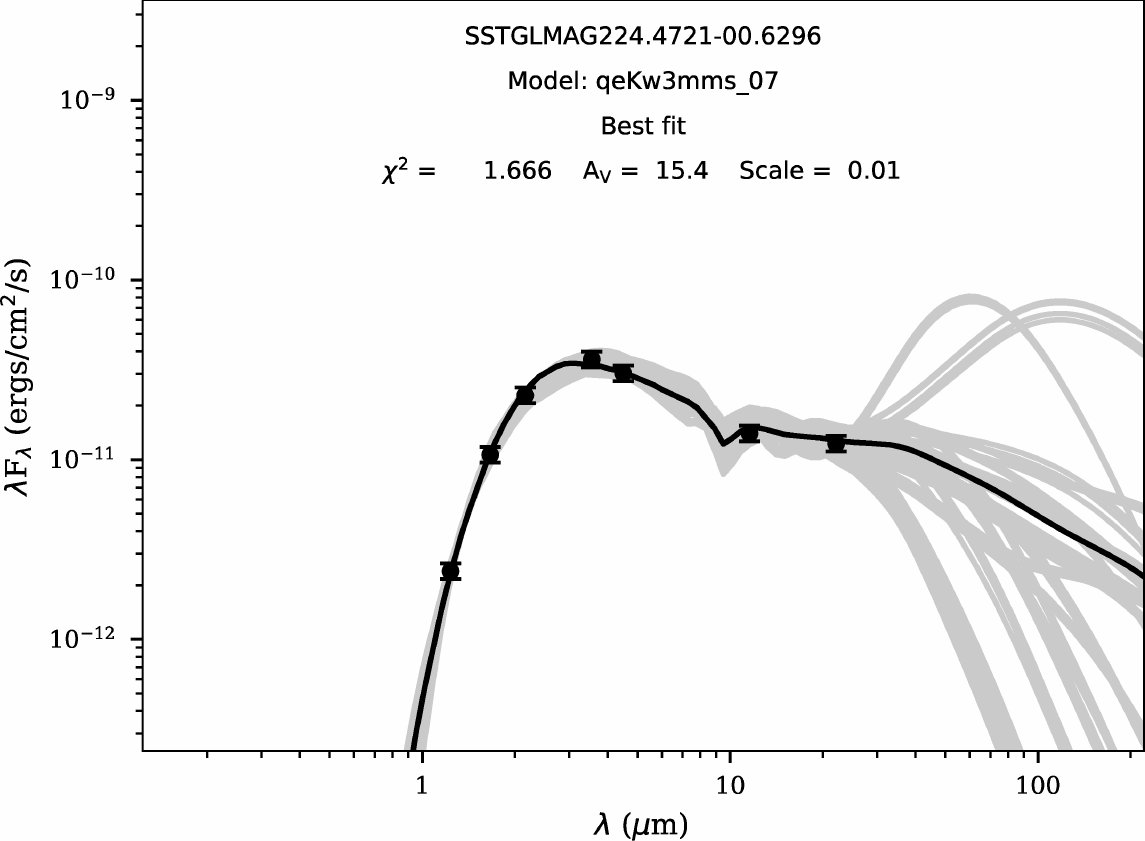}
\hfill
\includegraphics[width=0.32\textwidth]{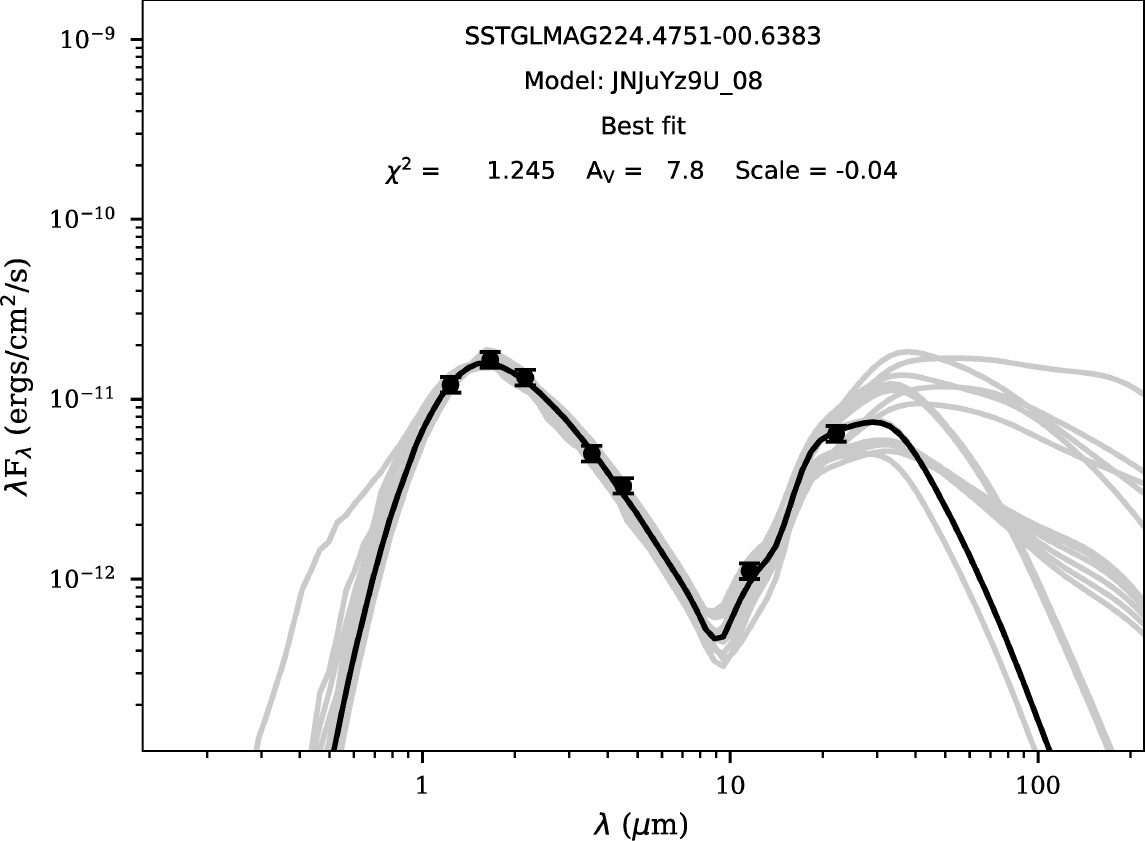} \par
\vspace{2mm}
\includegraphics[width=0.32\textwidth]{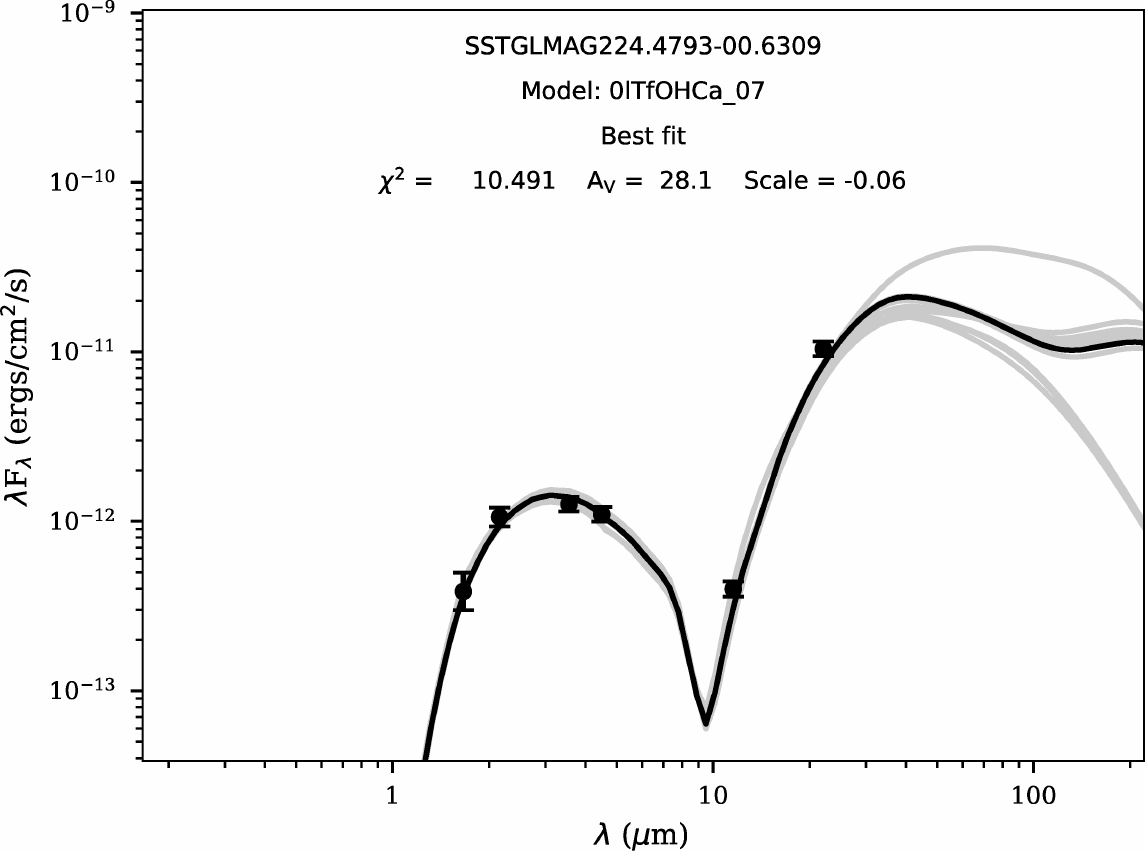}
\hfill
\includegraphics[width=0.32\textwidth]{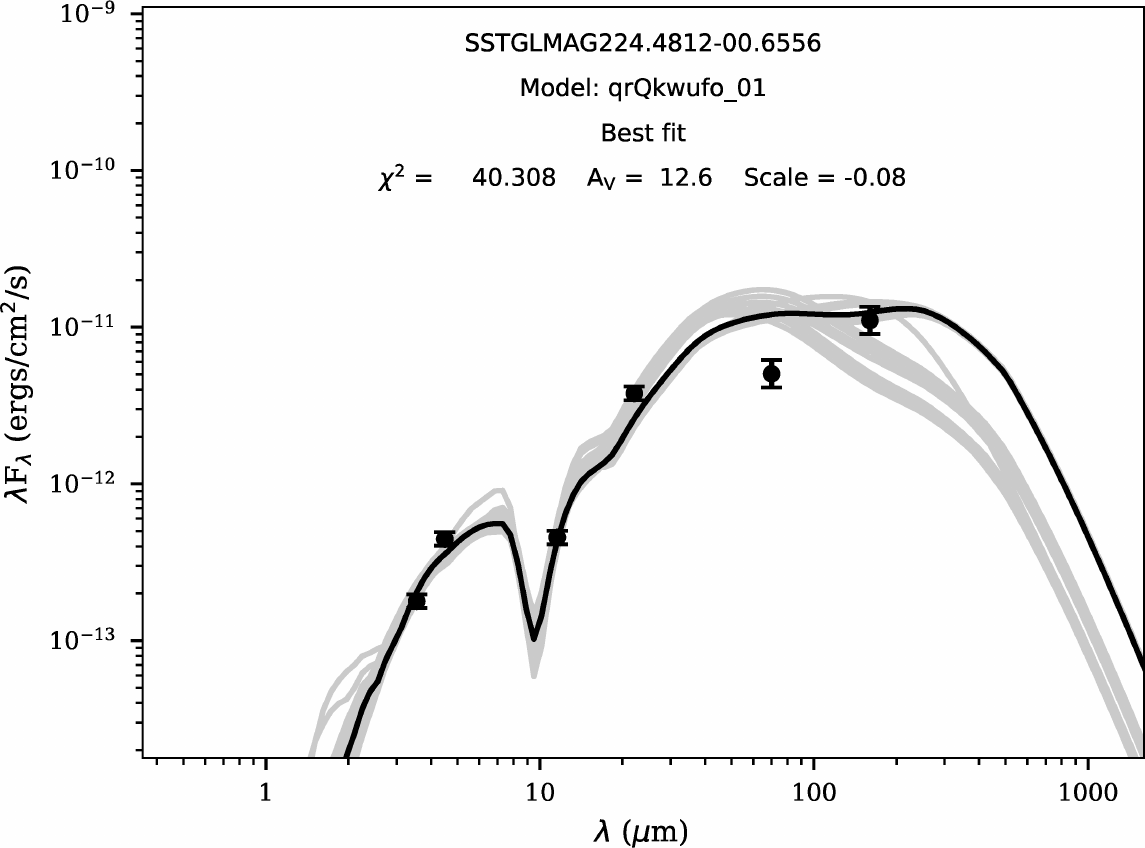}
\hfill
\includegraphics[width=0.32\textwidth]{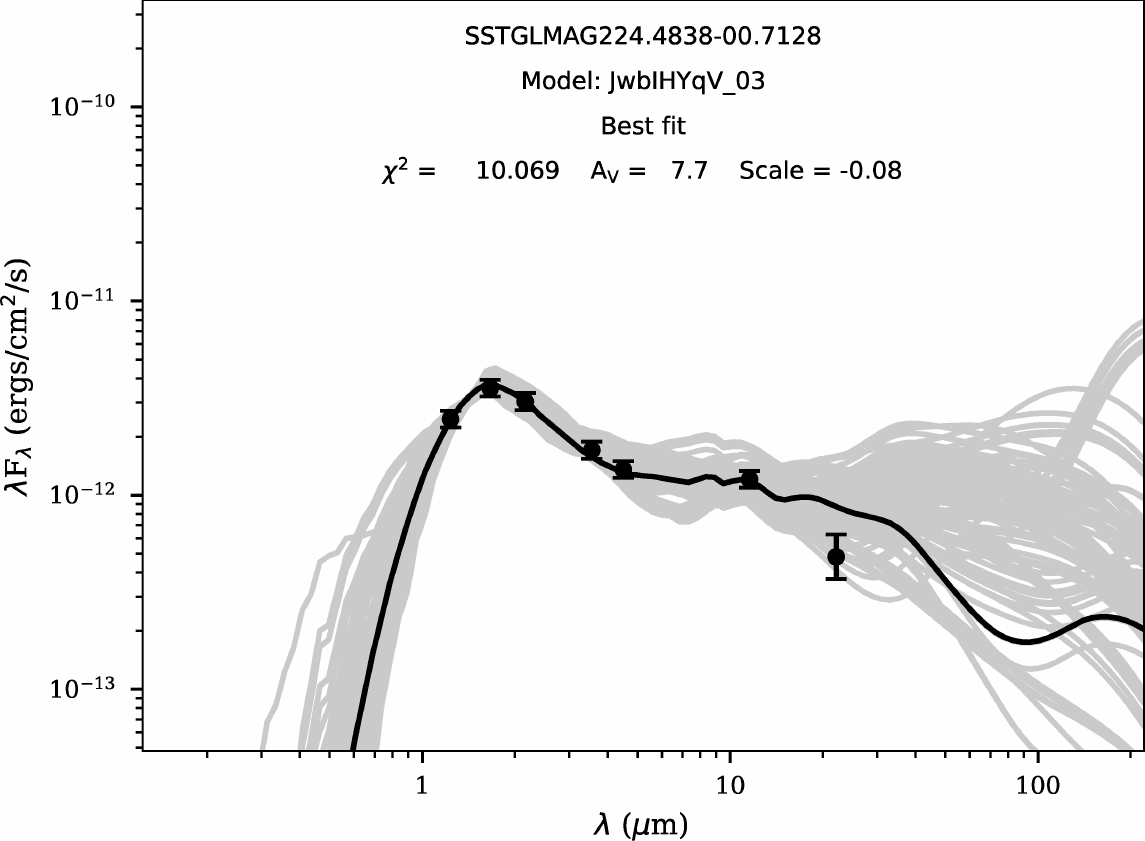} \par
\vspace{2mm}
\includegraphics[width=0.32\textwidth]{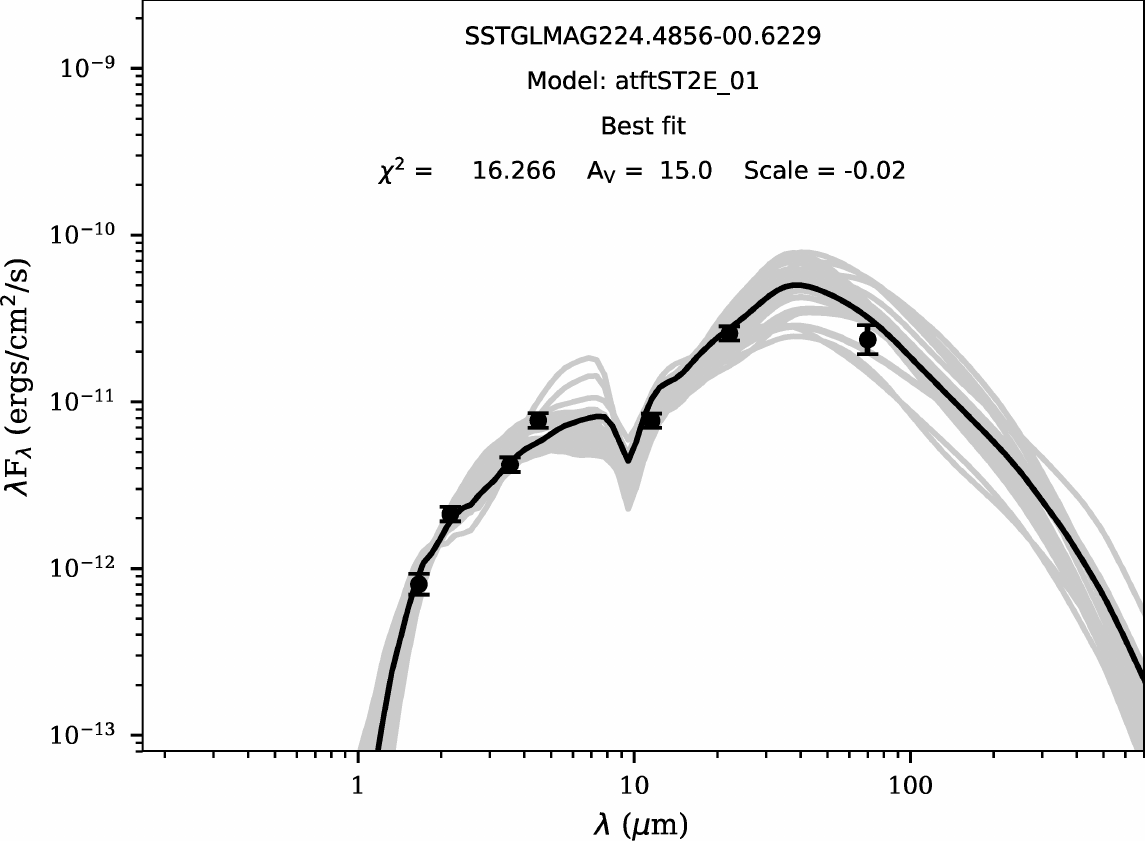}
\hfill
\includegraphics[width=0.32\textwidth]{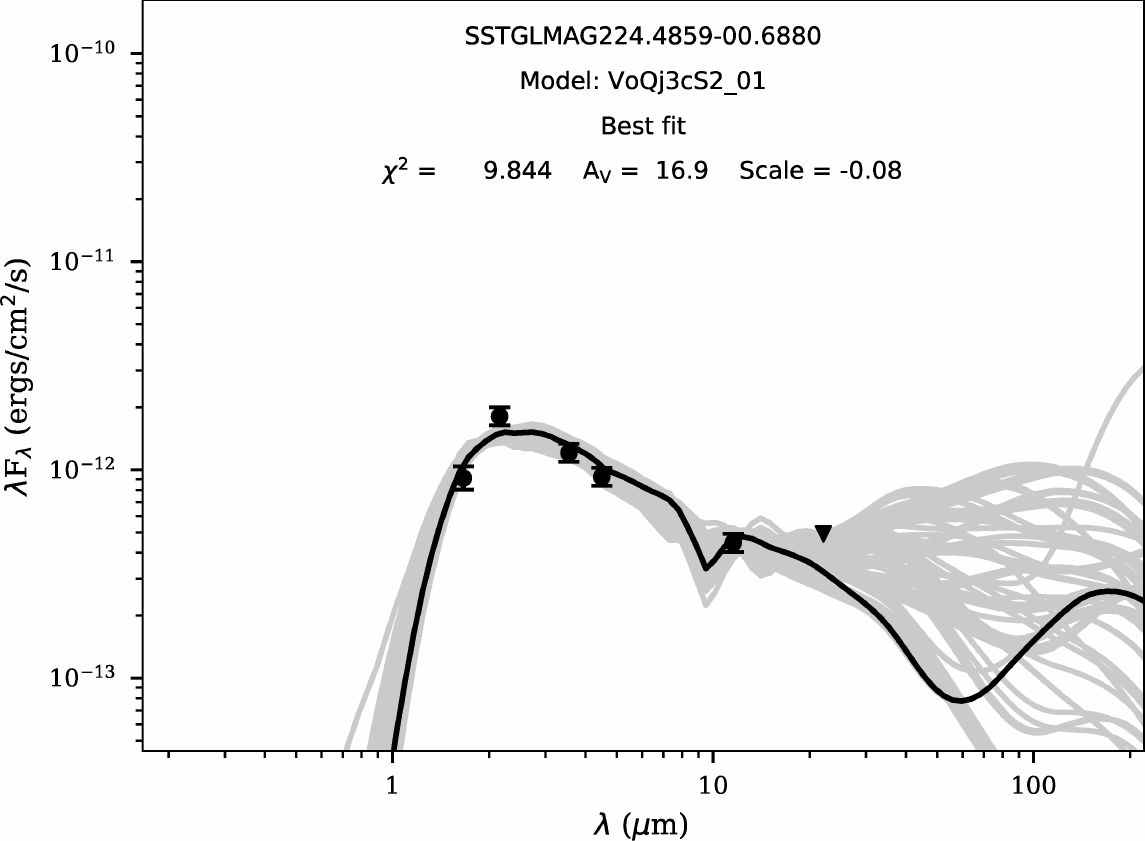}
\hfill
\includegraphics[width=0.32\textwidth]{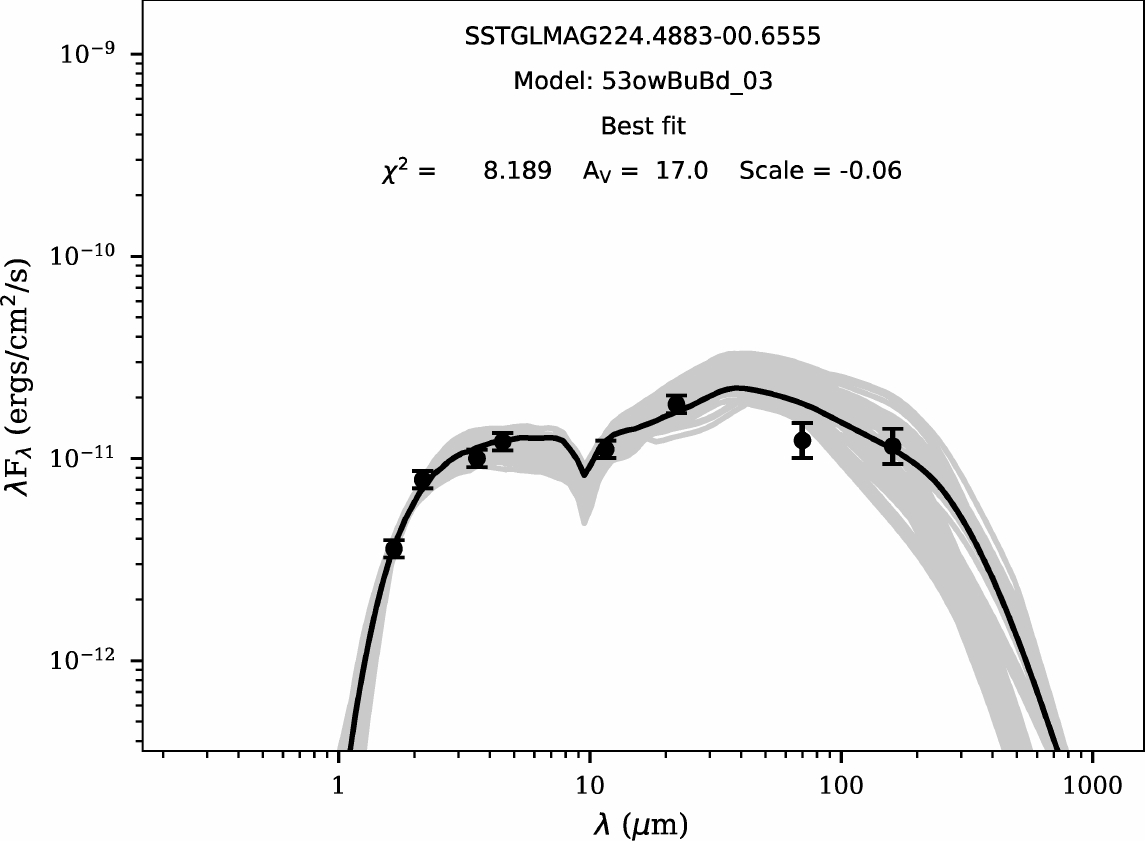} \par
\vspace{2mm}
\includegraphics[width=0.32\textwidth]{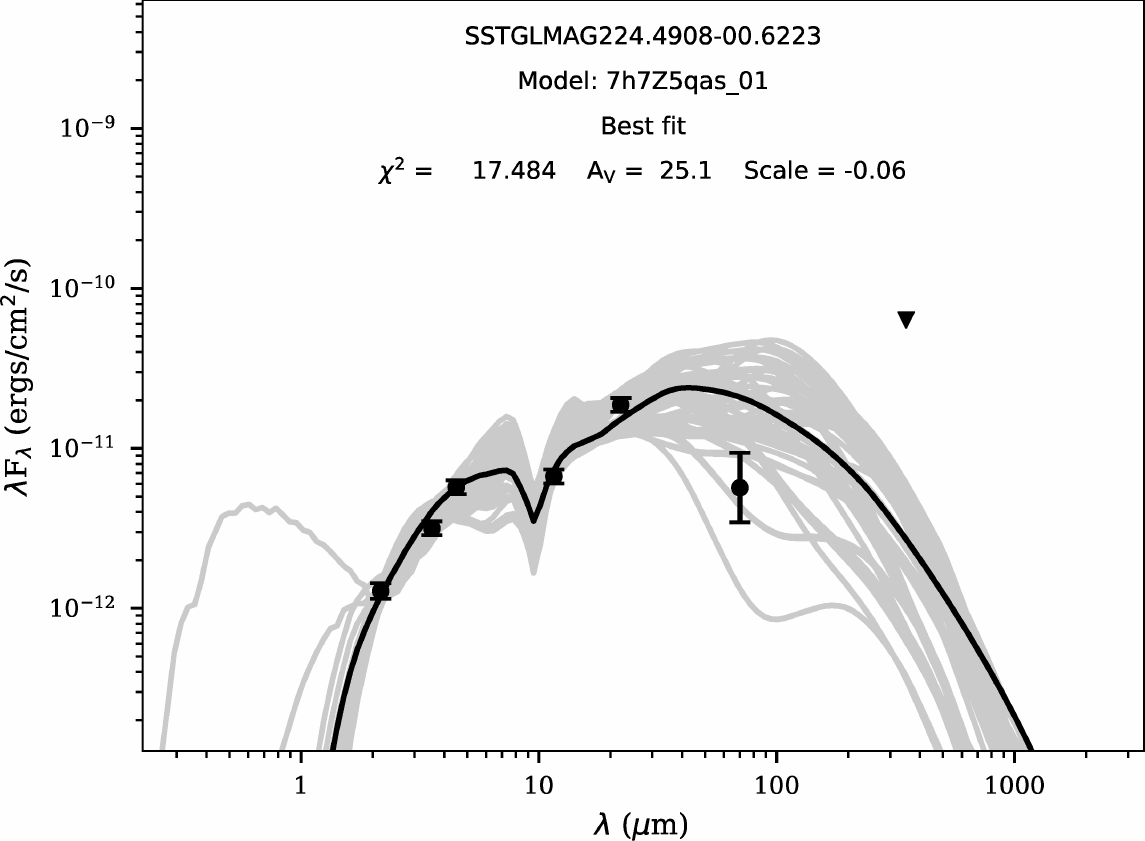}
\hfill
\includegraphics[width=0.32\textwidth]{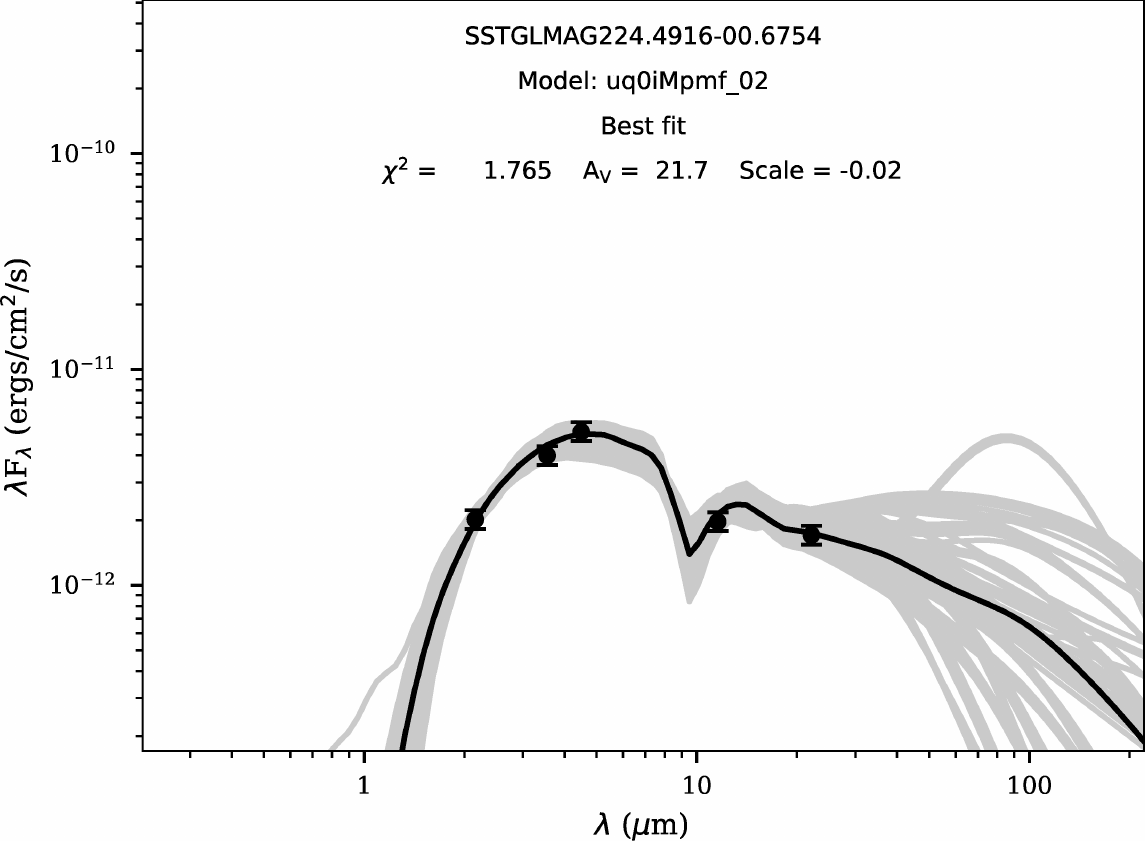}
\hfill
\includegraphics[width=0.32\textwidth]{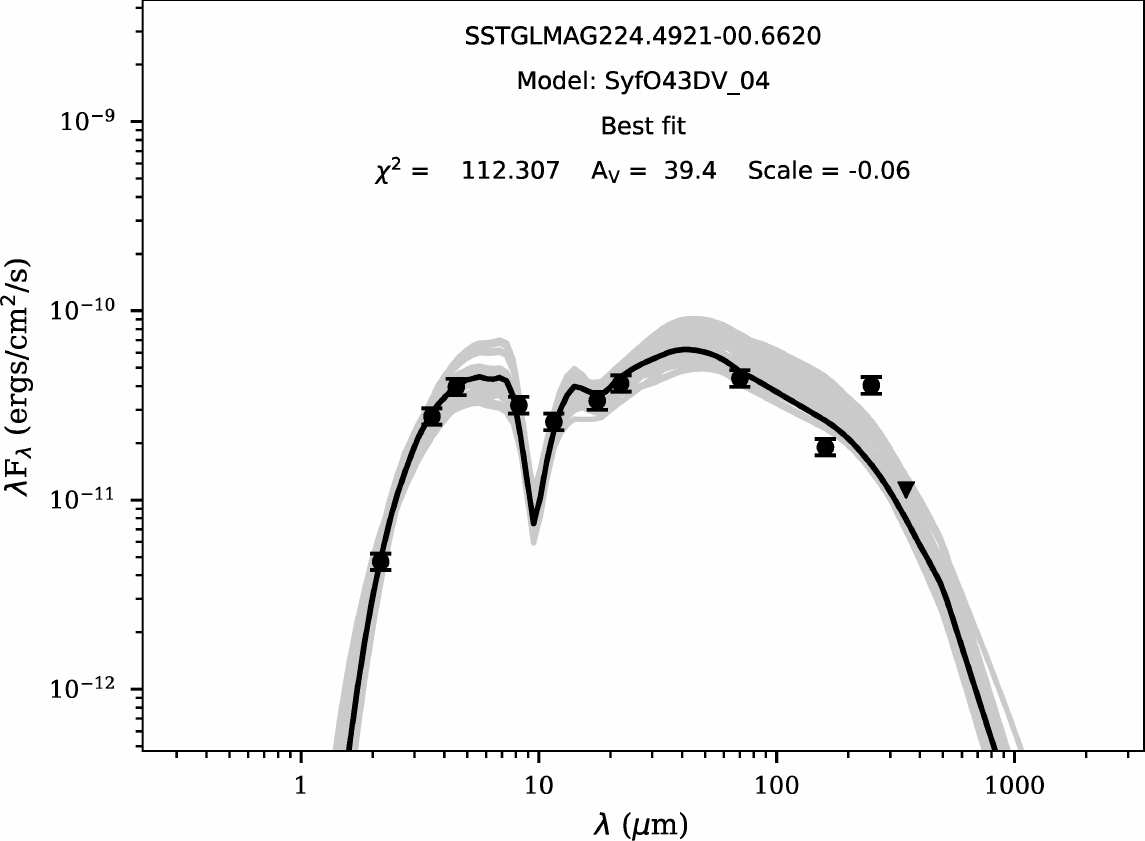} \par
\vspace{2mm}
\includegraphics[width=0.32\textwidth]{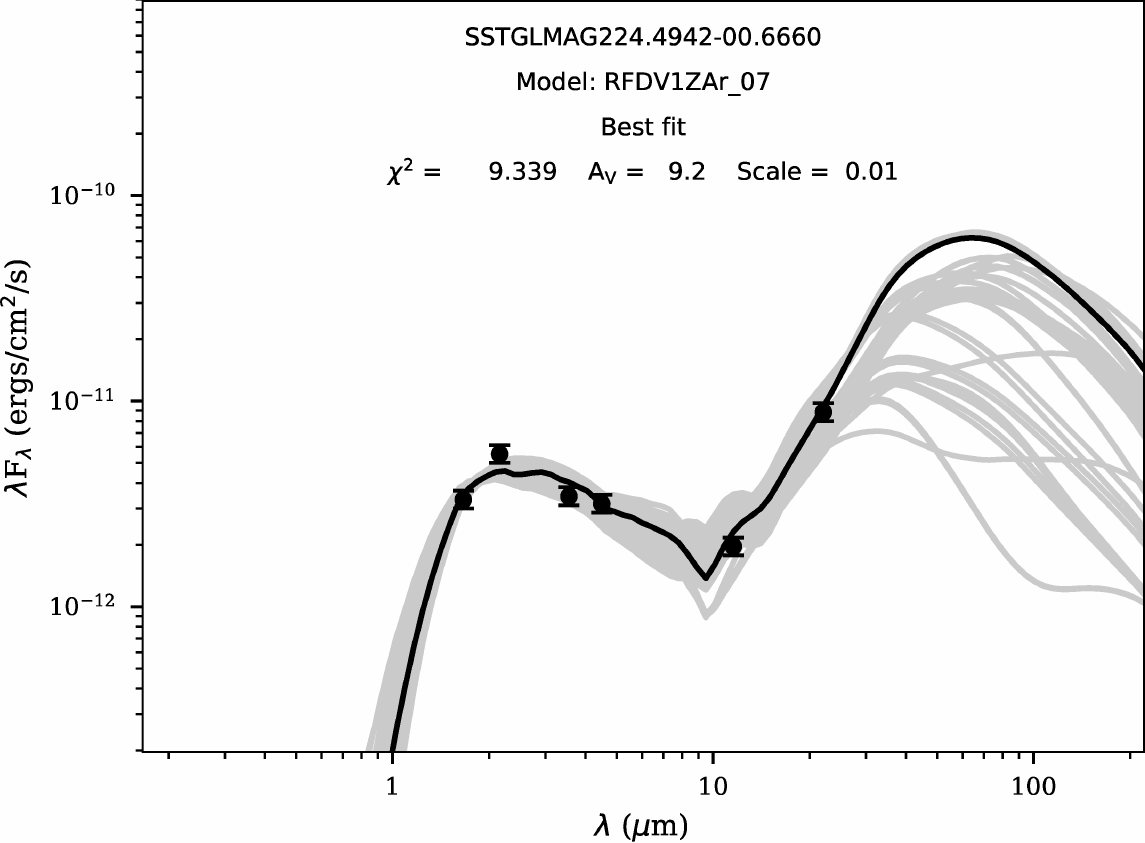}
\hfill
\includegraphics[width=0.32\textwidth]{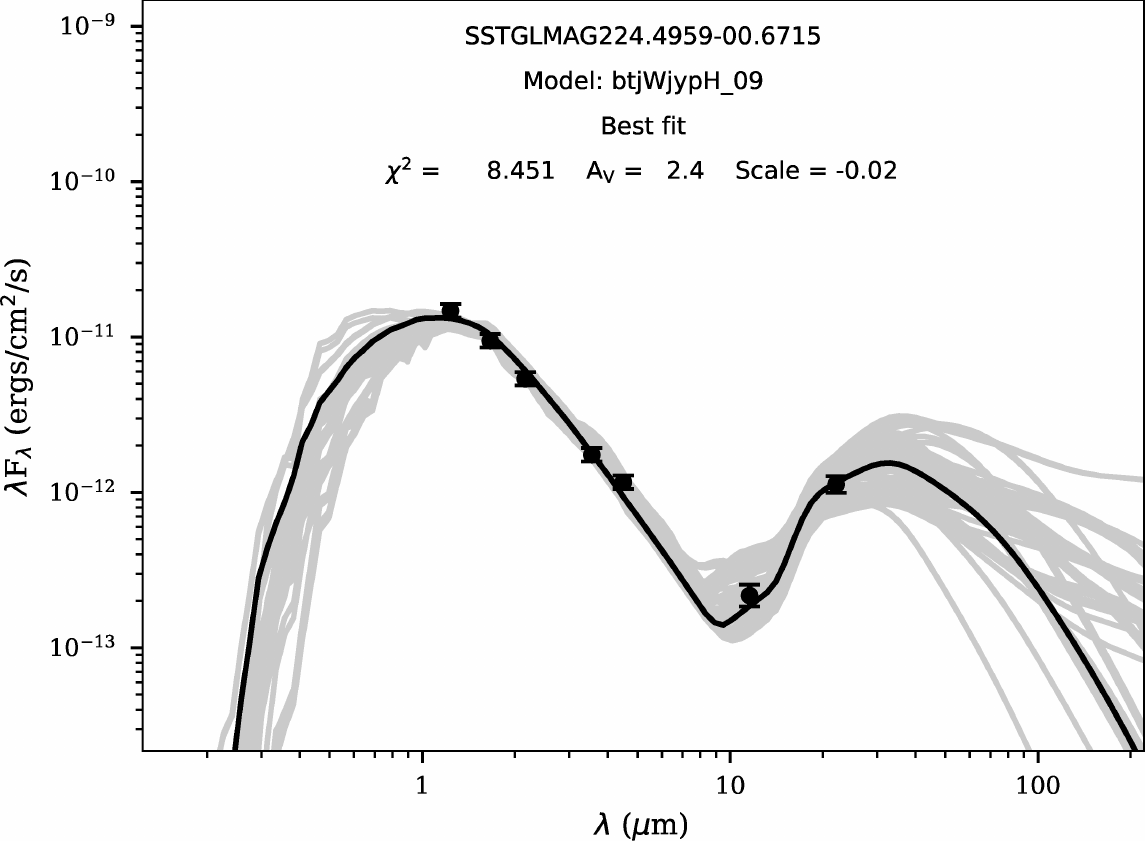}
\hfill
\includegraphics[width=0.32\textwidth]{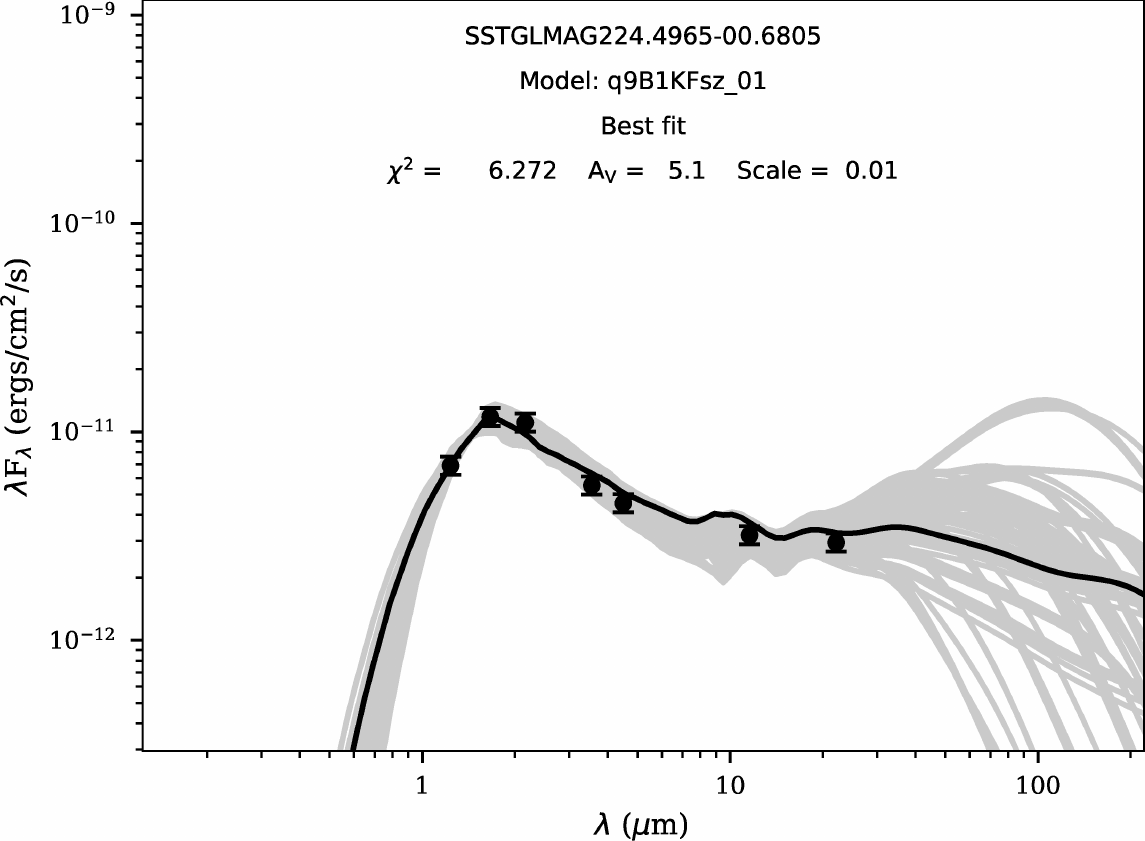}
\caption{Same as Fig.~\ref{f:SEDs1}  \label{f:SEDs11}}
\end{figure*}

\begin{figure*}
\includegraphics[width=0.32\textwidth]{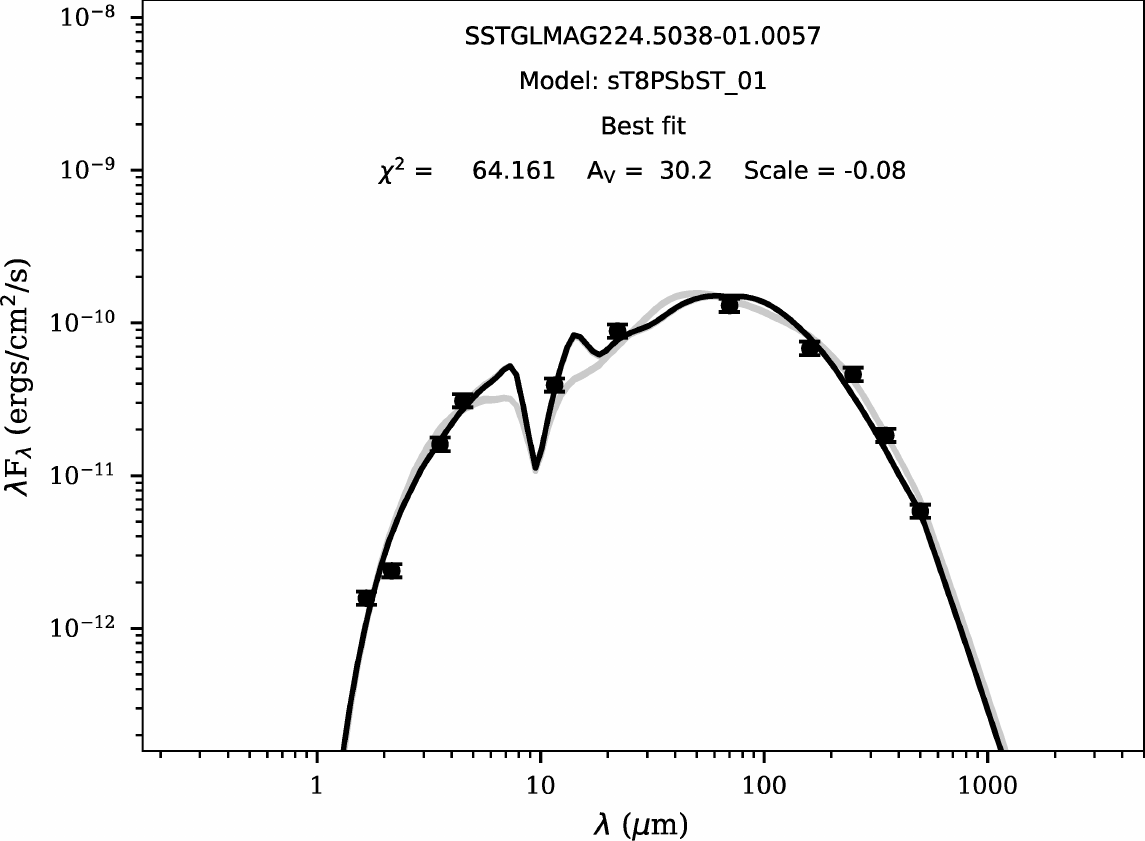}
\hfill
\includegraphics[width=0.32\textwidth]{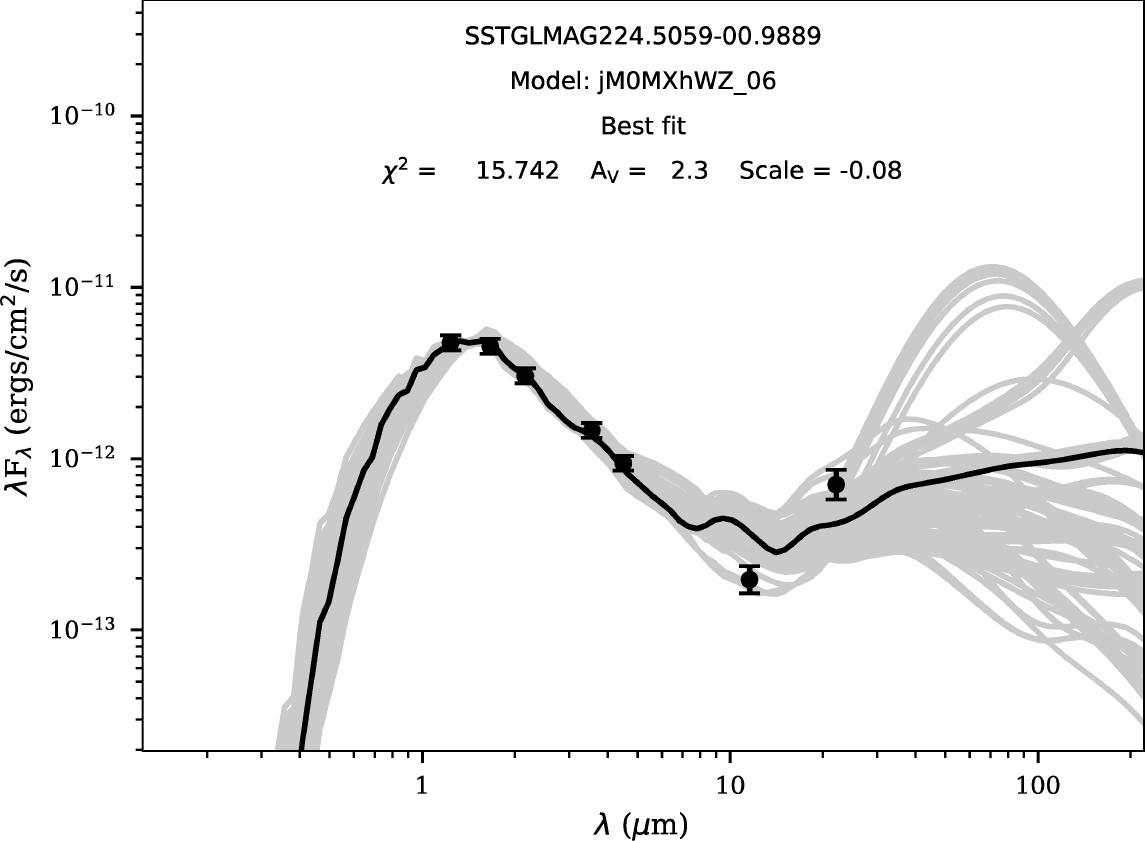}
\hfill
\includegraphics[width=0.32\textwidth]{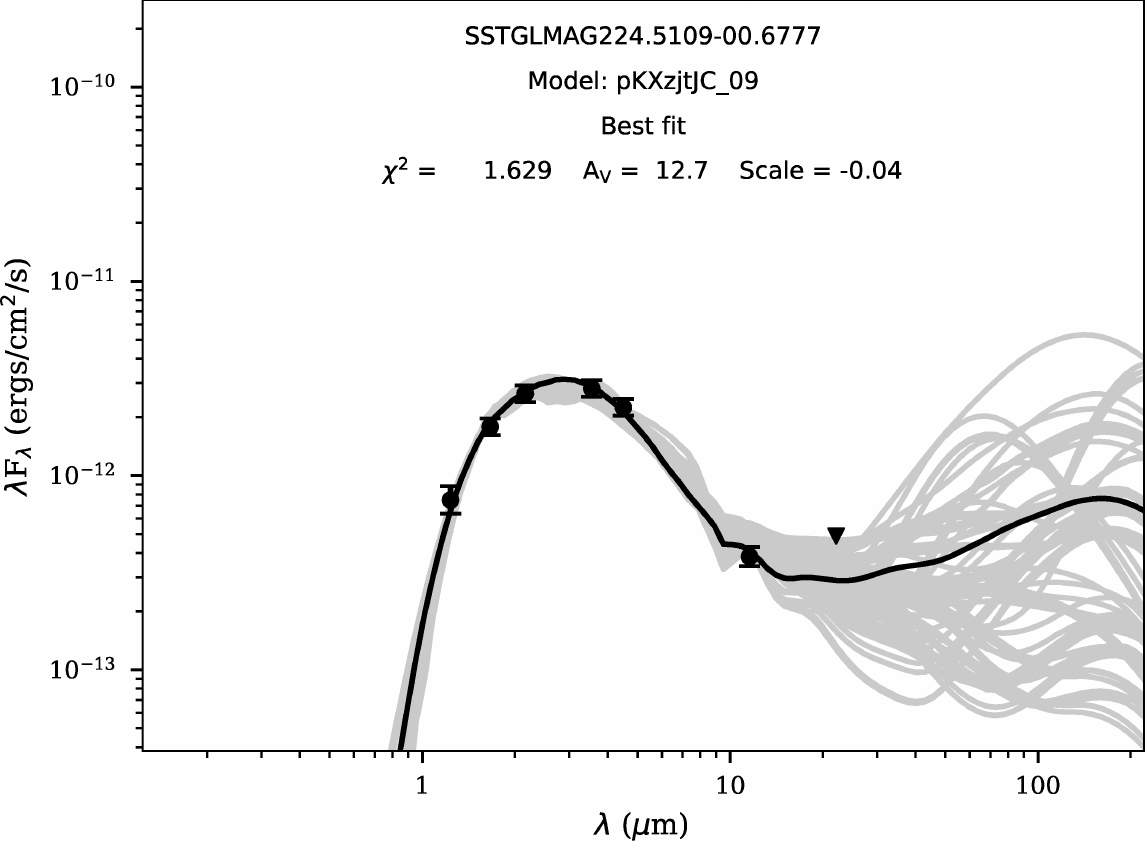} \par
\vspace{2mm}
\includegraphics[width=0.32\textwidth]{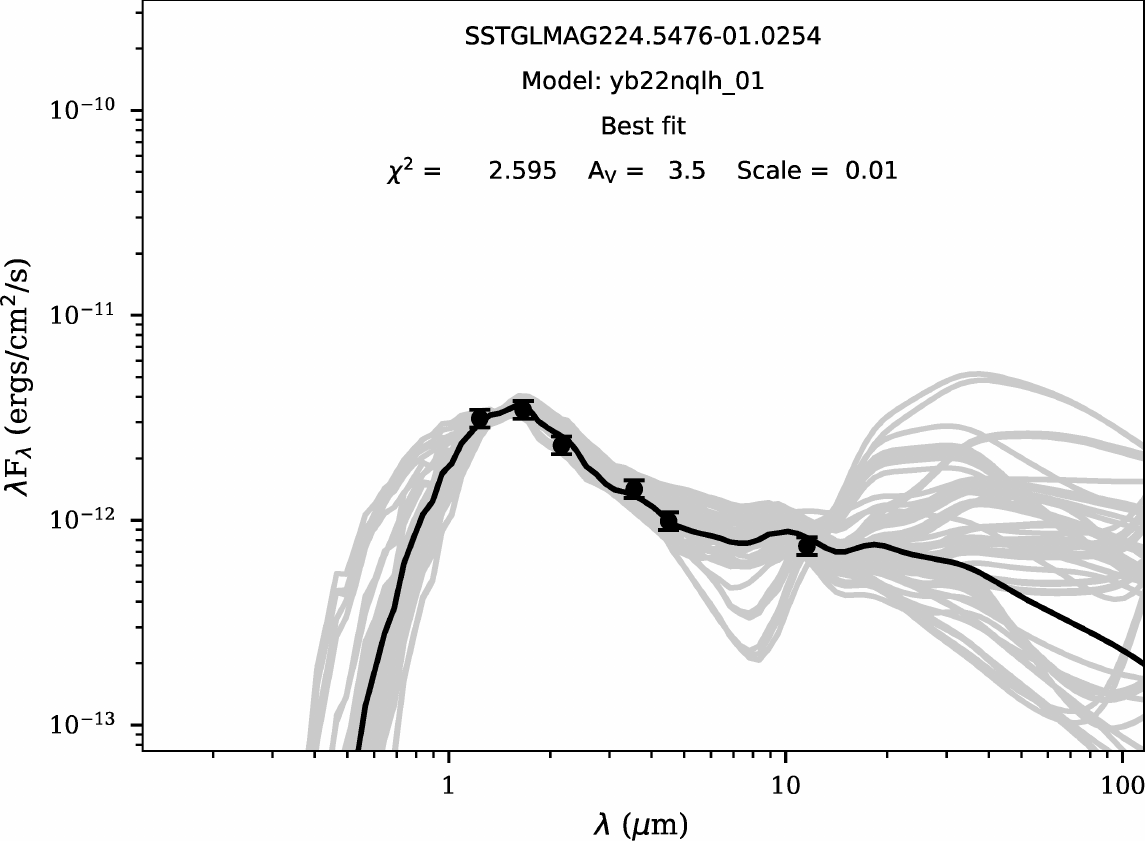}
\hfill
\includegraphics[width=0.32\textwidth]{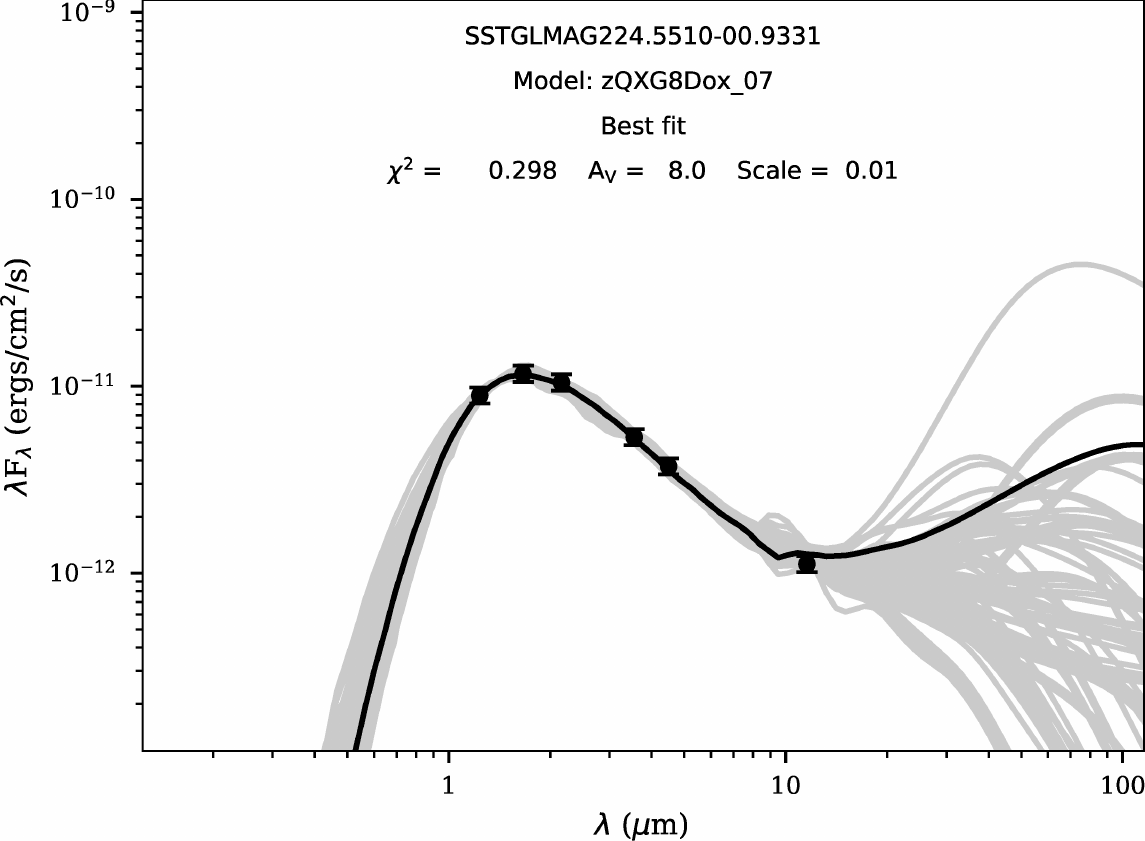}
\hfill
\includegraphics[width=0.32\textwidth]{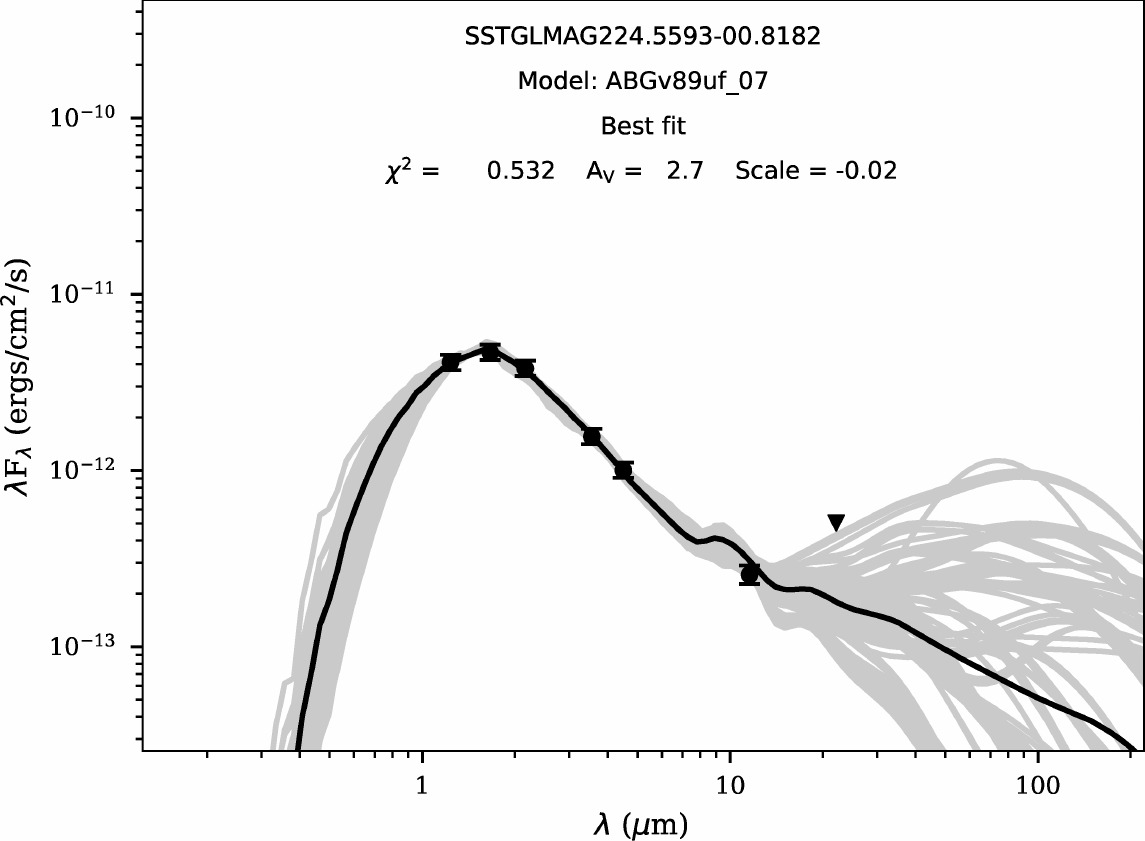} \par
\vspace{2mm}
\includegraphics[width=0.32\textwidth]{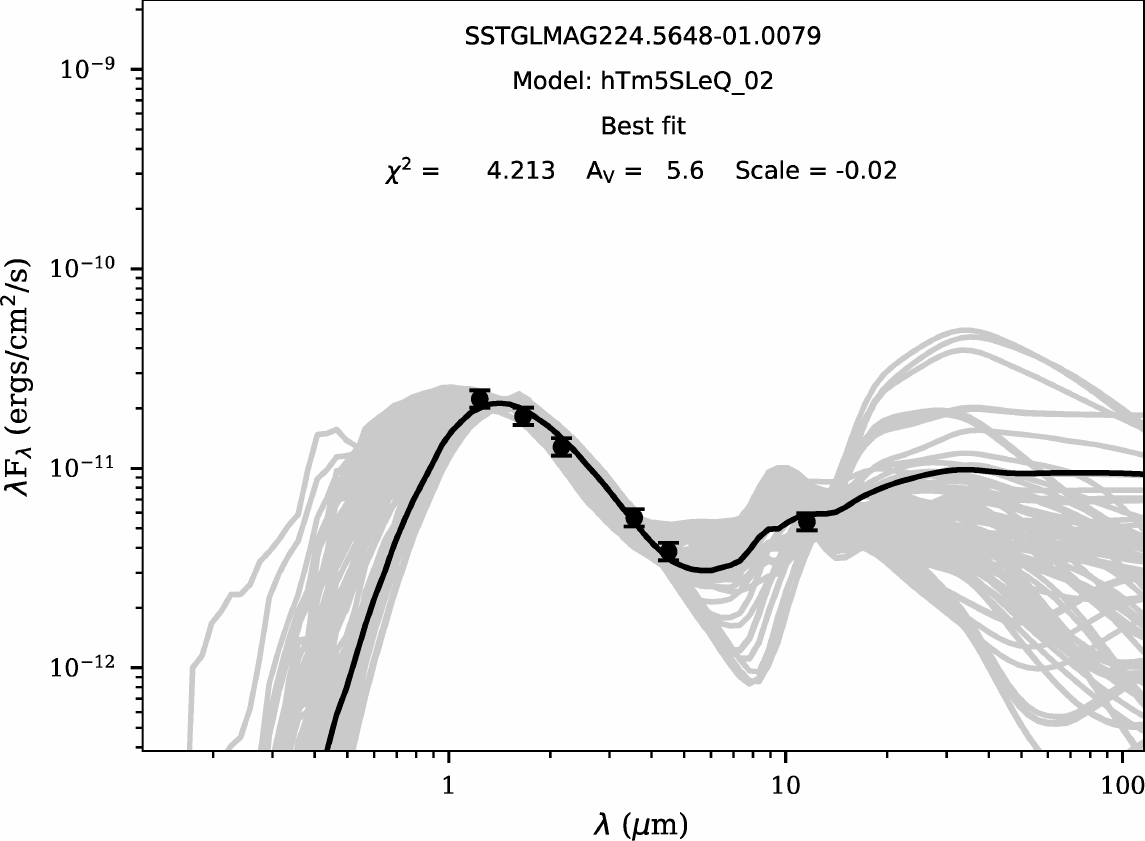}
\hfill
\includegraphics[width=0.32\textwidth]{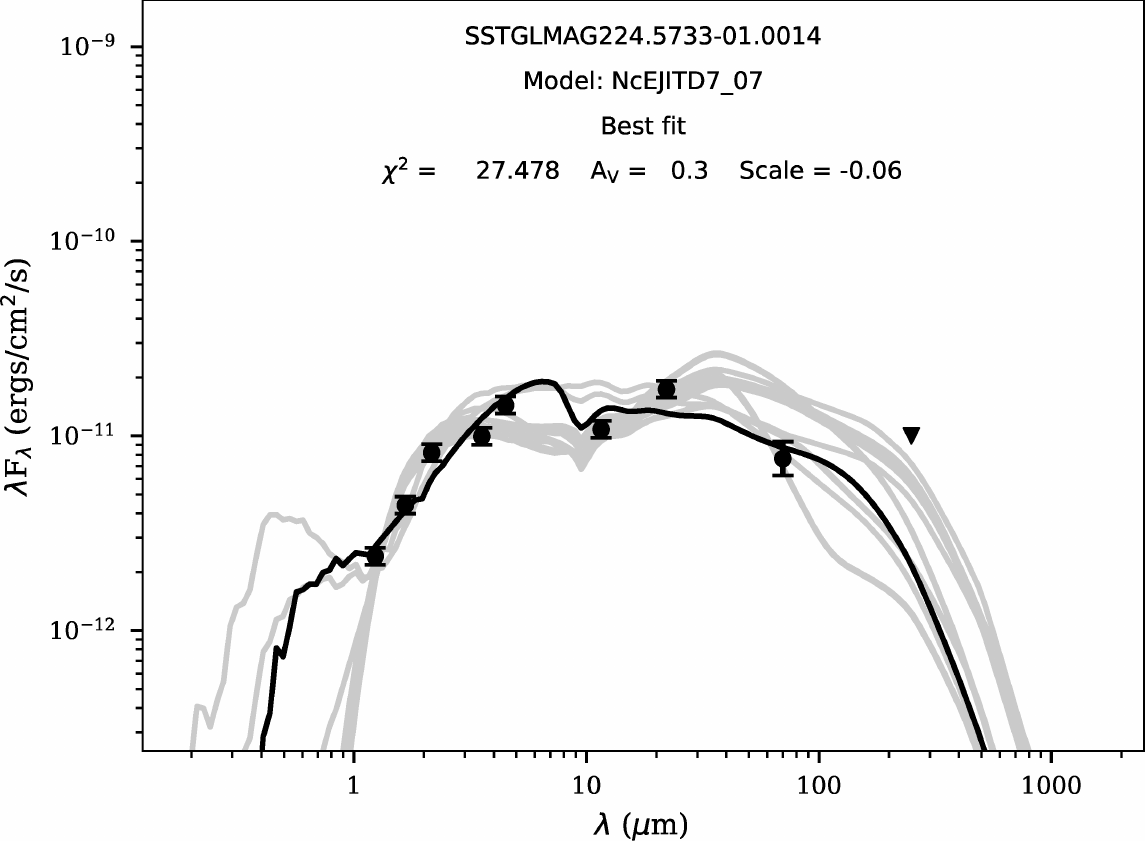}
\hfill
\includegraphics[width=0.32\textwidth]{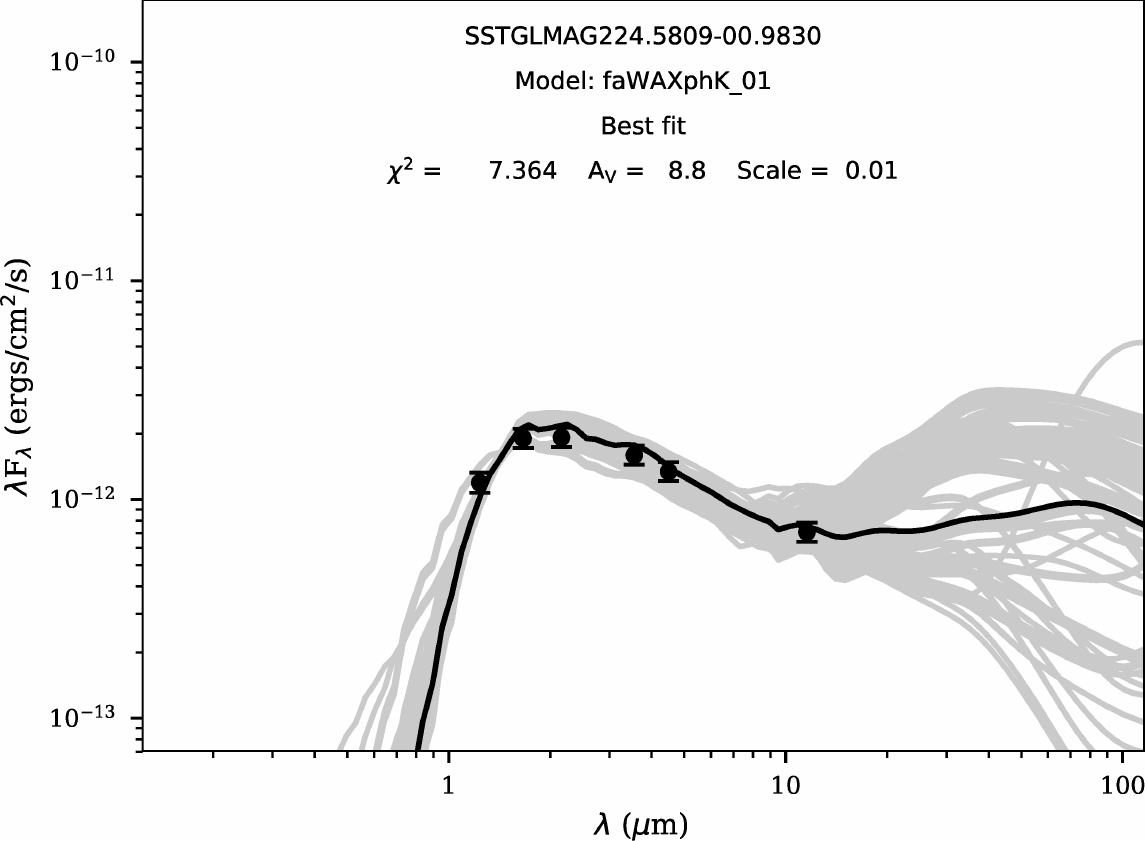} \par
\vspace{2mm}
\includegraphics[width=0.32\textwidth]{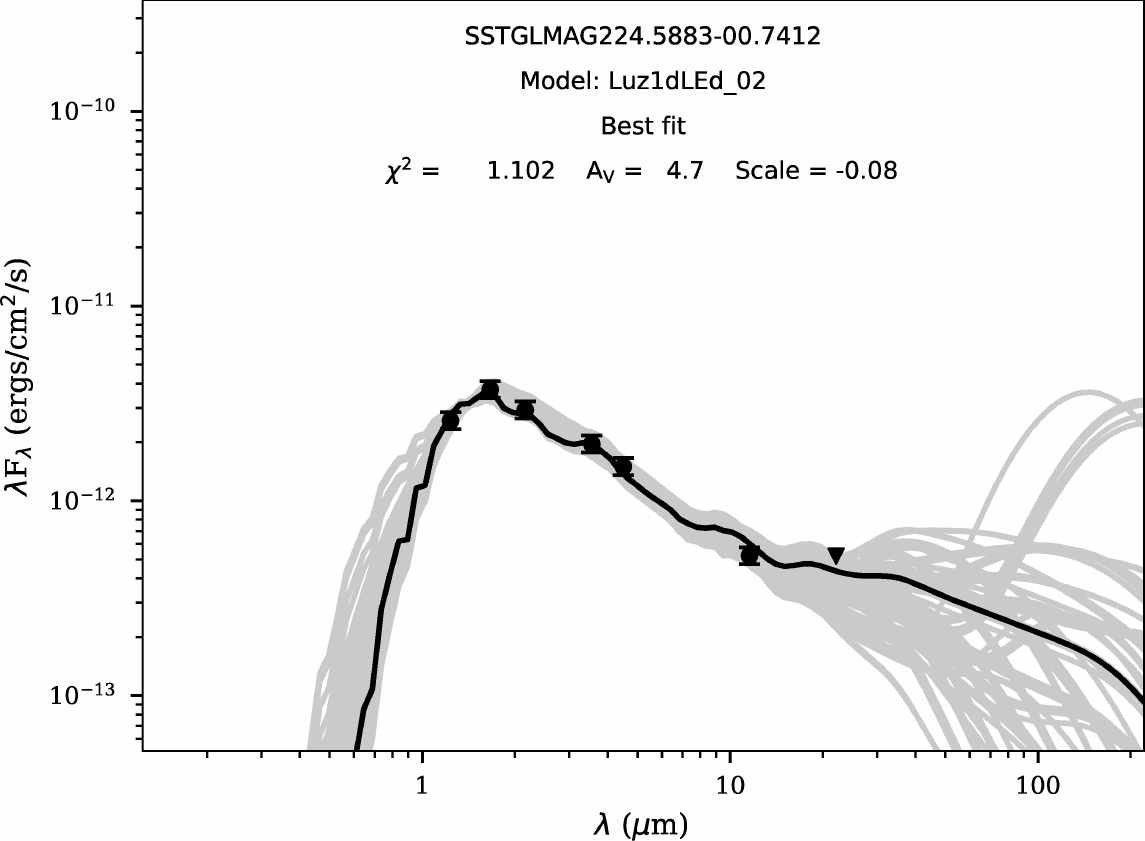}
\hfill
\includegraphics[width=0.32\textwidth]{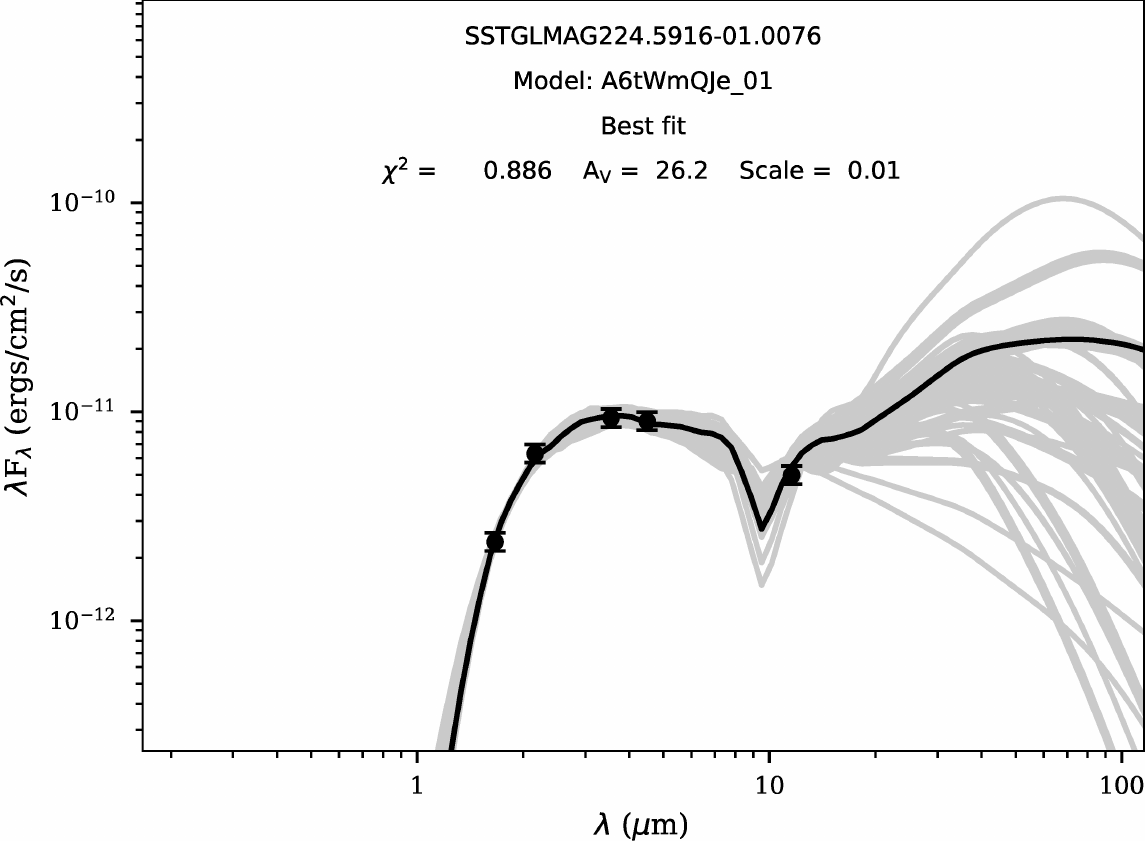}
\hfill
\includegraphics[width=0.32\textwidth]{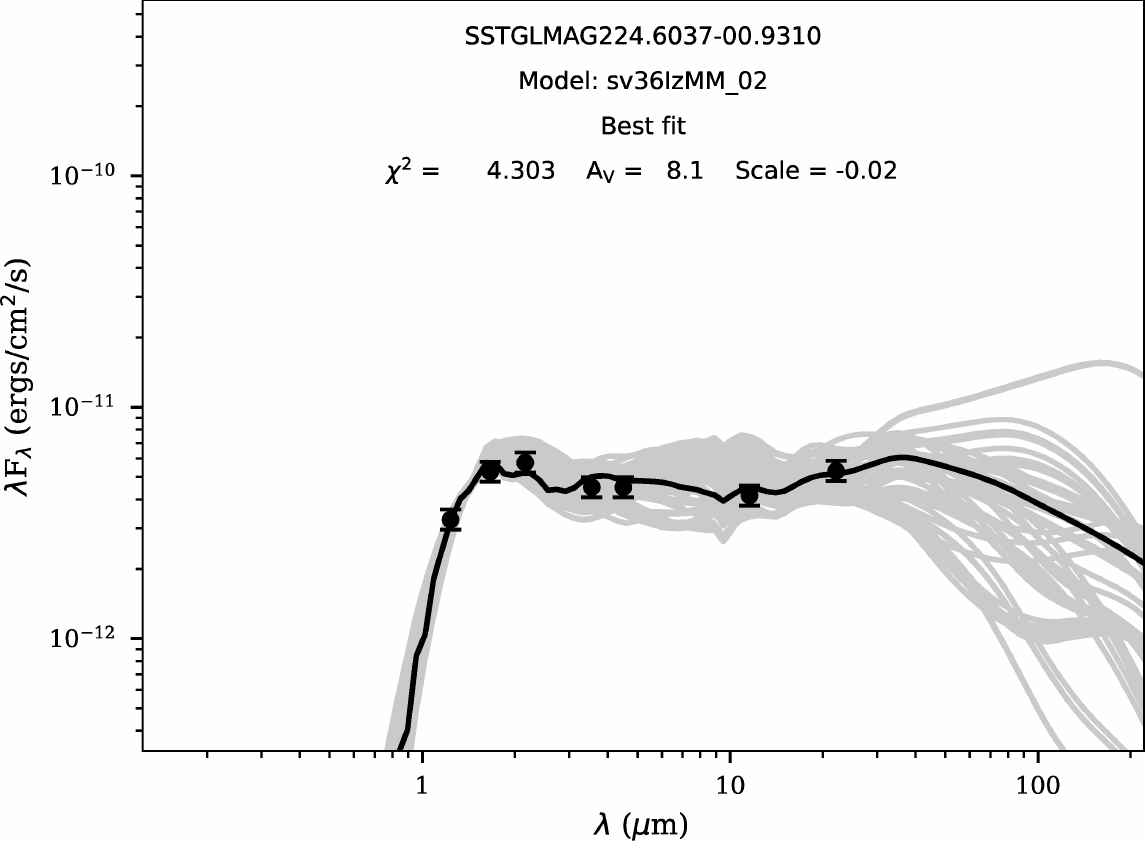} \par
\vspace{2mm}
\includegraphics[width=0.32\textwidth]{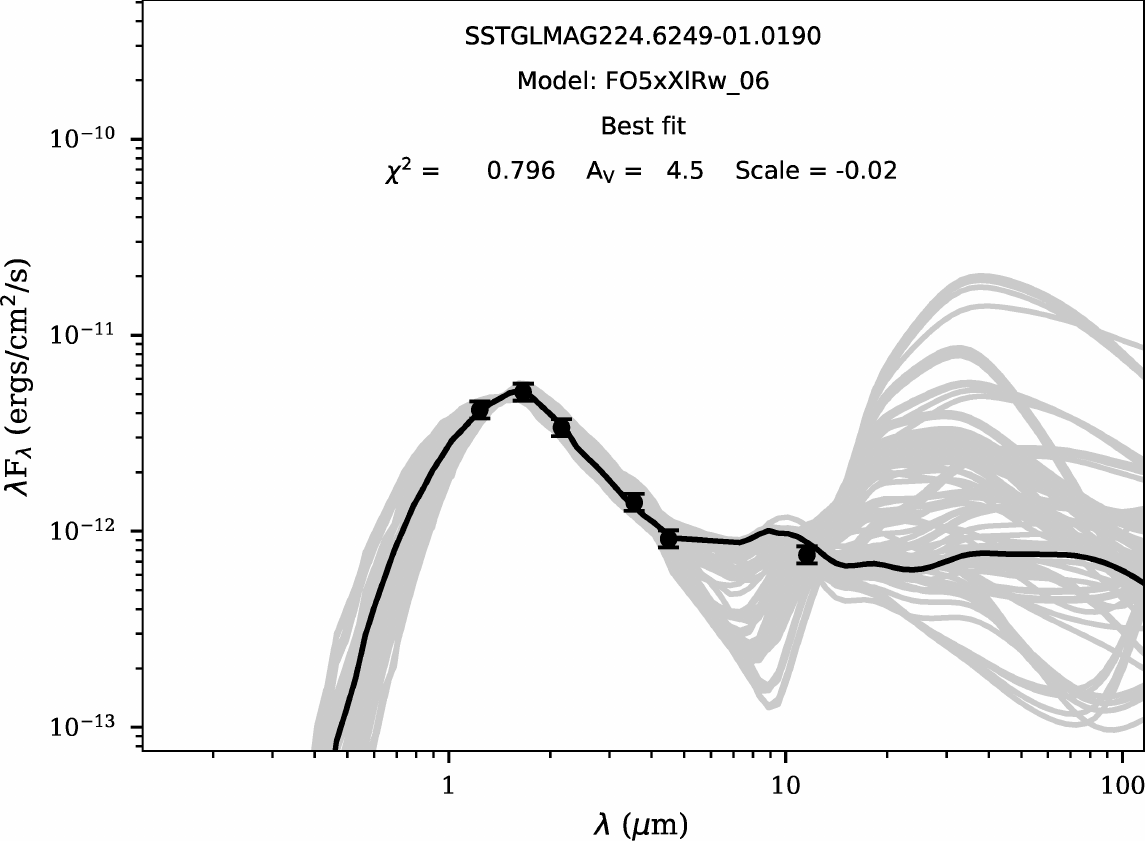}
\hfill
\includegraphics[width=0.32\textwidth]{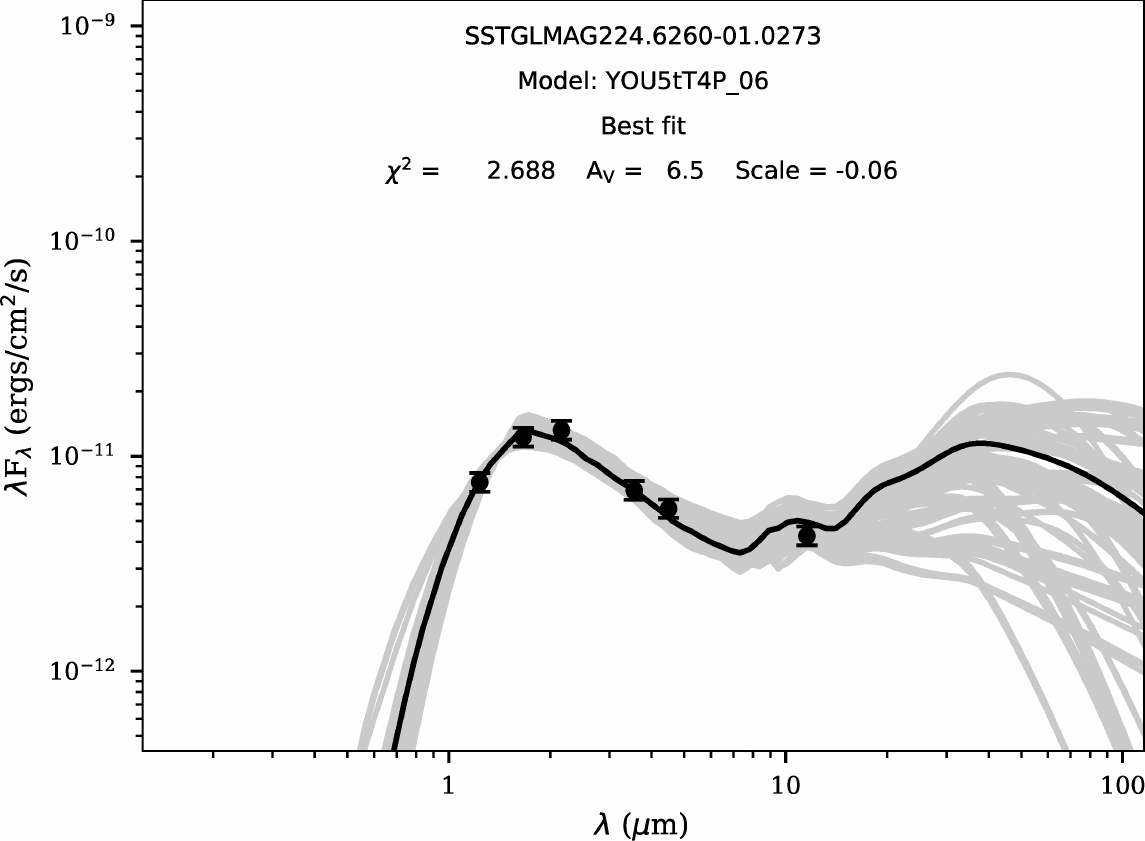}
\hfill
\includegraphics[width=0.32\textwidth]{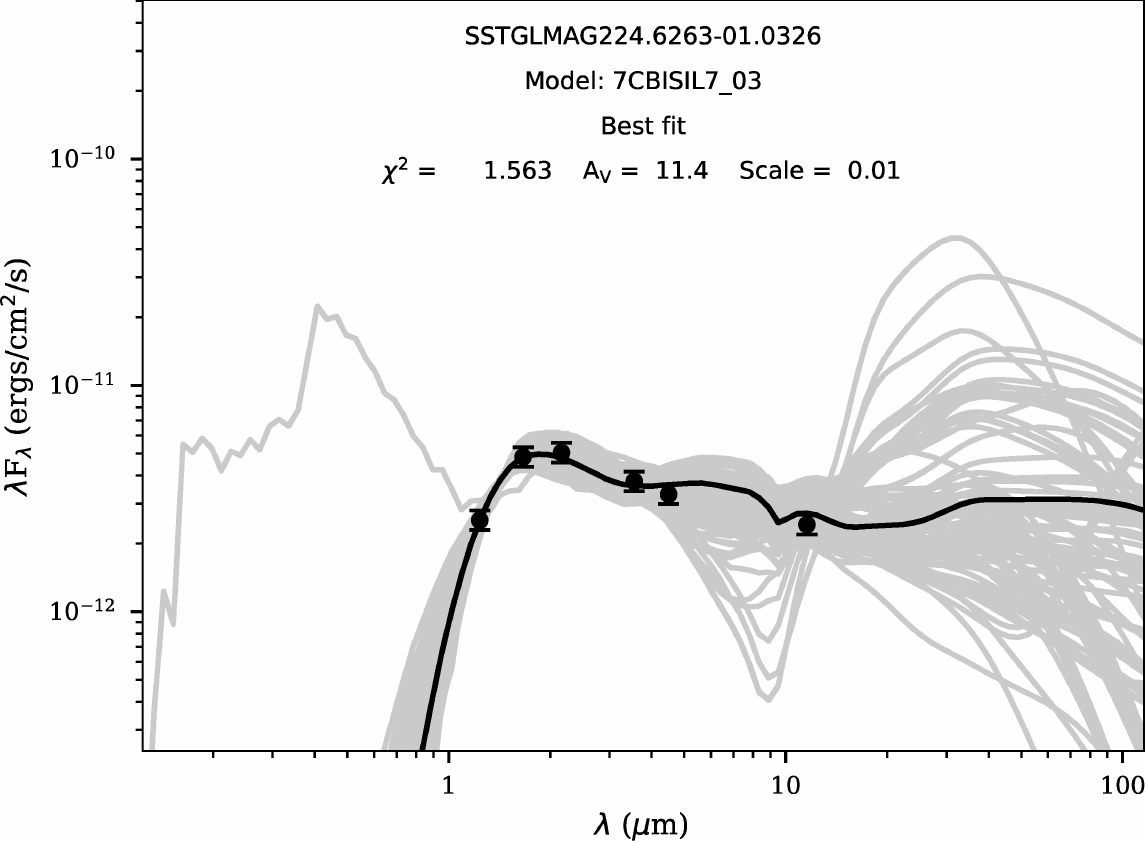}
\caption{Same as Fig.~\ref{f:SEDs1}  \label{f:SEDs12}}
\end{figure*}

\begin{figure*}
\includegraphics[width=0.32\textwidth]{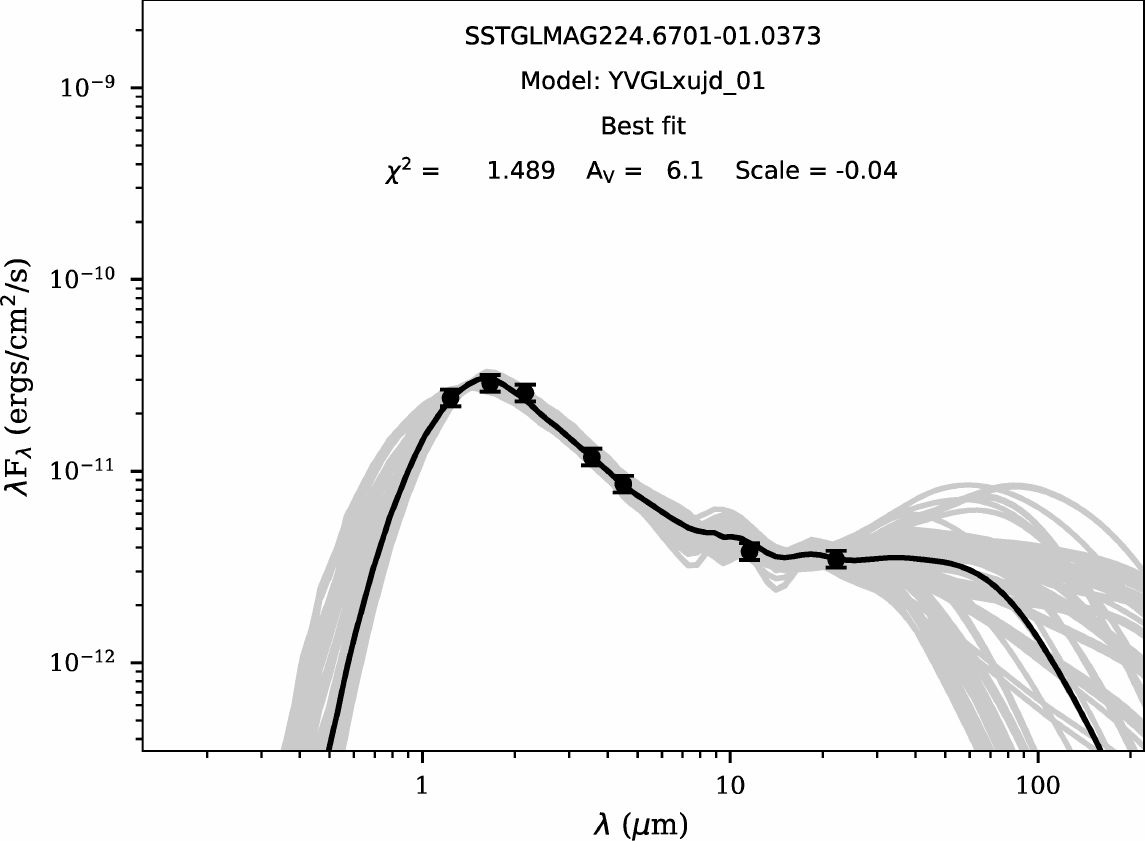}
\hfill
\includegraphics[width=0.32\textwidth]{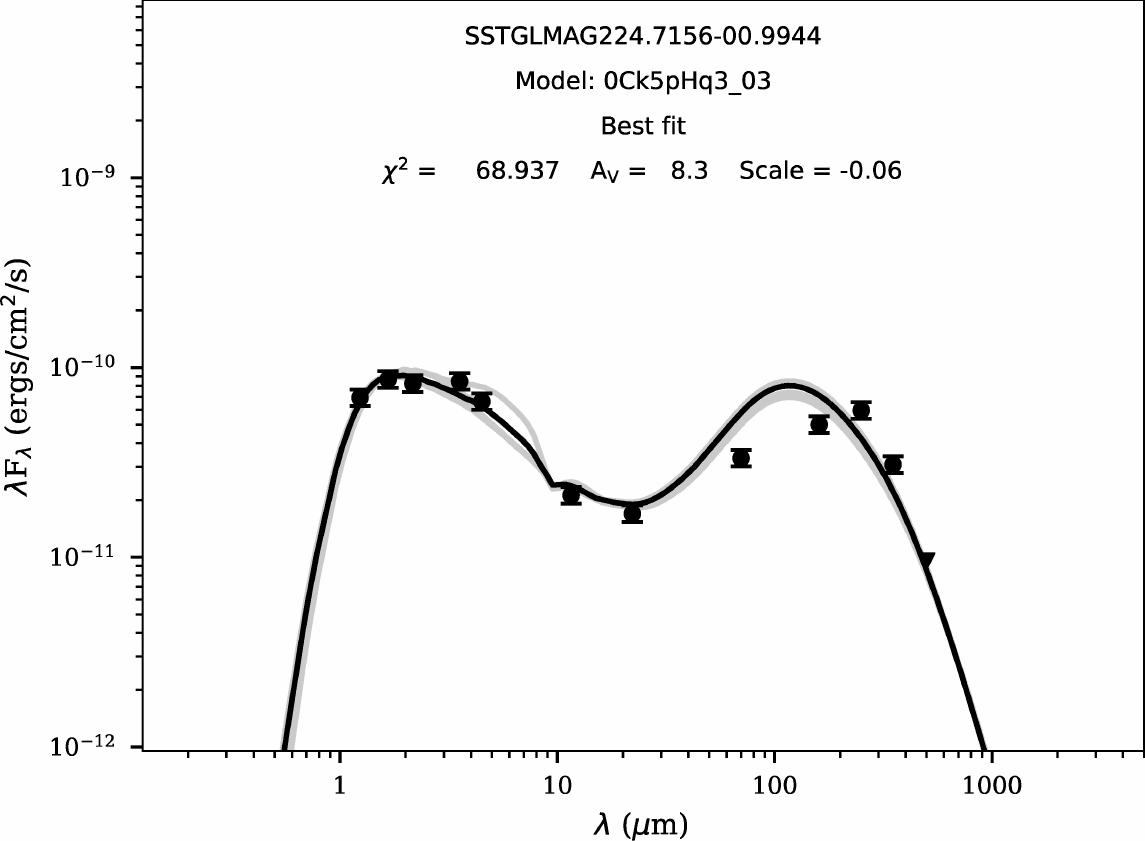}
\hfill
\includegraphics[width=0.32\textwidth]{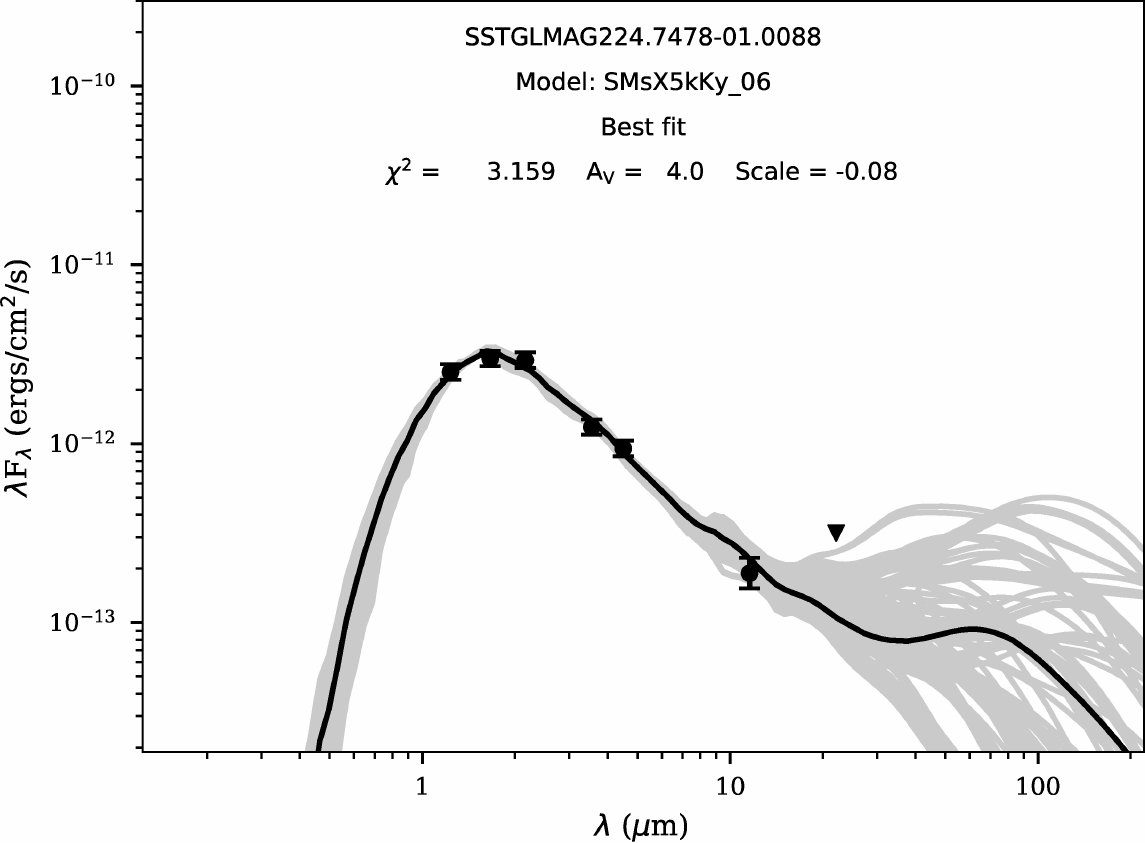} \par
\vspace{2mm}
\includegraphics[width=0.32\textwidth]{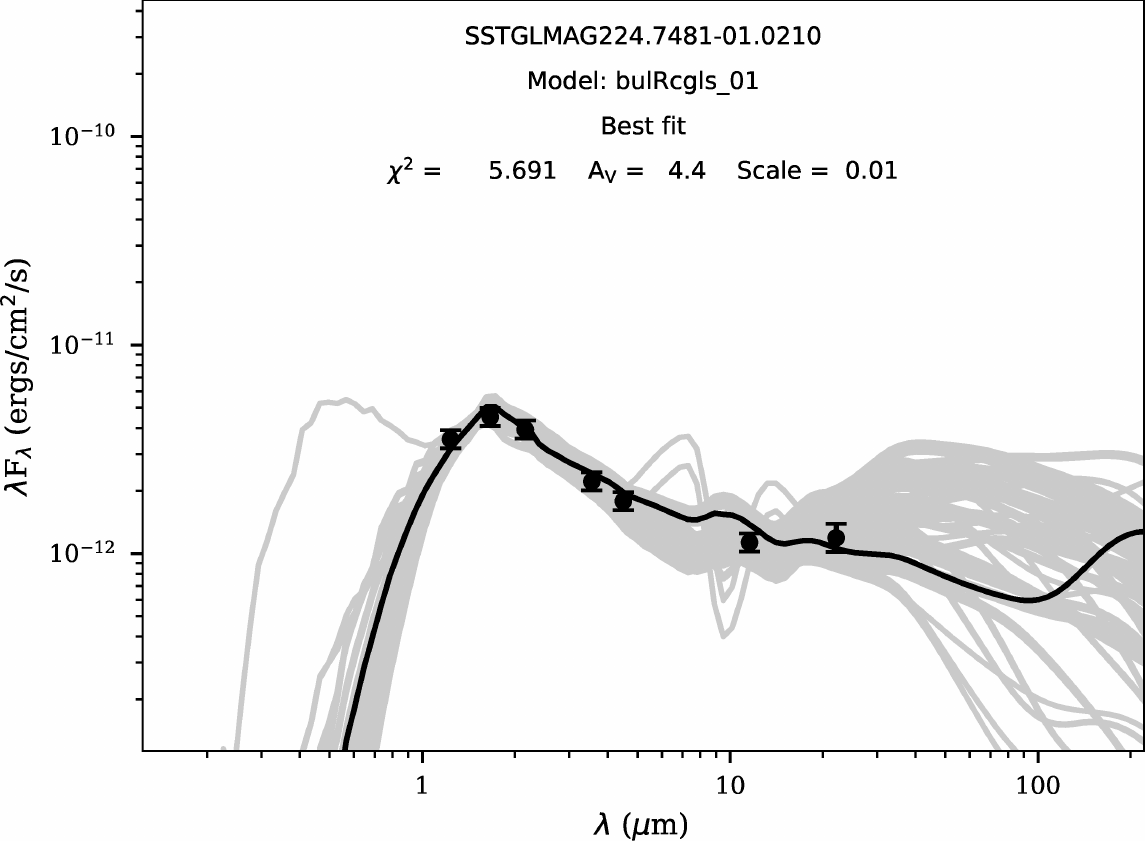}
\hfill
\includegraphics[width=0.32\textwidth]{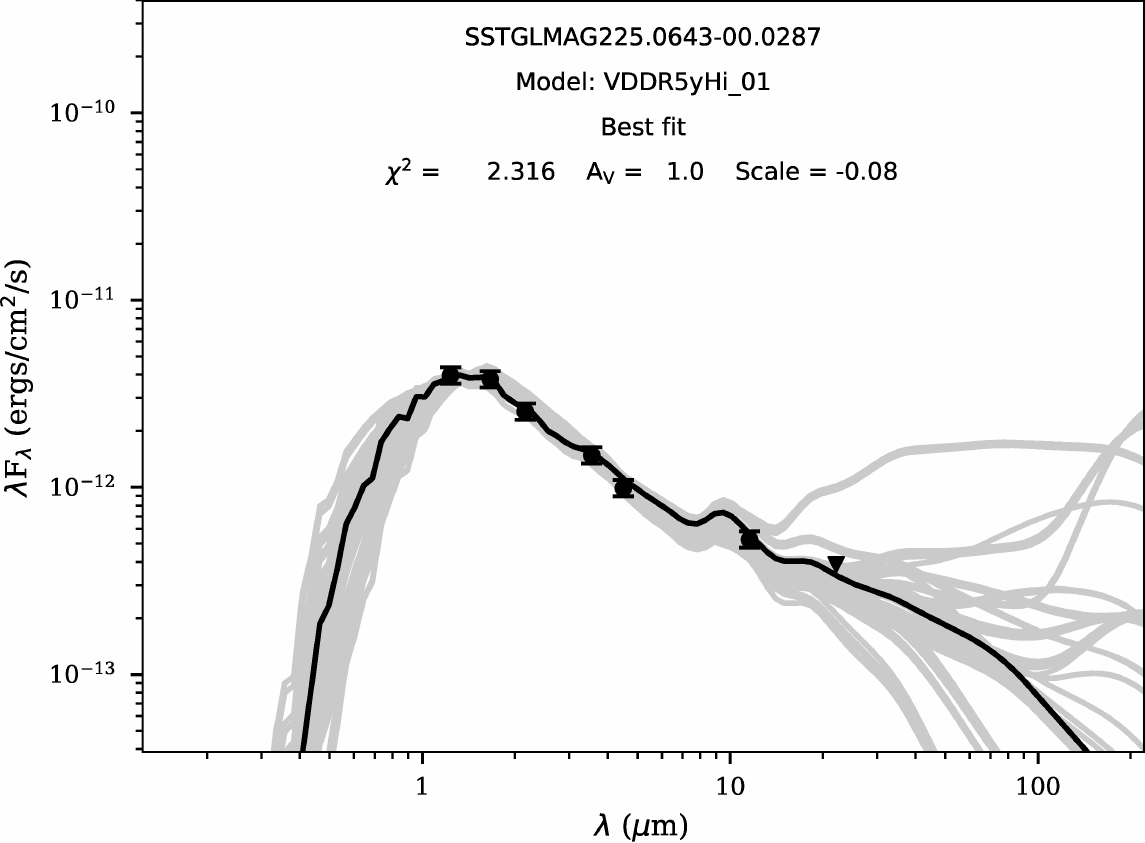}
\hfill
\includegraphics[width=0.32\textwidth]{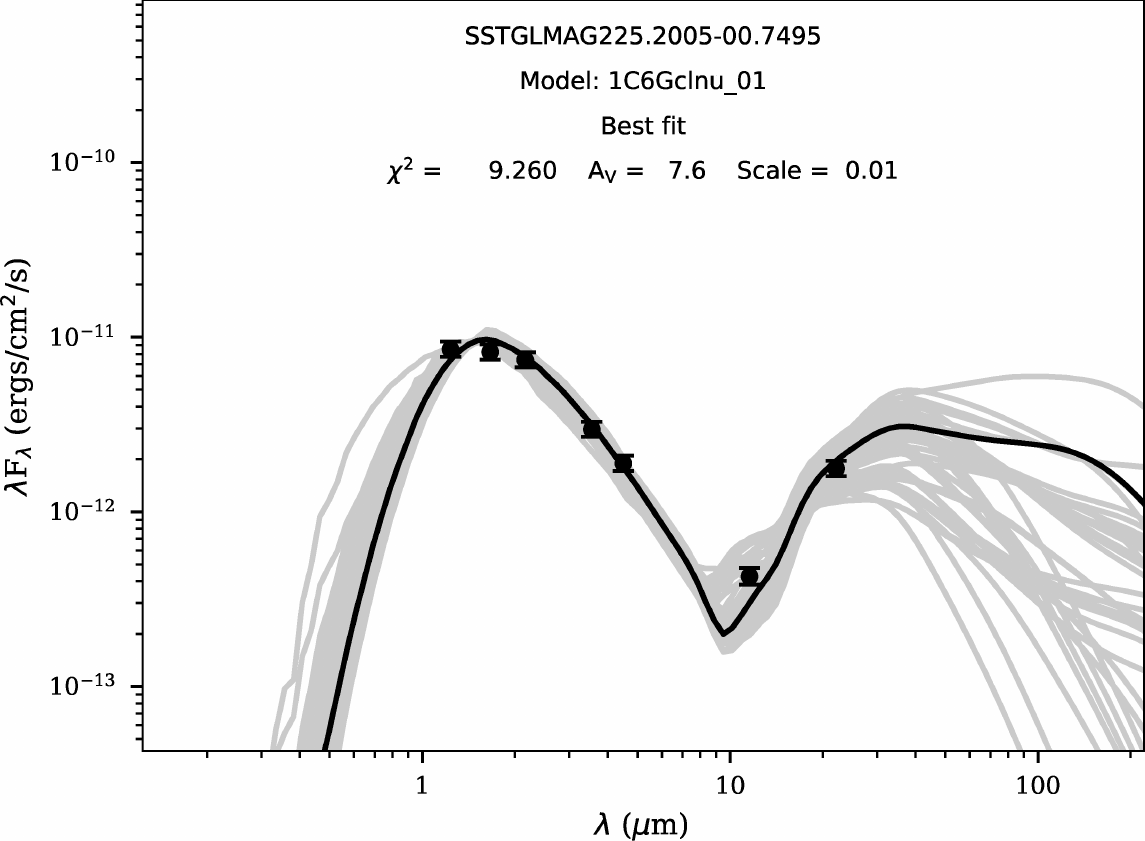} \par
\vspace{2mm}
\includegraphics[width=0.32\textwidth]{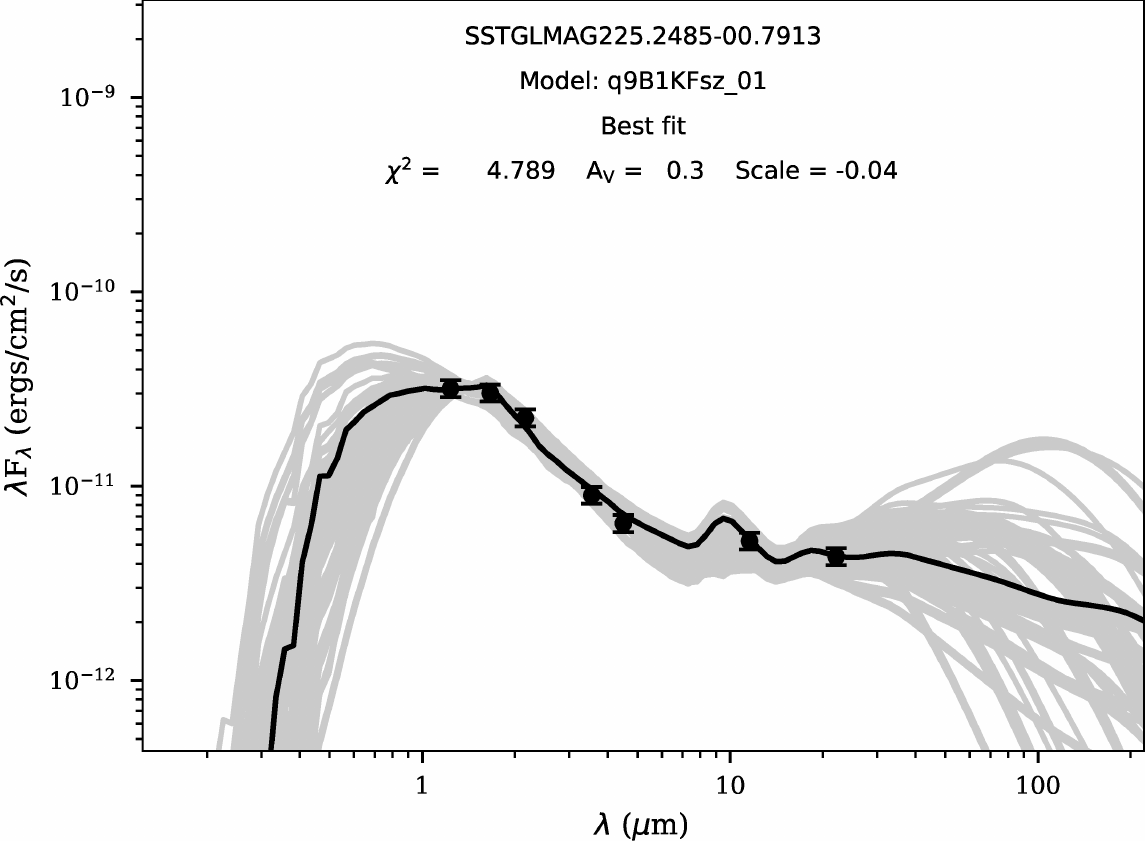}
\hfill
\includegraphics[width=0.32\textwidth]{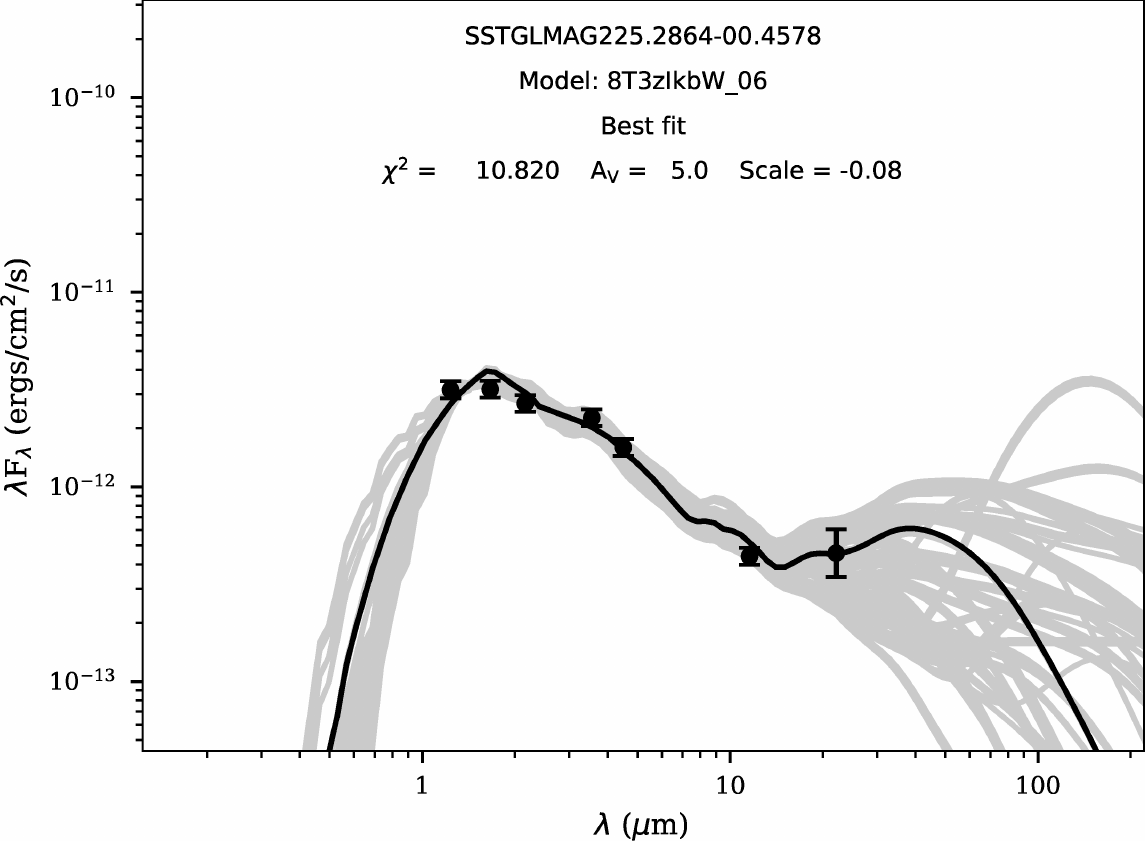}
\hfill
\includegraphics[width=0.32\textwidth]{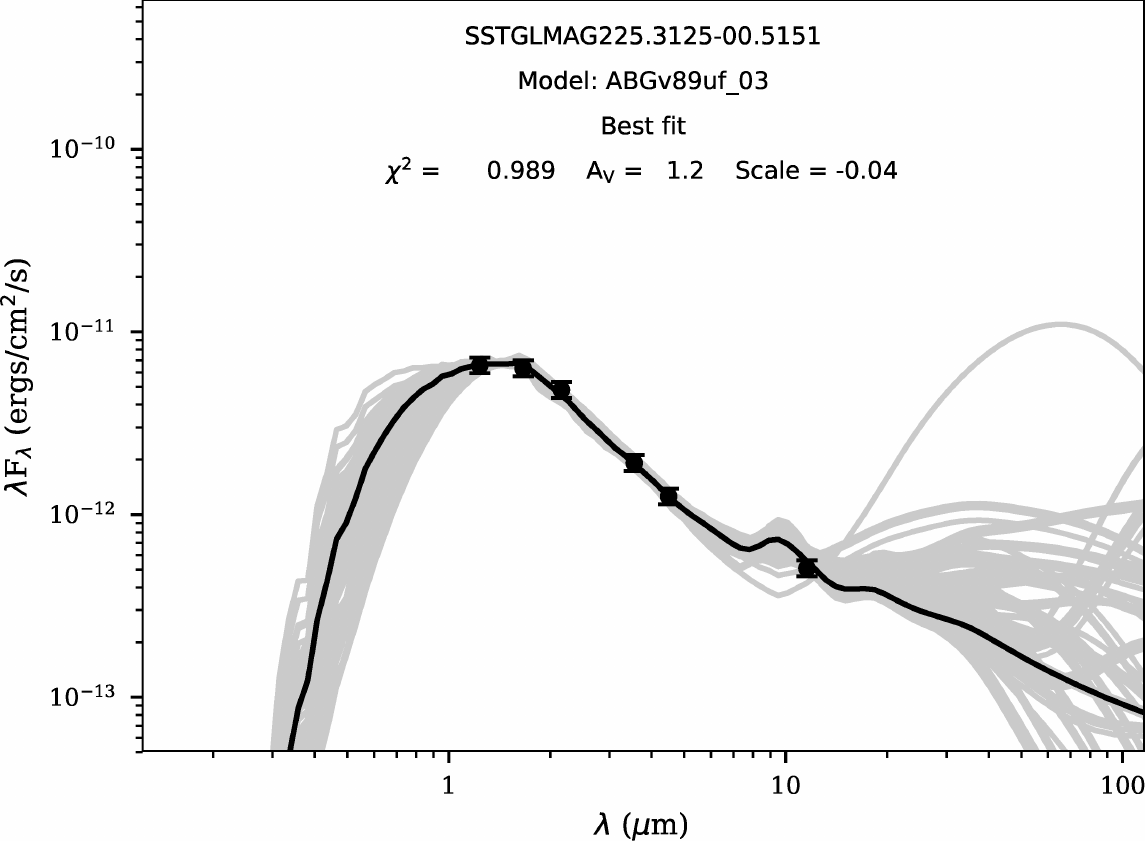} \par
\vspace{2mm}
\includegraphics[width=0.32\textwidth]{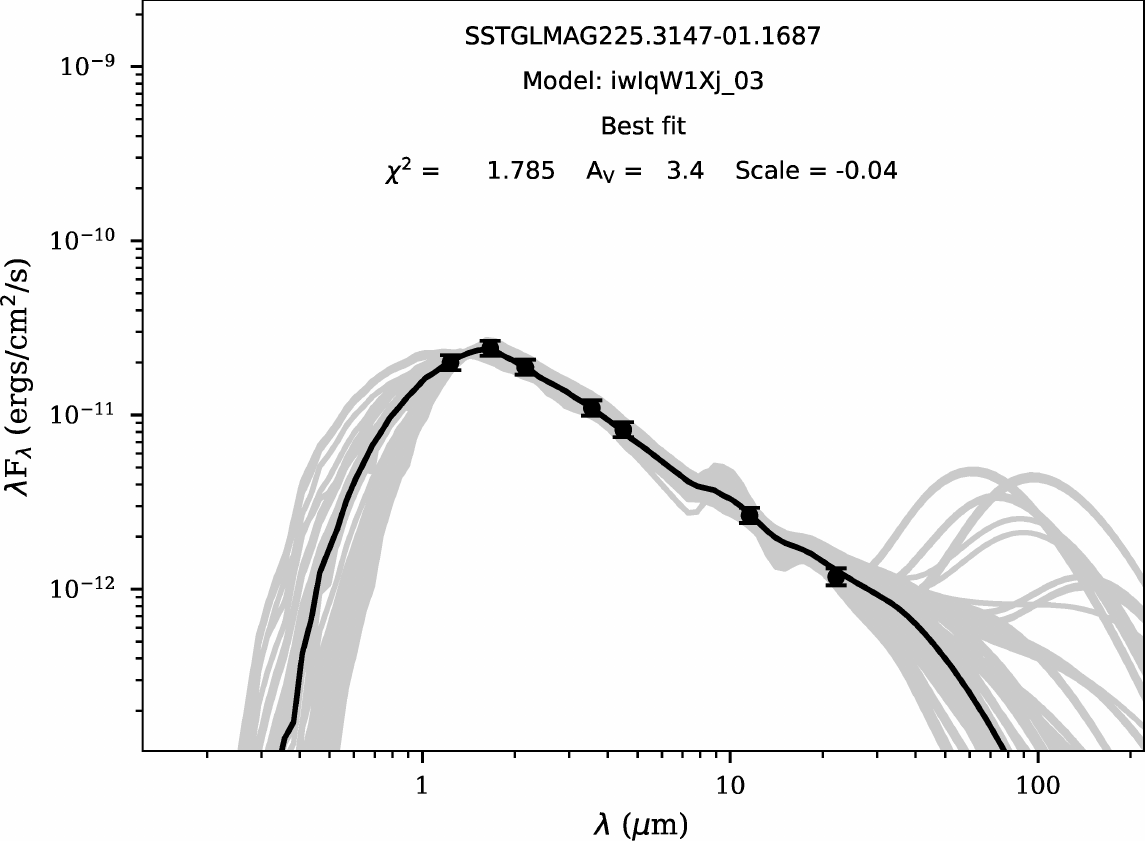}
\hfill
\includegraphics[width=0.32\textwidth]{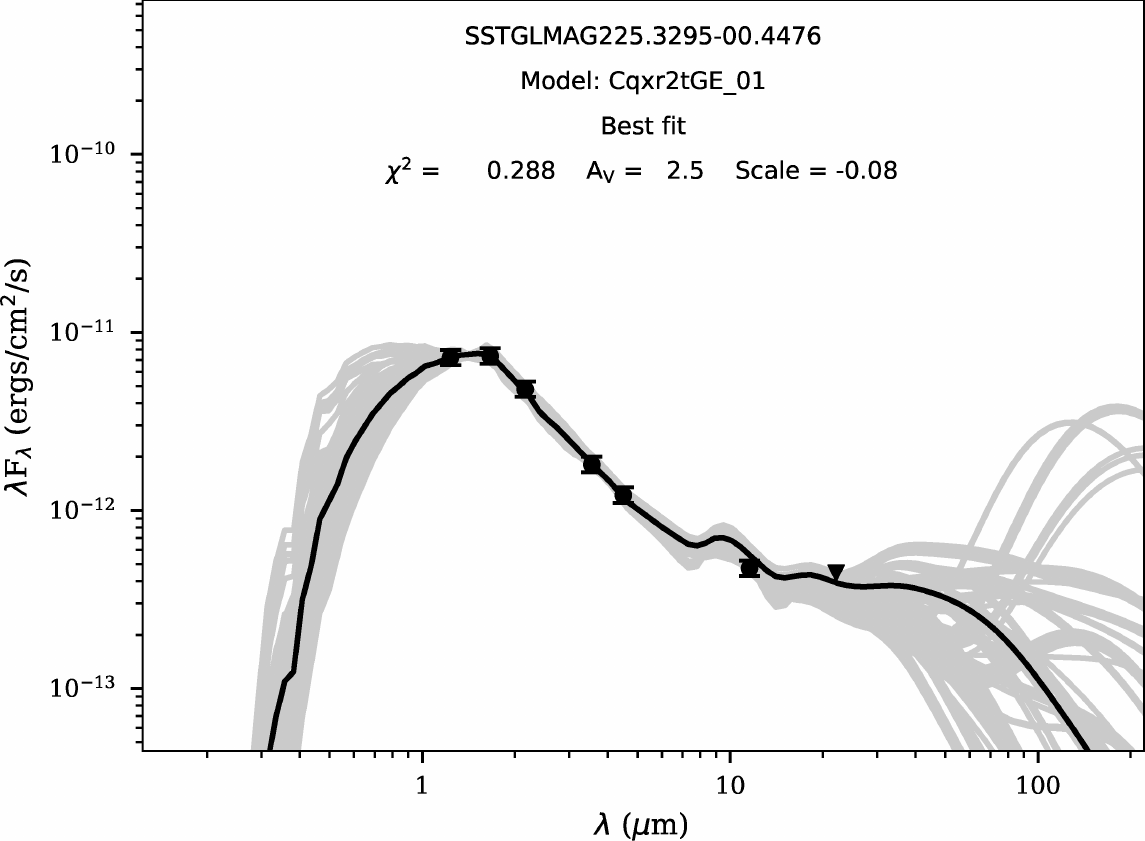}
\hfill
\includegraphics[width=0.32\textwidth]{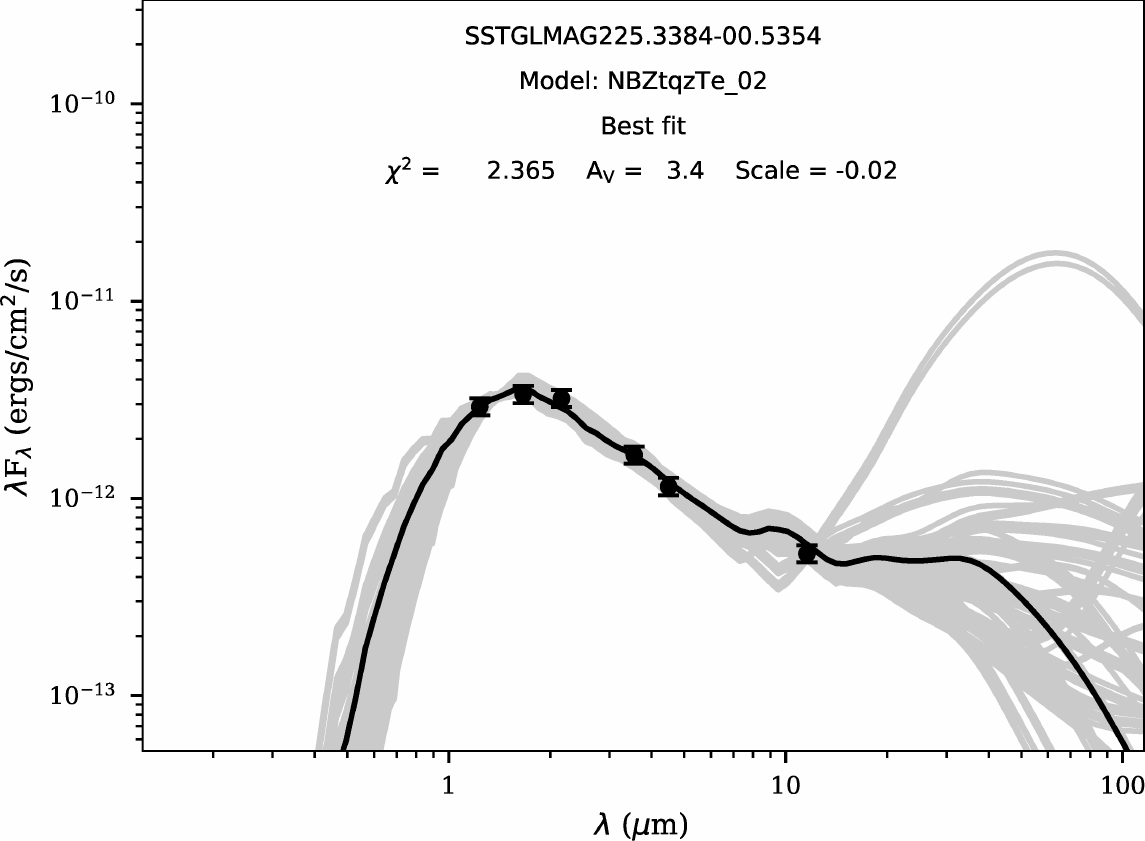} \par
\vspace{2mm}
\includegraphics[width=0.32\textwidth]{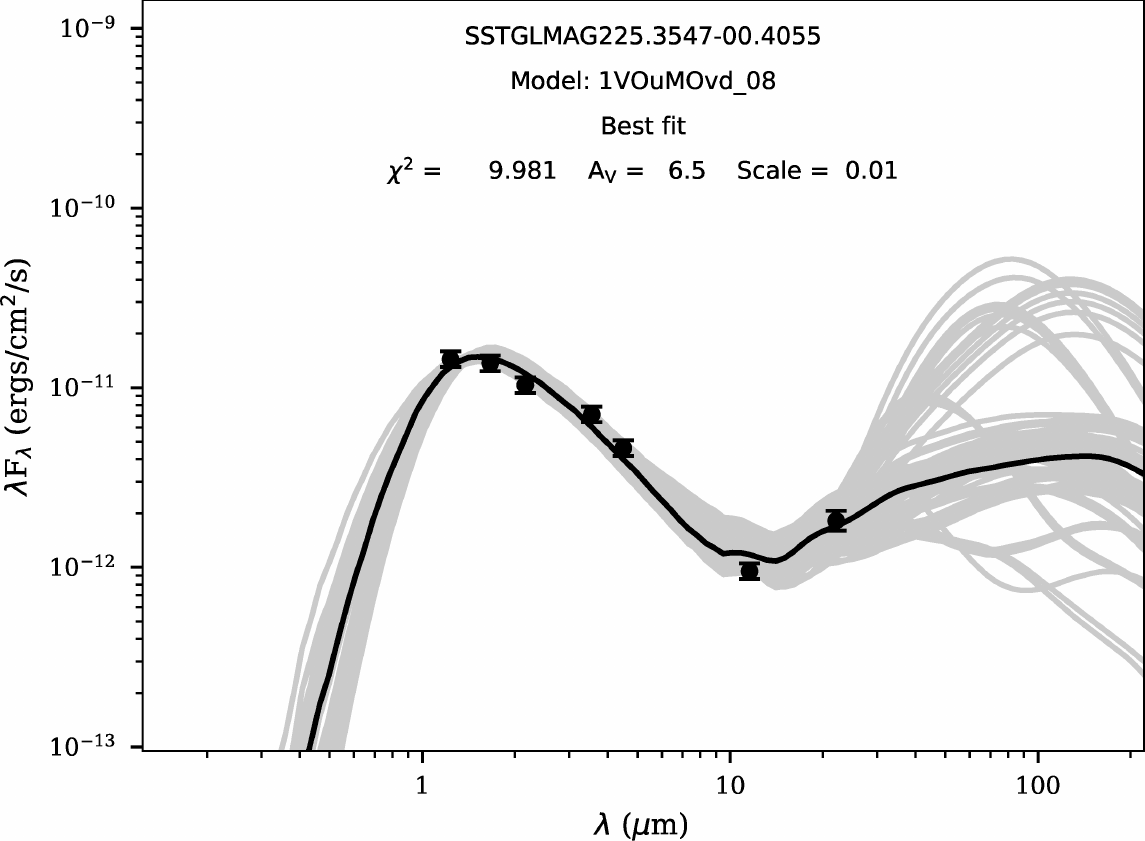}
\hfill
\includegraphics[width=0.32\textwidth]{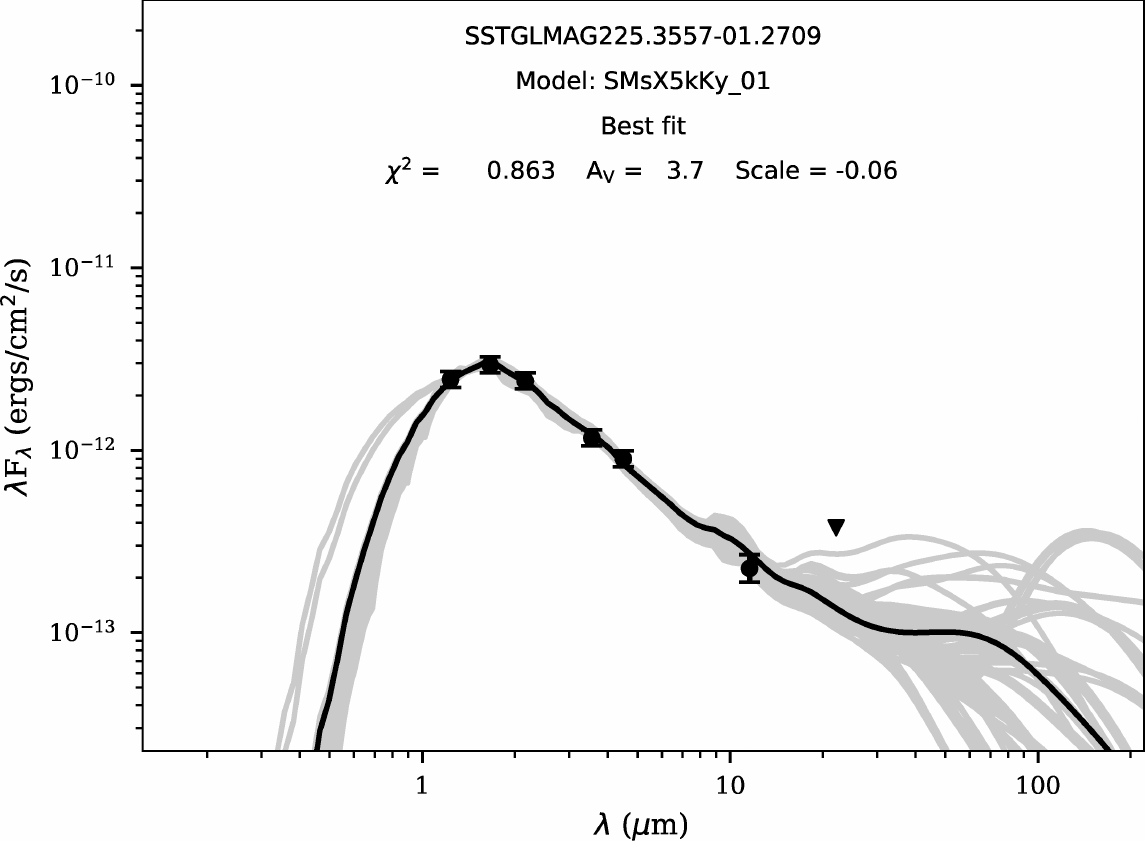}
\hfill
\includegraphics[width=0.32\textwidth]{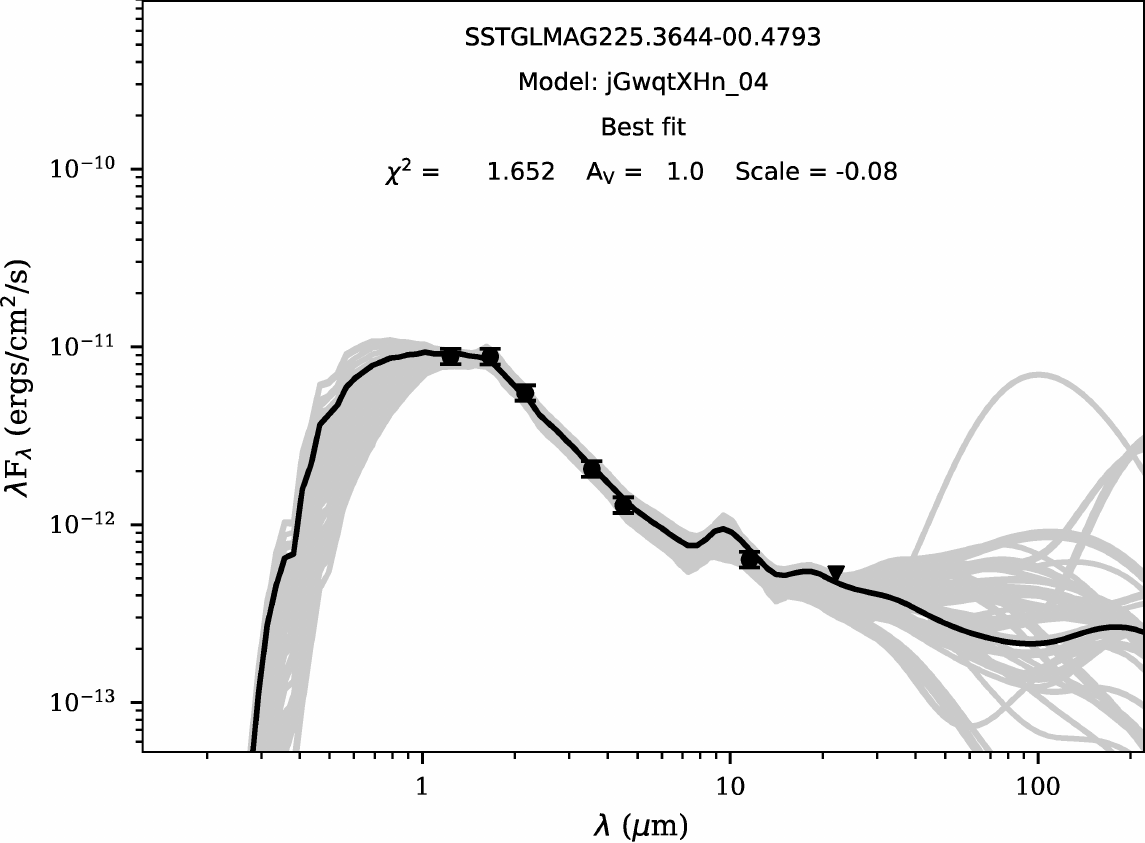}
\caption{Same as Fig.~\ref{f:SEDs1}  \label{f:SEDs13}}
\end{figure*}

\begin{figure*}
\includegraphics[width=0.32\textwidth]{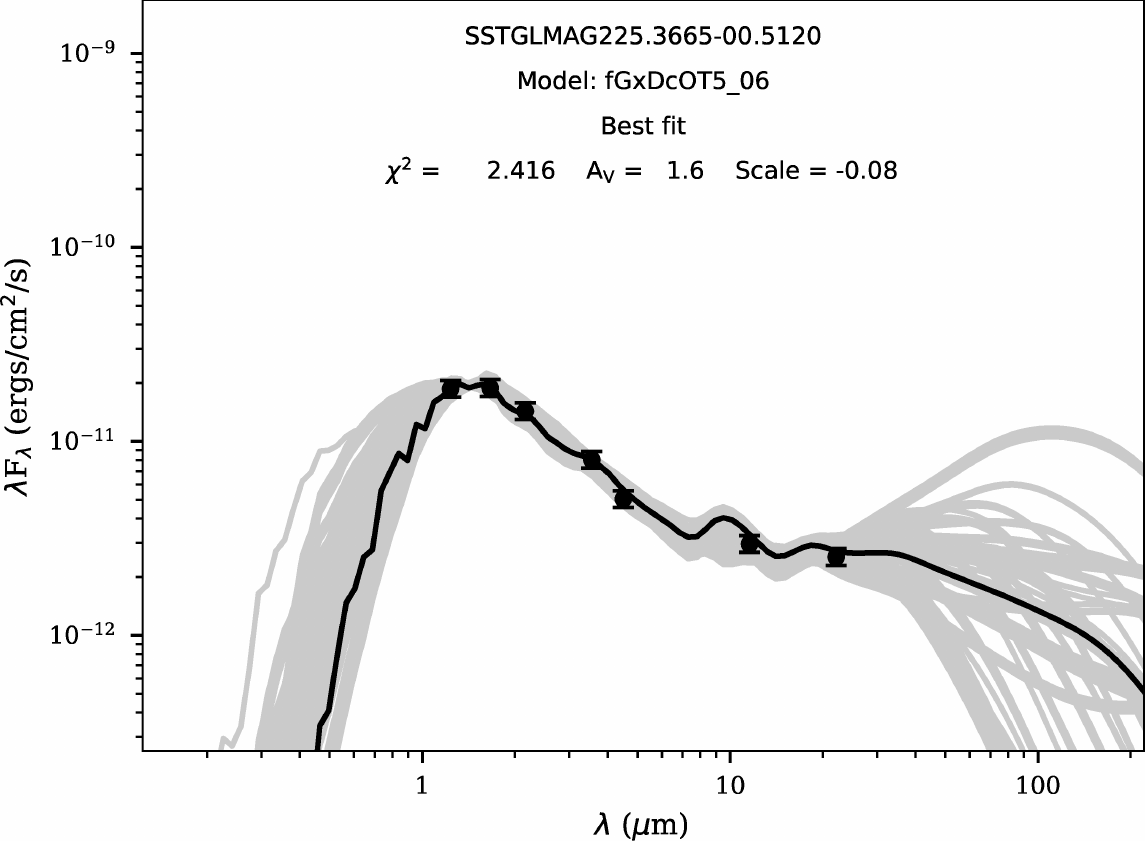}
\hfill
\includegraphics[width=0.32\textwidth]{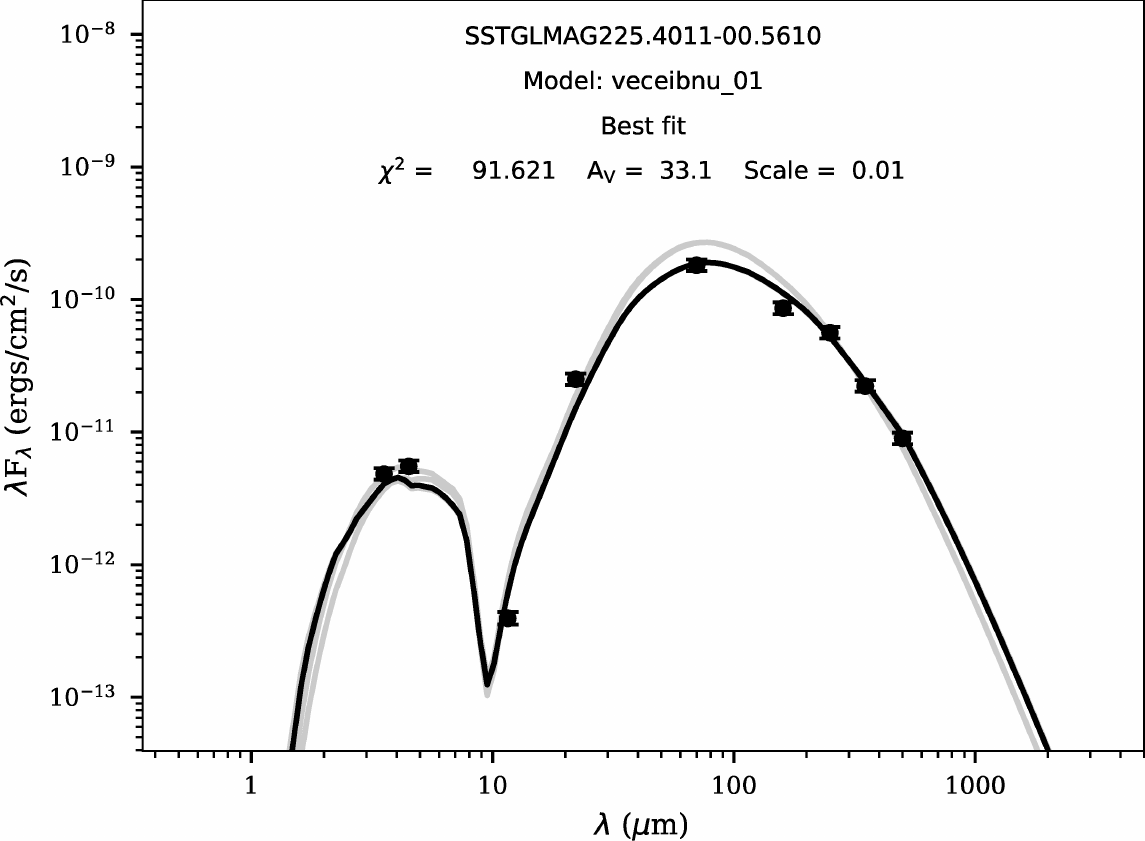}
\hfill
\includegraphics[width=0.32\textwidth]{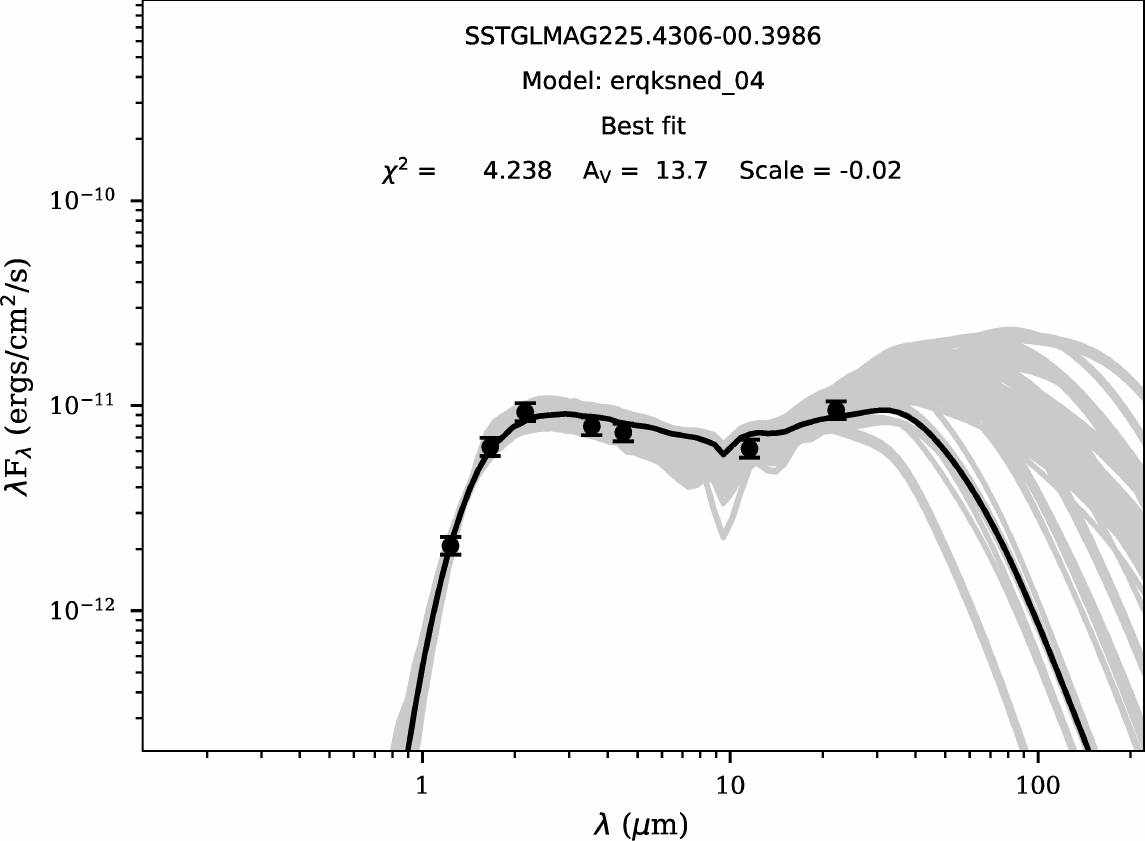} \par
\vspace{2mm}
\includegraphics[width=0.32\textwidth]{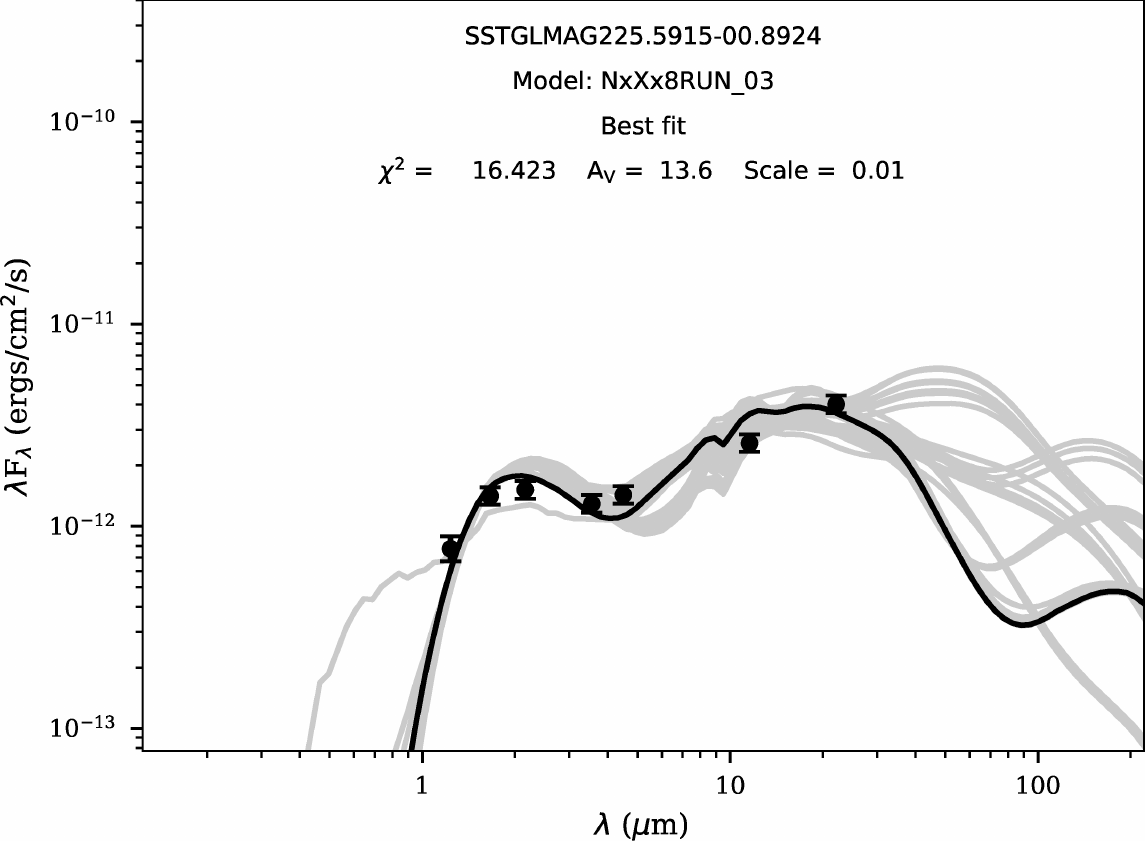}
\hfill
\includegraphics[width=0.32\textwidth]{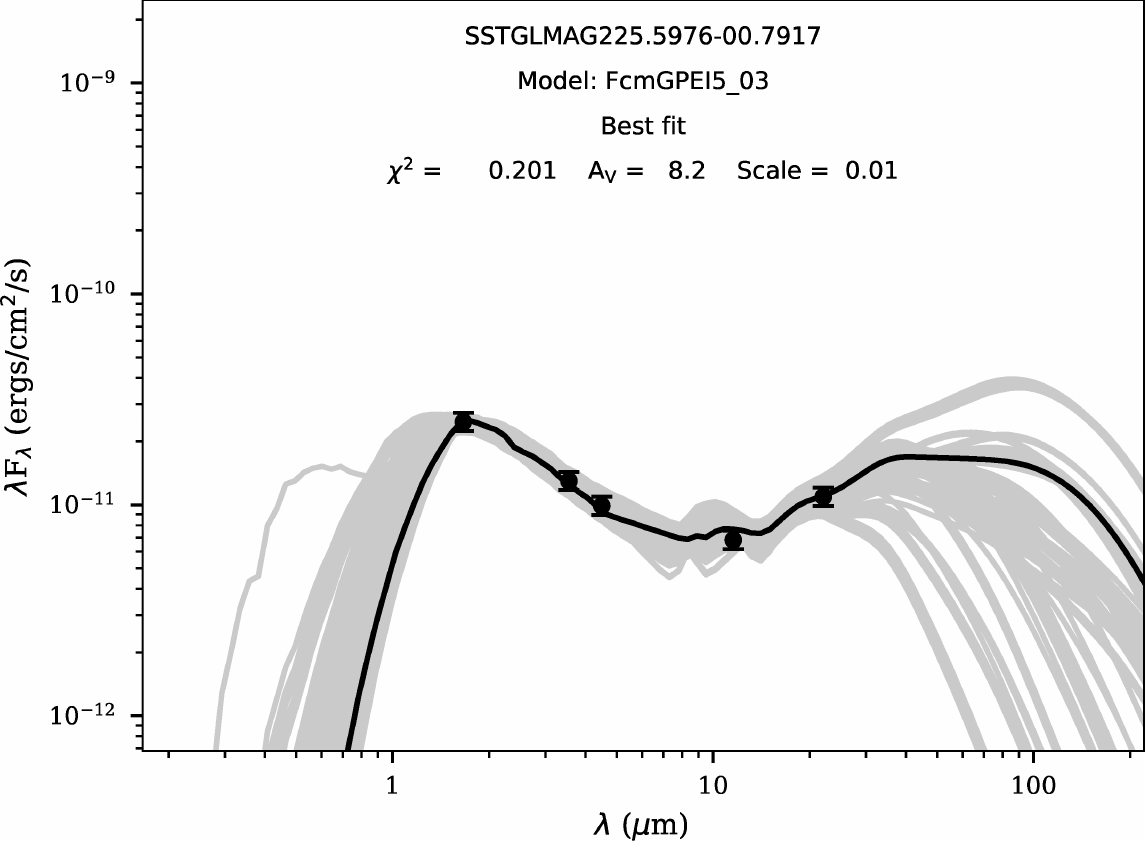}
\hfill
\includegraphics[width=0.32\textwidth]{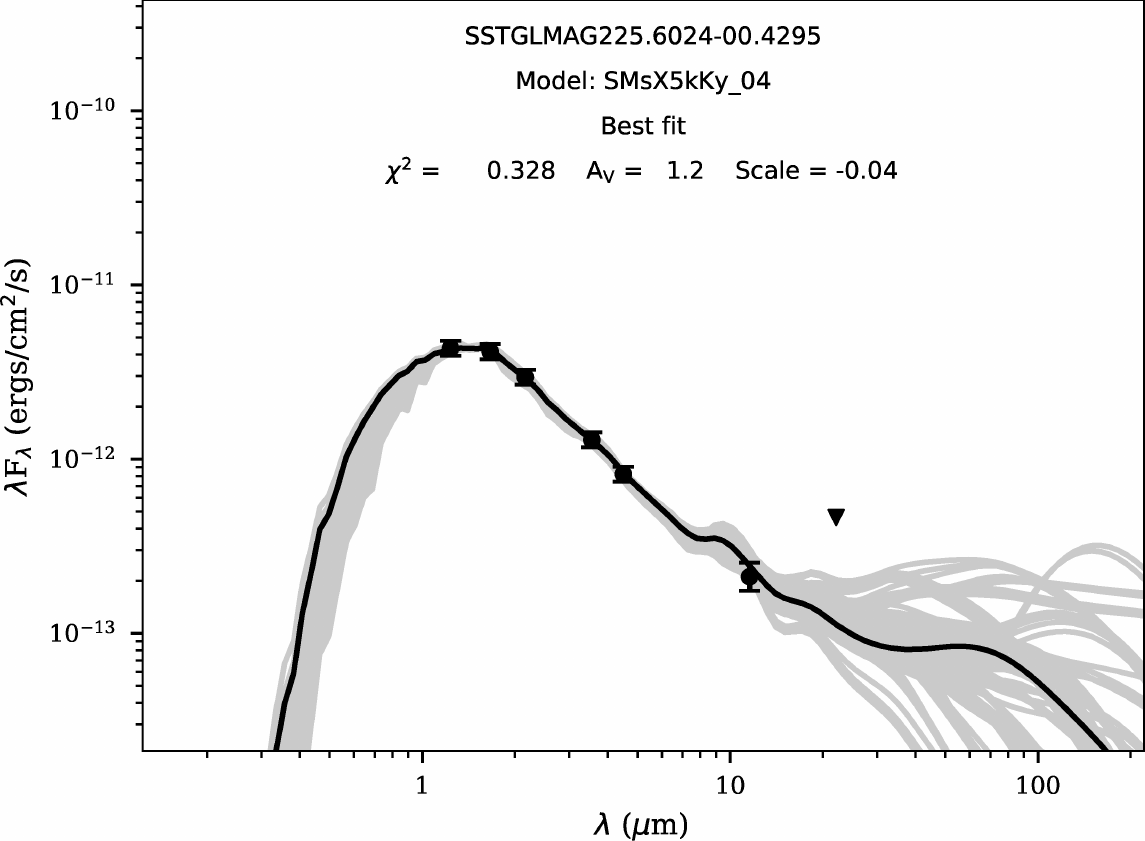} \par
\vspace{2mm}
\includegraphics[width=0.32\textwidth]{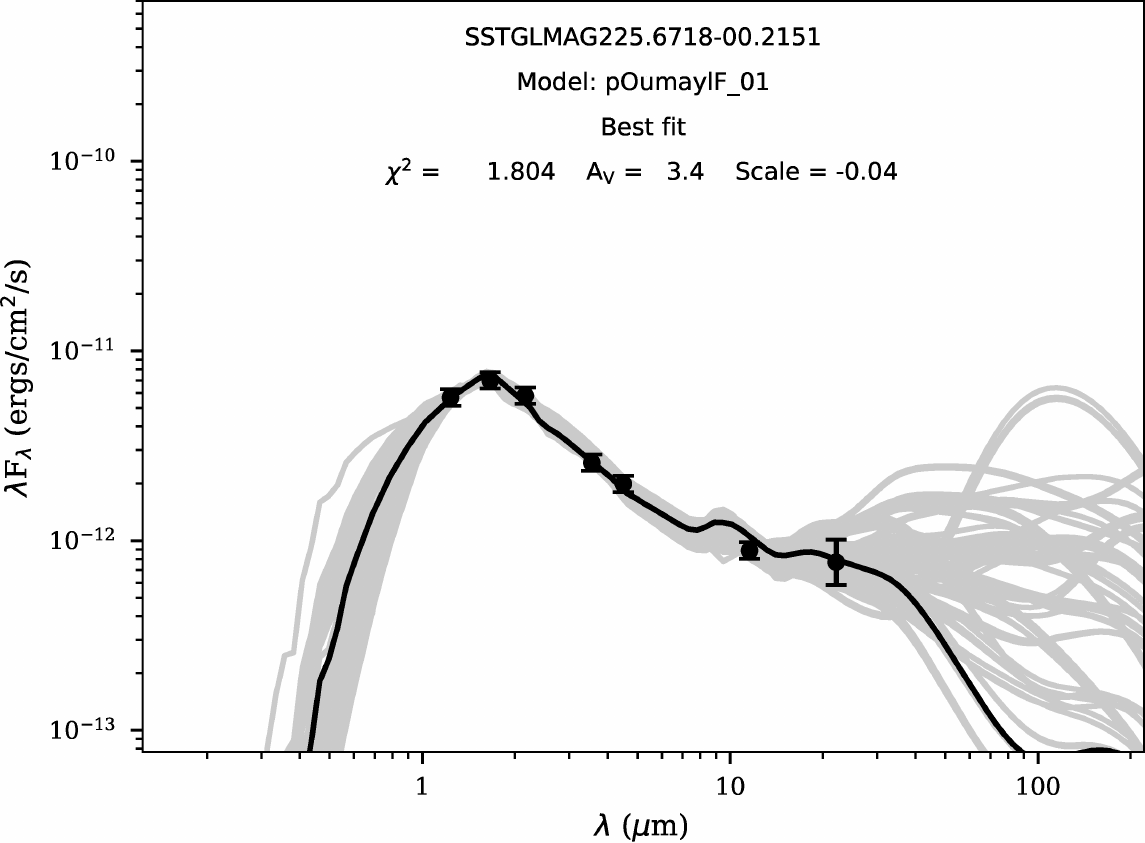}
\hfill
\includegraphics[width=0.32\textwidth]{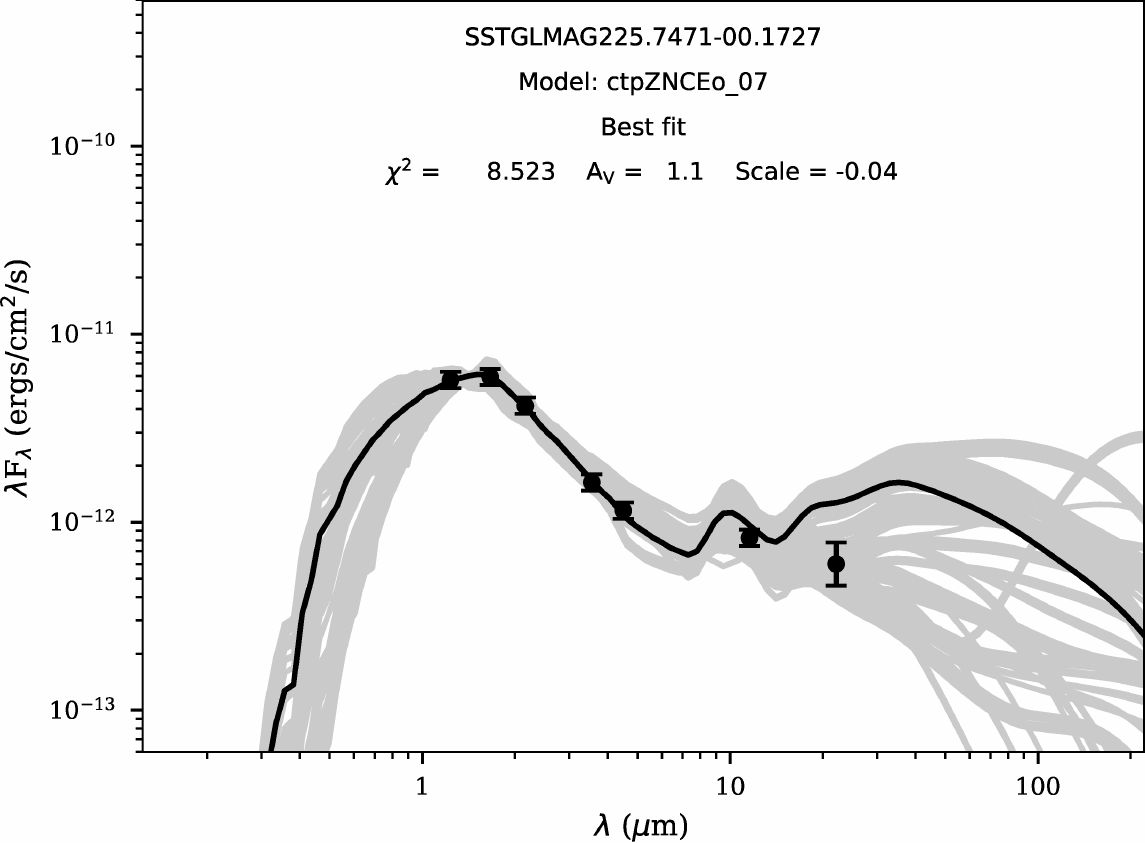}
\hfill
\includegraphics[width=0.32\textwidth]{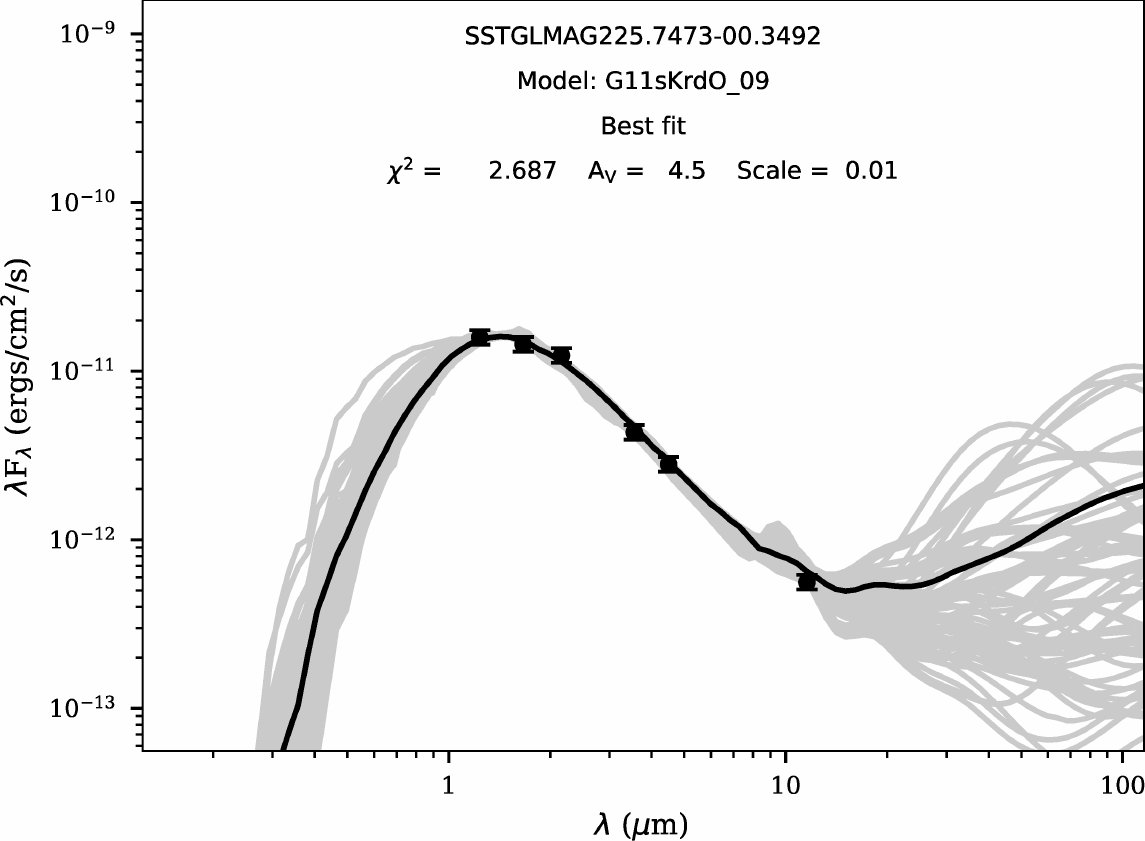}
\caption{Same as Fig.~\ref{f:SEDs1}  \label{f:SEDs14}}
\end{figure*}

\section{A Comparison Between the YSO Candidate Lists}
\label{s:F16comp}

We matched the final list of YSO candidates to the \citet{fischer2016} catalog using the AllWISE designations. Out of the 93 YSO candidates from \citet{fischer2016} located in CMa--$l224$, 20 AllWISE sources are not in our final list of YSO candidates. 

Two out of 20 WISE sources were not detected by {\it Spitzer} (J070724.82$-$094712.0, J070927.40$-$102921.0), and one source was resolved into multiple {\it Spitzer} sources (J070810.47$-$102720.3); one of these sources is in the GLIMPSE360 catalog with a position offset $>$3$''$ from the WISE position. 

For 12 WISE sources selected as YSO candidates in \citet{fischer2016}, the distance between the {\it Spitzer} and AllWISE catalog source was larger than 1$''$ and the AllWISE match was removed 
(J070613.51$-$095837.9, J070801.11$-$102427.5, 
J070818.24$-$102933.9, J070822.75$-$102916.3, 
J070856.28$-$102910.4, J070923.27$-$102747.4,	
J070923.42$-$102818.8, J070924.77$-$102534.0, 
J071012.84$-$103213.0, J071023.74$-$103327.1,	
J071024.84$-$103247.7, J071306.61$-$111803.8).
\begin{packed_enum}
\item[--] After removing the AllWISE photometry, 5 out of 12 {\it Spitzer} sources had only 3.6 and 4.5 $\mu$m photometry left; four of these sources are included in our [3.6][4.5]-only source list of `possible YSO candidates' (J070822.75$-$02916.3,  J070923.42$-$102818.8,  J071306.61$-$111803.8, J070613.51$-$095837.9). One source (J070856.28-102910.4) does not have the GLIMPSE360 Catalog counterpart and was removed from the list. 
\item[--] Four out of 12 sources did not fulfill the 2MASS-{\it Spitzer} YSO selection criteria and where removed from the list (J070801.11-102427.5, J070924.77-102534.0, J070923.27-102747.4, J071012.84-103213.0). 
\item[--] One source was well-fit with the stellar photosphere model and removed (J070818.24-102933.9).
\item[--] Two sources are on our final YSO candidates list (J071024.84-103247.7, J071023.74-103327.1).
\end{packed_enum}

For 5 WISE sources selected as YSO candidates in \citet{fischer2016}, the distance between the {\it Spitzer} and AllWISE catalog source was less than 1$''$ and a combined {\it Spitzer} and AllWISE photometry was used in the analysis.
\begin{packed_enum}
\item[--] Two out of 5 sources were removed as background galaxy candidates (J070857.81-102906.5, J071006.22-103140.2).
\item[--] Three sources are on the final list of YSO candidates, but the AllWISE matches/photometry were removed based on the ext\_flg criterion (J071222.96-111625.1, J070920.64-102817.4, J070925.85-102907.7). 
\end{packed_enum}

In summary, out of 20 \citet{fischer2016} YSO candidates not matched to our final list of YSO candidates using AllWISE designation, five are classified as YSO candidates based on the 2MASS and {\it Spitzer} photometry only, and four are on the [3.6][4.5]-only list of possible YSO candidates. Seven GLIMPSE360 sources that were initially matched to AllWISE sources were removed from the list based on the 2MASS and GLIMPSE360 photometry. One source that became an [3.6][4.5]-only source after removing the AllWISE photometry in the initial selection of YSO candidates was removed due to the lack of the GLIMPSE360 Catalog counterpart. Three YSO candidates from \citet{fischer2016} do not have a clear {\it Spitzer} counterpart.

\end{document}